\documentclass[a4paper,12pt,spanish,twoside,openright]{report}
\usepackage[utf8]{inputenc}
\usepackage[a4paper,right=30mm,left=40mm,top=45mm,bottom=30mm]{geometry}

\usepackage[T1]{fontenc}
\usepackage[spanish]{babel}
\usepackage{graphicx}
\graphicspath{ {images/} }
\usepackage{ulem}
\usepackage{caption}
\usepackage{subcaption}
\usepackage[hidelinks]{hyperref}

\usepackage[usenames]{color}
\usepackage{xcolor}
\usepackage{xurl}

\usepackage{multirow}

\usepackage{listings}
\definecolor{codegreen}{rgb}{0,0.6,0}
\definecolor{codegray}{rgb}{0.5,0.5,0.5}
\definecolor{codepurple}{rgb}{0.58,0,0.82}
\definecolor{backcolour}{rgb}{0.95,0.95,0.92}
\lstdefinestyle{mystyle}{
    backgroundcolor=\color{backcolour},   
    commentstyle=\color{codegreen},
    keywordstyle=\color{magenta},
    numberstyle=\tiny\color{codegray},
    stringstyle=\color{codepurple},
    basicstyle=\ttfamily\footnotesize,
    breakatwhitespace=false,         
    breaklines=true,                 
    captionpos=b,                    
    keepspaces=true,                 
    numbers=left,                    
    numbersep=5pt,                  
    showspaces=false,                
    showstringspaces=false,
    showtabs=false,                  
    tabsize=2
}
\usepackage{natbib}
\usepackage{apalike}
\usepackage{multirow}
\usepackage{array} 

\usepackage{changepage}   

\usepackage{pdfpages}

\begin{document}

\begin{titlepage}
   \begin{center}
        \includegraphics[width=0.3\textwidth]{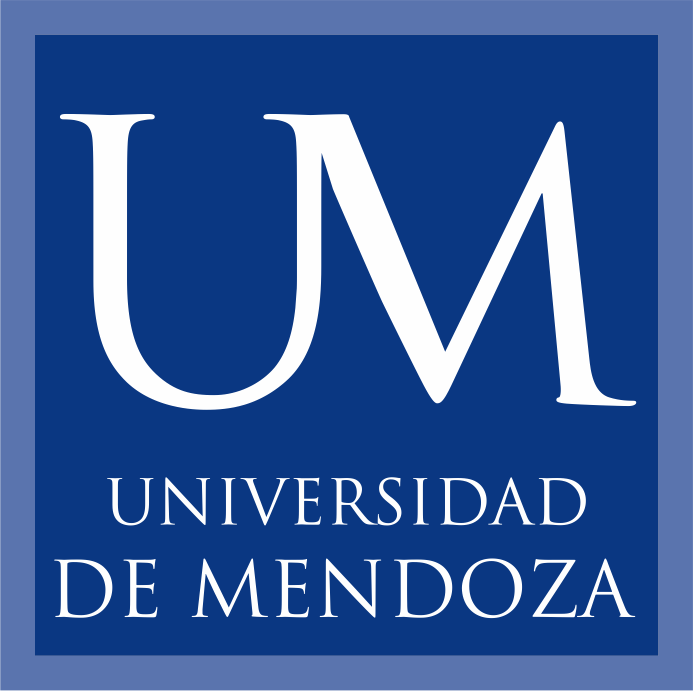}
        
        \vspace*{1.5cm}
        
        \textbf{\Large UNIVERSIDAD DE MENDOZA}
        
        \vspace*{0.5cm}
        
        \textbf{\Large FACULTAD DE INGENIERÍA}

        \vspace{2cm}
       
        \textbf{\LARGE TESIS DOCTORAL}

        \vspace{2.5cm}
        
        \textbf{\Huge Investigación sobre acceso, uso y exploración efectiva de datos observacionales y bibliográficos astronómicos a partir de la sonorización}
        
        \vfill

        \textbf{\large Autora: Bioing. Johanna Casado}
        
        \vspace*{0.5cm}
        
        \textbf{\large Directora: Dra. Beatriz García}
        
        \vspace{0.5cm}

    \end{center}
            
    \large Mendoza, 2023
            
    \vspace{0.8cm}
     
\end{titlepage}

\chapter*{Dedicatoria y agradecimientos}
Durante el transcurso de escritura de esta tesis me prometí a mi misma que este capítulo no sería un mero formalismo, sino que deseo utilizarlo para expresar sinceramente lo que siente mi corazón. De más está decir que estoy segura que me quedarán nombres de personas e instituciones olvidados, espero incluir a todas las personas con las que he interactuado en estos largos seis años, todos han sido una pieza fundamental para que esta disertación haya llegado a esta instancia.

Mi cerebro a discutido sobre el primer lugar de esta dedicatoria y claramente les expreso que no he llegado a un consenso, por lo que el primer lugar está compartido entre mi directora y mi familia (en primer lugar pareja e hijos, y en segundo lugar mi suegra y mis padres).

A la Dra. Beatriz García, quien claramente hizo posible que hoy esté presentando esta tesis. Fue quién planteó la necesidad en un inicio, la cual llegó a mi en el momento que estaba terminando mi trabajo final de grado. Con solo leer el título me tiré a la pileta. Sinceramente, Beatriz tuvo que tolerar mis incertidumbres y locuras, creo que se debe acordar de las veces que renuncié a la beca incluso antes de saber que CONICET la iba a aprobar. Recuerdo ese día que recibí el mail de aprobación y el mensaje de Beatriz diciendo que debíamos conversar. Les cuento que durante esa charla le dije que aceptaba la beca con la condición de que no iba a subirme a un avión (obvio que después viaje en avión). Para no hacerla más larga, en resumen quiero agradecerle su paciencia y perseverancia, porque en estos seis años me ha dedicado un tiempo invaluable, ha seguido de cerca todo mi desarrollo y me ha empujado a mejorar día a día como profesional.

A mi pareja, hija e hijo, las palabras no alcanzan, creo que todos saben lo que es tolerar la montaña rusa emocional de un integrante de la familia (y sinceramente creo que la definición montaña rusa se queda corta al describir las emociones que acompañan una tesis doctoral). Recuerdo que antes de iniciar la beca doctoral le dije a mi pareja `Bueno, ahora tenemos que decidir, porque no voy a tener hijos si empiezo ahora el doctorado'. Me río con solo recordarlo, acá estamos yo terminando el doctorado y nosotros con una hija (Isabella) e hijo (Santiago). En resumen, gracias por todos estos años de apoyo incondicional, por tolerar todos esos momentos donde llevé las frustraciones de trabajo a casa, gracias por estar siempre.

A mi suegra, pido disculpas a mi mamá, pero es que viviendo a 250 km de mi mamá y papá, sinceramente sin este ángel que me dio la vida no me encontraría en el lugar que estoy hoy. Mi suegra hizo posible que pueda concentrarme en mi trabajo mientras ella ha cuidado a mis hijos, porque ver el amor que ella le da a mis hijos cada día hizo posible que dedique horas de calidad a mi trabajo. Sinceramente, gracias. A mi mamá y papá, porque más allá de estar a 250 km de distancia son los que han hecho posible que yo hoy sea la persona que soy, se que sacrificaron mucho y que siempre dieron lo máximo para que yo hoy sea la persona que soy, los quiero inmensamente. A mi hermana, por su eterno apoyo moral, escucha y presencia. Solo decirte Romi que siempre has estado en el momento que más necesitaba, gracias. A mi hermano, que si bien cada uno camina su camino se que siempre está ahí si levanto el teléfono. También al resto de mi familia de sangre y política, ¡gracias!.

Formalmente, en este punto quiero agradecer a CONICET por la Beca Doctoral, haciendo posible que me dedique exclusivamente al desarrollo de mi tesis doctoral. Quiero resaltar en este punto el agradecimiento por las facilidades que existen hoy en día para las maternidades, hace poco me enteré que son recientes y quería expresar que son sumamente valiosas. A ITeDA Mendoza por ser el instituto en donde he llevado adelante mi investigación y me ha brindado todo el apoyo necesario durante estos años.

A la Universidad de Mendoza, por ser algo así como una casa para mí, es aquí donde he crecido académicamente y donde también hoy tengo el agrado de trabajar. Particularmente, agradecer al Bioing. Gastón Jaren quien al momento de iniciar esta tesis era el director del Instituto de Bioingeniería, y quien me envió el mail con la propuesta de beca hecha por la Dra. Beatriz García, gracias por participar de intermediario al inicio de esta tesis y por creer en mi; al Dr. Ricardo Cabrera que fue la primer persona perteneciente a CONICET que conocí y me abrió las puertas del instituto INBIOMED cuando aún era estudiante; a la Dra. Cristina Parraga, que me lee desde el cielo, más allá de las diferencias que pudimos tener admiro su carrera y su calidad como persona; al Dr. Alfredo Iglesias por la oportunidad de formar parte de la Facultad de Ingeniería y todo el apoyo para hacer crecer el Instituto de Bioingeniería; a la Dra. Vanesa Bazzocchini, la persona con la que inicié en el mundo de la investigación, gracias de corazón por todo lo que me enseñaste; al Dr. Lucas Iacono, quien me brindó un espacio para trabajar en el proyecto del robot seguidor de línea y me aconsejó al iniciar mi camino de doctorando. Adicionalmente, quiero agradecer a cada una de las personas que formaron y forman parte de mi día a día desde hace tantos años en la Universidad de Mendoza (Grupo de Robótica, INBIOMED, DICyT de la mano de la Dra. Parraga, Instituto de Bioingeniería), soy feliz trabajando y estudiando en este lugar con todos ustedes (incluye a todos los que no he llegado a nombrar). También por si aún no queda explícito, a todos los estudiantes con los que he trabajado o colaborado en sus trabajos finales, y con los que aún estamos trabajando.

En cuanto a las instituciones con las que he tenido el placer de cruzarme en estos años, agradezco a la Oficina de Astronomía para el Desarrollo (OAD) localizada en Sudáfrica que fue el primer lugar que me abrió sus puertas en el extranjero haciendo posible que trabajara con la Dra. Wanda Díaz Merced durante un mes. Allí conocí al Dr. Kevin Govender (Director), la Dra. Vanessa McBride (astrónoma) y Nuhaah Solomon, quienes estuvieron atentos en todo momento e hicieron de la estadía una hermosa experiencia. Particularmente, agradecer a la Dra. Wanda Díaz Merced quien me dedicó un tiempo invaluable durante los primeros tres años de trabajo. Gracias por ser mi guía en el desarrollo de interfaces accesibles y centradas en el usuario, por enseñarme a ver mas allá de la discapacidad, desde la inclusión (fue quien me enseñó el término diversidad funcional) y por guiarme en el análisis cualitativo de grupos focales.

Otra institución que me abrió sus puertas fue la Universidad de Southampton, donde estuve trabajando tres semanas y se realizaron las pruebas con usuarios del software desarrollado. Particularmente, agradecerle al Dr. Poshak Gandhi quien gestionó la visita, donde además se otorgó una computadora Apple como herramienta para el desarrollo, y con quien se ha mantenido contacto luego para la publicación de un paper. A Southampton Sight quienes recibieron nuestra propuesta de probar una herramienta nueva y accedieron amablemente, brindando su tiempo y experiencia al testear la herramienta. Adicionalmente, agradecerle al proyecto REINFORCE, sus integrantes y a la Unión Europea (agente financiador), ya que en relación con este proyecto se trabajaron los últimos tres años de esta tesis y se tuvo la posibilidad de sonorizar nuevos tipos de datos.

A las Bioing. Aldana Palma y Julieta Carricondo, quienes me acompañaron en los primeros pasos de esta tesis doctoral. Hermosas tardes trabajando en sonorización cuando incluso no tenía para nada claro lo que eso significaba, espero que se acuerden siempre de nuestro totem Mingus y de la biocueva. A Lucía, Paloma y Sofía, quienes siendo estudiantes de la carrera de Bioingeniería de la Universidad de Mendoza contribuyeron con el análisis de normativas, ¡muchas gracias!. A Alexis Mancilla, Javier Malla y el Dr. Angel Cancio por ayudarme a integrarme esos primeros años de la beca doctoral, siempre es difícil llegar a un lugar nuevo pero ellos me hicieron sentir pertenencia. Al Dr. Agustín Cobos, a quien conocí años después de empezar a trabajar pero que siempre está atento, de verdad valoro tu `ya no llegas', siempre me hace reír. A Gonzalo De La Vega con quien hemos compartido mucho desde que iniciamos el desarrollo de la web de sonoUno. Al Dr. Pablo Teruya por su constante apoyo y guía, sobre todo en la etapa final que tanto me ha costado. Al Dr. Diego Tramontina, en parte responsable de que haya ingresado a CONICET, quien apareció misteriosamente cuando no sabía por donde empezar a escribir para darme su consejo, los misterios de la vida, porque claramente lo escuche y acá estamos.

A mis compañeros de cursado de las materias obligatorias y amigas/amigos: Adriana Neri, Emiliano Aparicio, Noelia Robles, Orlando Deluigi, Patricio Gonzalez y Luis Arce. Gracias por todo ese apoyo compartido para transitar ese camino difícil que es el cursado. Particularmente, a Emiliano Aparicio, con quien más allá del cursado nuestra relación de amistad nos llevó a un apoyo mutuo y a salidas al cine siguiendo las películas de Marvel y DC Comics.

Quiero también agradecer a las personas con las que he compartido grandes momentos de aprendizaje en sonorización, como son el Dr. Jeffrey Cooke y Jeff Hannan (desarrolladores del software StarSound), con quienes se ha mantenido contacto durante todo el desarrollo y han probado gentilmente la herramienta sonoUno. Al Dr. Paul Green quien ayudó en muchas mejoras de sonoUno y utilizó la herramienta dentro de un desarrollo propio (Sensing the Dynamic Universe). Al Prof. Carlos Morales Socorro quien amablemente utilizó la versión web de sonoUno en las aulas del IES José Frugoni Pérez, La Rocha, España, como soporte a la enseñanza. Al Dr. Theodoros Avgitas quien gentilmente ha trabajado con el equipo de sonoUno para realizar la sonorización de datos de muongrafía, y actualmente propuso incorporar a sonoUno en una instalación que llevarán a cabo en el museo `Jules Verne' de Nantes, Francia. A Gary Hemming, Dr. Massimiliano Razzano y Dr. Francesco Di Renzo por todo su apoyo y proveer los datos para la sonorización de Glitches, provenientes de la detección de ondas gravitacionales. A la Dra. Christine Kourkoumelis, MSc. Stelios Vourakis y Dr. Stelios Angelidakis por toda su ayuda y proveer archivos de datos para el desarrollo de la sonorización de partículas del Gran Colisionador de Hadrones.

Al Dr. Pierre Chanial, quien se incorporó estos últimos años al desarrollo de sonoUno con una propuesta de unificación de librerías y recursos mediante un servidor. A Natasha Bertaina y Belén Olivera, quienes también han contribuido al proyecto sonoUno desde sus trabajos finales de grado.

A todas aquellas personas que han probado la herramienta a lo largo de estos años y nos han brindado sus elogios y críticas constructivas. También a las personas que se tomaron el tiempo de probar la herramienta en nuevos entornos como el caso donde instalaron sonoUno en el sistema operativo `Centos' y en `Raspberry Pi'. A aquellas personas que han llevado a sonoUno a las aulas, como el caso expuesto en la sección \ref{sect:uso_canarias}. A quienes se han ofrecido a contribuir desde su experiencia, recientemente recibimos la oferta para traducir el programa y los manuales a otros idiomas. A todos ellos, y también a aquellos que me haya olvidado mencionar, quiero agradecerles por contribuir a que sonoUno y su concepto crezcan. Particularmente, también quiero agradecer a todos los que lean este trabajo, amé escribir cada parte de él.

\chapter*{Acknowledgements}

During the course of writing this thesis, I promised myself that this chapter would not be a mere formality, I want to use it to sincerely express what my heart feels. Needless to say that I am sure I will misplace some people and institutions' names, I hope to be able to include all the people with whom I have interacted in these long six years, they have all been a fundamental piece for this final stage of my thesis.

The first place in this text produced a discussion in my mind, and I couldn't reach a consensus, so the first place is shared by my PhD director and my family (my daughter, son, and husband; then my mother-in-law and my parents).

To Dr. Beatriz García, who clearly contribute to making this thesis possible. She brings the need to me at the beginning when I was finishing my bioengineering career. When I read the title I knew that I want to investigate this topic. Sincerely, Beatriz tolerates all my uncertainties and crisis, even when I quit before the CONICET grant resolution. I remember that after CONICET approve my grant, I told to Beatriz that I will continue with the investigation but I wasn't going to travel by plane (of course I traveled by plane during my PhD). In summary, I want to thank her patience and persistence, because during these six years, she dedicates invaluable time to following my work and she pushes me to be a better professional.

To my husband, daughter, and son, the words are not enough, I think all of you know what is an emotional hurricane at home (and I think a hurricane is not enough to describe the emotions during a PhD). I remember that at the beginning I told him `Now we have to decide because I don't want to have kids during my PhD'. I laugh remembering it, here we are, I am finishing my PhD and we have a daughter (Isabella) and a son (Santiago). In summary, thanks for all these years of unconditional support, for tolerating all that moments of frustration, and thank for always being there.

To my mother-in-law, apologies to my mom, but living 250 km from my parents, without this angel that life give to me, I wouldn't be able to reach this moment. My mother-in-law makes it possible for me to concentrate on my work at the time she looks after my kids because seeing the love she gives to them each day makes it possible for me to dedicate valuable hours to my work, thanks from all of my heart. To my mom and dad, because they make it possible for me to follow my dreams and became the person I am, I know they sacrifice a lot and always gave me all they could and more, I love you so much. To my sister, for all her unconditional support and presence. To my brother, to always be there at a phone call distance. Also to the rest of my family and in-laws, thanks.

Formally, at this point I want to acknowledge CONICET for the grant, making possible a complete dedication to my PhD. I want to point out the considerations that the organization gave to motherhood, I know that this facility is recent and I want to express their importance. To ITeDA Mendoza for being the Institute where I carry out my investigation and for all their support.

To the University of Mendoza, for being something like a home to me, here I grew up academically and it is where I work today as a professor. Particularly, thanks to BioEn. Gastón Jarén, who was the director of the Bioengineer Institute at the beginning of my PhD and who sent me the email with the proposal, thanks for believing in me; Dr. Ricardo Cabrera, the first person related to CONICET that I met, he opens the doors to the INBIOMED institute when I was an engineering student; to Dr. Cristina Parraga, who reads me from heaven, I admire her career and her kindness; to Dr. Alfredo Iglesias for the opportunity to work on the University and all his support to the Bioengineering institute; to Dr. Vanesa Bazzocchini, with her, I initiate my research path, thanks from my heart for all your training; to Dr. Lucas Iacono, who gave me the place to work with a line follower robot and advice me at the beginning of my PhD. Additionally, I want to thank each person that was and is part of each work day since I began studying at the University of Mendoza (Robotic group, INBIOMED, DICyT at hand of Dr. Parraga, Bioengineer Institute), I'm really happy working and studying in this place with all of you (including all the names I don't write). Also, just in case, I include in this acknowledgment all students with whom I have worked or collaborated, even these days.

Regarding the institutions with which I have had the pleasure of coming across in these years, I am grateful to the Office of Astronomy for Development (OAD) located in South Africa, which was the first place that opened its doors to me abroad, making it possible for me to work with Dr. Wanda Díaz Merced for a month. There I met Dr. Kevin Govender (Director), Dr. Vanessa McBride (Astronomer), and Nuhaah Solomon, who were attentive at all times and made the stay a beautiful experience. In particular, I would like to thank Dr. Wanda Díaz Merced, who dedicated invaluable time to me during the first three years of work. Thank you for being my guide in the development of accessible end user centred interfaces, for teaching me to see beyond disability, from inclusion (he was the one who taught me the term functional diversity), and for guiding me in the qualitative analysis of focus groups.

Another institution that opened its doors to me was the University of Southampton, where I worked for three weeks and tested the software developed with users. Particularly, thanks to Dr. Poshak Gandhi who managed the visit, where an Apple computer was also given as a tool for development, and with whom contact has been maintained later for the publication of a paper. To Southampton Sight who received our proposal to test a new tool and kindly agreed, giving their time and experience testing the tool. Additionally, thank the REINFORCE project, its members, and the European Union (financing agent), since in relation to this project the last three years of this thesis were worked on and there was the possibility of sonifying new types of data.

To BioEn. Aldana Palma and Julieta Carricondo, who accompanied me in the first steps of this doctoral thesis. Beautiful afternoons working on sonification when I wasn't even sure what that meant, I hope you always remember our Mingus totem and the `Biocueva'. To Lucía, Paloma, and Sofía, who during their Bioengineering studies at the University of Mendoza, contributed to the analysis of the ISO standard. To Alexis Mancilla, Javier Malla, and Dr. Angel Cancio for helping me to integrate those first years of the doctoral fellowship, it is always difficult to reach a new place but they made me feel like I belong. To Dr. Agustín Cobos, whom I met some years after starting the work but who is always attentive, I really appreciate your `you no longer arrive', he always makes me laugh. To Gonzalo De La Vega with whom we have shared a lot since we started developing the sonoUno website. To Dr. Pablo Teruya for his constant support and guidance, especially in the final stage which has cost me so much. To Dr. Diego Tramontina, partly responsible for me joining CONICET, who mysteriously appeared when I didn't know where to start writing to give me his advice, the mysteries of life because I clearly heard him and here we are.

To my classmates and friends: Adriana Neri, Emiliano Aparicio, Noelia Robles, Orlando Deluigi, Patricio Gonzalez, and Luis Arce. Thank you for all that shared support to travel that difficult path that are the courses. Particularly, to Emiliano Aparicio, with whom beyond the course our friendship relationship led us to mutual support and going to the cinema following Marvel and DC Comics.

I also want to thank the people with whom I have shared great learning moments in sonification, such as Dr. Jeffrey Cooke and Jeff Hannan (StarSound software developers), with whom I have maintained contact throughout the development and have kindly tested the sonoUno tool. To Dr. Paul Green who helped in many improvements of sonoUno and used the tool within his own development (Sensing the Dynamic Universe). To Prof. Carlos Morales Socorro who kindly used the web version of sonoUno in the classrooms of the IES José Frugoni Pérez, La Rocha, Spain, as a teaching support. To Dr. Theodoros Avgitas who has kindly worked with the sonoUno team to perform the sonorization of muongraphy data, and currently proposed to incorporate sonoUno in an installation that they will carry out at the `Jules Verne' museum in Nantes, France. To Gary Hemming, Dr. Massimiliano Razzano, and Dr. Francesco Di Renzo for all their support and for providing the data for the Glitches sonification, which ones come from the detection of gravitational waves. To Dr. Christine Kourkoumelis, MSc. Stelios Vourakis and Dr. Stelios Angelidakis for all their help and for providing data files for the development of the Large Hadron Collider particle sonification.

To Dr. Pierre Chanial, who has joined the development of sonoUno in recent years with a proposal to unify libraries and resources through a server. To Natasha Bertaina and Belén Olivera, who have also contributed to the sonoUno project since their final degree projects.

To all those people who have tried the tool over the years and have given us their praise and constructive criticism. Also to the people who took the time to test the tool in new environments such as the case where they installed sonoUno on the `Centos' operating system and on `Raspberry Pi'. To those people who have taken sonoUno to the classroom, as the case described in the section \ref{sect:uso_canarias}. To those who have offered to contribute from their experience, we recently received the offer to translate the program and manuals into other languages. To all of them, and also to those I forgot to mention, I want to thank you for helping sonoUno and its concept grow. In particular, I also want to thank everyone who reads this work, I loved writing every part of it.

\chapter*{Resumen}

El análisis de datos en las ciencias del espacio se realiza desde hace años exclusivamente de forma visual, pese a que la mayor cantidad de datos pertenece a porciones no visibles del espectro electromagnético. Esto, por un lado limita el estudio de lo desconocido a las posibilidades de resolución actual de las pantallas, y por otro, excluye a un grupo de personas que presentan algún tipo de discapacidad visual. Teniendo en cuenta lo mencionado, y que las personas con algún tipo de discapacidad se encuentran con muchas barreras para lograr estudios académicos y puestos de trabajo estables, se enfoca la presente investigación en nuevas modalidades de acceso a los datos, pero teniendo en cuenta la accesibilidad e inclusión de personas con diversidad funcional desde el principio.

Se ha evidenciado que la percepción multimodal (uso de más de un sentido) puede ser un buen complemento para la exploración visual y el entendimiento de los datos científicos complejos. Esto es especialmente cierto para datos astrofísicos, compuestos de una suma de diferentes modos oscilatorios resultando en el complejo arreglo de datos final. Esta propuesta se centra en la habilidad humana de adaptarse a los datos y a la interacción con el sonido, con el fin de analizar conjuntos de datos y producir una aplicación orientada a nivelar las posibilidades de acceso a la información en el campo de desempeño de la física y la astronomía (aunque la herramienta es aplicable también a cualquier tipo de datos en archivos de 2 o más columnas (.txt o .csv)) para personas con discapacidad. Además, propone el estudio de capacidades científicas y tecnológicas para la generación de herramientas con un enfoque novedoso, centrado en el usuario y orientadas a: un problema social específico, el uso de lenguajes de programación en software libre y el diseño de infraestructura para mejorar la inclusión.

\tableofcontents

\listoffigures

\listoftables

\chapter*{Introducción}
\addcontentsline{toc}{chapter}{Introducción}

Al iniciar la presente investigación y comenzar a utilizar la palabra `accesible', la cual deriva de la palabra accesibilidad, siempre se la relacionó con la habilidad de producir algo que pudiera ser utilizado por personas con diversidad funcional sin que significara mayor dificultad para una u otra. Sin embargo, rápidamente, al tener relación con personas con discapacidad y comenzar a profundizar en el tema, esa primera definición de accesibilidad fue cambiando.

Analizando el término `discapacidad' y `accesibilidad', durante los últimos 50 años la investigación sobre accesibilidad dejó de centrarse en la discapacidad como un problema médico o deficiencia que debía repararse, dando un giro hacia un constructo más social \citep{macketal2021}. Así, se comenzó a diferenciar entre la discapacidad como una condición física o biológica y la discapacidad como construcción social y ambiental. En el caso de esta tesis, el texto estará centrado en esta última definición de discapacidad, teniendo en cuenta todas las capacidades que tiene una persona y buscando reducir al mínimo las limitaciones que podría llegar a experimentar, principalmente en el caso de acceso a una interfaz humano-computadora. En este sentido es interesante pensar, además, en las dificultades a las que también se enfrentan las personas que se define (o autodefinen) como no discapacitadas, para abordar los desafíos que se detallan en el presente trabajo. Otro término que se utilizará ampliamente es `inclusión' que hace referencia a la necesidad de que el sistema, herramienta o entorno que se está desarrollando se ajuste a las necesidades o capacidades de todas las personas. Se debe tener en cuenta que este trabajo no busca investigar nuevas modalidades de acceso a los datos para personas con discapacidad, sino que busca investigar sobre estas nuevas modalidades promoviendo la inclusión en los espacios científicos y buscando mejorar el despliegue actual de los datos en astronomía y astrofísica, para asegurar el acceso universal.

En cuanto a la realidad laboral de las personas con discapacidad en Argentina, existen diferentes normativas que buscan reglamentar y promover la inclusión como ser: la ``Ley 22.431 - Sistema de protección integral de los discapacitados'', la Ley que reglamenta el certificado único de discapacidad, la Ley que reglamenta el trasporte público adaptado e incluso la Ley 26.378 que aprueba la ``Convención sobre los Derechos de las Personas con Discapacidad y su protocolo facultativo'' elaborado y publicado por las Naciones Unidas. Sin embargo, datos del Instituto Nacional de Estadísticas y Censos de la República Argentina \citep{indec2018} muestran una realidad preocupante: solo un 32\% de las personas mayores a 14 años que sufren una o más discapacidades tienen un empleo. Otro inconveniente con el que se encuentran las personas con discapacidad es el acceso a la educación, un 20\% de las personas con discapacidad de 15 años o más han dejado sus estudios con nivel primario incompleto. De los que lograron terminar la escuela primaria un 46\% no logró finalizar la secundaria, un 20\% logró terminarla y solo un 13\% accedió a algún tipo de educación superior. Se considera que tanto para el ámbito académico como laboral, uno de los problemas fundamentales es la falta de herramientas y modalidades de presentación de la información que promuevan y generen inclusión.

Una realidad que llamó la atención, es que las prácticas de intercambio de información en todos los aspectos de las ciencias STEM, incluida la preparación académica y especialmente en matemáticas en la educación superior y las ciencias en las aulas escolares, pueden impedir que la mayoría de las personas participen, por ejemplo en el campo de la astronomía. Durante los últimos 100 años, la humanidad ha sido testigo de un cambio de mentalidad general hacia la naturalización del derecho humano inherente a la participación equitativa e igualitaria en las líneas actuales de investigación científica, como así también en educación y difusión de las ciencias.

Una extensa y profunda búsqueda bibliográfica sugirió que la falta de experimentos de percepción (aplicados al análisis de datos astronómicos), la brecha de desarrollo causada por los enfoques del conocimiento en cada momento de la historia de la ciencia y de la humanidad, el desempeño y los estilos de producción en el mercado laboral a medida que las máquinas se integraron en la investigación, acompañado por el lento desarrollo de las tecnologías de audio (en comparación con las tecnologías visuales) y su aplicación a los factores humanos, contribuyeron a una disminución de las formas asociadas con otros tipos de experiencia sensorial y al predominio de formas uniformes de transmitir y analizar la información científica en general y astronómica en especial. En ese sentido, \citet{wandatesis2013} y \citet{mercedetal2011} investigaron si las modalidades visuales eran mejores que otras modalidades sensoriales para la exploración de datos astrofísicos, a través de experimentos de percepción y encontraron que el uso del sonido aumentaba la sensibilidad del astrónomo experto frente a eventos en los datos que por naturaleza son de muy baja relación señal/ruido, o bien ambiguos para el ojo humano. 

A través de la historia, no hay evidencia experimental (para información astronómica) que respalde por qué la astronomía profesional dejó de usar otras modalidades sensoriales para el despliegue de datos bajo análisis, máximo si se tiene en cuenta que la mayor parte del espectro electromagnético (su principal objeto de estudio) se encuentra fuera del rango visible de longitudes de onda. Esto lleva a la pregunta de por qué la física y la astronomía han seguido utilizando las mismas transacciones e interacciones uniformes, digitales, cognitivas, expresivas y reflexivas. Esto puede deberse a la falta de evidencia experimental (relacionada con las ventajas de uso de modalidades no visuales), a la curva de aprendizaje, al lento progreso de la resolución de audio o los métodos de aprendizaje uniformes (educación superior), entre otros. Sin embargo, ciertamente está claro ahora que la mayoría de las personas videntes usan sus ojos para dar sentido a su mundo, y esta es la modalidad sensorial humana predominante, que se utiliza para dar sentido consciente al conocimiento astronómico.

La presente disertación expone la investigación realizada sobre el uso del despliegue multimodal (a través de sonido como anexo a la visualización) de datos para evaluar la accesibilidad y poder diseñar, programar y testear un software de exploración multimodal de datos de las ciencias del espacio. El trabajo que se presenta, se basa en la hipótesis de que ``mediante la presentación multimodal de datos astronómicos y astrofísicos se podría mejorar las técnicas actuales de despliego y reconocimiento de rasgos especiales en dichos datos, promoviendo y facilitando la inclusión en el campo de la investigación''. Mediante esta investigación se estaría dando respuesta a los científicos que expresan la necesidad de nuevas formas de exploración de datos que permitan análisis sin una manipulación previa \citep{wandatesis2013}, también a la necesidad de generar herramientas y prácticas que permitan la inclusión de personas con discapacidad en los ámbitos académicos y laborales y, en parte, a los planteos relacionados con que no existe evidencia de las mejoras que significaría una aproximación multisensorial al análisis de datos científicos.

La realidad actual de las ciencias espaciales nos muestra que la nueva generación de telescopios en tierra y en el espacio, y detectores especiales como centelladores o detectores Cherencov, utilizan técnicas experimentales avanzadas para la captura de datos astrofísicos (energía electromagnética y partículas), con precisión y resolución temporal mejoradas. Sin embargo, aún se encuentran con limitaciones en el análisis de datos, algunas de ellas son:
\begin{enumerate}
    \item \textit{La ausencia de técnicas eficientes con bases adaptativas para detallar y analizar datos no lineales y no estacionarios}: la definición de la base es dependiente de los datos, suponiendo una base definida a posteriori, esto implica un enfoque muy diferente al análisis de datos matemáticos regulares \citep{bendat1993}. Usando la sonorización y la percepción humana, la base podría ser definida a posteriori.
    \item \textit{La limitación en la utilización de la presentación visual de los datos}: éstos generalmente contienen mucha más información que la que puede ser efectivamente desplegada usando las tecnologías disponibles actualmente.  
    \item \textit{La habilidad humana para interpretar los datos del espacio extendido, limitada por la visualización y la perceptualización de los datos}: actualmente se trabajan en esta limitación filtrando (priorizando) los datos para desplegar sólo la información que se cree es importante para el problema que se estudia, esto lleva a hacer algunas suposiciones sobre el resultado, se estima que algunos descubrimientos pueden haberse perdido.
    \item \textit{Los métodos estándar de análisis de datos también hacen a la astronomía inaccesible} para personas física o sensorialmente discapacitadas (ciegos, orientados de forma auditiva, o quienes sufren desórdenes cognitivos). Esta investigación persigue reducir esta brecha.
\end{enumerate}

En búsqueda de mejorar las limitaciones expuestas anteriormente y salvar la falta de inclusión encontrada, se proponen los siguientes objetivos generales para este trabajo:
\begin{itemize}
    \item Investigar sobre nuevas modalidades de acceso a los datos astronómicos y astrofísicos, promoviendo la inclusión en los espacios científicos y buscando mejorar la forma de despliegue actual.
    \item Producir herramientas y documentación orientada a mejorar el acceso a los datos promoviendo la inclusión.
\end{itemize}

Con el fin de lograr los objetivos generales planteados, se proponen los siguientes objetivos específicos:
\begin{itemize}
    \item Revisión bibliográfica sobre el tema y de los software existentes para sonorización de datos.
    \item Estudio de normativas vigentes para la realización de programas e interfaces accesibles, y producción de una herramienta basada en software libre (por ejemplo Python).
    \item Realización de un grupo focal integrado por personas con y sin discapacidad visual, para investigación sobre usabilidad y accesibilidad del programa desarrollado,
    \item Rediseño de la interfaz humano computadora a partir de la investigación previa; planteo de posibles mejoras e inclusión de funcionalidades al programa, en base a los resultados del grupo focal y el intercambio con los usuarios finales.
    \item Recuperación de datos de bases de datos especificas (por ejemplo: SIMBAD, NASA, SDSS, ADSAbs) para las pruebas técnicas de la herramienta de despliegue multimodal y pruebas con usuarios.
    \item Publicación de los resultados y transferencia a la comunidad.
\end{itemize}

En base a los objetivos planteados y con el objetivo de describir de la mejor forma posible la investigación y desarrollo realizados durante esta tesis, se ha dividido el escrito en siete (7) capítulos. En primer lugar se describe la base y punto de partida de la investigación y desarrollo abordados, encontrando en el Capítulo \ref{cap:marco_teorico} información sobre discapacidad, el diseño centrado en el usuario y el actual enfoque multisensorial en las ciencias del espacio. Complementando lo mencionado, el Capítulo \ref{cap:estado_del_arte} describe las herramientas de sonorización existentes, haciendo una distinción entre las que existían antes de 2017 y las que han sido desarrolladas a lo largo de estos años de forma paralela con el desarrollo propio de esta tesis.

Los siguientes cinco (5) capítulos describen prácticamente de forma cronológica, los temas abordados. El Capítulo \ref{cap:analisis_normativo} contiene el análisis normativo realizado al iniciar el trabajo donde se evaluaron las herramientas existentes en 2017 y su accesibilidad, también se aplicó un estudio de accesibilidad a tres bases de datos astronómicas, lo que decantó en un documento con recomendaciones de accesibilidad y el primer diseño de interfaz gráfica para el software sonoUno (descripto a lo largo de los últimos cuatro (4) capítulos y producto final de esta tesis). El Capítulo \ref{cap:sonouno_v1} describe el desarrollo de la primer versión de sonoUno, disponible para Windows, MacOS y Ubuntu.

Una vez que se logró la primer versión del software de sonorización, se realizó la primer prueba con usuarios de forma presencial en la Universidad de Southampton, Reino Unido. Dicho trabajo se describe en el Capítulo \ref{cap:fg_completo} junto con las pruebas realizadas a distancia con contactos de diferentes partes del mundo; para éstas últimas, las comunicaciones se realizaron por email y a través de una encuesta volcada en Google Form. Adicionalmente, al final del Capítulo \ref{cap:fg_completo} se presentan dos casos donde se utiliza el programa sonoUno para sonorización de datos astronómicos.

Durante el desarrollo de esta tesis, además de trabajar sobre la interfaz gráfica y su accesibilidad, se desarrollaron nuevas sonorizaciones de datos de grandes facilidades astrofísicas y bases de datos. En el Capítulo \ref{cap:reinforce} se detallan los desarrollos sobre sonorización de partículas del Gran Colisionador de Hadrones, imágenes y datos de muongrafía. Además, se agregan ejemplos de aplicación del software a espectros de galaxias, estrellas variables y rayos cósmicos. En este mismo capítulo se describe la versión web de la herramienta y los primeros pasos para lograr una plataforma de sonorización de datos que unifique las diferentes versiones del programa.

Finalmente, el Capítulo \ref{cap:sonouno_vfinal} se dedicó a describir la página web con información dedicada al proyecto sonoUno, allí se encuentran también los manuales de instalación y de usuario, los accesos a la herramienta, una sección de noticias y una galería que muestra diferentes datos y sus sonorizaciones. Se incluye en el Capítulo \ref{cap:sonouno_vfinal} un apartado con el primer entrenamiento en sonorización realizado durante 2022 dentro del equipo de trabajo y un apartado que establece las bases para futuros desarrollos e investigaciones.

Desde 2017 que se inició esta investigación, se han ido realizando transferencias a la comunidad junto con publicaciones científicas que le fueron dando pautas al equipo de trabajo de que se estaba dirigiendo en la dirección correcta. Se ha participado de numerosos congresos, donde se presentaron los avances en el desarrollo de la propuesta (se incluyen las publicaciones en las secciones correspondientes), desde 2019 se participó en un proyecto de colaboración internacional (proyecto REINFORCE, descripto en el Capítulo \ref{cap:reinforce}) y se recibió invitación para participar del Audible Universe Workshop 1 y 2, un encuentro internacional que desde su primer edición en 2021 se ha realizado de forma anual convocando a especialistas en sonorización de diferentes áreas de estudio.

\chapter{Marco teórico}
\label{cap:marco_teorico}

\section{Introducción}

En la actualidad, el astrónomo con discapacidad visual, independientemente de su nivel profesional, público interesado, estudiante o experto, no tiene acceso a la misma cantidad y calidad de información que un astrónomo vidente. Las matemáticas y la física también son ciencias que se enseñan de manera unisensorial, suponiendo que el estudiante sea vidente, y hay muy poca investigación sobre cómo enseñar usando diferentes modalidades sensoriales. La falta de libros de texto y materiales apropiados para que las personas con discapacidad visual aprendan conceptos matemáticos y científicos de una manera eficiente, útil y eficaz, que no dependa de la memorización, es un ejemplo de falta de acceso.

Dicha falta de acceso en los diferentes niveles de aprendizaje, se mantienen aún cuando se registraron los primeros desarrollos de interfaces de sonorización para el análisis de datos hace unos 20 años. Esto puede tener relación con que muchas de las interfaces creadas para despliegues multisensoriales, entre ellas interfaces de sonorización que buscan mejorar la accesibilidad, se centran en funcionalidades estereotipo o funcionalidades que se creen usables. Esto se hace dejando fuera del ciclo de diseño y desarrollo la experiencia de usuarios y especialmente cómo los usuarios con discapacidad se comunican con la interfaz y experimentan su discapacidad. 

Si bien, áreas profesionales como las ciencias sociales, estudios de discapacidad, educación, profesionales de HCI y guías de accesibilidad (por mencionar algunas), comentan sobre la imposibilidad de crear una interfaz que se adapte a las necesidades de cada usuario o con un diseño universal, poco se habla de interfaces que se centren en las capacidades y experiencias de las personas que la están utilizando, y principalmente interfaces que le ofrezcan a personas con discapacidad autonomía digital. Esta autonomía no resuelve las cosas por el usuario sino que le permite adaptar activamente la interfaz y modificar el despliegue a sus propias necesidades. Un enfoque centrado en el usuario desde el inicio del desarrollo, combinado con un despliegue e interacciones multisensoriales ofrece una primer aproximación al paradigma que se propone, permitiendo la incorporación de nuevas modalidades sensoriales y brindando la posibilidad de elegir la forma de interactuar con la información.

En el presente Capítulo se describirá la situación actual de las personas con discapacidad, documentado en gran parte por las Naciones Unidas. Adicionalmente, se plasmará el marco de referencia correspondiente al diseño de interfaces centrado en el usuario y la sonorización de datos astronómicos y astrofísicos. Al converger aquí estos dos campos de investigación, se decidió que cada uno de ellos tenga su sección correspondiente. Se concluye con una reflexión sobre percepción visual y auditiva, mostrando como estos campos de investigación convergen en esta tesis.

\section{Sistemas digitales y accesibilidad}

La sostenibilidad de los sistemas digitales depende de lograr diseños, ya sea HCI (interfaces humano computadora), UX (experiencia de usuario) y UCD (diseño centrado en el usuario), que mejoren la calidad de vida y fomenten la realización humana \citep{eganbenyon2017,knowlesetal2016}. El informe sobre desarrollo humano de las Naciones Unidas en el año 2015 \citep{undp2015} aborda la realización humana sobre la base del acceso al trabajo en términos de desarrollo humano y su sostenibilidad. El informe destaca: ``el vínculo entre el desarrollo humano y el trabajo no es automático'' \citep{undp2015}, dado que la calidad del trabajo es una dimensión importante para garantizar que el trabajo mejore el desarrollo humano. También menciona que algunos trabajos son muy perjudiciales para el desarrollo humano, sobre todo aquellos que violentan los derechos humanos. La falta de autonomía con la que se encuentran las personas con discapacidad a la hora de utilizar herramientas digitales, hace que se vean obligadas a seguir una carrera profesional definida por obstáculos potenciales, en lugar de elegirla en base a sus intereses.

El Informe sobre Desarrollo Humano de las Naciones Unidas del año 2016 \citep{undp2016} establece que debido a que algunos grupos sociales (minorías étnicas, pueblos indígenas, personas con discapacidad) son sistemáticamente discriminados y, por lo tanto, excluidos, se necesitan medidas específicas para que puedan lograr resultados equitativos en materia de desarrollo humano. El informe también establece que:

\begin{adjustwidth}{1.27cm}{}
Pese a la gran diversidad de identidades y necesidades, los grupos en situación de marginación, como las minorías étnicas, los pueblos indígenas, las personas con discapacidad, las personas que viven con el VIH/SIDA y las personas lesbianas, gais, bisexuales, transgénero e intersexuales, a menudo se enfrentan a limitaciones similares, como la discriminación, el estigma social y el riesgo de sufrir daños. Sin embargo, cada grupo tiene también necesidades especiales que hay que satisfacer para poderse beneficiar de los progresos en materia de desarrollo humano. \citep[p.14]{undp2016}
\end{adjustwidth}

\noindent Las necesidades específicas de los científicos con discapacidades dependen de la disciplina. No es suficiente proporcionar adaptaciones, sino que es necesario crear las herramientas que alienten, apoyen y promuevan el máximo rendimiento de cada individuo. La inclusión y las facilidades son fundamentales para empoderar a las personas y posibilitarles que vivan de forma independiente, tengan un empleo y contribuyan a la sociedad.

El informe sobre desarrollo humano actual, que data del año 2021 \citep{undp2021}, hace referencia al rol del ser humano y las presiones planetarias, definiendo un nuevo horizonte que ubica como uno de los ejes principales la equidad: ``Si la \textbf{equidad}, la innovación y la gestión se convierten en elementos centrales de lo que significa llevar una buena vida, el ser humano podrá prosperar y se aliviarán las presiones planetarias'' \citep[p.8]{undp2021}. La equidad, sobre todo para personas con alguna discapacidad es una necesidad y sobre todo un derecho, al cual la sociedad no está respondiendo de manera coherente y sostenida. Si bien se reconoce que las políticas han tomado acción para que esto mejore, las estadísticas siguen siendo preocupantes. El último reporte sobre discapacidad y desarrollo publicado en 2018 por las Naciones Unidas, indica que la relación entre empleo y población de las personas con discapacidad de 15 años o más es 50\% menor que para las personas sin discapacidad \citep{ddr2018}. Adicionalmente, los salarios para las personas con discapacidad son más bajos, y por si fuera poco, el 32\% de las personas con discapacidad que tienen un empleo consideran que el espacio de trabajo dificulta o no es accesible. 

El trabajo digno facilita un sentido de realización, interacción social y consolidación. Permite que las personas utilicen sus habilidades, adquieran y utilicen nuevas, aumenten su red social y fortalezcan su espíritu emprendedor. Es de suma importancia brindar el acceso y las herramientas para que personas con diversidad funcional puedan ejercer su libre voluntad a la hora de elegir una carrera y un trabajo, generar un ambiente de trabajo accesible dentro del campo de la investigación de datos astronómicos y astrofísicos puede ser un primer paso para lograr tal propósito. Cabe aclarar que se mencionó a personas con ``diversidad funcional'', por ser un término que hace foco en las capacidades de las personas y no así en su deficiencia o discapacidad.

\subsection{Accesibilidad a los datos de las ciencias espaciales}

El Grupo de Trabajo sobre Accesibilidad y Discapacidad (WGAD) de la Sociedad Astronómica Estadounidense publicó en 2017 un informe titulado ``Recomendaciones de Accesibilidad para Revistas'' \citep{wandaetal2017} que puede tomarse como una señalización para la accesibilidad formal, preparado después de la recopilación de sugerencias de accesibilidad con observaciones relacionadas con la astronomía y revisión por parte de la comunidad de astrónomos profesionales (astrónomos con puestos posdoctorales y más de 4 artículos logrados como primeros autores luego de obtener la carrera de grado). El documento enfatizó claramente la importancia y la necesidad del diseño centrado en el usuario y las evaluaciones de usabilidad de las herramientas y recursos disponibles, llegando a la conclusión de que las discapacidades y/o los estilos de aprendizaje deben tenerse en cuenta al diseñar tecnologías y sistemas de información. 

Los astrónomos discapacitados, a través del documento desarrollado en el marco del WGAD muestran que es posible contribuir en un nivel profesional buscando generar y promover la accesibilidad en publicaciones científicas. Es de suma importancia crear las herramientas necesarias que permitan la realización de investigaciones expertas por parte de las personas con discapacidad, ya sea de aparición tardía o congénita. En un nivel diferente de experiencia, es importante que los estudiantes con discapacidades de nivel escolar y universitario puedan acceder a la misma cantidad y calidad de información que sus compañeros y tener éxito en las transiciones a través de los niveles académicos. La inclusión sólo se logrará si se brindan las mismas oportunidades a todas las personas desde el inicio de su camino hacia su vida profesional.

Concentrando el análisis en los diferentes aspectos de la vida profesional de un astrónomo o astrónoma, se debe tener en cuenta que la forma de acceder tanto a datos de objetos públicos como a bibliografía científica es a través de bases de datos web, sean estas bases de datos de bibliografía como el caso de ADSABS (\textit{Astrophysics Data System}), o bases de datos de objetos como el caso de: SIMBAD (\textit{Set of Indications, Measurements, and Bibliography for Astronomical Data}), SDSS (\textit{Sloan Digital Sky Survey}) o la base de datos del Observatorio Pierre Auger (\textit{Auger Open Data}). Si bien estos últimos años algunas de las plataformas web mencionadas han tenido actualizaciones, un estudio realizado por \citet{casadoetal2018} mostró poca accesibilidad a dichas bases de datos utilizando lectores de pantalla y siguiendo las recomendaciones de la norma WCAG 2.0 (\textit{Web Content Accessibility Guidelines}).

La falta de existencia de un modelo de trabajo formal y evaluaciones robustas de accesibilidad a las bases de datos, ha afectado severamente el desarrollo de herramientas para el tratamiento de los datos y revisión de literatura que puedan ser utilizados por las personas discapacitadas. Si bien existen plataformas digitales para evaluar accesibilidad, las mismas no son abordadas correctamente, los desarrolladores u organizaciones que los utilizan buscan conocer rápidamente el resultado sin comprender que el análisis de accesibilidad es un proceso complejo que debe ser desarrollado desde el inicio y en conjunto con la herramienta. Debe ser considerado que los evaluadores automáticos hasta el momento evalúan el código fuente buscando un problema, por lo que no serían capaces de detectar problemas en el discurso que transmite la web, problemas de acceso que pueda presentar un lector de pantalla o problemas en la disposición de los elementos propios de la web, entre otras cosas.

\citet{nunezetal2019} realiza una revisión de bibliografía identificando los métodos utilizados para evaluar accesibilidad de páginas web, allí se identifican tres métodos de evaluación: (1) herramientas automáticas; (2) evaluación de expertos; (3) testeo con usuarios. Este trabajo muestra que el método más utilizado es el (1), y en segundo lugar estaría el (3); pero en este último caso, no se especifica si el testeo con usuarios se refuerza con un análisis de confiabilidad que valide los hallazgos. 

Esta falta de claridad en la forma de abordar la accesibilidad de herramientas digitales, en general somete al usuario a lo que ha sido decidido y asumido por los desarrolladores, decantando en lo que el usuario puede o no puede hacer (limitándolo y dejándolo sin opción para decidir). En el caso de la inclusión de discapacidad, este enfoque y el desarrollo de ``tecnologías accesibles'', puede interpretarse como un intento basado en el modelo social de discapacidad en búsqueda de una rápida respuesta a la evidente exclusión socioeconómica que sufren las personas con discapacidad. Tales intentos, sesgados por obtener esta rápida respuesta, ignoran cómo cada usuario experimenta su discapacidad en su vida cotidiana, y por ende, cómo cada persona experimenta la discapacidad en frente de las interfaces hombre-máquina. \citet{ghai2002} afirma en su reporte, que tales prácticas desmotivan a las personas con discapacidad, porque las construcciones a su alrededor amenazan con invisibilizar completamente a la persona.

Evaluaciones de accesibilidad solidas y holísticas son necesarias, ya que se evidencia que las personas con discapacidades se acostumbran a las interacciones dolorosas y físicamente extenuantes impuestas por las interfaces actuales, dadas por su estructura y dialógica. Las pautas de accesibilidad del consorcio web (\href{https://www.w3.org/standards/webdesign/accessibility}{\underline{\textcolor{blue}{link}}}), así como otras normativas similares, deben verse como un punto de partida para la accesibilidad, pero debe tenerse en cuenta que pueden presentar un criterio de éxito débil y, como consecuencia, no proporcionar un acceso equitativo en todos los aspectos de la interacción. Las prácticas mencionadas hasta aquí comúnmente solo se refieren al usuario sin sugerir granularidad y no toma en cuenta la interacción con el usuario final.

Adicionalmente a las normativas existentes, las recomendaciones de accesibilidad publicadas en revistas pueden servir como punto de partida para que los administradores de bases de datos y desarrolladores de software tengan en cuenta la diversidad de formas de comunicación entre la interfaz y personas con y sin discapacidad, todo esto con el objetivo de lograr un diseño sostenible y centrado en el usuario. Los enfoques de diseño de acceso digital actuales tienen un punto de vista limitado y no permiten una amplia diversidad de percepciones, desempeño y estilos de aprendizaje, y sus posibles descubrimientos posteriores.

Reforzando lo mencionado y dirigiéndonos al diseño centrado en el usuario, \citet{stanfill2015} considera una interfaz digital como un predictor de discurso y se enfoca en el despliegue web teniendo en cuenta audiencias sin discapacidad. Este autor sostiene que las suposiciones integradas en las interfaces son normativas, no son basadas en evaluaciones exhaustivas y tampoco en indicativos de experiencias de usuarios. De acuerdo con esta premisa, los sitios digitales son basados en la asunción superficial de usuarios que aceptan o rechazan características. En el caso de la discapacidad, el acceso a la ruta que se debe seguir para tener éxito en una tarea, en una interfaz digital, está predeterminado, los usuarios deben adaptarse a rutas y acciones prefijadas para completar una tarea. No hay duda de que las personas con discapacidad encuentran mayor inclusión sin importar la dificultad, sin embargo, retomando lo mencionado por \citet{ghai2002} si no tomamos acción para incluir a las personas con diversidad funcional en los desarrollos seguiremos invisibilizando su persona y construyendo herramientas poco usables y poco útiles.

\subsection{Diseño centrado en el usuario}

Tal como ya se mencionó, el campo de la astronomía depende principalmente de los datos y, dados los altos costos computacionales de la gestión de datos y el aumento en las tasas y el tamaño de muestreo, el enfoque actual se inclina más hacia la reducción de los costos computacionales y la mejora de la calidad de los datos. El resultado es una visualización de funcionalidades e interacción no centrada en el usuario sino en los datos. El término ``funcionalidades'' incluye, entre otros, los métodos de diseño, entrada, procesamiento y salida del sistema. 

\begin{figure}[ht]
    \centering
    \includegraphics[width=1\textwidth]{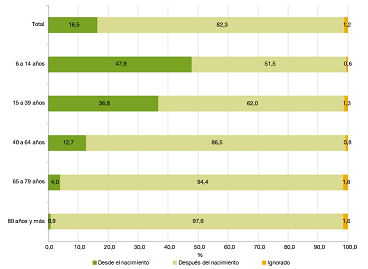}
    \caption{Imagen extraída del ``Estudio Nacional sobre el Perfil de las Personas con Discapacidad'', sección 8.1, p.62 \citep{indec2018}.}
    \label{fig:origen_discap}
\end{figure}

Esta característica de los sistemas en el campo de la astronomía, y según el WGAD, corta los vínculos del astrónomo discapacitado profesional con el sistema científico y su actividad como investigador. Para entender el problema de manera amplia, se debe considerar y aceptar que la discapacidad no es exclusivamente congénita y de nacimiento, sino que un gran porcentaje de ellas son de inicio tardío y pueden adquirirse a raíz de una enfermedad o accidente. El informe del Instituto Nacional de Estadísticas y Censos de la República Argentina \citep{indec2018} expone una gráfica que se muestra aquí en la Figura \ref{fig:origen_discap}, donde se evidencia como a partir de los 14 años el porcentaje de personas que adquieren la discapacidad después del nacimiento aumenta abruptamente, pasando a ser este mayoritario. Lo que nos lleva a que cualquier profesional podría adquirir una discapacidad, quedando excluido del ámbito científico al que pertenece debido a la falta de accesibilidad en herramientas y sistemas digitales.

La era digital con su continua competencia en el desarrollo de interfaces no establece como requisito para los desarrolladores una formación o educación continua en el campo del diseño centrado en el usuario (UCD por sus siglas en inglés), experiencia humano computadora (HCE) o interacción humano computadora (HCI). Por ende, el comportamiento humano, la diversidad funcional y la forma en que las personas actúan no son tomadas en cuenta a la hora del diseño de software. Esto puede ser uno de los factores que lleva a los desarrolladores a interpretar la accesibilidad como algo que se logra siguiendo un manual, una normativa o normalizando el comportamiento a una sola modalidad sensorial (como ser el caso de la ceguera), cuando en realidad las discapacidades se presentan por lo general en conjunto con otras patologías (según el \citet{indec2018}, en personas de 6 años o más, el 41\% de la población Argentina con discapacidad presenta más de una).

En base a lo que se viene exponiendo, la pregunta de cómo construir y lograr la inclusión por parte de las arquitecturas digitales e interfaces utilizadas en astronomía no ha sido investigada de manera sistemática y profunda. La inclusión de personas funcionalmente diversas en los desarrollos relacionados a astronomía no ha sido una práctica común, pese a que su colaboración contribuiría a analizar y eventualmente comprobar la efectividad e impacto de los desarrollos, en conjunto a su funcionalidad y uso a través de los años. Hasta el momento solo se conocen cuatro software que trabajan con o incluyen a personas con discapacidad (contando el desarrollo propio de este trabajo), en la sección \ref{sect:desktop_astronomia} se describen dos software que incluyen personas con discapacidad visual en el grupo de desarrollo, xSonify y starSound; luego, en la sección \ref{sect:web_astronomia} un desarrollo reciente menciona un proyecto donde trabajan temas de inclusión.

Ampliando la búsqueda a otras ciencias y aplicaciones, se encuentran trabajos que buscan lograr y fomentar los diseños centrados en el usuario como \citet{grundyetal2020}, cuyo trabajo consiste en modelar el comportamiento de los mismos, manteniendo actualizaciones con nuevos reportes de los consumidores y ayudando así a lograr diseños más centrados en el destinatario final. No queda del todo claro si pretenden prescindir del usuario una vez que se tenga ese modelo, lo que sería un error, dado que las necesidades de los mismos cambian al cambiar el entorno, el tiempo y la situación, entre otras cosas.

Por su lado, \citet{niespelayo2010} remarcan la necesidad de incluir especialistas en factores humanos para mejorar la comunicación entre usuarios y desarrolladores. En dicho trabajo llevan a cabo un caso de estudio con un desarrollo relacionado con cuidado médico, en un principio se realizan reuniones entre los desarrolladores y destinatarios, demostrando las dificultades que encuentran los primeros a la hora de interpretar algunos pedidos o críticas (por ejemplo, los desarrolladores no comprendían la necesidad de los consumidores en acceder a la lista de medicamentos detallada en lugar de sintetizada). Acto seguido, se realizaron intercambios con los especialistas en factores humanos, quienes mediante un análisis cognitivo de los requerimientos pudieron trasmitirle a los desarrolladores la razón que respaldaba el pedido de los consumidores. Esto resalta la importancia de un enfoque centrado en el usuario y de los equipos de investigación multidisciplinarios.

Recientemente, \citet{huynhetal2021} menciona limitaciones que encuentran las personas que usan los programas para poder reportar errores o problemas en las herramientas que utilizan, por ejemplo una persona con discapacidad puede tener problemas para reportar un error cuando la plataforma para reportarlo no es accesible. Para ello el trabajo propone una herramienta que permite reportar errores y problemas de forma más asertiva y con campos que requieren la información detallada que necesitan los desarrolladores. Los mencionados ejemplos confirman lo que se expone desde el inicio de este trabajo: para que un desarrollo sea accesible, usable y eficiente, debe ser centrado en el usuario desde el inicio.

Reforzando lo que se expone anteriormente, \citet{ritteretal2014} explica en su trabajo que muchas áreas como la interacción humano-computadora (HCI), la experiencia del usuario (UX), el diseño centrado en el usuario (UCD) o el diseño de sistemas centrados en el ser humano (HCSD o UCSD, si se emplea la palabra usuario en lugar de humano) están preocupadas por el objetivo de mejorar la interacción de las personas con los sistemas y las computadoras, solo difieren en los métodos utilizados para abordar este objetivo. Esta heterogeneidad con la que nos encontramos por un lado dificulta la toma de decisiones por parte del desarrollador en cuanto a cual método seguir. Por otro lado, la oferta de métodos poco validados y que prometen resultados en corto tiempo hace confusa y poco atractiva la adopción de la técnica del diseño centrado en el usuario desde el inicio. Debe tenerse en cuenta que ésta técnica necesita de equipos interdisciplinarios (desarrolladores, expertos en astronomía y astrofísica (especialmente para el caso de estudio de este trabajo), expertos en diseño de interfaces, expertos en técnicas de análisis cualitativo y cuantitativo) y sobre todo tiempo, necesario para llevar adelante múltiples encuentros con los destinatarios de la herramienta.

\section{Enfoque multisensorial para el análisis de datos}
\label{sect:enfoque_multisensorial}

En ciencias como la astronomía y la astrofísica, en las cuales se centra la aplicación de la investigación objetivo de esta tesis, los científicos interactúan de forma permanente con datos numéricos, en general representados de forma visual. Estas interacciones implican una respuesta que se relaciona mayormente con la percepción e interpretación de patrones o eventos aislados. En la actualidad, dichas interacciones están limitadas por las herramientas de análisis de datos disponibles y por la resolución de los elementos de despliego como pantallas. 

Teniendo en cuenta que el número de datos obtenidos es cada vez  más grandes y se adquieren con mayor rapidez, la necesidad de darles sentido y poder mejorar el método de análisis de los mismo también se incrementa. La investigación realizada por \citet{wandatesis2013} durante su tesis doctoral, mediante un estudio de percepción, demostró que el despliegue multisensorial de los datos puede mejorar la detección de señales, especialmente si se trata de datos astronómicos. Ello nos permite inferir que la sonorización de dichos datos en conjunto con la visualización de los mismos contribuiría a una mejor comprensión de los resultados, además permitiría a personas con diversidad funcional el análisis de datos científicos.

\subsection{Manejo de datos en ciencias del espacio}

Los Macrodatos (Big Data) son un tema candente estos días y particularmente relevante en astrofísica, donde los avances continuos en tecnología conducirán a cada vez más grandes conjuntos de datos en el futuro. Actualmente las tecnologías disponibles simplemente no están a la altura de recolectar, procesar y dar sentido a tanta información en tiempo real. \citet{ibm2012} ha propuesto el uso de sistemas cognitivos definidos como capaces de aprender a partir de su interacción con los datos y los seres humanos, en principio debido a que continuamente se programan a sí mismos. Estos sistemas cognitivos si se aplican a diferentes entornos, no evalúan el riesgo de tener el usuario al margen de la actividad misma de investigación, y no lo mantienen informado acerca de las incertidumbres que cada sistema, proceso o herramienta pueden generar. La sonorización plantea la opción de evaluación de riesgos (en relación a la toma decisiones) y proporciona al analista un control total sobre los datos y la exploración, lo cual conduce a la toma de decisiones. Los sistemas cognitivos se basan en el uso de formas de onda y aprendizaje automático, sin embargo el cerebro humano sigue siendo la más poderosa herramienta para la lectura de grandes conjuntos de datos. En ese sentido, debe ser un ser humano quien examina y decide sobre el análisis de la información en cuestión.

Basados en el uso de sonido como complemento de prototipos de visualización y teniendo en cuenta las técnicas de percepción, un estudio basado en la técnica de grupo focal realizado en el CfA-Harvard (Center for Astrophysics), fue de los primeros que se centraron en el usuario a partir de la evaluación de las necesidades de los físicos al realizar el análisis de datos \citep{wandatesis2013}. Los astrofísicos que participaron en el estudio, informaron el deseo de escuchar los datos ``tan despojados como sea posible de manipulación, antes de cualquier reducción o procesamiento'', así como la necesidad de disminuir la incertidumbre (por ejemplo, ``necesitamos saber si el código está haciendo algo poco confiable con los datos''). No parece que los sistemas cognitivos puedan lograr esto último, ya que impondrán formas de onda de datos con el riesgo de enmascarar eventos que podrían ser de importancia, aumentando la incertidumbre y la carga de trabajo. 

Los especialistas en procesamiento de datos en ciencias del espacio acuerdan que  las herramientas actuales o los sistemas cognitivos automatizados no serán un reemplazo para el cerebro humano \citep{hassanetal2013}. Por otro lado, aunque las computadoras pueden hacer el mundo más accesible para las personas discapacitadas, la naturaleza visual de la computación tal como es implementada usualmente, la falta de recursos y la falta de modelos a seguir con personas discapacitadas, hacen que sea difícil para éstas participar en cursos de ciencia y verse como potenciales científicos. Esto es particularmente cierto en astronomía, donde a pesar de la naturaleza intrínsecamente invisible de más del 90\% del universo, hacemos un gran esfuerzo para transformar los datos en imágenes para analizarlas visualmente, bloqueando efectivamente el acceso a personas orientadas con el sonido. 

También sabemos que mientras algunas disciplinas científicas son dependientes de la integración de la ciencia y el pensamiento computacional, pocas lo son tanto como la astronomía. Todos los esfuerzos se han dirigido a fomentar el desarrollo de arquitecturas de alto rendimiento (por ejemplo CPUs multinúcleo y tarjetas gráficas potentes), interoperatividad (diferentes aplicaciones pueden funcionar simultáneamente en conjuntos de datos compartidos) y flujos de trabajo colaborativo, permitiendo a varios usuarios trabajar simultáneamente, intercambiando información y experiencias de visualización. Sin embargo, nuevas técnicas para el despliego de datos e información, como la integración de la perceptualización multimodal (vista y oído) para la exploración y análisis de los datos de la física espacial, son poco exploradas.

\subsection{Sonorización para el análisis de datos}

\subsubsection{Primeros indicios en el campo de las ciencias espaciales}

La sonorización como complemento a la visualización de datos no es ampliamente utilizada, quizás debido a la falta de una base teórica sólida y escasos análisis de percepción. A pesar de ello, el uso del sonido para analizar los datos de la ciencia espacial no es una idea nueva. Si analizamos el tema desde el punto de vista de los desarrollos a lo largo de la historia, podemos afirmar que, por ejemplo, el telégrafo condujo a grandes avances en telecomunicaciones, mayormente en lo que hace a la comunicación de una señal vocal de un punto a otro en tiempo real. Dichos avances y la investigación en estática e interferencia para mejorar la comunicación llevó a grandes descubrimientos científicos, basados en estudios sobre percepción auditiva. 

El uso de la percepción multisensorial para el análisis de datos en ciencias del espacio ha sido documentado desde hace tiempo, como por ejemplo el uso de audio para estudiar impulsos atmosféricos expuesto por \citet{preece1894} y \citet{barkhausen1919}. Los orígenes de esos descubrimientos (utilizando herramientas desarrolladas en el campo de las telecomunicaciones), se relacionan con los avances tecnológicos que estaban comenzando a despegar. Algunos términos sensoriales fueron aceptados en aquel momento para describir los fenómenos escuchados, como son: emisión en modo de silbido, zumbido, pío (canto de los pájaros). En 1982, se reprodujeron datos como sonido para detectar micrometeoroides que impactan en la Voyager 2 cuando atraviesa los anillos de Saturno \citep{scarfetal1982}. En este trabajo fue utilizada una de las primeras herramientas basadas en Beatnik/JavaScript, se probaron varias técnicas de sonorización de campo magnético y mediciones de plasma para la magnetopausa. 

Desde la década de 1970 y durante unos cuarenta años, \citet{gurnettetal2005} ha utilizado la sonorización de datos espaciales para producir los primeros intentos documentados del uso del sonido relacionado con analizar, desentrañar y transmitir la información contenida en los datos de diferentes misiones espaciales, como las sondas Cassini y la ya mencionada Voyager. Esta documentación es muy importante ya que es fácil confundir a los pioneros en este tema con otras personas que utilizaban la sonorización a finales de los 80.

Pese a estos antecedentes de los '80 y '90, con el avance de nuevas tecnologías, en lugar de maximizarse esta multisensorialidad, el enfoque de nuevas aplicaciones se concentró en despliegues y herramientas casi exclusivamente basadas en la percepción y capacidad visual. Esta tendencia, deja a un gran número de personas con discapacidad visual excluidas \citep{shakespeare2017}, sin posibilidad de acceso a profesionalizarse en ciencias y entre ellas en astronomía.

\subsubsection{Primeros registros de sonorizaciones con programas o Scripts}

El interés en la técnica de sonorización dentro de la comunidad de física espacial ha ido en aumento desde que el primer prototipo de sonorización de datos de la física espacial fue lanzado en septiembre de 2005 \citep{candeyetal2005}, por la división de heliofísica del Centro de Vuelo Espacial Goddard (GSFC) de la NASA. Dicha herramienta se sustentó en el interés por utilizar parámetros de sonido para realizar una exploración más detallada de datos de plasma y partículas del viento solar. Simultáneamente, los científicos involucrados en el proyecto sugirieron que las técnicas de sonorización podrían incrementar la detección de evidencias en la búsqueda de planetas extra solares, a partir de datos adquiridos por el telescopio Kepler \citep{laughlin2006}. Greg Laughlin, astrónomo de Santa Cruz, California, desarrolló una herramienta de reducción de datos para este fin. La herramienta incluye sonorización (así como periodogramas) y cuando las órbitas son keplerianas, se pueden escuchar inclusive los armónicos. La aplicación incluye enlaces a varios datos de exoplanetas y de esta manera permite visualizar y sonorizar diversos sistemas.

En este sentido, es interesante mencionar que en 2011 Batalha, de la Universidad Estatal de San José en California, y miembro clave del Equipo Científico de Kepler, reprodujo datos sonorizados durante el coloquio del Centro Smithsonian de Astrofísica de Harvard \citep{batalha2011}, para Kepler 10b, un exoplaneta pequeño (1,4 de radio terrestre). Para poder establecer las propiedades del planeta, debieron realizar astrosismología a la estrella, por lo que en los datos sonorizados es posible escuchar diferentes frecuencias que pueden indicar la presencia de un exoplaneta. La herramienta no permite que el usuario mapee parámetros como el timbre, el volumen o el tono, tampoco es posible explorar los datos a medida que se sonorizan. \citet{gurnett2012}, por su parte ha usado la sonorización de datos espaciales para analizar y transmitir la información contenida en los datos de diferentes misiones, por ejemplo Cassini. 

Investigadores que trabajan con datos de exoplanetas detectados por el telescopio Kepler han propuesto utilizar otro prototipo de sonorización: el software xSonify \citep{candeyetal2006}. En este caso, se buscaba tratar cada punto fotométrico como una nota, con el brillo determinando el tono y la duración de cada nota asociada con el tiempo probable de un tránsito típico. El programa xSonify es el primer desarrollo que proporciona una interfaz donde el usuario puede ingresar los datos, visualizarlos y sonorizarlos. Adicionalmente, muestra un panel de configuraciones, permitiendo al usuario elegir los parámetros que mejor se ajusten a sus necesidades. Cuenta también con un lector de pantalla propio que puede activarse o desactivarse, y otras características de accesibilidad como etiquetas en los elementos de la interfaz (se puede encontrar una descripción mas detallada en la sección \ref{sect:desktop_astronomia}).

El equipo ``Stereo-solar wind'' de la Universidad de California, en Berkeley, también implementó una herramienta de sonorización para transmitir información sobre parámetros caracterizados en datos de viento solar \citep{bithellmorales2012}. El software proporciona cinco formas diferentes de sonorizar los datos con la capacidad de cambiar la frecuencia fundamental utilizada. Actualmente, han puesto a disposición cuatro gráficos de datos de flujo de energía de composición iónica para helio (He), carbono (C), magnesio (Mg) y hierro (Fe) del satélite ``Advance Composition Explorer'' (ACE) de la NASA. Esta herramienta está destinada a la divulgación y educación.

Para los procesos no lineales, la fundamental de la forma de onda de los datos puede cambiar o, como en el caso de los plasmas turbulentos (como en el viento solar), la fundamental puede no ser el componente más bajo de la forma de onda. \citet{alexanderetal2011} realizó la sonorización de los datos de viento solar para el departamento de ingeniería y ciencias atmosféricas de la Universidad de Michigan. En este caso, el objetivo principal de los investigadores era tratar de escuchar información que sus ojos podrían haber pasado por alto en los datos de velocidad del viento solar y densidad de partículas recopilados por el satélite ACE de la NASA. El trabajo fue presentado en la reunión ICAD 2010 en Washington DC, el autor utilizó un alto nivel de licencia artística. La sonorización no llevó a los investigadores a nuevos descubrimientos, pero reconocieron el poder de la sonorización para el análisis de plasmas solares. 

Otros ejemplos que podemos mencionar son: \textit{'See-Through-Sound'}, un proyecto dedicado a convertir imágenes en sonido para permitir a personas con discapacidad visual detectar objetos a su alrededor \citep{henriquesetal2014}; \textit{R-Scuti}, una instalación audio-visual que propone la conversión de datos astronómicos de la base de datos AAVSO (por sus siglas en ingles ``American Association of Variable Star Observers'') en un entorno de exhibición, utilizando grabaciones de observaciones de estrellas variables \citep{laurentizetal2021}; y un \textit{laboratorio de ciencia accesible}, investigación que resalta la necesidad de enfoques multimodales en la enseñanza y entornos de aprendizaje, presentando principios para el diseño de material inclusivo basado en el diseño centrado en el usuario y diseño universal para ambientes de enseñanza \citep{reynagaetal2020}.

El proyecto ``Sensing the Dynamic Universe'' del Centro de Astrofísica de la Universidad de Harvard, realizó una implementación teórica de un protocolo de diseño centrado en el usuario. Utilizando una plataforma web para educación y divulgación (puede acceder al link \href{https://lweb.cfa.harvard.edu/sdu/}{\underline{\textcolor{blue}{aquí}}}), su enfoque se basa en brindar un soporte individualizado que permita a los usuarios ajustar su despliegue web de forma dinámica. Dicha tarea se plantea basados en elementos conocidos como atajos de teclado, paletas de colores para ajustes de contraste, opciones para cambiar el tamaño de letra y exploración en el uso de la plataforma para la prevención de errores. Todas las configuraciones pueden ser habilitadas o deshabilitadas por el usuario. Este proyecto brinda autonomía durante el aprendizaje, trabajando en la flexibilidad, elecciones, y el conocimiento de experiencias de personas, permitiéndole al participante utilizar su experiencia digital. Dentro de la información proporcionada por la página web, se encuentran videos de sonorizaciones realizadas con una adaptación del software sonoUno, desarrollo producto de esta tesis (más información en la sección \ref{sect:uso_SDU} del Capítulo \ref{cap:fg_completo}).

Enfocándonos en otro tipo de investigación, \citet{asquith2010} desarrolló la herramienta para la sonorización de datos obtenidos del Gran Colisionador de Hadrones (LHC). El ``LHC Sound'', fue  lanzado en 2010 para sonorizar los datos del LHC, representa datos reales y simulados utilizando instrumentos musicales y su aplicación es exclusivamente en divulgación. Cabe destacar que solo se ha podido acceder a videos o sonidos, y datan de esa fecha, por lo que se considera que no ha tenido actualizaciones.

Otro ejemplo de sonorización de datos es el esfuerzo por representar de manera multimodal ondas gravitacionales. Estas ondas y/o sus fuentes, como los agujeros negros, pueden no verse a simple vista. Las ondas gravitacionales son una deformación del espacio-tiempo solo detectable a partir de las grandes instalaciones de LIGO  (en Estados Unidos) y VIRGO (en Italia). El equipo del Observatorio de Ondas Gravitacionales con Interferómetro Láser (LIGO, por su nombre en inglés) de la Universidad de Syracuse expresó la necesidad de herramientas para convertir eventos ultra pequeños en algo perceptible mediante sonidos \citep{monrclairuniversity}. Por ejemplo, Levin y su grupo han sonorizado predicciones para ondas gravitacionales desde pares de agujeros negros, representando los sonidos de su colisión \citep{explorations2011}. De esta manera, LIGO detectó el primer evento de ondas gravitacionales en la historia, en el año 2015 \citep{LIGO2016}. Está documentado que la onda se detectó como una señal de chirrido, que duró más de 0,2 segundos y aumentó en frecuencia y amplitud en aproximadamente 8 ciclos de 35 Hz a 250 Hz. La señal está en el rango audible, aunque es difícil de detectar debido a la baja relación señal/ruido. Se ha descrito que la señal se parece al "gorjeo" de un pájaro; LIGO tiene un centro de ciencia abierta donde las personas pueden descargar datos de audio "confiables". Adicionalmente, LIGO tiene relación con un grupo de trabajo llamado ``Black Hole Hunter'' quienes han desarrollado una actividad disponible en la web para clasificar sonidos de ondas gravitacionales \citep{blackholehunter}.

Desde 2019, el equipo de trabajo formado durante y con el objetivo del desarrollo del software sonoUno, se unió al proyecto REINFORCE-GA 872859 (con el soporte del ``EC Research Innovation Action'' bajo el programa H2020 SwafS-2019-1). En el marco de esta colaboración internacional se tiene contacto con el Observatorio de Ondas Gravitacionales VIRGO y la Organización Europea para la Investigación Nuclear (CERN), entre otras. Para las últimas instancias de la investigación de esta tesis se trabajó en la sonorización de Ondas Gravitacionales, Glitches (nombre que reciben las diferentes fuentes de ruido que se encuentran en el ambiente y en los instrumentos de medición utilizados para la detección de las ondas gravitacionales), partículas (generadas en el Gran Colicionador de Hadrones (LHC), entre las cuales encontramos electrones, protones, muones, fotones y fotones convertidos) y muongrafía (método en el cual utilizan los muones producto de la lluvia de partículas secundarias producidas por los rayos cósmicos para analizar diferentes espacios, como por ejemplo el interior de las pirámides). 

En la página web de \href{https://www.sonouno.org.ar/reinforce-demonstrators/}{\underline{\textcolor{blue}{sonoUno}}} se encuentra un apartado con algunos ejemplos de sonorización de Glitches e información sobre las diferentes sonorizaciones de datos. Adicionalmente, en el Capítulo \ref{cap:reinforce} se describen en detalle las herramientas de sonorización realizadas y las actualizaciones de sonoUno.

\subsubsection{Software de sonorización de datos que permiten ingreso de archivos}

En los últimos años, se ha encontrado evidencia de herramientas que buscan ser más generales aceptando como entrada el arreglo de datos y produciendo la sonorización de los mismos. Algunas de estas herramientas son: Sonification Sandbox \citep{sandbox2007}, MathTrax \citep{mathtraxweb}, xSonify (que ya fue mencionado anteriormente debido a su relación con NASA) \citep{wandaetal2011}, Sonifyer \citep{sonifyer2008}, Sonipy \citep{worraletal2007}, Planethesizer \citep{riber2018}, StarSound \citep{cookeetal2017} y recientemente STRAUSS \citep{strauss2021}. Los primeros cuatro de ellos no han sido actualizados y por ello probablemente sus librerías estén obsoletas, por lo que pueden presentar problemas con sus funcionalidades. En particular, Sonifyer fue diseñado para sonorizar datos de electroencefalografía, demostrando de esta manera que otras ramas de la ciencia también se ocupan de la detección o despliegue multimodal. MathTrax tiene fines exclusivamente educativos. Una descripción más detallada de cada uno de ellos se brindará en el Capítulo \ref{cap:estado_del_arte}, sección \ref{sect:software_independientes}.

Acerca de la interfaz gráfica de usuario (GUI) que presentan estos programas, Sonipy y STRAUSS no tiene una; en cuanto a los demás, muestran una GUI compleja con muchos elementos. En algunos casos, presentan ventanas emergentes y en otros, pestañas para separar las diferentes funcionalidades. Estas últimas características, obliga a los usuarios a cambiar entre ventanas, un tema desaconsejado por especialistas en accesibilidad y diseño centrado en el usuario.

Los mencionados hasta ahora son programas de escritorio que deben ser instalados en la computadora, con las complejidades que resulten de los diferentes sistemas operativos, lo que para una persona con discapacidad pueden resultar insalvable. Es interesante mencionar que en casos de discapacidad visual severa, el usuario utiliza lectores de pantalla que facilitan la navegación por la interfaz digital. Un problema que se encuentra aquí es que no todos los desarrollos aceptan ser utilizados por los lectores de pantalla disponibles gratuitamente. Se ampliará el tema de lectores de pantalla en el Capítulo \ref{cap:estado_del_arte}, sección \ref{sect:lectores_pantalla}.

Debido a la complejidad que conlleva la instalación de programas, y la necesidad de herramientas multiplataforma, hace unos años se comenzaron a evidenciar desarrollos de programas de sonorización de datos que funcionan en la web. Este es el caso de herramientas como: TwoTone \citep{twotoneweb}, TimeWorkers \citep{chafe2019}, Sonification Blocks \citep{sonifblocks2017} y Web Sandbox \citep{websandbox2017}. En el caso de TwoTone, solo hay tutoriales y videos disponibles y se muestra un marco en el que el usuario puede abrir un conjunto de datos y sonorizarlo con diferentes configuraciones de sonido. En el caso de TimeWorkers, el usuario debe descargar una carpeta zip y cambiar algunas configuraciones del navegador para ejecutar el programa. Sonification Blocks presenta una interfaz de arrastrar y soltar que permite la generación de sonido con diferentes configuraciones. Finalmente, Web Sandbox, actualmente llamado ``Highcharts Sonification Studio'', basado en el software Sonification Sandbox, permite al usuario producir gráficos auditivos y visuales a partir de ecuaciones y abrir conjuntos de datos csv (archivos de datos separados por comas), además de descargar el trabajo actual en diferentes formatos. Todas las herramientas mencionadas hasta el momento que sean comparables con la propuesta de esta tesis serán descriptas con mayor detalle en el Capítulo \ref{cap:estado_del_arte}, donde se detallará el Estado del Arte en el tema.

Es necesario destacar, que los recursos previamente descriptos han sido desarrollados intuitivamente por científicos espaciales que necesitan explorar los datos de una manera diferente, pero aún así son desarrollos centrados en los datos, no en el usuario. El resultado de un diseño con estas características, o sea, no centrado en los usuarios finales, es el diseño de prototipos bien intencionados, pero de baja fidelidad, que no logran la accesibilidad y la sostenibilidad. Los prototipos presentados no permiten al científico espacial reducir la incertidumbre impuesta por la arquitectura del sistema, lo que se relaciona con la necesidad de secuencias de comandos en tiempo real y el manejo intuitivo de las funcionalidades del prototipo. Estas herramientas no consideran las estrategias de adaptación de la audiencia objetivo, haciendo difícil la navegación en la GUI y extremadamente difícil la arquitectura de la información.

\subsection{Algunos comentarios sobre percepción sensorial}

Hasta el momento se ha comentado sobre accesibilidad y la realidad que viven las personas con discapacidad. Adicionalmente, se ha hablado sobre la técnica de sonorización y como esta podría ayudar a generar mayor inclusión y mejoras en el análisis de datos incluso para personas sin discapacidad. Sin embargo, si bien los fines de esta tesis no profundizan sobre las técnicas de percepción, es necesario describir un marco de referencia que permita comprender por que se propone la sonorización como complemento a la visualización.

Los humanos estamos continuamente experimentando, explorando y monitoreando información mediante señales sensoriales de múltiples modalidades. De acuerdo con \citet{massaroetal1993} y \citet{carson2007} se producen dos etapas entre estímulo y respuesta: los sentidos y la decisión. El estímulo es descifrado para producir un evento sensorial como salida. Esta salida es información que se usa para tomar una decisión que dará lugar a una respuesta. La respuesta es entonces la salida del proceso sensorial y no del estímulo. \citet{wandatesis2013} realizó experimentos en los cuales el sonido como un anexo a la visualización de datos, es visto como un proceso del sistema de entrada fisiológico. El sujeto interpreta el estímulo (sesgo de convicción) y elige cómo responder (sesgo de decisión). El ``sesgo de convicción'' tiene que ver con el proceso sensorial y el ``de decisión'', con el resultado \citep{ohare1991}. El análisis de datos de las ciencias espaciales involucra un proceso de percepción que determina el siguiente estado del mismo. Una percepción temprana de cualquier señal presente en los datos ampliará las probabilidades de encontrar eventos en los mismos.

\citet{baldassi2006} comprobó que el desorden de percepción visual conduce no solo hacia un incremento en los errores de juicio, sino también a un incremento en la fuerza de la señal percibida y la confianza de la decisión en pruebas erróneas. En estos estudios, se prueba y justifica una predicción directa: en un ambiente desordenado, estímulos erróneamente percibidos deben ser vistos a mayor intensidad de señal que cuando el objetivo se presenta en forma aislada. El mismo principio puede ser ampliado hacia decisiones mucho más complejas como la detección de señales incrustadas en ruido en datos astrofísicos. En esta ciencia, se analizan datos con carga cognitiva de ruido, que no solo afecta el juicio y confianza cuando se identifica la señal, sino que al mismo tiempo, provocan mecanismos de atención que inhiben la sensibilidad periférica \citep{schnepsetal2007}. Ha sido probado por las imágenes de resonancia magnética \citep{schwartzetal2005} que la carga de trabajo en la periferia del campo visual suprime la percepción en el centro y que la carga de trabajo en el centro, afecta la detección en la periferia. Esto sugiere que el centro y la periferia visual interfieren cuando se los carga con tareas de atención \citep{wegerinhoff2006,martinezcondeetal2006,posnetetal1994,ruccietal2007}. En el caso de la astrofísica, la señal puede estar en cualquier parte de los datos (periferia o centro). Por lo que, el complemento mediante otra modalidad sensorial a la visualización de los datos podría ser de ayuda en la toma de decisiones durante el análisis de dichos datos.

Sin lugar a dudas, las herramientas y prototipos presentados subrayan la necesidad que tienen los científicos espaciales de formas alternativas centradas en el usuario para analizar los datos de la ciencia espacial, utilizando la percepción multimodal para aumentar las detecciones en los datos e incluso para identificar si una señal es real o no. Esto debe verse como un esfuerzo por complementar la tecnología en constante desarrollo, con las habilidades humanas, y como un esfuerzo real para mostrar cómo las ciencias espaciales se toman en serio el permitir la diversidad de estilos de percepción, lo que también puede llevar a un incremento en el número de descubrimientos.

\section{Conclusiones}

En las secciones previas se describió la triste realidad de las personas con discapacidad, cómo se encuentran con barreras a la hora de capacitarse y profesionalizarse. Adicionalmente, se muestra como ha aumentado el uso de sonorización como complemento a la visualización de datos astronómicos y astrofísicos. Al comenzar a explorar los datos de forma multisensorial, se abre una puerta a las personas que han perdido alguna de las modalidades sensoriales para que puedan explorar la misma información que personas con todas las modalidades sensoriales sanas. Al final de la sección \ref{sect:enfoque_multisensorial}, se agregan algunos comentarios sobre la percepción sensorial, lo que refuerza la utilidad de la herramienta desarrollada tanto para personas con discapacidad como para las que no presentan una. 

En la presente investigación, y en base al marco de referencia teórico presentado, demostraremos el impacto del diseño de una nueva herramienta de análisis, centrada en el usuario desde el inicio. Para ello se recurrió a la posibilidad de interacción que brindan los grupos focales y la mejora en el diseño atendiendo a las recomendaciones del usuario final. Esta aproximación está en acuerdo con la propuesta para la eliminación de las barreras funcionales, que impiden una verdadera inclusión.

\chapter{Estado del arte}
\label{cap:estado_del_arte}

\section{Introducción}

El número de proyectos que utilizan sonorización para representar datos astrofísicos ha crecido enormemente durante los últimos 10 años. \citet{zanella2022}, luego de un workshop de sonorización realizado en agosto de 2021, elaboró un repositorio con los software existentes a diciembre de ese año, lo cual arroja un resultado de 98 proyectos desarrollados desde 1962, muchos de ellos discontinuados, con falta de documentación o sin evidencias de aplicaciones en ciencias. Casi el 80\% de los proyectos de sonorización han sido realizados entre 2011 y 2021. Hasta 2017, fecha en la cual se inició  el presente trabajo cuyo resultado o producto final es el sonoUno (un software de sonorización de datos astronómicos centrado en el usuario), solo el 50\% de los software incluidos en el mencionado repositorio habían iniciado su desarrollo.

Teniendo en cuenta lo descripto anteriormente, es importante destacar que en el estado del arte se describirán tanto software anteriores como contemporáneos al desarrollo de sonoUno. Programas anteriores, con aplicación en astronomía y astrofísica, fueron los que se utilizaron como referencia previa (xSonify, Sonification Sandbox y MathTrax) y con los cuales se trabajó el estudio normativo que se presentará en el Capítulo \ref{cap:analisis_normativo}. Software contemporáneos, que fueron diseñados en paralelo con la investigación de esta Tesis, tales como StarSound (con cuyos desarrolladores se mantiene intercambio), Astronify y Afterglow, entre otros, serán descriptos principalmente para establecer similitudes y diferencias existentes entre ellos y sonoUno. Se incluirá también una herramienta de sonorización que comenzó su desarrollo en 2021 llamada STRAUSS, resulta importante porque es un paquete de Python enfocado en la sonorización de datos en astronomía \citep{strauss2021}. Se debe recordar que en general y como mencionamos en el Capítulo \ref{cap:marco_teorico}, los programadores muy pocas veces producen recursos centrados en el usuario y, en muy pocas ocasiones, con fines que van más allá de la educación o difusión de disciplinas científicas.

Teniendo en cuenta que en esta disertación se realiza un diseño centrado en el usuario y enfocado en la inclusión de personas con diversidad funcional, el presente capítulo comienza con la descripción de las herramientas `lectores de pantalla'. Dichas aplicaciones son fundamentales para las personas ciegas o con baja visión que pretenden acceder al uso de otras herramientas de computación. Cabe agregar que éstas últimas generalmente no tienen en cuenta a los lectores de pantalla, no describen su uso en conjunto, no se analiza su usabilidad y por supuesto, no se estudia el impacto de la falta de integración entre los lectores de pantalla y los recursos informáticos. 

El presente capítulo, luego de presentar los lectores de pantalla, continua con la descripción de los software de sonorización pertinentes a la propuesta de esta tesis, desarrollados para ser instalados o funcionar en el escritorio de una computadora. Dichos programas son diferenciados en tres categorías: aquellos que aceptan archivos de datos genéricos (como ser tablas de datos en un formato específico) sin aplicación conocida en astronomía; aquellos que aceptan archivos de datos genéricos pero hay información de su aplicación en el campo de la astronomía o astrofísica (aquí se ubicaría el software desarrollado en esta tesis); desarrollos o scripts programados para un tipo de datos específico de las ciencias espaciales (estos suelen ser el caso de desarrollos relacionados con telescopios especiales o facilidades como el Gran Colisionador de Hadrones, que pueden producir datos relacionados con detección de partículas elementales, un tema que también se aborda en la presente tesis, derivado del diseño de sonoUno).

Por último y en una sección aparte, el Capítulo \ref{cap:estado_del_arte} describe algunas herramientas de sonorización que cuentan con una interfaz web. Aquí también se divide dichas interfaces entre las que se aplican a datos fuera de las ciencias del espacio o presentan otros fines (como por ejemplo uso de sonido para educación), y aquellas que tienen ejemplos de aplicación en astronomía o astrofísica.

\section{Lectores de pantalla}
\label{sect:lectores_pantalla}

Los lectores de pantalla son herramientas que se instalan en el sistema operativo como cualquier otro programa; su función es leer todo lo que aparece en la pantalla generando acceso al contenido para personas con problemas de visión. En principio, pueden acceder a las configuraciones de sistema y avisar (o informar) cuando se abre o cierra alguna aplicación, pero se busca que puedan describir también los programas y aplicaciones que se utilizan en las interfaces digitales. Esta última tarea es muy compleja, debido a que depende de la arquitectura de cada programa, muchas de ellas no permiten el acceso de segundas aplicaciones, negando así que el lector de pantalla pueda realizar su trabajo.

El objetivo principal de un lector de pantalla es ofrecer acceso y control sobre un sistema computacional, por ejemplo para personas con discapacidad visual \citep{evansBlenkhorn2008}. Para el objeto de esta tesis, se tendrán en cuenta los lectores de pantalla que describen la interfaz mediante sonido, no aquellos que utilizan el sistema braille. Los primeros son ampliamente utilizados, debido a que no necesitan conocimientos previos como sucede en el caso del lenguaje Braille, y no tienen mayores costos adicionales ya que se reproduce el sonido con el parlante de la computadora u otro dispositivo.

Una persona con discapacidad visual utiliza las interfaces digitales mediante teclados, por lo que el modo de navegación será con teclas que permitan navegar entre elementos y mediante atajos de teclado. Un lector de pantalla junto con las configuraciones de sistema, permiten setear todos estos parámetros para que la persona pueda seleccionar y ajustar el modo de navegación a sus necesidades. El funcionamiento de los lectores de pantalla se basa en describir el elemento de la interfaz que tiene foco (con esto referimos al elemento donde nos encontramos posicionados en la pantalla, por ejemplo cuando se está escribiendo en un editor de texto, el foco está en el cursor). Allí, el lector de pantalla no solo describe el elemento sino también su funcionalidad y opciones (por ejemplo, si se está en un reproductor de música y se tiene el foco en el botón Play/Pausa, el lector de pantalla indicará que se encuentra posicionado en un botón, que el mismo tiene la función de reproducir y poner pausa al sonido, y que el estado actual del botón es Pausa indicando que si se presiona se procederá a reproducir el sonido).

En la presente tesis se utilizarán lectores de pantalla disponibles para los sistemas operativos Windows, Ubuntu y MacOS, con la finalidad de evaluar la accesibilidad de los software de sonorización de datos existentes a 2017 (Sonification Sandbox, MathTrax y xSonify, durante el análisis descripto en el Capítulo \ref{cap:analisis_normativo}) y el programa que se desarrolló durante este trabajo. Para el sistema operativo Windows algunas de las opciones de lector de pantalla son el Narrador integrado con el sistema, NVDA (de uso gratuito), JAWS (pago) y Dolphin (pago). En el caso de Ubuntu se puede descargar el lector de pantalla Orca. Finalmente, MacOS tiene su propio lector de pantalla que viene instalado dentro de las características de accesibilidad del sistema operativo, llamado VoiceOver. Se utilizaron los lectores de pantalla NVDA, Orca y VoiceOver para realizar el testeo de los programas en cada uno de los sistema operativo (un ejemplo del uso de NVDA en windows puede visualizarse en el siguiente \href{https://youtu.be/OCnK_NkYXIU}{\underline{\textcolor{blue}{link}}}).

\section{Software independiente de sonorización de datos}
\label{sect:software_independientes}

Debido a que el número de trabajos de sonorización es extenso y no todos son pertinentes al marco de trabajo del presente tema, a continuación se describen: los desarrollos utilizados como base para esta tesis (xSonify, Sonification Sandbox y MathTrax); otros que presentan características similares pero se encuentran desactualizados y trabajos contemporáneos al desarrollo de la presente investigación. Concerniente a estos últimos, se incluirá aquellos que presentan alguna relación o incumbencia, ya sea porque buscan generar accesibilidad, se estableció algún contacto con los desarrolladores o buscan mejorar el análisis de datos en investigación científica.

\begin{figure}[!ht]
	\centering
	\includegraphics[width=1\textwidth]{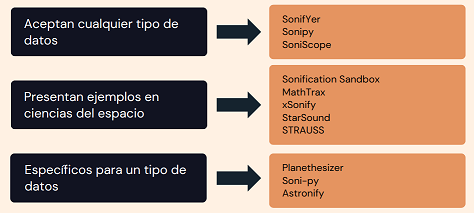}
	\caption{Cuadro que muestra las tres categorías adoptadas y los programas que fueron ubicados en cada una de ellas.}
	\label{fig:cuadro_soft_escritorio}
\end{figure}

La totalidad de los software de sonorización que se describen aquí fueron divididos en tres categorías (ver Figura \ref{fig:cuadro_soft_escritorio}), las primeras dos contienen en su mayoría programas con interfaz gráfica que aceptan un archivo de datos como entrada. Esto hace posible su uso en diferentes ciencias siempre y cuando se cumpla con el formato de entrada. La diferencia entre estas dos primeras categorías está dado por la aplicación o no de los programas en las ciencias del espacio.

Por otro lado, en la tercer categoría se describen tres herramientas que presentan objetivos específicos como sonorizar un conjunto de datos o los datos de un instrumento de las ciencias del espacio, los cuales son: Planethesizer, Soni-py y Astronify. La decisión de no incluir todos los desarrollos hechos para un conjunto de datos específico, se tomó en base a que el desarrollo planteado en esta disertación se basa en una plataforma que permita el ingreso de datos de diversa procedencia y realice un despliegue multimodal (no solo auditivo) de los mismos, pudiendo utilizarse en el ámbito profesional de investigación.

Cabe destacar que los desarrollos centrados en Arte no han sido incluidos en el presente capítulo y por lo tanto no forman parte del estado del arte actual. Esta tesis investiga la sonorización como producto que representa fielmente los datos de entrada, sin ningún tipo de modificación o manejo de los mismos. No se busca hacer un sonido agradable al oído o que produzca sentimientos, se busca poder hacer ciencia utilizando la sonorización como complemento a la visualización.

\subsection{Software que aceptan cualquier tipo de datos}

\subsubsection{SonifYer}

SonifYer es un software de sonorización que ha sido desarrollado para encefalografía por el Grupo de Investigación en Sonorización de la Universidad de las Artes de Berna. Si bien escapa de lo que es el análisis de datos en astronomía, se incluye aquí porque permite ingresar un set de datos independiente de su procedencia. Los ejes principales de su desarrollo fueron \citep{sonifyer2008}: usabilidad (indican que debe poder utilizarlo tanto un profesional como un principiante), facilidad en la instalación (comparan aquí con los sistemas `plug and play') y distribución (permitir portabilidad de las sonorizaciones).

Al ser un programa diseñado para ser usado en encefalografía, donde se tienen múltiples canales de datos que deben poder ser comparados entre sí, una potencialidad de este software es que permite importar diferentes archivos de datos y se pueden configurar hasta 32 canales. Esto es de interés en astronomía, donde para algunos tipos de datos se requieren múltiples gráficos y sonorizaciones, para poder compararlos entre sí.

\begin{figure}[!ht]
    \centering
    \includegraphics[width=1\textwidth]{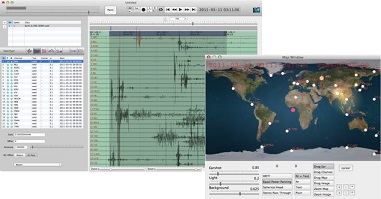}
    \caption{Captura de pantalla del programa SonifYer extraída de su página web \citep{sonifyerweb}.}
    \label{fig:sonifyer}
\end{figure}

Una de sus desventajas es que sólo ha sido diseñado para sistemas operativos MacOS, esto hace difícil su uso en lugares donde no se tiene disponibilidad de dicho sistema operativo. Además, presenta una interfaz gráfica compleja con múltiples ventanas y muchos elementos. En la Figura \ref{fig:sonifyer} se muestran la ventana principal a la izquierda y una ventana `mapa' a la derecha. La ventana principal despliega los datos, indicando una tabla con canales y su información a la izquierda y otro panel a la derecha con las gráficas por canal. El panel con las opciones de sonorización está en la parte superior. La ventana `mapa' indica la ubicación del dato seleccionado en la ventana principal, en el caso de la Figura \ref{fig:sonifyer} son datos que provienen de diferentes partes del mundo (si los datos fueran de encefalografía la ventana `mapa' mostraría una cabeza con la posición de cada electrodo). Cabe destacar aquí que las múltiples ventanas, así como un número alto de funcionalidades en un mismo despliego, son características difíciles de describir y utilizar con lectores de pantalla.

Lamentablemente, la última actualización del programa fue en el año 2013. En 2019 se anunció en el sitio web de este desarrollo que el software ya no es compatible con las nuevas versiones de MacOS, por lo que no puede ser utilizado. A principios de 2023, al intentar ingresar a la página web del desarrollo se evidenció que no está disponible actualmente.

\subsubsection{Sonipy}

\citet{worraletal2007} lo describe como un desarrollo de código abierto, modular, que busca integrar librerías y trabajos existentes o desarrollos propios cuando fuera necesario, con el objetivo de producir investigación en sonorización y despliegue auditivo. Los lineamientos del trabajo se basan en: 
\begin{itemize}
    \item Integrabilidad: que se pueda utilizar en diferentes ámbitos y ciencias; 
    \item Flexibilidad: busca integrar desarrollos existentes, no suplantarlos o superponerse a ellos; 
    \item Extensibilidad: en el caso particular de que ningún módulo responda a las necesidades del proyecto deberá ser desarrollado;
    \item Accesibilidad: donde hacen referencia a la accesibilidad a los módulos existentes que se integrarían en el desarrollo, no a que el desarrollo sea accesible para usuarios con diversidad funcional;
    \item Portabilidad: que pueda utilizarse en diferentes plataformas;
    \item Disponibilidad: hace referencia a que el software debe poder ser usado con las menores restricciones posibles;
    \item Durabilidad: se necesita compromiso de parte de la comunidad para que proyectos de estas características sobrevivan en el tiempo.
\end{itemize}

Este desarrollo cuenta con una publicación y una página web propia \citep{sonipyweb}, aunque no se encuentra en la documentación forma de descargarlo o dirección donde el código esté alojado. Otro detalle es que la última fecha de actualización del sitio web es 2009. Aún así se considera de valor ya que las bases que describen la propuesta son interesantes y extensas en comparación con otras contribuciones. Los autores expresan incluso la importancia de un desarrollo modular, y la inclusión de las librerías y desarrollos existentes dentro de un marco de trabajo más amplio.

\subsubsection{SoniScope}

SoniScope es un programa que comenzó su desarrollo recientemente, el trabajo que lo describe data de 2022 \citep{soniscope}. Esta herramienta produce la sonorización de gráficos de dispersión, la diferencia con otros recursos es la forma en que selecciona que datos sonorizar. Se basa en la referencia `\textit{auscultar con un estetoscopio}' para describir el elemento que utilizan para seleccionar el rango de datos que llaman `lente visual' (traducido del ingles `\textit{visual lens}'); el mismo es una forma geométrica (por ejemplo, circulo o cuadrado) que tiene su centro en el puntero y se puede cambiar de tamaño con una barra deslizante o con la ruedita del puntero. La finalidad es que sólo se sonoricen los datos que se encuentren dentro de esta forma geométrica, permitiendo que la lente visual se utilice como dispositivo de auscultación (comparándolo nuevamente con la metáfora del estetoscopio).

Es un método novedoso  que recién se comienza a desarrollar. No se encuentran evidencias que se esté teniendo en cuenta al usuario final en el desarrollo, o que se esté trabajando en un marco de diseño centrado en la inclusión y la accesibilidad.

\subsection{Software que aceptan cualquier tipo de datos y ejemplos en las ciencias del espacio}
\label{sect:desktop_astronomia}

\subsubsection{Sonification Sandbox}
\label{sect:desktop_astronomia_sandbox}

El programa Sonification Sandbox es una herramienta que ha sido desarrollada por el Laboratorio de Sonorización del Departamento de Psicología del Instituto de Tecnología de Georgia \citep{sandboxgeorgialab}. La herramienta permite importar los datos en formato de tabla (CSV), los grafica y produce la sonorización de los mismos.

\begin{figure}[p]
    \centering
    \begin{subfigure}[b]{0.49\textwidth}
         \centering
         \includegraphics[width=\textwidth]{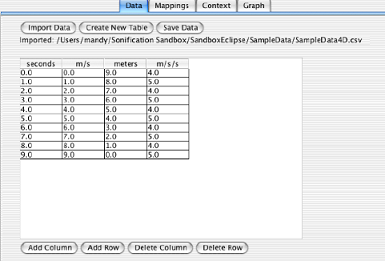}
         \caption{Pestaña datos}
         \label{fig:sandbox_datos}
     \end{subfigure}
     \hfill
    \begin{subfigure}[b]{0.49\textwidth}
         \centering
         \includegraphics[width=\textwidth]{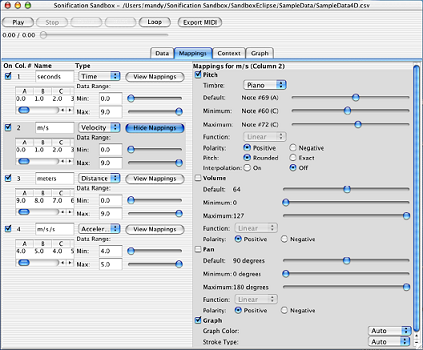}
         \caption{Pestaña mapeo}
         \label{fig:sandbox_mapeo}
     \end{subfigure}
     \hfill
    \begin{subfigure}[b]{0.49\textwidth}
         \centering
         \includegraphics[width=\textwidth]{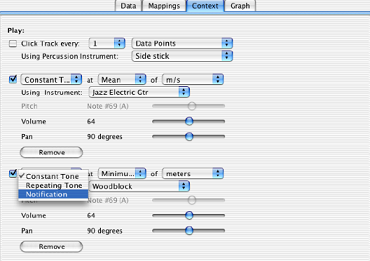}
         \caption{Pestaña contexto}
         \label{fig:sandbox_contex}
     \end{subfigure}
     \hfill
    \begin{subfigure}[b]{0.49\textwidth}
         \centering
         \includegraphics[width=\textwidth]{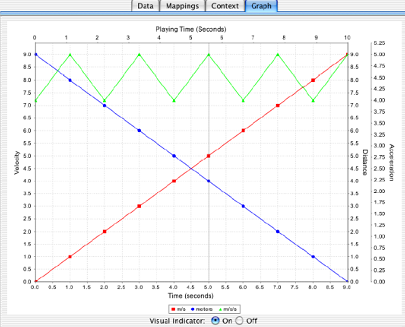}
         \caption{Pestaña gráfico}
         \label{fig:sandbox_graph}
     \end{subfigure}
    
    \caption{Imagen con cuatro capturas de pantalla del programa Sonification Sandbox que muestra cada una de las pestañas. Las imágenes fueron extraídas de \citet{sandboxgeorgialab}.}
    \label{fig:sandbox_2}
\end{figure}

El desarrollo está hecho en Java, es multiplataforma, pero dependiente de la versión de Java en la que está programado. Para la generación de sonido se utilizó el paquete JavaSound obteniendo una salida en formato MIDI. Una vez sonorizados los datos, el programa permite guardar la imagen, el audio, los datos por si se les ha aplicado algún cambio y un video con la sonorización. En cuanto a la configuración del sonido, el programa ofrece una sección con parámetros ajustables para que el usuario pueda elegir como producir la sonorización. En la Figura \ref{fig:sandbox_mapeo} se observan las configuraciones de sonido entre las cuales se puede seleccionar por ejemplo, si sonorizar con cambios de tono (notas) o de volumen: entre los parámetros de tono se puede seleccionar el instrumento y valores de nota mínima y máxima; mientras que entre los parámetros de volumen se puede seleccionar volumen mínimo y máximo, y polaridad.

Sonification Sandbox ha sido mantenido con diversas actualizaciones a través de los años, la primer publicación fue en 2003. En 2007, \citet{sandbox2007} describen la arquitectura y la interfaz gráfica de la versión de escritorio. Finalmente, en el año 2017 se anuncia la versión web del desarrollo, que continúa vigente en la actualidad y se describirá más adelante en la sección \ref{sect:web_astronomia_sandbox}.

\begin{figure}[p]
    \centering
    \includegraphics[width=1\textwidth]{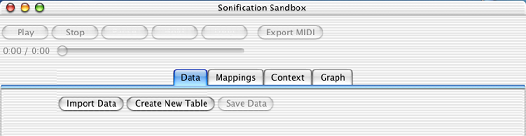}
    \caption{Captura de pantalla de la versión de escritorio del programa Sonification Sandbox. La imagen fue extraída de \citet{sandboxgeorgialab}.}
    \label{fig:sandbox_1}
\end{figure}

En cuanto a la versión de escritorio de Sonification Sandbox, que es la pertinente a esta sección, tuvo su última actualización en diciembre de 2014, siendo compatible con Java 7. En la Figura \ref{fig:sandbox_1} se muestra la interfaz gráfica de usuario (GUI) con la pestaña Datos seleccionada. La GUI presenta botones para controlar la reproducción en la parte superior y luego una sección con ventanas superpuestas que pueden seleccionarse utilizando las etiquetas con su nombre: datos (\textit{Data}), mapeo (\textit{Mappings}), contexto (\textit{Context}) y gráfico (\textit{Graph}).

La Figura \ref{fig:sandbox_2} muestra la vista de las cuatro pestañas que pueden seleccionarse. Siendo la Figura \ref{fig:sandbox_datos} la pestaña de datos luego de importar un archivo, allí se muestra: la dirección desde donde se obtuvo el archivo, los datos en formato tabla y en la parte inferior presenta botones que permiten agregar y eliminar filas y columnas.

La Figura \ref{fig:sandbox_mapeo} muestra la sección de mapeo o configuración de sonido con todas sus partes, al inicio solo se muestra la parte de la izquierda que permite seleccionar a qué columna de datos se le va a aplicar la configuración de sonido. En esa sección a la izquierda se cuenta con algunas configuraciones genéricas, cuando se selecciona alguna de las opciones (en la Figura \ref{fig:sandbox_mapeo} está seleccionada la columna 2) aparece a la derecha un panel con más configuraciones específicas que permiten, entre otras cosas, seleccionar el instrumento para la sonorización de dicha columna.

La Figura \ref{fig:sandbox_contex} expone la sección contexto, en dicho lugar se pueden configurar y agregar los sonidos que dan contexto a la sonorización. Estos sonidos son aquellos que indiquen el cruce de alguno de los ejes, si la reproducción es cíclica puede ser un sonido que indique el final de la gráfica antes de que vuelva a comenzar nuevamente desde el inicio, y así cualquier otro sonido que le de contexto a la reproducción del gráfico.

La Figura \ref{fig:sandbox_graph} presenta la pestaña gráfico, allí se muestran en el gráfico con diferente color cada una de las columnas seleccionadas. Además, durante la reproducción se despliega una barra horizontal que indica la posición del punto o puntos que están siendo sonorizados en un momento dado.

Sonification Sandbox es uno de los programas de sonorización que presenta aplicación en astronomía pero no restringe el tipo de datos de entrada, haciendo posible su uso en otras ciencias. Además, permite una amplia configuración de parámetros de sonido y gráfico que brindan al usuario la posibilidad de ajustar la salida a sus necesidades. Cabe mencionar que este programa no tiene registro de haber realizado un diseño centrado en el usuario, pero si menciona dentro de sus lineamientos la premisa de ser usable y la ventaja de acercar la astronomía a personas con discapacidad visual.

\subsubsection{MathTrax}

MathTrax es un programa desarrollado por la NASA para educación inicial y media, con el objetivo de graficar ecuaciones, simulaciones físicas y archivos de datos. Es interesante dentro de los límites de esta disertación porque realiza tanto el despliegue visual como sonoro de dichos datos, con la idea de generar accesibilidad e incluir a personas con discapacidad visual en el estudio de STEMS \citep{mathtraxweb}. Cabe destacar que la última actualización registrada es de diciembre de 2008.

\begin{figure}[p]
    \centering
    \includegraphics[width=1\textwidth]{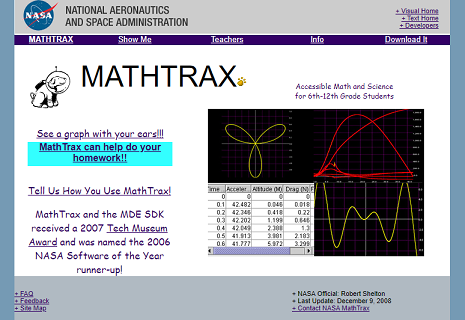}
    \caption{Captura de pantalla de la versión visual de la web de MathTrax.}
    \label{fig:mathtrax_web}
\end{figure}

La Figura \ref{fig:mathtrax_web} muestra una captura de pantalla de la página web del desarrollo, la misma tiene una versión visual y otra de solo texto para ser más accesible a lectores de pantalla. Allí se puede encontrar información sobre el programa, explicaciones sobre su uso y la opción de descarga. MathTrax contaba con una versión para Windows y otra para MacOS, actualmente la versión de Windows ya no está disponible. Lo interesante es que la versión para MacOS es el archivo `.jar', el cual puede ejecutarse en otros sistemas operativos teniendo la versión y archivos de java adecuados.

\begin{figure}[p]
    \centering
    \includegraphics[width=1\textwidth]{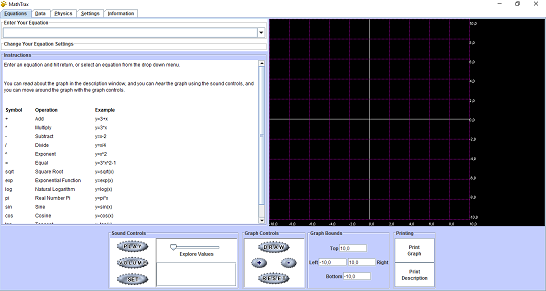}
    \caption{Captura de pantalla de la interfaz de MathTrax cuando recién se abre la aplicación.}
    \label{fig:mathtrax_gui}
\end{figure}

La Figura \ref{fig:mathtrax_gui} presenta una captura de pantalla de la interfaz gráfica del programa, con la pestaña ecuaciones seleccionada. La interfaz consta de cinco pestañas: 1) \textit{Ecuaciones}, que permite seleccionar entre ecuaciones con un menú desplegable o escribir la propia, luego de seleccionada la función, la misma es graficada en el panel de la derecha; 2) \textit{Datos}, que permite cargar datos en tabla formato txt, visualizarlos y sonorizarlos; 3) \textit{Física}, contiene dos pestañas internas, una de ellas permite mediante ecuaciones y parámetros realizar la trayectoria de una montaña rusa, la otra pestaña posibilita simular el lanzamiento de un cohete; 4) \textit{Configuraciones}, aquí se tiene la opción de configurar la parte del gráfico ya que las configuraciones de sonido se encuentran junto con las opciones de reproducción en la parte inferior de la ventana; 5) \textit{Información}, por último esta pestaña contiene información sobre el programa, sus desarrolladores e información de contacto.

\begin{figure}[p]
    \centering
    \includegraphics[width=1\textwidth]{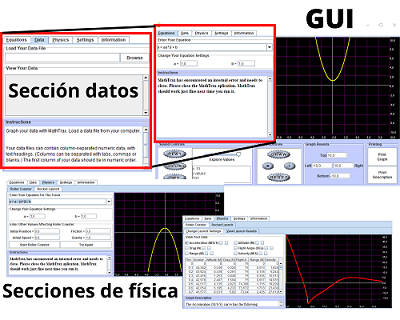}
    \caption{Combinación de imágenes que muestran las vistas de las primeras tres pestañas del programa Mathtrax.}
    \label{fig:mathtrax_comp}
\end{figure}

En la Figura \ref{fig:mathtrax_comp} se muestra un arreglo de imágenes que son capturas de pantalla de las diferentes pestañas descriptas. Por ejemplo las primeras dos pestañas comparten la sección gráfico (panel derecho) y opciones de reproducción (panel inferior), cambiando solamente el panel de la izquierda donde una permite el ingreso de la ecuación y la otra el ingreso de datos. En cuanto a la tercer sección, ésta sólo comparte la parte inferior con los botones de reproducción (en la Figura \ref{fig:mathtrax_comp} no fue incluido el panel inferior en las secciones de física por un tema de espacios), ya que la sección de gráfico está presente durante la simulación pero no así mientras se configuran las variables.

\begin{figure}[p]
    \centering
    \begin{subfigure}[b]{0.49\textwidth}
         \centering
         \includegraphics[width=\textwidth]{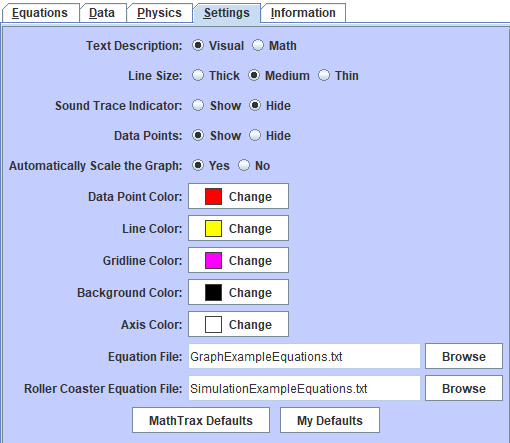}
         \caption{Pestaña configuraciones}
         \label{fig:mathtrax_set}
     \end{subfigure}
     \hfill
    \begin{subfigure}[b]{0.4\textwidth}
         \centering
         \includegraphics[width=\textwidth]{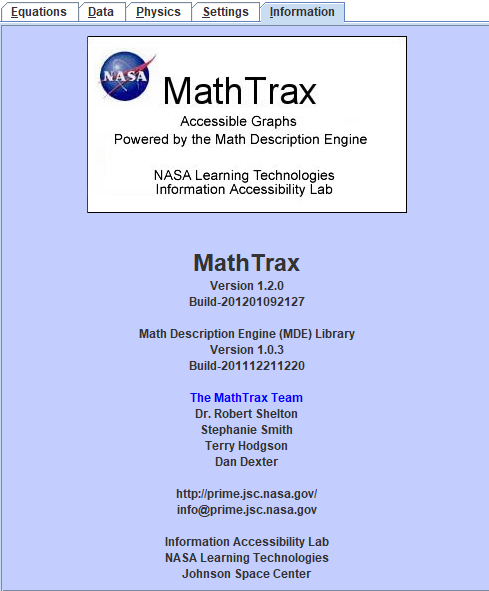}
         \caption{Pestaña información}
         \label{fig:mathtrax_info}
     \end{subfigure}
    \caption{Captura de pantalla de las últimas dos pestañas del programa MathTrax.}
    \label{fig:mathtrax_3y4}
\end{figure}

La Figura \ref{fig:mathtrax_3y4} contiene las captura de pantalla de las dos últimas pestañas del programa, la pestaña que permite la configuración del gráfico (ver Figura \ref{fig:mathtrax_set}) y la pestaña que presenta la información sobre el programa (ver Figura \ref{fig:mathtrax_info}). En cuanto a las configuraciones, dicha pestaña solo contiene las configuraciones del gráfico, como se mencionó anteriormente, la configuración de sonido se encuentra ubicada en el botón `Set' debajo de los botones `Play' y `Volume'. Por su lado, la pestaña información enuncia la versión del programa, todo lo relacionado con sus autores y filiación, junto con la dirección web y mail de contacto.

Si bien Mathtrax fue concebido para hacer accesible el estudio de STEMs para estudiantes de grado, la sección que permite ingresar datos en formato tabla hace posible su uso con datos astronómicos o astrofísicos que cumplan con ese formato. Esta característica, junto con que es un desarrollo hecho por la NASA, posibilitó incluirlo en la categoría de programas con ejemplos en ciencias del espacio. El diseño de la interfaz es similar a Sonification Sandbox, pero es destacable que MathTrax cuenta con atajos de teclado para cambiar entre pestañas, dado que por lo general estas pestañas no son accesibles a la navegación por teclado y el lector de pantalla.

\subsubsection{xSonify}

El software xSonify \citep{candeyetal2006} fue desarrollado inicialmente por Anton Schertenleib en lenguaje Java; Robert Candey y Wanda Díaz-Merced continuaron mejorando la herramienta. Este software busca brindar acceso al análisis de datos de física espacial para personas con y sin discapacidades visuales. xSonify es una aplicación multiplataforma de código abierto que genera un despliegue auditivo y un gráfico con un archivo de datos de dos columnas importado por el usuario. En este programa se puede configurar el tono, volumen y ritmo del despliegue sonoro. El usuario puede exportar el archivo MIDI con el resultado de la sonorización.

\begin{figure}[p]
    \centering
    \includegraphics[width=1\textwidth]{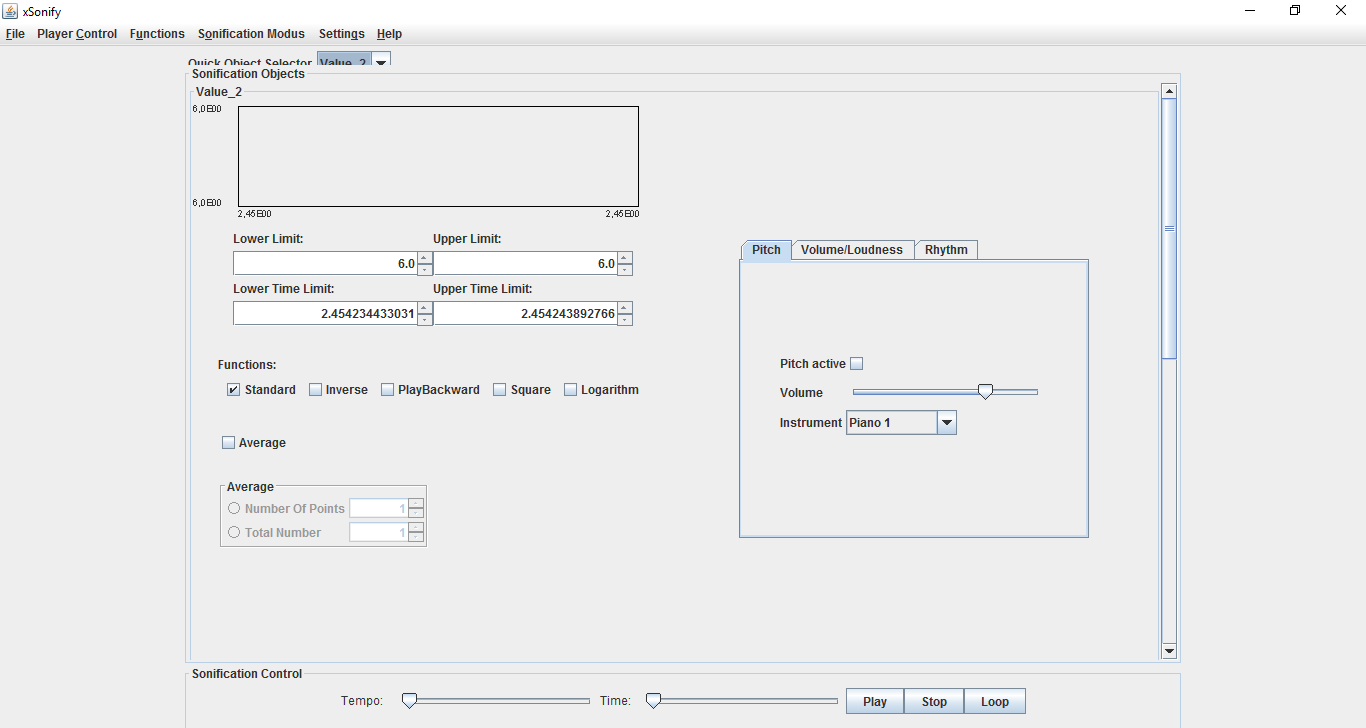}
    \caption{Captura de pantalla del programa xSonify al iniciarse.}
    \label{fig:xsonify}
\end{figure}

En la siguiente actualización, cerca del año 2011 \citep{wandaetal2011}, algunas de las mejoras que se incorporaron permitieron sonorizar datos de más de dos columnas con un despliego visual por columna de datos a sonorizar. Esto se logra con datos donde todas las columnas son dependientes de la primera, luego utiliza un despliegue como el observado en el interior de la ventana de la Figura \ref{fig:xsonify} por cada columna a sonorizar. En dicha ventana se puede seleccionar los parámetros de sonido a utilizar con cada columna y cuales columnas sonorizar en la reproducción. Otro de los aspectos importante de este software es la posibilidad de analizar grandes conjuntos de datos, algo muy importante en el área de investigación de las ciencias del espacio.

Una particularidad que diferencia a este proyecto de los demás es que fue uno de los primeros en incluir a una astrónoma con discapacidad visual dentro del equipo de desarrollo. xSonify tiene un lector de pantalla propio que puede activarse o desactivarse, sin anular las características propias del sistema (esto quiere decir que con el lector de pantalla del xSonify desactivado puede utilizarse el lector de pantalla disponible en el sistema operativo). Esto último tiene relación con dejar que el usuario decida que característica es la que mejor se ajusta a sus necesidades y una consecuencia clara de incluir especialistas en accesibilidad en el desarrollo. Es preciso mencionar que la última actualización que se registra es de 2014.

\subsubsection{StarSound}

\begin{figure}[p]
    \centering
    \includegraphics[width=1\textwidth]{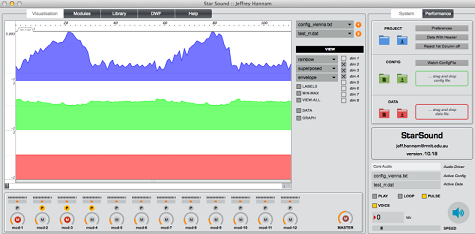}
    \caption{Captura de pantalla del programa StarSound extraída de un articulo publicado en 2022 \citep{starsound2022}}
    \label{fig:starsound}
\end{figure}

Es un software de sonorización para datos astronómicos y astrofísicos, tales como curvas de luz y espectrogramas, fue desarrollado por Jeffrey Cooke, Jeffrey Hannan y Gary Foran (astrónomo con discapacidad visual). Acepta como entrada datos formato tabla de dos columnas (1D), que pueden ser graficados en un eje de coordenadas x-y. Contiene una interfaz gráfica de usuario que puede verse en la Figura \ref{fig:starsound}, y un sintetizador de sonido con una interfaz que permite manipular sus parámetros (disponible en otra de las pestañas, la cual se puede seleccionar en la parte superior de la interfaz mostrada en la Figura \ref{fig:starsound}).

Este desarrollo ha tenido en cuenta características de accesibilidad durante su diseño, los parámetros del sintetizador de sonido pueden ser configurados en un archivo de texto que puede abrirse y modificarse utilizando un lector de pantalla. Además, comenta la existencia de atajos de teclado, mediante los cuales puede ser utilizada su interfaz gráfica. Sin embargo, esta interfaz presenta un gran número de elementos, los cuales se relacionan con una gran carga de memoria para el usuario, haciendo difícil su utilización por personas que se están iniciando en las ciencias de sonorización o astronomía.

Este programa es contemporáneo al desarrollo de esta tesis, con bases y lineamientos similares, e incluye a un investigador con discapacidad visual dentro de su proyecto. Uno de los primeros trabajos que describe el desarrollo de este software \citep{cookeetal2017} tuvo colaboración de miembros del equipo de trabajo vinculado con esta tesis. Sin embargo, la diferencia principal reside en dónde puso el foco de atención cada desarrollo, siendo StarSound un programa que se enfocó en las técnicas de sonorización, con una interfaz que despliega herramientas similares a las que contiene un sintetizador de sonido; mientras que la presente contribución está más orientada al diseño centrado en el usuario, teniendo en cuenta una investigación que involucra intercambio con usuarios finales y grupos de investigación. No obstante, se mantiene un canal de comunicación activo con las personas involucradas en el desarrollo de StarSound.

\subsubsection{STRAUSS}

Es un desarrollo reciente, presenta sus primeras actualizaciones en GitHub hace dos años \citep{straussGithub}. Es un paquete de Python, desarrollado con la finalidad de mejorar el actual despliegue visual de datos y la accesibilidad. Presenta la posibilidad de ser utilizado por personas sin conocimiento en sonorización o programación, mediante ejemplos predefinidos y utilizando las configuraciones por defecto. Sin embargo, presenta documentación que permite su uso en un nivel más avanzado permitiendo modificar los parámetros de sonorización.

Una aplicación directa de este desarrollo es en el proyecto ``Audio Universe'' en el cual recientemente han publicado un tour por el sistema solar, para el cual han utilizado la herramienta STRAUSS (para más información pueden acceder al siguiente \href{https://www.audiouniverse.org/}{\underline{\textcolor{blue}{link}}}). \citet{strauss2021} explica como se desarrolló dicho tour utilizando diferentes instrumentos para cada planeta, junto con la configuración de sonido para representar la distancia y posición de cada objeto.

\subsection{Software desarrollados específicamente para un tipo de datos}

\subsubsection{Planethesizer}

Este desarrollo fue creado para la sonorización de exoplanetas a partir de las variaciones que presenta la curva de luz de la estrella que orbitan \citep{riber2018}. En este caso utilizan también sintetizadores de sonido que pueden ser configurados por el usuario, y un complejo sistema diseñado para realizar un despliegue visual bien sincronizado con el despliegue sonoro. \citet{riber2018} describe en su trabajo que han logrado sonorizar dos sistemas de exoplanetas además del Trappist-1, que es con el cual prueban la funcionalidad del programa.

No se han encontrado publicaciones adicionales sobre este programa, y la página web a la cual redirecciona para descargas tiene fecha de subida del archivo en 2017. Este software tiene como principios el despliegue multimodal de datos, que el usuario pueda controlar el sonido, y el desarrollo de una interfaz intuitiva, entre otras. 

Los principios que promueven los desarrollos de herramientas para la sonorización en ciencias del espacio son muy similares, aunque hasta el momento ninguno cuenta con publicaciones relacionadas a técnicas cualitativas para evaluar la interfaz o estudios de percepción, tampoco presentan planificación para llevarlas a cabo.

\subsubsection{Soni-py}

El desarrollo Soni-py que se describe en esta sección no se debe confundir con el previamente descripto Sonipy, son independientes y no presentan relación entre sí, así lo describen \citet{soni-pyGithub} en su repositorio de GitHub. En este caso, el desarrollo es un algoritmo escrito en su mayoría en Python, que permite sonorizar gráficos de dispersión.

Soni-py no cuenta con interfaz gráfica, se puede utilizar directamente desde el entorno de Python importando la librería. Se debe indicar el arreglo de datos a sonorizar (x e y), los parámetros de frecuencia (ya que la frecuencia mínima y máxima son seteadas por el usuario) y tiempo (cuenta con una duración de dos segundos por defecto pero puede ser modificado); luego de ejecutado el comando, puede salvarse el archivo de sonido en wav. \citet{soni-py2021} exponen su uso para sonorizar curvas de luz de supernova.

\subsubsection{Astronify}

Este trabajo se basa en sonorizar datos de observaciones hechas por telescopios, particularmente este desarrollo ha sido llevado a cabo por \citet{astronify} operado por AURA, el tercer satélite del Sistema de Observación Terrestre de la NASA.

Astronify es un paquete que está en activo desarrollo en la actualidad, sonoriza datos de series de tiempo, específicamente curvas de luz. Es una librería escrita en Python que puede importarse en este entorno de desarrollo y utilizarse para realizar la sonorización. Su página web cuenta con el link a GitHub, videos que describen y muestran la sonorización, tutoriales, material multimedia y un juego. Éste último tiene dos niveles y presenta diferentes características de una curva de luz.

Este es el primer programa, de los descriptos en esta tesis, que cuenta con una web descriptiva específica del proyecto, donde cuenta con herramientas como material audio-visual y un juego. Dicho juego podría ser comparable con una actividad de entrenamiento, que sirva para instruir a la comunidad en el análisis multimodal de curvas de luz. Sin embargo, no se encuentra documentación que indique si realmente esta es la intención de este proyecto y si está realmente tomando registro de las contribuciones de los usuarios que responden a dicho juego.

\section{Software con interfaz web para sonorización de datos}
\label{sect:sorftware_web}

Durante los últimos 5 años, producto de las dificultades que presentan los software independientes en cuanto a su instalación en sus versiones de escritorio, se comenzó a hablar sobre desarrollos de sonorización que funcionan en plataformas web. En un inicio, se encontraban sonorizaciones ya generadas de datos específicos. Sin embargo, en la actualidad se encuentran plataformas que permiten cierta interacción por parte del usuario, algunas sólo permiten modificaciones de parámetros y otras incluso el ingreso de datos.

\begin{figure}[!ht]
	\centering
	\includegraphics[width=1\textwidth]{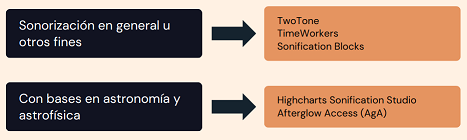}
	\caption{Cuadro que muestra las dos categorías adoptadas y las aplicaciones que fueron ubicadas en cada una de ellas.}
	\label{fig:cuadro_soft_web}
\end{figure}

A continuación se describen algunas de estas plataformas, haciendo la distinción entre las que tienen base o surgieron desde el ámbito de la astronomía y aquellas que no, o que cuentan con una finalidad diferente al análisis de datos (ver Figura \ref{fig:cuadro_soft_web}).

\subsection{Interfaz web para sonorización de datos en general u otros fines}

\subsubsection{TwoTone}

\begin{figure}[!ht]
    \centering
    \begin{subfigure}[b]{0.7\textwidth}
         \centering
         \includegraphics[width=\textwidth]{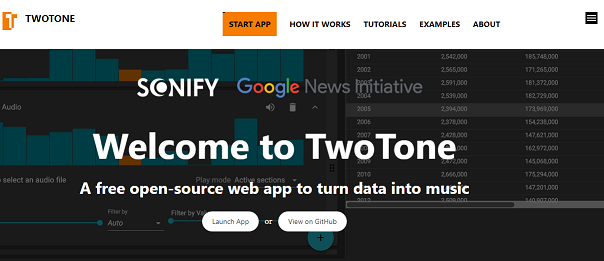}
         \caption{Página inicial con información general}
         \label{fig:twotoneinit}
     \end{subfigure}
     \hfill
    \begin{subfigure}[b]{0.7\textwidth}
         \centering
         \includegraphics[width=\textwidth]{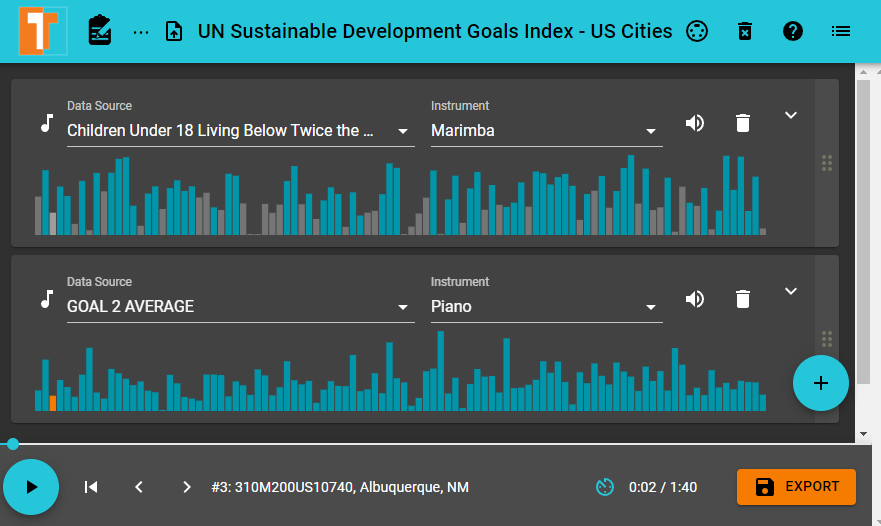}
         \caption{Aplicación web de sonorización}
         \label{fig:twotonegui}
     \end{subfigure}
    \caption{Capturas de pantalla de la página web inicial y de la aplicación web del software TwoTone.}
    \label{fig:twotone}
\end{figure}

Este desarrollo web busca hacer accesibles datos civiles (como por ejemplo producción de miel en Estados Unidos, o disminución en la abundancia de insectos en Dinamarca durante 22 años) a través de sonorización. Actualmente, el desarrollo es un proyecto de la empresa Sonify (Audio - Datos - Tecnologías emergentes) y cuenta con el soporte de `Google News Initiative' \citep{twotoneweb}. En la Figura \ref{fig:twotoneinit} se presenta la página web inicial del proyecto con un mensaje de bienvenida, un botón que da acceso a la app y un menú en la parte superior que da acceso a información y tutoriales.

La interfaz web cuenta con pocos elementos, lo que en cierto modo facilitaría su navegación (ver Figura \ref{fig:twotonegui}). Al inicio presenta un tutorial, donde van surgiendo ventanas que explican las secciones principales, tales como la relacionada con reproducción, la carga de datos, como así también las que permiten agregar más de un conjunto de datos, la parte de configuraciones y cómo guardar el sonido. Una vez terminado el tutorial, o salteándolo con el botón `Skip', se puede abrir un conjunto de datos, precargados o propios, para proceder con la sonorización. TwoTone no genera un gráfico con los datos, solo muestra un gráfico de barra  representativo de la nota MIDI que corresponde a cada punto del arreglo de datos a reproducir.

Algo importante a destacar, que comenta la empresa en la página web, es que actualmente tienen un proyecto en curso denominado `Narración basada en datos: Whichita' (traducido del ingles `Data-Driven Storytelling: Whichita'), dicha iniciativa tiene un año de antigüedad y trabajan con periodistas y personas con discapacidad visual en la búsqueda de nuevos enfoques para la comunicación de datos a través de sonido.

Este tipo de proyectos donde se involucra a las personas en el desarrollo de nuevas herramientas para mejorar la comunicación y el análisis tienen un gran potencial. Generalmente, surgen soluciones más centradas en el usuario, y por ende, más usables y eficientes. El caso de esta herramienta, desde el punto de vista de los lineamientos del presente trabajo, no permite un análisis de datos y tampoco un complemento entre el despliegue visual y auditivo. Aún así se considera de interés por su trabajo con personas con discapacidad, su actualidad y por el uso de sonorización para transmitir información.

\subsubsection{TimeWorkers} 

Es un entorno web que permite ingresar un archivo de datos, graficarlo y sonorizarlo \citep{chafe2019}. Está desarrollado en Java y se plantea como una herramienta de uso gratuito y código abierto. \citet{chafe2019} menciona también que descargando las herramientas, puede utilizarse de forma local, sin necesidad de tener una conexión a internet para producir la sonorización.

No cuenta con una interfaz gráfica que permita interactuar con los datos y los parámetros de configuración. Solo cuenta con un espacio para cargar los datos, que luego se convierte en el espacio donde se muestra el gráfico, y los botones de reproducción (reproducir y detener). Actualmente, la \href{https://ccrma.stanford.edu/~cc/sonify/}{\underline{\textcolor{blue}{página web}}} redirecciona a un espacio con información sobre sonorizaciones específicas, actividades de workshop y un link donde se da acceso al instructivo, el código y una interfaz que permite ejecutar código en un entorno web (\href{https://github.com/ccrma/webchuck}{\underline{\textcolor{blue}{WebChucK IDE}}}).

\subsubsection{Sonification Blocks}

\citet{sonifblocks2017} definen a Sonification Block como una herramienta para enseñar programación a estudiantes de secundaria, con la motivación de aprender a programar mientras están generando música. En este caso se utiliza la sonorización para motivar a estudiantes en el aprendizaje. La interfaz de Sonification Block es bastante simple y la técnica que utilizan es a través de bloques, análogo a la herramienta `Scratch' desarrollada por el Instituto Tecnológico de Massachusetts (MIT) para enseñar programación a niños de primaria.

\begin{figure}[p]
    \centering
    \includegraphics[width=1\textwidth]{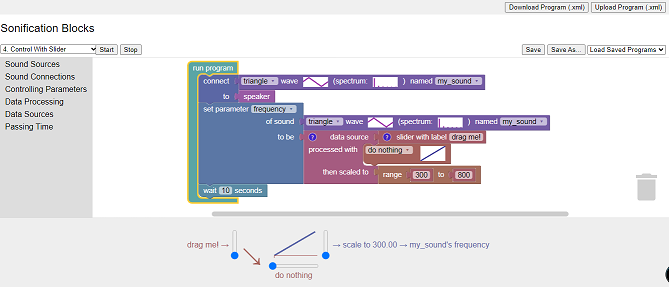}
    \caption{Captura de pantalla de la interfaz web de Sonification Blocks con un programa de ejemplo llamado: `4. Control with Slider'}
    \label{fig:sonifblocks}
\end{figure}

La Figura \ref{fig:sonifblocks} muestra la interfaz web de este programa, en la parte superior derecha se puede cargar o descargar el proyecto en formato `xml', lo que ofrece portabilidad y comodidad para que estudiantes puedan trabajar tanto en casa como en la escuela. Otra alternativa que ofrece es guardar el programa en el buscador que se está utilizando, lo cual se puede hacer con los botones ubicados debajo de los mencionados anteriormente. Dirigiendo la atención a los bloques de programación, a la izquierda se encuentra debajo del título un menú desplegable con ejemplos que muestran las diferentes características y opciones que ofrece la herramienta, en la Figura \ref{fig:sonifblocks} se muestra el ejemplo `4. Control with Slider' que muestra un programa que reproduce un sonido, despliega la curva en la parte inferior y tres sliders que permiten modificar parámetros de dicho sonido. Para reproducir o detener el sonido se utilizan los botones que están al lado del menú desplegable de ejemplos. Por último, cada ítem que se encuentra a la izquierda, debajo del menú desplegable de ejemplos, muestra bloques divididos en esas categorías que permiten programar un sonido desde cero.

Es una herramienta interesante y muestra la utilidad que tiene la sonorización para diferentes ámbitos del conocimiento. El despliegue multisensorial permite transmitir y adquirir el conocimiento utilizando diferentes vías de comunicación (vista y oído en el caso particular de este trabajo). Sonification Blocks muestra como un despliegue multisensorial puede atraer la atención de estudiantes para desarrollar capacidades que necesitan para su buen desempeño escolar.

\subsection{Software que tienen sus bases en astronomía y astrofísica}
\label{sect:web_astronomia}

\subsubsection{Highcharts Sonification Studio}
\label{sect:web_astronomia_sandbox}

La versión web de Sonification Sandbox, llamada `Highcharts Sonification Studio' \citep{sandboxwebtool} fue desarrollada con cuatro objetivos como guía \citep{websandbox2017}: (1) maximizar la accesibilidad; (2) ser una herramienta fácil de utilizar para principiantes sin reducir su alta performance; (3) maximizar la portabilidad; y (4) maximizar su utilidad. Se debe recordar que este programa tiene una versión de escritorio desarrollada previamente y descripta en la sección \ref{sect:desktop_astronomia_sandbox}.

\begin{figure}[p]
    \centering
    \includegraphics[width=1\textwidth]{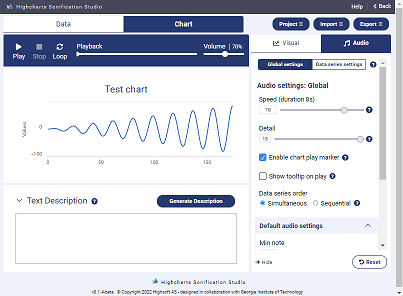}
    \caption{Captura de pantalla de la versión web de Sonification Sandbox. Se muestra la pestaña `Chart' que muestra el gráfico y las opciones de reproducción.}
    \label{fig:sandbox_web}
\end{figure}

La Figura \ref{fig:sandbox_web} muestra la interfaz web, que presenta funcionalidades similares a la versión de escritorio. Cuenta con dos pestañas, una para los datos y la otra contiene todos los elementos para la visualización y sonorización. En esta segunda pestaña se encuentran los botones de reproducción, el gráfico y un panel a la derecha con dos pestañas más para poder alternar entre las configuraciones de sonido y de gráfico.

Es notable que este proyecto es el único que ha mantenido actualizaciones por casi 20 años. Si bien la aplicación de escritorio no ha tenido nuevas actualizaciones, los esfuerzos se han concentrado en el despliegue web. Aún así, es uno de los desarrollos comparable con el trabajo que se ha realizado durante esta tesis.

\subsubsection{Afterglow Access (AgA)}

AgA fue desarrollado por Skynet en la Universidad de Carolina del Norte – Chapel Hill como parte del proyecto STEM+C de la Fundación Nacional de Ciencias (NSF) (NSF-1640131), investigación que respalda el compromiso multisensorial de estudiantes con discapacidad visual y videntes para promover el aprendizaje integrado de astronomía e informática (también conocido en inglés como: Innovators Developing Accessible Tools for Astronomy o IDATA) \citep{afterglowweb}. El objetivo principal de este desarrollo es el análisis de imágenes astronómicas mediante sonorización. La herramienta fue desarrollada por un equipo multidisciplinario (estudiantes de secundaria, de grado, astrónomos, ingenieros de software, investigadores en educación, entre otros) e incluyo a personas con y sin discapacidad visual. Sus primeras publicaciones fueron posteriores al inicio de esta tesis, entre 2018 y 2019.

\begin{figure}[!ht]
    \centering
    \includegraphics[width=1\textwidth]{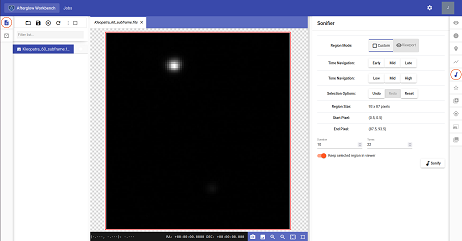}
    \caption{Captura de pantalla de la interfaz web de Afterglow Access}
    \label{fig:afterglow}
\end{figure}

La Figura \ref{fig:afterglow} muestra la interfaz web de la herramienta, en la parte superior izquierda de la ventana permite cargar archivos de imágenes en formato `fits', guardar archivos que estén en el programa, cerrar archivos abiertos y refrescar la lista. En la parte central de la ventana es donde se muestran las imágenes, en la parte inferior se muestra información y opciones de configuración sobre la imagen desplegada. Luego, a la derecha, hay un menú con iconos que contienen las funcionalidades del programa. En dicho menú se encuentra la configuración de imagen donde se muestra también el histograma, la información del archivo de datos, la opción para añadir marcadores, la opción para sonorizar y otras propias del análisis de imágenes astronómicas.

Este desarrollo es uno de los pocos que cuenta con estudios e intercambio con usuarios desde el inicio del trabajo. \citet{afterglow2020} describe el proceso realizado para lograr una herramienta centrada en el usuario. Entre las técnicas utilizadas estuvieron modelos de roles, ejercicios con preguntas como `¿Qué?-Si' en búsqueda de un flujo creativo de ideas, contacto directo o indirecto a través de GitHub entre usuarios, diseñadores y desarrolladores, entre otras. Esto marca un antecedente claro y publicado, en el campo de la astronomía, de la necesidad e importancia del diseño centrado en el usuario para poder crear herramientas útiles, usables y eficientes para la sonorización  de datos.  

\section{Conclusiones}

Teniendo en cuenta los programas presentados en este capítulo, se evidencia que aún cuando existe un número considerable de herramientas que producen sonorización, muy pocas tienen una interfaz gráfica que le permita al usuario controlar el proceso de producción de sonido y a la vez la visualización de datos. De las que cumplen con tal requisito y presentan aplicación en el campo de la astronomía y astrofísica, solo cuatro (contando a sonoUno, el desarrollo de esta Tesis) han incluido personas con discapacidad visual en el diseño. Afinando un poco más el detalle, sólo una herramienta además de sonoUno ha realizado intercambio con usuarios durante su desarrollo y solo cuenta con una versión web.

Si ahora tenemos en cuenta que la presente investigación se inició en 2017, el único programa que contaba con un integrante con discapacidad visual era xSonify. Adicionalmente, los únicos programas de sonorización que permitían al usuario manipulación de parámetros de configuración y contaban con aplicación en las ciencias del espacio era Sonification Sandbox y MathTrax. Es por ello que solo se hizo hincapié en el uso de lectores de pantalla inicialmente en estos tres programas, descubriendo que la mayoría de los elementos no son detectados por los lectores de pantalla de acceso gratuito (NVDA, Orca y Voice Over). Es preciso también destacar que, para dicha fecha, ningún proyecto mencionaba la necesidad de incluir al usuario final en el desarrollo, ni la importancia de las técnicas de diseño centrado en el usuario. Tampoco se enfocaban en el estudio de normas que permitieran la comparación entre las distintas propuestas en lo que hace a usabilidad de estas herramientas, todas ellas cuestiones que se consideraron centrales en el desarrollo de esta investigación.

\chapter{Análisis normativo para el primer diseño de interfaz}
\label{cap:analisis_normativo}

\section{Introducción}

En este capítulo se desarrollará el análisis normativo basado en la norma ISO 9241-171:2008 titulada `Guías en accesibilidad de software' y la norma WCAG 2.0 titulada `Guías de accesibilidad para el contenido web', aplicado a las diferentes herramientas para sonorización y acceso a los datos relacionados con astronomía y astrofísica. Como primer paso en cada análisis se describirá la metodología, el análisis realizado, los resultados obtenidos y se cerrará con las conclusiones parciales de cada investigación (sección \ref{sect:iso_seccioncompleta} y sección \ref{sect:w3c_seccioncompleta}).

En cuanto al marco de referencia y la justificación de la elección de cada una de las normas se elaboró una sección previa a las que describen el análisis (sección \ref{sect:marco_teorico_normativas}). En la misma se detallan normas existentes relacionadas al desarrollo de software y páginas web con foco en la accesibilidad. Luego, para cerrar este capítulo se describen dos consecuencias inmediatas de estos análisis, que son una serie de recomendaciones publicadas y el diseño de la interfaz gráfica de usuario del software sonoUno.

Este capítulo evidencia la importancia que tienen las normas, las limitaciones que presentan y el valor que tiene realizar este análisis previo al desarrollo de software. Las limitaciones de las normas, deben entenderse desde el punto de vista que es un texto que refleja una realidad determinada y sintetizada. Esto último se desprende de comprender que la discapacidad y la realidad que vive una persona con diversidad funcional, escapa a lo que se puede expresar en un texto. Es por ello que estas normas deben tomarse como indica su título, como `guías', y no como una `receta'. Lo expresado refuerza la necesidad que enfatiza esta tesis en todo momento, se debe incluir a las personas con diversidad funcional en el desarrollo de cualquier herramienta desde el inicio.

\section{Normativas ISO y W3C}
\label{sect:marco_teorico_normativas}

Resulta necesario describir, aún de forma resumida, las normativas ISO (``Internacional Organization for Standardization'') y W3C (``World Wide Web Consortium'') existentes referentes al desarrollo de software con foco en la accesibilidad. Se pretende dar un marco de referencia, justificando, por un lado, la elección de la norma ISO 9241-171:2008 para el análisis de software accesible que se utilizará en un análisis posterior (detallado en la sección \ref{sect:iso_seccioncompleta}), y además, la elección de la norma WCAG 2.0 para el análisis de accesibilidad de bases de datos astronómicas (detallado en la sección \ref{sect:w3c_seccioncompleta}).

\subsection{Normas para el desarrollo de herramientas y sistemas}
\label{sect:normas_iso}

La accesibilidad busca que tanto productos, como servicios y facilidades sean factibles de ser usadas por personas con discapacidad. La norma ISO 13407 ``Proceso de diseño centrado en el usuario para sistemas interactivos (traducido del inglés: \textit{Human-centred design processes for interactive systems})'', actualmente ha sido revisada por la norma \textbf{ISO 9241-210:2019} ``Ergonomía de las interacciones humano-sistema - Parte 210: Proceso de diseño centrado en el usuario para sistemas interactivos (traducido del inglés: \textit{Ergonomics of human-system interaction - Part 210: Human-centred design for interactive systems})''. Según esta norma el término ``usabilidad'' es definido como un producto, servicio o instalación que puede ser utilizado con efectividad, eficiencia y satisfacción por usuarios específicos en un contexto de uso dado. Adicionalmente a lo expuesto por la norma, existe un documento que redacta una convención para los derechos de las personas con discapacidad \citep{un2021file} que ha sido firmado por una gran cantidad de países (entre ellos Argentina) para asegurar igualdad de condiciones y trato para personas con discapacidad con respecto a personas que no presentan o documentan discapacidad.
 
Existen diferentes normas, además de la mencionada, que describen partes del proceso para diseños usables y accesibles. A continuación se mencionan algunas de ellas:
\begin{itemize}
    \item La norma \textbf{ISO 9241-11:2018} ``Ergonomía de la interacción hombre-sistema - Parte 11: Usabilidad: Definiciones y conceptos'' (traducido del inglés: \textit{Ergonomics of human-system interaction — Part 11: Usability: Definitions and concepts}) provee un marco de referencia para entender el término usabilidad, aplicándolo a diferentes entornos y sistemas. 
    \item La norma \textbf{ISO 11064} ``Diseño ergonómico de centros de control'' (traducido del inglés: \textit{Ergonomic design of control centres}) provee un marco de trabajo para la creación de centros de control ergonómicos, con el objetivo de minimizar el potencial de errores humanos. 
    \item La norma \textbf{ISO 14915} ``Ergonomía de software para interfaces de usuario multimedia'' (traducido del inglés: \textit{Software ergonomics for multimedia user interfaces}), actualmente retirada y sin registro de norma que la suplante. 
    \item La norma \textbf{ISO/IEC/IEEE 26514:2022} ``Ingeniería de sistemas y software — Diseño y desarrollo de información para usuarios'' (traducido del inglés: \textit{Systems and software engineering — Design and development of information for users}) describe el proceso para desarrollar la documentación e información que necesitan los usuarios durante el uso de un programa. 
    \item La norma \textbf{ISO/TR 16982:2002} ``Ergonomía de la interacción hombre-sistema: métodos de usabilidad que respaldan el diseño centrado en el ser humano'' (traducido del inglés: \textit{Ergonomics of human-system interaction — Usability methods supporting human-centred design}) describe métodos para abordar la usabilidad en sistemas centrados en el ser humano. 
    \item La norma \textbf{ISO 9241-220:2019} ``Ergonomía de la interacción hombre-sistema — Parte 220: Procesos para habilitar, ejecutar y evaluar el diseño centrado en el ser humano dentro de las organizaciones'' (traducido del inglés: \textit{Ergonomics of human-system interaction — Part 220: Processes for enabling, executing and assessing human-centred design within organizations}) se enfoca en el diseño centrado en el usuario desde el punto de vista organizacional, no aborda nuevos métodos o técnicas de diseño centrado en el ser humano. 
    \item Por último, la norma ISO/TS 16071:2003 ``Ergonomía de la interacción hombre-sistema: orientación sobre accesibilidad para interfaces hombre-computadora'' (traducido del inglés: \textit{Ergonomics of human-system interaction — Guidance on accessibility for human-computer interfaces}), ha sido revisada por la norma \textbf{ISO 9241-171:2008} ``Ergonomía de la interacción hombre-sistema. Parte 171: Orientación sobre la accesibilidad del software'' (traducido del inglés: \textit{Ergonomics of human-system interaction — Part 171: Guidance on software accessibility}), esta última describe buenas prácticas para el diseño de un software accesible.
\end{itemize}

Entre todas las normas ISO mencionadas, se destaca que la única norma que tiene en cuenta la accesibilidad al sistema por parte de personas con discapacidad es la norma \textbf{ISO 9241-171:2008}, por lo que se decidió como primer estudio en esta tesis analizar los alcances y objetivos de esta norma. La misma pretende proveer herramientas para el diseño de software con el fin de alcanzar el máximo nivel de accesibilidad. Esta reglamentación responde a los requerimientos sociales y legislativos de alcanzar la accesibilidad, lo que aplica a software que se utiliza en espacios públicos, en casa, en el trabajo y en educación.

\subsection{Normativa enfocada en accesibilidad web}
\label{sect:normas_w3c}

En cuando a lo que páginas web se refiere, el \citet{w3cstandard} desarrolla normativas para asegurar el crecimiento de la web a largo plazo. Dentro de la categoría de estándares de accesibilidad presenta varias recomendaciones, borradores de recomendaciones y notas de grupo, entre las cuales se encuentran:
\begin{itemize}
    \item ATAG 2.0: Pautas de accesibilidad para herramientas de creación de contenido;
    \item WAI-ARIA 1.1: Guía de implementación enfocada a requerimientos semánticos que deben tener los objetos para su óptima relación con tecnologías de asistencia (como lectores de pantalla);
    \item WCAG 2: Pautas de accesibilidad para el contenido Web;
    \item UAAG: Pautas de accesibilidad para los agentes de usuario (por ejemplo navegadores).
\end{itemize}

Se decidió realizar el primer análisis con las WCAG 2.0, debido a que este estándar describe diferentes criterios de éxito con los cuales se puede evaluar cuán accesible es una página web. Se debe considerar que este análisis fue realizado en el año 2017 al inicio de esta tesis doctoral, con la finalidad de evaluar cuán accesible es para el usuario buscar y descargar el contenido necesario para hacer un análisis de datos astronómico o astrofísico con el software de sonorización que se desarrolla en este trabajo. En la actualidad esta norma cuenta con un reconocimiento de ISO (ISO/IEC 40500), y nuevas versiones disponibles. Se realizó una comparación con la versión WCAG 2.2 evidenciándose pequeños cambios en algunos de los ítem o criterios de éxito que no modifican significativamente la conclusión general obtenida por el estudio realizado en 2017.

\section{Análisis normativo ISO}
\label{sect:iso_seccioncompleta}

\subsection{Introducción}

Existe una falta de referencias que cuestionen los desarrollos de software actuales y traten de explicar el problema detrás de la escasez de accesibilidad. Comprender por qué las herramientas actuales no son accesibles es crucial para no repetir y propagar los mismos problemas de una herramienta a otra. Conjuntamente, se considera que la evaluación de una norma que describe prácticas para hacer un software accesible, es un primer paso para construir un marco de referencia para el desarrollo de un software centrado en el usuario final desde el principio.

En base a los estándares existentes, enfocándonos en el diseño centrado en el usuario y sobre todo en la accesibilidad (como se detalla en la sección \ref{sect:normas_iso}), se decidió aplicar un análisis en base a la norma ISO 9241-171:2008 a los software de sonorización de datos disponibles en 2017 (Sonification Sandbox, MathTrax y xSonify - previamente descriptos en la sección \ref{sect:desktop_astronomia}). El criterio de selección se basó en que tuviesen una interfaz gráfica que permitiera al usuario la configuración de parámetros, que posibilitara importar datos en formato tabla (realizando el despliegue gráfico y sonoro) y que presentaran aplicaciones en astronomía.

Durante el estudio de la norma y una evaluación preliminar que hicieron dos de los integrantes del grupo de análisis que llevó a cabo este estudio, se detectó que los criterios mediante los que la norma propone evaluar accesibilidad en los software se podían interpretar de diferentes maneras, dependiendo de la experiencia y conocimiento del evaluador. Debido a esto, se decidió realizar el estudio con un grupo más amplio, incluyendo personas con y sin discapacidad, y con diferentes niveles de pericia en el análisis de datos astronómicos en general. 

\subsection{Metodología}
\label{sect:iso_metodologia}

La decisión sobre el uso de la norma ISO 9241-171:2008 para realizar un análisis comparativo entre grupos de profesionales con y sin discapacidad, surgió de la necesidad de establecer un marco común para el análisis de software. Se priorizó en todo momento las características de accesibilidad presentes en los software, y se amplió el grupo de evaluadores considerando que las recomendaciones de este estándar en particular podrían interpretarse subjetivamente. En este sentido, el trabajo con la ISO 9241 permitió determinar la efectividad, o no, del uso de la propia norma.

\begin{figure}[!ht]
    \centering
    \includegraphics[width=1\textwidth]{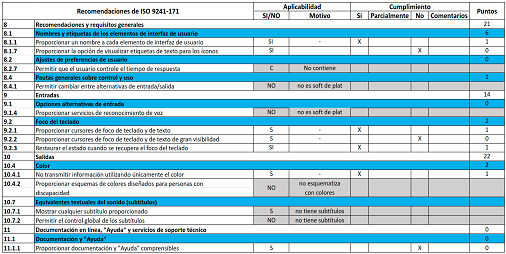}
    \caption{Se muestra un ejemplo con algunos ítem del análisis en xSonify utilizando el apéndice C de la norma ISO 9241-171:2008.}
    \label{fig:iso_anexoC_ejemplo}
\end{figure}

Considerando el texto completo de la norma ISO 9241-171:2008, el apéndice C de la misma contiene una lista de verificación con las recomendaciones que se deben considerar a la hora de evaluar accesibilidad (la Figura \ref{fig:iso_anexoC_ejemplo} muestra un ejemplo con algunos de los ítem de la norma; sino en el Apéndice \ref{ap:A_anexoc_enblanco} se adjunta una copia de la tabla presentada por la norma). En la descripción de la norma detalla los alcances de cada uno de dichos ítem y algunos ejemplos. Para los fines de la evaluación propuesta, se utilizó el mencionado apéndice para comprobar si los software bajo análisis tuvieron en cuenta estas recomendaciones, en particular enfocadas en la accesibilidad para personas ciegas o con baja visión. En función de llevar a cabo esta tarea se conformaron dos grupos: 

\begin{enumerate}
    \item Un primer grupo formado por tres estudiantes de bioingeniería en el mismo año de cursado (identificados como participante 1, 2 y 5), un estudiante de doctorado en ingeniería con mención en bioingeniería (participante 3) y un doctor en ingeniería con especialidad en computación y comunicaciones (participante 4);
    \item Un segundo grupo integrado por un estudiante de grado de física (discapacidades múltiples congénitas y de inicio tardío), un estudiante de doctorado (discapacidades congénitas; ciego/con deficiencia visual), dos informáticos (discapacidades congénitas y de inicio temprano) y un astrónomo e informático con discapacidad (discapacidad de inicio temprano), todos usuarios de herramientas de accesibilidad.
\end{enumerate}

En un principio, el primer grupo con formación en informática pero poca o nula experiencia en accesibilidad, utilizó el apéndice C de la norma para comprobar si los software bajo análisis cumplían con cada ítem enumerado. Luego, utilizando las tablas conformadas por cada uno de los evaluadores, se realizó un informe con lo expresado respecto a cada punto de la norma y fue enviado al segundo grupo, el cual contaba con experiencia en accesibilidad más allá de su condición física y sensorial. Este segundo grupo evaluó no sólo el software, sino también la norma en sí.

\subsubsection{Grupo 1}

El primer grupo de evaluadores, con una formación centrada en computación e informática, pero no en accesibilidad, llevó adelante el análisis de forma individual con sólo dos encuentros virtuales pactados. En el primer encuentro se presentó la norma ISO 9241-171, donde para familiarizarse con ella se explicó y discutió cada punto; en la segunda reunión se acordó cómo realizar el análisis y se puntualizó en el significado de ``Aplica''/ ``No aplica'' y ``Cumple''/``Cumple parcialmente''/``No cumple'', etiquetas que tienen la finalidad de categorizar cada criterio de la norma. Al finalizar este segundo encuentro, se definieron cinco categorías siguiendo la guía estándar para posicionar cada ítem de la norma ISO: 

\begin{enumerate}
    \item ``Cumple'': cuando el software satisface completamente la declaración del ítem ISO;
    \item ``Cumple parcialmente'': cuando el software obedece al menos una parte de la declaración ISO;
    \item ``No cumple'': cuando el software no cumple, o cumple escasamente las declaraciones ISO;
    \item ``N/A'' (No aplica): cuando los evaluadores consideren que la declaración ISO del ítem no aplica al software;
    \item Sin información: cuando el evaluador no comenta el ítem (la norma contempla algunos ítem que podrían no ser aplicados al programa y marca cada uno de ellos en la tabla proporcionada como anexo C en dicha norma).
\end{enumerate}

Luego de estas dos reuniones, cada evaluador realizó la valoración en sus computadoras y en el momento oportuno, para que no hubiera interferencia entre evaluadores. Incluso con estas dos charlas previas, los resultados fueron muy heterogéneos, como mostraremos en las próximas secciones. Cabe aclarar aquí que los evaluadores 3 y 4 (la estudiante de doctorado y el doctor en ingeniería) habían realizado previamente un curso sobre diseño de software centrado en el usuario y tenían experiencia en programación de aplicaciones. En cuanto a las tres estudiantes de bioingeniería, solo tenían conocimientos de programación en el marco de su carrera.

Para llevar a cabo el análisis y evaluar los ítem del Apéndice C de la norma ISO (Apéndice \ref{ap:A_anexoc_enblanco}), se descargaron los tres software (Sonification Sandbox, MathTrax y xSonify) de las páginas web de los desarrollos, y se probaron en MacOS Mojave y Windows 10. Se decidió realizar la prueba en estos sistemas operativos por disponibilidad y sencillez en la instalación de los tres programas evaluados, la instalación de los paquetes y versiones de java necesarias para que funcionaran los programas en Ubuntu es compleja. A lo largo de la prueba, este primer grupo de analistas acordó utilizar la almohadilla táctil de la computadora y el teclado de la notebook como entradas, y la pantalla junto con los parlantes integrados como salidas. Los lectores de pantalla utilizados fueron: NVDA para Windows y Voice Over para MacOS. La salida táctil (ítem 10.9; algunos ítem internos indican: 10.9.1 No transmitir la información únicamente mediante salida táctil, 10.9.3 Permitir ajustar las salidas táctiles) no fue evaluada porque no se dispone de las herramientas específicas para hacerlo. Finalmente, para completar la tabla del anexo C de la norma, se siguieron las instrucciones allí expresadas: primero, se indicó si el ítem era aplicable o no y luego, en caso de ser aplicable, se señala si el programa cumple de forma total, parcial o no cumple con la recomendación.

Una vez realizado el análisis de accesibilidad de los software, se utilizó el coeficiente alfa de Krippendorff \citep{hayeskrip2007} para evaluar la confiabilidad de estos resultados parciales. Dicho coeficiente fue seleccionado porque permite estimar la confiabilidad entre más de dos evaluadores cuando existen datos faltantes (lo cual fue importante debido a que hubieron algunos ítems sin evaluar). El cálculo se realizó siguiendo el trabajo de \citet{krip2011}, donde se desarrollan los pasos para calcular dichos coeficientes y se menciona la disponibilidad de macros existentes para el programa ‘IBM SPSS Statistics’ desarrolladas por Andrew Hayes. En el trabajo de \citet{swert2012} se muestran diferentes capturas de pantalla con el uso de las macros para calcular el coeficiente Alfa de Krippendorff (KALPHA), por lo que se siguió este procedimiento utilizando las macros ya existentes para calcular el coeficiente específico para este estudio.

\subsubsection{Grupo 2}

Respecto del segundo grupo, conformado por personas con experiencia en accesibilidad como campo de trabajo y/o simultáneamente con discapacidad, la metodología de trabajo fue diferente. En este caso, se utilizaron las tablas y descripciones previamente realizadas por el Grupo 1 como punto de partida para evaluar el cumplimiento del software bajo análisis y la exhaustividad de la norma ISO.

Los integrantes del Grupo 2 definieron previamente los criterios de evaluación que se detallan a continuación:

\begin{enumerate}
    \item Definir los objetivos de la evaluación: lograr el acceso al análisis de datos.
    \item Definir poblaciones objetivo para la evaluación de necesidades y servicios:
    \begin{itemize}
        \item Personas con discapacidad, en personas específicas que son ciegas y deficientes visuales (pueden ser solo b/VI o b/VI en combinación con otras discapacidades)
    \end{itemize}
    \item Criterios de evaluación: Importancia de los elementos evaluados utilizando ISO para lograr el objetivo de análisis de datos.
    \begin{itemize}
        \item Definir la importancia de cuán crucial es que el elemento evaluado por ISO exista o no en la interfase digital para lograr el objetivo final definido anteriormente.
        \item Definir el análisis de datos como la efectividad de los pasos concatenados para la exploración sensorial en el despliegue de los datos (en este caso, limitado a visual y auditiva) y una mayor exploración matemática de los mismos.
        \item Definir pasos concatenados como: para acceder a la característica evaluada el usuario tiene que llevar:
        \begin{enumerate}
            \item Pasos homogéneos: aquellos que comprenden pasos no compuestos para realizar el análisis ISO siguiendo los puntos del apéndice C,
            \item Pasos de alta granularidad (agregación de pasos) que aumentan la complejidad de la interacción necesaria para tener éxito (tener que realizar pasos polifacéticos para lograr el elemento evaluado del documento ISO), es decir, "¿necesito realizar una iteración homogénea?".
        \end{enumerate}
    \end{itemize}
\end{enumerate}

Se realizaron reuniones reiteradas donde los participantes discutieron si cumplían o no con cada ítem en particular, agregando comentarios en cada caso sobre las limitaciones de cada punto.

\subsection{Resultados}

Una vez organizados los grupos y definida la metodología, cada integrante del Grupo 1 realizó su parte del trabajo de forma independiente y los datos obtenidos fueron recogidos por los autores. Luego los integrantes del Grupo 2 realizaron su parte del trabajo.

Consecuencia de la primera etapa, realizada por el grupo uno, se recibió una planilla de excel por evaluador y por software (se incluyen las planillas obtenidas en el Apéndice \ref{ap:A_anexoc_planillasXeval}, agrupadas por evaluador). Con dichos datos, por un lado se preparó un documento de texto por razones de accesibilidad y con el fin de enviarlo al grupo dos, y por el otro, se crearon las tablas (incluidas aquí desde Cuadro \ref{tab:byeval_mathtrax_mac} a \ref{tab:byeval_ss_win}) para mostrar los resultados parciales del grupo 1. Adicionalmente, se utilizaron las macros descriptas anteriormente para calcular el coeficiente de confiabilidad KALPHA. 

En la segunda etapa, se recibió el mismo archivo de texto enviado para el análisis, pero en este caso con los comentarios de los participantes del grupo dos sobre el software y sobre la norma (se adjunta este documento en el Apéndice \ref{ap:A_anexoc_grupo2}). Se debe recordar que el análisis del estándar se ejecutó para los tres software mencionados y en los dos sistemas operativos, MacOS y Windows (no en Linux porque no todo el software analizado corre bajo esta plataforma).

\subsubsection{Grupo 1}

\begin{table}[t]
    \centering
    \begin{tabular}{| l | c | c | c | c |}
        \hline
        Resultados & Eval. 1 & Eval. 2 & Eval. 3 & Eval. 4 \\ \hline
        Cumple & 64 & 65 & 40 & 40 \\
        Cumple parcialmente & 28 & 21 & 31 & 31 \\
        No cumple & 29 & 25 & 14 & 14 \\
        N/A & 17 & 23 & 55 & 55 \\
        Sin información & 12 & 6 & 0 & 0 \\ \hline
        Total & 140 & 140 & 140 & 140 \\ \hline
    \end{tabular}
    \caption{Resultados finales de los evaluadores del G1 para el software Mathtrax en MacOS.}
    \label{tab:byeval_mathtrax_mac}
\end{table}

\begin{table}[t]
    \centering
    \begin{tabular}{| l | c | c | c | c |}
        \hline
        Resultados & Eval. 1 & Eval. 2 & Eval. 3 & Eval. 4 \\ \hline
        Cumple & 59 & 53 & 40 & 40 \\
        Cumple parcialmente & 14 & 17 & 20 & 20 \\
        No cumple & 32 & 40 & 22 & 22 \\
        N/A & 24 & 24 & 58 & 58 \\
        Sin información & 11 & 6 & 0 & 0 \\ \hline
        Total & 140 & 140 & 140 & 140 \\ \hline
    \end{tabular}
    \caption{Resultados finales de los evaluadores del G1 para el software xSonify en MacOS.}
    \label{tab:byeval_xsonify_mac}
\end{table}

\begin{table}[t]
    \centering
    \begin{tabular}{| l | c | c | c | c |}
        \hline
        Resultados & Eval. 1 & Eval. 2 & Eval. 3 & Eval. 4 \\ \hline
        Cumple & 60 & 56 & 37 & 37 \\
        Cumple parcialmente & 16 & 13 & 23 & 23 \\
        No cumple & 35 & 41 & 25 & 25 \\
        N/A & 19 & 24 & 55 & 55 \\
        Sin información & 10 & 6 & 0 & 0 \\ \hline
        Total & 140 & 140 & 140 & 140 \\ \hline
    \end{tabular}
    \caption{Resultados finales de los evaluadores del G1 para el software Sonification Sandbox en MacOS.}
    \label{tab:byeval_ss_mac}
\end{table}

\begin{table}[t]
    \centering
    \begin{tabular}{| l | c | c | c | c | c |}
        \hline
        Resultados & Eval. 1 & Eval. 2 & Eval. 3 & Eval. 4 & Eval. 5 \\ \hline
        Cumple & 63 & 63 & 32 & 32 & 90 \\
        Cumple parcialmente & 18 & 23 & 36 & 36 & 7 \\
        No cumple & 30 & 25 & 17 & 17 & 30 \\
        N/A & 17 & 23 & 55 & 55 & 13 \\
        Sin información & 12 & 6 & 0 & 0 & 0 \\ \hline
        Total & 140 & 140 & 140 & 140 & 140 \\ \hline
    \end{tabular}
    \caption{Resultados finales de los evaluadores del G1 para el software MathTrax en Windows.}
    \label{tab:byeval_mathtrax_win}
\end{table}

\begin{table}[t]
    \centering
    \begin{tabular}{| l | c | c | c | c | c |}
        \hline
        Resultados & Eval. 1 & Eval. 2 & Eval. 3 & Eval. 4 & Eval. 5 \\ \hline
        Cumple & 59 & 53 & 34 & 34 & 70 \\
        Cumple parcialmente & 14 & 17 & 24 & 24 & 13 \\
        No cumple & 31 & 40 & 24 & 24 & 49 \\
        N/A & 24 & 24 & 58 & 58 & 8 \\
        Sin información & 12 & 6 & 0 & 0 & 0 \\ \hline
        Total & 140 & 140 & 140 & 140 & 140 \\ \hline
    \end{tabular}
    \caption{Resultados finales de los evaluadores del G1 para el software xSonify en Windows.}
    \label{tab:byeval_xsonify_win}
\end{table}

\begin{table}[t]
    \centering
    \begin{tabular}{| l | c | c | c | c | c |}
        \hline
        Resultados & Eval. 1 & Eval. 2 & Eval. 3 & Eval. 4 & Eval. 5 \\ \hline
        Cumple & 60 & 56 & 30 & 30 & 76 \\
        Cumple parcialmente & 16 & 13 & 29 & 29 & 12 \\
        No cumple & 35 & 41 & 26 & 26 & 43 \\
        N/A & 19 & 24 & 55 & 55 & 3 \\
        Sin información & 10 & 6 & 0 & 0 & 6 \\ \hline
        Total & 140 & 140 & 140 & 140 & 140 \\ \hline
    \end{tabular}
    \caption{Resultados finales de los evaluadores del G1 para el software Sonification Sandbox en Windows.}
    \label{tab:byeval_ss_win}
\end{table}

\begin{figure}[p]
    \centering
    \includegraphics[width=0.8\textwidth]{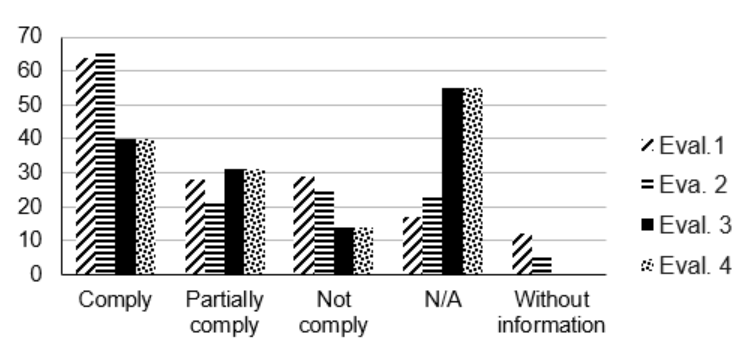}
    \caption{Gráfico de barras de los resultados por evaluador para el software MathTrax en MacOS.}
    \label{fig:iso_byeval_mathtrax_mac}
\end{figure}

\begin{figure}[p]
    \centering
    \includegraphics[width=0.8\textwidth]{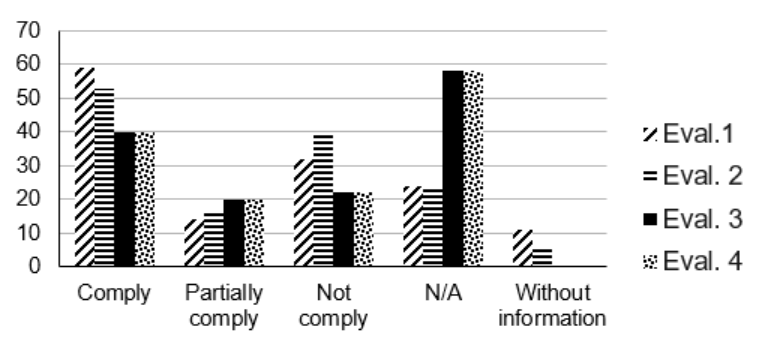}
    \caption{Gráfico de barras de los resultados por evaluador para el software xSonify en MacOS.}
    \label{fig:iso_byeval_xsonify_mac}
\end{figure}

\begin{figure}[p]
    \centering
    \includegraphics[width=0.8\textwidth]{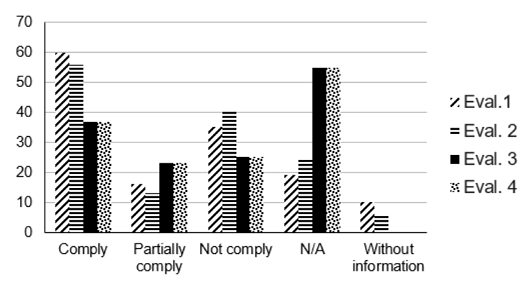}
    \caption{Gráfico de barras de los resultados por evaluador para el software Sonification Sandbox en MacOS.}
    \label{fig:iso_byeval_ss_mac}
\end{figure}

Los resultados finales obtenidos por el Grupo 1 para el software analizado se detallan en los Cuadros \ref{tab:byeval_mathtrax_mac} a \ref{tab:byeval_ss_win}. La Figura \ref{fig:iso_byeval_mathtrax_mac} muestra los resultados para distintos evaluadores para MathTrax en MacOS, se puede apreciar que en el mejor de los casos el estándar no alcanza el 50\% de cumplimiento (el valor de cumplimiento obtenido es cercano a 70 teniendo en cuenta un total de 140 ítems); también se nota una marcada diferencia en las categorías `cumple' y `sin información' entre los evaluadores 1 y 2, en contraste con los evaluadores 3 y 4. La Figura \ref{fig:iso_byeval_xsonify_mac} muestra los resultados de los evaluadores para xSonify en MacOS, este caso presenta menor cumplimiento en general y la diferencia entre los evaluadores 1 y 2, en contraste a los evaluadores 3 y 4 permanece. La Figura \ref{fig:iso_byeval_ss_mac} muestra los resultados de los evaluadores para Sonification Sandbox en MacOS, las diferencias entre evaluadores en cuanto al cumplimiento del estándar son similares, pero aún así existen; es apreciable que las categorías `no aplicable (N/A)' y `sin información' son casi iguales a las del software xSonify.

\begin{figure}[p]
    \centering
    \includegraphics[width=1\textwidth]{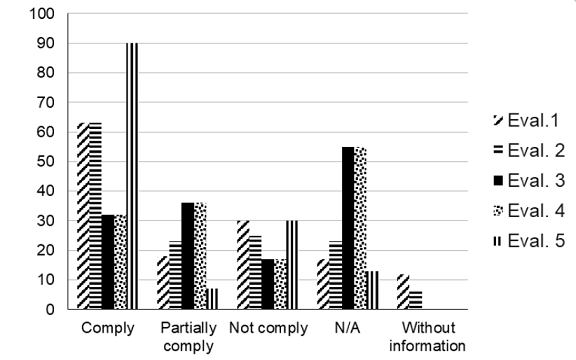}
    \caption{Gráfico de barras de los resultados por evaluador para el software MathTrax en Windows.}
    \label{fig:iso_byeval_mathtrax_win}
\end{figure}

\begin{figure}[p]
    \centering
    \includegraphics[width=1\textwidth]{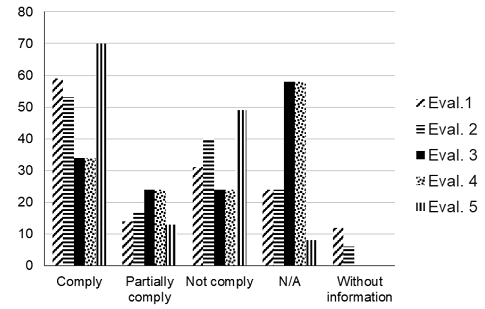}
    \caption{Gráfico de barras de los resultados por evaluador para el software xSonify en Windows.}
    \label{fig:iso_byeval_xsonify_win}
\end{figure}

\begin{figure}[p]
    \centering
    \includegraphics[width=1\textwidth]{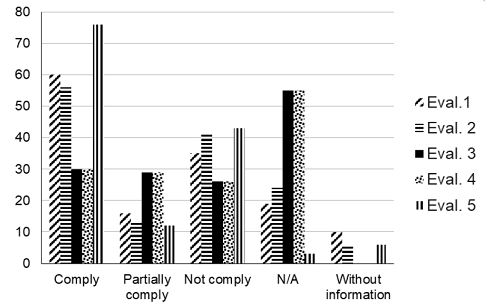}
    \caption{Gráfico de barras de los resultados por evaluador para el software Sonification Sandbox en Windows.}
    \label{fig:iso_byeval_ss_win}
\end{figure}

Acerca de la Figura \ref{fig:iso_byeval_mathtrax_win} (resultados por evaluador para MathTrax en Windows), la Figura \ref{fig:iso_byeval_xsonify_win} (resultados por evaluador para xSonify en Windows) y la Figura \ref{fig:iso_byeval_ss_win} (resultados por evaluador para Sonification Sandbox en Windows), los niveles de cumplimiento de los primeros 4 evaluadores son similares a los obtenidos para el sistema operativo macOS, pero hay una marcada diferencia con el Eval. 5. Estos hallazgos, además de la diferencia previamente detectada entre los evaluadores 1-2 y 3-4, refuerzan la teoría planteada al inicio del análisis sobre la posibilidad de que la interpretación de la norma presente subjetividad y tenga una fuerte relación con la experiencia y conocimiento del evaluador.

Más allá de que todos los evaluadores coincidan en que varios ítem de la normativa no se cumplen, cabe destacar la heterogeneidad de los resultados entre los diferentes evaluadores. Esto puede deberse a las diferentes experiencias que tienen los evaluadores en términos de accesibilidad y en el desarrollo de software, pero también puede estar relacionado con cómo cada persona interpreta los diferentes puntos del estándar. Esto último sería preocupante, porque si bien el estándar busca ser una guía para el desarrollo de herramientas accesibles, muchos desarrolladores lo toman como condición suficiente para decir que su herramienta es accesible, sin serlo completamente.

El presente estudio demuestra la importancia de seguir una norma reforzando la necesidad de que se debe ser consciente que no es un criterio suficiente, sino sólo un peldaño más o una plataforma de la cual partir para llegar a la inclusión en lo que a desarrollo de software concierne.

\subsubsection{Análisis de confiabilidad - Grupo 1}

\begin{table}[t]
    \centering
    \begin{tabular}{|c|c|c|c|c|c|}
        \hline
        \multicolumn{3}{|c|}{Windows} & \multicolumn{3}{|c|}{MacOS} \\ \hline
        MathTrax & SS* & xSonify & MathTrax & SS* & xSonify \\
        0.4116 & 0.4481 & 0.4861 & 0.5391 & 0.5697 & 0.6109 \\ \hline
    \end{tabular}
    \caption{Valores del coeficiente Krippendorff alpha para cada software en cada sistema operativo. Se contó con cinco evaluadores en Windows y cuatro en MacOS. (*SS: Sonification Sandbox)}
    \label{tab:valores_kalpha_confiabilidad}
\end{table}

Observando la heterogeneidad de los resultados que emergieron de la aplicación de las normas, se procedió a realizar el análisis de confiabilidad mediante el coeficiente KALPHA. En el Cuadro \ref{tab:valores_kalpha_confiabilidad} se muestra el valor obtenido para cada uno de los programas en cada uno de los sistemas operativos. Se puede observar que el coeficiente es muy bajo (tener en cuenta que un valor de 0,75 o superior se considera confiable \citep{hayeskrip2007}); la mejor puntuación es para xSonify con un coeficiente de 0,61. 

\begin{figure}[p]
    \centering
    \begin{subfigure}[b]{0.49\textwidth}
         \centering
         \includegraphics[width=\textwidth]{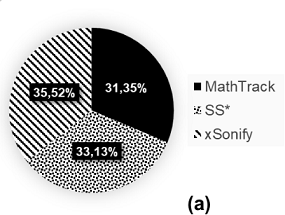}
         \caption{MacOS.}
         \label{fig:iso_confiabilidad_macos}
     \end{subfigure}
     \hfill
    \begin{subfigure}[b]{0.49\textwidth}
         \centering
         \includegraphics[width=\textwidth]{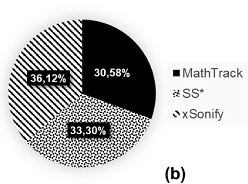}
         \caption{Windows.}
         \label{fig:iso_confiabilidad_windows}
     \end{subfigure}
    \caption{Gráfico de torta representando los valores obtenidos en el análisis de confiabilidad KALPHA para cada software, se realiza un gráfico de torta para cada sistema operativo. (*SS: Sonification Sandbox)}
    \label{fig:iso_confiabilidad_kalpha}
\end{figure}

La figura \ref{fig:iso_confiabilidad_kalpha} muestra un gráfico de torta representando los valores del coeficiente alfa de Krippendorff para cada software en cada sistema operativo; se puede observar la poca diferencia entre los valores correspondientes a cada sistema operativo. Los coeficientes de confiabilidad en porcentaje son muy similares entre sí. Estos resultados pueden deberse, como se mencionó anteriormente, a la dificultad de interpretar la norma en partes específicas, pero también a que el estándar aborda situaciones de accesibilidad en general para diferentes tipos de herramientas y al aplicarlo en un programa específico el resultado se torna dependiente de la interpretación de cada término o elemento en sí.

También se puede observar en los valores que cuando aumenta el número de evaluadores (en el caso del análisis en Windows) el coeficiente de confiabilidad disminuye en lugar de aumentar. Esto nos da la pauta de la influencia que tiene la interpretación de cada evaluador en cada punto.

Es interesante mencionar que este tipo de estudio no cuenta con antecedentes publicados, de manera que el aporte, que resulta original, puede ser de ayuda para diseñar actualizaciones de los estándares de referencia.

\subsubsection{Grupo 2}

Teniendo en cuenta que los evaluadores del primer grupo no tenían experiencia en accesibilidad y la baja fiabilidad de los resultados, el análisis de los puntos del estándar fue realizado por un segundo grupo con experiencia en accesibilidad. Este grupo partió del análisis realizado por el primer grupo y examinó punto por punto la norma, realizando comentarios sobre cada uno de estos ítems.

Algunos de los resultados de análisis más relevantes se resumen a continuación, se detalla punto por punto el texto de la recomendación ISO 9241-171:2008 del ítem en particular, el análisis del G1 y lo expuesto por el G2 (el documento con el análisis del G2 completo se incluye en el Apéndice \ref{ap:A_anexoc_grupo2}).

\begin{itemize}
    \item \textbf{Ítem 8.1.1 - Asignar un nombre a cada elemento de la interfaz gráfica de usuario (GUI).}

        \underline{La norma expresa}: “El software debe asociar un nombre identificativo a cada elemento de la GUI, excepto cuando el nombre sea redundante”
        
        \underline{Análisis de G1 - Evaluadores 1, 2 y 5}:
        \begin{itemize}
            \item MathTrax - Windows y MacOS (Cumple - 100\%)
            \item xSonify - Windows y MacOS (Cumple - 100\%)
            \item Sonification Sandbox - Windows y MacOS (Cumple - 100\%)
        \end{itemize}
        
        \underline{Análisis de G1 - Evaluadores 3 y 4}:
        \begin{itemize}
            \item Sonification sandbox - Windows y MacOS (Cumplimiento parcial - 50\%): Al mirar cada elemento de la interfaz, se nota que falta el nombre del control deslizante de navegación de datos, el resto de los elementos están todos identificados con texto.
            \item MathTrax - Windows (Cumplimiento parcial - 50\%): Al observar cada elemento de la interfaz, el gráfico no tiene nombre. Además, en `Física-Lanzamiento de cohete-Ver resultados de lanzamiento-Ver sus datos' hay cuadros combinados sin nombre, que solo aparecen en el sistema operativo Windows.
            \item MathTrax - MacOS (Cumple - 100\%): Cuando se mira cada elemento de la interfaz, se puede ver que el gráfico tiene una etiqueta, pero no dice explícitamente que es un gráfico. Aún así, se considera que cumple con este criterio.
            \item xSonify - Windows y MacOS (Cumple - 100\%): Al mirar cada elemento de la interfaz se nota que el gráfico no indica ser un gráfico. Aún así, se considera que cumple con este criterio.
        \end{itemize}
        
        \underline{Análisis de G2}:
        \begin{itemize}
            \item La tecnología de asistencia debería funcionar en Windows para usar los aspectos que aparecen en los elementos de la GUI: (1) ``La interfaz asume que la persona tiene orientación y que la persona puede percibir de manera autónoma que la etiqueta se refiere a un gráfico, no debería ser así''; (2) ``Uso de sonido y gráficos, reconocimiento de funcionalidades, orientación en la interfaz en cuanto a la identificación de las funcionalidades que se muestran (gráfica visual y auditiva) y en cuanto a la navegación, este ítem es fundamental para el inicio, avance y culminación de una tarea de análisis. Esto también es muy necesario para aprender a usar el sistema y para llevar a cabo el objetivo principal del programa que es traducir los datos en sonido''.
        \end{itemize}
    
    \item \textbf{Ítem 8.1.3 - Dar nombres únicos en contexto.}
    
        \underline{La norma expresa}: ``Cada nombre de un elemento de la interfaz de usuario especificado por los desarrolladores de software debe ser único en su contexto''.
        
        \underline{Análisis de G1 - Todos los evaluadores}: 
        \begin{itemize}
            \item MathTrax - Windows y MacOS (Cumple - 100\%)
            \item xSonify - Windows y MacOS (Cumple - 100\%)
            \item Sonification Sandbox - Windows y MacOS (Cumple - 100\%)
        \end{itemize}
        
        \underline{Análisis de G2}: Los evaluadores concluyen que no cumple, G1 debería haber evaluado si cumple con las herramientas de accesibilidad, como los lectores de pantalla.
    
    \item \textbf{Ítem 8.1.5 - Mostrar nombres.}
    
        \underline{La norma expresa}: ``Si un elemento de la interfaz de usuario tiene una representación visual y no forma parte de los componentes estándar de la interfaz de usuario, el software debe mostrar su nombre a los usuarios (ya sea de forma predeterminada o a pedido del usuario)''.
        
        \underline{Análisis de G1 - Todos los evaluadores}:
        \begin{itemize}
            \item Sonification Sandbox - Windows y MacOS (Cumple - 100\%)
            \item MathTrax - Windows y MacOS (Cumple - 100\%)
            \item xSonify - Windows y MacOS (Cumple - 100\%): este software no contiene iconos y expone los nombres de todos los elementos mostrados con espacios de texto.
        \end{itemize}
        
        \underline{Análisis de G2}: Los revisores identifican que este parámetro se refiere a etiquetas visibles de los elementos, no se refiere simétricamente a muestras auditivas o táctiles. El parámetro de identificación se considera de suma importancia para la orientación, el desempeño eficiente del trabajo, la reducción de obstáculos con la arquitectura y la reducción de la carga cognitiva.
        
    \item \textbf{Ítem 9.2.1 - Proporcionar cursores de foco de teclado y texto.}
        
        \underline{La norma expresa}: ``El software debe proporcionar un cursor de foco de teclado que indique visualmente qué elemento de la interfaz de usuario tiene foco de teclado en un momento dado, así como un cursor de texto que indique la ubicación del foco dentro de un elemento de texto''.
        
        \underline{Análisis de G1 - Evaluadores 1 y 2}:
        \begin{itemize}
            \item Sonification Sandbox - Windows y MacOS (Cumple - 100\%)
            \item MathTrax - Windows y MacOS (Cumple - 100\%)
            \item xSonify - Windows y MacOS (Cumple - 100\%)
        \end{itemize}
        
        \underline{Análisis de G1 - Evaluadores 3 y 4}:
        \begin{itemize}
            \item Sonification Sandbox - Windows (Cumple parcialmente - 50\%): Algunos elementos tienen el foco de teclado y cursor, especialmente si se puede hacer clic con el mouse. La tabla también lo tiene.
            \item Sonification Sandbox - MacOS (Cumple parcialmente - 50\%): Algunos elementos tienen el foco de teclado y cursor. Se evalúa con Voice Over (lector de pantalla de MacOS) desactivado, ya que este cuenta con su propio cursor de foco de teclado.
            \item MathTrax - Windows (Cumple parcialmente - 50\%): Presenta el cursor de enfoque del teclado pero no el de texto.
            \item MathTrax - MacOS (Cumple parcialmente - 50\%): tiene enfoque de teclado solo en algunos elementos y no tiene un cursor de texto.
            \item xSonify: Windows y MacOS (Cumple parcialmente - 50\%): presenta el enfoque del teclado en algunos elementos. Tiene un cursor de texto en cuadros de texto no editables, este software no contiene cuadros de texto editables.
        \end{itemize}
        
        \underline{Análisis de G1 - Evaluador 5}:
        \begin{itemize}
            \item MathTrax - Windows y MacOS (Cumple - 100\%)
            \item Sonification Sandbox - Windows (Cumple parcialmente - 50\%): en algunos casos no presenta la opción.
            \item Sonification Sandbox - MacOS (Cumple - 100\%)
            \item xSonify - Windows (Cumple parcialmente - 50\%): en algunos casos no presenta la opción.
            \item xSonify - MacOS (Cumple - 100\%)
        \end{itemize}
        
        \underline{Análisis de G2}:
        \begin{itemize}
            \item ``Se debe asegurar que los prototipos en su arquitectura definan el foco de atención dentro de un teclado, mouse y demás periféricos. De esta forma, el software proporcionará un cursor de foco de atención que indica de forma visual, sonora y táctil manteniendo los rangos de activación y alertas, mencionando qué elemento de la interfaz de usuario tiene el foco de atención del teclado en un momento dado; así como un cursor de texto que especifica la ubicación del foco dentro de un texto o elemento gráfico. El despliego visual del foco de atención debe ocurrir simultáneamente y en sincronía con otras alertas multisensoriales''.
            \item ``Se debe tener en cuenta que para las personas con discapacidad, perder el foco de atención puede significar perder la orientación en la escena prototipo. Por ejemplo, una persona con dislexia que depende de la sincronización de un mecanismo de atención visual y auditivo para desambiguar el texto, pierde la orientación si los dos elementos no están sincronizados''.
        \end{itemize}
        
    \item \textbf{Ítem 9.3.8 - Permitir a los usuarios deshabilitar la repetición de clave.}
    
        \underline{La norma expresa}: ``El software responsable del repetidor de tecla debe permitir a los usuarios desactivar esta función''.
        
        \underline{Análisis de G1 - Evaluadores 1, 2 y 5}:
        \begin{itemize}
            \item MathTrax - Windows y MacOS (No cumple - 0\%): no permite una configuración mínima de tiempo de pulsación de tecla.
            \item xSonify - Windows y MacOS (No cumple - 0\%): no tiene esta opción.
            \item Sonification Sandbox - Windows y MacOS (No cumple - 0\%): no dispone de esta opción.
        \end{itemize}
        
        \underline{Análisis de G1 - Evaluadores 3 y 4}: Este ítem no se aplica a ningún software porque no son responsables de la activación de repetición de clave.
        
        \underline{Análisis de G2}: Referente a la evaluación de G1 (evaluadores 3 y 4) ``aunque la norma habla de que se active desde el sistema estándar o plataforma, se debe considerar el esfuerzo físico que requieren las personas con problemas de movilidad. Este punto de la norma se refiere al software y no al sistema operativo, por lo que los árbitros, reconociendo la falta de investigación y las dificultades para implementar estos aspectos en el software, dicen que se cumple con un 25\%''.
    
    \item \textbf{Ítem 9.3.10 - Proveer atajos de teclado.}
        
        \underline{La norma expresa}: ``El software debe proporcionar atajos de teclado para funciones de uso frecuente. Es decir: el usuario puede presionar `Ctrl + C' para copiar, `Ctrl + V' para pegar o `Ctrl + P' para imprimir.
        
        \underline{Análisis de G1 - Evaluador 1}: 
        \begin{itemize}
            \item MathTrax - Windows y MacOS (Cumple - 100\%)
            \item xSonify - Windows y MacOS (Cumple parcialmente - 50\%): en algunos casos, esta opción no funciona, sin embargo, con el lector de pantalla de MacOS sí.
            \item Sonification Sandbox - Windows y MacOS (Cumple parcialmente - 50\%): no todos los comandos se pueden activar a través del teclado.
        \end{itemize}
        
        \underline{Análisis de G1 - Evaluador 2}:
        \begin{itemize}
            \item MathTrax - Windows y MacOS (Cumple parcialmente - 50\%): tiene la opción pero no funciona.
            \item Sonification Sandbox - Windows y MacOS (Cumple parcialmente - 50\%): tiene la opción, pero no funciona. No todos los comandos se pueden manejar con el teclado.
            \item xSonify - Windows y MacOS (Cumple parcialmente - 50\%): tiene la opción, pero no funciona.
        \end{itemize}
        
        \underline{Análisis de G1 - Evaluadores 3 y 4}:
        \begin{itemize}
            \item Sonification Sandbox - Windows y MacOS (Cumple parcialmente - 50\%): proporciona algunos atajos de teclado, pero no se consideran suficientes.
            \item MathTrax: Windows y MacOS (Cumple parcialmente - 50\%): proporciona algunos atajos de teclado, pero no se consideran suficientes.
            \item xSonify - Windows y MacOS (Cumple - 100\%)
        \end{itemize}
        
        \underline{Análisis de G1 - Evaluador 5}:
        \begin{itemize}
            \item MathTrax - Windows (Cumple - 100\%)
            \item xSonify - Windows (Cumple parcialmente - 50\%): no proporciona accesos directos para todas las funciones.
            \item Sonification Sandbox - Windows (No cumple - 0\%): no permite la opción.
        \end{itemize}
        
        \underline{Análisis de G2}: ``Este punto es muy importante porque refuerza la tendencia de separar una discapacidad de otra (discapacidades que ocurren de manera aislada). De acuerdo con la descripción del artículo, si la persona solo accede usando el teclado, y no se puede acceder a todas las funciones con el teclado, ¿de qué otra manera se le proporcionará al cliente acceso a todo el escenario del software?''.
    
    \item \textbf{Ítem 10.3.3 - Ajustar la escala y la disposición de los elementos de la interfaz de usuario en función de los cambios en el tamaño de fuente.}
        
        \underline{La norma expresa}: ``El software debe ajustar la escala y la disposición de los elementos de la interfaz de usuario según sea necesario para tener en cuenta los cambios de tamaño en el texto incrustado o asociado”.
        
        \underline{Análisis de G1 - Evaluadores 1 y 2}:
        \begin{itemize}
            \item MathTrax - Windows y MacOS (Cumple - 100\%)
            \item xSonify - Windows y MacOS (No cumple - 0\%): no se permite modificar la escala.
            \item Sonification Sandbox - Windows y MacOS (No cumple - 0\%): no se permite modificar la escala ni el tamaño de letra.
        \end{itemize}
        
        \underline{Análisis de G1 - Evaluadores 3 y 4}: Este ítem no se considera aplicable porque ninguno de los programas permite cambiar el tamaño de fuente.
        
        \underline{Análisis de G1 - Evaluador 5}: 
        \begin{itemize}
            \item MathTrax - Windows (No cumple - 0\%): no permite cambiar el tamaño de letra.
            \item xSonify - Windows (No cumple - 0\%): no permite cambiar el tamaño de letra.
            \item Sonification Sandbox - Windows (No cumple - 0\%): no permite cambiar el tamaño de letra.
        \end{itemize}
        
        \underline{Análisis de G2}: Los evaluadores consideran que este ítem elimina la responsabilidad del programador. Estos elementos de implementación automática deben observarse con mucho cuidado, se supone que el software tiene el mecanismo explicado anteriormente para interactuar con el elemento estándar sin causar cambios en la plataforma u otros programas.
    
    \item \textbf{Ítem 10.6.1 - Usar patrones de tono en lugar del valor del tono para transmitir información}
        
        \underline{La norma expresa}: ``Al transmitir información de sonido, el software debe usar patrones de tono temporales o basados en frecuencia en lugar de usar un solo tono o volumen absoluto. Es decir: en un servicio de teleconferencia, un par de tonos de alto a bajo (en lugar de solo un tono bajo) indica que una persona está desconectada''.
        
        \underline{Análisis de G1 - Evaluadores 1, 2 y 5}:
        \begin{itemize}
            \item MathTrax - Windows y MacOS (Cumple - 100\%)
            \item xSonify - Windows y MacOS (Cumple - 100\%)
            \item Sonification Sandbox - Windows y MacOS (Cumple - 100\%)
        \end{itemize}
        
        \underline{Análisis de G1 - Evaluadores 3 y 4}:
        \begin{itemize}
            \item Sonification Sandbox: Windows y MacOS (no aplicable): este elemento se considera no aplicable porque el software no utiliza tonos para indicar cambios y, por definición, la sonorización depende de los datos.
            \item MathTrax - Windows y MacOS (Cumple - 100\%): en este caso, se considera compatible aunque utiliza un solo tono porque se produce durante la sonorización de datos para indicar ejes cruzados.
            \item xSonify: Windows y MacOS (no aplicable): este elemento se considera no aplicable porque el software no utiliza tonos para indicar cambios y, por definición, la sonorización depende de los datos.
        \end{itemize}
        
        \underline{Análisis de G2}: ``Este punto no menciona que cada alarma debe ser reconocible, asociativa y más o menos única para cada evento en particular. Asimismo, no considera que existan tonos no aptos para personas neurodiversas. Los árbitros se sorprenden al detectar que la norma aparentemente indica a los desarrolladores que las discapacidades ocurren de manera aislada''.
        
    \item \textbf{Ítem 10.9.3 - Permite ajustar salidas táctiles.}
    
        \underline{La norma expresa}: ``El software debería permitir a los usuarios ajustar los parámetros de la salida táctil para evitar molestias, dolor o lesiones. Es decir: un usuario con percepción táctil reducida puede regular individualmente el límite superior de la salida táctil de un sistema con retorno forzado''.
        
        \underline{Análisis de G1 - Todos los evaluadores}: Este ítem no se considera aplicable porque los evaluadores no cuentan con las herramientas táctiles para comprobarlo.
        
        \underline{Análisis de G2}: ``Este artículo es de suma importancia debido a su relevancia para las personas sordociegas, neurodiversas y que necesitan alertas multisensoriales (entre otras). Los evaluadores encuentran que la norma ha sido muy limitada en el aspecto táctil y su combinación multimodal''.
    
\end{itemize}

Este análisis punto por punto de la ISO 9241-171:2008-Apéndice C llevado a cabo por el G2 condujo a la siguiente conclusión expuesta por dicho grupo: \textit{``La evaluación parece dirigir a los desarrolladores a pensar que la mayoría de las discapacidades ocurren de forma aislada, lo que lleva a dichos desarrolladores a aislar posibles estrategias que complican las soluciones futuras. Por eso es dudosa su recopilación de datos''}. Adicionalmente, les sorprendió mucho que la norma no considere las entradas y salidas multimodales; además, que no tenga en cuenta la sincronización automática en el sistema para personas sordas, a pesar de que mantiene una presencia a través de ejemplos de las necesidades de ayuda a esta población en varios puntos. No se consideran sistemas autónomos similares a TTY (modo teléfono de texto) con bidireccionalidad para el acceso a ambos lados, por ejemplo, una persona sorda que se comunica por una aplicación de teléfono fijo o móvil o por sistemas informáticos tradicionales a una persona ciega con problemas de movilidad.

\subsection{Análisis y discusión}

La norma ISO utilizada expresa en su introducción que brinda orientación para las actividades de diseño centradas en el ser humano con el objetivo de maximizar la accesibilidad, además aclara que podría aplicarse durante toda la vida útil de un sistema informático. En este sentido, para llevar a cabo este estudio se aplicó el estándar ISO 9241-171:2008 a tres software de sonorización, con aplicación en el campo de la astronomía, para determinar el nivel de accesibilidad de cada uno.

De acuerdo a los resultados del Grupo 1 (personas con antecedentes en computación e informática, no en accesibilidad), se detecta una heterogeneidad entre los resultados obtenidos por los evaluadores; se puede apreciar en las Figuras \ref{fig:iso_byeval_mathtrax_mac} a \ref{fig:iso_byeval_ss_win} una marcada diferencia en las columnas relacionadas al cumplimiento de la norma, también sucede algo similar en las relacionadas con los ítem que no son aplicables. Para cuantificar la diferencia detectada entre evaluadores y estimar la confiabilidad del análisis realizado, se aplicó el coeficiente alfa de Krippendorff. Aún cuando el valor de coeficiente no indica confiabilidad nula, confirma que los evaluadores no están de acuerdo en su análisis y arroja un resultado que tiende a indicar poco nivel de confianza en los datos obtenidos.

En este punto del estudio, se consideraron dos supuestos: (1) aumentar el número de evaluadores con experiencia en computación e informática aumentaría la confiabilidad; (2) la diferencia de conocimiento y experiencia entre los evaluadores los lleva a una interpretación diferente del estándar. El segundo supuesto se basa en las similitudes observadas entre los evaluadores 3 y 4 (un estudiante de doctorado y un doctor recién graduado), frente a las similitudes entre los evaluadores 1 y 2 (estudiantes de bioingeniería en el mismo año de cursado). Se debe recordar que los evaluadores 3 y 4, más allá de no tener experiencia en desarrollo de software accesible, habían realizado previamente un curso de diseño centrado en el usuario. Estas diferencias entre evaluadores se consideraron positivas para analizar esta primer hipótesis de que el estándar podía interpretarse de forma heterogénea, y llevaron a una primer confirmación de esta premisa. Sin embargo, y reforzado por el bajo nivel de confiabilidad, se decidió incluir expertos en accesibilidad para complementar el análisis.

En esa segunda etapa, y teniendo en cuenta que los archivos .pdf, las tablas y las ventanas emergentes (necesarias para probar un software y puntuar cada ítem del estándar) son de difícil acceso para las personas con discapacidad, se decidió que el segundo grupo analizara los ítem partiendo del análisis previo realizado por el Grupo 1. Para ello, se envió al Grupo 2 un documento de texto con la información de la norma y del análisis del Grupo 1, descripto para cada ítem ISO como se indica en la sección \ref{sect:iso_metodologia}. Sorprendentemente, en este segundo análisis realizado por expertos en accesibilidad, se detalla un claro desacuerdo por parte del Grupo 2, no solo con el Grupo 1, sino también con lo que indica y describe el estándar ISO. Mencionamos a continuación algunos ejemplos:

\begin{enumerate}
    \item \textbf{Punto 10.3.3}:
        
        \underline{La norma expresa}: ``La escala y la disposición de los elementos de la interfaz de usuario deben ser ajustadas por el software según sea necesario, para dar cuenta de los cambios de tamaño en el texto incrustado o asociado''.
        
        \underline{Comentario de evaluadores 3 y 4}: ``Este ítem no se considera aplicable porque ninguno de los software permite el cambio de tamaño de fuente''.
        
        \underline{Comentario del grupo 2}: ``Los evaluadores consideran que este ítem le quita responsabilidad al programador. Estos elementos de implementación automática deben observarse con mucho cuidado, se supone que el software tiene el mecanismo explicado anteriormente para interactuar con el elemento estándar sin causar cambios en la plataforma u otros programas''.
    
    \item \textbf{Punto 10.6.1}:
    
        \underline{La norma expresa}: ``Usar patrones de tono en lugar del valor del tono para transmitir información''.
        
        \underline{Comentario del grupo 2}: ``Este punto no menciona que cada alarma debe ser reconocible, asociativa y más o menos única para cada evento en particular. Asimismo, no considera que existen tonos no aptos para personas neurodiversas. Los evaluadores se sorprenden de que la norma aparenta decirle a los desarrolladores que las discapacidades ocurren de manera aislada''.
    
\end{enumerate}

Los resultados de los dos grupos de evaluación muestran que el texto del estándar puede ser interpretado de diferentes formas de acuerdo a la experiencia de cada persona. Esto es muy importante en el campo del desarrollo de software ya sean diseñadores, programadores, jefes de proyecto o usuarios finales (por mencionar algunas), ya que las personas involucradas siempre presentan diferentes antecedentes y niveles de experiencia. La heterogeneidad de los resultados, incluidos los comentarios del Grupo 2 (expertos en discapacidad y con experiencia en herramientas de accesibilidad), plantean que el estándar no es suficiente para poder decir que una herramienta es accesible. Siguiendo esta idea, la norma ISO podría tomarse como punto de partida, pero se necesita una mejor comprensión en el campo de la accesibilidad y de las personas involucradas.

Según la evidencia presentada, se recomienda fuertemente el contacto con los interesados que utilizarán la herramienta desde el principio del desarrollo, para que esté lo más centrado posible en el usuario, evitando suposiciones erróneas y asegurando que la herramienta sea utilizable en el largo plazo.

En cuanto a los software analizados en particular, todos presentan falta de cumplimiento de alrededor del 30 o 40 por ciento respecto de los postulados de la Norma, lo que sugiere que los mismos no sigue las recomendaciones del estándar de accesibilidad o diseño centrado en el usuario. Pero, incluso cuando el análisis arroja resultados similares para cada uno de ellos, \textbf{xSonify} presenta algunas características que hacen que sea más accesible y pertinente para nuestras futuras investigaciones frente a las otras dos opciones; algunas de estas características son: 
\begin{itemize}
    \item presenta todas las funcionalidades en las mismas ventanas (esto es muy importante, especialmente para usuarios con discapacidad visual);
    \item permite sonorizar múltiples columnas al mismo tiempo estableciendo una configuración de sonido particular para cada una (permite una fácil y ágil comparación entre datos en un archivo con diferentes columnas);
    \item presenta un lector de pantalla integrado y también permite utilizar el lector de pantalla nativo.
\end{itemize}

\section{Análisis normativo W3C}
\label{sect:w3c_seccioncompleta}

\subsection{Introducción}

En un principio, el programa xSonify presentaba comunicación con bases de datos, permitiendo a los usuarios realizar la búsqueda del dato a analizar escribiendo el nombre de identificación del objeto, sin tener que navegar una página web. En la actualidad ninguno de los software presenta este tipo de conexión, por lo que el usuario debe: buscar el objeto que desea estudiar en una base de datos, descargar el archivo, y recién ahí, si es que el formato del archivo descargado es compatible con el programa de análisis que utiliza, puede realizar el análisis. Por esta razón, en búsqueda de comprender mejor los retos a los que se enfrenta un científico con discapacidad en las ciencias espaciales, y complementando el estudio realizado a los software de análisis de datos en la sección \ref{sect:iso_seccioncompleta}, se plantea evaluar la accesibilidad a las bases de datos astronómicas (bases de las cuales una personas puede descargar información y datos astronómicos o astrofísicos para su posterior análisis).

\citet{boldyreff2002} menciona una serie de niveles relacionados entre sí que deben tenerse en cuenta en el momento de analizar accesibilidad (mencionándolos desde el interno al externo): funcionalidad, eficacia, usabilidad, estándares, efectos individuales, efectos grupales, efectos organizacionales y efectos sociales. Luego profundiza los cuatro niveles internos (correspondientes a temas técnicos) con las características de calidad enunciadas por la norma ISO/IEC 9126: 1991 (``Estándar de calidad sobre productos de software''): confiabilidad, eficiencia y mantenibilidad. De lo mencionado anteriormente surgen los siguientes niveles internos: confiabilidad, eficiencia, funcionalidad, usabilidad, eficacia, mantenibilidad y estándares.

Evaluar la accesibilidad web no es una tarea sencilla, se deben tener en cuenta desde los aspectos más técnicos hasta los aspectos sociales y la relación con el usuario final. Para los límites de este trabajo, y teniendo en cuenta que el objetivo de dimensionar la accesibilidad web tiene como finalidad conocer los problemas a los que se enfrentan las personas con discapacidad a la hora de desarrollarse académica y profesionalmente, se ha decidido comenzar analizando la accesibilidad de algunas bases de datos astronómicas como son: SIMBAD (\textit{Set of Indications, Measurements, and Bibliography for Astronomical Data}), SDSS (\textit{Sloan Digital Sky Survey}) y ADSABS (\textit{Astrophysics Data System}). Estas bases de datos permiten acceder a información sobre parámetros astrofísicos (como coordenadas, tipos espectrales y flujos), imágenes astronómicas y bibliografía, respectivamente. En esta sección se evaluará la accesibilidad a estas páginas web, que permiten acceder a datos previamente almacenados.

La evaluación de dichas páginas web se llevó a cabo desde el punto de vista de los estándares propuestos por el \citet{w3cstandard}, quién desarrolla normativas para asegurar el crecimiento de la web a largo plazo. Dentro de la categoría de accesibilidad presenta varias recomendaciones, borradores de recomendaciones y notas de grupo. En la sección \ref{sect:normas_w3c}, se detallan las normativas de W3C existentes, en este estudio particular se utilizará la norma \href{https://www.w3.org/TR/WCAG20/}{\underline{\textcolor{blue}{WCAG 2.0}}}.

\subsection{Metodología}

La web de W3C contiene información y los estándares que han ido desarrollando, entre ellos el WCAG 2.0. Además, presenta links con explicación de cada criterio, ejemplos de aplicación y, técnicas y fallas. La norma seleccionada presenta diferentes criterios de éxito que buscan ser una guía para analizar y generar mayor accesibilidad en los desarrollos web. Estos criterios de éxito se presentan en tres niveles de complejidad (A, AA y AAA). Debido a ello, la página puede presentar un nivel de accesibilidad nulo (si no cumple algún criterio de éxito nivel A), nivel A (si cumple todos los criterios de éxito nivel A y no cumple alguno nivel AA), nivel AA (si cumple todos los criterios de éxito AA y no cumple alguno nivel AAA) o nivel AAA (si cumple con todos los criterios de éxito nivel AAA).

Adicionalmente a las guías de accesibilidad detalladas para ser utilizadas en un análisis, dentro de W3C existe una lista de herramientas que analizan la accesibilidad de las páginas web, enumerando las fallas y beneficios encontrados, o bien indicando los criterios que deben ser analizados manualmente. Dentro de esa lista, teniendo en cuenta solo las que analizan WCAG 2.0 a fines de 2017, encontramos tres herramientas que realizan dicho análisis: AChecker, Examinator y Functional Accessibility Evaluator (FAE) 2.0. De los tres evaluadores utilizados en este estudio realizado a fines de 2017, en la actualidad solo se mantiene en funcionamiento el último mencionado \citep{fae2022}. Por tal motivo en el Apéndice \ref{ap:web_completo} se incluyen los reportes entregados por dichas páginas web al momento de realizado este estudio.

\subsection{Resultados}

En esta sección se presentarán los resultados obtenidos por las tres herramientas mencionadas, los cuales se han condensado en tablas que muestran las categorías adoptadas por cada programa y el número de ítem que se encuadra en cada categoría.

Para realizar el análisis utilizando AChecker se selecciona la pestaña ``Web Page URL'', dentro de opciones se marca ``WCAG 2.0 (Level AAA)'' y ``View by Guideline''. Luego en ``Address'' se coloca la url de la web a analizar y se presiona el botón ``Check It''. Una vez realizado el análisis, se despliegan los resultados en tres pestañas: problemas conocidos, problemas probables y problemas potenciales; y se muestra un enlace (``WCAG 2.0 (Level AAA)'') que permite acceder al listado de los criterios evaluados.

\begin{table}[t]
    \centering
    \begin{tabular}{ l c c c }
        \hline \hline
         & SIMBAD & SDSS & ADS \\ \hline
        Problemas conocidos & 86 & 8 & 0 \\
        Problemas probables & 0 & 0 & 0 \\
        Problemas potenciales & 192 & 233 & 30 \\ \hline
    \end{tabular}
    \caption{Resultados del análisis con ACHECKER.}
    \label{tab:w3c_achecker}
\end{table}

Achecker cuenta con un total de 168 criterios para evaluar de los establecidos por la norma WCAG 2.0. Cabe mencionar que al realizar el análisis, separa los problemas encontrados en problemas conocidos (aquellos que indudablemente presentan un problema de accesibilidad; por ejemplo no proveer texto alternativo para imágenes), problemas probables (sobre estos no se encontró ninguno) y problemas potenciales (por ejemplo, crear contenido que pueda ser representado de diferentes formas sin perder contenido). La lista completa con los problemas encontrados por página web se encuentra en el Apéndice \ref{ap:achecker}.

En la Tabla \ref{tab:w3c_achecker} se presentan los resultados obtenidos para la web principal de cada una de las bases de datos tenidas en cuenta en el presente trabajo. Es importante resaltar que el analizador se basa en el código fuente de la página web, por lo que se pueden repetir los errores, razón por la cual si sumamos los errores totales de cada página puede dar superior al número de criterios analizados. De esta forma, cada valor en la tabla representa la totalidad de errores encontrados en la web, no el número de criterios donde se ha encontrado un problema.

En el caso de FAE 2.0, se debe indicar la url a analizar, un título identificador de la página ingresada y elegir las características que mayor número de páginas nos permiten analizar:

\begin{itemize}
    \item Profundidad de evaluación: ``Include third-level pages''; 
    \item Seguir los enlaces en: ``Specified domain and all of its next-level subdomains'';
    \item Conjunto de reglas: ``HTML5 and ARIA Techniques''; 
    \item Límite de evaluación: ``25 pages''.
\end{itemize}

\begin{table}[t]
    \centering
    \begin{tabular}{l c c c}
        \hline \hline
         & SIMBAD & SDSS & ADS \\ \hline
        Violaciones & 8 & 5 & 5 \\
        Advertencias & 1 & 2 & 1 \\
        Chequeo Manual & 31 & 32 & 14 \\
        Aprobado & 7 & 24 & 1 \\
        Reglas no aplicadas & 85 & 69 & 111 \\ \hline
    \end{tabular}
    \caption{Resultados del análisis con FAE 2.0.}
    \label{tab:w3c_fae2.0}
\end{table}

El analizador proporciona un listado de estas reglas a revisar y los criterios de éxito de la WCAG 2.0 correspondientes. Más allá del límite de evaluación seleccionado (25 páginas) el número de páginas incluido en el análisis varía según lo que se analiza. Debido a que para el ADS solo se analiza la página principal, a los fines de este trabajo y para poder comparar con los demás analizadores utilizados, solo se muestran los resultados de dichas páginas principales (ver Cuadro \ref{tab:w3c_fae2.0}). En el caso de cada valor por página web, representa el número total de reglas que presenta violaciones, advertencias, que requieren chequeo manual, que han sido aprobadas o que no han sido aplicadas. Se encuentra el informe completo por página web analizada en el Apéndice \ref{ap:fae}.

\begin{table}[t]
    \centering
    \begin{tabular}{l c c c}
        \hline \hline
         & SIMBAD & SDSS & ADS \\ \hline
        Resultado general & 4.5(15) & 6.1(21) & 6.5(7) \\
        Limitación total para ver & 4.6(15) & 5.9(17) & 6.7(7) \\
        Limitación grave para ver & 4.3(14) & 6.4(19) & 7.1(7) \\
        Limitación de los miembros superiores & 4.9(10) & 5.7(13) & 4.8(5) \\
        Limitación para comprender & 3.6(9) & 5.9(13) & 5.9(4) \\
        Limitación por la edad & 4.9(11) & 6.4(18) & 7.2(7) \\ \hline
    \end{tabular}
    \caption{Resultados globales del análisis con EXAMINATOR (entre paréntesis se indica el número de pruebas).}
    \label{tab:w3c_examinator_globales}
\end{table}

Finalmente, EXAMINATOR solo requiere que se ingrese la url de la página a evaluar. La herramienta otorga un puntaje a los criterios que analiza (entre 1 y 10) y entrega un informe detallado del análisis. Además, realiza una ponderación de los criterios analizados, otorgando un puntaje general y unos específicos para ciertas limitaciones (ver Cuadro \ref{tab:w3c_examinator_globales}). En el Cuadro \ref{tab:w3c_examinator_puntuales} se muestra el número de pruebas realizadas y el estado de las mismas. Además, en el Apéndice \ref{ap:examinator} se incluyen los reportes por página web con los problemas encontrados.

\begin{table}[t]
    \centering
    \begin{tabular}{l c c c}
        \hline \hline
         & SIMBAD & SDSS & ADS \\ \hline
        Muy mal & 4 & 4 & 0 \\
        Mal & 5 & 2 & 3 \\
        Regular & 0 & 5 & 0 \\
        Bien & 0 & 1 & 0 \\
        Excelente & 6 & 9 & 4 \\
        Cantidad de pruebas & 15 & 21 & 7 \\ \hline
    \end{tabular}
    \caption{Resultados puntuales del análisis con EXAMINATOR.}
    \label{tab:w3c_examinator_puntuales}
\end{table}

\subsection{Análisis y discusión}

Acudiendo al detalle de los resultados de los programas de análisis de accesibilidad, se observa que entre los principales criterios no cumplidos se encuentran la falta de texto alternativo a las imágenes, secuencias de foco de teclado mal ordenadas, imágenes con porcentajes de contraste que no son adecuados (según el establecido por la norma), detalles en el código que pueden incumplir el criterio de éxito ``2.3 Convulsiones'' y falta de mecanismos para manipular grandes bloques de información. En el análisis llevado a cabo se puede concluir que las páginas web que dan acceso a las bases de datos no son accesibles: el no cumplimiento de uno solo de los criterios ya sería descalificador del recurso en términos de la norma.

Según AChecker la página web que presenta menores problemas de accesibilidad es ADS, siendo SIMBAD la que contiene mayores inconvenientes con respecto a los criterios de éxito.

Teniendo en cuenta los resultados presentados por FAE 2.0, en una inspección preliminar, aparentemente se obtiene un resultado similar. Sin embargo, si se presta atención a la cantidad de reglas no aplicables que detecta, se infiere que el porcentaje de criterios aplicados a ADS es mucho menor que el aplicado a las otras bases de datos. Se puede observar, además, que SDSS es la
base que presenta mayor cumplimiento de los criterios
analizados.

Examinator arroja como resultado que la página ADS es la que menos inconvenientes presenta, en concordancia con los resultados anteriores. Nuevamente se puede observar que SDSS es la que muestra más criterios cumplidos. Este analizador otorga además, puntuación
general de accesibilidad, donde se tienen en cuenta limitaciones como la falta de vista o problemas motores. Se evidencia que SIMBAD es la base menos accesible para personas con discapacidad.

Observando los criterios en los que fallaron las páginas, las tres tienen problemas con los siguientes lineamientos: ``1.3 Adaptable'', ``1.4 Distinguible'' y ``2.4 Navegable''. En cuanto a los demás criterios la mayoría debe evaluarse manualmente, como por ejemplo: ``1.2 Base de tiempo'', ``2.1 Teclado'' y ``2.2 Tiempo suficiente''. Resulta evidente que todos los criterios de éxito deberían reforzarse con un análisis manual.

\section{Resultados producto de los análisis normativos previos}

Los análisis normativos realizados en la sección previa dieron lugar a dos acciones inmediatas: la elaboración de recomendaciones para el desarrollo de herramientas accesibles, publicadas por el grupo de sonoUno en su página web oficial (ver sección \ref{sect:recomendaciones_publicadas}); y por otro lado el diseño de la interfaz gráfica en formato maqueta que se utilizó para el desarrollo del programa de sonorización sonoUno (ver sección \ref{sect:primer_diseño_GUI}).

\subsection{Recomendaciones de accesibilidad publicadas por el equipo de trabajo}
\label{sect:recomendaciones_publicadas}

En base a los análisis previos llevados a cabo con los diferentes estándar y la experiencia de diseño centrado en el usuario llevada a cabo desde el inicio del desarrollo del software sonoUno (llevado a cabo en el marco de esta tesis), se elaboró un documento con recomendaciones para el desarrollo de herramientas accesibles. El mismo consta de una lista de buenas prácticas que buscan priorizar una buena relación entre las interfaces humano-computadora y los usuarios finales, estas recomendaciones deben ser consideradas al momento de diseñar (y producir) una herramienta accesible. Cabe aclarar que estas recomendaciones también tienen contenido obtenido de la investigación con grupos focales realizada en el marco de esta tesis (este tema se explicará en el Capítulo \ref{cap:fg_completo}). Lo mencionado se debe a que este documento se actualizó luego de realizado dicho estudio, y la autora consideró pertinente describir aquí directamente la versión final de las recomendaciones de accesibilidad publicadas por el grupo de trabajo de sonoUno.

Las recomendaciones de accesibilidad, que se enumeran a continuación,  fueron escritas con el objetivo de promover la necesidad de buenas prácticas en inclusión y equidad que, normalmente, no forman parte de propuestas específicas de proyectos en ciencias o en propuestas relacionadas con desarrollos de software, búsqueda de datos, data mining o big data. Probablemente, los siguientes puntos puedan ser ampliados o mejorados, y deben ser considerados como un punto de partida para cada nuevo desarrollo:

\begin{enumerate}
    \item \textbf{La integración con tecnologías asistivas debe ser una prioridad desde el inicio de cualquier desarrollo o propuesta de trabajo o investigación en temas de software, asegurando que cada persona pueda utilizar la tecnología asistiva que mejor resuelva sus necesidades}. Para eso es muy importante preocuparse por la integración efectiva entre las tecnologías de asistencia y la interfaz gráfica de usuario (GUI) en desarrollo (ISO 9241-171:2008). El ítem 8.3.3, de las recomendaciones ISO indica: Evitar interferir con las funciones de accesibilidad (el software no debe deshabilitar o interferir con las funciones de accesibilidad de la plataforma).
    
    \item \textbf{Tratar de minimizar el tiempo para aprender una nueva herramienta y brindar independencia a las personas con discapacidad debe aceptarse como una buena práctica}. En este momento, aprender una nueva herramienta (una nueva tecnología de asistencia o una nueva interfaz) para personas con discapacidad significa mucho tiempo y ayuda de otras personas \citep{billahetal2017}. La arquitectura de back-end debe proporcionar una salida de front-end que permita una curva de aprendizaje baja hacia la identificación, interpretación y uso dinámico de funcionalidades; debe además, brindar independencia a las personas con discapacidades. Según el mandato de la ONU, los seres humanos tienen derecho a la igualdad, el sistema no debe etiquetar a las personas con discapacidad, sino que debe permitir que todos los usuarios usen el sistema para integrarse por igual.
    
    \item \textbf{Evitar ventanas emergentes}. \citet{bahr&ford2011} expresan: “los usuarios (participantes) consideraron las ventanas emergentes molestas y frustrantes y no las disfrutaron” [p.781]. Estos autores analizan la respuesta del usuario a las ventanas emergentes y destacan que debe haber una nueva forma de pensar sobre HCI. 
    El equipo de sonoUno propone el uso de paneles como alternativa a las ventanas emergentes, donde el usuario decide qué paneles mostrar. La técnica propuesta era mejor que las ventanas emergentes, pero el análisis mostró que todos los paneles abiertos al mismo tiempo son difíciles de navegar. Tal vez la navegación entre paneles con teclas de acceso directo solucione el problema, pero se requieren nuevos análisis con usuarios.
    
    \item \textbf{Las funcionalidades del software deben ser simples y presentar linealización entre sí}. En 2019 se realizó un estudio bajo la técnica de grupo focal en la Universidad de Southampton \citep{casadoFG2022}, los participantes detectaron la linealización del software sonoUno y lo compararon con otros software comerciales. Presentar un framework con una estructura familiar, lenguaje simple, consistencia entre vocabulario y linealización de funcionalidad, reduce la sobrecarga de memoria. Además, se concluyó que al minimizar el número de transacciones para realizar una tarea también se disminuye la sobrecarga de memoria.
    
    \item \textbf{La documentación, los tutoriales y los entrenamientos deben ser los temas principales}.
    La sección 11 de la norma ISO 9241-171:2008 describe el servicio de documentación, ayuda y soporte técnico en línea.
    La interfaz tiene que prever los aspectos de aprendizaje, recuperación de errores y soporte de ayuda. El back-end que maneja el soporte de ayuda, al momento de la visualización, debe seguir el foco principal de la interfaz para decidir, por ejemplo, cómo se organizan los elementos (en paralelo o en serie). 
    Se debe mantener la longitud y complejidad correspondientes, constancia uniforme de visualización, ser suficientemente descriptivo manteniendo un equilibrio con la existencia de palabras en lengua de signos y accesibilidad para personas con deficiencias/discapacidades de lectura y neurodiversas; todo lo mencionado busca evitar sobrecarga cognitiva, de memoria y convulsiones, entre otros factores. Al mismo tiempo, apuntan al complejo objetivo de tratar de mantener el aprendizaje, la ayuda y la recuperación de errores simples, mientras se abordan tantos aspectos de la interfaz como sea posible.
    Los participantes de los grupos focales \citep{casadoFG2022} expresan la necesidad de capacitarse sobre nuevas herramientas. Es crucial que se involucre al usuario desde el principio analizando en profundidad la información y siguiendo los procedimientos científicos de análisis de datos cualitativos para las decisiones con respecto a la interfaz. La situación ideal es que si esta parte está correctamente diseñada, los usuarios no necesitan que otras personas los ayuden permanentemente.
    
    \item \textbf{La interfaz gráfica debe ayudar al usuario a tener éxito y no sufrir fatiga, sobrecarga de memoria y cognitiva}, accediendo, realizando y completando las diferentes tareas \citep{casadoFG2022}.
    
    \item Para dar una imagen general y una imagen localizada de la interfaz: \textbf{la interfaz gráfica debe presentar diferentes opciones dinámicas para realizar y completar la tarea deseada por el usuario}, permitirle establecer dinámicamente diferentes configuraciones, decidir y cambiar a voluntad el camino seguido para realizar la tarea.
    Esto también se aplica para decidir la ruta de acción y para la recuperación de errores. Por ejemplo, en el caso de sonoUno, el usuario puede cambiar el timbre a voluntad para producir el sonido \citep{casadoFG2022}, y puede realizar cambios de modalidades de sonido en cada uno de los paneles de sonorización destinados al mapeo.
    
    \item \textbf{Deben evitarse los elementos innecesarios}, porque estos elementos confunden al usuario y congestionan la presentación visual. Es por eso que un análisis científico centrado en el usuario, detectando las visiones y expectativas de cada persona es de suma importancia.
    Se debe tener en cuenta que muchos elementos de la interfaz dificultan la navegación con tecnología de asistencia. Por ejemplo, en el caso de los lectores de pantalla, la tecnología de asistencia describe elemento por elemento y presenta la navegación en serie, por lo que pasar de una funcionalidad a otra dentro de una ventana con muchos elementos es muy difícil.
    Otro ejemplo es que las personas con algunas discapacidades de aprendizaje pueden enfrentar desafíos en cuanto a la percepción mental (tareas viso-espaciales) de patrones que pueden necesitar una referencia perceptiva en la interfase; según el modelo de código triple (verbal, simbólico y analógico) de estímulos cerebrales sensoriales, para contrastar el progreso en el análisis de la pantalla, para interpretar el progreso y apoyarse con el análisis.
    
    \item \textbf{La necesidad de precisión, aplicada al desempeño de tareas y al análisis de datos, es muy importante}. El usuario debe estar seguro de que está haciendo exactamente lo que pretende hacer y de que existen mecanismos para gestionar la incertidumbre; debe integrarse un modelo de gestión de la incertidumbre en la interfaz. Además, es muy importante que cada pantalla (auditiva, visual, táctil o combinación de cada una) esté en fase, sincronizada y activada a voluntad sin interferencia con la tecnología de asistencia, y que no exprese información desincronizada, desfasada y antagónica en la velocidad y complejidad decidida por el usuario.
    
    \item \textbf{Las pruebas con usuarios son muy importantes} \citep{metatleetal2015}, cada persona desarrolla diferentes técnicas para comunicarse y recibir información, además, cada una tiene diferentes enfoques del conocimiento y habilidades para procesar la información \citep{kavcic2005}.
    Un ejemplo de una técnica diferente para comunicarse con una interfaz es que las personas con problemas de motricidad fina necesitan elementos más grandes en la interfaz, porque señalar algo pequeño con un apuntador es muy difícil para ellos. Por otro lado, si los elementos que se muestran en la interfaz no están linealizados, obliga a las personas con discapacidades ortopédicas y neuromusculares a realizar movimientos físicos drásticos (por ejemplo, si `play' y `stop' están en diferentes lugares de la interfaz).
    Durante la investigación con grupo focal realizada por el equipo de trabajo \citep{casadoFG2022} se agregó una pregunta sobre la visualización 3D del gráfico desplegado por sonoUno y los participantes expresaron que esta forma alternativa de comunicación será útil quizás para personas que leen braille o personas con habilidades táctiles (este comentario se basa en que las personas que no leen braille, no tienen tan desarrollado su sentido del tacto). El equipo de sonoUno espera explorar más sobre esta suposición.
    
    \item \textbf{Es evidente que en muchos campos del abordaje de contenidos científicos de divulgación, los modelos táctiles son un apoyo para algunos públicos}. Tal es el caso de la exploración de estructuras de galaxias como lo llevan a cabo: Nicolas Bonne y ``The Tactile Universe'' \citep{tactileuniverse_web,bonneetal2018}; o Kimberly Arcand, con su obra ``Chandra Tactile Universe'' \citep{kimberlyetal_web}; alguna estrella definida \citep{madura2017}; o el descubrimiento táctil de las características de la superficie del planeta (A Touch of the Universe, \citet{ortiz_tactileuniv_web}), \citep{garciaetal2013,garciaetal2017}; entre otros ejemplos. 
    En estos días, muchas personas ciegas y con visión reducida (BVI, por sus siglas en ingles) no leen Braille; para ello hay que tener en cuenta que los textos escritos en ese formato no podrán ser utilizados por ellos. En este sentido, la información debe estar disponible a través de sonido, y los recursos habituales son los lectores de pantalla, que también deben evaluarse permanentemente.
    Este análisis es válido no solo para personas con discapacidad visual, el material propuesto debe ser adaptado para sordos, discapacitados motores, multidiscapacitados y neurológicamente diversos. Los individuos viven realidades múltiples que en muchos casos se combinan. Las discapacidades de una persona pueden ser permanentes o temporales, congénitas o adquiridas, reveladas o no reveladas.
    
    \item \textbf{Necesidad de explorar lo desconocido}. En general, las personas con discapacidad no tienen la oportunidad de explorar. Durante la sesión del grupo focal \citep{casadoFG2022} después del uso de sonoUno, las personas con discapacidad visual en diferentes aspectos de la alfabetización informática y la experiencia científica expresaron la necesidad de poder explorar los datos con precisión, certeza, efectividad, eficiencia y en un tiempo razonable.
    
    \item Relacionado con el punto anterior, \textbf{el usuario debe estar informado sobre todos los cambios} (los cambios pueden deberse a actualizaciones, elementos en la interfaz, elementos en los menús y en los submenús, ubicación y descripción de las funcionalidades a las que accede, escenario como un todo, etc.). Cualquier cambio, incluidos los mínimos, debe ser informado en cada despliegue sensorial (auditivo, visual, táctil o combinación de cada uno). Además, el usuario debe ser capaz de ubicarse y orientarse espacialmente en la interfaz.
    
    \item \textbf{No se deben hacer suposiciones}, cada suposición dejará atrás a personas con discapacidad o diferentes estilos de aprendizaje, quedando fuera de las consideraciones iniciales.
    Por ejemplo, el supuesto de que cada persona usa la computadora, excluye a las personas que no tienen esta capacidad \citep{casadoFG2022}. La suposición de que todos pueden interpretar la información mostrada siguiendo las mismas pistas multisensoriales deja fuera a las personas con sesgos sensoriales.
    
    \item \textbf{Se deben implementar enfoques multisensoriales y multiplataforma porque no todas las personas pueden acceder a todas las plataformas}. Por ejemplo, incluso si Linux es gratuito, las personas que usan la biblioteca de su universidad deben adaptarse a la plataforma disponible en este sitio. En este enfoque, las tecnologías de asistencia y sonoUno se han desarrollado teniendo en cuenta este marco.
    
    \item \textbf{La interfaz debe permitir a los usuarios deshacer y recordar sus acciones} (ISO 9241-171:2008).
    
    \item \citep{alonsoetal2008} enuncian algunos requisitos de usabilidad: que van desde la \textbf{adecuación de la tarea} (la tarea debe ser competente de acuerdo con las herramientas disponibles para el usuario, por ejemplo, se debe tener en cuenta que las personas con discapacidad visual están obligadas a comunicarse en serie con la computadora); \textbf{compensación dimensional} (la interfaz debe presentar equilibrio entre diferentes estilos sensoriales); \textbf{equivalencia de comportamiento} (todos los estilos sensoriales deben presentar la misma información, por ejemplo, todos los elementos visuales deben presentarse con la opción de un despliegue auditivo que proporcione la misma información); \textbf{evitar pérdidas semánticas} (se debe presentar información relevante en todos los estilos sensoriales); \textbf{independencia del dispositivo} (la interfaz debe funcionar con una amplia gama de tecnologías de asistencia y de sistemas operativos).
    El estudio anterior también concluyó que los usuarios novatos quieren toda la información, mientras que los usuarios expertos solo quieren lo que necesitan. La interfaz debe considerar la opinión del usuario y permitirle elegir qué y cómo mostrar la información.
    
    \item \citet{mullikenfalloon2018} realizaron un estudio sobre sitios web de bibliotecas con personas con discapacidad visual y algunas de las conclusiones fueron: la primera vez que visitan el sitio web consumen más tiempo porque tienen que explorar la página web; la inconsistencia de información entre diferentes bibliotecas web dificulta la tarea; y el usuario espera que la tecnología de asistencia (lector de pantalla en este caso) comunique toda la pantalla; entre otros. En las interfases digitales, \textbf{el usuario necesita ser informado de los cambios de escenario ya sean totales o parciales, orientación, movilidad, y debe poder ejercer su libre albedrío para decidir dinámicamente la ruta de interacción}.
    
    \item \textbf{Es de suma importancia evaluar las leyes de accesibilidad digital existentes}, la raíz de esas leyes, el rigor y la aplicación gubernamental de sus criterios de éxito para lograr todos los objetivos de este documento cumpliendo, a la vez, con esas leyes (hasta ahora se han encontrado 25 países con Leyes sobre los derechos de las personas con discapacidad).
    
    \item \textbf{El usuario tiene que guiar estrictamente el desarrollo; y no el pensamiento y las suposiciones de un equipo de desarrollo o las posibilidades de las plataformas de codificación}. El equipo de desarrollo tiene que encontrar una manera para que los usuarios guíen el desarrollo, logrando así que las personas con discapacidad tengan acceso a la misma cantidad y calidad de información que los pares sin discapacidad.
    
    \item \textbf{Es fundamental que las personas se sientan importantes, y que su opinión sea valorada por el desarrollador}, porque las personas con discapacidad están acostumbradas a no ser tenidas en cuenta.
    
    \item \textbf{Finalmente, hacer un diseño centrado en el usuario significa escucharlo y estar seguro de que la interfaz es usable}. No se deben hacer suposiciones, es una mejor práctica, y aunque no parezca conlleva menor tiempo, realizar un análisis de grupos focales o entrevistas con usuarios potenciales y analizar posibles soluciones.
    
\end{enumerate}

Para cerrar esta serie de recomendaciones se desea remarcar que cualquier persona puede desarrollar una discapacidad en cualquier momento de su vida. Las discapacidades pueden desarrollarse independientemente del patrimonio cultural, las personas de diferentes grupos raciales, etnias, grupos religiosos, miembros de la comunidad LGBTIQ+, diferentes géneros, identidades sexuales, grupos lingüísticos, edades, estatus socioeconómico, otros grupos subrepresentados y las intersecciones de cada una de ellas. Cualquier desarrollo debe consultar estricta y continuamente a diferentes miembros de la comunidad con y sin discapacidad buscando que sean protagonistas, propongan y supervisen la generación de los nuevos recursos y herramientas para la inclusión.

En esta era donde las computadoras son tan poderosas, los sistemas e interfaces que se ajustan a cada persona (incluso aquellas que presentan diversidad funcional) no deberían ser una utopía. Con este fin, la actualización de recomendaciones (como las que se presentan aquí), buenas prácticas y marcos de trabajo basados en estudios centrados en el usuario, además de técnicas de aprendizaje automático, deben ser promovidas.

\subsection{Primer diseño de interfaz centrada en el usuario}
\label{sect:primer_diseño_GUI}

En base al análisis del cumplimiento del estándar ISO 9241-171:2008, parte del análisis de las normativas web (WCAG 2.0) y las recomendaciones elaborados por el equipo de trabajo del software sonoUno, se diseñó una propuesta de nueva interfaz. Se buscó que esta interfaz permitiera al usuario, luego de ser implementada, manipular las diferentes funcionalidades para realizar las tareas de carga y sonorización de datos sin tener que hacer largos recorridos. Se describirá el proceso de diseño desde la primer propuesta realizada en papel, hasta la propuesta final que se implementa en la primer versión de sonoUno.

\subsubsection{Primer diseño propuesto}

\begin{figure}[ht!]
    \centering
    \includegraphics[width=1\textwidth]{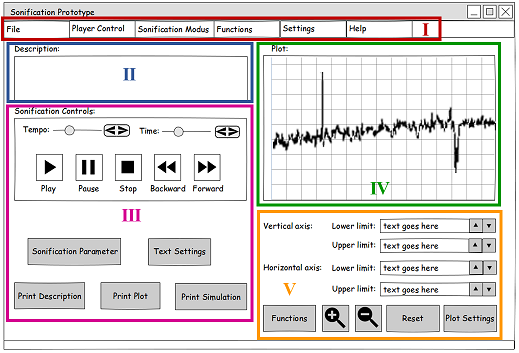}
    \caption{Propuesta de interfaz gráfica para el software sonoUno, diseñada a partir del análisis de normas realizado en el presente Capítulo e intercambios con usuarios durante la trayectoria de esta tesis doctoral.}
    \label{fig:propuesta_gui}
\end{figure}

La interfaz gráfica propuesta como primer aproximación, luego del primer análisis de software de sonorización con la norma ISO, se muestra en la Figura \ref{fig:propuesta_gui}. Este primer diseño buscó ofrecer al usuario la mayor cantidad de herramientas, tratando de no saturar el despliegue y tomando como referencia otros desarrollos análogos existentes en 2017 (como los ya descriptos Sonification Sandbox, MathTrax y xSonify). 

Inicialmente, y como puede observarse, se dividió la pantalla en dos grandes sectores, uno dedicado a las configuraciones de descripción y sonorización (II y III) y otro dedicado a la gráfica y sus características (IV y V). Se consideró que las configuraciones iniciales y el ingreso de datos se podía realizar a través de la barra de herramientas, contando cada funcionalidad con su atajo de teclado (I). A continuación se describirá cada una de las secciones disponibles en el menú:

\begin{itemize}
    \item \textbf{`Archivo' (File)} mostrará las opciones para crear una nueva plantilla, cargar el trabajo anterior, guardar lo que se está haciendo, importar datos, imprimir archivos (descripciones, gráficos y video de simulación) y salir del software. 
    \item \textbf{`Control de reproducción' (Player Control)} contendrá acciones similares a las de la sección III de la interfaz (recuadro magenta, ubicado en la parte inferior izquierda). Estas acciones se refieren a herramientas que adaptan la salida sonora a las necesidades del usuario y permiten la reproducción de la sonorización.
    \item \textbf{`Modo de sonorización' (Sonification Modus)} mostrará los parámetros de sonorización, a los que se puede acceder con el botón \textit{`Parámetros de sonorización' (Sonification Parameter)} disponible en la interfaz. Aparecerá una ventana emergente, la cual tendrá el foco del teclado mientras el usuario hace los ajustes necesarios, luego al cerrar la ventana el foco del teclado volverá al lugar donde estaba en la ventana principal antes de la interrupción. 
    \item \textbf{`Funciones' (Functions)} contendrá las funciones matemáticas que se podrán aplicar a los datos, para realizar un procesamiento. Se proyecta la posibilidad de que esta sección cuente con un puente a algún lenguaje de programación que permita una manipulación de datos más profunda.
    \item \textbf{`Configuraciones' (Settings)} permitirá al usuario tener cierto control sobre la forma visual de la interfaz, modificar colores, tamaños, tal vez agregar funciones que generalmente solo se usan en casos específicos. Se propone que el usuario, dentro del alcance del software, pueda modificar la interfaz gráfica agregando o quitando partes para hacerla más accesible a sus propias necesidades.
    \item \textbf{`Ayuda' (Help)}, aquí se debe describir las funcionalidades del software y cómo acceder a ellas, las modificaciones que se pueden realizar en la interfaz gráfica y cómo hacerlo, una descripción detallada de los iconos y botones del interfaz. También sería muy útil presentar en esta parte una forma de contacto con los desarrolladores o personal técnico, en caso de que haya problemas específicos que no estén contemplados en la documentación.
\end{itemize}

En cuanto a la parte visual del diseño, contiene una sección de descripción (Figura \ref{fig:propuesta_gui} - II, recuadro azul ubicado al centro a la izquierda) que coloca la información en texto que puede contener el archivo de datos a analizar, de esa forma se guarda en un lugar accesible donde el usuario puede leerlo o reproducirlo con un lector de pantalla tantas veces como sea necesario. El botón `Configuración de texto' (disponible en la sección III del diseño) permite modificar el formato del texto.

La sección control de sonorización (Figura \ref{fig:propuesta_gui} - III, recuadro magenta ubicado en la parte inferior izquierda), además del botón mencionado en el párrafo anterior, permite controlar la sonorización de datos que se realiza de forma sincrónica con la posición indicada en el gráfico. En la parte inferior de esta sección se encuentran los botones que permiten imprimir los resultados obtenidos.

Por su parte, la sección de la derecha contiene el gráfico (Figura \ref{fig:propuesta_gui} - IV, recuadro verde ubicado en la parte central derecha), y configuraciones referentes al mismo (Figura \ref{fig:propuesta_gui} - V, recuadro amarillo ubicado en la parte inferior derecha). Dentro de las opciones gráficas encontramos: botones para administrar los límites de ese gráfico (para que pueda seleccionar partes específicas de los datos), el botón `Funciones' (Functions) para procesar los datos si es necesario, botones para aumentar o disminuir el tamaño del gráfico, el botón `Reset' para volver a la configuración por defecto y `Configuraciones de gráfico' (Plot Settings) para modificar las características del mismo.

En este diseño de la interfaz existe una redundancia en cuanto a las rutas de acceso referente a un mismo proceso, esto tiene como finalidad que los usuarios puedan utilizar la forma que les resulte más accesible para realizar la tarea. Este detalle en el diseño es ampliamente utilizado para lograr una aplicación centrada en el usuario.

\subsubsection{Actualización del diseño para la nueva versión de sonoUno}

\begin{figure}[ht!]
    \centering
    \includegraphics[width=1\textwidth]{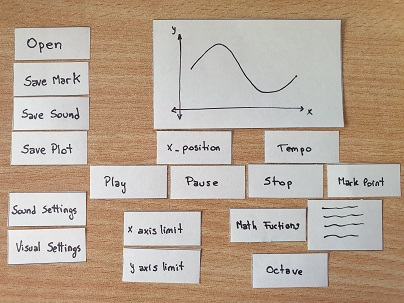}
    \caption{Proceso de rediseño de la interfaz para sonoUno, diseñada durante reuniones en un intercambio realizado en la oficina de IAU-OAD Sudáfrica [2018]}
    \label{fig:mock-up_gui}
\end{figure}

Luego de la primer versión de diseño de interfaz, se realizó un intercambio con diferentes investigadores en la Oficina de Astronomía para el Desarrollo (IAU-OAD) con sede en Sudáfrica en el año 2018. Durante dicha estadía se desarrolló un marco de trabajo especial y consensuado centrado en el usuario con discapacidad visual, del cual decantaron las recomendaciones presentadas en la sección \ref{sect:recomendaciones_publicadas} y una maqueta (Figura \ref{fig:mock-up_gui}) que permitió ir moviendo las funcionalidades para buscar el mejor diseño agrupándolas en secciones según su finalidad. La Figura \ref{fig:mock-up_gui} muestra el diseño final logrado con la maqueta, que sirvió de base para las actualizaciones del software.

\begin{figure}[ht!]
    \centering
    \includegraphics[width=1\textwidth]{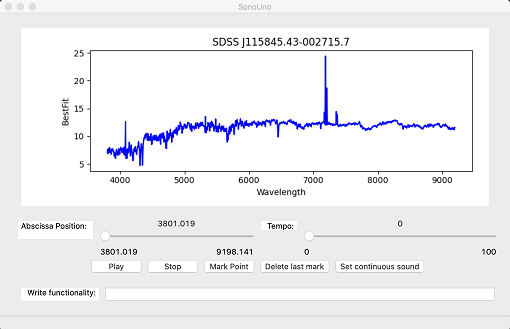}
    \caption{Actualización de la propuesta de interfaz gráfica para el software sonoUno, diseñada a partir del análisis de normas realizado en el presente Capítulo.}
    \label{fig:propuesta_final_gui}
\end{figure}

Con dicho diseño en mente se elaboraron los paneles que contendrían cada grupo de funcionalidades y se llegó al primer diseño de interfaz que puede observarse en la Figura \ref{fig:propuesta_final_gui}. Se detalla el proceso de elaboración del programa sonoUno en el Capítulo \ref{cap:sonouno_v1}. Luego, en el Capítulo \ref{cap:reinforce} se describen las actualizaciones realizadas al mismo y algunos ejemplos de uso con datos astronómicos y astrofísicos.

\section{Conclusiones}

Con el propósito de reconocer las necesidades de los usuarios y los problemas a contemplar para la sonorización y el desarrollo de nuevas herramientas para el análisis de datos usando sonido, dos grupos independientes de evaluadores fueron formados. Las personas involucradas contaban con diferentes antecedentes académicos y además, se involucró a personas con y sin discapacidad. Ambos grupos analizaron el cumplimiento y usabilidad de algunos software, siguiendo la norma ISO 9241-171:2008 sobre accesibilidad. Este análisis mostró la dificultad de aplicar este tipo de herramienta de manera objetiva: el análisis punto por punto de la norma ISO 9241-171:2008 mostró que cada analista puede tener un enfoque diferente de las recomendaciones y, en este sentido, los estudios de accesibilidad finales pueden presentar una parcialidad. 

Sin embargo, dicho estudio permitió inferir que incluso con baja accesibilidad (como se evidencia en la evaluación ISO), xSonify permite el análisis de diferentes conjuntos de datos a través del sonido y la visualización con una mejor interacción entre el usuario y la computadora (comparado con los otros dos desarrollos analizados). La principal diferencia de xSonfy es la posibilidad de tener todas las funcionalidades en el mismo espacio de trabajo, incluida la posibilidad de aplicar funciones matemáticas, por lo que se lo consideró un buen punto para iniciar nuevos desarrollos en la sonorización de datos científicos.

Por otro lado, teniendo en cuenta las herramientas evaluadoras de páginas web que siguen las recomendaciones de la norma W3C, las mismas resultaron adecuadas para el estudio preliminar de accesibilidad de bases de datos astronómicos, reducen el número de criterios que se deben analizar manualmente y entregan un informe detallado de los errores. Como estudio para evaluar la realidad a la que se enfrentan las personas con discapacidad a la hora de descargar información científica se considera suficiente. Sin embargo, si se quisiera promover el desarrollo de una base de datos para el acceso a la información científica, se deben analizar los criterios que quedaron excluidos, ya sea por su especificidad o por la necesidad de ser evaluados por un analista o por el usuario. Un ejemplo de un criterio que FAE 2.0 indicó que necesitaba revisión manual es que los videos que contenía la web debían tener descripción audible.

A partir del estudio presentado en este Capítulo se evidenció que los software disponibles para análisis de datos astronómico en 2017 no presentaban la accesibilidad que requerían los usuarios. Como las ciencias de la computación y la gestión de big data son parte del enfoque actual de los científicos para estudiar la naturaleza, un software centrado en el usuario, de código abierto, disponible para todas las plataformas, accesible incluso a través de Internet, que permite analizar datos con los mismos recursos que las herramientas tradicionales en astrofísica, parece ser el camino a seguir para un futuro nuevo diseño de software.

Este Capítulo también señala que un estándar no es suficiente para hacer que una herramienta sea accesible para personas con diversidad funcional. Se necesita capacitación sobre software accesible y retroalimentación constante con los usuarios finales para tener éxito. Una de las novedades de este trabajo de doctorado es que el actor principal del estudio es la voz de los usuarios y evaluadores con discapacidad, apoyada con una evaluación más profunda dirigida hacia un cambio de mentalidad, cambio social e institucional. La exploración multisensorial que aquí se propone no solo traerá personas con discapacidades al campo de las ciencias, sino que también aumentará la cantidad de descubrimientos por parte de científicos espaciales capacitados.

Cabe destacar que del estudio presentado en este capítulo se realizaron dos publicaciones en congresos nacionales (XXI Congreso Argentino de Bioingeniería (SABI-2017)) (Apéndice \ref{ap:A_papers}) \citep{sabi2017} y Reunión anual de la Asociación Argentina de Astronomía 2017 (Apéndice \ref{ap:B_proceeding}) \citep{casadoetal2018}, una en un congreso internacional (Simposio Internacional de Educación en Astronomía y Astrobiología (ISE2A)) (Apéndice \ref{ap:A_papers}) \citep{ise2a2017} y un paper publicado en revista (\textit{``International Journal of Sociotechnology and Knowledge Development''}) (Apéndice \ref{ap:A_papers}) \citep{ijskd2022}.

Los resultados presentados en el presente capítulo se utilizarán como base para todo el desarrollo posterior del software sonoUno. La investigación de usabilidad realizada a través de un grupo focal y las propuestas de nueva interfaz y sonorizaciones, que llevarán, como mostraremos, al desarrollo más allá de los datos astronómicos clásicos, pudiendo adaptar sus funcionalidades para el diseño de scripts que permitan sonorizar datos de grandes facilidades científicas, como son el caso del Gran Colisionador de Hadrones o los interferómetros para detección de ondas gravitacionales.

\chapter{Primera versión de sonoUno}
\label{cap:sonouno_v1}

\section{Introducción}

El ser humano por naturaleza explora el mundo a través de todos sus sentidos, sin embargo, existe un marcado predominio en el uso de visualizaciones para dar sentido a los conjuntos de datos en estudio. Este es el caso, incluso en astronomía, donde la mayoría de los datos bajo estudio están fuera del rango visible se encuentra evidencia sobre los beneficios del despliegue auditivo como complemento del despliegue visual \citep{wandatesis2013}.

Como ya se ha mencionado anteriormente, el uso de sonorización en astronomía existe desde hace años, algunos ejemplos recientes que aún no han sido mencionados son:
\begin{itemize}
    \item La sonorización del conjunto de datos astronómicos zCOSMOS, donde los autores describen el conjunto de datos y la estrategia de sonorización utilizada \citep{bardellietal2021};
    \item LightSound, un dispositivo electrónico que permite convertir la luz en sonido, se utiliza, principalmente,  para observar eclipses \citep{lightsound2020};
    \item Una plataforma de sonorización construida en colaboración con el equipo de ATLAS Outreach y una página web que permite al público en general escuchar experimentos en tiempo real \citep{cherstonetal2016};
    \item El software Quasar Spectroscopy Sound centrado en la sonorización de conjuntos de datos cosmológicos, la técnica utilizada es descripta por \citet{hansenetal2020};
    \item Un proyecto con la refuncionalización de dos aceleradores de partículas cósmicas donde se describió el uso de despliegue auditivo en conjunto con despliegue visual, el objetivo era hacer más accesibles los descubrimientos, la sonorización fue hecha con características musicales y desarrollada específicamente para este caso de estudio \citep{ohmetal2021};
    \item \citet{quintonetal2021} describe un diseño de sonorización de órbitas planetarias en cinturones de asteroides.
\end{itemize}

Sin embargo, en la mayoría de los casos, el mapeo llevado a cabo para obtener la sonorización del conjunto de datos ha sido definido por su creador y compartido como producto final. Es de interés para esta Tesis, mencionar la conferencia internacional en despliego auditivo (ICAD por sus siglas en inglés), la cuál tiene como finalidad, desde su creación en 1992, reunir a expertos multidisciplinarios que trabajan en el campo de la sonorización.

Dicha conferencia presenta un repositorio donde hay una gran cantidad de trabajos que \citet{andreopoulouYgoudarzi2021} agruparon en seis categorías: métodos de sonorización, herramientas/sistema de sonorización, estudios de revisión/opinión, estudios exploratorios, estudios de percepción/evaluación y otros. Esta revisión sistemática del repositorio ICAD destaca un alto porcentaje de artículos dedicados a métodos y herramientas de sonorización, en contraste con un bajo porcentaje de artículos dedicados a metodologías de diseño, estudios de percepción y métodos de evaluación. Es alarmante que los estudios de percepción presenten un crecimiento entre 2005-2009, pero disminuyan por debajo del 1\% al 2019; incluso cuando \citet{fergusonYbrewster2017} señalan la importancia de los estudios de percepción en los despliegues auditivos y reportan algunos parámetros psicoacústicos mencionando cómo las personas lo perciben.

\citet{supper2014} expresa preocupación acerca de que el sonido es un medio inmersivo y emocional en contraste con el despliegue visual, pero se debe tener en cuenta que aprendemos a estudiar conjuntos de datos visualmente y a separarlos de las pinturas artísticas prácticamente desde que nacemos; por lo que esta idea solo refuerza la necesidad de desarrollos de sonorización robustos y estudios de percepción para aprender cómo las personas entienden los parámetros acústicos, además de capacitaciones para aprender a escuchar y comprender conjuntos de datos. En la misma línea de Supper, \citet{neuhoff2019} también describe este desafío concerniente a la sonorización, compara a nivel perceptivo el despliegue visual y auditivo y, además, expone la existencia de un alto porcentaje de trabajos que utilizan la palabra música, entre todas las presentaciones del ICAD. ¿Es posible en solo 20 años de investigación y sin entrenamiento durante la etapa escolar, realmente separar y tratar de forma diferenciada a la música y la sonorización aplicada a la investigación?

En este sentido y teniendo en cuenta los crecientes ejemplos de sonorización en astrofísica, el software sonoUno tiene como objetivo proporcionar una plataforma de código abierto que permita a los usuarios abrir diferentes conjuntos de datos y explorarlos a través de un despliegue visual y auditivo, permitiendo ajustar la configuración visual y sonora para mejorar su percepción, ajustándose de esta forma al destinatario. Este proyecto está centrado en el usuario desde el principio y sigue los lineamientos presentados en el Capítulo \ref{cap:analisis_normativo}, donde se describió el estudio normativo realizado con software existentes al momento de inicio de este desarrollo en 2017, dando lugar al diseño de un marco de trabajo basado en dicho análisis y una búsqueda bibliográfica centrada en el uso de interfaces por parte de personas con discapacidad visual. Todo lo mencionado dio lugar al diseño de una interfaz gráfica de usuario en formato maqueta presentada en la sección \ref{sect:primer_diseño_GUI}. En el presente capítulo se describirá la estructura y el proceso de desarrollo del código que da funcionalidad al programa sonoUno, finalizando con ejemplos de uso con datos astrofísicos como rayos cósmicos y galaxias. Se agregan dos publicaciones hechas en congreso sobre este tema en el Apéndice \ref{ap:pub_sonouno_desktop} \citep{casadoetal2019,wdea2019}.

\section{Desarrollo de software}

El desarrollo propiamente dicho del software se dividió en diferentes etapas, la primera fue el diseño centrado en el usuario de la interfaz, cuya primera parte fue presentada en el Capítulo \ref{cap:analisis_normativo} (sección \ref{sect:primer_diseño_GUI}), por lo que aquí se continuará con la explicación partiendo de la maqueta de interfaz gráfica presentada y describiendo la interfaz funcional que permite la interacción del usuario con la herramienta y los datos. 

Otra etapa del desarrollo de software fue la planificación del diseño modular pensando tanto en las funcionalidades actuales como posibles funcionalidades a futuro, y la selección de herramientas y librerías con las cuales se trabajó. Finalmente, se realizó la programación de la primer versión de sonoUno, el detalle de las funcionalidades se explicará en la subsección correspondiente.

\subsection{Diseño centrado en el usuario}

\begin{figure}[ht!]
    \centering
    \includegraphics[width=1\textwidth]{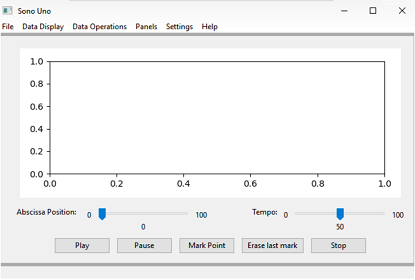}
    \caption{Interfaz gráfica de la versión prototipo de sonoUno, presentada en 2019.}
    \label{fig:cap4_initialwin}
\end{figure}

Partiendo del diseño de interfaz expuesto en la Figura \ref{fig:propuesta_gui}, la maqueta producto del análisis bibliográfico (ver Figura \ref{fig:mock-up_gui}) y todo el análisis normativo del Capítulo \ref{cap:analisis_normativo}, se materializó la interfaz gráfica de la Figura \ref{fig:cap4_initialwin}. Adicionalmente, se construyó la Tabla \ref{tab:agrupacion_funcionalidades} para categorizar las funcionalidades del programa en cuatro grupos principales, manteniendo una buena linearización de acuerdo a lo indicado en las recomendaciones de accesibilidad expuestas en la sección \ref{sect:recomendaciones_publicadas}. Cabe destacar que la versión de escritorio de SonoUno por el momento solo despliega el idioma inglés.

\begin{table}
    \centering
    \begin{tabular}{|c|c|c|c|}
        \hline
        \textbf{Despliegue} & \textbf{Operaciones} & \textbf{Configuraciones} & \textbf{Opciones} \\ 
         &  &  & \textbf{de I/O} \\ \hline \hline
        Posición de & Selección de & Sonido & Abrir/Importar \\
        abscisas & límites &  &  \\ \hline
        Tempo & Función & Gráfico & Guardar \\
         & cuadrática &  & las marcas \\ \hline
        Reproducir & Función &  & Guardar \\
         & logaritmo &  & el sonido \\ \hline
        Pausa & Invertir eje & & Guardar \\
         & de ordenada & & el gráfico \\ \hline
        Marcar punto & Función & & Borrar todas \\
         & promedio & & las marcas \\ \hline
        Borrar última & Octave & & Cerrar \\
        marca & & & \\ \hline
        Detener & & & \\
         & & & \\ \hline
    \end{tabular}
    \caption{Agrupación de funcionalidades por similitud teniendo en cuenta el objetivo o aplicación de cada una.}
    \label{tab:agrupacion_funcionalidades}
\end{table}

Cuando se abre la aplicación, la primer ventana que se muestra (Figura \ref{fig:cap4_initialwin}) contiene un menú con seis elementos y las funcionalidades correspondientes al despliegue de datos. En cuanto al menú, el mismo presenta seis elementos, que se corresponden con los grupos de funcionalidades definidos en la Tabla \ref{tab:agrupacion_funcionalidades}: `Despliegue' se corresponde con `Data Display'; `Operaciones' con `Data Operations'; `Configuraciones' con `Settings'; y `Opciones de I/O' con `File'; los dos restantes no tienen correspondencia directa, pero el menú `Panels' permite abrir y cerrar cada uno de los cuatro paneles, y finalmente el menú `Help' permite acceder a la ayuda y manual referente al programa.

En cuanto a los elementos (como botones y gráfico) que se muestran en la ventana, todos corresponden al grupo `Despliegue' y se encuentran ubicados en un panel (elemento de la interfaz gráfica que permite agrupar diferentes componentes de interacción como los botones, y a su vez puede ser mostrado u ocultado). Las demás funcionalidades están ubicadas en los paneles correspondientes y se encuentran inicialmente ocultas, pueden ser mostradas por el usuario con el atajo de teclado correspondiente o desde el menú, ya sea seleccionando la funcionalidad deseada o habilitando el panel desde `Panels'. La razón por la que inicialmente solo se muestran las funcionalidades correspondientes al despliegue visual y sonoro es la necesidad de una interfaz inicial simple y con los elementos básicos, de esta forma no se satura al usuario con información y se le permite ir explorando y aumentando el nivel de complejidad de la interfaz de acuerdo a sus necesidades.

\begin{figure}[ht!]
    \centering
    \includegraphics[width=1\textwidth]{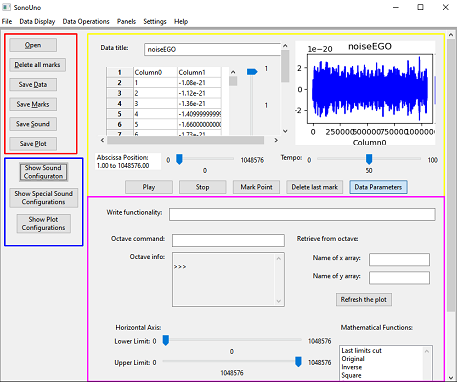}
    \caption{Interfaz gráfica con todos los paneles principales habilitados y recuadrados con distintos colores.}
    \label{fig:cap4_gui_allpanels}
\end{figure}

\begin{figure}[ht!]
    \centering
    \includegraphics[width=1\textwidth]{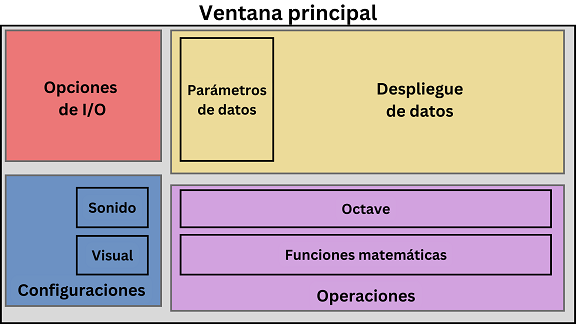}
    \caption{Diagrama con la disposición de los paneles existentes en sonoUno.}
    \label{fig:cap4_diagrama_paneles}
\end{figure}

La Figura \ref{fig:cap4_gui_allpanels} muestra la interfaz gráfica con todos los paneles principales desplegados y algunos de los paneles secundarios (se les llama así a los paneles que se encuentran dentro de paneles, por ejemplo el panel `Data parameters' o `Parámetros de datos' que se incluye dentro del panel principal `Data display' recuadrado en amarillo). Luego, la Figura \ref{fig:cap4_diagrama_paneles} muestra un diagrama con todos los paneles disponibles en la interfaz gráfica de sonoUno, dichos paneles serán explicados en cuatro subsecciones a continuación.

\subsubsection{Panel `Despliegue'}

Este panel contiene los elementos de reproducción, visibles al iniciar el programa, y los elementos referentes a los datos ingresados en un subpanel específico llamado `Parámetros de datos'. En las Figuras \ref{fig:cap4_gui_allpanels} y \ref{fig:cap4_diagrama_paneles} está ubicado en la parte superior derecha, corresponde al recuadro amarillo. 

En cuanto a los elementos referentes a los datos, los mismos incluyen el título (que puede modificarse), la grilla que muestra los datos (con la opción de modificar títulos de columnas) y la selección de ejes a graficar teniendo como opción todas las columnas del archivo de datos. Se ubicó esta sección en un panel aparte para poder ocultarse y mostrarse según la necesidad del usuario, sin embargo, se mantuvo dentro del panel principal de `Despliegue' debido a que es totalmente dependiente de los datos. La descripción detallada sobre la función de cada elemento será presentada en la sección \ref{sect:cap4_descrip-func}.

\subsubsection{Panel `Opciones de I/O'}

Este panel contiene los elementos de entrada y salida en formato botones, con la misma funcionalidad y nombre que los encontrados en el menú `File' o `Archivo'. Se considera importante incluir estos botones porque personas con movilidad reducida pueden acceder mejor a botones, los menú requieren motricidad fina y sin movimientos espasmódicos. En las Figuras \ref{fig:cap4_gui_allpanels} y \ref{fig:cap4_diagrama_paneles} está ubicado en la parte superior izquierda, corresponde al recuadro rojo. 

\subsubsection{Panel `Operaciones'}

Este panel contiene los elementos que permiten aplicar operaciones matemáticas predefinidas, recorte de ejes y la posibilidad de utilizar comandos de octave. En las Figuras \ref{fig:cap4_gui_allpanels} y \ref{fig:cap4_diagrama_paneles} está ubicado en la parte inferior derecha, corresponde al recuadro rosa. Particularmente, en la Figura \ref{fig:cap4_diagrama_paneles} se puede observar el panel principal `Operaciones' y los dos paneles secundarios que lo componen.

La funcionalidad de Octave se colocó en un panel particular debido a la especificidad de su tarea, que posibilita escribir comandos de octave en el cuadro de texto etiquetado como `Octave command' (el segundo cuadro dentro del panel `Operaciones') y los envía para ser procesados en el entorno de octave. Durante el proceso de trabajo en octave, en el cuadro de texto no editable `Octave info' (tercer cuadro dentro del panel `Operaciones') se muestran todos los mensajes devueltos por octave ante el comando ejecutado. Al terminar el trabajo con los datos en octave, se deben devolver los datos a graficar indicando qué datos corresponden al eje x y cuáles al el eje y, para lo que debe utilizarse el espacio a la derecha de los cuadros de texto etiquetado como `Retrieve from octave'. Debido a que este espacio solo será utilizado por personas interesadas en usar octave para el procesamiento de datos es que se decidió incluirlo en un panel oculto, el cual puede mostrarse utilizando el atajo de teclado o el menú.

Por otro lado, en un panel aparte se colocan las dos barras deslizables para modificar el máximo y mínimo valor de x (permitiendo así seleccionar un rango específico del eje de abscisas para sonorizar), junto con una lista de funciones matemáticas predefinidas que pueden utilizarse en los datos (por ejemplo: función inversa, cuadrática y suavizado, entre otras). Estas últimas se incluyeron en un mismo panel por simplicidad para el usuario y porque todas producen un cambio inmediato en el despliegue visual y sonoro de los datos.

\subsubsection{Panel `Configuraciones'}

Este panel contiene los elementos de configuraciones tanto sonoras como visuales. En las Figuras \ref{fig:cap4_gui_allpanels} y \ref{fig:cap4_diagrama_paneles} está ubicado en la parte inferior izquierda, corresponde al recuadro azul. Debido a que el sonido cuenta con configuraciones más sencillas como cambios en los niveles máximos y mínimos de frecuencia o volumen, y cambios en la selección de instrumento, se consideró dividir las configuraciones de sonido en dos paneles secundarios. Mientras que los cambios visuales del gráfico son solo referentes al despliegue visual de datos, por lo que se incluye un solo panel secundario. El programa en su versión actual aún no permite modificaciones visuales de la interfaz gráfica a realizar por el usuario, el diseño se adapta a las configuraciones del sistema operativo donde se corre el sonoUno (por ejemplo, si se usa con la configuración de vista nocturna, el fondo de la interfaz gráfica es negro o gris oscuro). Se detallaran de forma extensa las opciones de configuración en la sección \ref{sect:cap4_descrip-func}.

\subsection{Diseño modular e implementación}

\begin{figure}[ht!]
    \centering
    \includegraphics[width=0.7\textwidth]{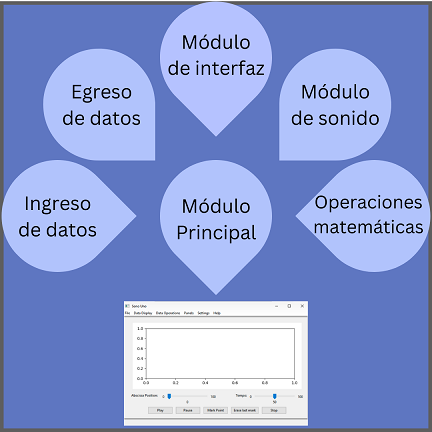}
    \caption{Diseño modular de sonoUno.}
    \label{fig:cap4_dismodular}
\end{figure}

Una vez diagramada la interfaz gráfica y decidido su diseño, se comenzó con la programación del software sonoUno. Pensando en un proyecto que permitiera incorporación de personal a largo plazo para su desarrollo y mantenimiento, se realizó un diseño modular desde el principio, siguiendo la experiencia compartida en documentos referentes al programa xSonify. Este tipo de diseño en desarrollo de software permite que múltiples desarrolladores trabajen en diferentes secciones del programa, siendo luego su integración más ágil y práctica. El diseño modular elaborado para esta primer parte del desarrollo se muestra en la Figura \ref{fig:cap4_dismodular}.

Si se observa el Cuadro \ref{tab:agrupacion_funcionalidades}, al compararlo con este diseño modular se evidencia que los módulos están íntimamente relacionados con la agrupación de funcionalidades. El módulo \textit{`Ingreso de datos'} permite la búsqueda en el sistema de un archivo de datos formato tabla (con extensión txt o csv) para enviar luego al módulo principal el arreglo con dichos datos listo para ser desplegados y utilizados. El módulo \textit{`Operaciones matemáticas'} incluye el código necesario para llevar a cabo las transformaciones en los datos, recibiendo del módulo principal el arreglo de datos y devolviendo el mismo transformado. El módulo \textit{`Egreso de datos'} permite guardar los datos trabajados, el gráfico mostrado en la interfaz, el sonido y las coordenadas de las marcas que pueden hacerse sobre los datos. El \textit{`Módulo de sonido'} incluye la generación de sonido a partir de un arreglo de datos que es enviado por el módulo principal, y todas las posibles configuraciones del mismo. El \textit{`Módulo de interfaz'} comprende todo el diseño de la interfaz gráfica y los eventos para hacerla funcional. En cuanto al \textit{`Módulo principal}, el mismo hereda al módulo de interfaz, para poder otorgar funcionalidad a cada evento de interfaz al recibir la información del usuario y de cada uno de los demás módulos. De esta forma, mediante la integración de todos los módulos mencionados se obtiene la interfaz gráfica funcional de sonoUno.

Consecuentemente de lo expuesto hasta ahora, para el desarrollo de sonoUno se utilizó el lenguaje de programación \href{https://www.python.org/}{\underline{\textcolor{blue}{Python}}} por ser ampliamente utilizado, de acceso libre y promueve el desarrollo colaborativo. Muchos desarrolladores escriben y mantienen librerías en dicho lenguaje que además son ampliamente utilizadas. Cabe destacar que durante los años de desarrollo se ha ido actualizando a las últimas versiones de Python y las correspondientes librerías, lo que en varias ocasiones conllevó cambios en el código y hasta cambio de librerías. En este sentido, se utilizaron y, en algunos casos, se siguen utilizando las librerías desarrolladas en Python que detallamos a continuación:

\begin{itemize}
    \item Interfaz gráfica: 
    \begin{itemize}
        \item \href{https://www.wxpython.org/}{\underline{\textcolor{blue}{wxPython}}}: es un kit de herramientas para interfaz gráfica multiplataforma, es bastante completa en cuanto a los elementos que contiene y presenta una documentación extensa y detallada. Toda la interfaz gráfica de sonoUno se programó utilizando esta librería y se ha continuado usando hasta la última versión presentada en el año 2022.
        \item \href{https://matplotlib.org/}{\underline{\textcolor{blue}{Matplotlib}}}: esta librería provee las herramientas para crear gráficos estáticos, animados e interactivos, permite desplegar el gráfico en un elemento Panel que provee wxPython. De esta forma se logra graficar los datos y modificar dicho gráfico, siempre de acuerdo a las instrucciones del usuario dentro del mismo despliegue de interfaz. Se continúa utilizando en la última versión del programa.
    \end{itemize}
    \item Manejo de datos: los dos paquetes utilizados para el manejo de datos permiten el ingreso y manipulación de archivos en arreglos formato columna o tablas, aplicar operaciones matemáticas, modificaciones de forma y orden, entre otras cosas. El motivo por el que se usa uno o el otro fue por la simplicidad que presenta cada uno en diferentes tareas y la documentación en línea que se ha encontrado sobre cada problema específico a resolver. Ambas librerías siguen utilizándose hasta la actualidad.
    \begin{itemize}
        \item \href{https://numpy.org/}{\underline{\textcolor{blue}{Numpy}}}: particularmente se considera que permite un mejor manejo del array para la comunicación con la etapa de sonido y graficado.
        \item \href{https://pandas.pydata.org/}{\underline{\textcolor{blue}{Pandas}}}: en este caso se estima que el ingreso de archivos es mejor manejado y es más cómodo el uso de DataFrame para manipular los datos.
    \end{itemize}
    \item Sonido: en este caso, para la primer versión se utilizó sonidos MIDI, por eso se describen estas librerías aquí, sin embargo no se continuó con su uso en las futuras versiones debido a problemas de instalación y portabilidad.
    \begin{itemize}
        \item \href{https://bspaans.github.io/python-mingus/}{\underline{\textcolor{blue}{Mingus}}}: es un paquete de sonido multiplataforma que permite la generación, reproducción y guardado de música formato MIDI.
        \item \href{https://www.fluidsynth.org/}{\underline{\textcolor{blue}{FluidSynth}}}: es un sintetizador de sonido basado en las especificaciones `SoundFont', no presenta una interfaz gráfica para ser utilizada pero cuenta con APIs que facilitan su uso. En este caso se utiliza la librería Mingus para establecer la comunicación con FluidSynth.
    \end{itemize}
    \item Operaciones con Octave: con la finalidad de realizar operaciones con los datos en el entorno de Octave se utiliza la librería de python `oct2py' que provee un puente de comunicación entre ambos entornos.
\end{itemize}

Utilizando las librerías descriptas y los diseños elaborados se desarrolló el código en Python dividido en diferentes scripts (al menos uno por módulo). Se proveerá el código fuente del programa sonoUno en su versión final del año 2022 y se completará su descripción en el Capítulo \ref{cap:sonouno_vfinal}. En cuanto a la versión descripta aquí, además de ser ejecutada desde el código fuente utilizando el script principal llamado sonoUno, se elaboró un instalador para Windows que permitía instalar el programa y generaba un acceso directo en el escritorio. Lamentablemente, por un tema de incompatibilidad de herramientas, no se pudo mantener este instalador para las últimas versiones; en la actualidad la forma de utilizar el programa es a través de su código fuente o en su versión web.

\subsection{Descripción de funcionalidades}
\label{sect:cap4_descrip-func}

En esta subsección se describirán todas las posibilidades que tiene el programa, desde el ingreso de datos y el despliegue, hasta las operaciones matemáticas y configuraciones que se pueden realizar. Cabe destacar que las funcionalidades descriptas aquí son las que presentaba el programa al momento de realizar las pruebas con usuarios en el año 2019, las herramientas u opciones agregadas luego de esa fecha serán descriptas en los capítulos correspondientes. La organización de esta subsección será por grupos de funcionalidades (ver Cuadro \ref{tab:agrupacion_funcionalidades}).

\subsubsection{Opciones de I/O}

\begin{enumerate}

    \item \textit{Abrir un archivo}
    
    Una vez abierto el programa, existen tres formas de llegar a la ventana del buscador del sistema para abrir un archivo de datos (ver Figura \ref{fig:open_options}): (1) desde el menú archivo con el ítem `Abrir' (`Open'); (2) desde el menú paneles, ítem archivo (`Panels -> File'), que despliega el panel contenedor del botón `Abrir' (`Open') y elementos que muestran datos del archivo (ver Figura \ref{fig:open_elements}); (3) con el atajo de teclado `Ctrl+O' indicado al lado del ítem `Abrir' o en la documentación. 
    
    Una vez en la ventana del buscador de sistema (ver Figura \ref{fig:open_buscador}) se puede acceder a la carpeta donde se encuentran los datos y seleccionarlos para ser abiertos. En la Figura \ref{fig:open_guiwithdata} se puede observar la interfaz inicial con un conjunto de datos desplegados.
    
    Por otro lado, ligado a esta funcionalidad se despliega en el panel de entrada/salida una grilla (ver Figura \ref{fig:open_elements}) donde se muestra información sobre los datos abiertos. En la Figura \ref{fig:param_functions} se muestran algunas de las opciones que sonoUno ofrece en cuanto al despliegue de los datos, teniendo en cuenta que hasta el momento se despliegan dos columnas de los datos donde una depende de la otra. El panel de parámetros de datos permite cambiar el título de los datos (cambiándolo también en el despliegue gráfico pero sin alterar el archivo importado); permite ingresar títulos de columna si los datos no contaban con ellos, o cambiarlos si es que los datos ya tenían título de columna (ver Figuras \ref{fig:param_anychange} y \ref{fig:param_columntitle}); por último, permite elegir qué columna mostrar en el eje x y cuál en el eje y (ver Figuras \ref{fig:param_ejeyFlux} y \ref{fig:param_ejeyBestfit}).
    
    \begin{figure}[p]
        \centering
        \begin{subfigure}[b]{0.50\textwidth}
             \centering
             \captionsetup{justification=centering}
             \frame{\includegraphics[width=\textwidth]{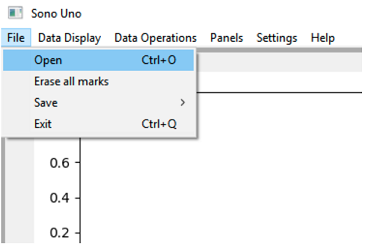}}
             \caption{Ítem `Open' existente en el menú archivo.}
             \label{fig:open_menuop}
         \end{subfigure}
         \hfill
        \begin{subfigure}[b]{0.49\textwidth}
             \centering
             \captionsetup{justification=centering}
             \frame{\includegraphics[width=\textwidth]{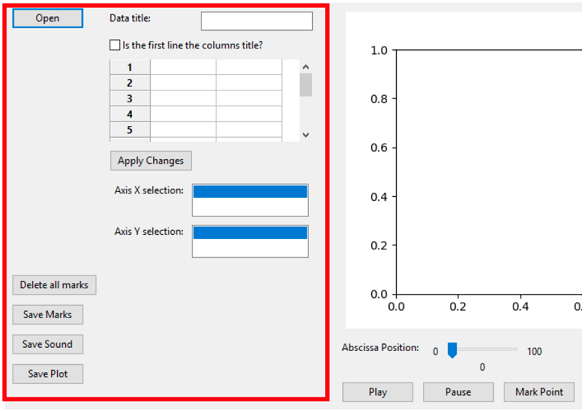}}
             \caption{Elementos de ingreso de datos disponibles en el panel de I/O.}
             \label{fig:open_elements}
         \end{subfigure}
         \hfill
        \begin{subfigure}[b]{0.60\textwidth}
             \centering
             \captionsetup{justification=centering}
             \frame{\includegraphics[width=\textwidth]{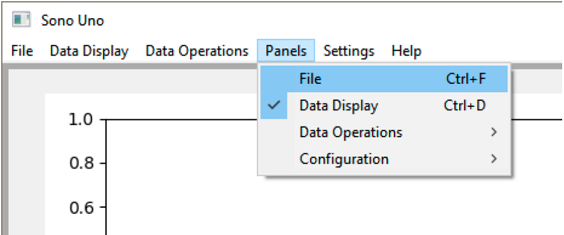}}
             \caption{Ítem `File' existente en el menú paneles, que permite visualizar el panel de I/O.}
             \label{fig:open_menupanel}
         \end{subfigure}
        \caption{Diferentes formas de acceder al ingreso de datos en el software sonoUno.}
        \label{fig:open_options}
    \end{figure}
    
    \begin{figure}[p]
        \centering
        \begin{subfigure}[b]{0.45\textwidth}
             \centering
             \captionsetup{justification=centering}
             \frame{\includegraphics[width=\textwidth]{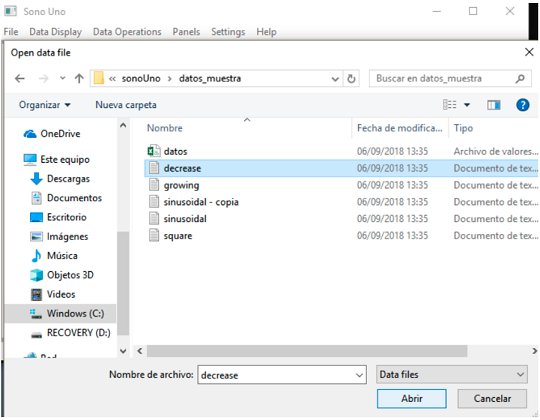}}
             \caption{Buscador de archivo del sistema operativo.}
             \label{fig:open_buscador}
         \end{subfigure}
         \hfill
        \begin{subfigure}[b]{0.53\textwidth}
             \centering
             \captionsetup{justification=centering}
             \frame{\includegraphics[width=\textwidth]{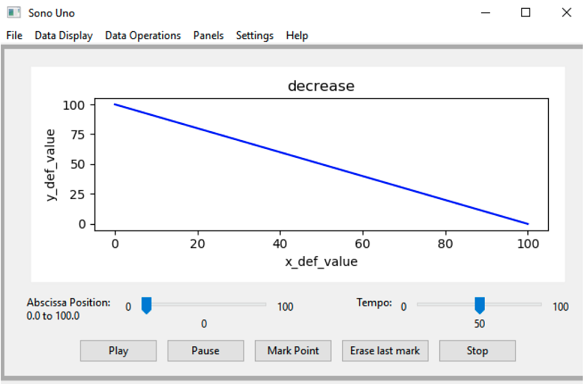}}
             \caption{Interfaz gráfica con los datos desplegados.}
             \label{fig:open_guiwithdata}
         \end{subfigure}
         
        \caption{Acciones posteriores a presionar el botón `Abrir' (`Open').}
        \label{fig:open_procedimiento}
    \end{figure}
    
    \begin{figure}[p]
        \centering
        \begin{subfigure}[b]{0.45\textwidth}
             \centering
             \captionsetup{justification=centering}
             \frame{\includegraphics[width=\textwidth]{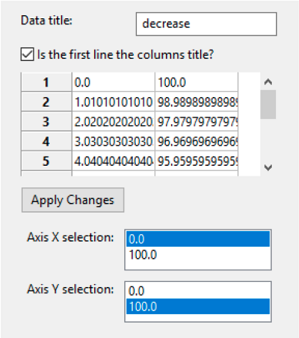}}
             \caption{Vista con los datos recién ingresados.}
             \label{fig:param_anychange}
         \end{subfigure}
         \hfill
        \begin{subfigure}[b]{0.43\textwidth}
             \centering
             \captionsetup{justification=centering}
             \frame{\includegraphics[width=\textwidth]{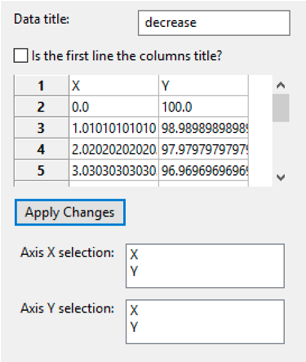}}
             \caption{Se ha agregado el título a las columnas.}
             \label{fig:param_columntitle}
         \end{subfigure}
         \hfill
        \begin{subfigure}[b]{0.49\textwidth}
             \centering
             \captionsetup{justification=centering}
             \frame{\includegraphics[width=\textwidth]{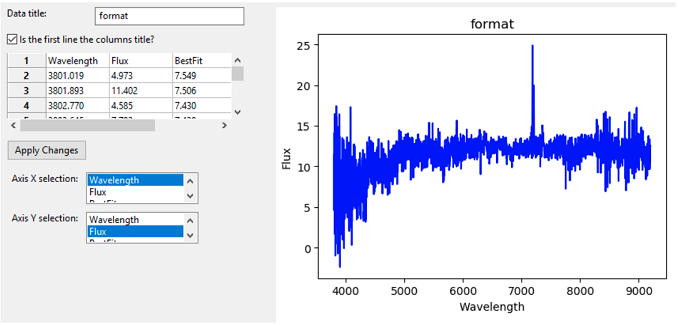}}
             \caption{Datos con la columna Flux seleccionada para el eje y.}
             \label{fig:param_ejeyFlux}
         \end{subfigure}
         \hfill
        \begin{subfigure}[b]{0.49\textwidth}
             \centering
             \captionsetup{justification=centering}
             \frame{\includegraphics[width=\textwidth]{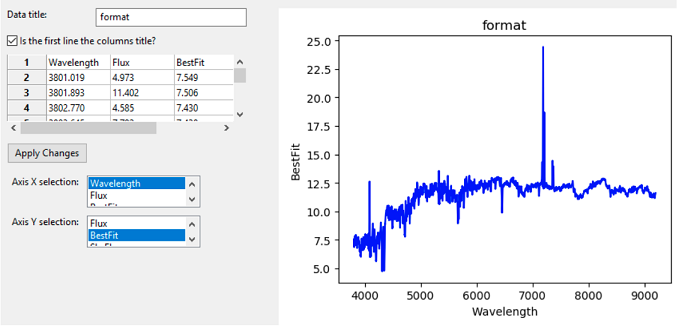}}
             \caption{Datos con la columna BestFit seleccionada para el eje y.}
             \label{fig:param_ejeyBestfit}
         \end{subfigure}
        \caption{Muestra de diferentes funcionalidades que ofrece el panel de parámetros de datos: (a) y (b) muestran que se puede incluir títulos de columnas de ser necesario; (c) y (d) muestra que se puede seleccionar las columnas a representar en cada eje.}
        \label{fig:param_functions}
    \end{figure}

    Una vez abiertos los datos en la interfaz de sonoUno, puede iniciarse la sonorización con el botón `Play'. Al iniciar la reproducción de sonido, una línea vertical de color rojo aparecerá en el área del gráfico indicando la posición que está siendo sonorizada (ver Figura \ref{fig:abscisa_slider}).

    \item \textit{Borrar todas las marcas hechas en los datos}
    
    Esta funcionalidad se encuentra en la sección de `Archivo' (`File'), la Figura \ref{fig:deletemarks_buttons} muestra las formas de acceder a ella: por un lado, el ítem que se encuentra en el menú `Archivo' (Figura \ref{fig:deletemarks_filemenu}); por otro, el botón alojado en el panel `Archivo' (Figura \ref{fig:deletemarks_filepanel}).
    
    El programa sonoUno permite salvar puntos que son de interés para la persona que está analizando los datos (se marcan utilizando una línea vertical de color negro en el punto marcado), para luego poder extraerlos en una tabla si fuera necesario. El botón descripto aquí permite eliminar todas estas marcas previamente realizadas en los datos con una sola acción, la Figura \ref{fig:deletemarks_plotview} muestra un antes (Figura \ref{fig:deletemarks_guiwith}) y un después (Figura \ref{fig:deletemarks_guiwithout}) de la interfaz al presionar el elemento `Borrar todas las marcas' (`Erase all marks').
    
    \begin{figure}[p]
        \centering
        \begin{subfigure}[b]{0.49\textwidth}
             \centering
             \captionsetup{justification=centering}
             \frame{\includegraphics[width=\textwidth]{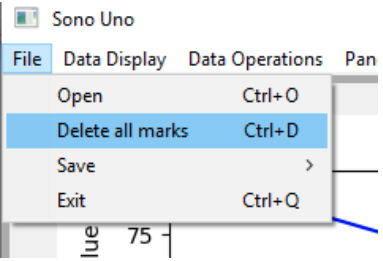}}
             \caption{Ubicación de la funcionalidad en el menú `File'.}
             \label{fig:deletemarks_filemenu}
         \end{subfigure}
         \hfill
        \begin{subfigure}[b]{0.49\textwidth}
             \centering
             \captionsetup{justification=centering}
             \frame{\includegraphics[width=\textwidth]{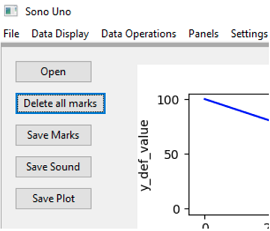}}
             \caption{Ubicación de la funcionalidad en el panel `Archivo'.}
             \label{fig:deletemarks_filepanel}
         \end{subfigure}
         
        \caption{Formas de acceder a la funcionalidad `Borrar todas las marcas' (`Erase all marks').}
        \label{fig:deletemarks_buttons}
    \end{figure}
    
    \begin{figure}[p]
        \centering
        \begin{subfigure}[b]{0.49\textwidth}
             \centering
             \captionsetup{justification=centering}
             \frame{\includegraphics[width=\textwidth]{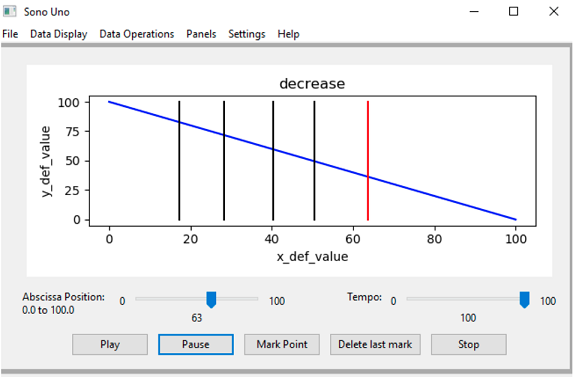}}
             \caption{Interfaz con varias marcas hechas en los datos (líneas negras).}
             \label{fig:deletemarks_guiwith}
         \end{subfigure}
         \hfill
        \begin{subfigure}[b]{0.49\textwidth}
             \centering
             \captionsetup{justification=centering}
             \frame{\includegraphics[width=\textwidth]{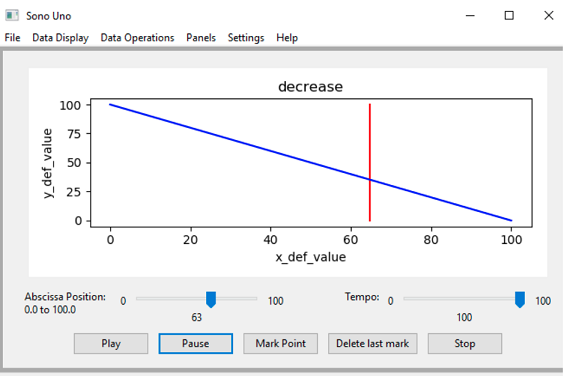}}
             \caption{Interfaz luego de presionar el botón `Borrar todas las marcas'.}
             \label{fig:deletemarks_guiwithout}
         \end{subfigure}
         
        \caption{Antes y después en el gráfico de la acción para borrar todas las marcas.}
        \label{fig:deletemarks_plotview}
    \end{figure}
    
    \item \textit{Salvar archivos}
    
    El programa permite salvar tres tipos de archivos: las coordenadas de las marcas hechas en los datos, el sonido y el gráfico. La Figura \ref{fig:savemarks_elementsygui} muestra las dos formas de acceder al elemento de interfaz para guardar las marcas, una mediante el submenú `Guardar' (`Save') localizado en el menú `Archivo' (`File') (ver Figura \ref{fig:savemark_filemenu}) y la otra desde el botón `Guardar marcas' (`Save marks') en el panel `Archivo' (ver Figura \ref{fig:savemark_filepanel}). Cuando las marcas son guardadas, las mismas son borradas del gráfico, en las Figuras \ref{fig:savemarks_guiwith} y \ref{fig:savemarks_guiwithout} se muestra la interfaz antes y después del proceso de salvar las marcas en los datos. 
    
    Algo que se debe tener en cuenta es que al presionar el botón `Guardar' (`Save') (para cualquiera de los tres tipos de datos que se pueden salvar) se abre una ventana que permite decidir donde se desean guardar los datos y elegir el nombre para el archivo a guardar.
    
    En cuanto al proceso de guardar el sonido, la Figura \ref{fig:savesound_elements} muestra las dos formas de acceder a la funcionalidad. De forma similar, la Figura \ref{fig:saveplot_elements} muestra los elementos que permiten guardar el gráfico. Por último, la Figura \ref{fig:output_all} muestra una captura de pantalla de la carpeta que contiene los tres archivos guardados (Figura \ref{fig:output_files}) y cada uno de los archivos abiertos con la aplicación correspondiente para su tipo de archivo.
    
    \begin{figure}[p]
        \centering
        \begin{subfigure}[b]{0.49\textwidth}
             \centering
             \captionsetup{justification=centering}
             \frame{\includegraphics[width=\textwidth]{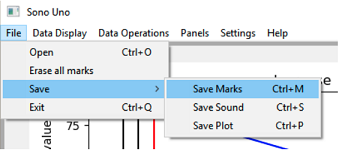}}
             \caption{Ubicación de la funcionalidad en el submenú `Guardar' (`Save').}
             \label{fig:savemark_filemenu}
         \end{subfigure}
         \hfill
        \begin{subfigure}[b]{0.49\textwidth}
             \centering
             \captionsetup{justification=centering}
             \frame{\includegraphics[width=0.6\textwidth]{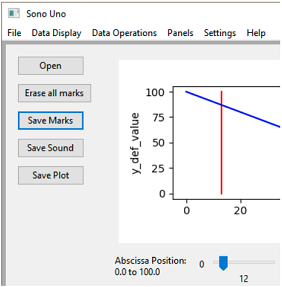}}
             \caption{Ubicación de la funcionalidad en el panel `Archivo'.}
             \label{fig:savemark_filepanel}
         \end{subfigure}
         \hfill
        \begin{subfigure}[b]{0.49\textwidth}
             \centering
             \captionsetup{justification=centering}
             \frame{\includegraphics[width=\textwidth]{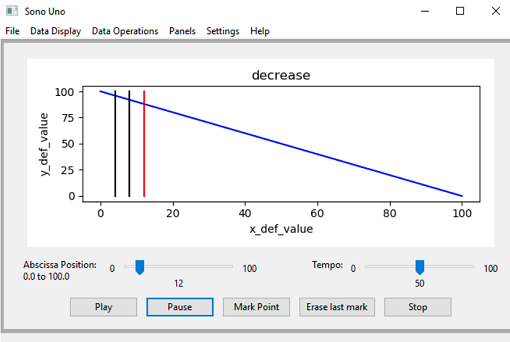}}
             \caption{Interfaz gráfica con dos marcas hechas en los datos.}
             \label{fig:savemarks_guiwith}
         \end{subfigure}
         \hfill
        \begin{subfigure}[b]{0.49\textwidth}
             \centering
             \captionsetup{justification=centering}
             \frame{\includegraphics[width=\textwidth]{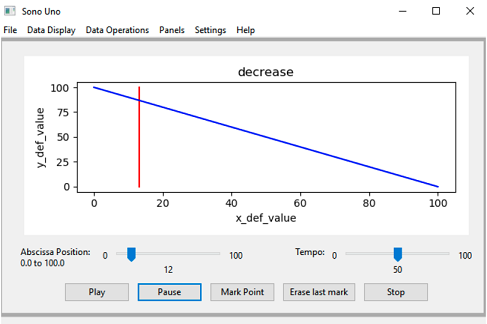}}
             \caption{Interfaz gráfica luego de guardar las marcas.}
             \label{fig:savemarks_guiwithout}
         \end{subfigure}
         
        \caption{Formas de acceder a la funcionalidad `Guardar marcas' (`Save Marks') y también, un antes y después de la interfaz gráfica al guardar las marcas.}
        \label{fig:savemarks_elementsygui}
    \end{figure}
    
    \begin{figure}[p]
        \centering
        \begin{subfigure}[b]{0.49\textwidth}
             \centering
             \captionsetup{justification=centering}
             \frame{\includegraphics[width=\textwidth]{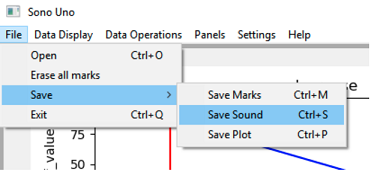}}
             \caption{Ubicación de la funcionalidad en el submenú `Guardar' (`Save').}
             \label{fig:savesound_filemenu}
         \end{subfigure}
         \hfill
        \begin{subfigure}[b]{0.49\textwidth}
             \centering
             \captionsetup{justification=centering}
             \frame{\includegraphics[width=0.6\textwidth]{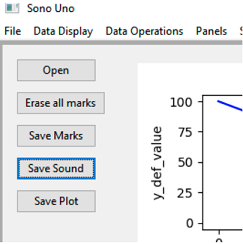}}
             \caption{Ubicación de la funcionalidad en el panel `Archivo'.}
             \label{fig:savesound_filepanel}
         \end{subfigure}
         
        \caption{Dos formas de acceder a la funcionalidad `Guardar sonido' (`Save Sound').}
        \label{fig:savesound_elements}
    \end{figure}
    
    \begin{figure}[p]
        \centering
        \begin{subfigure}[b]{0.49\textwidth}
             \centering
             \captionsetup{justification=centering}
             \frame{\includegraphics[width=\textwidth]{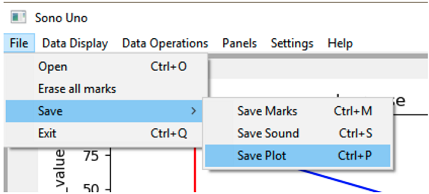}}
             \caption{Ubicación de la funcionalidad en el submenú `Guardar' (`Save').}
             \label{fig:saveplot_filemenu}
         \end{subfigure}
         \hfill
        \begin{subfigure}[b]{0.49\textwidth}
             \centering
             \captionsetup{justification=centering}
             \frame{\includegraphics[width=0.6\textwidth]{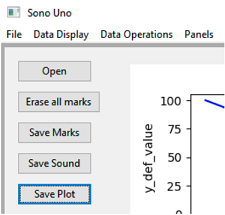}}
             \caption{Ubicación de la funcionalidad en el panel `Archivo'.}
             \label{fig:saveplot_filepanel}
         \end{subfigure}
         
        \caption{Dos formas de acceder a la funcionalidad `Guardar gráfico' (`Save Plot').}
        \label{fig:saveplot_elements}
    \end{figure}
    
    \begin{figure}[p]
        \centering
        \begin{subfigure}[b]{0.49\textwidth}
             \centering
             \captionsetup{justification=centering}
             \frame{\includegraphics[width=0.7\textwidth]{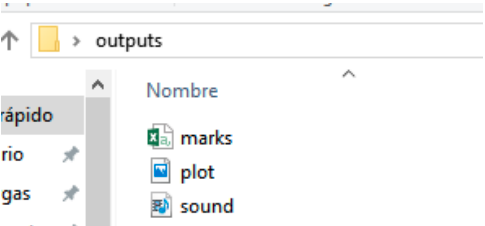}}
             \caption{Archivos guardados previamente.}
             \label{fig:output_files}
         \end{subfigure}
         \hfill
        \begin{subfigure}[b]{0.49\textwidth}
             \centering
             \captionsetup{justification=centering}
             \frame{\includegraphics[width=\textwidth]{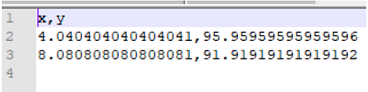}}
             \caption{Archivo de texto con las coordenadas de las dos marcas salvadas.}
             \label{fig:output_marks}
         \end{subfigure}
         \hfill
        \begin{subfigure}[b]{0.49\textwidth}
             \centering
             \captionsetup{justification=centering}
             \frame{\includegraphics[width=0.7\textwidth]{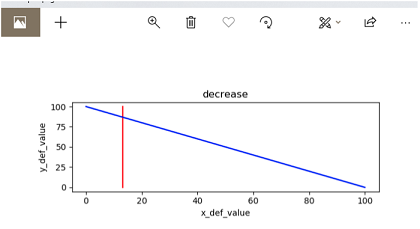}}
             \caption{Archivo del gráfico abierto con el visualizador de imágenes.}
             \label{fig:output_plot}
         \end{subfigure}
         \hfill
        \begin{subfigure}[b]{0.49\textwidth}
             \centering
             \captionsetup{justification=centering}
             \frame{\includegraphics[width=\textwidth]{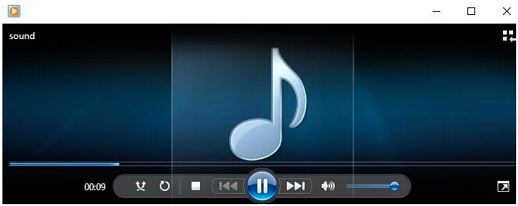}}
             \caption{Archivo de sonido abierto con el reproductor de música.}
             \label{fig:output_sound}
         \end{subfigure}
         
        \caption{Archivos de salida guardados con el programa sonoUno.}
        \label{fig:output_all}
    \end{figure}
    
    \item \textit{Salir del programa}
    
    La última funcionalidad incluida dentro del grupo de entrada y salida, es el cierre del programa. La figura \ref{fig:exit_elements} muestra las opciones existentes para cerrar el programa: se puede ejecutar desde la `X' en la esquina superior derecha (Figura \ref{fig:exit_x}), desde el ítem correspondiente en el menú `Archivo' (`File') y con el atajo de teclado `Ctrl+Q'.
    
    \begin{figure}[p]
        \centering
        \begin{subfigure}[b]{0.49\textwidth}
             \centering
             \captionsetup{justification=centering}
             \frame{\includegraphics[width=0.8\textwidth]{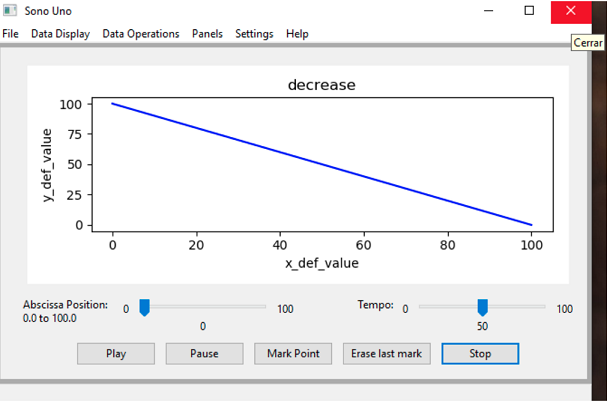}}
             \caption{Ubicación de la `X' para cerrar el programa}
             \label{fig:exit_x}
         \end{subfigure}
         \hfill
        \begin{subfigure}[b]{0.49\textwidth}
             \centering
             \captionsetup{justification=centering}
             \frame{\includegraphics[width=0.6\textwidth]{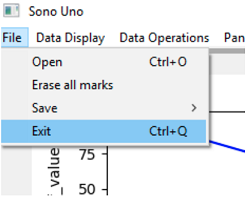}}
             \caption{Ubicación del ítem `Cerrar' (`Exit') en el menú `Archivo' (`File').}
             \label{fig:exit_menu}
         \end{subfigure}
         
        \caption{Opciones disponibles para cerrar el programa.}
        \label{fig:exit_elements}
    \end{figure}
    
\end{enumerate}

\subsubsection{Despliegue}

\begin{enumerate}
    \item \textit{Posición de abscisas}
    
    Dentro de las funcionalidades de despliegue, se encuentra el elemento deslizable que permite seleccionar la posición del cursor de reproducción dentro de la gráfica. La Figura \ref{fig:abscisa_elements} muestra las formas de acceder a dicho elemento: mediante un ítem en el menú `Despliegue' (`Data Display') (Figura \ref{fig:abscisa_menu}), utilizando directamente el elemento deslizable en el panel `Despliegue' (indicado con un cuadro rojo en la Figura \ref{fig:abscisa_slider}) o bien utilizando el atajo de teclado `Shift+A'.
    
    En el caso de este elemento, para poder modificar la posición en el eje x se requiere utilizar el elemento deslizable, por lo que las opciones de menú y atajo de teclado en este caso lo que hacen en ubicar el foco de teclado en dicho elemento. Luego para poder modificar la posición de la barra se puede utilizar el puntero o las flechas en el teclado.
    
    \begin{figure}[p]
        \centering
        \begin{subfigure}[b]{0.49\textwidth}
             \centering
             \captionsetup{justification=centering}
             \frame{\includegraphics[width=0.95\textwidth]{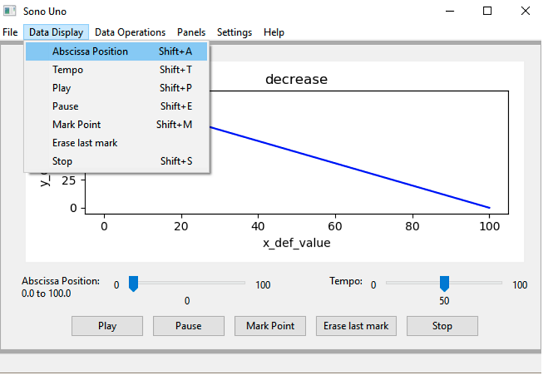}}
             \caption{Ubicación del ítem para seleccionar el elemento deslizable de abscisa.}
             \label{fig:abscisa_menu}
         \end{subfigure}
         \hfill
        \begin{subfigure}[b]{0.49\textwidth}
             \centering
             \captionsetup{justification=centering}
             \frame{\includegraphics[width=\textwidth]{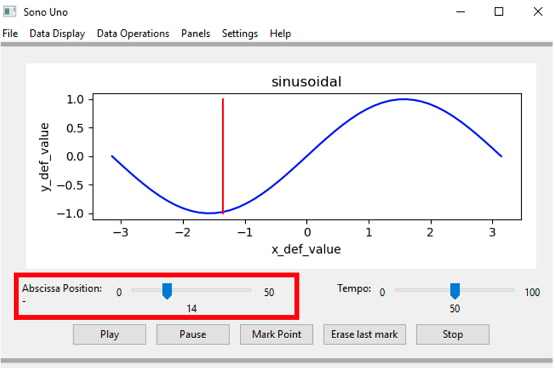}}
             \caption{Ubicación del elemento deslizable abscisa en la interfaz gráfica.}
             \label{fig:abscisa_slider}
         \end{subfigure}
         
        \caption{Opciones disponibles para acceder al elemento que permite modificar la posición en x.}
        \label{fig:abscisa_elements}
    \end{figure}
    
    \item \textit{Selección de tempo}
    
   Otro elemento deslizable disponible en el grupo de elementos de `Despliegue' es el que permite seleccionar el tempo para la reproducción del sonido. La Figura \ref{fig:tempo_elements} muestra las formas de acceder a dicho elemento: mediante un ítem en el menú `Despliegue' (`Data Display') (Figura \ref{fig:tempo_menu}), utilizando directamente el elemento deslizable en el panel `Despliegue' (indicado con un cuadro rojo en la Figura \ref{fig:tempo_slider}) o bien utilizando el atajo de teclado `Shift+T'.
    
    Análogamente al caso anterior, para poder modificar el tempo se requiere utilizar el elemento deslizable, por lo que las opciones de menú y atajo de teclado en este caso lo que hacen en ubicar el foco de teclado en dicho elemento. Luego para poder modificar la posición de la barra se puede utilizar el puntero o las flechas en el teclado.
    
    \begin{figure}[p]
        \centering
        \begin{subfigure}[b]{0.49\textwidth}
             \centering
             \captionsetup{justification=centering}
             \frame{\includegraphics[width=0.8\textwidth]{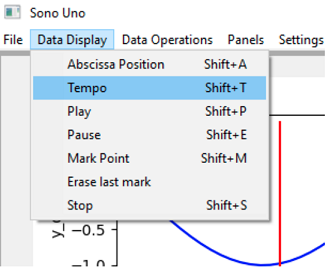}}
             \caption{Ubicación del ítem para seleccionar el elemento deslizable de tempo.}
             \label{fig:tempo_menu}
         \end{subfigure}
         \hfill
        \begin{subfigure}[b]{0.49\textwidth}
             \centering
             \captionsetup{justification=centering}
             \frame{\includegraphics[width=\textwidth]{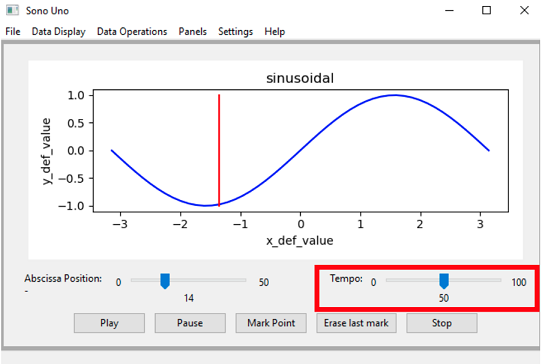}}
             \caption{Ubicación del elemento deslizable tempo en la interfaz gráfica.}
             \label{fig:tempo_slider}
         \end{subfigure}
         
        \caption{Opciones disponibles para acceder al elemento que permite modificar el tempo para la reproducción de sonido.}
        \label{fig:tempo_elements}
    \end{figure}
    
    \item \textit{Botón de reproducción}
    
    Es el primer botón en la línea inferior de botones y se encarga de iniciar la reproducción de los datos, donde se percibe la sonorización y el avance del cursor en el gráfico. La Figura \ref{fig:play_elements} muestra las formas de acceder a dicha funcionalidad: mediante el ítem ubicado en el menú `Despliegue' (`Data Display') (Figura \ref{fig:play_menu}), usando el botón `Play' en el panel `Despliegue' (Figura \ref{fig:play_button}, el recuadro azul que lo rodea indica que tiene el foco de teclado), o bien utilizando el atajo de teclado `Shift+P'.
    
    A diferencia de los dos casos anteriores, en el caso de los elemento botón, que producen la acción al ser presionados, todas las formas de acceso realizan directamente la acción. Por ejemplo, al presionar la combinación de teclas `Shift+P' se inicia la reproducción de los datos sin tener que presionar adicionalmente el botón.
    
    \begin{figure}[p]
        \centering
        \begin{subfigure}[b]{0.49\textwidth}
             \centering
             \captionsetup{justification=centering}
             \frame{\includegraphics[width=0.8\textwidth]{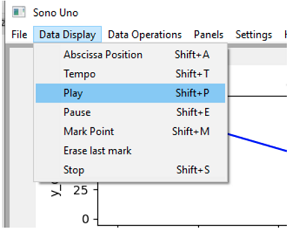}}
             \caption{Ubicación del ítem para seleccionar el elemento de reproducción.}
             \label{fig:play_menu}
         \end{subfigure}
         \hfill
        \begin{subfigure}[b]{0.49\textwidth}
             \centering
             \captionsetup{justification=centering}
             \frame{\includegraphics[width=\textwidth]{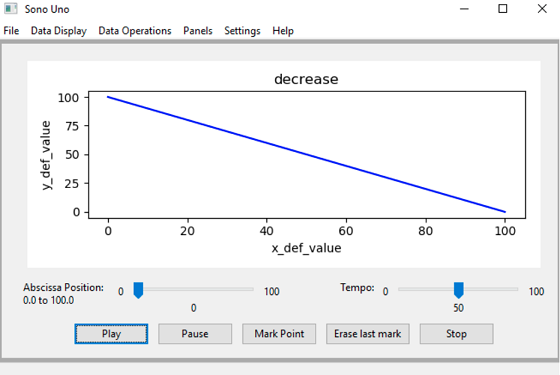}}
             \caption{Ubicación del botón `Play' en la interfaz gráfica.}
             \label{fig:play_button}
         \end{subfigure}
         
        \caption{Opciones disponibles para acceder al elemento para iniciar la reproducción.}
        \label{fig:play_elements}
    \end{figure}
    
    \item \textit{Botón de pausa}
    
    Es el segundo botón en la línea inferior de botones y se encarga de pausar la reproducción de los datos, permitiendo detener la reproducción para poder retomarla luego en el lugar donde fue pausada. La Figura \ref{fig:pausa_elements} muestra las formas de acceder a dicha funcionalidad: mediante el ítem ubicado en el menú `Despliegue' (`Data Display') (Figura \ref{fig:pausa_menu}), usando el botón `Pause' en el panel `Despliegue' (Figura \ref{fig:pausa_button}, el recuadro azul que lo rodea indica que tiene el foco de teclado), o bien utilizando el atajo de teclado `Shift+E'.
    
    \begin{figure}[p]
        \centering
        \begin{subfigure}[b]{0.49\textwidth}
             \centering
             \captionsetup{justification=centering}
             \frame{\includegraphics[width=0.8\textwidth]{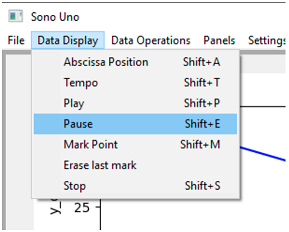}}
             \caption{Ubicación del ítem para seleccionar el elemento `Pause'.}
             \label{fig:pausa_menu}
         \end{subfigure}
         \hfill
        \begin{subfigure}[b]{0.49\textwidth}
             \centering
             \captionsetup{justification=centering}
             \frame{\includegraphics[width=\textwidth]{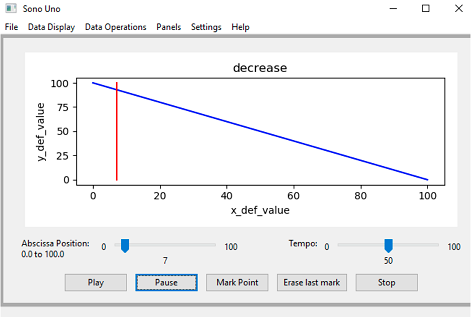}}
             \caption{Ubicación del botón `Pause' en la interfaz gráfica.}
             \label{fig:pausa_button}
         \end{subfigure}
         
        \caption{Opciones disponibles para acceder al elemento que permite pausar la reproducción.}
        \label{fig:pausa_elements}
    \end{figure}
    
    \item \textit{Botón para marcar un punto en el gráfico}
    
    Es el tercer botón en la línea inferior de botones y se encarga de realizar una marca en el gráfico, en el punto en el cual se encontraba el cursor de gráfico (línea vertical roja) al ser presionado el botón. Adicionalmente a la marca en el gráfico, se guardan las coordenadas del punto marcado para que el usuario si desea pueda guardarlas posteriormente en un archivo formato tabla. La Figura \ref{fig:mark_elements} muestra las formas de acceder a dicha funcionalidad: mediante el ítem ubicado en el menú `Despliegue' (`Data Display') (Figura \ref{fig:mark_menu}), usando el botón `Marcar punto' (`Mark Point') en el panel `Despliegue' (Figura \ref{fig:mark_button}, el recuadro azul que lo rodea indica que tiene el foco de teclado), o bien utilizando el atajo de teclado `Shift+M'.
    
    \begin{figure}[p]
        \centering
        \begin{subfigure}[b]{0.49\textwidth}
             \centering
             \captionsetup{justification=centering}
             \frame{\includegraphics[width=0.8\textwidth]{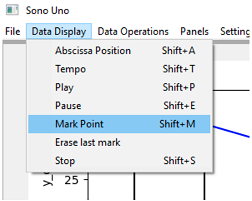}}
             \caption{Ubicación del ítem para marcar un punto en la gráfica.}
             \label{fig:mark_menu}
         \end{subfigure}
         \hfill
        \begin{subfigure}[b]{0.49\textwidth}
             \centering
             \captionsetup{justification=centering}
             \frame{\includegraphics[width=\textwidth]{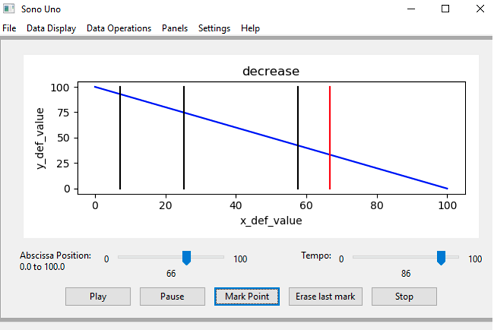}}
             \caption{Ubicación del botón `Mark Point' en la interfaz gráfica.}
             \label{fig:mark_button}
         \end{subfigure}
         
        \caption{Opciones disponibles para acceder al elemento que permite marcar un punto en la gráfica.}
        \label{fig:mark_elements}
    \end{figure}
    
    \item \textit{Botón para borrar la última marca realizada}
    
    El cuarto botón en la línea inferior de botones  se encarga de borrar la última marca realizada en el gráfico, eliminando también las coordenadas que fueron guardadas sobre esa marca. Un detalle sobre este elemento es que puede borrar todas las marcas realizadas en los datos, siempre borrando una a la vez y desde la última realizada hacia atrás hasta llegar a la primera ejecutada. 
    
    La Figura \ref{fig:deletemark_elements} muestra las formas de acceder a dicha funcionalidad: mediante el ítem ubicado en el menú `Despliegue' (`Data Display') (Figura \ref{fig:deletemark_menu}), usando el botón `Borrar última marca' (`Erase last mark') en el panel `Despliegue' (Figura \ref{fig:deletemark_button}, el recuadro azul que lo rodea indica que tiene el foco de teclado), o bien utilizando el atajo de teclado `Shift+D'.
    
    \begin{figure}[p]
        \centering
        \begin{subfigure}[b]{0.49\textwidth}
             \centering
             \captionsetup{justification=centering}
             \frame{\includegraphics[width=0.8\textwidth]{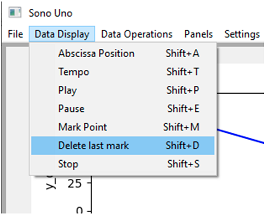}}
             \caption{Ubicación del ítem para borrar la última marca realizada en el gráfico.}
             \label{fig:deletemark_menu}
         \end{subfigure}
         \hfill
        \begin{subfigure}[b]{0.49\textwidth}
             \centering
             \captionsetup{justification=centering}
             \frame{\includegraphics[width=\textwidth]{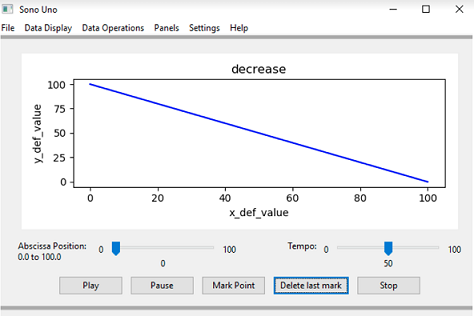}}
             \caption{Ubicación del botón `Erase last mark' en la interfaz gráfica.}
             \label{fig:deletemark_button}
         \end{subfigure}
         
        \caption{Opciones disponibles para acceder al elemento que permite borrar la última marca realizada en el gráfico.}
        \label{fig:deletemark_elements}
    \end{figure}
    
    \item \textit{Botón para detener la reproducción}
    
    El último botón en la línea inferior de botones se encarga de detener la reproducción de los datos, reiniciando la reproducción y dejando el cursor en el inicio de los datos (invisible en la interfaz gráfica). La Figura \ref{fig:stop_elements} muestra las formas de acceder a dicha funcionalidad: mediante el ítem ubicado en el menú `Despliegue' (`Data Display') (Figura \ref{fig:stop_menu}), usando el botón `Detener' (`Stop') en el panel `Despliegue' (Figura \ref{fig:stop_button}, el recuadro azul que lo rodea indica que tiene el foco de teclado), o bien utilizando el atajo de teclado `Shift+S'.
    
    \begin{figure}[p]
        \centering
        \begin{subfigure}[b]{0.49\textwidth}
             \centering
             \captionsetup{justification=centering}
             \frame{\includegraphics[width=0.8\textwidth]{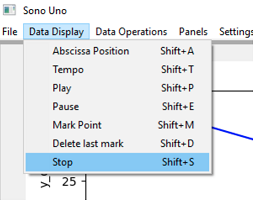}}
             \caption{Ubicación del ítem para seleccionar el elemento `Detener' (`Stop').}
             \label{fig:stop_menu}
         \end{subfigure}
         \hfill
        \begin{subfigure}[b]{0.49\textwidth}
             \centering
             \captionsetup{justification=centering}
             \frame{\includegraphics[width=\textwidth]{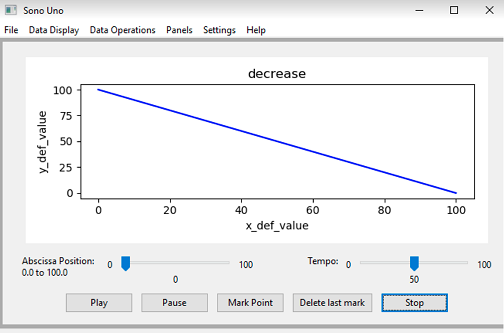}}
             \caption{Ubicación del botón `Detener' (`Stop') en la interfaz gráfica.}
             \label{fig:stop_button}
         \end{subfigure}
         
        \caption{Opciones disponibles para acceder al elemento que permite detener la reproducción.}
        \label{fig:stop_elements}
    \end{figure}
    
\end{enumerate}

\subsubsection{Operaciones}

\begin{enumerate}
    \item \textit{Selección de rango en el eje de abscisas}
    
    Esta funcionalidad consta de dos elementos deslizables que permiten seleccionar, una el valor mínimo y otra el valor máximo del eje x, permitiendo de esa forma ajustar el eje a un rango de datos específico. Se eligen elementos deslizables porque permiten modificar el valor moviendo una etiqueta entre el valor mínimo y máximo de los datos. Se debe tener en cuenta que los elementos deslizables muestran la cantidad de valores disponibles en el eje x, es por ello que siempre comienza en 0 independientemente del archivo de datos que se ingrese.
    
    La Figura \ref{fig:cutx_elements} muestra los elementos que se relacionan con la función de seleccionar un rango específico de los datos. En cuanto a la forma de mostrar los elementos deslizables en la interfaz gráfica hay tres: utilizando el ítem en el submenú `Límite horizontal' (`Horizontal Limit') para límite superior o inferior (Figura \ref{fig:cutx_menumath}), habilitar directamente el panel de deslizadores y funciones matemáticas (recuadro rojo en la Figura \ref{fig:cutx_sliders}) desde el submenú `Operaciones' (`Data Operations') dentro del menú `Paneles' (`Panels') (Figura \ref{fig:cutx_menupanel}), o bien utilizando los atajos de teclado correspondientes (`Ctrl+Shift+X' para mostrar el panel y establecer el foco de teclado en el elemento deslizable del límite inferior en x; `Ctrl+Shift+W' para mostrar el panel y establecer el foco de teclado en el elemento del límite superior; `Ctrl+X' para habilitar directamente el panel de deslizables y funciones matemáticas).
    
    La Figura \ref{fig:cutx_ejemplo} muestra desplegados los datos de una función decreciente que presenta límite inferior en 0 y superior en 100, además, se muestra el panel de deslizadores y funciones matemáticas habilitado. En el ejemplo de dicha Figura se ha realizado una selección de rango posicionando el valor mínimo de x en el valor 20 y el máximo en el valor 80, puede apreciarse tanto en los elementos deslizables como en el eje x del despliegue gráfico.
     
    \begin{figure}[p]
        \centering
        \begin{subfigure}[b]{0.49\textwidth}
             \centering
             \captionsetup{justification=centering}
             \frame{\includegraphics[width=0.8\textwidth]{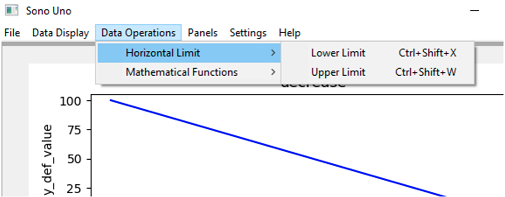}}
             \caption{Submenú `Límite horizontal' (`Horizontal Limit') dentro del menú `Operaciones' (`Data Operations') que contiene los elementos de selección de rango.}
             \label{fig:cutx_menumath}
         \end{subfigure}
         \hfill
        \begin{subfigure}[b]{0.49\textwidth}
             \centering
             \captionsetup{justification=centering}
             \frame{\includegraphics[width=\textwidth]{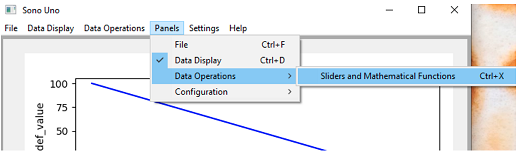}}
             \caption{Submenú `Operaciones' (`Data Operations') dentro del menú `Paneles' (`Panels') que permite mostrar u ocultar el panel con las barras deslizables y funciones matemáticas.}
             \label{fig:cutx_menupanel}
         \end{subfigure}
         \hfill
        \begin{subfigure}[b]{0.49\textwidth}
             \centering
             \captionsetup{justification=centering}
             \frame{\includegraphics[width=\textwidth]{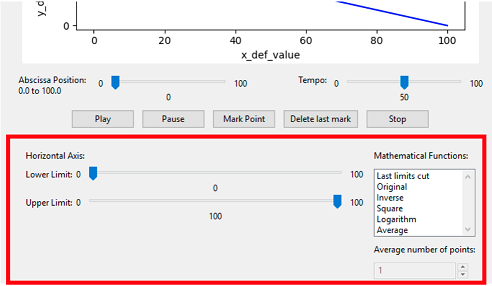}}
             \caption{Panel que contiene las barras deslizables y funciones matemáticas predefinidas (recuadro rojo).}
             \label{fig:cutx_sliders}
         \end{subfigure}
         \hfill
        \begin{subfigure}[b]{0.49\textwidth}
             \centering
             \captionsetup{justification=centering}
             \frame{\includegraphics[width=0.8\textwidth]{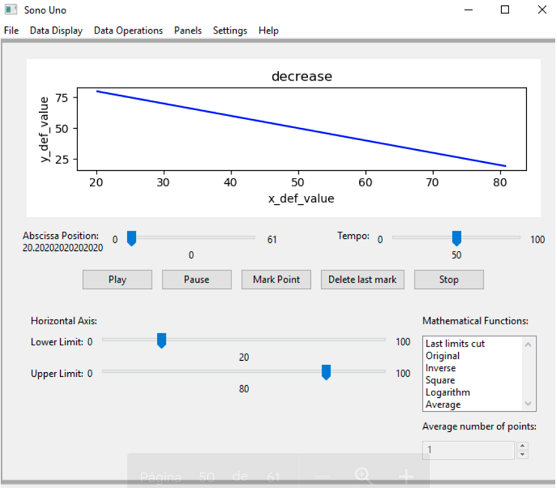}}
             \caption{Ejemplo de selección de rango en el eje horizontal, en este caso de un conjunto de 100 datos se ha seleccionado entre el valor 20 y 80.}
             \label{fig:cutx_ejemplo}
         \end{subfigure}
         
        \caption{Diferentes capturas de pantalla de los elementos correspondientes a la selección de rangos en el eje de abscisas.}
        \label{fig:cutx_elements}
    \end{figure}
    
    \item \textit{Funciones matemáticas predefinidas}
    
    Dentro del sonoUno, se programaron seis operaciones matemáticas que pueden aplicarse a los datos presionando un botón, entre ellas encontramos: volver a los datos originales, mostrar el último rango de datos seleccionado, invertir el eje y, aplicar una función cuadrática, un logaritmo o un suavizado (indicado como promedio en las capturas de pantalla) a los datos.
    
    La Figura \ref{fig:mathop_elements} muestra la forma de acceder a dichas funciones matemáticas en el menú (Figura \ref{fig:mathop_menu}), donde en la mayoría de los casos se realiza la acción al seleccionar la función. El caso particular donde no se cumple esta premisa es la función suavizado, ya que al presionar dicho ítem se habilita un cuadro que permite seleccionar el número de datos con los que se realiza el promedio para el suavizado que por defecto es 1 (elemento inferior en la Figura \ref{fig:mathop_operaciones}). La Figura \ref{fig:mathop_operaciones} muestra el elemento lista que despliega todas las operaciones que pueden realizarse a los datos y la caja donde se indica el valor para la función suavizado (esta caja solo se activa al seleccionar dicha función), el cuadro que se muestra en esta Figura se encuentra posicionado a la derecha de los elementos deslizables que permiten seleccionar un rango en eje x. Otra forma de acceder a estos elementos de operaciones matemáticas es habilitando el panel de deslizables y funciones matemáticas como indica la Figura \ref{fig:cutx_menupanel}. Por último, la Figura \ref{fig:mathop_examples} muestra capturas de pantalla donde se ha aplicado cada una de las funciones matemáticas mencionadas.
    
    \begin{figure}[ht!]
        \centering
        \begin{subfigure}[b]{0.49\textwidth}
             \centering
             \captionsetup{justification=centering}
             \frame{\includegraphics[width=\textwidth]{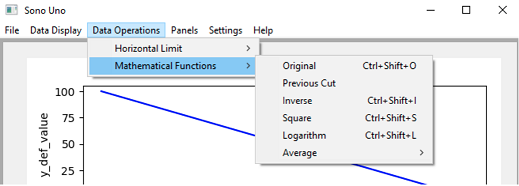}}
             \caption{Ítems con las funciones matemáticas disponibles dentro del submenú `Funciones matemáticas' (`Mathematical Functions') en el menú `Operaciones' (`Data Operations').}
             \label{fig:mathop_menu}
         \end{subfigure}
         \hfill
        \begin{subfigure}[b]{0.49\textwidth}
             \centering
             \captionsetup{justification=centering}
             \frame{\includegraphics[width=0.5\textwidth]{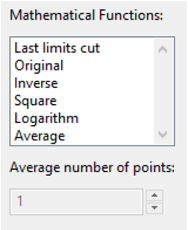}}
             \caption{Elemento lista donde se muestran las operaciones matemáticas que pueden seleccionarse.}
             \label{fig:mathop_operaciones}
         \end{subfigure}
         
        \caption{Listado con las operaciones matemáticas predefinidas con las que cuenta esta versión de sonoUno.}
        \label{fig:mathop_elements}
    \end{figure}
    
    \begin{figure}[p]
        \centering
        \begin{subfigure}[b]{0.49\textwidth}
             \centering
             \captionsetup{justification=centering}
             \frame{\includegraphics[width=\textwidth]{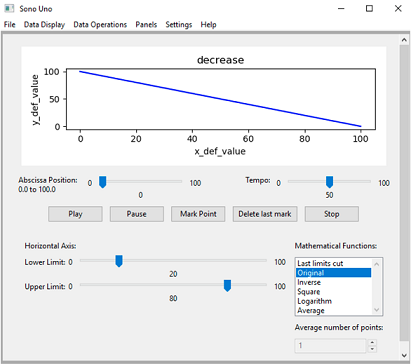}}
             \caption{Original.}
             \label{fig:mathop_original}
         \end{subfigure}
         \hfill
        \begin{subfigure}[b]{0.49\textwidth}
             \centering
             \captionsetup{justification=centering}
             \frame{\includegraphics[width=\textwidth]{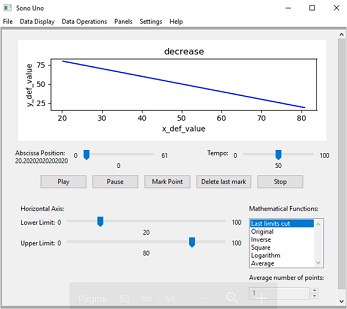}}
             \caption{Último rango seleccionado.}
             \label{fig:mathop_lastlimit}
         \end{subfigure}
         \hfill
        \begin{subfigure}[b]{0.49\textwidth}
             \centering
             \captionsetup{justification=centering}
             \frame{\includegraphics[width=\textwidth]{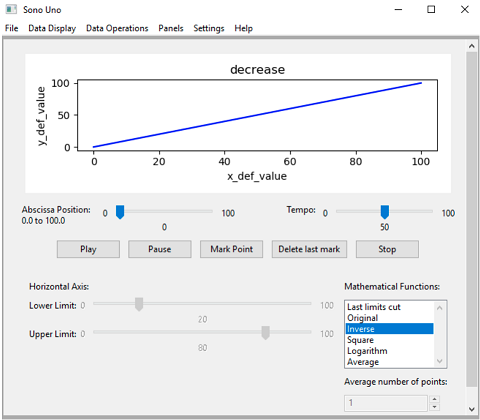}}
             \caption{Invertir eje y.}
             \label{fig:mathop_inverse}
         \end{subfigure}
         \hfill
        \begin{subfigure}[b]{0.49\textwidth}
             \centering
             \captionsetup{justification=centering}
             \frame{\includegraphics[width=\textwidth]{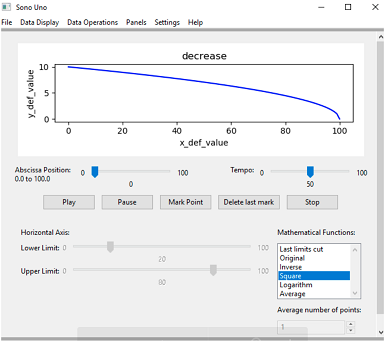}}
             \caption{Función Cuadrática.}
             \label{fig:mathop_square}
         \end{subfigure}
         \hfill
        \begin{subfigure}[b]{0.49\textwidth}
             \centering
             \captionsetup{justification=centering}
             \frame{\includegraphics[width=\textwidth]{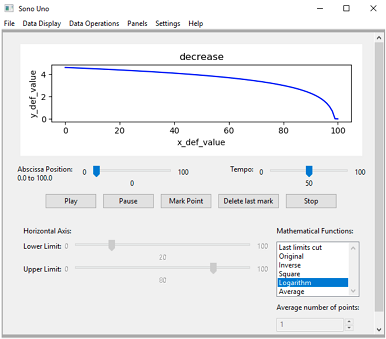}}
             \caption{Función logarítmica}
             \label{fig:mathop_log}
         \end{subfigure}
         \hfill
        \begin{subfigure}[b]{0.49\textwidth}
             \centering
             \captionsetup{justification=centering}
             \frame{\includegraphics[width=\textwidth]{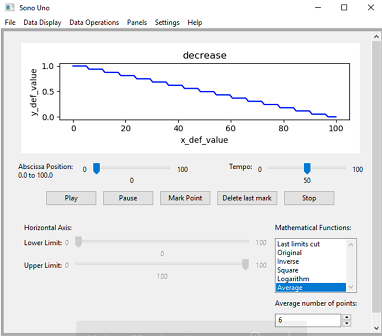}}
             \caption{Suavizado}
             \label{fig:mathop_smooth}
         \end{subfigure}
         
        \caption{Ejemplos de aplicación de cada una de las funciones predefinidas al conjunto de datos ingresado, correspondientes a una función lineal decreciente.}
        \label{fig:mathop_examples}
    \end{figure}
    
    \item \textit{Puente con Octave}
    
    Además de las funciones matemáticas que pueden aplicarse con un botón, al ser un programa que busca permitir el análisis de datos astronómicos y astrofísicos, se evaluó la necesidad de incluir herramientas que permitan ese análisis y adaptarse a las diferentes necesidades de análisis de los científicos que utilizarían la herramienta. Por tal razón, como octave es una herramienta ampliamente utilizada en astrofísica, se decidió incorporar esta herramienta desde el inicio del desarrollo del software.
    
    Es posible establecer un puente de comunicación entre el entorno de trabajo de Python y el entorno de Octave a partir de la librería de Python `oct2py'. Se utilizó dicha librería para establecer la comunicación y se diseñó e implementó los elementos que pueden observarse en el centro del recuadro rosa en la Figura \ref{fig:cap4_gui_allpanels} para facilitar la interacción con el usuario. 
    
    Para aplicar comandos de octave a los datos puede utilizarse un cuadro de texto etiquetado como `Comando de octave', donde luego el mensaje que devuelve octave se muestra en el cuadro de texto no editable que se encuentra debajo. Una vez que se realizan las operaciones deseadas en el entorno de octave, debe indicarse a SonoUno que arreglos de datos debe pedirle a octave para graficar como eje x y eje y (para lo que se utilizan los cuadros de texto y el botón colocados a la derecha de los cuadro de texto de octave).
    
    Cabe aclarar que esta sección de octave no estuvo presente en la versión de sonoUno que se probó con Grupo Focal en la Universidad de Southampton (ver Capítulo \ref{cap:fg_completo}). Sin embargo, se consideró de valor incluirlo en esta descripción de la primer versión de sonoUno porque se implementó en los dos meses posteriores a dicha prueba de grupo focal, previo al análisis de los datos obtenidos con dicha herramienta de análisis cualitativa.
    
\end{enumerate}

\subsubsection{Configuraciones}

\begin{enumerate}
    \item \textit{Acceso a las configuraciones en general}
    
    En el caso de las configuraciones, hasta el momento sonoUno dispone de configuraciones de sonido y de gráfico solamente. Debido a que ambos tipos de configuraciones son diferentes y constan de diferentes elementos se diseñó un panel que contiene un botón para cada tipo de configuración, y dicho botón (o su atajo de teclado) será el encargado de mostrar y ocultar los paneles de configuración de sonido y de gráfico.
    
    La Figura \ref{fig:config_elemgenerales} muestra las formas de acceder a las configuraciones: mediante el menú `Paneles' (`Panels') (Figura \ref{fig:config_menupanel}) haciendo visible el panel, con el menú `Configuraciones' (`Configuration') (Figura \ref{fig:config_menuconfig}) utilizando los ítems de cada submenú, o bien utilizando el atajo de teclado `Ctrl+C' para mostrar el panel de configuraciones (este atajo de teclado será cambiado en futuras actualizaciones por superponerse con un atajo de teclado del sistema). Dentro de la misma Figura \ref{fig:config_panel} se muestra el panel configuraciones con las secciones de sonido y gráfico sin seleccionar.
    
    \begin{figure}[p]
        \centering
        \begin{subfigure}[b]{0.54\textwidth}
             \begin{subfigure}[b]{\textwidth}
                 \centering
                 \captionsetup{justification=centering}
                 \frame{\includegraphics[width=\textwidth]{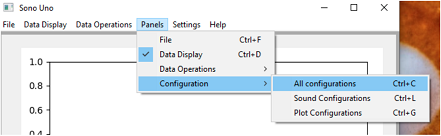}}
                 \caption{Ítem `Todas las configuraciones' (`All configurations') seleccionado en el submenú `Configuraciones' (`Configuration') del menú `Paneles' (`Panels').}
                 \label{fig:config_menupanel}
             \end{subfigure}
             \begin{subfigure}[b]{\textwidth}
                 \centering
                 \captionsetup{justification=centering}
                 \frame{\includegraphics[width=\textwidth]{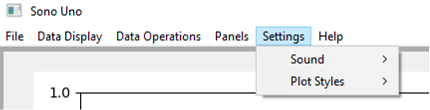}}
                 \caption{Submenús del menú `Configuraciones' (`Configuration') que corresponden a los elementos específicos de configuraciones de sonido y gráfico.}
                 \label{fig:config_menuconfig}
             \end{subfigure}
         \end{subfigure}
         \hfill
        \begin{subfigure}[b]{0.45\textwidth}
             \centering
             \captionsetup{justification=centering}
             \frame{\includegraphics[width=0.85\textwidth]{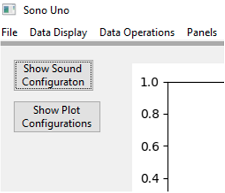}}
             \caption{Panel de configuraciones con los botones de cada sección de configuración sin presionar.}
             \label{fig:config_panel}
         \end{subfigure}

        \caption{Formas de mostrar el panel de configuraciones y vista de los botones correspondientes a cada tipo de configuraciones sin presionar.}
        \label{fig:config_elemgenerales}
    \end{figure}
    
    \item \textit{Configuraciones de sonido}
    
    Al momento de esta versión de sonoUno, donde se utilizaban sonidos MIDI con diferentes instrumento, la configuración de sonido que se permitía era la selección del instrumento con el cual se generaría la sonorización. La Figura \ref{fig:soundconfig_general} muestra por un lado la forma de acceder al panel de configuraciones de sonido (ver Figuras \ref{fig:soundconfig_menupanel} y \ref{fig:soundconfig_menuconfig}), y por otro, el panel de configuraciones de sonido con la lista de instrumentos posibles (Figura \ref{fig:soundconfig_panel}).
    
    \begin{figure}[p]
        \centering
        \begin{subfigure}[b]{0.54\textwidth}
             \begin{subfigure}[b]{\textwidth}
                 \centering
                 \captionsetup{justification=centering}
                 \frame{\includegraphics[width=\textwidth]{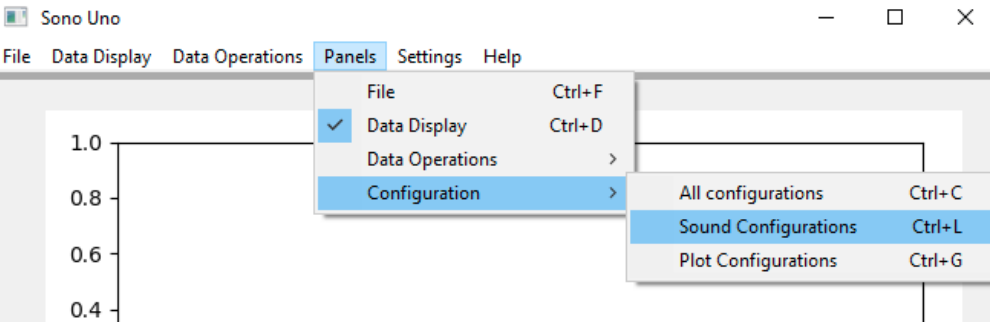}}
                 \caption{Ítem `Configuraciones de sonido' (`Sound Configurations') seleccionado en el submenú `Configuraciones' (`Configuration') del menú `Paneles' (`Panels').}
                 \label{fig:soundconfig_menupanel}
             \end{subfigure}
             \begin{subfigure}[b]{\textwidth}
                 \centering
                 \captionsetup{justification=centering}
                 \frame{\includegraphics[width=\textwidth]{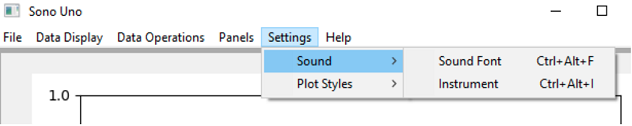}}
                 \caption{Submenú `Sonido' (`Sound') del menú `Configuraciones' (`Settings') con los elementos `Fuente de sonido' (`Sound Font') e `Instrumento' (`Instrument').}
                 \label{fig:soundconfig_menuconfig}
             \end{subfigure}
         \end{subfigure}
         \hfill
        \begin{subfigure}[b]{0.45\textwidth}
             \centering
             \captionsetup{justification=centering}
             \frame{\includegraphics[width=\textwidth]{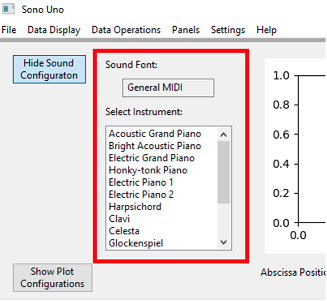}}
             \caption{Panel de configuraciones de sonido con el elemento que permite realizar la elección de instrumento.}
             \label{fig:soundconfig_panel}
         \end{subfigure}

        \caption{Formas de mostrar el panel de configuraciones de sonido y recuadrado en rojo el panel que contiene dichos elementos.}
        \label{fig:soundconfig_general}
    \end{figure}
    
    \item \textit{Configuraciones de gráfico}
    
    Las Figuras \ref{fig:plotconfig_menupanel} y \ref{fig:plotconfig_menuconfig} muestran las dos formas de acceder al panel de configuraciones de gráfico, mediante el menú `Paneles' (`Panels') y sino, el correspondiente a `Configuraciones' (`Settings').
    
    En cuanto a las configuraciones de gráfico que permite sonoUno, se utilizan las configuraciones disponibles en la librería `matplotlib', para lo que se genera un panel con elementos del estilo lista mostrando las opciones para el estilo de línea, tipo de marcador, color y grilla (por defecto desactivada). Dicho panel se muestra en el recuadro rojo de la Figura \ref{fig:plotconfig_panel}. Particularmente, en el caso de la grilla, cuenta con opciones de configuración propias que aparecen cuando el usuario habilita la opción para mostrar la grilla en el gráfico (ver Figura \ref{fig:plotconfig_grid}).
    
    La Figura \ref{fig:plotconfig_style_ejemplo} presenta diferentes combinaciones de estilo de gráfico, por ejemplo, la Figura \ref{fig:plotconfig_style_color} despliega un gráfico con línea sólida, marcador tipo punto y color rojo.
    
    \begin{figure}[p]
        \centering
        \begin{subfigure}[b]{0.49\textwidth}
             \centering
             \captionsetup{justification=centering}
             \frame{\includegraphics[width=0.9\textwidth]{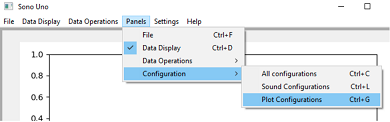}}
             \caption{Ítem `Plot Configurations' seleccionado en el submenú `Configuration' del menú `Panels'.}
             \label{fig:plotconfig_menupanel}
         \end{subfigure}
         \hfill
        \begin{subfigure}[b]{0.49\textwidth}
             \centering
             \captionsetup{justification=centering}
             \frame{\includegraphics[width=\textwidth]{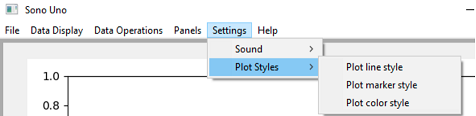}}
             \caption{Submenú `Plot Styles' del menú `Settings' con los elementos propios de configuración.}
             \label{fig:plotconfig_menuconfig}
         \end{subfigure}
         \hfill
        \begin{subfigure}[b]{0.49\textwidth}
             \centering
             \captionsetup{justification=centering}
             \frame{\includegraphics[width=0.8\textwidth]{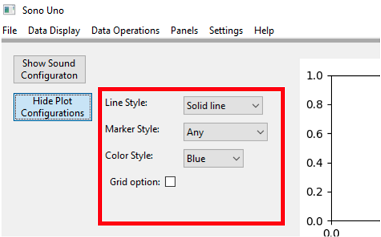}}
             \caption{Panel con los elementos de configuración de gráfico, la grilla en este caso está desactivada.}
             \label{fig:plotconfig_panel}
         \end{subfigure}
         \hfill
        \begin{subfigure}[b]{0.49\textwidth}
             \centering
             \captionsetup{justification=centering}
             \frame{\includegraphics[width=\textwidth]{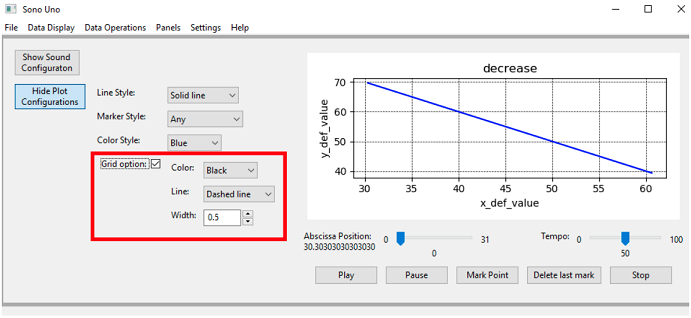}}
             \caption{Elementos de configuración de grilla indicados con un recuadro rojo, se muestran luego de activar la grilla.}
             \label{fig:plotconfig_grid}
         \end{subfigure}
         
        \caption{Muestra las formas de habilitar el panel de configuración de gráfico y los elementos de configuración disponibles.}
        \label{fig:plotconfig_general}
    \end{figure}
    
    \begin{figure}[p]
        \centering
        \begin{subfigure}[b]{\textwidth}
             \centering
             \captionsetup{justification=centering}
             \frame{\includegraphics[width=0.6\textwidth]{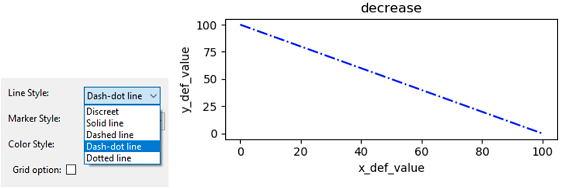}}
             \caption{Selección de estilo de línea `guión-punto'.}
             \label{fig:plotconfig_style_line}
         \end{subfigure}
         \hfill
        \begin{subfigure}[b]{\textwidth}
             \centering
             \captionsetup{justification=centering}
             \frame{\includegraphics[width=0.6\textwidth]{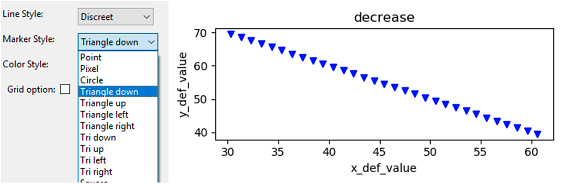}}
             \caption{Selección de marcador triángulo hacia abajo.}
             \label{fig:plotconfig_style_marker}
         \end{subfigure}
         \hfill
        \begin{subfigure}[b]{\textwidth}
             \centering
             \captionsetup{justification=centering}
             \frame{\includegraphics[width=0.6\textwidth]{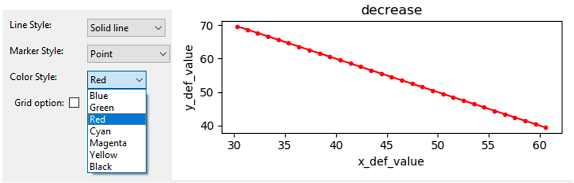}}
             \caption{Selección de color rojo.}
             \label{fig:plotconfig_style_color}
         \end{subfigure}
         
        \caption{Diferentes ejemplos de estilos de gráfico seleccionados con los elementos de configuración disponibles en sonoUno.}
        \label{fig:plotconfig_style_ejemplo}
    \end{figure}
    
\end{enumerate}

\section{Resultados}

En esta sección se mostrarán casos de uso con datos reales de astronomía y astrofísica. Si bien el programa se puede utilizar con cualquier conjunto de datos en formato de columnas, como por ejemplo la función decreciente utilizada para mostrar las funcionalidades en la sección anterior, el objetivo principal de esta tesis se basa en el acceso, uso y exploración de datos astronómicos usando la sonorización como complemento de la visualización, asegurando una aproximación multisensorial o multimodal a dichos datos.

\begin{figure}[ht!]
    \centering
    \begin{subfigure}[b]{0.49\textwidth}
        \centering
        \captionsetup{justification=centering}
        \frame{\includegraphics[width=\textwidth]{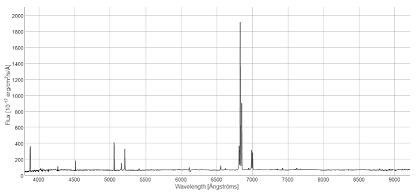}}
        \caption{Imagen obtenida de la base de datos del SDSS, de donde provienen los datos del espectro galáctico desplegado.}
        \label{fig:cap4_result_sdss_basedatos}
    \end{subfigure}
    \hfill
    \begin{subfigure}[b]{0.49\textwidth}
        \centering
        \captionsetup{justification=centering}
        \frame{\includegraphics[width=\textwidth]{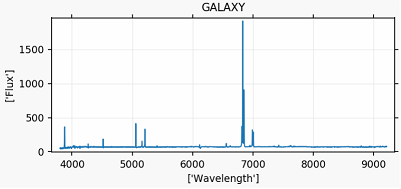}}
        \caption{Gráfico obtenido a partir de los datos descargados del SDSS, procesados con sonoUno.}
        \label{fig:cap4_result_sdss_sonouno}
    \end{subfigure}
         
    \caption{Comparación de las gráficas desplegadas por el SDSS y el programa sonoUno.}
    \label{fig:cap4_result_sdss}
\end{figure}

\begin{figure}[ht!]
    \centering
    \begin{subfigure}[b]{0.49\textwidth}
        \centering
        \captionsetup{justification=centering}
        \frame{\includegraphics[width=\textwidth]{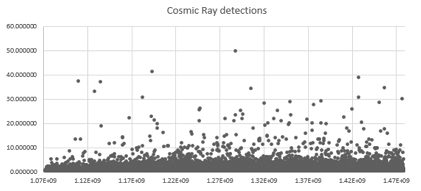}}
        \caption{Gráfico obtenido a partir de los datos de rayos cósmicos públicos descargados, procesados con hoja de cálculo.}
        \label{fig:cap4_result_auger_excel}
    \end{subfigure}
    \hfill
    \begin{subfigure}[b]{0.49\textwidth}
        \centering
        \captionsetup{justification=centering}
        \frame{\includegraphics[width=\textwidth]{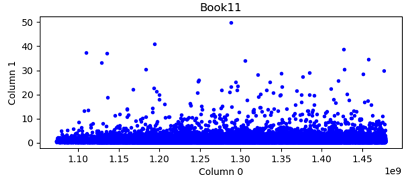}}
        \caption{Gráfico obtenido a partir de los datos descargados de `Auger open access Data'  con sonoUno.}
        \label{fig:cap4_result_auger_sonouno}
    \end{subfigure}
         
    \caption{Comparación de las gráficas obtenidas para datos públicos de rayos cósmicos, provenientes del Observatorio Pierre Auger, en Malargüe, usando una hoja de cálculo y el programa sonoUno.}
    \label{fig:cap4_result_auger}
\end{figure}
    
Los datos que se muestran en esta sección provienen de la base de datos Sloan Digital Sky Survey y de la base de datos del observatorio Pierre Auger instalado al sur de Mendoza, Argentina. Estos datos de una Galaxia y de Rayos Cósmicos, respectivamente, se utilizarán para corroborar el proceso de graficación del software, comparando el gráfico provisto por la bases de datos de referencia o un programa de uso masivo como ser una hoja de cálculo, con el gráfico obtenido en sonoUno. 

La Figura \ref{fig:cap4_result_sdss} muestra la gráfica obtenida de los datos de una galaxia identificada como \href{http://skyserver.sdss.org/dr15/en/tools/quicklook/summary.aspx?}{\underline{\textcolor{blue}{`SDSS J151806.13+424445.0'}}}, en la Figura \ref{fig:cap4_result_sdss_basedatos} se observa la gráfica proporcionada por la base de datos y en la Figura \ref{fig:cap4_result_sdss_sonouno} la gráfica elaborada por el programa sonoUno. De la comparación entre ambas gráficas se puede inferir que son iguales, lo que confirmaría una correcta producción de la gráfica en sonoUno.

En el caso de los datos descargados de la página de \href{https://labdpr.cab.cnea.gov.ar/opendata/data.php}{\underline{\textcolor{blue}{Pierre Auger}}}, el Observatorio provee de ciertos gráficos de datos pero no el que se muestra en la Figura \ref{fig:cap4_result_auger}, por ello para la comparación se procedió a realizar el ploteo de puntos con un programa de hoja de cálculo. La Figura \ref{fig:cap4_result_auger} muestra la comparación de la gráfica obtenida en la hoja de cálculo y la generada por sonoUno, evidenciándose que ambas gráficas son iguales.

Cabe destacar, que gracias a las configuraciones de gráfico disponibles en sonoUno, se puede ajustar la gráfica de acuerdo al tipo de datos desplegado. Por ejemplo, en el caso de la Figura \ref{fig:cap4_result_sdss} que representa datos del espectro de una galaxia, relacionando la longitud de onda con el flujo, los mismos deben representarse de forma continua. En cambio, para el caso de rayos cósmicos, que se manifiestan como eventos registrados por los detectores (en este caso detectores Cherencov), el gráfico debe realizarse con puntos, que relacionan Energía del rayo cósmico primario en función del tiempo de detección (ver Figura \ref{fig:cap4_result_auger}). Esta capacidad de sonoUno en su interfaz gráfica, permite obtener un gráfico representativo de los datos que se están desplegando y en relación con el análisis que de desea hacer.

En cuanto al sonido generado, se comprueba que el mismo tiene mayor contenido de tonos bajos (correspondientes a los valores mínimos que son mayoría en ambos datos) y menos contenido de tonos altos (correspondientes a los picos o valores cercanos al valor máximo) (se puede acceder a los sonidos en los siguientes links: \href{https://www.sonouno.org.ar/wp-content/uploads/sites/9/2021/12/SDSS-J151806.13424445.0_flux_fmax2003.wav}{\underline{\textcolor{blue}{galaxia}}} y rayos cósmicos \href{https://drive.google.com/file/d/1skXWhcQvocmXJ_7n5RyEWSPQVKm2r7f5/view?usp=sharing}{\underline{\textcolor{blue}{completo}}} (con variación de notas de piano y mayor tempo) y \href{https://www.sonouno.org.ar/wp-content/uploads/sites/9/2021/10/auger_public_2016_11_12_range1285973511-1294550092_sound.wav}{\underline{\textcolor{blue}{recorte}}} (con un sonido de piano pero menor tempo)), lo que ayuda a la detección de señales que pueden confundirse con el ruido, teniendo entonces la posibilidad de explorar detección de rasgos que débilmente se impongan sobre dicho ruido, una de las premisas en torno de la percepción señales y de esta investigación: el uso de más de un sentido implica mayor índice de detección de rasgos significativos o patrones en los datos.

\section{Conclusiones}

Se presenta en este capítulo la primera versión de sonoUno, exponiendo las bases de su diseño consecuencia del Capítulo \ref{cap:analisis_normativo} y todas las funcionalidades que se lograron implementar para esta fase. Esta versión fue presentada y subida a la plataforma GitHub en el año 2019, habiendo logrado un software de sonorización de datos que permitía ingresar archivos de datos en formato columnas, graficarlos, sonorizarlos, aplicar ciertas funciones matemáticas y ajustar parámetros de configuración como instrumento y color de gráfico.

Es preciso aclarar que se logró esta primera versión de interfaz centrada en el usuario desde su concepción, permitiendo la integración de características de diseño como la integración de funcionalidades por paneles. Ésta última posibilita un diseño con pocos elementos a primera vista, pero gran potencial y número de funcionalidades que pueden mostrarse a voluntad del usuario.

Por otro lado, el diseño modular y la metodología de desarrollo de software utilizada permitió a los integrantes del equipo trabajar sobre los diferentes módulos de forma autónoma e integrada a la vez. De esa forma se logró un desarrollo efectivo en el tiempo esperado. Este tipo de diseño aplicado ha permitido la inclusión de nuevos módulos al programa en los años posteriores de desarrollo, trabajando con colaboradores de diferentes lugares.

En base a los resultados mostrados, es preciso destacar que el software realiza un despliegue confiable y que se corresponde con el sonido reproducido. Se detecta que la resolución sonora que ofrece los rangos disponibles en la librería de sonido no es suficiente para conjuntos de datos que presentan gran dispersión, siendo este un aspecto que se ha mejorado en las versiones posteriores (se profundizará sobre este tema en el Capítulo \ref{cap:reinforce}).

Los siguientes pasos luego de esta primera implementación se relacionaron con pruebas de software con futuros usuarios que serán descriptas en el próximo capítulo (Capítulo \ref{cap:fg_completo}). Es importante tener en cuenta que esta herramienta tiene la potencialidad de permitir el análisis de funciones y datos de diferentes orígenes, por parte de estudiantes y profesionales con o sin discapacidad visual. En el caso de escuelas que trabajan especialmente con alumnos ciegos o con visión reducida, o que tienen integrados a estos alumnos en clases de instituciones educativas tradicionales, este recurso puede ser una herramienta de valor en el proceso de enseñanza-aprendizaje.

Pensado originalmente como una herramienta para acceder y sonorizar datos astronómicos, este desarrollo muestra que es posible generar mayor accesibilidad a cualquier tipo de datos de una forma complementaria a las actuales, brindando la posibilidad de análisis mutimodal de dichos datos, asegurando mayor independencia del usuario respecto del recurso de tutores y otros dispositivos hápticos.

\chapter{Pruebas con usuarios}
\label{cap:fg_completo}

\section{Introducción}

Las tecnologías digitales tienen una capacidad sustancial para mejorar la inclusión o, por el contrario, para excluir aún más a las personas. En las últimas dos décadas, las organizaciones internacionales han priorizado un compromiso serio con el hecho de que personas con discapacidad tengan acceso al mundo digital. La Organización de las Naciones Unidas (ONU) consagró, como un derecho humano, el acceso digital a la información (artículo 19 Declaración Universal de los Derechos Humanos-UDHR \citep{un2021declar}, y Convención sobre los Derechos de las Personas con Discapacidad (CHRPD) \citep{un2021file} anexo g, artículo 2, artículo 4, artículo 9 apartados b, f, g, h, artículo 21, artículo 24). En los últimos años se ha comenzado a evidenciar un crecimiento de iniciativas, movimientos y esfuerzos para igualar el acceso a los recursos, acciones que condujeron a la concreción del manifiesto de Disability Interaction/eXperience, DIX \citep{holloway2019}. Además, la ONU profundizó la acción posible en este tema, al establecer en el artículo 24 de la CHRPD inciso e, que: ``(e) Se brindan medidas efectivas de apoyo individualizado en entornos que maximicen el desarrollo académico y social, en consonancia con el objetivo de la plena inclusión''. Aquí, ``individualizado'' se refiere a una acción específica para el individuo, teniendo en cuenta las perspectivas individuales y sociales de cada uno \citep{who2002}. De todas maneras, debe tenerse en cuenta que cuando se habla de inclusión y en particular de tecnologías para la inclusión, las aproximaciones pueden ser muy diversas, tanto en recursos como en atención a las necesidades de los usuarios finales. Finalmente, la inclusión puede abordar diversos aspectos que impiden al ser humano su pleno desarrollo en áreas diversas, como las que se trabajan en ciencias. Por ello, en esta tesis, nos concentramos en la posibilidad de introducir la sonorización como una herramienta inclusiva y a la vez novedosa para el análisis de los datos en el amplio sentido de la multimodalidad.

Aunque existen muchas iniciativas de programas para sonorización de datos, ninguna ha sido capaz de mejorar la situación de las personas con discapacidad, en pos de permitirles estar a la par de la población en general. Particularmente, en el caso de la educación, que es el primer paso hacia la profesionalización: la historia se repite, dejando a los discapacitados en posiciones desventajosas. En cuanto a los trabajos en ciencias, las personas con discapacidad (especialmente las discapacidades congénitas, es decir, adquiridas desde el nacimiento) aún no pueden desempeñarse adecuadamente y, en la mayoría de los casos, no pueden trabajar como investigadores. Esto no se debe a una falta de esfuerzo o a los logros de los individuos, sino a la reticencia en contratar a personas con discapacidad \citep{vornholtetal2018}, la mayoría de las veces por prejuicio, siendo lo anterior especialmente cierto para las personas con discapacidades congénitas,  como por ejemplo: discapacidades sensoriales, ortopédicas, neurodiversas y de aprendizaje. La situación es un tanto diferente para las personas con discapacidades de inicio tardío, en este caso ya tienen una habilidad adquirida y en algunos casos cuentan con cierta estabilidad laboral; sin embargo, aún así sufren despidos y les es difícil la reinserción laboral.

En términos de accesibilidad a las interfaces digitales, el paradigma del Diseño Centrado en el Usuario desde el inicio del desarrollo propuesto,  puede beneficiar a las personas con discapacidad, eliminando barreras y personalizando la experiencia del usuario; por ejemplo, al interactuar con tecnologías digitales, un usuario con ceguera congénita debe tener la libertad de elegir el orden de navegación; hasta ahora, su única opción es la navegación secuencial. 

Es preocupante y cuestionable que costosos algoritmos de aprendizaje automático, principalmente destinados a promover el consumo, estén constantemente monitoreando nuestras respuestas, adaptando opciones y reconfigurando el algoritmo para que se ajuste a nuestros intereses o necesidades, pero que estos algoritmos casi no se utilice para adaptar las interfaces de accesibilidad al usuario, haciendo que las mismas estén más centradas en el usuario con discapacidad.

La implementación y desarrollo actual del diseño tecnológico, junto a las evaluaciones de usabilidad con técnicas como los grupos focales, siguen normas internacionales de estandarización como la WCAG, W3C, WAI-ARIA y similares (mencionadas en el Capítulo \ref{cap:analisis_normativo}). Si bien estas son normas importantes, no pueden tener en cuenta la realidad cambiante de cada individuo y sus necesidades, tomar dichas normas como un manual irrefutable puede conducir a perjudicar la voz de los usuarios discapacitados. Enfoques centrados sólo en normativas de accesibilidad conllevan el riesgo de terminar muy lejos de las necesidades reales del usuario, otorgando adicionalmente una falsa seguridad de éxito a los desarrolladores.

Consideremos el siguiente ejemplo: una evaluación de usabilidad de una interfaz donde los usuarios más experimentados en el uso de interfaces digitales pueden navegar sin problemas, pero a un usuario discapacitado que no puede hacerlo se le dice ``otras personas lograron hacerlo''. En dicho caso, según el sociólogo y colaborador de derechos humanos de la ONU, \citet{shakespeare2017}, con tal comentario se le está diciendo al usuario discapacitado: ``tu situación es falsa''. Por lo general, según dicho autor, las fallas en el diseño y la arquitectura del sistema terminan echando la culpa a los discapacitados: ``¡El problema es que eres diferente!''. Del mismo modo, el usuario con discapacidad puede sentir que el sistema no lo toma en cuenta y que no le queda más remedio que ceder. Ignorar las capacidades y habilidades de cada individuo, dejándolos sin acceso a herramientas e información, es una violación de los Derechos Humanos de las Personas con discapacidad \citep{un2021file}. Lo mencionado puede evidenciarse en un análisis del pensamiento social, el cual, por lo general, se corresponde con el pensamiento de los desarrolladores de tecnologías digitales, donde se encuentra la tendencia a creer que es imposible crear una interfaz que se ajuste a todos.

La UDHR \citep{un2021declar} y la CHRPD \citep{un2021file} exigen el desarrollo/implementación de un marco que se centre en preguntas como las enunciadas a continuación: ``¿Cómo permito que el usuario utilice esta interfaz para desempeñarse de manera óptima, sin tener que cambiar drásticamente su situación particular? ¿Qué se implementará para maximizar la experiencia del usuario, sin someterlo a curvas de aprendizaje y transacciones que puedan perjudicar su autoestima y excluirlo aún más?''. La Organización Mundial de la Salud con su clasificación internacional de Funcionamiento, Discapacidad y Salud, promueve el acercamiento a la realidad de cada persona tratándola como una variación y no como una anormalidad.

La comunidad de especialistas en HCI y las ciencias de la computación están cambiando la forma de acercarse a la ``discapacidad'' del individuo, centrándose en cambio en un marco integrador que considera: la discapacidad individual y su funcionamiento (en contexto), y las capacidades de individuo; es decir, un enfoque basado en la capacidad y no en la incapacidad. \citet{holloway2019} propone un enfoque y un marco de integración de la discapacidad en la interacción humano-computadora y el mundo digital. De una manera muy inteligente, pide la integración de tecnologías para atender a las personas con discapacidad, no exige que las personas con discapacidad tengan opciones limitadas o decidan. Teniendo en cuenta que la discapacidad es multidimensional, se considera que el marco de interacción de la discapacidad propuesto en 2019 \citep{holloway2019} carece de una pregunta: ¿cómo integro tecnologías que implican actualizaciones sostenibles, no solo estrategias a corto plazo?. Si el objetivo es impactar el desarrollo social del individuo para que la discapacidad sea considerada de manera general, necesitamos crear tecnologías que no sólo se adapten a las habilidades individuales, sino que también evolucionen. Es decir, que progresen a medida que los estilos de interacción se transformen con el tiempo y, en última instancia, predigan y se adapten a los estilos de interacción que pueden surgir en el futuro. ¿Cómo diseñamos tecnologías que permitan a las personas funcionar con lo que ya tienen/traen y tal como son?

Considerando lo mencionado y los objetivos específicos de esta tesis, el presente capítulo aporta al tercer objetivo específico, el cual enuncia: ``Realización de un grupo focal integrado por personas con y sin discapacidad visual, para investigación sobre usabilidad y accesibilidad del programa desarrollado''. Así es que, teniendo en cuenta que esta sección presenta un análisis de grupo focal donde se ha probado sonoUno, se han adoptado las definiciones de usabilidad, eficiencia y eficacia de la norma ISO 9241-11:2018, las cuales se detallan a continuación:
\begin{enumerate}
    \item \textit{Usabilidad}: grado en que un sistema, producto o servicio puede ser utilizado por usuarios específicos para lograr objetivos específicos con eficacia, eficiencia y satisfacción en un contexto de uso específico.
    \item \textit{Eficiencia}: recursos utilizados en relación con los resultados obtenidos.
    \item \textit{Eficacia}: precisión y exhaustividad con la que los usuarios alcanzan los objetivos especificados.
\end{enumerate}

Adicionalmente al análisis con grupo focal, se realizaron intercambios por correo electrónico con personas que trabajan en astronomía, comentándoles sobre actualizaciones del programa y solicitando que instalen y usen la herramienta con el fin de recabar datos de usuarios expertos (sección \ref{sect:mails_completa}). Por último, en este capítulo también se mencionan dos casos de uso, donde dos de las personas contactadas por correo electrónico decidieron utilizar sonoUno en diferentes entornos.

\section{Pruebas con Grupo Focal}
\label{sect:FG_completo}

Esta sección describe el planteo, desarrollo y análisis de resultados, de un grupo focal que examinó el programa sonoUno. En este contexto, se describen los hallazgos realizados con esta técnica tanto positivos, como aspectos a mejorar para hacer el programa más accesible y usable. Producto de este análisis se publicó un paper en la revista \textit{``American Journal of Astronomy and Astrophysics''} \citep{casadoFG2022}, el cual se adjunta en el Apéndice \ref{ap:FG}.

\subsection{Metodología adoptada para el análisis de grupo focal}
\label{sect:method_FG}

Con la primera versión del software sonoUno funcionando, se utilizó la técnica de grupo focal para investigar su efectividad y usabilidad por parte de personas con y sin discapacidad. La práctica fue llevada a cabo en la ciudad de Southampton, Reino Unido, durante abril de 2019. Adicionalmente a lo enunciado, se utilizó el intercambio con usuarios para consultarles sobre la calidad del sonido y su opinión sobre esta nueva propuesta de enfoque multimodal para el análisis de datos astrofísicos.

El encuentro e intercambio se basó en una serie de preguntas y una lista de tareas diseñadas con tres niveles de complejidad (bajo, medio y alto). Las sesiones se dividieron en dos espacios que se realizaron el mismo día, con un descanso intermedio que permitió a los participantes asimilar e interactuar (hablar entre ellos, por ejemplo) sobre el primer uso del software. La primera sesión se sustentó en una pregunta de introducción, la presentación de la herramienta a los participantes, una lista de tareas que debían seguir con diferentes niveles de complejidad, y terminó con preguntas sobre la efectividad de la herramienta (dos de ellas sobre la usabilidad del software, una sobre el sonido y la última sobre la posibilidad de analizar datos con esta herramienta). 

En cuanto a la segunda sesión, después de un descanso de 15 minutos, la primera tarea fue probar el programa con un conjunto de datos astronómicos descargados de la base de datos \href{http://skyserver.sdss.org/dr15/en/tools/quicklook/summary.aspx?ra=179.8036794828000&dec=-00.5238033275900}{\underline{\textcolor{blue}{Sloan Digital Sky Survey}}}, las personas podían usar el programa como quisieran. Luego de esta primera tarea, la sesión continuó con más preguntas: cinco se centraron en la usabilidad del programa, una en el sonido y dos para su opinión sobre el enfoque multimodal de los datos y si lo utilizarían en su trabajo (las preguntas y la lista de tareas están disponibles en el Apéndice \ref{ap:FG}, luego de la publicación en la revista `American Journal of Astronomy and Astrophysics').

Los participantes fueron invitados por correo electrónico y contacto personal. Entre las personas contactadas, nueve aceptaron participar en el grupo focal y fueron divididos en cuatro grupos según si trabajaban en astrofísica y su diversidad funcional:
\begin{enumerate}
    \item \underline{Grupo A}: formado por 4 personas que no trabajan con datos astrofísicos en su día a día (2 con baja visión, 2 totalmente ciegas). Los cuatro utilizan el lector de pantalla (NVDA) con el software y 2 de ellos utilizan adicionalmente la lupa; las herramientas utilizadas se encontraban preinstaladas en la computadora.
    \item \underline{Grupo B}: 1 persona sin deficiencia sensorial, no es astrónomo de profesión, su carrera de formación es informática.
    \item \underline{Grupo C}: 1 persona con baja visión con formación en astronomía (profesionalmente). Esta persona utilizó el lector de pantalla junto con la lupa.
    \item \underline{Grupo D}: 3 personas sin deficiencia sensorial, formación profesional en astronomía (Rango de edad: 22-27).
\end{enumerate}

Debido a limitaciones de tiempo y disponibilidad de espacios, los grupos B y C no pudieron ampliarse. Es extremadamente difícil reunir muchas personas con discapacidad visual que trabajen activamente en el campo de la astrofísica (o cualquier otro campo). No son muchas las personas que adquieren la discapacidad luego de obtener el grado y luchan contra las barreras impuestas para seguir ejerciendo su profesión. Por otro lado, las personas con discapacidad visual congénita o de inicio temprano, no cuenta con herramientas que le permitan estudiar y desarrollarse en el ámbito académico y de investigación. Disminuir (y por que no, extinguir) estas barreras es uno de los factores motivadores de la presente investigación, y uno de los objetivos de la herramienta en desarrollo. En cuanto al Grupo B, compuesto por una sola persona, se consideró importante tener alguien sin discapacidad que no trabajara en astrofísica en este estudio, inclusive del análisis decantan importantes recomendaciones que refuerzan declaraciones hechas por los otros grupos.

Aún con las limitaciones que presentan los grupos conformados, se consideró que eran suficientes para obtener primeras respuestas a nuestras preguntas de investigación, que son: 1) ¿El software en desarrollo es usable, efectivo y eficiente?; 2) ¿El programa permite el análisis de datos a través del sonido?; 3) ¿Cuál es la opinión de los participantes con respecto al análisis multimodal de datos astrofísicos?

En cuanto a los encuentros con los diferentes grupos, todas las reuniones se llevaron a cabo de forma análoga y se grabaron en formato audio, previo consentimiento de los participantes. El proyecto también obtuvo la aprobación por parte del comité de ética de la Universidad de Southampton en Southampton, Reino Unido, antes de su inicio (referencia ERGO II 48331), toda la documentación fue elaborada en conjunto y fue presentada por dicha Universidad. Una vez finalizadas las reuniones de los grupos focales, las grabaciones de audio fueron transcritas para su posterior análisis y destruidas después de un período de un año.

\subsection{Análisis de las transcripciones}
\label{sect:analisis_transcripciones}

El análisis de las transcripciones fue realizado por dos miembros del equipo de trabajo, utilizando la técnica de ‘condensación sistemática de texto’ \citep{malterud2012}. Dicha técnica se basa en recorrer el texto de las traducciones con la finalidad de concretar el contenido, pudiendo definir al final un número específico de categorías representativas. 

En primer lugar, se realizan encuentros para acordar los pasos a seguir y el como abordar el análisis, en este caso se acordó recorrer el texto lo más libre de preconcepciones posibles, extrayendo frases representativas que se repitan durante la entrevista y/o entre los participantes. A partir de entonces, se continuó por separado recorriendo las transcripciones, hasta que se definieron ``afirmaciones'' representativas. Tras un primer análisis, se realizó una nueva reunión para discutir y consensuar las afirmaciones y las posibles categorías que se consideraron características de las contribuciones de los voluntarios en el Grupo Focal. Finalmente, trabajando de forma individual, se categorizaron cada una de las afirmaciones o declaraciones extraídas, revisando nuevamente si las categorías eran suficientes y representativas. 

Para identificar a los participantes se utilizó un código con el objetivo de mantener su anonimato. A continuación, al explicar las categorías identificadas se presentarán frases textuales de las transcripciones. Dado que el grupo focal fue desarrollado en inglés y este texto está en castellano, primero y entre comillas simple se indicará la frase en español, seguido con comillas dobles y entre paréntesis la frase original en inglés; debe tenerse en consideración que hay traducciones que no son literales, debido a que se tiene en cuenta el contexto de la conversación. Siguiendo el camino descripto, las categorías definidas fueron:
\begin{enumerate}
    \item \underline{Sobrecarga de memoria}:

    Determinada por afirmaciones referentes a la necesidad de realizar muchas transacciones para alcanzar su objetivo. Ejemplos de afirmaciones de los participantes incluyen: A4 durante la prueba principal \textit{`mi instinto cuando intento pausar es volver a play en lugar de reiniciar' (``my instinct when trying to pause playback is instead to restart'')} (en la versión utilizada de sonoUno, Reproducir y Pausa están asignados a diferentes botones); A2 siguiendo la instrucción del moderador (`Debes seleccionar un instrumento' (``You have to select an instrument'')) responde: \textit{`¿La fuente de sonido?' (``The sound font?'')} (aquí A2 se perdió en el espacio de texto anterior a seleccionar instrumento, el cual solo indicaba que el software usaba una fuente de sonido determinada; este elemento innecesario desvió la atención de A2 y contribuyó a la sobrecarga de memoria). Además, las afirmaciones sobre la falta de consistencia también contribuyen a la sobrecarga de la memoria, refiriéndose a la confusión sobre las diferencias entre la visualización gráfica y auditiva: B expresó \textit{`cuando vi el gráfico original, con la línea continua, no esperé escuchar notas individuales' (``when I saw the original graph, with the continuous line, I don’t expect to hear individual notes'')}. Además, hubo comentarios positivos con respecto a esta categoría: B expresó \textit{`todo fue bastante obvio' (``everything was quite obvious'')} reforzando la presencia de consistencia entre otras aplicaciones y sonoUno; en concordancia, A4 declaró \textit{`Estos son los mismos comandos que se encuentran en las opciones predeterminadas en el Bloc de notas' (``These are the same commands that are the default options in Notepad'')}; en el caso de C, la expresión durante una búsqueda de funcionalidad fue \textit{`Entonces, ese es el tercero, sí, está bien' (``So, that’s the third, yeap, ok'')}.

    \item \underline{Necesidad de información}:

    Categoría determinada por las respuestas de los participantes e indicaciones directas sobre la necesidad de que sonoUno los mantenga informados sobre cada acción realizada de forma multimodal. Particularmente, B expresó \textit{`Si marcas un punto, ¿cuál es el valor de la longitud de onda?' (``If you mark a point, what is the wavelength value?'')}; C1 también comentó refiriéndose al programa \textit{`¿Y me dice cuál es mi rango seleccionado?' (``And does it tell me what my selected range is?'')}. Además, surge la necesidad de confirmación después de una acción, D2 comentó \textit{`pero espero ver algo aquí cuando presione marcar punto' (``but I expect to see something here when I push mark point'')}; el problema en esta versión era que la marca negra se colocaba detrás de la barra roja y el usuario solo podía verla luego de que continuaba la reproducción. Asimismo, asegurar la comunicación con la tecnología de asistencia y la efectividad en comunicar los mensajes dentro de sonoUno debe ser un enfoque importante; en este sentido, A4 expresa \textit{`dice 25 slider [...] ¿qué significa slider?' (``it says 25 slider [...] what’s slider mean?'')}. Sobre la tecnología de asistencia y la importancia de proporcionar las mismas herramientas a las que las personas están acostumbradas, A2 destaca \textit{`[...] necesito que me lo recuerden, pero eso es porque prefiero tener la información en braille a mi lado [...]' (``[...] I need to be reminded, but that’s because I prefer to have the information in braille besides me [...]'')}. Adicionalmente, A4 y A1 comentaron sobre la necesidad de precisión y retroalimentación de sonoUno: A4 indicó \textit{`[...] pero para cuando, si escuchas, escuchas el pico, para el momento en que presionas la tecla de atajo [...]'} (aquí luego hizo una expresión indicando que para cuando se llegaba a presionar la tecla ya había pasado el pico y el cursor estaba en otro punto) \textit{(``[...] but by the time, if you listen, you hear the peak, by the time you get the shortcut key [...]'')}; agregando, A1 expresó \textit{`Sí, yo estaba preocupado [...]'} (haciendo referencia a que también había notado lo mismo que A4) \textit{(``Yea, I was concerned [...]'')}.

    \item \underline{Necesidad de poder elegir}:

    En este caso, se encuentra definida por el discurso de los participantes donde destacaron que sonoUno ofrece a los usuarios con discapacidad la posibilidad de explorar los datos y controlar las interacciones de principio a fin: C expresó \textit{`[...] libertad también, sentí que era mucho más capaz de acceder a los pequeños datos que quiero hacer' (``[...] freedom as well, I felt like I was a lot more able to access the bitty data that I want to do'')}; A1 también comentó: \textit{`Creo que en las circunstancias en las que trabajas con personas que no tienen, literalmente, ninguna visión, entonces es bastante valioso porque te da acceso a algo a lo que de otro modo no tendrías acceso, también información a la que no tienen acceso' (``I think in the circumstances where you are working with people who have no, literally no vision, then it’s quite valuable because it gives you access to something that you otherwise wouldn’t have access to, so information that you wouldn’t have access to'')}. En contraste, los participantes sin discapacidad visual declararon la necesidad de más funciones que permitan trabajar con los datos y la visualización gráfica: D2 dijo \textit{`Necesito algunas cosas interactivas, realmente analizar en la línea de comando' (``I need some interactive things, really analysing on the command line'')}; en cuanto a B \textit{`[...] estas barras, me gustaría ponerles números específicos' (``[...] these bars, I would like to put specific numbers on'')} refiriéndose a las barras deslizantes que permiten cambiar los límites del eje x en el gráfico. Además, algunos comentarios remarcan la necesidad de controlar la interacción y la necesidad de precisión de la acción que realiza el usuario, junto con la posibilidad de ajustar la configuración de la interfaz: por ejemplo, A2 preguntó durante la segunda prueba después del descanso \textit{`Entonces, ¿es esto piano otra vez?' (``So, is this a piano again?'')}; B señaló \textit{`[...] toma mucho tiempo reproducir el gráfico completo' (``[...] take a lot of time to play the full graph'')}; B, D1, D2 y D3 expresaron \textit{`[...] hacer más interactivo el gráfico' (``[...] make the plot more interactive'')}; C declaró: \textit{`Creo que tener la capacidad de cambiar el instrumento o el tiempo es realmente importante' (``I think having the ability to change the instrument or the time is really important'')}.

    \item \underline{Necesidad de entrenamiento}:

    Esta categoría se encuentra definida por las respuestas que señalan la necesidad de ayuda al inicio usando la herramienta, en este caso por personas con discapacidad visual: A3 y A2 intercambiaron \textit{`[...] si alguien te hubiera mostrado cómo abrir diferentes carpetas, y como es la estructura de carpetas, ¿hubiera sido más fácil?' (``[...] if somebody could demonstrate how to open different folders, or the folder structure and how it looks, would it have been easier?''), `Probablemente si' (``Probably, yes'')}; también, A1 complementó \textit{`[...] pero en esta etapa ciertamente necesitamos gente alrededor' (``[...] but at this stage we certainly need people around'')}, refiriéndose a la versión de sonoUno probada. Además, el discurso de los participantes también indicó la necesidad de capacitación sobre la técnica general de sonorización: B mencionó \textit{`[...] no creo que pueda detectar las pequeñas tendencias' (``[...] I don’t think that I could pick out the small trends'')}; D2 complementó con \textit{`no era tan fácil identificar las cosas, solo por el sonido' (``it was not that easy to identify the things, just by sound'')}; A4 resaltó esto con una pregunta \textit{`Quiero decir, si tienes cuatro imágenes, ¿cómo sabes que la línea de hidrógeno de esta imagen está en la misma posición que la línea de hidrógeno en la segunda que analizas, para saber que se está moviendo a la misma velocidad o a la misma distancia? porque en este momento lo estás haciendo visualmente, no puedo mirar la imagen, ¿cómo una persona ciega mira dos imágenes de audio?' (``I mean if you’ve got four pictures, how do you know that the hydrogen line of this picture is in the same position as the hydrogen line on the second one that you analyse, to know that is moving at the same speed or the same distance? ’cause at the moment you are doing it visually, I can’t look at the picture, how does a blind person look at two audio pictures?'')}.

    \item \underline{Aspectos sociales}:

    Esta categoría está formada por las declaraciones de los participantes donde se evidencia que las personas con discapacidad cargan en ellos mismos la culpa de algún mal funcionamiento del software, resaltando además la incomprensión que se tiene sobre las discapacidades: C expresó \textit{`[...] mi única queja sería la imposibilidad de aumentar el tempo aún más de lo que se puede de momento, porque \textbf{soy impaciente}' (``[...] my only complaint would be the inability to increase the tempo even more than you can at the moment, because I’m impatient'')}; B comentó \textit{`en cuanto a las personas ciegas, ¿cómo pueden usar la interfaz?' (``for blind people, how they can use the interface?'')}; D3 afirmó en respuesta a cualquier ventaja del software \textit{`por razones científicas no sé, por razones de divulgación mucho' (``for scientific reasons I don’t know, for outreach reasons a lot'')}. Además, las respuestas de las personas con discapacidad señalaron la importancia de los desarrollos centrados en el usuario que les permitan llevar a cabo investigaciones científicas: C indicó que \textit{`eventualmente haría que el tipo de investigación que tenía que hacer fuera aún más fácil, mucho más fácil de lo que es posible ahora, permitiéndome investigar cosas que antes me resultaban muy difíciles' (``eventually it would make the kind of research that I had to do even easier, much much easier than is possible now, allowing me to research things that I would found very difficult before'')}.
    
\end{enumerate}

\subsection{Análisis de confiabilidad}

A partir de todo el proceso de análisis, se construyeron dos tablas teniendo como encabezado las categorías, donde cada analista clasificó los enunciados por categoría y luego se midió la concordancia entre ambos mediante el coeficiente kappa \citep{cohen1960}. El trabajo de \citet{lombardetal2002} destaca la importancia del análisis de contenido y la confiabilidad; allí explican la confiabilidad entre codificadores como una medida del grado en que los diferentes jueces tienden a estar de acuerdo. En cuanto a \citet{hallgren2012}, específicamente sobre el coeficiente kappa, menciona que se usa para variables nominales y “solo son adecuadas para diseños completamente cruzados con exactamente dos codificadores” \citep[p.6]{hallgren2012}. Por lo que, dado que el análisis decantó en variables nominales (denominadas en este trabajo como ``categorías'') y todo el estudio fue realizado por dos codificadores, se utilizó el coeficiente kappa para cuantificar la confiabilidad del estudio.

\begin{table}
	\centering
	\caption{Matriz de proporciones de afinidad obtenidas a partir de la tabla de categorías de cada evaluador \citep{cohen1960}. Se utiliza para identificar las categorías los números con los que fueron presentadas en la sección anterior.}
	\label{tab:cap5_agreem_matrix}
	\begin{tabular}{cccccccc} 
		\hline
		\multicolumn{2}{c}{\multirow{2}{*}{Nº de categoría}} & \multicolumn{5}{c}{Eval. A} &  \\
		 & & 1 & 2 & 3 & 4 & 5 & P$_{iB}$ \\
		\hline
		\multirow{5}{*}{Eval. B} & 1 & 0.107 & 0 & 0 & 0 & 0 & 0.107 \\
		 & 2 & 0 & 0.178 & 0 & 0 & 0 & 0.178 \\
		 & 3 & 0 & 0.071 & 0.286 & 0.036 & 0.036 & 0.429 \\
		 & 4 & 0.036 & 0 & 0 & 0.107 & 0 & 0.143 \\
		 & 5 & 0 & 0 & 0 & 0 & 0.143 & 0.143 \\
		 & P$_{iA}$ & 0.143 & 0.249 & 0.286 & 0.143 & 0.179 & 1.00 \\
		\hline
	\end{tabular}
\end{table}

En cuanto al método de análisis de confiabilidad con el coeficiente kappa, \citet{simwrig2005,mchugh2012} describen paso a paso como se lleva a cabo el análisis de confiabilidad. Dicha técnica se ajusta a trabajos realizados con datos nominales y a cargo de dos codificadores. La Tabla \ref{tab:cap5_agreem_matrix} muestra una matriz de afinidad, a partir de la cual se calcula el coeficiente kappa: k = 0,768. Siguiendo los criterios de \citet{cohen1960}, el valor kappa (k) ilustra la proporción de afinidad después de eliminar la concordancia azarosa, y se calcula como:

\begin{equation}
    k = \frac{po-pc}{1-pc}
	\label{eq:cap5_kappa}
\end{equation}

donde $po$ es la proporción de unidades en las cuales los jueces tienen afinidad y $pc$ es la proporción de unidades en las cuales la concordancia es azarosa.

Empleando \ref{eq:cap5_kappa} con la matriz de afinidad (Tabla \ref{tab:cap5_agreem_matrix}), el coeficiente kappa es 0.768. Según \citet{mchugh2012}, un kappa igual a 1 representa concordancia perfecta y un valor entre .60-.79 es moderado. Teniendo en cuenta que 0,76 está cerca del umbral superior, se considera que el coeficiente indica una buena afinidad entre los codificadores, lo que implica que los resultados obtenidos en la presente investigación son fiables.

\subsection{Resultados}

Siguiendo las categorías definidas previamente (ver sección \ref{sect:analisis_transcripciones}), se resaltarán algunos hallazgos importantes durante la técnica de condensación de texto en cada una de ellas.

\begin{enumerate}
    \item \underline{Sobrecarga de memoria}:

    En esta categoría todos los participantes reconocen la linealización de funcionalidades que presenta sonoUno, adicionalmente, A4 destaca el hecho de que un menú era redundante. Con la finalidad de profundizar sobre el comentario de A4, se preguntó a otros cuatro participantes sobre este punto y todos estuvieron de acuerdo con el grupo A. El menú en cuestión es el menú `Paneles' donde su única funcionalidad es mostrar u ocultar paneles, tarea que también puede cumplirse utilizando el botón pertinente en los demás menús; por ejemplo, el panel de configuraciones de sonido puede mostrarse desde el menú `Paneles' y desde el menú `Configuraciones'. 
    
    Se encontró que los botones separados de reproducción y pausa confunden al usuario (estas funcionalidades están en los mismos botones en la última versión publicada). 
    
    Con respecto a la consistencia, B esperaba escuchar un sonido continuo cuando la trama presenta una línea continua (el sonido predeterminado actual es discreto), esto presenta una inconsistencia entre el despliegue visual y auditivo. 
    
    Esta categoría demuestra la importancia de la linealización de funcionalidades y la consistencia con otras aplicaciones ampliamente utilizadas, todos los participantes en primer lugar buscaron las funcionalidades genéricas donde se suponía que debían estar siguiendo su experiencia previa con aplicaciones de computadora. Se logró evidenciar que las funcionalidades genéricas de sonoUno están en los lugares esperados y son fáciles de encontrar (por ejemplo guardar, abrir, reproducir, detener).

    \item \underline{Necesidad de información}:

    Todos los participantes (100\%) expresaron la necesidad de información y confirmación sobre sus acciones (algunos comentarios parafraseados: ``¿Dónde estoy? ¿Dónde están los picos? ¿Estoy apuntando el pico correctamente?''; ``Sería bueno saber el rango seleccionado''; ``Sería bueno saber el valor de la longitud de onda del punto marcado y la posición exacta en abscisas''). En este sentido se debe asegurar una buena comunicación con el usuario y con la tecnología de asistencia. Específicamente, se hizo evidente una dependencia de la tecnología de asistencia durante la sesión del grupo focal: la falta de sincronización entre el prototipo y el lector de pantalla obligó a los participantes con discapacidad visual (55.5\% del total, 100\% de los discapacitados visuales) a realizar acciones que no forman parte de la tarea principal, para hacer que todo funcione. Un ejemplo de ello fue preguntar ``¿Qué significa el control deslizante?'', esto se debe principalmente a no poder indicarle al lector de pantalla el texto específico que debe leer al llegar a dicho objeto. SonoUno usa las configuraciones de accesibilidad nativas y se puede usar con diferentes lectores de pantalla, lo que en muchos casos no permite ajustes tan específicos como acotaciones o definiciones. En resumen, se resalta la necesidad de una mejor comunicación entre las diferentes aplicaciones y las herramientas de tecnología de asistencia.

    \item \underline{Necesidad de poder elegir}:

    Esta categoría surge de los participantes con discapacidad (grupo A y C, 100\% de los participantes en estos grupos), ya que existen pocas herramientas que les permitan hacer selecciones y explorar. En general, las que lo permiten, requieren que el usuario realice una gran cantidad de transacciones y conlleva una alta curva de aprendizaje. Si bien, en el caso de sonoUno durante la sesión de grupo focal los participantes expresan que esta herramienta les permite explorar y elegir entre diferentes opciones según lo deseen (ver definición de esta categoría en la sección anterior para citas textuales) esto no es común especialmente en el campo de investigación. 
    
    En esta categoría, también surgió la necesidad de contar con más configuraciones de sonido (``una variedad de instrumentos que son notablemente diferentes''; ``para agregar sonido continuo también''); necesidad de mayor precisión (``Me gustaría poner números específicos'', refiriéndose a los controles deslizantes de corte en `x'); necesidad de retornar sobre algunas posiciones para volver a escuchar la sonorización; la posibilidad de cambiar los tamaños de los paneles; y la posibilidad de ajustar la configuración de la fuente, entre otros.

    \item \underline{Necesidad de entrenamiento}:

    Durante la sesión de grupo focal con los diferentes grupos, surgió la necesidad de capacitación tanto en el uso del prototipo como en la técnica de sonorización. La sonorización aplicada para analizar conjuntos de datos es un nuevo enfoque, por lo tanto, las personas expresan la necesidad de orientación para ganar confianza en la técnica, junto con pautas para garantizar que la mayoría de las personas entiendan los mismos datos de manera coherente. 
    
    En el caso del prototipo, fue evidente que las personas durante el segundo período tenían mucha más confianza en el uso del software; esto nos mostró que el software presenta una curva de aprendizaje rápida. Tal vez, una combinación de cursos de capacitación para sonorización usando sonoUno podría ser un buen punto de partida, fortaleciendo la técnica y haciendo que sonoUno sea más robusto y centrado en el usuario.

    \item \underline{Aspectos sociales}:
    
    Esta categoría evidencia como el constructo social perjudica a las personas con discapacidad, constructo que las personas sin discapacidad desconocen, omiten o mal interpretan. Durante la sesión, las personas sin discapacidad consideraron la herramienta principalmente como un programa de divulgación; ninguno pensó/expresó que las personas con discapacidad puedan hacer ciencia (D3 afirmó en respuesta a la consulta sobre alguna ventaja del software ``por razones científicas no sé, por razones de divulgación mucho''). Por otro lado, las personas con discapacidad expresaron su gratitud y echaron sobre sus propios hombros la culpa de los errores del prototipo (ver la definición de esta categoría en la sección \ref{sect:analisis_transcripciones}). 
    
    Esto evidencia lo lejos que se está de la inclusión, pone en relieve un gran malentendido con respecto a la accesibilidad: cada persona con diversidad funcional, teniendo la herramienta adecuada, debería poder hacer lo que desee si tiene la voluntad. El otro posible enemigo de esta afirmación es la mentalidad de los desarrolladores, si los mismos no están capacitados para considerar las necesidades de personas funcionalmente diversas en el desarrollo de las herramientas desde el principio, escuchando, evaluando e integrando soluciones que se adapten a sus necesidades, la producción de una herramienta usable y útil es muy difícil.
    
\end{enumerate}

En cuanto a las preguntas de investigación expresadas en la sección \ref{sect:method_FG} para este estudio en particular, en respuesta a la primera de ellas (``¿El software en desarrollo es usable, efectivo y eficiente?'') se evidencia que el programa cumple con su objetivo principal (mostrar visualmente y sonorizar datos astronómicos). Sin embargo, presenta una interfaz muy simple, sin las herramientas que un astrónomo utiliza hoy en día en su profesión, además de cierta complejidad para personas con discapacidad visual.

En cuanto a la segunda pregunta de esta investigación de grupo focal (``¿El software permite el análisis de datos a través de sonorización?''), todos los participantes lograron identificar correctamente (a través de sonido) una función decreciente en la primera tarea de datos que se les asignó. En una segunda tarea, detectaron correctamente tres picos que corresponden a tres líneas de emisión espectral. Es importante resaltar que los participantes sin discapacidad expresaron que más allá de poder detectar estos patrones, por el momento no creen poder analizar datos a través del sonido, en correspondencia con \citet{supper2012}. Pero nuestros hallazgos no cierran la puerta a la sonorización, el discurso general obtenido de los participantes deja en evidencia la falta de experiencia en el uso del sonido y la novedad de la técnica. Todos los participantes dijeron que utilizarían el software en el futuro si tuvieran la oportunidad. Volviendo a la sonorización, los participantes con discapacidad visual también mencionaron la importancia del entrenamiento para el uso de la técnica en análisis de datos. De ello se concluye que diseñar una formación adecuada es de suma importancia para poder empezar a hablar de análisis de datos a través de sonorización, más allá del uso de este software en particular. Es muy importante tener presente que su contraparte sensorial, la exploración visual, se enseña desde el inicio de la etapa escolar.

En cuanto a la última pregunta (``¿Cuál es la opinión de los participantes sobre el análisis multimodal de datos astrofísicos?''): todos los participantes expresaron que se trata de una técnica novedosa y prometedora ya que permite que personas actualmente excluidas puedan explorar datos astronómicos. Uno de los participantes (grupo C) expresó que con esta técnica podría continuar con su investigación, postergada por su discapacidad visual (``Creo que me podría servir, para continuar con mi investigación'') (``Me permitiría investigar cosas que me hubieran resultado muy difíciles antes'').

Como reflexión final, el estudio demuestra una marcada necesidad de poder explorar los datos de forma autónoma y eficaz. Los participantes sin discapacidad mostraron una marcada preocupación por la posibilidad de poder manipular sus datos con esta nueva herramienta, sin dicha manipulación expresan que no se puede considerar un análisis de datos. En cuanto a los participantes con discapacidad, destacaron la posibilidad de una mayor autonomía en el análisis de datos (algo que hasta el momento ninguna herramienta les permite), pero siguen preocupados por la confiabilidad y precisión del programa.

\subsection{Conclusiones parciales}
\label{sect:fg_conclusiones}

La generalización que se evidencia durante el análisis de contextos de uso (ya sean circunstancias o entorno presente) de tecnología digital es muy peligrosa. Incluso cuando es ampliamente evaluado y analizado este contexto, muchas personas aún no pueden adaptarse y hacer frente a la interfaz digital lograda. Las respuestas de los participantes de los grupos focales dejan ver que están abrumados por las interacciones digitales, se sienten minimizados, excluidos, sin voz, inadecuados y con la responsabilidad de aligerar un aire pesado producido sistemáticamente por las presiones de que deben encajar. Fue evidente durante la sesión, observando sus comentarios adicionales y el lenguaje corporal, que las personas con discapacidades están acostumbradas a agradecer incluso cuando la herramienta no es útil y tienden a culparse por los errores o problemas que experimentan con la herramienta, incluso cuando el error no tiene relación con ellos. Además, no expresan ningún tipo de decepción o comentario negativo directamente, todas las recomendaciones que hicieron se basaron en un comentario positivo o incluso en una gratitud.

Si bien es cierto que es un desafío que un programa se adapte a las necesidades de todos para crear una única pantalla estática, también es muy cierto que es posible utilizar con dicho fin mecanismos que están siendo empleados regularmente para anticipar cómo piensan y reaccionan los humanos, usando datos personales para ajustar publicidades en muchos casos no deseadas. Estos mismos mecanismos podrían usarse para crear un entorno que permita a las personas depositar su experiencia (su contexto) y adaptarse con poco o ningún esfuerzo a la pantalla, de modo que  puedan sentirse reconocidas y empoderadas para producir o terminar una tarea con su propio máximo potencial.

Los resultados que se presentan refuerzan la idea de que la interfaz debe ayudar al usuario a tener éxito sin sufrir fatiga o sobrecarga cognitiva y de memoria. Durante el desarrollo de la interfaz de usuario, no se deben hacer suposiciones (por ejemplo: la suposición de que cada persona usa habitualmente la computadora con el teclado y mouse convencional excluye a las personas que no tienen esta capacidad; la suposición de que todos pueden interpretar la información mostrada siguiendo las mismas pistas multisensoriales deja fuera a las personas con sesgos sensoriales). La mejor práctica aquí es realizar análisis de entrevistas o grupos focales de alta granularidad con usuarios potenciales y considerar posibles soluciones centradas en ellos.

Las personas han expresado en varias ocasiones que sonoUno les permite explorar conjuntos de datos de una manera nueva (aún en su primera versión limitada). El ser humano, por naturaleza, necesita explorar. Si las herramientas informáticas modernas hacen posible la creación de interfaces que permitan a las personas con discapacidad autonomía y exploración para trabajar con datos digitales, ¿por qué se duda en desarrollarlas a gran escala? ¿Por qué es tan difícil para los desarrolladores trabajar en equipo con personas con discapacidad? Las personas funcionalmente diversas necesitan herramientas para ser autónomas, herramientas que les aseguren precisión, certeza, eficacia y eficiencia en su trabajo y en igualdad de condiciones.

Lograr una herramienta universal puede parecer una utopía, pero el error es pensar en pequeño; se necesita un enfoque y un esfuerzo más ambicioso: al igual que las piezas de un gran rompecabezas encajan, cada tecnología de asistencia puede formar un componente del gran diseño, todo detrás del mismo protocolo de comunicación (lo que significa que la comunicación no solo es verbal, sino también entre softwares o APIs). En este escenario ideal, las personas pueden elegir cualquier tecnología de asistencia y programa informático que deseen, o que se adapte a sus necesidades, y hacer que funcione. Puede que no se esté tan lejos de hacerlo realidad, de la misma manera que matemáticas y gramática forman parte de la capacitación desde la infancia, se debe comenzar a hablar sobre el diseño centrado en el usuario y el enfoque multisensorial para el análisis de datos.

\section{Colegas y usuarios contactados a distancia}
\label{sect:mails_completa}

Desde el inicio de este trabajo se ha mantenido una relación con personas en el ámbito de la astronomía que están realizando desarrollos similares, o bien que están interesados en aplicar la sonorización en sus ámbitos laborales. En primer lugar se debe mencionar a la Dra. Wanda Díaz Merced, quien nos ayudó durante lo primeros cuatro años de trabajo, guiando el desarrollo para lograr un eficiente despliegue y comprobando la accesibilidad de la herramienta en cada paso.

Otro caso que es deseable destacar, es el intercambio que se ha mantenido con el equipo de desarrollo del programa StarSound, habiendo comenzado en 2018, luego de una conferencia donde mostraron su primera versión. Al compartir un objetivo similar se intercambiaron con los desarrolladores los primeros resultados en sonorización de sonoUno, resultando útil para definir las herramientas y la forma de encarar la sonorización. Durante el intercambio resaltaron lo novedoso de realizar sonorización en Python, debido a que en 2018 aún no existían muchos paquetes que permitieran realizar producción de sonido y despliego del mismo en una interfaz. A lo largo de 2019, ya con una primer versión de sonoUno y buscando mejorar el despliegue de sonido, se intercambió nuevamente experiencias con el equipo de trabajo. Amablemente sugirieron opciones para continuar utilizando FluidSynth o incluso cambiar a librerías que producen sonido sin utilizar el protocolo MIDI. El problema que continuó persistiendo fue que la mayoría de las librería contemplan la generación de un archivo de sonido, ya sea MIDI o wav, pero no tienen soporte para la reproducción de sonido nota por nota, o punto por punto. Fue decisión del equipo de desarrollo de sonoUno priorizar este estilo de reproducción para mantener la sincronicidad entre el despliegue gráfico y el despliegue sonoro, podría decirse que es uno de los pilares de sonoUno.

Es también destacable la ayuda recibida de la Universidad de Southampton, Reino Unido, donde el Dr. Poshak Gandhi junto con la Dra. Wanda Díaz Merced y la Dra. Beatriz García, aseguraron las condiciones para cumplir con todos los requerimientos que permitieron realizar el encuentro con Grupos Focales (descripto en el Capítulo \ref{cap:fg_completo}). Esto brindó la posibilidad de evaluar la herramienta y establecer un plan de mejoras para mantener el enfoque centrado en el usuario, planteado desde el inicio del trabajo. Adicionalmente, durante este encuentro fue donada al proyecto una computadora Mac, de gran utilidad para mantener un desarrollo multiplataforma, pudiendo así testear la herramienta en este sistema operativo con sus herramientas de accesibilidad nativas.

Además de las mencionadas, durante 2020 y 2021 se mantuvieron intercambios con dos grupos de trabajo que buscaban aplicar el desarrollo sonoUno en sus proyectos. Uno de estos fue el proyecto `\textit{Sensing the Dynamic Universe}' (ver sección \ref{sect:uso_SDU}) iniciado por el Dr. Paul Green y la Dra. Wanda Díaz Merced, contando posteriormente con otros colaboradores para la generación de material y página web. El otro proyecto fue un `\textit{Taller de Introducción a la Sonorización Científica}' (ver sección \ref{sect:uso_canarias}) realizado por integrantes de la Asociación Astronómica y Educativa de Canarias “Henrietta Swan Leavitt” (AAEC) y alumnos del IES José Frugoni Pérez.

Finalmente, desde 2019 hasta la fecha se forma parte de una colaboración internacional que tiene como objetivo involucrar y promover la cooperación de los ciudadanos con los centros de investigación, buscando que contribuyan activamente en el desarrollo de nuevos conocimientos para las necesidades de la ciencia y la sociedad. En este proyecto denominado REINFORCE (Infraestructura de investigación para ciudadanos en Europa; financiado por European Union's Horizon 2020 SWAFS ``Science with and for Society''), se contribuyó mediante el desarrollo de la versión web de sonoUno y algunos scripts de sonorización atendiendo las necesidades de datos particulares, como es el caso de: imágenes, partículas generadas aquí en la Tierra en el Gran Colisionador de Hadrones (LHC) y muones provenientes del espacio (se profundiza sobre estos desarrollos en el Capítulo \ref{cap:reinforce}).

Teniendo como referente a las personas que se fueron conociendo estos años y a través de las mencionadas colaboraciones, para las últimas dos versiones de sonoUno se contactó por correo electrónico a estos grupos de especialistas relacionados con la astronomía en diversos campos. Se les solicitó que instalaran sonoUno, lo utilizaran en proyectos e investigaciones personales, realizando actividades de análisis de datos propios de sus prácticas diarias, y que compartieran con el equipo de desarrollo de sonoUno comentarios sobre su uso y recomendaciones, para lo cual se les compartió una encuesta utilizando Google Forms. Se realizó una publicación en el Simposio de la Unión Astronómica Internacional (IAU367) sobre las conclusiones del intercambio con usuarios y las actualizaciones realizadas en sonoUno que se describirán a continuación \citep{iau367} (se adjunta el trabajo publicado en el Apéndice \ref{ap:pub-y-mails}).

\subsection{Metodología de contacto}

A principios de 2020 se realizó el primer contacto con usuarios por email, luego de la sesión de grupo focal a inicios de 2019 (descripto en la sección \ref{sect:FG_completo}), se incluye el email enviado para invitarlos a probar el programa en el Apéndice \ref{ap:pub-y-mails}, el primer email corresponde a este primer contacto. En la versión del software 3.0, se agregaron cambios como: la unificación del botón `Play' y `Pausa'; la traducción a Python 3 con el consecuente cambio de librería de sonido (esto último se debió a que Mingus no estaba disponible para Python 3, la nueva librería de sonido elegida es pygame, la cual no necesita paquetes adicionales). 

Este primer contacto a distancia se realizó con trece (13) personas, la mayoría astrónomos, entre ellos se obtuvo respuesta de ocho (8) personas. Es decir, que respondió comentando sobre su experiencia el 61\% de las personas, los comentarios recibidos fueron respondidos a través del mismo email enviado, no completaron el formulario. En general las respuestas fueron positivas, elogiando los avances sobre el programa, de las ocho (8) personas que respondieron solo cuatro (4) de ellas comentaron sobre el proceso de instalación. En cuanto a esto último, dos (2) (de los cuatro (4) que reportaron la instalación) tuvieron aún problemas con la instalación, uno de ellos con la librería wxPython en Ubuntu y el otro con el programa octave en Mac. Sin embargo, ambos pudieron superar las dificultades y lograr que el programa funcionara en sus computadoras. Se brindará un resumen con las recomendaciones de los usuarios en la sección \ref{sect:usuarios_recomend}.

Luego de implementar nuevas mejoras que se estaban desarrollando a inicios del 2020, sumando mejoras que decantaron del intercambio con usuarios por email, en Julio del 2020 se volvió a enviar una nueva versión de sonoUno (v3.1) a las mismas trece (13) personas y a un nuevo grupo. En cuanto a las personas que contactamos por segunda vez, solo dos (2) de ellas respondieron. Es destacable que con una de las personas que respondió en esta ocasión, se mantuvo contacto sobre nuevas mejoras en sonoUno hasta fines de 2020, debido a que utilizaría el desarrollo para mostrar videos de sonorizaciones en una página web que estaban desarrollando (ver sección \ref{sect:uso_SDU}).

En cuanto al nuevo grupo al cual se le mandó esta nueva versión de sonoUno, estuvo conformado por los colaboradores en el proyecto REINFORCE, se envió el email a través de la lista de colaboradores y se obtuvo respuesta de seis (6) colegas. Estas respuestas fueron tres (3) por email, respondiendo al pedido y sin detalles sobre la instalación o uso del programa. Las tres (3) respuestas restantes fueron mediante la encuesta de google form, los tres pudieron instalar el programa en sus computadoras (una Mac y dos Linux/Ubuntu), solo uno reportó problemas en la instalación por librerías faltantes. En cuanto al programa en principio comentaron que no lo utilizaría para su trabajo en el día a día, pero fue motivado por el hecho de que sonoUno aún no presenta las funcionalidades necesarias para que puedan utilizarlo en sus propias investigaciones (uno de ellos trabaja con datos de muongrafía y otro de los participantes con Glitches). Sin embargo, todos expresaron al final de la encuesta que lo utilizarían en un futuro y que lo recomendarían. Otro tema destacable fue que todos recomendaron que el programa se pueda utilizar desde la línea de comandos.

Las sugerencias de los usuarios recopiladas hasta el momento y que se encuentran en la planificación para próximas actualizaciones se detallará en la sección \ref{sect:usuarios_recomend}.

\subsection{Recomendaciones de los usuarios}
\label{sect:usuarios_recomend}

En este apartado se especifican las recomendaciones hechas tanto por los usuarios del grupo focal, como también por los intercambios realizados por email. Se enumeran las recomendaciones de todos los intercambios hechos hasta el momento para evidenciar de mejor forma aquellas que se repiten de una prueba a la otra, dándoles así mayor prioridad al pedido. En cuanto a la sesión de grupo focal, las sugerencias principales fueron las siguientes: 

\begin{enumerate}
    \item Ubicar el botón Play y Pausa en el mismo botón, análogo a los reproductores de audio (Grupo A y C);
    \item Algunos de los datos que corresponden a la línea decreciente usan la misma nota (Todos los grupos), esto fue debido a una falta de resolución en el sonido;
    \item Uno de los paneles de la interfaz gráfica, no estaba en la sección Panel del menú (Grupo B);
    \item El control deslizante debería mostrar el valor numérico del punto donde se encuentra el cursor (Grupo D), en la primer versión se mostraba la posición del array;
    \item Una pregunta que se repitió fue ¿Por qué el software solo muestra las primeras diez filas del conjunto de datos? (Grupo D), en esa primer versión se había decidido que fuera así por la demora que representa cargar el arreglo de datos completo;
    \item Para cambiar de instrumento tienen que pulsar Stop y volver a reproducir (Todos los grupos), sino no se actualiza la configuración seleccionada.
\end{enumerate}

Por otro lado, durante el intercambio de correos electrónicos con especialistas en astronomía y astrofísica, se repitieron algunas de las sugerencias descriptas en la lista anterior. Adicionalmente, se tuvieron las siguientes sugerencias: 

\begin{enumerate}
    \item Mayor rapidez del sonido, es decir, poder reproducir los datos más rápido;
    \item Poder escribir directamente las frecuencias mínima y máxima del sonido.
    \item En las barras deslizables de los límites del sonido, poder introducir una frecuencia mínima que sea MAYOR que la frecuencia máxima, de esa forma se puede representar una curva de luz de una estrella en magnitud de forma que valores altos a nivel numérico, que corresponden a bajo brillo, se escuchen con sonido grave. Y al revés, valores bajos de magnitud, que corresponden a altos brillos, con frecuencias altas. En ese sentido se debería también cambiar el gráfico.
    \item Representar la distancia entre dos puntos de x con el tiempo entre notas, para poder percibir con el sonido cuando dos puntos de x están más cerca o más lejos (hasta el momento se sonorizaban los puntos con el mismo intervalo de tiempo).
    \item Graficar y sonorizar dos o tres columnas en función de  x al mismo tiempo.
    \item Poder habilitar una forma automática de que al finalizar la sonorización vuelva al inicio de la reproducción (función loop).
\end{enumerate}

Para definir los pasos a seguir en la actualización de sonoUno se evaluaron las diferentes recomendaciones de acuerdo al marco de trabajo de sonoUno, con el fin de analizar si se correspondían con el planteo inicial de sonoUno. Luego, se tuvieron en cuenta la complejidad de implementación, el aporte de cada una en cuanto a la accesibilidad de sonoUno y si era factible de realizar por el equipo de trabajo.

\subsection{Mejoras de sonoUno basadas en las recomendaciones de usuarios}
\label{sect:mails_mejoras}

En base a la lista de recomendaciones de usuarios expuesta en la sección \ref{sect:usuarios_recomend}, se consideró que todas podían  ser implementadas en sonoUno para hacerlo más centrado en el usuario, sin perder de vista que es un programa de sonorización de datos. Como primer paso, las recomendaciones se dividieron en 3 grupos: 
\begin{itemize}
    \item \underline{Interfaz}: incluye aquellas recomendaciones que están relacionadas con la interfaz y no requieren un procesamiento o cambio de la estructura interna del código.
    \item \underline{Sonido}: comprende las recomendaciones relacionadas con la producción de la sonorización.
    \item \underline{Implementación compleja}: puede contener una mejora en sonorización, operaciones de datos y cambios gráficos que requieran mejorar aspectos en la estructura interna del algoritmo que pueden afectar varias funcionalidades del software.
\end{itemize}

Las modificaciones de interfaz seleccionadas no presentaban complejidad y se implementaron por completo. Por ejemplo: `Play' y `Pausa' ahora están en el mismo botón; el panel `Parámetros de datos' se cambió a la sección de paneles en el menú para seguir una adecuada linealización de tareas; cuando se presiona el botón `Marcar punto', la línea de marca aparece encima de la línea de posición, siendo visible su efecto en el momento que se presiona; y, todos los elementos en la interfaz muestran una descripción en un cuadro de texto cuando se pasa el puntero sobre ellos.

En cuanto a las recomendaciones de sonido, la primera fue un error descubierto durante la sesión del grupo focal donde para poder escuchar el instrumento seleccionado se debía detener la reproducción y volver a reproducir el sonido. Ahora los parámetros de sonido se pueden cambiar sin presionar ningún botón adicional, es decir que al seleccionar el instrumento si se está reproduciendo el sonido, este cambia automáticamente y continua reproduciendo con el nuevo instrumento seleccionado. La segunda recomendación de sonido se basó en la resolución, debido a que con la librería anterior se utilizaban notas MIDI (en función de agregar contexto se destaca que previo a la traducción a Python3, usando la librería Mingus para generar sonido MIDI, este restringe la cantidad de notas a las notas reales del instrumento, solo alrededor de 70 notas en el instrumento Piano, por ejemplo). Con la finalidad de solucionar este inconveniente con la resolución y en paralelo a la actualización a Python3, se comenzó a utilizar la librería Pygame para generar y reproducir sonido, modificando los parámetros de sonido directamente en su forma de onda. El problema de resolución se solucionó temporalmente, pero la forma de incluir sonidos MIDI quedó en agenda, debido a que es un sonido más familiar y armónico para los usuarios y por lo tanto, más demandado.

Continuando con las recomendaciones sobre el sonido, con respecto al tiempo que demora la reproducción de los datos, la limitación la impone el tempo mínimo que permite la interfaz gráfica de usuario, y por lo tanto el tiempo de reproducción no se puede modificar. En cambio, se propone reducir la cantidad de elementos del conjunto de datos, realizando un promedio entre los valores más cercanos; esta actualización se agregó pudiendo utilizarse con un comando desde un espacio de texto incluido en la nueva versión de sonoUno, pero debe probarse en el próximo contacto con los usuarios finales para evaluar su efectividad. La última recomendación sobre el sonido fue, que el tiempo entre notas debe coincidir con la distancia de los puntos en el arreglo de datos, de acuerdo a eso al iniciar la reproducción el programa mide el espacio entre cada par de puntos del eje `x' y calcula la duración en tiempo que debe existir en forma de  silencio entre cada nota.

En última instancia, en el grupo de implementaciones complejas, primero se trabajó con las barras deslizantes, que se modificaron para mostrar el valor real del conjunto de datos y no la posición en la matriz. Una vez logrado eso y debido a la similitud de la tarea, se habilitó la posibilidad de escribir el valor deseado para llevar el cursor en `x' arriba del control deslizante. Acerca de la cuadrícula de datos que en la versión anterior solo mostraba los primeros diez valores, se investigó una forma de mostrar todo el conjunto de datos sin congelar la interfaz gráfica de usuario, en este momento el conjunto de datos se muestra en páginas de 100 valores indicando la cantidad de páginas total que se utilizan para mostrar el arreglo completo. De esta forma, solo se cargan de a 100 valores a la vez, lo cual se logra hacer en menos de un segundo.

En cuanto a las actualizaciones que requieren un cambio considerable de la interfaz de usuario, ya sea modificando la vista actual o agregando más elementos interactivos, se decidió implementarlas en un formato de prueba, implementando primero estas actualizaciones con un comando específico que se escribirá en un cuadro de texto destinado a tal fin en la interfaz, y así poder probarlas antes de incluir cambios considerables en la interfaz gráfica de usuario que generen una mayor curva de aprendizaje y una mayor carga de memoria. Este enfoque permite no solo garantizar que la funcionalidad sea útil, sino también realizar intercambios de usuarios para decidir la forma óptima de incluirla en la interfaz gráfica de usuario manteniendo el concepto centrado en el usuario. 

Estas actualizaciones finales incluyen una función de bucle para la reproducción de los datos (porque algunos conjuntos de datos, como las curvas de luz, se representaron mejor con esta funcionalidad), y una función para mostrar y sonorizar más de una columna en el mismo gráfico o en gráficos diferentes; esta última en una etapa de prueba que no permite configurar parámetros de sonido o de gráfico aún. 

Por último, la recomendación de generar varios archivos de audio con un comando a partir de diferentes archivos de datos en una carpeta se realizó un año más tarde. Aún cuando esta funcionalidad no se relaciona directamente con el análisis multimodal de datos, fue solicitada por la mayoría de usuarios que probaron el programa en los últimos dos años. Esta tarea se logró generando un Script para sonorizar desde la línea de comandos que se agregó en la misma carpeta donde se encuentra el script para desplegar la interfaz gráfica. El primer enfoque, busca todos los archivos csv o txt en una carpeta (ubicación dada por el usuario) y guarda el sonido y la gráfica (si se indica) de las dos primeras columnas de cada archivo, en la misma carpeta. Siguiendo en esta línea, todos los nuevos planteamientos de sonorización realizados en sonoUno se han desarrollado con estos dos formatos, un script para ejecutar en modo bash y luego la integración en el framework gráfico de sonoUno.

\subsubsection{Interfaz web}

Atendiendo a recomendaciones de los usuarios y con el fin de asegurar el uso amplio y global del software, se planteó la necesidad del desarrollo de una página web y una versión web del sonoUno que, además, asegurara e acceso a través de múltiples plataformas y redujera los problemas de instalación, que se evidencias en la versión escritorio de esta herramienta. La tarea de ``traducción'' del software  se inició a principios de 2020 y se desarrolló en el marco del proyecto REINFORCE (GA 872859, con el soporte de la EC Research Innovation Action bajo el programa H2020 SwafS-2019-1). Adicionalmente a lo que se necesitaba en dicha colaboración internacional, ya se contaba con reportes de la complejidad de instalar una aplicación completa con todas sus dependencias, incluso cuando los usuarios solo necesitaban funciones sencillas que podrían ser mostradas en una página web. Es por ello que se decidió responder al desafío de desarrollar una interfaz web, con menos funcionalidades que el programa base de Python, pero que ayudaría a superar estas limitaciones.

Una interfaz web implica el uso de HTML, JavaScript y CSS para realizar la programación, lo que da como resultado diferentes bases de código para las versiones nativa y web, esto requiere un esfuerzo de programación adicional. Además, el acceso a la web también se logra a través de un navegador web, que introduce una capa de software intermedia, especialmente cuando se trata de la interfaz de usuario, con sus propias reglas y estándares. La implementación en cuanto a la programación se puede dividir en front-end y back-end. El front-end es la parte más fundamental, ya que proporciona la presentación de datos real y las funciones de sonido. Esta sección brinda la parte interactiva de la experiencia, que también es la más atractiva para los usuarios, tiene un diseño de alto contraste que se puede personalizar a través de CSS y también aprovecha las características ARIA de los navegadores web modernos que permiten personalizar el flujo de trabajo del lector de pantalla. Algunas de las limitaciones de la interfaz web provendrán de las velocidades de procesamiento más lentas en el navegador web y algunas limitaciones en la producción de sonido. Cabe destacar que el diseño de los elementos de interfaz gráfica siguen el mismo marco de trabajo que la versión de escritorio, logrando así menor carga de trabajo para las personas que ya conocen el programa y manteniendo el diseño centrado en el usuario.

Por otro lado, la parte de back-end del desarrollo web ha sido planificada para realizar trabajos que dependen de capacidades de software más complejas, de modo que el usuario no necesite instalar paquetes de software tan grandes y complejos localmente. Probablemente, el problema se encuentre al tratar de trabajar con grandes conjuntos de datos en la versión web, debido a la complejidad de transferir archivos grandes desde el dispositivo local al back-end. Se describirá de forma amplia el desarrollo web en el Capítulo \ref{cap:reinforce}, sección \ref{sect:cap6_webcompleta}.

\subsection{Conclusiones parciales}

Sobre esta sección de contacto por correo electrónico y consecuentes actualizaciones, podemos concluir que el desarrollo propuesto a logrado un mayor nivel de diseño centrado en el usuario. SonoUno tiene el potencial de mejorar el trabajo científico trayendo otros estilos sensoriales al análisis de datos, se ha logrado dar respuesta a la mayoría de las recomendaciones hechas por las personas que han dado su devolución. Cabe destacar que se mantuvo siempre como prioridad el hecho de que sonoUno posibilite el análisis de datos a personas con diversidad funcional, se tuvo en cuenta que ninguna de las actualizaciones comprometa el hecho de que el programa permite a personas con discapacidad visual explorar datos y tomar decisiones por sí mismos. 

Esta sección muestra la importancia de un enfoque centrado en el usuario, lo que permite un desarrollo versátil y ajustable a las necesidades de las personas. Resulta evidente que este enfoque debe ser tomado en cuenta incluso desde la ciencia computacional, la ingeniería y el diseño de software, entre otros. Esto le permitiría a los desarrolladores tener en cuenta al usuario final desde el principio, haciendo que las herramientas sean más usables, eficientes y útiles, teniendo en cuenta la accesibilidad.

\section{Casos exitosos de uso}
\label{sect:casos_uso}

\subsection{Proyecto `\textit{Sensing the Dynamic Universe}}
\label{sect:uso_SDU}

Durante el segundo semestre de 2020 y como parte de la colaboración con el proyecto `Sensing the Dynamic Universe' (SDU) se mantuvo comunicación con el Dr. Paul Green (Centro de Astrofísica, Harvard \& Smithsonian). Varios miembros del grupo de trabajo en sonoUno mantuvieron intercambios donde el objetivo era la producción de la página web \href{https://lweb.cfa.harvard.edu/sdu/}{\underline{\textcolor{blue}{`Sensing the Dynamic Universe'}}} alojada en la Universidad de Harvard, concebida desde el inicio para ser una página accesible que permitiera a personas con y sin discapacidad conocer qué es una estrella variable y profundizar en el tema desde un enfoque multisensorial.

\begin{figure}[ht!]
        \centering
        \begin{subfigure}[b]{1\textwidth}
             \centering
             \captionsetup{justification=centering}
             \frame{\includegraphics[width=0.7\textwidth]{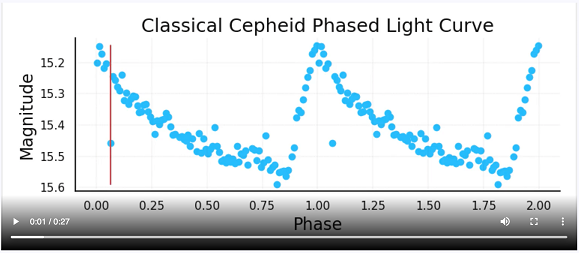}}
             \caption{Cefeida.}
             \label{fig:cap5_cepheid_phase}
         \end{subfigure}
         \hfill
        \begin{subfigure}[b]{1\textwidth}
             \centering
             \captionsetup{justification=centering}
             \frame{\includegraphics[width=0.7\textwidth]{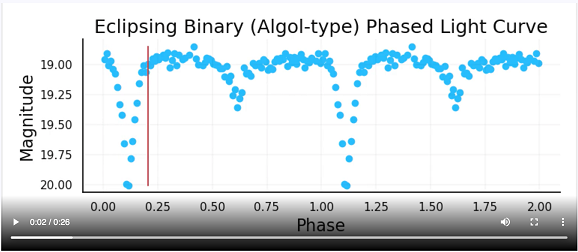}}
             \caption{Binaria Eclipsante.}
             \label{fig:cap5_eclip_phase}
         \end{subfigure}
        \caption{Gráficas de curvas de luz en fase, para dos tipos de estrellas variables. Las imágenes son capturas de pantalla de los videos de sonorización disponibles en la página web del SDU (\href{https://lweb.cfa.harvard.edu/sdu/cepheids.html}{\textcolor{blue}{\underline{Cefeida}}}; \href{https://lweb.cfa.harvard.edu/sdu/eclipsingbinaries.html}{\textcolor{blue}{\underline{Binaria Eclipsante}}}).}
        \label{fig:cap5_sdu_phase}
\end{figure}

Desde el equipo de desarrollo de sonoUno se aportó con varias actualizaciones del software, como ser respetar el tiempo entre notas (explicado en la sección \ref{sect:mails_mejoras}), importante para representar la diferencia de tiempo entre dos eventos o sucesos. En el caso de estrellas variables, esto es de suma importancia, ya que las curvas de luz representan la cantidad de luz recibida por el instrumento en un cierto momento. Una forma de representar las curvas de luz es mediante un diagrama de fase donde se representan dos períodos completos, la Figura \ref{fig:cap5_sdu_phase} representa los diagramas de fase de dos tipos de estrella variable: Cefeida (Figura \ref{fig:cap5_cepheid_phase}) y Binaria Eclipsante (Figura \ref{fig:cap5_eclip_phase}). 

\begin{figure}[ht!]
        \centering
        \begin{subfigure}[b]{1\textwidth}
             \centering
             \captionsetup{justification=centering}
             \frame{\includegraphics[width=0.75\textwidth]{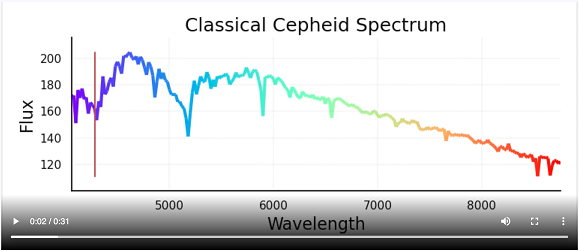}}
             \caption{Cefeida.}
             \label{fig:cap5_cepheid_spectrum}
         \end{subfigure}
         \hfill
        \begin{subfigure}[b]{1\textwidth}
             \centering
             \captionsetup{justification=centering}
             \frame{\includegraphics[width=0.75\textwidth]{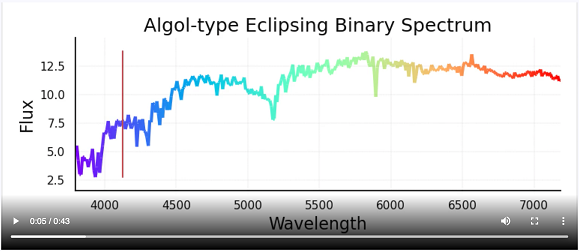}}
             \caption{Binaria Eclipsante.}
             \label{fig:cap5_eclip_spectrum}
         \end{subfigure}
        \caption{Gráficas del espectro para dos tipos de estrellas variables. Las imágenes son capturas de pantalla de los videos de sonorización disponibles en la página web del SDU (\href{https://lweb.cfa.harvard.edu/sdu/cepheids.html}{\textcolor{blue}{\underline{Cefeida}}}; \href{https://lweb.cfa.harvard.edu/sdu/eclipsingbinaries.html}{\textcolor{blue}{\underline{Binaria Eclipsante}}}).}
        \label{fig:cap5_sdu_spectrum}
\end{figure}

Adicionalmente, se realizaron modificaciones a sonoUno para poder realizar los videos del espectro de dichas estrellas variables que se muestran en la plataforma web (Figura \ref{fig:cap5_sdu_spectrum}) (vale la pena destacar que todas estas mejoras están presentes y disponibles en el repositorio en GitHub, en una bifurcación del código hecha por \href{https://github.com/joepalmo/sonoUno}{\textcolor{blue}{\underline{Joe Palmo}}}). En cuanto a las actualizaciones, estas fueron mayormente visuales y en la producción del sonido para que se perciba más continuo, todo ello necesario para representar datos espectrales de estrellas variables en particular, debido a que se coloreó la gráfica para que siguiera el color correspondiente a la longitud de onda que representa el espectro de la estrella. En cuanto al sonido, la representación de los espectros de estas estrellas corresponde a una curva continua, por lo que debe representarse con sonido continuo.

\begin{figure}[ht!]
    \centering
    \includegraphics[width=1\textwidth]{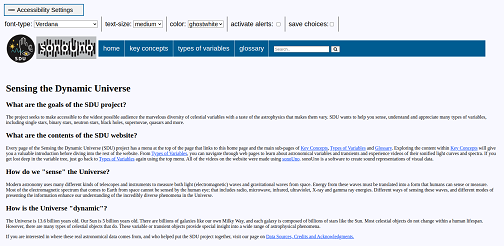}
    \caption{Captura de pantalla de la interfaz web \textit{`Sensing the Dynamic Universe'}.}
    \label{fig:cap5_sdu}
\end{figure}

En lo que hace a este desarrollo, puede encontrarse información general y un video demostrativo en la \href{https://pweb.cfa.harvard.edu/research/sensing-dynamic-universe}{\textcolor{blue}{\underline{web}}} del Centro de Astrofísica, Harvard \& Smithsonian. Además, cuentan con un espacio particular dentro de la web de la Universidad de Harvard (hacer click \href{https://lweb.cfa.harvard.edu/sdu/}{\textcolor{blue}{\underline{aquí}}}), la página presenta una sección de configuraciones de accesibilidad, donde se puede modificar el despliegue visual, activar alertas y guardar la interacción. Así mismo, se puede observar que el diseño es simple y no presenta mayores elementos, lo que hace a la navegación con el lector de pantalla más fácil (ver Figura \ref{fig:cap5_sdu}).

\subsection{Taller de sonorización dictado por el IES José de Frugoni Pérez en España}
\label{sect:uso_canarias}

A fines del año 2021 una nueva aproximación al uso del sonoUno surgió a partir del contacto con Carlos Morales Socorro, profesor de matemáticas en el IES José Frugoni Pérez, La Rocha, España. La propuesta fue utilizar la versión web del software sonoUno en las aulas para estudiar gráficas en general a partir de sonido. A principios de 2022, durante el primer taller de `Introducción a la Sonorización', seis participantes con diferente grado de discapacidad visual lograron reconocer características globales de funciones utilizando sonoUno (ver \href{https://astronomiayeducacion.org/taller-de-introduccion-a-la-sonificacion/?cn-reloaded=1&cn-reloaded=1}{\textcolor{blue}{\underline{nota}}}).

\begin{figure}[ht!]
    \centering
    \includegraphics[width=1\textwidth]{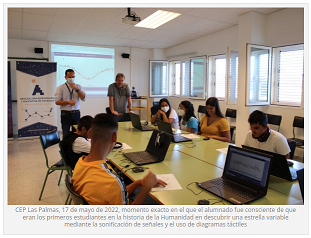}
    \caption{Imagen extraída de la \href{https://astronomiayeducacion.org/taller-2-de-sonificacion-descubriendo-el-universo/?cn-reloaded=1}{\textcolor{blue}{\underline{noticia}}} publicada en la web de la AAEC sobre el segundo taller de sonorización realizado en España.}
    \label{fig:cap5_canarias}
\end{figure}

Meses después, en el marco del Día Internacional de la Luz, se realizó allí mismo el segundo taller de sonorización, donde, utilizando sonoUno, seis estudiantes ciegos lograron descubrir una estrella variable utilizando la sonorización (Figura \ref{fig:cap5_canarias}) (noticia publicada en la \href{https://astronomiayeducacion.org/taller-2-de-sonificacion-descubriendo-el-universo/?cn-reloaded=1}{\textcolor{blue}{\underline{web}}} de la Asociación Astronómica y Educativa de Canarias “Henrietta Swan Leavitt” (AAEC) y en \href{https://twitter.com/aaecastro/status/1527991640924868608}{\textcolor{blue}{\underline{Twitter}}}). En la nota publicada en AAEC, Carlos expone que se ha comenzado a utilizar a sonoUno ampliamente en las aulas de la escuela como soporte a la enseñanza, un hito sumamente importante para este desarrollo que demuestra la importancia de esta herramienta y toda la investigación y desarrollo de técnicas que la acompañan.

\section{Conclusiones}

Durante el presente capítulo se ha logrado demostrar como un diseño centrado en el usuario desde el inicio y con un sólido marco de trabajo, permite el desarrollo y mejora de una herramienta usable, que ha sido elegida por investigadores a nivel internacional para ser incorporada en sus proyectos (sección \ref{sect:casos_uso}). 

En general, de la entrevista de grupo focal (sección \ref{sect:FG_completo}) se pudo concluir que la herramienta permitía el análisis de datos para personas con discapacidad visual, se lograron mejorar varias características de accesibilidad siguiendo los comentarios de los usuarios al hacer la herramienta más centrada en el usuario. Sin embargo, durante ese mismo intercambio, al ser testeada por personas sin discapacidad, se remarcó que la herramienta era muy simple y no presentaba mayores características que les permitiera realizar un análisis de datos complejos.

En base a estos últimos comentarios, aprovechando que el equipo de desarrollo contaba con una lista de investigadores interesados en tener novedades sobre el desarrollo y la pandemia de COVID-19 que se atravesó entre 2020 y 2021, se decidió realizar un intercambio con estos contactos vía correo electrónico. El objetivo principal fue seguir desarrollando la herramienta, haciéndola más centrada en el usuario, poniendo atención al investigador sin discapacidad sin perder de vista las características de accesibilidad ya logradas en sonoUno. Esto último resulta de gran importancia, debido a que sonoUno se plantea como un programa que permita el análisis multisensorial de los datos por personas con y sin discapacidad. Cabe destacar que sonoUno pretende unir a las personas con y sin discapacidad, permitiendo que utilicen una misma herramienta.

Estos intercambios con usuarios videntes, no sólo permitieron un diseño más centrado en un usuario investigador y asociado con datos específicos a ser estudiados (curvas de luz, espectros, entre otros), sino también, dieron lugar a casos de uso donde se eligió a sonoUno como herramienta de despliegue de datos. Adicionalmente, estos intercambios y en particular la relación con el proyecto REINFORCE de la Unión Europea, permitió expandir las fronteras de sonoUno, logrando funcionalidades que permiten sonorizar datos no tradicionales presentados en tablas de dos o más columnas, sino partículas y una primera aproximación a la sonorización de imágenes. Por su importancia en lo que hace a innovación, este tema se profundizará en el Capítulo \ref{cap:reinforce}.

Finalmente, es muy importante siempre tener presente que se debe abordar la causa de la inequidad en primera instancia, para luego poder eliminarla o encontrar una forma alternativa que permita la realización de las personas, cualquiera sean sus habilidades y capacidades. Hay muchos casos en los que eliminar barreras es fácil si desde el principio se hizo un enfoque centrado en el usuario con un buen diseño.

\chapter{Sonorización de datos de grandes facilidades astrofísicas y grandes bases de datos}
\label{cap:reinforce}

\section{Introducción}

Como se mencionó brevemente en el Capítulo \ref{cap:fg_completo}, desde 2019 hasta inicios de 2023 se formó parte de la colaboración \href{https://reinforceeu.eu/}{\textcolor{blue}{\underline{REINFORCE}}}. Durante el desarrollo de ese proyecto, uno de los objetivos principales de la colaboración fue el despliegue multisensorial de datos astrofísicos, particularmente datos de partículas del Gran Colisionador de Hadrones (LHC), datos de muografía e imágenes de Glitches (ruido instrumental o ambiental presente en el equipo VIRGO) obtenidas en el Observatorio Europeo de Ondas Gravitacionales (EGO). Con tal fin se generaron nuevos códigos de sonorización para estos datos que se presentan en un formato diferentes al ya disponible en sonoUno (datos en formato columnas donde una de ellas debe poder tomarse como variable independiente y otra como variable dependiente). Se profundizará sobre este tema en la sección \ref{sect:cap6_sonif_reinforce}. 

En cuanto a la versión de escritorio de sonoUno, luego de las pruebas con usuarios descriptas en el Capítulo \ref{cap:fg_completo}, se realizaron varias actualizaciones. Las mismas se describen en la sección \ref{sect:cap6_vescritorio}, mencionando las características agregadas y en algunos casos las limitaciones encontradas y propuestas a futuro. Adicionalmente, en la sección \ref{sect:cap6_vescritorio} se muestra el uso de tres tipos de datos astronómicos en esta versión de sonoUno.

Otro elemento clave en la participación dentro del proyecto REINFORCE fue el inicio del desarrollo de la versión web del software de sonorización de datos sonoUno. Solo a modo de recordatorio, sonoUno es un paquete de software centrado en el usuario que permite a personas con diferentes estilos sensoriales explorar y analizar datos científicos presentados en formato tabla; a través de una representación visual y de sonorización sincronizadas. Éste mismo concepto de la versión escritorio, fue llevado a la nueva interfaz web que se describe en detalle, en la sección \ref{sect:cap6_webcompleta}.

Además del trabajo que se menciona anteriormente, está en marcha un proyecto con el grupo de trabajo de sonorización, para unificar el código base de las dos versiones de sonoUno, ayudando así a optimizar el flujo de trabajo del desarrollo. Para lograr este nuevo enfoque se propone utilizar un servidor, que permita ofrecer el servicio de sonorización tanto a sonoUno como a otras plataformas que deseen utilizarlo. Se profundizará sobre este tema en la sección \ref{sect:cap6_server}.

\section{Sonorizaciones específicas de grandes facilidades astrofísicas que trabajan en la línea de multimensajero}
\label{sect:cap6_sonif_reinforce}

Dentro de la colaboración REINFORCE, los desarrollos y trabajos de investigación se encontraban divididos en diferentes equipos de trabajo. A continuación se enumeran los diez equipos de trabajo que formaron parte del proyecto:

\begin{itemize}
    \item WP1: Gestión de proyectos
    \item WP2: Ciencia ciudadana (estrategia de compromiso)
    \item WP3-6: Demostrador de ciencia ciudadana a gran escala

    \begin{itemize}
        \item WP3: Cazadores de ruido de ondas gravitacionales
        \item WP4: Exploradores de aguas profundas
        \item WP5: Buscadores de nuevas partículas en el Gran Colisionador de Hadrones (LHC)
        \item WP6: Estudios interdisciplinares con Geociencia y Arqueología
    \end{itemize}

    \item WP7: Incrementando los sentidos, incrementando la inclusión
    \item WP8: Actividades de compromiso participativo
    \item WP9: Evaluación de impacto
    \item WP10: Sensibilización y sostenibilidad
\end{itemize}

El grupo de trabajo en el cual se encontraba la sonorización de datos, entre otras actividades, fue el WP7. El compromiso asumido en cuanto a la sonorización fue posibilitar que los demostradores desarrollados por los grupos de trabajo WP3, WP4, WP5 y WP6, pudieran ser utilizados por una amplia audiencia, incluso aquella con discapacidad visual. En cuanto a los mencionados demostradores, se basaron en proyectos desarrollados en la plataforma \href{https://www.zooniverse.org/}{\underline{\textcolor{blue}{Zooniverse}}}, la cual permite incluir información general y diferentes actividades para que todas las personas puedan hacer ciencia.

Los datos de cada demostrador (particulares de este proyecto y puestos en funcionamiento por los WP3, WP4, WP5 y WP6) necesitaron una atención específica para poder ser sonorizados. Los datos del WP3, en específico los datos de Glitch (un ruido transitorio que puede ser producido por el dispositivo o por el entorno y se superpone a la onda gravitacional haciendo complicada su detección), pueden ser sonorizados en sonoUno utilizando los datos crudos, pero es preferible utilizar las imágenes para clasificarlos, sobre todo porque fue la forma de presentar el demostrador en la plataforma \href{https://www.zooniverse.org/projects/reinforce/gwitchhunters}{\underline{\textcolor{blue}{Zooniverse}}}. Un ejemplo concreto puede observarse en la galería web de sonoUno (\href{https://www.sonouno.org.ar/glitch-1126409678-84375/}{\underline{\textcolor{blue}{link}}}), allí se despliega la imagen de un Glitch, el video de la sonorización de la imagen y la sonorización e imagen extraídas de sonoUno, sonorizando las dos tablas componentes de ese Glitch. Se describirá el desarrollo de la sonorización de imagen en la sección \ref{sect:cap6_imgsonif}.

\begin{figure}[ht!]
        \centering
        \begin{subfigure}[b]{1\textwidth}
             \centering
             \captionsetup{justification=centering}
             \frame{\includegraphics[width=0.7\textwidth]{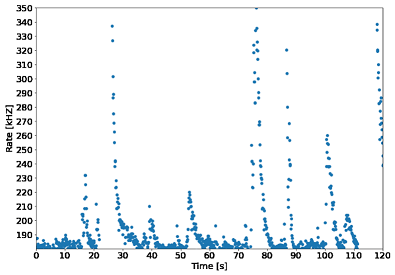}}
             \caption{Imagen extraída del proyecto `Deep Sea Explorer' en Zooniverse, biolominiscencia 1: Conteo de picos (\href{https://www.zooniverse.org/projects/reinforce/deep-sea-explorers/classify}{\underline{\textcolor{blue}{link}}}).}
             \label{fig:cap6_wp4-1}
         \end{subfigure}
         \hfill
        \begin{subfigure}[b]{1\textwidth}
             \centering
             \captionsetup{justification=centering}
             \frame{\includegraphics[width=1\textwidth]{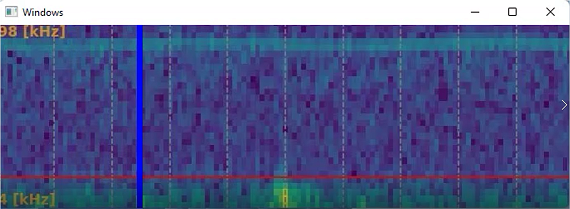}}
             \caption{Captura de pantalla de una imagen (ubicada en \href{https://www.zooniverse.org/projects/reinforce/deep-sea-explorers}{\underline{\textcolor{blue}{bioacústica 1}}}) sonorizada con el código para sonorización de imágenes de sonoUno (\href{https://youtu.be/pkiGdZu5gEo}{\underline{\textcolor{blue}{link}}}).}
             \label{fig:cap6_wp4-2}
         \end{subfigure}
        \caption{Imágenes tomadas de ambas secciones en el demostrador Zooniverse `Deep Sea Explorer'.}
        \label{fig:cap6_wp4}
\end{figure}

Sobre el WP4, trabajan con datos de neutrinos en el contexto del Telescopio de Neutrinos de kilómetros cúbicos (\href{https://www.km3net.org/}{\underline{\textcolor{blue}{KM3NeT}}}). La forma de estudiar los neutrinos es a través de su interacción con el medio, debido a que son partículas neutras (como su nombre lo indica) y de masa muy reducida, que interactúan de manera débil con la materia y no es posible estudiarlas directamente. En este caso particular se estudian a través de la radiación Cherenkov, que se produce cuando el neutrino atraviesa el agua, a una velocidad mayor que la de la luz en ese mismo medio. Otro dato particular a tener en cuenta, es que este telescopio se encuentra sumergido en el mar mediterráneo. De forma similar al grupo de trabajo anterior, estos detectores reciben no solo la información de la radiación Cherenkov, sino también de fuentes de ruido. El tipo de ruido que se encontrará aquí es producido por fuentes de sonido y fuentes de luz (se debe tener en cuenta que cualquier ser vivo tiene la capacidad de producir luz). En el caso del sonido, este se relaciona con mamíferos marinos y debe ser detectado como ruido para poder filtrarlo de la señal obtenida, este tipo de ruido lo denominan ``bioacústico''. En cuanto al ruido producido por la luz emitida por seres vivos en el fondo del mar, se denomina ``bioluminiscencia''. Ambos clasificadores pueden encontrarse en la página de Zooniverse de este equipo de trabajo (\href{https://www.zooniverse.org/projects/reinforce/deep-sea-explorers}{\underline{\textcolor{blue}{link}}}). En cuanto a la contribución desde la sonorización, las curvas de bioluminiscencia pueden graficarse en sonoUno ya que son tablas de dos columnas, es más, las gráficas localizadas en el clasificador fueron obtenidas con sonoUno (ver Figura \ref{fig:cap6_wp4-1}). Para el caso de bioacústica, la fuente es un sonido y con el se genera una imagen, un audiograma, de todas formas para sonorizar dicha imagen puede ser utilizado el código de sonorización de imágenes que se presentará en la sección \ref{sect:cap6_imgsonif} (la Figura \ref{fig:cap6_wp4-2} muestra un ejemplo y puede observarse un video en el siguiente \href{https://youtu.be/pkiGdZu5gEo}{\underline{\textcolor{blue}{link}}}).

\begin{figure}[!ht]
    \centering
    \includegraphics[width=1\textwidth]{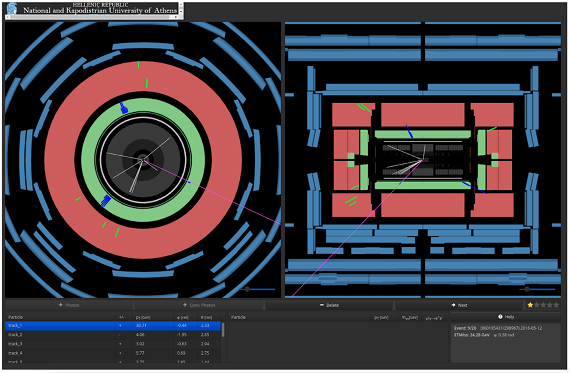}
    \caption{Muestra el despliego del evento 860195431 en el programa HYPATIA. A la izquierda una vista transversal, y a la derecha una vista longitudinal del detector; se indica con rosado en ambas vistas la partícula seleccionada.}
    \label{fig:cap6_lhcdemonstrator}
\end{figure}

En el caso del WP5, se trabaja con datos de partículas obtenidas en el Gran Colisionador de Hadrones (LHC). En el marco del proyecto REINFORCE el demostrador `Nueva búsqueda de partículas en el CERN' disponible en \href{https://www.zooniverse.org/projects/reinforce/new-particle-search-at-cern}{\underline{\textcolor{blue}{Zooniverse}}}, además de la detallada explicación sobre el proyecto del WP5, dirige a los usuarios al programa HYPATIA (la Figura \ref{fig:cap6_lhcdemonstrator} muestra un ejemplo con un evento desplegado). En dicho programa es donde se les permite clasificar las partículas generadas en diferentes eventos en el LHC, visualmente en las gráficas del detector, al ser seleccionada una partícula, se colorea de rosado para que la persona pueda identificarla y clasificarla. Resulta evidente de esta gráfica y de la naturaleza de este tipo de datos, que para sonorizarlos se necesita un enfoque diferente al que hemos descripto hasta el momento. En este caso, no se tiene una tabla de dos columnas representable en una gráfica 2D, y sonorizar la imagen completa no aportaría mayor información para diferenciar entre partículas. En la sección \ref{sect:cap6_lhc_sonif} se detallarán los pasos de diseño y desarrollo del código para sonorizar partículas.

Finalmente, el WP6 trabaja con tomografía muónica, es decir que utilizan muones provenientes del espacio para realizar una tomografía aplicable a Geociencia y Arqueología. En cuanto a la fuente, los muones, se debe tener en cuenta que cuando una partícula de rayos cósmicos entra en contacto con la atmósfera, reacciona con los átomos de las moléculas presentes en el aire, produciendo partículas secundarias que, a la vez, continúan interactuando hasta llegar al suelo. Dicho fenómeno se conoce como lluvia atmosférica extendida de partículas secundarias, en las que se encuentran, según su abundancia: muones, electrones y fotones, además de otras partículas elementales y antimateria. 

\begin{figure}[ht!]
    \centering
        \begin{subfigure}[b]{0.53\textwidth}
             \centering
             \captionsetup{justification=centering}
             \frame{\includegraphics[width=1\textwidth]{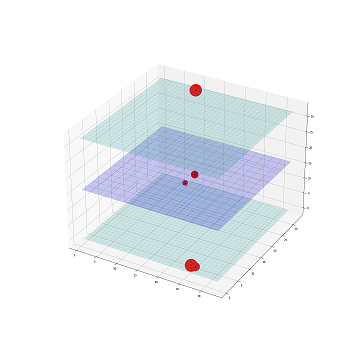}}
             \caption{Representación 3D.}
             \label{fig:cap6_muon_linea_3D}
         \end{subfigure}
         \hfill
        \begin{subfigure}[b]{0.445\textwidth}
             \centering
             \captionsetup{justification=centering}
             \frame{\includegraphics[width=1\textwidth]{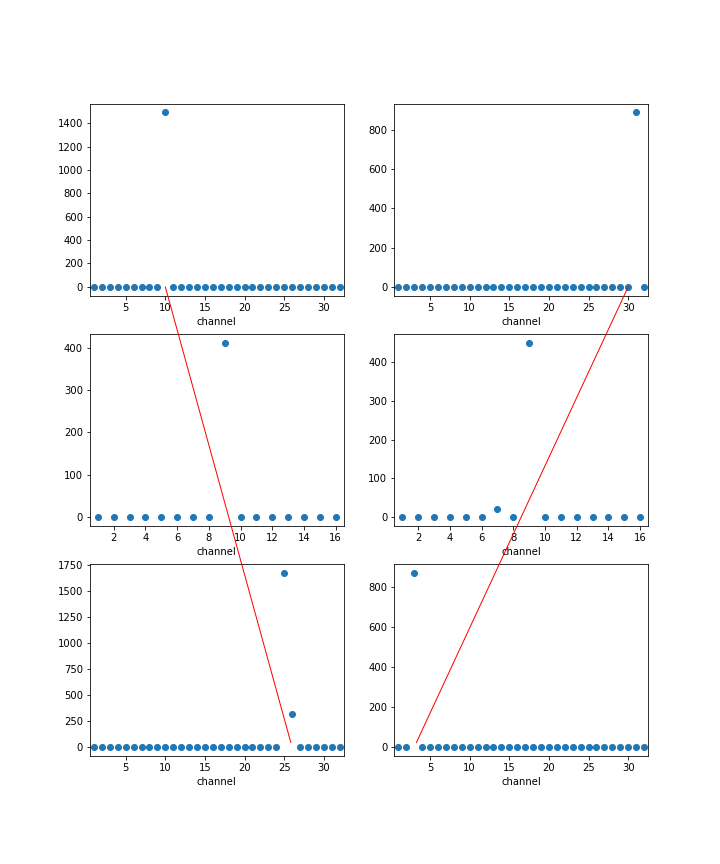}}
             \caption{Representación 1D.}
             \label{fig:cap6_muon_linea_1D}
         \end{subfigure}

    \caption{Ambas vistas del evento 5762520\_3538527 que representa la existencia de un muón. Se incluye el \href{https://youtu.be/EYhcdyO2w2I}{\underline{\textcolor{blue}{link}}} a un video con la sonorización de este evento.}
    \label{fig:cap6_muon_linea}
\end{figure}

\begin{figure}[ht!]
    \centering
         \begin{subfigure}[b]{0.53\textwidth}
             \centering
             \captionsetup{justification=centering}
             \frame{\includegraphics[width=1\textwidth]{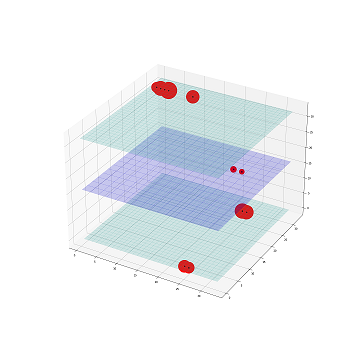}}
             \caption{Representación 3D.}
             \label{fig:cap6_muon_nolinea_3D}
         \end{subfigure}
         \hfill
        \begin{subfigure}[b]{0.445\textwidth}
             \centering
             \captionsetup{justification=centering}
             \frame{\includegraphics[width=1\textwidth]{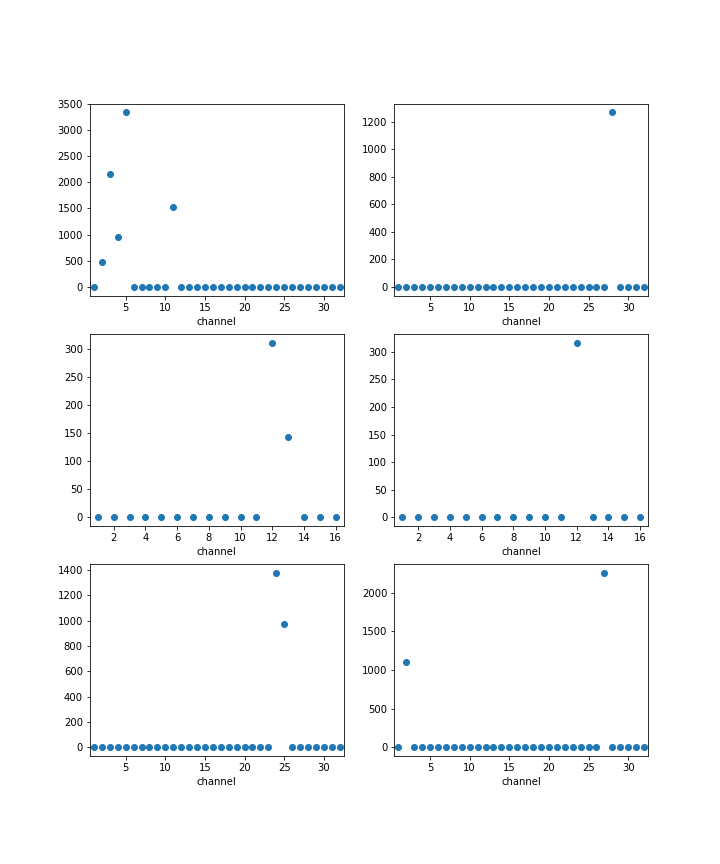}}
             \caption{Representación 1D.}
             \label{fig:cap6_muon_nolinea_1D}
         \end{subfigure}
         
    \caption{Ambas vistas del evento 5793328\_1297611 donde no se percibe la existencia de un muón. Se incluye el \href{https://youtu.be/MwZ0_EEuD_g}{\underline{\textcolor{blue}{link}}} a un video con la sonorización de este evento.}
    \label{fig:cap6_muon_nolinea}
\end{figure}

Habiendo aclarado de donde proviene la fuente, lo que resta mencionar es cómo se utilizan para realizar esta tomografía. Los muones producidos por lluvia de rayos cósmicos están ampliamente estudiados, se conoce muy bien su distribución y cómo interactúan. Es por ello, que colocando un detector de muones luego de que estos atraviesen un objeto, es posible estimar la densidad media de la materia atravesada por esta partícula, 204 veces más masiva que el electrón, y por tanto, muy penetrante. Un ejemplo de aplicación de este método es el descubrimiento de una cámara en una pirámide, debido a que la densidad del aire es mucho menor a la densidad del material de construcción. Ahora bien, en el caso del demostrador de tomografía muónica, lo que buscan es poder clasificar entre diferentes archivos de datos, identificando cuál representa la existencia de un muón y cuales no, a través de la determinación de su trayectoria, utilizando 3 centelladores ubicados uno sobre el otro: los muones atravesarán las 3 capas de detectores sin desviación, permitiendo ``dibujar'' la trayectoria del muón (la Figura \ref{fig:cap6_muon_linea} muestra las gráficas correspondiente a un caso de existencia de muón; la Figura \ref{fig:cap6_muon_nolinea} incluye las gráficas donde se aprecia que no puede trazarse una línea recta entre las capas del detector, por lo que no existe un muón). Si bien, los muones son partículas, la forma de mostrar y analizar los datos es completamente diferente a la descripta para el LHC, por lo que requirió un nuevo enfoque que se describe en la sección \ref{sect:cap6_muon_sonif}.

\subsection{Sonorización de imágenes}
\label{sect:cap6_imgsonif}

Esta técnica se comienza a desarrollar principalmente para el grupo de trabajo que estudia Glitches (WP3), sin embargo, luego es utilizada para sonorizar imágenes de datos de bioluminiscencia y bioacústica (WP4). Adicionalmente, también había surgido en intercambios previos la posibilidad de sonorizar imágenes de otros tipos de datos, por ejemplo, se consultó a la autora la posibilidad de sonorizar agujeros negros durante el intercambio en Reino Unido para realizar las pruebas con usuarios.

Debido a que ha sido una consulta constante, se pretende con este primer enfoque, plantear una línea de investigación que estudie la forma de sonorizar una imagen que contiene datos en al menos tres dimensiones. Dicho estudio excede los límites de esta tesis, pero sí se plantea aquí un primer enfoque.

\begin{figure}[!ht]
    \centering
    \includegraphics[width=1\textwidth]{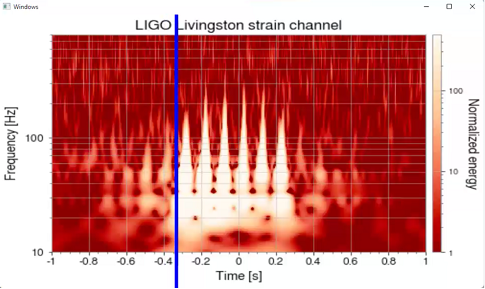}
    \caption{Muestra la imagen de un Glitch que contiene una barra vertical azul indicando la posición sonorizada (\href{https://www.sonouno.org.ar/glitch-1126409678-84375/}{\underline{\textcolor{blue}{link a video}}}).}
    \label{fig:cap6_glitch_primertest}
\end{figure}

Esta técnica de sonorización ha sido desarrollada en Python, utilizando la librería openCV para la manipulación de imágenes y la librería de sonido de sonoUno para el proceso de sonorización. Para esta primer aproximación se realiza una sonorización de los niveles de gris que contiene la imagen por columna (por ejemplo, en la Figura \ref{fig:cap6_glitch_primertest} puede observarse una imagen de Glitch con una línea vertical azul, dicha línea de color indica los pixeles que están siendo sonorizados).

\begin{figure}[!ht]
    \centering
    \includegraphics[width=1\textwidth]{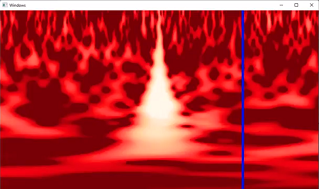}
    \caption{Muestra la imagen de un Glitch del tipo blip, la cual no tiene cuadrícula y tampoco los bordes de ejes. Se aprecia la barra vertical azul indicando la posición sonorizada (\href{https://youtu.be/x6Abb2Ekb8s}{\underline{\textcolor{blue}{link}}}).}
    \label{fig:cap6_glitch_blip}
\end{figure}

En cuanto al proceso específico que sigue el algoritmo, en primer lugar utiliza la librería OpenCV para convertir la imagen a escala de grises:

\begin{lstlisting}[language=Python]
    img = cv.imread(path)
    img = cv.resize(img, (960, 540))
    gray_img = cv.cvtColor(img, cv.COLOR_BGR2GRAY)
\end{lstlisting}

Con la imagen en escala de grises, dentro de un bucle que permite recorrer todas las columnas de la imagen, se continúa con la tarea de medir el nivel de gris en la columna a sonorizar (se va a sonorizar la columna específica durante cada interacción del bucle, y al mismo tiempo se almacena el valor para guardar la sonorización en un archivo wav). A continuación se muestra cómo se evalúa el nivel de gris, sumando todas las filas de dicha columna y dividiendo por el número de filas. Adicionalmente, se divide por el máximo nivel de gris de toda la imagen para obtener un valor representativo del nivel de gris entre 0 y 1:

\begin{lstlisting}[language=Python]
    column = gray_img[:, j:j+1]
    for i in range(0,column.shape[0]):
        if i==0:
            suma = int(column[i,0])
        else:
            suma = suma + int(column[i,0])
    value = (suma/column.shape[0])/gray_img.max()
\end{lstlisting}

En cuanto a los parámetros de sonido, se considera el valor de brillo (blanco) correspondiente al tono más alto y el valor más oscuro (negro) al tono más bajo (silencio). Los demás parámetros como instrumento y límites de frecuencia de sonido por el momento se configuran de forma fija al inicio del código, pero se pretende agregar una forma de indicarlo a través de línea de comandos a futuro. El código completo incluye las líneas propias de la visualización con OpenCV, el bucle para recorrer las columnas de la imagen, la selección de tecla para terminar la reproducción antes de tiempo y las líneas propias de la sonorización que utilizan el valor calculado anteriormente:

\begin{lstlisting}[language=Python]
    sound.make_sound(value, 1)
    time.sleep(0.02)
\end{lstlisting}

El código toma como parámetro de ingreso el path donde se encuentra la imagen en la computadora, por lo que se puede sonorizar cualquier tipo de imágenes en los formatos permitidos por OpenCV (por nombrar algunos: jpg, png, bmp, dib, jpeg, tiff). El primer tipo de sonorización de Glitch que se realizó fue la que se muestra en la Figura \ref{fig:cap6_glitch_primertest}, el video está disponible en la galería web de sonoUno (\href{https://www.sonouno.org.ar/glitch-1126409678-84375/}{\underline{\textcolor{blue}{link}}}); pero se percibió que al sonorizar la imagen completa del Glitch, también se estaba sonorizando la cuadrícula, los ejes y los títulos. Es por ello que posteriormente, para este tipo de datos, se realizó una prueba sin la cuadrícula y sin los bordes, obteniendo la Figura \ref{fig:cap6_glitch_blip} (\href{https://youtu.be/x6Abb2Ekb8s}{\underline{\textcolor{blue}{link}}} al video).

Al poder sonorizar diferentes imágenes con este algoritmo, también se probó su eficacia en los datos de neutrinos (WP4), lográndose así sonorizar las imágenes obtenidas en el eje de bioacústica (un video de ejemplo se encuentra en el siguiente \href{https://youtu.be/pkiGdZu5gEo}{\underline{\textcolor{blue}{link}}}).

Aún siendo una sonorización sencilla, en este tipo de imágenes y para estos datos permite detectar los rasgos que se necesitan. Sin embargo, utilizando los diferentes combinaciones de parámetros de sonido y la potencialidad de la librería OpenCV para el análisis de imágenes, esta podría ser una basta línea de investigación con el objetivo de lograr una sonorización de imágenes que sea representativa y comparable con el contenido de la imagen.

\subsection{Sonorización de partículas del LHC}
\label{sect:cap6_lhc_sonif}

La Figura \ref{fig:cap6_lhcdemonstrator} muestra una captura de pantalla del programa HYPATIA con un evento desplegado. Para poder discriminar entre partículas, originadas por los haces de protones que son acelerados a casi la velocidad de la luz antes de hacerlos colisionar, el detector presenta diferentes subsistemas de detectores: el detector interno, los calorímetros, el espectrómetro de muones y los sistemas de imanes.

\begin{itemize}
    \item \textit{Detector interno}: se encuentra alrededor del tubo portador del haz del LHC, es lo primero que se encuentran las nuevas partículas generadas. Está formado por tres subdetectores diferentes que registran con precisión el paso de las partículas producidas. En el caso de las partículas neutras, como son los fotones, atraviesan este detector sin dejar rastros.
    \item \textit{Calorímetros}: construidos de materiales pesados (como acero y plomo) buscan detectar tanto partículas cargadas como neutras, logrando que estas interactúen y depositen toda su energía, permitiendo su medición. Existen dos tipos de calorímetro en ATLAS:
    \begin{itemize}
        \item \textit{Electromagnético}: representado con color verde en la Figura \ref{fig:cap6_lhcdemonstrator}, es el encargado de absorber y permitir la medición de energía depositada por electrones, positrones y fotones. Los hadrones energéticos logran traspasar este calorímetro depositando únicamente una fracción de su energía.
        \item \textit{Hadrónico}: representado con color rojo en la Figura \ref{fig:cap6_lhcdemonstrator}, posibilita la medición de la energía total depositada por hadrones, tales como los protones y neutrones. A la altura de esta capa se logran detener casi todas las partículas incidentes, posibilitando medir su energía.
    \end{itemize}
    \item \textit{Espectrómetro de muones}: encargado de detectar muones energéticos, ya que como tienen la capacidad de atravesar el detector casi sin interactuar, depositan muy poca energía en los detectores previos.
    \item \textit{Sistema de imanes}: responsable de curvar la trayectoria de las partículas cargadas en el detector interno y el espectrómetro muónico, permitiendo medir con precisión sus momentos en base a la curvatura de la trayectoria.
\end{itemize}

Los neutrinos son las únicas partículas que atraviesan el detector sin interactuar, sin embargo, se puede medir su energía indirectamente a partir del desequilibrio de su momento en el plano perpendicular al eje del haz.

Ahora bien, conociendo la estructura del detector y como detecta las diferentes partículas, se resumirán las características que presentan cada una y como se caracterizan. Las partículas de interés son: electrón, fotón, fotón convertido y muón, además de partículas que no encajan entre las anteriores y que se llaman desconocidas. Cada una de estas partículas describen una trayectoria característica:

\begin{enumerate}
    \item \textit{Electrón}: representado por una trayectoria en el detector interno (línea en la primera área gris) que apunta a un clúster (depósito de energía indicado de color azul) en el calorímetro electromagnético (área verde);
    \item \textit{Fotón}: representado por un clúster en el calorímetro electromagnético pero sin rastro en el detector interno;
    \item \textit{Fotón convertido}: representado por dos trayectorias muy cercanas en el detector interno, que apunta a un clúster en el calorímetro electromagnético;
    \item \textit{Muón}: representado por una larga trayectoria que atraviesa todas las capas del detector, podría atravesar un clúster, pero no siempre deja ese rastro de energía;
    \item \textit{Desconocido}: cualquier otra representación que no se ajuste a las anteriores.
\end{enumerate}

En un primer momento y conociendo que una de las columnas del archivo de datos indica que tipo de partícula se está representando, se pensó en indicar con diferentes sonidos los tipos de partículas a sonorizar. Sin embargo, con esa representación no se estaría permitiendo a la persona explorar los datos y poder clasificar, en su lugar, se le estaría dando la respuesta de antemano. Por lo expresado se continuó tomando en consideración diferentes visualizaciones de partículas, extrayendo las características representativas en cada capa del detector. Se recayó en las siguientes características:

\begin{enumerate}
    \item La presencia o ausencia de la trayectoria en el detector interior;
    \item La presencia o ausencia del clúster en el calorímetro electromagnético;
    \item Finalmente, tener en cuenta que el muón traspasa todo el detector, presentando una larga trayectoria.
\end{enumerate}

Con esto en mente, la siguiente pregunta fue ¿Cómo producir un sonido que represente cada situación lo suficientemente general como para usarlo con cada partícula y lo suficiente como para poder discriminar entre cada partícula?. Después de varias reuniones entre los WP5 y WP7, la propuesta final de sonorización consistió en sonorizar cada trayectoria de partículas del evento con la siguiente configuración de sonido:

\begin{enumerate}
    \item En primer lugar el sonido de una tick-mark (sonido característico que corresponde a un mensaje definido y claro, por ejemplo una bocina indica un llamado de atención) representada en este caso por un sonido bip. En este caso es indicativo del centro del detector y el comienzo del viaje de la partícula (este sonido está presente en cada sonorización de partículas).
    \item Para representar la trayectoria o la ausencia de la misma en el detector interior:
    
    \begin{itemize}
        \item Un sonido continuo con una frecuencia específica (nota de piano `D6', valor 1174.66Hz) para representar cada trayectoria individual presente en el detector interno (duración de una trayectoria simple en el detector interno: 2s); en el caso de un muón, el sonido continuo presenta mayor tiempo de duración para representar una trayectoria más larga que excede el detector interno (duración del sonido de un muón: 4s).
        \item En el caso de dos trayectorias, se sonorizan dos sonidos continuos con dos frecuencias diferentes (frecuencia 1 de la nota de piano `D6', valor 1174,66Hz; frecuencia 2 de la nota de piano `C6', valor 1046.50) (la duración de estas trayectorias en el detector interior: 2s).
        \item En caso de ausencia de trayectoria, se sonoriza un silencio de 2s.
    \end{itemize}
    
    \item Una segunda tick-mark representa el final del detector interno y el comienzo del calorímetro electromagnético (se utiliza una nota de piano `F7', valor 2793.82Hz; duración 1ms)

    \begin{itemize}
        \item En el caso de los muones, esta tick-mark se sonoriza a los 2 segundos del inicio del sonido continuo, lo que indica que el muón pasa del detector interno al calorímetro y continúa su trayectoria.
    \end{itemize}
    
    \item Cuando existe un clúster en el calorímetro electromagnético, se reproduce una compilación específica de sonidos cortos para representarlo, y el volumen del mismo está relacionado con la energía del clúster (menos energía-menor volumen; más energía-mayor volumen).

    \begin{itemize}
        \item En el caso de los muones, el clúster se sonoriza junto con el sonido de la trayectoria en el tiempo correspondiente (justo después de la marca que indica la transición al calorímetro).
    \end{itemize}
\end{enumerate}

\begin{figure}[ht!]
    \centering
         \begin{subfigure}[b]{1\textwidth}
             \centering
             \captionsetup{justification=centering}
             \frame{\includegraphics[width=1\textwidth]{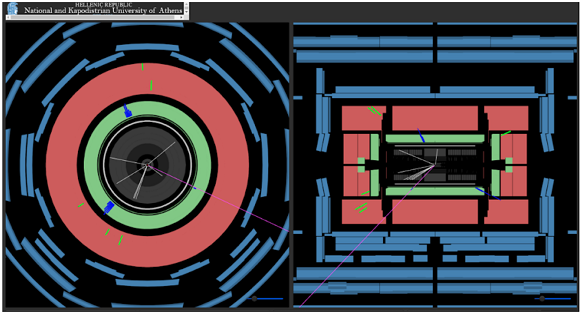}}
             \caption{HYPATIA}
             \label{fig:cap6_lhc_hypatia}
         \end{subfigure}
         \hfill
        \begin{subfigure}[b]{1\textwidth}
             \centering
             \captionsetup{justification=centering}
             \frame{\includegraphics[width=1\textwidth]{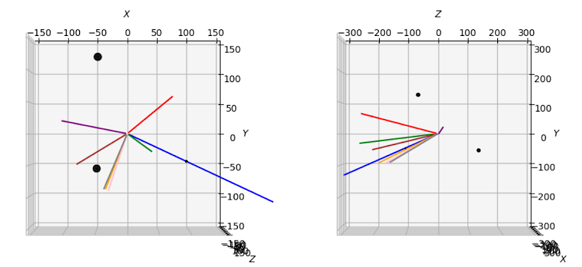}}
             \caption{SonoUno}
             \label{fig:cap6_lhc_sonouno}
         \end{subfigure}
         
    \caption{Se muestra el mismo evento: 860195431; desplegado en \ref{fig:cap6_lhc_hypatia} por HYPATIA y en \ref{fig:cap6_lhc_sonouno} por el script desarrollado.}
    \label{fig:cap6_lhc_comparacion}
\end{figure}

Siguiendo el método descripto, se puede utilizar el script desarrollado para generar la sonorización a partir de un archivo de texto en formato columna (siempre que respete las posiciones de columnas de los archivos proporcionados por el WP5), por lo que serviría incluso para producir la sonorización de las grandes cantidades de eventos que existen en el LHC sin modificaciones o trabajo extra. Por esta razón, existen dos versiones, una para utilizar en línea de comandos que trabajaría de forma autónoma generando los sonidos de cada evento y almacenándolos en una carpeta con el mismo nombre del archivo que le da origen. La otra versión del script realiza la misma tarea pero mostrando el complemento visual por pantalla (se puede observar un video de ejemplo en el siguiente \href{https://youtu.be/XMaYIJkJIHg}{\underline{\textcolor{blue}{link}}}). Adicionalmente, en la Figura \ref{fig:cap6_lhc_comparacion} se puede observar el mismo evento representado en HYPATIA y con el script de sonoUno, se muestran con el script las mismas vistas desplegadas por HYPATIA para poder realizar la comparación visual entre ambas plataformas.

Un detalle que debe tenerse en cuenta, es que el algoritmo en primer lugar busca un separador ubicado en el archivo de texto que indica la separación entre eventos, logrando así obtener los diferentes eventos de un archivo identificados por su nombre. Luego continúa con la tarea de separar entre trayectorias y clúster, para continuar con el despliegue y sonorización. Esta característica permite que en un mismo archivo txt puedan encontrarse un gran número de eventos, y el código entregará un archivo de sonido (el gráfico es opcional) por cada evento que registre en dicho archivo de texto con el nombre del evento como título.

\subsection{Sonorización de datos de moungrafía}
\label{sect:cap6_muon_sonif}

Si bien el muón es una partícula, como las que se encuentran en el LHC, la tomografía muónica pretende informar cuando existe y cuando no, por lo que en este caso se debe investigar una forma de despliegue tal que muestre la información y permita a la persona clasificar la existencia o no de un muón. No se puede utilizar el enfoque de sonorizar una trayectoria (como se hace en el LHC), debido a que no se conoce previamente si los datos contienen o no esta trayectoria, lo que se conoce es los lugares del detector donde hubo depósito de energía.

\begin{figure}[ht!]
    \centering
        \begin{subfigure}[b]{0.53\textwidth}
             \centering
             \captionsetup{justification=centering}
             \frame{\includegraphics[width=1\textwidth]{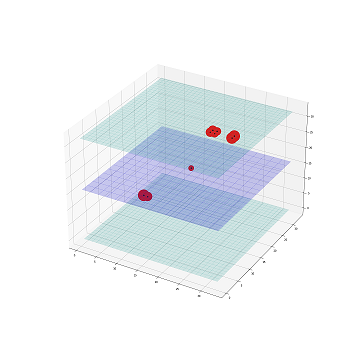}}
             \caption{Representación 3D.}
             \label{fig:cap6_muon_dudoso_3D}
         \end{subfigure}
         \hfill
        \begin{subfigure}[b]{0.445\textwidth}
             \centering
             \captionsetup{justification=centering}
             \frame{\includegraphics[width=1\textwidth]{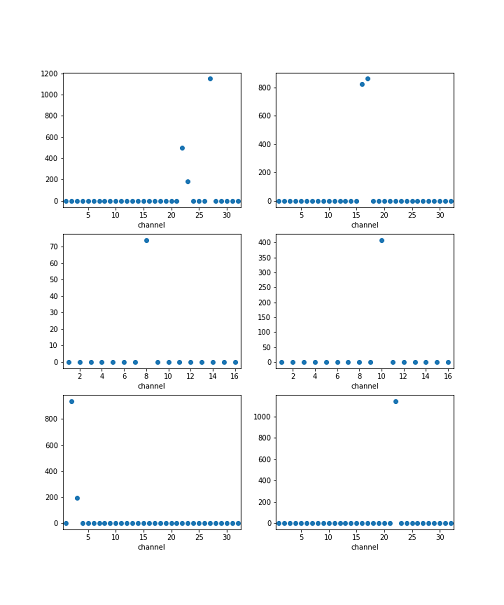}}
             \caption{Representación 1D.}
             \label{fig:cap6_muon_dudoso_1D}
         \end{subfigure}

    \caption{Ambas vistas de un evento que representa la existencia de un muón pero el algoritmo no puede representar la línea porque existe depósito de energía en otras partes del detector.}
    \label{fig:cap6_muon_dudoso}
\end{figure}

Para ilustrar mejor la información que se tiene de base para la sonorización, se recurre a las imágenes de los tres casos posibles que se presentan en esta etapa inicial. En primer lugar comentar que la representación 3D muestra con recuadros inclinados de color cada una de las tres capas del detector y con puntos rojos los depósitos de energía; en cuanto a la representación 1D, muestra las vistas `x' e `y' en cada columna de gráficos y mediante las filas representa las tres capas del detector (un detalle a tener en cuenta es que el detector del medio presenta 16 canales en lugar de 32 como los superior e inferior). Las Figuras \ref{fig:cap6_muon_linea} y \ref{fig:cap6_muon_nolinea} presentan los dos casos opuestos, el más sencillo donde se puede trazar una línea entre los depósitos de energía en el detector; el otro, donde los depósitos de energía no coinciden entre sí, por lo que no puede graficarse una línea.

Ahora bien, la Figura \ref{fig:cap6_muon_dudoso} muestra un caso donde se aprecia la existencia de un muón, pero el algoritmo no pudo trazar una línea entre detectores porque hay depósitos de energía que no corresponden a dicha partícula. Es necesario aclarar que el algoritmo mencionado es el utilizado por el WP6 para generar la gráfica 1D, hasta el momento solo se ha descripto la información disponible previo al diseño de la sonorización.

Conociendo las características del archivo de datos y la forma en que debe ser desplegado para permitir a los ciudadanos realizar una clasificación, se procede a enumerar las características a tener en cuenta para el diseño de sonorización:

\begin{itemize}
    \item Se debe poder diferenciar entre los sucesos de cada capa: tal vez sonorizando una capa del detector luego de la otra a intervalos fijos de tiempo porque se encuentran a la misma distancia;
    \item Debe poder localizarse el lugar donde se producen los depósitos de energía: tal vez sonorizando solo los lugares donde se produce un depósito y respetando un silencio donde no hubo depósito de energía;
    \item Surge la pregunta aquí, ¿cómo se puede sonorizar la posición en la imagen 3D de forma intuitiva? 
    \item El último detalle a tener en cuenta, es que debe poder relacionarse cada una de las capas a través de la sonorización, permitiéndole a la persona percibir si los depósitos de energía están alineados o no.
\end{itemize}

Teniendo en cuenta estas cuatro necesidades, se comenzó con el diseño de la sonorización, en primer lugar para evitar el problema de la sonorización 3D y debido a que este tipo de datos cuenta con una representación 1D, se resuelve sonorizar esta última representación. Evaluando la cantidad de datos por detector se evidencia que cuenta con un máximo de 32 puntos, por lo que se podrían utilizar notas de piano para representar cada canal, sonorizando las notas de piano de los canales que tienen depósito de energía. Con estas premisas se realizó la primer aproximación de sonorización, la cual dio buenos resultados, al sonorizar una capa luego de la otra y haber mapeado los canales como si fueran un teclado de piano, se logra establecer una correlación entre las diferentes capas del detector. De esta forma se comprobó que el planteo era prometedor.

De esta primera prueba se descubrió que era mejor sonorizar las capas del detector de arriba hacia abajo, debido a que de esta forma se corresponde con la forma en la que se detecta el muón. Siguiendo, se realizó una prueba configurando los canales a 16 teclas de piano (es decir dos canales por tecla en la capa superior e inferior y un canal por tecla en la capa del medio), donde la relación sonora entre los depósitos de energía de cada capa se percibía mejor. Se pueden observar dos videos de ejemplo en los siguientes link: \href{https://youtu.be/269lsyC_bGg}{\underline{\textcolor{blue}{existencia de muón}}} y \href{https://youtu.be/1nDW8XGV1PQ}{\underline{\textcolor{blue}{caso dudoso donde no existe muón}}}.

Una vez realizadas las actualizaciones y configuraciones mencionadas al script, se logró la primer versión de sonorización de muones. La correspondencia entre canales de cada detector y notas de piano se muestra en el Cuadro \ref{tab:cap6_piano_muonchanel}. De esta forma, cuando un conjunto de datos presenta depósito de energía en más de un canal, se sonoriza una composición de notas de piano, correspondientes a los canales que presentan energía (ejemplo de esto se puede percibir en la sonorización de las Figuras \ref{fig:cap6_muon_nolinea} y \ref{fig:cap6_muon_dudoso}).

\begin{table}[ht!]
    \centering
    \caption{Contiene la configuración de notas de piano por canal de los detectores de muongrafía}
    \label{tab:cap6_piano_muonchanel}
    \begin{tabular}{|c|c|c|c|}
        \hline
        \multirow{2}{2cm}{Nota piano} & \multicolumn{3}{|c|}{Canales de cada detector} \\ \cline{2-4}
         & Detector Superior & Detector Central & Detector Inferior \\ \hline
         `A3' & Canal 1 y 2 & Canal 1 & Canal 1 y 2 \\ \hline
         `B3' & Canal 3 y 4 & Canal 2 & Canal 3 y 4 \\ \hline
         `C4' & Canal 5 y 6 & Canal 3 & Canal 5 y 6 \\ \hline
         `D4' & Canal 7 y 8 & Canal 4 & Canal 7 y 8 \\ \hline
         `E4' & Canal 9 y 10 & Canal 5 & Canal 9 y 10 \\ \hline
         `F4' & Canal 11 y 12 & Canal 6 & Canal 11 y 12 \\ \hline
         `G4' & Canal 13 y 14 & Canal 7 & Canal 13 y 14 \\ \hline
         `A4' & Canal 15 y 16 & Canal 8 & Canal 15 y 16 \\ \hline
         `B4' & Canal 17 y 18 & Canal 9 & Canal 17 y 18 \\ \hline
         `C5' & Canal 19 y 20 & Canal 10 & Canal 19 y 20 \\ \hline
         `D5' & Canal 21 y 22 & Canal 11 & Canal 21 y 22 \\ \hline
         `E5' & Canal 23 y 24 & Canal 12 & Canal 23 y 24 \\ \hline
         `F5' & Canal 25 y 26 & Canal 13 & Canal 25 y 26 \\ \hline
         `G5' & Canal 27 y 28 & Canal 14 & Canal 27 y 28 \\ \hline
         `A5' & Canal 29 y 30 & Canal 15 & Canal 29 y 30 \\ \hline
         `B5' & Canal 31 y 32 & Canal 16 & Canal 31 y 32 \\ \hline
    \end{tabular}
\end{table}

Adicionalmente a la sonorización, como el objetivo de esta tesis es la investigación de nuevas modalidades de acceso a los datos, centrándonos en un despliegue multimodal de los datos, se incluyeron cambios en el despliegue gráfico sincronizados con el despliegue auditivo. Es así que, cada punto que se está sonorizando cambia de color y adopta un color de rojo a violeta correspondiente a la frecuencia del sonido, esto es: rojo para baja frecuencia y violeta para alta frecuencia. El arreglo de colores se detalla en el Cuadro \ref{tab:cap6_color_muonchanel}. Una vez que el punto fue sonorizado se deja de color negro (nuevamente esta configuración puede apreciarse en los videos correspondientes a las imágenes).

\begin{table}[ht!]
    \centering
    \caption{Contiene la configuración de color por canal de los detectores de muongrafía}
    \label{tab:cap6_color_muonchanel}
    \begin{tabular}{|c|c|c|c|}
        \hline
        \multirow{2}{2cm}{Color} & \multicolumn{3}{|c|}{Canales de cada detector} \\ \cline{2-4}
         & Detector Superior & Detector Central & Detector Inferior \\ \hline
         `Rojo' & 1 a 4 & 1 y 2 & 1 a 4 \\ \hline
         `Naranja' & 5 a 8 & 3 y 4 & 5 a 8 \\ \hline
         `Amarillo' & 9 a 12 & 5 y 6 & 9 a 12 \\ \hline
         `Oliva' & 13 a 16 & 7 y 8 & 13 a 16 \\ \hline
         `Verde' & 17 a 20 & 9 y 10 & 17 a 20 \\ \hline
         `Celeste' & 21 a 24 & 11 y 12 & 21 a 24 \\ \hline
         `Azul' & 25 a 28 & 13 y 14 & 25 a 28 \\ \hline
         `Violeta' & 29 a 32 & 15 y 16 & 29 a 32 \\ \hline
    \end{tabular}
\end{table}

Si bien se debe seguir perfeccionando este script para lograr representar los datos de muongrafía de la mejor forma posible, se considera una primera aproximación exitosa, la cual permite realizar una primera clasificación, solo debiendo profundizar en algunos casos que aún son dudosos para detectar la presencia del muón con el sonido.

\section{Sonorizaciones de grandes bases de datos}
\label{sect:cap6_vescritorio}

Más allá de la sonorización de datos obtenidos de grandes facilidades astrofísicas, desde el inicio de esta tesis se han utilizado archivos de bases de datos para desarrollar sonoUno. Por lo que en esta sección se describirán las bases de datos utilizadas y los tipos de datos sonorizados.

Previamente a la descripción de las bases de datos, se detallará la interfaz de escritorio con todas las actualizaciones que se incluyeron producto de las pruebas con usuarios descriptas en el Capítulo \ref{cap:fg_completo}. Luego, se incluirán tres subsecciones donde se mostrará el uso de la interfaz de escritorio con datos astronómicos: espectro de galaxia (base de datos \href{https://www.sdss4.org/}{\underline{\textcolor{blue}{Sloan Digital Sky Survey}}}), rayos cósmicos (base de datos de \href{https://opendata.auger.org/}{\underline{\textcolor{blue}{Pierre Auger}}}) y curvas de luz (base de datos \href{https://asas-sn.osu.edu/variables}{\underline{\textcolor{blue}{ASAS-SN Variable}}}).

\subsection{Actualizaciones en la versión escritorio}

Los contactos con los usuarios finales, durante la investigación de grupo focal y algunas consultas de pruebas de sonoUno enviadas por correo electrónico (donde las preguntas se centran en algunos aspectos de usabilidad y si se necesita alguna funcionalidad nueva), han ayudado a resolver errores y actualizaciones en sonoUno haciéndolo más centrado en el usuario. Una de estas actualizaciones, brevemente descriptas en el Capítulo \ref{cap:fg_completo}, fue el cambio de la librería de sonidos a pygame. Esta decisión fue tomada debido a su mantenimiento y por ser una de las pocas librerías que presenta herramientas para reproducir sonido por tonos posibilitando atender las necesidades de la interfaz gráfica de usuario (permitir la interacción del usuario mientras el sonido se está reproduciendo), además de guardar el mismo sonido en un formato de archivo como wav. Esta acción también soluciona un error encontrado con los instrumentos continuos usando la biblioteca Mingus durante la sesión de grupo focal.

Durante las pruebas de sonoUno realizadas por usuarios contactados por email donde utilizaron sus propios conjuntos de datos, algunos reportan un problema con los puntos de datos importados como NaN. El tipo de dato NaN para librerías como Numpy y Pandas, representa espacios vacíos en un arreglo de datos, por ejemplo casos donde una columna tiene más información que otra. En la primer versión de sonoUno, dicho tipo de datos se trataba de la misma forma que 0, es decir que consideraba que era un dato más, lo que era un error. Para la nueva actualización, se trabajó utilizando la librería Numpy para detectar los conjuntos de datos con valores NaN, si se encuentra en una posición donde le siguen valores numéricos, no lo grafica ni sonoriza; y si después de este no hay otro número, el software lo toma como el final de la reproducción. 

Otro problema detectado, también referente a la reproducción de archivos de datos, surgió al sonorizar un conjunto de datos que contiene el mismo valor durante todo el eje `y' (o la columna a reproducir). Eso se debía a que la función que mapea dicho eje a valores entre 0 y 1, utiliza una función de normalización donde resta el valor mínimo y máximo de dicha columna en el denominador, por lo que en el caso donde todos los valores del eje `y' son iguales, dicha función intentaba dividir por 0. En la nueva versión, antes de la normalización se comprueba si el valor máximo y mínimo de la columna es el mismo o no: si son iguales, sonoUno reproduce el mismo tono durante todos los valores.

Particularmente, durante diferentes pruebas con curvas de luz de estrellas variables, se señaló otro error, sonoUno reproduce todos los puntos de datos con el mismo silencio entre ellos (aunque entre ambos puntos la distancia sea mayor), dando la sensación de que todos los puntos están a la misma distancia. En la versión actual, sonoUno corresponde el silencio entre tonos con el espacio entre puntos de datos, un ejemplo de esta funcionalidad se puede ver en los videos disponibles en la Galería de datos de sonoUno (\href{https://www.sonouno.org.ar/comic-rays-2005-to-2017/}{\underline{\textcolor{blue}{link aquí}}}). Además, se señaló la importancia de una función de bucle para analizar algunos conjuntos de datos como ser las curvas de luz, esta función se agregó como un comando (`playloop()') que debe escribirse en el cuadro de texto correspondiente. Se considerará el agregado de un elemento tipo botón en trabajos futuros, realizando el correspondiente análisis con usuarios.

Algunos usuarios finales durante las pruebas, notan que el sonido guardado no mantiene los mismos parámetros de sonido establecidos por ellos durante la sonorización. Esto se resolvió junto con otra actualización importante, la reproducción del punto donde la barra roja señala cuando cambia su posición (la última se recomendó varias veces durante las pruebas). En la versión actual al cambiar de posición se sonoriza el punto que se marcó con la barra de estado o en la posición indicada por teclado, lo que además permite avanzar o retroceder en los datos con las teclas de flechas escuchando cada posición en los datos.

Siguiendo múltiples consultas y pedidos, se diseñó y presentó un primer enfoque para el despliegue (sonorización y visualización) de datos con múltiples columnas en el mismo gráfico. Para ello se tuvo en cuenta que se pueden desplegar hasta cuatro o cinco sonorizaciones simultáneas pudiendo aún distinguirlas entre sí, para la sonorización se utilizan cambios de instrumento, idealmente de diferentes tipos (por ejemplo uno de viento, otro de cuerda y otro de percusión). Actualmente, sonoUno cuenta con la posibilidad de sonorizar tres columnas de forma simultanea con parámetros predefinidos que no pueden ser modificados por el usuario (ver sección \ref{sect:cap6_escritorio_sloan}, Sonorización de espectro de galaxia). Esta no es una tarea trivial y cuenta con muy poca documentación al respecto, para avanzar en este tema será necesario combinar mejoras en la producción de sonido, pruebas de percepción, entrenamientos y análisis con usuarios para incluir los elementos necesarios a la interfaz.

Otro tema que se consultó en numerosas ocasiones es la cantidad de datos o el tamaño del archivo a sonorizar. En este caso, se realizaron pruebas con archivos de dos columnas y dos millones de filas, obteniendose la sonorización sin problemas en la interfaz. El detalle surge al intentar guardar el archivo de sonido, donde se requiere mayor procesamiento para generar el archivo wav y se encontraron largas demoras. Otro punto a tener en cuenta es el tiempo que conllevará escuchar la sonorización completa, ya que hasta el momento se respeta el tiempo de sonorización de la interfaz.

Durante todo el proceso que permitió detectar errores y necesidades para el despliegue de datos, se utilizaron tanto datos patrones como reales, obtenidos de diferentes instrumentos. Todos los datos utilizados en estas pruebas han sido datos astronómicos obtenidos de dominios públicos (base de datos Sloan Digital Sky Survey (SDSS) y Observatorio Pierre Auger, por ejemplo) o colaboraciones internacionales (como Proyecto REINFORCE (GA 872859) e intercambio con usuarios que utilizaron curvas de luz). En este sentido, en la próxima sección se muestran ejemplos de uso con espectro de galaxia, rayos cósmicos y curvas de luz.

\subsection{Base de datos ``Sloan Digital Sky Survey''}
\label{sect:cap6_escritorio_sloan}

La base de datos SDSS cuenta con una extensa cantidad de imágenes, espectroscopia en el rango visible y espectroscopia en el infrarrojo, entre otros. Dichos datos han sido obtenidos principalmente por el telescopio propio del proyecto situado en el observatorio Apache Point, Nuevo México, Estados Unidos (\textit{``Sloan Foundation 2.5m Telescope''}). Ya en la etapa 4 (es la web utilizada para las pruebas durante esta tesis) y etapa 5 (publicada a finales de 2022) se han utilizado también el \textit{``Irénée du Pont Telescope''} ubicado en la región de Atacama en la cordillera de Los Andes, Chile, obteniendo así imágenes del hemisferio sur; y debido a que muchas estrellas son muy brillantes para poder observarse con los telescopios mencionados, se ha incluido el \textit{NMSU 1-meter telescope} ubicado en el observatorio Apache Point. Esta base de datos es operada por una gran cantidad de científicos colaboradores, de diferentes instituciones y ubicados en diferentes partes del mundo (se puede obtener información más precisa sobre la colaboración en su \href{https://www.sdss.org/collaboration/}{\underline{\textcolor{blue}{página web}}}).

En cuanto a los datos descargados para esta tesis, se utilizó una herramienta a la cual se accedía desde SDSS desde un link denominado \textit{``vista rápida''}, sirve para localizar diferentes galaxias y sus espectros (se puede acceder a dicha herramienta en el siguiente \href{https://skyserver.sdss.org/dr18/VisualTools/navi}{\underline{\textcolor{blue}{link}}}). Allí se puede \textit{navegar} una imagen y seleccionar con el puntero las diferentes galaxias o estrellas, pudiendo acceder a la \textit{vista rápida} del objeto seleccionado. 

\begin{figure}[ht!]
    \centering
    \begin{subfigure}[b]{1\textwidth}
         \centering
         \captionsetup{justification=centering}
         \frame{\includegraphics[width=1\textwidth]{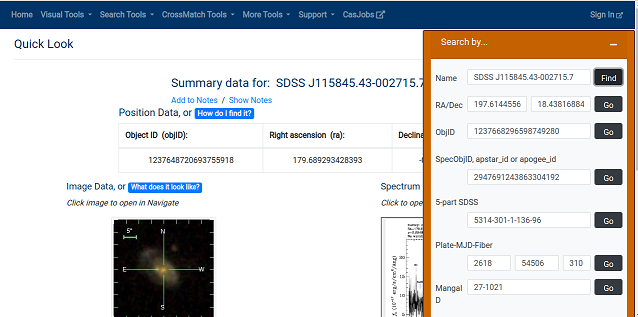}}
         \caption{Búsqueda de objeto.}
         \label{fig:cap6_basedatos_sdss_busqueda}
     \end{subfigure}
     \hfill
    \begin{subfigure}[b]{1\textwidth}
         \centering
         \captionsetup{justification=centering}
         \frame{\includegraphics[width=1\textwidth]{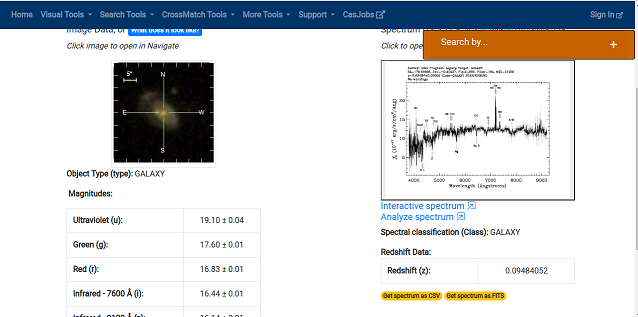}}
         \caption{Descarga de csv.}
         \label{fig:cap6_basedatos_sdss_download}
    \end{subfigure}
         
    \caption{Capturas de pantalla de la herramienta \textit{vista rápida} de la base de datos SDSS.}
    \label{fig:cap6_basedatos_sdss_quicklook}
\end{figure}

Los datos se acceden desde la herramienta \textit{SkyServer}, en la misma herramienta visual mencionada anteriormente (\href{https://skyserver.sdss.org/dr18/VisualTools/quickobj}{\underline{\textcolor{blue}{vista rápida}}}) se puede buscar el objeto si se conoce su nombre, coordenadas, ID del objeto, entre otros parámetros. Esto es útil porque en la mayoría de los casos se conoce los parámetros del objeto. Tomando como ejemplo la galaxia \textit{SDSS J115845.43-002715.7}, que se muestra en la \href{https://www.sonouno.org.ar/galaxy-sdss-j115845-43-002715-7/}{\underline{\textcolor{blue}{galería}}} de sonoUno con su correspondientes sonorizaciones, se pueden tomar su nombre o sus coordenadas ecuatoriales absolutas, ascensión recta y declinación (RA/Dec) que se proporcionan al inicio y buscarla en la base de datos (ver Figura \ref{fig:cap6_basedatos_sdss_busqueda}).

Si bien la web de SDSS ha sido mejorada y modificada a través de estos años, su accesibilidad con lectores de pantalla sigue siendo un tema a mejorar. Se logra evidenciar que los elementos de interfaz están mejor descriptos, sin embargo el discurso que se presenta con el lector de pantalla sigue careciendo de orden y sentido, las cosas se presentan en algunos casos de forma inversa y siguen habiendo elementos difíciles o imposibles de acceder con lectores de pantalla.

Continuando ahora con el archivo de datos, al descargar el espectro de galaxia, el mismo contiene cuatro columnas: la primera correspondiente a la longitud de onda, que se utilizara como coordenada de abscisas (en el caso de estos datos el rango de longitud de onda es el del espectro visible); la segunda corresponde al flujo, que representa la cantidad de luz que llega al instrumento en cada longitud de onda; la tercera es el mismo flujo pero se le ha aplicado algún filtro o suavizado para disminuir el ruido; y la cuarta es la intensidad de luz del cielo en cada longitud de onda, presenta información de fondo que se percibe en todas las observaciones e incluye el efecto atmosférico (tenga en cuenta que en el caso de SDSS son observaciones realizadas desde tierra).

\begin{figure}[ht!]
    \centering
    \begin{subfigure}[b]{0.49\textwidth}
         \centering
         \captionsetup{justification=centering}
         \frame{\includegraphics[width=0.6\textwidth]{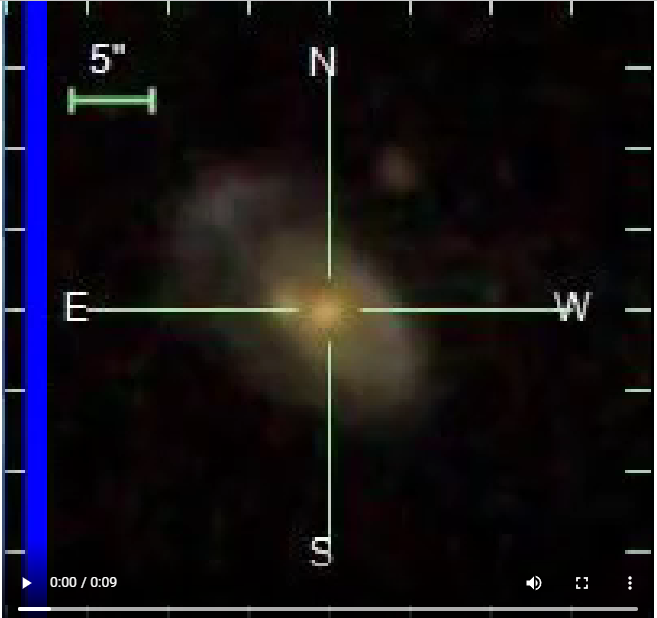}}
         \caption{\href{https://www.sonouno.org.ar/wp-content/uploads/sites/9/2022/05/J115845.43-002715.7_video.mp4}{\underline{\textcolor{blue}{Video}}}.}
         \label{fig:cap6_basedatos_sdss_galaxia1}
     \end{subfigure}
     \hfill
    \begin{subfigure}[b]{0.49\textwidth}
         \centering
         \captionsetup{justification=centering}
         \frame{\includegraphics[width=1\textwidth]{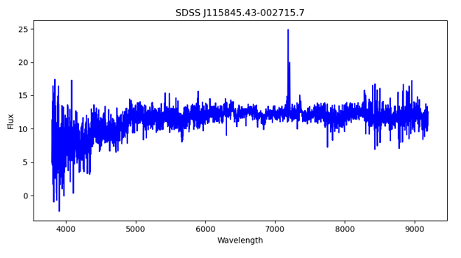}}
         \caption{Flujo (\href{https://www.sonouno.org.ar/wp-content/uploads/sites/9/2021/12/SDSS-J115845.43-002715.7_flux_fmax2003.wav}{\underline{\textcolor{blue}{sonido}}}).}
         \label{fig:cap6_basedatos_sdss_galaxia2}
    \end{subfigure}
    \hfill
    \begin{subfigure}[b]{0.5\textwidth}
         \centering
         \captionsetup{justification=centering}
         \frame{\includegraphics[width=1\textwidth]{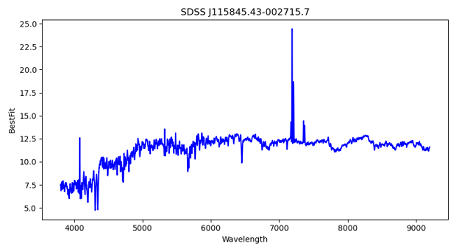}}
         \caption{Flujo con mejor ajuste (\href{https://www.sonouno.org.ar/wp-content/uploads/sites/9/2021/12/SDSS-J115845.43-002715.7_bestfit_fmax2003.wav}{\underline{\textcolor{blue}{sonido}}}).}
         \label{fig:cap6_basedatos_sdss_galaxia3}
     \end{subfigure}
     \hfill
    \begin{subfigure}[b]{0.49\textwidth}
         \centering
         \captionsetup{justification=centering}
         \frame{\includegraphics[width=1\textwidth]{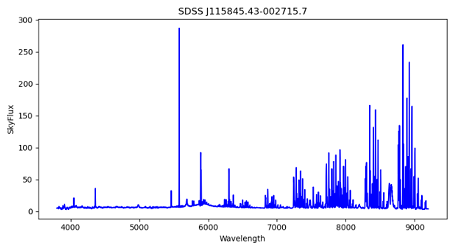}}
         \caption{Flujo de fondo (\href{https://www.sonouno.org.ar/wp-content/uploads/sites/9/2021/12/SDSS-J115845.43-002715.7_skyflux_fmax2003.wav}{\underline{\textcolor{blue}{sonido}}}).}
         \label{fig:cap6_basedatos_sdss_galaxia4}
    \end{subfigure}
         
    \caption{Imágenes obtenidas a partir de los datos del espectro de galaxia SDSS J115845.43-002715.7, en cada caso la variable independiente es la longitud de onda.}
    \label{fig:cap6_basedatos_sdss_galaxia}
\end{figure}

En la página web dedicada a este objeto con las sonorizaciones (link \href{https://www.sonouno.org.ar/galaxy-sdss-j115845-43-002715-7/}{\underline{\textcolor{blue}{aquí}}}), disponible en la web de sonoUno, se pueden observar los datos del objeto al inicio y luego el material audiovisual. Se presenta un video con la sonorización de la imagen de la galaxia, continuando luego con la imagen y audio de cada una de las sonorizaciones (ver Figura \ref{fig:cap6_basedatos_sdss_galaxia}).

\begin{figure}[!ht]
    \centering
    \includegraphics[width=1\textwidth]{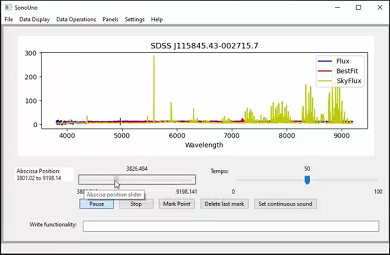}
    \caption{Captura de pantalla del video donde se muestra la sonorización de múltiples columnas, puede acceder al video en el siguiente \href{https://youtu.be/5TTdcgeDaPw}{\underline{\textcolor{blue}{link}}}.}
    \label{fig:cap6_basedatos_sdss_sonounoalldata}
\end{figure}

Dado que este tipo de datos contiene tres columnas que dependen todas de la misma variable independiente, pueden mostrarse todas en un mismo gráfico (ver Figura \ref{fig:cap6_basedatos_sdss_sonounoalldata}). El video al cual se puede acceder desde la descripción de la Figura \ref{fig:cap6_basedatos_sdss_sonounoalldata} muestra la primer aproximación a la sonorización de múltiples columnas. Hasta el momento se conoce que el cerebro humano puede distinguir entre dos y cuatro sonidos provenientes de diferentes fuentes al mismo tiempo, pudiendo discriminar entre ellos la información que contienen. Si bien por la práctica sabemos que discriminamos mayor cantidad de sonidos, se debe recordar que en este caso hablamos de cantidad de sonidos que se puedan analizar de forma simultanea. Si bien se logró realizar esta primer aproximación, no se ha presentado aún formalmente dentro de las funcionalidades de sonoUno debido a dos razones: (1) se debe realizar un nuevo diseño para incluir la configuración de parámetros para cada una de las gráficas, con sus consecuentes pruebas con usuarios; (2) se debe profundizar en los estudios de percepción y entrenamiento para entender mejor como el oído y cerebro humano interpreta este tipo de sonorización.

\subsection{Base de datos ``Pierre Auger - Open Data''}

Los rayos cósmicos son partículas con alta energía que viajan con gran velocidad, cuando estas partículas del espacio exterior impactan con la atmósfera, se produce lo que se conoce como lluvia atmosférica extendida de partículas secundarias. El observatorio Pierre Auger, ubicado en Malargüe, provincia de Mendoza, se dedica a medir las lluvias de partículas que se producen por el ingreso de estos rayos cósmicos (se puede obtener más información en su \href{https://visitantes.auger.org.ar/#}{\underline{\textcolor{blue}{página web}}}).

Si bien el lugar de trabajo donde se desarrolló esta tesis tiene relación directa con el observatorio Pierre Auger, por ser un desarrollo de acceso libre y para que todos los usuarios puedan utilizar el conjunto de datos, se han utilizado para las pruebas los datos de rayos cósmicos disponibles en la web \textit{``Pierre Auger Observatory Open Data''} (\href{https://labdpr.cab.cnea.gov.ar/opendata/}{\underline{\textcolor{blue}{link}}}). A la fecha, la última publicación de datos fue en febrero de 2021, sin embargo se utilizará aquí como ejemplo la liberación anterior con datos entre 2005-2017. 

Previo a mostrar los ejemplo, cabe destacar que el archivo csv \citep{augerData} contiene toda la información referente a cada detección realizada por los detectores, contiene en total 79 columnas (el detalle de lo que refiere a cada columna puede encontrarse en la web de los \href{https://labdpr.cab.cnea.gov.ar/opendata/data.php}{\underline{\textcolor{blue}{datos}}}). En el caso que se presenta aquí, se mostrarán los eventos ocurridos y su energía, por lo que se utilizarán como eje `x' el tiempo (\textit{gpstime}, columna 3) y como eje `y' la energía del evento (\textit{sd\_energy}, columna 11).

\begin{figure}[ht!]
    \centering
    \begin{subfigure}[b]{0.49\textwidth}
         \centering
         \captionsetup{justification=centering}
         \frame{\includegraphics[width=1\textwidth]{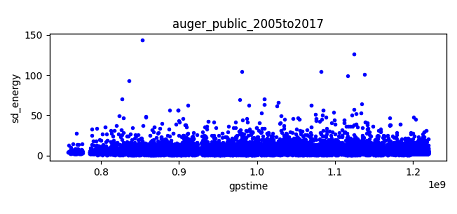}}
         \caption{Archivo completo (\href{https://www.sonouno.org.ar/wp-content/uploads/sites/9/2021/12/auger2017_sound_fulldata.wav}{\underline{\textcolor{blue}{sonido}}}).}
         \label{fig:cap6_basedatos_auger_data1}
     \end{subfigure}
     \hfill
    \begin{subfigure}[b]{0.49\textwidth}
         \centering
         \captionsetup{justification=centering}
         \frame{\includegraphics[width=1\textwidth]{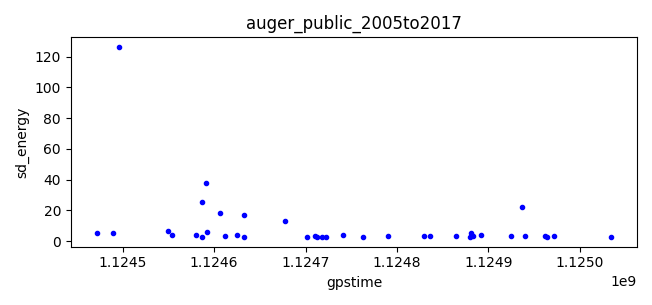}}
         \caption{Rango específico (\href{https://www.sonouno.org.ar/wp-content/uploads/sites/9/2021/12/auger2017_sound_1124471483-1125034238.wav}{\underline{\textcolor{blue}{sonido}}}).}
         \label{fig:cap6_basedatos_auger_data2}
    \end{subfigure}
         
    \caption{Capturas de pantalla de sonoUno con los datos de \textit{Pierre Auger Observatory Open Data} 2005-2017, se muestra la energía de cada evento con respecto al tiempo. En el caso de \ref{fig:cap6_basedatos_auger_data2} el rango del gráfico es: 1124471483-1125034238.}
    \label{fig:cap6_basedatos_auger_data}
\end{figure}

En el caso de los datos de rayos cósmicos se utilizó la configuración de gráfico de sonoUno que permite solo trazar los puntos sin línea que los una (ver Figura \ref{fig:cap6_basedatos_auger_data}), esto se debe a que son eventos que ocurren en un determinado tiempo. El ejemplo que se percibe en la Figura \ref{fig:cap6_basedatos_auger_data}, está disponible en la galería de sonoUno, en la sección de rayos cósmicos (se puede acceder a este ejemplo en el siguiente \href{https://www.sonouno.org.ar/comic-rays-2005-to-2017/}{\underline{\textcolor{blue}{link}}}). En dicha página web se muestran los datos completos con su sonorización al inicio, continuando con dos cortes diferentes y su correspondiente sonorización. Se incluye además, cerca de cada imagen, la opción para descargar el archivo de datos.

Es importante destacar la importancia para este conjunto de datos que tiene una de las actualizaciones que se hizo en sonoUno, mediante la sonorización se puede relacionar el tiempo que paso entre un evento y el otro debido al silencio que se percibe entre un dato y el otro. Se incluye aquí un \href{https://youtu.be/3GR4gcx92kU}{\underline{\textcolor{blue}{link}}} a un video donde se muestra la sonorización del mismo rango de datos presentado en la Figura \ref{fig:cap6_basedatos_auger_data2}, allí se puede observar como en los espacios sin datos avanza la línea roja sin reproducir sonido.

\subsection{Base de datos ``ASAS-SN''}

Las estrellas variables presentan la característica de que su brillo varía en el tiempo, lo que puede deberse a características propias de la estrella (intrínsecas, como lo son las variables pulsantes, eruptivas o cataclísmicas) o a características externas (extrínsecas, como lo son las eclipsantes). En el caso de este trabajo se utilizaron ejemplos de estrella variable intrínseca, del tipo pulsante, específicamente del tipo cefeida (la característica de este tipo de estrellas es que su período es proporcional a su luminosidad). Además, se sonorizó un ejemplo de estrella variable extrínseca, del tipo eclipsante, que son estrellas binarias cuyo plano de su órbita coincide con la dirección de observación, por lo que se observa una estrella pasar frente a la otra produciendo eclipses (el instrumento observa una disminución en la cantidad de luz recibida).

\begin{figure}[ht!]
    \centering
    \begin{subfigure}[b]{0.49\textwidth}
         \centering
         \captionsetup{justification=centering}
         \frame{\includegraphics[width=1\textwidth]{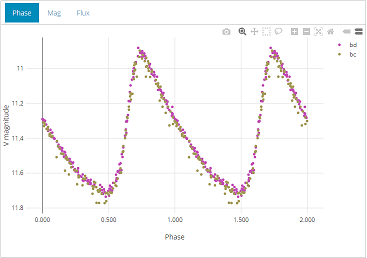}}
         \caption{Cefeida, \href{https://asas-sn.osu.edu/variables/753bdd73-38a7-5e43-b6c0-063292c7f28d}{\underline{\textcolor{blue}{base de datos}}}}
         \label{fig:cap6_basedatos_asas_cefeidaweb}
     \end{subfigure}
     \hfill
    \begin{subfigure}[b]{0.49\textwidth}
         \centering
         \captionsetup{justification=centering}
         \frame{\includegraphics[width=1\textwidth]{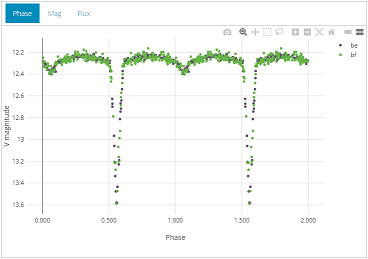}}
         \caption{Binaria eclipsante, \href{https://asas-sn.osu.edu/variables/dfa51488-c6b7-5a03-abd4-df3c28273250}{\underline{\textcolor{blue}{base de datos}}}}
         \label{fig:cap6_basedatos_asas_eclipsanteweb}
    \end{subfigure}
    \hfill
    \begin{subfigure}[b]{0.49\textwidth}
         \centering
         \captionsetup{justification=centering}
         \frame{\includegraphics[width=1\textwidth]{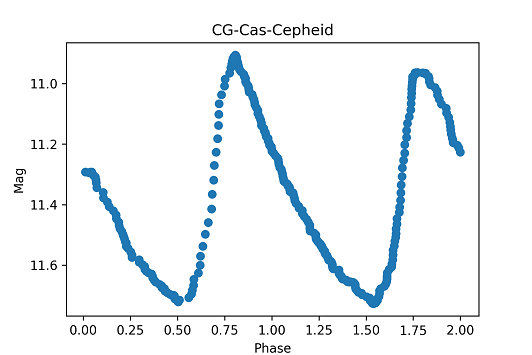}}
         \caption{Captura de sonoUno (\href{https://www.sonouno.org.ar/wp-content/uploads/sites/9/2023/01/CG-Cas-Cepheid.csv_sound.wav}{\underline{\textcolor{blue}{sonido}}}).}
         \label{fig:cap6_basedatos_asas_cefeidasonouno}
     \end{subfigure}
     \hfill
    \begin{subfigure}[b]{0.49\textwidth}
         \centering
         \captionsetup{justification=centering}
         \frame{\includegraphics[width=1\textwidth]{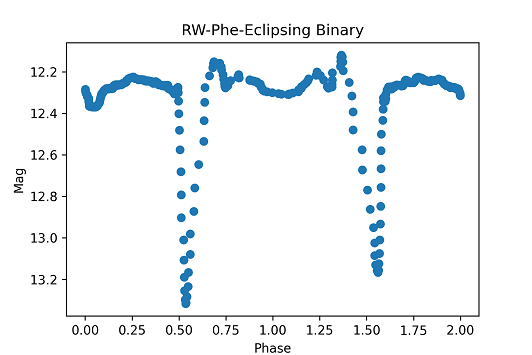}}
         \caption{Captura de sonoUno (\href{https://www.sonouno.org.ar/wp-content/uploads/sites/9/2023/01/RW-Phe-Eclipsing-Binary.csv_sound.wav}{\underline{\textcolor{blue}{sonido}}}).}
         \label{fig:cap6_basedatos_asas_eclipsantesonouno}
    \end{subfigure}
         
    \caption{Capturas de pantalla obtenida de la base de datos y de sonoUno con los datos descargados de la base de datos ASAS-SN. En el caso de las imágenes obtenidas de sonoUno se agrega el link a sonido en la descripción.}
    \label{fig:cap6_basedatos_asas}
\end{figure}

La base de datos \textit{``All-Sky Automated Survey for Supernovae''} (\href{https://asas-sn.osu.edu/variables}{\underline{\textcolor{blue}{ASAS-SN}}}) ofrece una base de datos de estrellas variables de acceso libre, donde se puede descargar un archivo formato tabla (extensión `csv') con los datos de observación particulares de la estrella seleccionada. Fue de esta base de datos de la cual se seleccionaron las siguientes estrellas variables para sonorizar con sonoUno:

\begin{itemize}
    \item Cefeida: ASASSN-V J000059.21+605732.5 / CG Cas (\href{https://asas-sn.osu.edu/variables/753bdd73-38a7-5e43-b6c0-063292c7f28d}{\underline{\textcolor{blue}{link}}} a la base de datos)
    \item Binaria eclipsante: ASASSN-V J003016.19-462759.5 / RW Phe (\href{https://asas-sn.osu.edu/variables/dfa51488-c6b7-5a03-abd4-df3c28273250}{\underline{\textcolor{blue}{link}}} a la base de datos)
\end{itemize}

Un detalle a tener en cuenta con los datos obtenidos de las estrellas variables mencionadas, es que no tienen directamente una columna con los valores de fase, debido a que existen diferentes formas de obtener esos valores de fase dependiendo de lo que se desea graficar. En el caso de este trabajo, se realizó el cálculo de fase teniendo en cuenta la época basada en la `Fecha Juliana Heliocéntrica' que indica la misma base de datos (\textit{`Epoch (HJD)'}) y siguiendo la Ecuación \ref{eq:cap6_asas_phase}.

\begin{equation}
    \phi = \frac{t-t0}{P}
    \label{eq:cap6_asas_phase}
\end{equation}

\begin{equation}
    \phi = \frac{t-2457412.70647}{4.3652815}
    \label{eq:cap6_asas_phasecefeida}
\end{equation}

\begin{equation}
    \phi = \frac{t-2458053.49761}{5.4134367}
    \label{eq:cap6_asas_phasebinaria}
\end{equation}

Como los parámetros de época y período son diferentes para cada una de las estrellas variables seleccionadas en la Ecuación \ref{eq:cap6_asas_phasecefeida} se puede observar el cálculo de fase para la estrella Cefeida y en la Ecuación \ref{eq:cap6_asas_phasebinaria} el correspondiente a la estrella Binaria Eclipsante. Finalmente, utilizando este cálculo de fase como variable independiente y la magnitud como variable dependiente se obtuvieron los gráficos presentados en las Figuras \ref{fig:cap6_basedatos_asas_cefeidasonouno} y \ref{fig:cap6_basedatos_asas_eclipsantesonouno}.

Si bien el cálculo mencionado puede realizarse de forma externa, para posteriormente ingresar la tabla a sonoUno y realizar la sonorización; en su lugar se realizó un script que está disponible en el repositorio de sonoUno con el nombre \textit{`sonify\_bash\_lightcurve.py'} (\href{https://github.com/sonoUnoTeam/sonoUno/blob/master/sonoUno/sonify_bash_lightcurve.py}{\underline{\textcolor{blue}{link}}}). Dicho código permite cargar en una variable el valor de época y período referente a los datos (actualmente de forma manual en el código), para luego indicar por bash el directorio donde se ubica el archivo a sonorizar. El script automáticamente guarda el sonido en esa misma carpeta con el mismo nombre de archivo y el agregado `\_sound'; opcionalmente, se le puede indicar al ejecutar el código que guarde el gráfico, en cuyo caso también lo guardará en la misma carpeta y con el mismo nombre adicionando `\_plot'. Los gráficos y sonidos que se presentan en las Figuras \ref{fig:cap6_basedatos_asas_cefeidasonouno} y \ref{fig:cap6_basedatos_asas_eclipsantesonouno} fueron obtenidos con el script que se describió aquí.

\section{Versión Web de sonoUno}
\label{sect:cap6_webcompleta}

La versión web de sonoUno ha sido desarrollada utilizando el marco de trabajo de la versión escritorio. Más allá de que la programación de la misma no fue realizada por la disertante, el marco de trabajo que se utilizó es el presentado en los Capítulos \ref{cap:analisis_normativo} y \ref{cap:fg_completo}, y en las pruebas y características de accesibilidad también se contribuyó. Esta nueva versión presenta las siguientes funcionalidades:

\begin{itemize}
    \item Selección de rangos específicos de datos en el eje `x';
    \item Marcar y guardar puntos de interés en los datos;
    \item Aplicar funciones matemáticas predefinidas (por ejemplo, logaritmo y cuadrado);
    \item Modificar las configuraciones de sonido;
    \item Modificar las configuraciones del gráfico.
\end{itemize}

Las diversas funcionalidades del software, así como el despliegue gráfico, se describen en esta sección. Sobre este trabajo y el desarrollo de sonoUno se han publicado tres proceedings en diferentes reuniones científicas que se adjuntan en el Apéndice \ref{ap:pub-webyreinforce} \citep{hcii2021,hcii2022,rmxac2022}.

\subsection{Herramientas utilizadas}
\label{sect:herramientas_web}

El entorno de desarrollo que se utiliza para la versión basada en la web es Angular. Esta elección se hizo porque promueve la modularidad y facilita el trabajo colaborativo. Al estar basada en la web, a esta versión de sonoUno se puede acceder desde la mayoría de las plataformas utilizando un buscador. 

En cuanto a la accesibilidad de sus elementos, se ha utilizado el protocolo ARIA para mejorar el flujo de trabajo de lectores de pantalla, como ya se mencionó en la sección \ref{sect:mails_mejoras} del Capítulo \ref{cap:fg_completo}, se profundizará en la sección \ref{sect:cap6_accesibilidadweb}.

En cuanto al código propiamente dicho, se ha utilizado HTML, JavaScript y CSS, debido a que son herramientas necesarias para el desarrollo de una interfaz web.

\subsection{Accesibilidad de la interfaz web}
\label{sect:cap6_accesibilidadweb}

En el caso de sonoUno web es el estándar ARIA el que proporciona los medios e impulsa la implementación de características de accesibilidad. Esta herramienta permite la inclusión de etiquetas diseñadas especialmente para lectores de pantalla, las cuales especifican el texto, la prioridad de lectura, y la capacidad de ``atrapar'' la navegación dentro de un contexto determinado cuando es necesario (un ejemplo de esto sería al emitir un mensaje de alerta).

La navegación para los usuarios con problemas de visión se lleva a cabo principalmente utilizando la tecla TAB o los cursores, y consiste en saltar de un elemento a otro, volviendo al inicio una vez que se ha pasado por todos los elementos legibles. Esto da como resultado que la presencia de muchos elementos resulte en un largo recorrido que además de tiempo conlleva una gran carga cognitiva para el usuario. Para abordar este problema, se agregaron paneles plegables que engloban a las funcionalidades por categorías, dejando solo visibles en la página de inicio el gráfico, los botones de reproducción y dichos paneles. Las categorías siguen el criterio de la interfaz de escritorio, siendo estas: operaciones matemáticas, configuraciones de sonido y configuraciones de gráfico.

Paralelamente, muchos usuarios prefieren utilizar la interfaz con atajos de teclado que permiten un acceso rápido a las funciones de uso común. A medida que el usuario se familiariza con este método de interacción, no se necesitan más ciclos de elementos, lo que mejora en gran medida la velocidad con la que se pueden realizar las tareas, sobre todo para personas con discapacidad visual.

El tema de atajos de teclado no es menor, se debe tener especial cuidado al configurarlos, ya que existen combinaciones estándar definidas y otras de uso común. Estas combinaciones preexistentes no deben reemplazarse (o sea, no debe utilizarse un atajo de teclado reservado para definir una nueva tarea), porque podría interrumpirse el uso normal del navegador web.

\subsection{Síntesis de sonido en la Interfaz Web}

Una de las principales diferencias con la versión desarrollada en Python y la implementación web, es el uso de JavaScript para implementar el back-end de síntesis de sonido. En el caso de este trabajo, se seleccionó la biblioteca Tone.js. Esta biblioteca permite la generación de tonos puros, de la misma manera que lo hace la versión Python de sonoUno, lo que garantiza que se puedan usar frecuencias precisas para representar valores variables, manteniendo un despliegue de sonido análogo en ambas interfaces.

Al usar Tone.js, la ejecución de la síntesis de tonos individuales se lleva a cabo de forma asíncrona, por lo que no resulta práctico avanzar con la sonorización mientras se espera que termine la producción del siguiente tono. En su lugar, el método adoptado fue programar todos los tonos con anticipación, indicando que los temporizadores se detengan o destruyan si el usuario selecciona comandos de `Pausa' o `Detener'. Esta porción de código también se encarga de actualizar la posición del marcador vertical y el control deslizante de abscisas.

\subsection{Despliegue de la interfaz web}

\begin{figure}[ht!]
        \centering
        \begin{subfigure}[b]{1\textwidth}
             \centering
             \captionsetup{justification=centering}
             \frame{\includegraphics[width=1\textwidth]{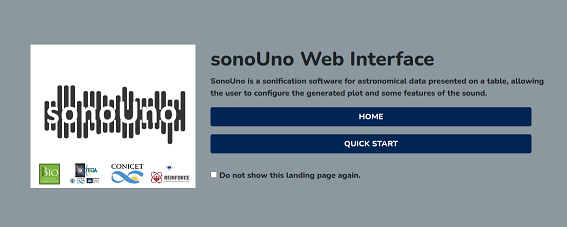}}
             \caption{Página inicial.}
             \label{fig:cap6_paginicial_web}
         \end{subfigure}
         \hfill
        \begin{subfigure}[b]{0.54\textwidth}
             \centering
             \captionsetup{justification=centering}
             \frame{\includegraphics[width=1\textwidth]{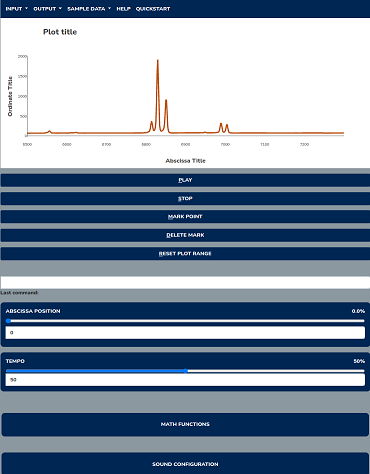}}
             \caption{Interfaz web.}
             \label{fig:cap6_guiweb}
         \end{subfigure}
         \hfill
         \begin{subfigure}[b]{0.45\textwidth}
             \centering
             \captionsetup{justification=centering}
             \frame{\includegraphics[width=0.98\textwidth]{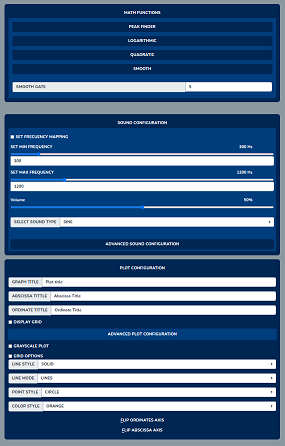}}
             \caption{Paneles desplegados.}
             \label{fig:cap6_paneles}
         \end{subfigure}
         
        \caption{Interfaz web con su página de inicio, el despliegue de funcionalidades que se aprecia al ingresar y luego los paneles de operaciones matemáticas y configuraciones desplegados.}
        \label{fig:cap6_guiweb_3view}
    \end{figure}

\begin{figure}[ht!]
        \centering
        \begin{subfigure}[b]{0.54\textwidth}
             \centering
             \captionsetup{justification=centering}
             \frame{\includegraphics[width=1\textwidth]{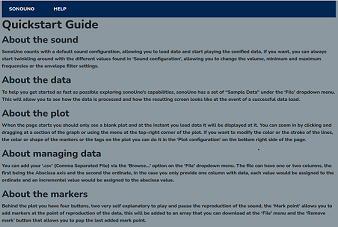}}
             \caption{Guía de inicio rápido.}
             \label{fig:cap6_quickstart_web}
         \end{subfigure}
         \hfill
        \begin{subfigure}[b]{0.45\textwidth}
             \centering
             \captionsetup{justification=centering}
             \frame{\includegraphics[width=0.97\textwidth]{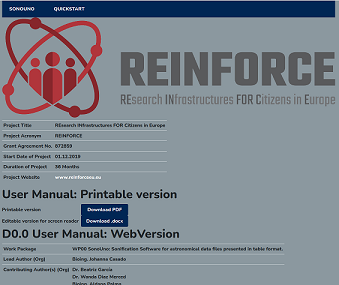}}
             \caption{Manual de sonoUno web.}
             \label{fig:cap6_help_web}
         \end{subfigure}
         
        \caption{La versión web de sonoUno ofrece una guía rápida, además del manual de usuario disponible en el botón Ayuda.}
        \label{fig:cap6_quickstart_y_ayuda}
    \end{figure}

La interfaz web de sonoUno recibe al usuario con un mensaje inicial donde se describe brevemente la herramienta y se muestran los logos correspondientes (Figura \ref{fig:cap6_paginicial_web}). En conjunto, esta página de inicio tiene dos botones que permiten ingresar directamente al programa (`HOME'), o acceder a una guía rápida que muestra brevemente una explicación de las funcionalidades del programa (`QUICK START') (Figura \ref{fig:cap6_quickstart_web}).

\begin{figure}[ht!]
        \centering
        \begin{subfigure}[b]{0.49\textwidth}
             \centering
             \captionsetup{justification=centering}
             \frame{\includegraphics[width=0.7\textwidth]{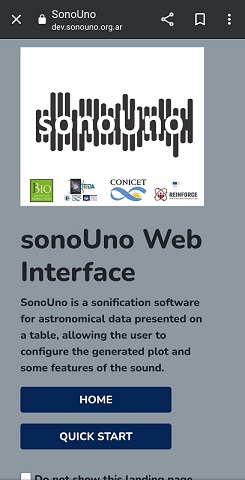}}
             \caption{Página inicial.}
             \label{fig:cap6_paginicial_celu}
         \end{subfigure}
         \hfill
        \begin{subfigure}[b]{0.49\textwidth}
             \centering
             \captionsetup{justification=centering}
             \frame{\includegraphics[width=0.7\textwidth]{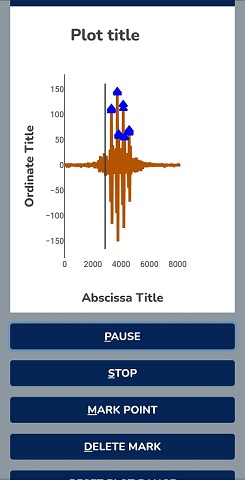}}
             \caption{Interfaz web.}
             \label{fig:cap6_guicelu}
         \end{subfigure}
         
        \caption{Interfaz web abierta desde un teléfono celular, se muestra la página de inicio y el despliegue de funcionalidades que se aprecia al ingresar a HOME.}
        \label{fig:cap6_guicelu_2view}
    \end{figure}

En la interfaz del programa (Figura \ref{fig:cap6_guiweb}) se pueden observar las mismas funcionalidades, respetando un diseño similar al diseño de la aplicación de escritorio. Cabe destacar que se utiliza el mismo marco de trabajo que en la versión de escritorio, solo que adaptado en este caso a un despliegue web, que además se ajusta a la web en el celular o dispositivos móviles (como tabletas) (Figura \ref{fig:cap6_guicelu_2view}). La página muestra cuatro secciones diferenciadas:

\begin{enumerate}
    \item \textit{La barra de navegación}, que brinda acceso a las funcionalidades de entrada/salida, datos de muestra y acceso a la guía rápida y el manual. El diseño de este elemento se basó en un estudio de caso de usuario realizado en la Universidad de Southampton con personas videntes y ciegas, utilizando la versión de escritorio de sonoUno. Los resultados destacaron la necesidad de una interfaz simple que solo muestre las funcionalidades principales en el marco principal.
    \item \textit{El área de despliegue}, que incluye el gráfico y los botones de control: Play/Pausa, Detener, Marcar punto, Eliminar marca y Reiniciar el gráfico. 
    \item \textit{El área de herramientas} que proporciona lo siguiente:
    \begin{itemize}
        \item Un cuadro de texto, que permite ejecutar los elementos de interfaz (como botones) ingresando el comando en formato texto. Esta mejora, también presente en la interfaz de escritorio permite menor carga cognitiva para el usuario, teniendo que realizar menor número de acciones para realizar una tarea.
        \item Dos controles deslizantes, con los que se puede controlar la posición de abscisas y el tempo;
        \item Una sección de funciones matemáticas, que permite la selección de funciones matemáticas simples para aplicar a los datos, como una función cuadrática o una función que marca automáticamente los picos en el gráfico;
    \end{itemize}
    \item \textit{El área de configuraciones}, donde incluimos tanto las configuraciones de sonido, como las de gráfico, más allá de que en la interfaz parezcan dos secciones separadas. Espacialmente se colocaron separadas para que fueran fácilmente identificadas tanto en el despliegue visual como con el lector de pantalla.
    \begin{itemize}
        \item La sección de configuración de sonido contiene herramientas con las que modificar la forma en que se procesa el audio;
        \item La sección de configuración del gráfico permite al usuario modificar la forma en que se muestran los datos en el gráfico.
    \end{itemize}
\end{enumerate}

El despliegue gráfico cuenta con tres paneles que pueden desplegarse o colapsarse, con el objetivo de no saturar la ventana principal con funcionalidades. En la Figura \ref{fig:cap6_paneles} se muestran los tres paneles desplegados con sus funcionalidades, estos son: (1) funciones matemáticas (correspondiente al área de herramientas); (2) configuraciones de sonido (correspondiente al área de configuraciones); y (3) configuraciones de gráfico (correspondiente al área de configuraciones).

En cuanto a las opciones de ayuda en esta versión, se diseñó una página de ayuda rápida donde se describe en general lo que puede hacer la herramienta (Figura \ref{fig:cap6_quickstart_web}). Adicionalmente, en el botón `HELP' se encuentra el manual de la versión web desplegado como texto en una página web (Figura \ref{fig:cap6_help_web}), se puede encontrar en la parte inicial de esa nueva página dos botones que permiten descargar el manual en formato `pdf' y archivo de texto.

Algo que es importante destacar es que la versión web de sonoUno se puede utilizar en cualquier dispositivo móvil (Figura \ref{fig:cap6_guicelu_2view}). Esto se logró cambiando el diseño del sitio web, haciéndolo extensible a todos los tamaños de pantalla, incluidos dispositivos móviles o de escritorio. Esto da como resultado una experiencia consistente mientras se mantiene la interfaz accesible y el esfuerzo de actualizaciones se realiza sobre un mismo código fuente.

\subsubsection{Uso de datos de ejemplo e ingreso de datos propios}

\begin{figure}[ht!]
        \centering
        \begin{subfigure}[b]{1\textwidth}
             \centering
             \captionsetup{justification=centering}
             \frame{\includegraphics[width=0.7\textwidth]{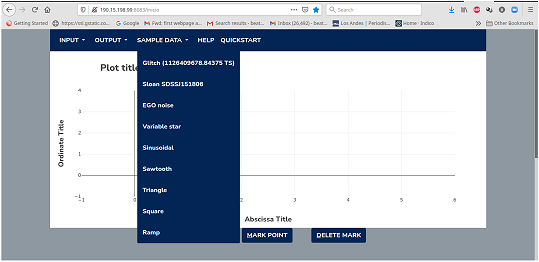}}
             \caption{Datos de ejemplo.}
             \label{fig:cap6_web_sampledata}
         \end{subfigure}
         \hfill
        \begin{subfigure}[b]{1\textwidth}
             \centering
             \captionsetup{justification=centering}
             \frame{\includegraphics[width=0.7\textwidth]{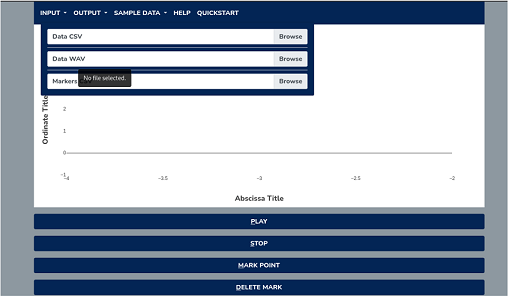}}
             \caption{Abrir datos propios.}
             \label{fig:cap6_web_inputdata}
         \end{subfigure}
         \hfill
        \begin{subfigure}[b]{1\textwidth}
             \centering
             \captionsetup{justification=centering}
             \frame{\includegraphics[width=0.7\textwidth]{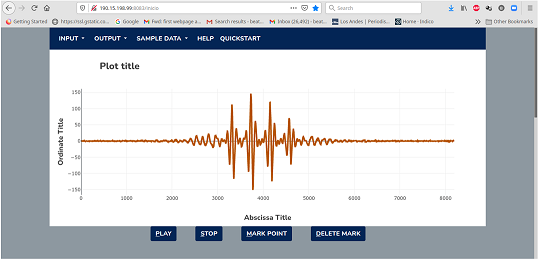}}
             \caption{Datos de Glitch seleccionado y desplegado.}
             \label{fig:cap6_web_gwdata}
         \end{subfigure}
         
        \caption{Métodos de entrada para ingresar datos en sonoUno web.}
        \label{fig:cap6_web_openwebsystem}
    \end{figure}

Los usuarios pueden graficar y sonorizar datos rápidamente con las funcionalidades proporcionadas a través de la página de inicio, en la opción `Datos de muestra' o `Entrada' en la barra de navegación (Figura \ref{fig:cap6_web_openwebsystem}). En el caso de `Datos de muestra', al seleccionar cualquiera de las opciones de este submenú (Figura \ref{fig:cap6_web_sampledata}), se carga automáticamente el conjunto de datos seleccionado en el área de gráfico y se encuentra disponible la sonorización para ser iniciada por el usuario (Figura \ref{fig:cap6_web_gwdata}).

La funcionalidad de uso de datos personalizados dentro de la aplicación se ha ampliado, comparado con la función de escritorio, para permitir que las personas carguen archivos CSV y WAV de gran tamaño. Los usuarios ahora también pueden cargar archivos de marcas hechas en los datos, guardados previamente a través de esta interfaz.

Al presionar el botón Reproducir, se muestra en el gráfico una línea vertical que se mueve de izquierda a derecha a lo largo del área de trazado, indicando el progreso a lo largo del eje x. Al mismo tiempo, se reproduce la sonorización, con variaciones de tono correspondientes con la amplitud de la gráfica, que coinciden con el punto de datos en la posición en el eje x. Es posible saltar a diferentes puntos de los datos moviendo la barra vertical de progreso hacia adelante y hacia atrás, con la barra deslizable `Posición de abscisa'. 

\subsubsection{Entrada de texto de valores junto con controles deslizantes}

Durante el desarrollo web, y teniendo en cuenta los pedidos de usuarios en ambas versiones, se realizaron mejoras en relación con la accesibilidad de la interfaz web. El usuario aún puede seleccionar frecuencias de sonido entre 300 y 1200 Hz mediante la barra deslizante, pero ahora también se puede hacer mediante la entrada directa de texto, en la parte superior derecha de cada barra deslizable (Figura  \ref{fig:cap6_paneles}). Con esta mejora, el usuario puede elegir frecuencias precisas de forma sencilla sin tener que perder tiempo ajustando los dos controles deslizantes. Dicha mejora está disponible también para el tempo y la posición de abscisas (Figura \ref{fig:cap6_guiweb}).

\subsubsection{Opción de línea de comandos}

De forma análoga a la opción de escritorio, se ha agregado la funcionalidad de ingreso de texto para poder realizar la acción deseada escribiendo el comando indicado. En lugar de tener que seleccionar una función a través de una lista desplegable o con movimientos precisos del puntero (ciertas discapacidades motoras no pueden utilizar un puntero para navegar la interfaz), los usuarios ahora pueden escribir el nombre de la función directamente (Figura \ref{fig:cap6_guiweb} debajo de `Reiniciar rango de gráfico').

\subsubsection{Inversión de ejes}

Durante una de las pruebas de software, realizadas de forma previa al uso de la versión web en Canarias, España (ver sección \ref{sect:uso_canarias}), uno de los usuarios resaltó la necesidad de invertir el eje de ordenadas en el gráfico, invirtiendo a su vez la frecuencia de sonorización. Esto es especialmente necesario para la representación de curvas de luz, ya que el gráfico se forma al representar la magnitud con respecto a la longitud de onda, pero esta magnitud en los valores más bajos corresponde a mucho brillo (lo que debería estar representado en la zona alta del gráfico y sonorizarse con frecuencias tendientes a sonidos agudos (siendo el máximo valor de frecuencia 3000Hz, y el valor máximo por defecto 1200Hz). Por el contrario, los valores de magnitud más altos, corresponden a poco brillo lo que debería representarse en la parte baja del gráfico y con frecuencias más graves (siendo el mínimo valor 0Hz, y el valor mínimo por defecto 300Hz).

\begin{figure}[p]
        \centering
        \begin{subfigure}[b]{1\textwidth}
             \centering
             \captionsetup{justification=centering}
             \frame{\includegraphics[width=1\textwidth]{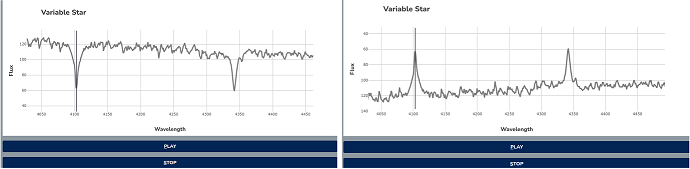}}
             \caption{Comparación de gráficos antes y después de invertir ejes.}
             \label{fig:cap6_web_flipaxisplot}
         \end{subfigure}
         \hfill
        \begin{subfigure}[b]{1\textwidth}
             \centering
             \captionsetup{justification=centering}
             \frame{\includegraphics[width=1\textwidth]{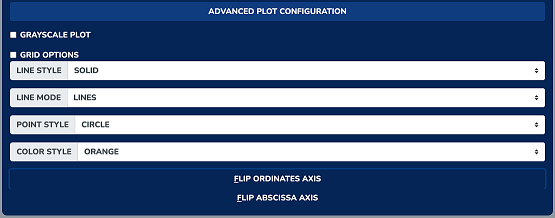}}
             \caption{Botones para invertir ejes.}
             \label{fig:cap6_web_flipaxisbuttons}
         \end{subfigure}
         
        \caption{Inversión de ejes en el sonoUno web, los botones se encuentran dentro de configuraciones avanzadas de gráfico.}
        \label{fig:cap6_web_flipaxis}
    \end{figure}

Invertir los ejes de coordenadas, para poder llevar a cabo la tarea mencionada es algo que se puede hacer ahora en las versiones de sonoUno (ver Figura \ref{fig:cap6_web_flipaxis}). Dichas inversiones de ejes ajustan de forma automática las frecuencias para que la sonorización siga representando el gráfico desplegado.

\subsubsection{Buscador de picos en la señal}

Esta función le permite marcar los picos en el gráfico automáticamente. Para los efectos de esta función, un pico se define por el valor que es mayor que los valores anterior y siguiente, además de las comprobaciones con la condición de estar por encima del promedio del gráfico, más un valor calculado por el diferencial entre el más alto y el más bajo, multiplicado por el porcentaje de sensibilidad.

\begin{figure}[p]
    \centering
    \includegraphics[width=0.5\textwidth]{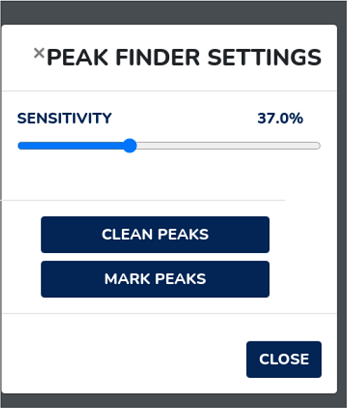}
    \caption{Cuadro de diálogo con la configuración del buscador de picos.}
    \label{fig:cap6_web_peakfinder}
\end{figure}

La ventana de configuración (Figura \ref{fig:cap6_web_peakfinder}) cuenta con: un control deslizante que permite configurar la sensibilidad, donde se puede seleccionar el valor que mejor funciona para los datos actuales; un botón para borrar los picos marcados; y un botón para agregarlos como marcadores en el gráfico.

Para cerrar la ventana modal, puede hacer clic en el botón ``Cerrar'' en la parte inferior izquierda, en la `X' junto al título, o bien, puede presionar la tecla Escape en su teclado.

\subsubsection{Ingreso de sonido como dato}

\begin{figure}[ht!]
    \centering
    \includegraphics[width=0.5\textwidth]{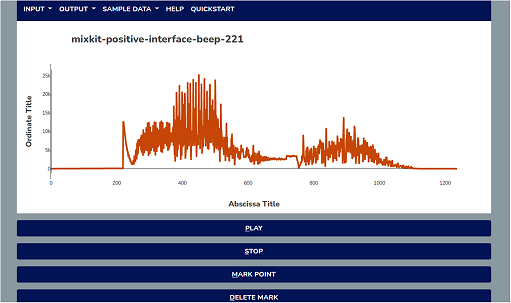}
    \caption{Ejemplo de un archivo de audio abierto en la interfaz web de sonoUno.}
    \label{fig:cap6_web_soundIN}
\end{figure}

La versión de sonoUno web permite el despliegue y análisis de un archivo de sonido (.WAV) (Figura \ref{fig:cap6_web_soundIN}). En este caso, la visualización en el área de gráfico es la envolvente de la intensidad del sonido, calculada como la media cuadrática.

\section{Hacia una plataforma de sonorización unificada}
\label{sect:cap6_server}

Como parte de la versión 2 del entregable de sonoUno (D7.2) en el marco del proyecto REINFORCE, se avanzó hacia el logro de dos objetivos:

\begin{itemize}
    \item Hacer que el software sea más versátil en el manejo de sonorizaciones, o sea ir más allá de una sola serie de tiempo (descripto en la sección \ref{sect:cap6_sonif_reinforce});
    \item Proporcionar una sonorización como servicio, utilizando un servidor de plataforma.
\end{itemize}

Esto agregará complejidad a la tarea, pero permitirá a largo plazo unificar las dos versiones existentes (escritorio y web), minimizando los esfuerzos de actualizaciones, ya que no será necesario actualizar la herramienta en dos lenguajes de programación diferente.

Dentro de WP7, se investigó la novedosa idea de incrustar en el navegador una aplicación Web Server Gateway Interface (WSGI) que actúa como el servidor sonoUno usando WebAssembly. Este enfoque habría maximizado la reutilización de código entre el servidor real de sonoUno, utilizado por los investigadores para cálculos intensivos o para trabajos por lotes, y los servidores locales de las computadoras de los científicos ciudadanos, usuarios de Zooniverse. Habría tenido la ventaja de preservar los recursos de los centros informáticos institucionales. Se hubiese podido construir un prototipo utilizando el micro-marco Flask, en el que las solicitudes podrían manejarse localmente. Sin embargo, la tecnología que permitía ejecutar Python dentro de un navegador (Pyodide) no era lo suficientemente estable en todos los navegadores populares como para justificar ir más allá de la fase de creación de prototipos, incluso cuando el enfoque pudo parecer muy prometedor en un futuro no muy lejano. Es por eso, que en su lugar, se entregaron dos piezas de software para satisfacer la necesidad encontrada:

\begin{itemize}
    \item La biblioteca de Python sonoUno, que es lo suficientemente versátil como para ejecutarse en muchos entornos diferentes, incluido un navegador, beneficiándose de la experiencia que obtuvimos con Pyodide;
    \item El servidor sonoUno, que deriva de una arquitectura clásica cliente-servidor.
\end{itemize}

\subsection{La librería sonoUno}

Se encuentra disponible en el índice de paquetes de Python (\href{https://pypi.org/project/sonounolib/}{\underline{\textcolor{blue}{link}}}), está destinada a ser utilizada como el primer bloque de construcción para sonorizar los datos de los experimentos científicos. Las guías de usuario y de referencia del proyecto están disponibles en el siguiente \href{https://pchanial.gitlab.io/sonouno-library/}{\underline{\textcolor{blue}{link}}}.
Esta librería proporciona herramientas genéricas para importar archivos de audio, transformar las ondas sonoras, exportarlas y reproducirlas. Pretende ser la única fuente de herramientas de sonorización para todos los labores de sonoUno. Permite la unificación de todos ellos, al menos en lo que respecta a la parte algorítmica del proceso de sonorización.

La clase más importante de esta librería es la clase `Track'. Almacena información intrínseca a las ondas sonoras:

\begin{itemize}
    \item La frecuencia de muestreo (por defecto: 44100 Hz);
    \item La amplitud máxima (por defecto: 1);
    \item La onda de sonido como una matriz numpy float64.
\end{itemize}

También resguarda un marcador de tiempo \textit{cue\_write} que indica cuándo ocurrirá la próxima escritura en la pista de audio. Las pistas administran su propio buffer de datos, que cambia de tamaño automáticamente a medida que se le agregan más sonidos.

La librería actualmente cuenta con tres \textit{back-ends} para reproducir sonidos:

\begin{itemize}
    \item En navegadores, a través de los proyectos PyScript y Pyodide, que permiten ejecutar Python de forma nativa sin transpilación, a través de WebAssembly. Se agregó la funcionalidad para inyectar archivos WAVE como BLOB (objetos binarios grandes) en la etiqueta de audio HTML5. Para experimentar con la ejecución de Python (pero también muchas librerías compiladas en C, como Numpy) en un navegador y reproducir sonidos, hay una demostración disponible en la Figura \ref{fig:cap6_server_soundej1}.

\begin{figure}[!ht]
    \centering
    \includegraphics[width=1\textwidth]{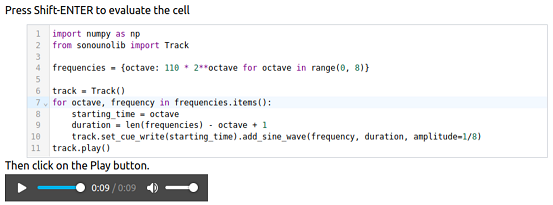}
    \caption{Muestra el uso de la librería sonoUno en un navegador, puede correrse con el siguiente \href{https://pchanial.gitlab.io/sonouno-library/demo_pyscript.html}{\underline{\textcolor{blue}{link}}}}
    \label{fig:cap6_server_soundej1}
\end{figure}
    
    \item En Jupyter Notebook, se utilizan instancias de audio de IPython como widgets por parte de Jupyter. En Linux, este entorno se puede probar con un Jupyter Notebook dockerizado:

\begin{lstlisting}
    $ docker run --network host --rm pchanial/sonounolib:0.4.0
\end{lstlisting}

    Luego se muestran las instrucciones y se puede acceder al servidor de laboratorio de Jupyter en un navegador copiando y pegando una URL con el formato:
    
\begin{lstlisting}
	http://127.0.0.1:8888/lab?token=bc03475af361693e02 cfff472ae54cb879be49b2e6d500c6
\end{lstlisting}
    
    Para ejecutar la demostración, seleccione la notebook \textit{demo.ipynb} en el panel izquierdo, debe aparecer una ventana similar a la que muestra la Figura \ref{fig:cap6_server_soundej2}. 

\begin{figure}[!ht]
    \centering
    \includegraphics[width=1\textwidth]{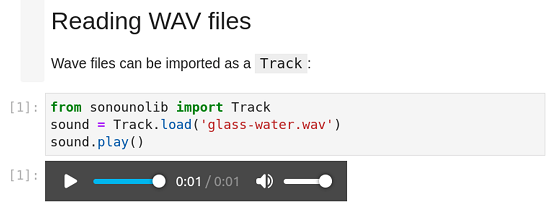}
    \caption{Muestra el uso de la librería sonoUno siguiendo los pasos de Jupyter Notebook}
    \label{fig:cap6_server_soundej2}
\end{figure}
    
    A diferencia del caso anterior, el código en Python no es ejecutado por el navegador, sino por un kernel Jupyter externo, que ejecuta un intérprete CPython.
    
    \item En todos los demás casos, la API multiplataforma \textit{PortAudio} se utiliza para reproducir sonidos. Por ejemplo, en un sistema operativo \textit{Debian} o \textit{Ubuntu}, se debe instalar el paquete `libportaudio2'.
\end{itemize}

\subsubsection{\textbf{Ejemplo 1:} Reproducción de una onda sinusoidal}

Para agregar una onda sinusoidal a la pista, se requiere especificar:

\begin{itemize}
    \item la frecuencia de oscilación, en Hertz;
    \item la duración, en segundos;
    \item y la amplitud, relativa a la amplitud máxima de la pista.
\end{itemize}

\begin{lstlisting}[language=Python]

    from sonounolib import Track
    track = Track().add_sine_wave(
        440, duration=2, amplitude=1/4)
    track.play()
\end{lstlisting}

\subsubsection{\textbf{Ejemplo 2}: Reproducción de una superposición de ondas sinusoidales}

Para superponer otras ondas sinusoidales generadas a la pista, uno tiene que rebobinar la escritura de entrada al momento en que comienzan las nuevas ondas sinusoidales:

\begin{lstlisting}[language=Python]

    import numpy as np
    from sonounolib import Track
    frequencies = {octave: 110 * 2**octave for octave in range(0, 8)}
    track = Track()
    for octave, frequency in frequencies.items():
    	starting_time = octave
    	duration = len(frequencies) - octave + 1	track.set_cue_write(starting_time).add_sine_wave(frequency, duration)
    track.play()
\end{lstlisting}

\subsubsection{\textbf{Ejemplo 3:} reproduciendo notas}

La notación científica de tono se puede utilizar para tocar notas. Por ejemplo: las notas C4, F\#4 y G$\flat$4 pueden ser referenciadas por las cuerdas 'C4', 'F\#4' y 'Gb4'.

\subsection{Motivación para el servidor sonoUno: Integración con Zooniverse}

\begin{figure}[!ht]
    \centering
    \includegraphics[width=1\textwidth]{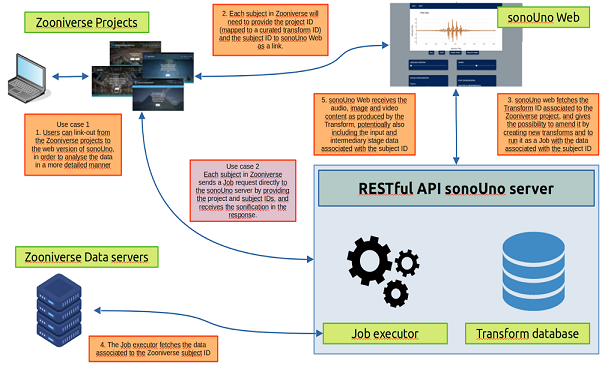}
    \caption{Arquitectura para un servidor RESTful API que permite la conexión de los demostradores de REINFORCE en Zooniverse a sonoUno.}
    \label{fig:cap6_server_concept}
\end{figure}

Además del trabajo llevado a cabo para crear una librería que pueda usarse en muchos contextos diferentes, en relación con sonoUno, también se ha estudiado una forma de integrar completamente sonoUno en los demostradores de Zooniverse. Se ha implementado la arquitectura que se diseñó siguiendo un análisis de requisitos que se presenta en la Figura \ref{fig:cap6_server_concept}.

A través de esta arquitectura, es posible vincular conjuntos de datos de eventos individuales en Zooniverse directamente a sonoUno y viceversa. Ha sido necesario montar un servidor no previsto inicialmente en el proyecto, para atender las peticiones de la interfaz de programación de aplicaciones (API), así como la propia definición de la API, pero ha aportado un valor añadido importante. Con su implementación, los usuarios pueden escuchar la sonorización de un conjunto de datos en Zooniverse, cabe destacar que la sonorización ha sido generada previamente y almacenada en la base de datos del servidor. Se pretendía seguir trabajando para proporcionar a los usuarios de Zooniverse un enlace a la web de sonoUno para modificar el tono, el timbre y el tempo, entre otros, haciéndola más accesible.

Lamentablemente, no se logró establecer la comunicación con Zooniverse para implementar este enfoque. Sin embargo, se plantea continuar investigando la posibilidad de utilizar este desarrollo para la unificación de interfaces gráficas de sonoUno y para ofrecer la sonorización utilizada en sonoUno como servicio para otras plataformas que lo deseen.

\subsection{Descripción del servidor de sonoUno}

El proyecto está alojado en el siguiente \href{https://doi.org/10.5281/zenodo.7717567}{\underline{\textcolor{blue}{link}}} y la documentación inicial se encuentra en el README de \href{https://github.com/sonoUnoTeam/sonoUno-server}{\underline{\textcolor{blue}{GitHub}}}. Las API del servidor sonoUno se describen utilizando OpenAPI v3 en el siguiente \href{http://api.sonouno.org.ar/redoc}{\underline{\textcolor{blue}{link}}}. 

Para comenzar, se ha puesto a disposición un Jupyter Notebook dockerizado `demo\_client.ipynb'. Utiliza las API del servidor de producción para iniciar sesión como usuario de prueba, crear una transformación, ejecutar un trabajo y reproducir el archivo de audio resultante.

\begin{lstlisting}
    $ docker run --network host --rm pchanial/sonouno-server-demo
\end{lstlisting}

El stack tecnológico es el siguiente:

\begin{itemize}
    \item Docker componse para la organización de contenedores;
    \item FastAPI para la aplicación web Python 3.10;
    \item MongoDB para la base de datos de usuarios, transformaciones y trabajos;
    \item MinIO, un almacén de objetos compatible con S3, para mostrar las salidas de sonorización;
    \item Nginx para mostrar la aplicación;
    \item Tokens JWT para la gestión de identidades y accesos.
\end{itemize}

\begin{figure}[ht!]
    \centering
        \begin{subfigure}[b]{0.8\textwidth}
             \centering
             \captionsetup{justification=centering}
             \frame{\includegraphics[width=1\textwidth]{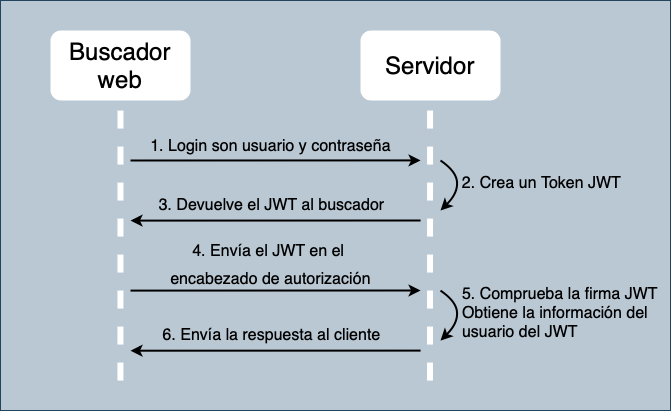}}
             \caption{Buscador web}
             \label{fig:cap6_server_webworkflow}
         \end{subfigure}
         \hfill
        \begin{subfigure}[b]{0.8\textwidth}
             \centering
             \captionsetup{justification=centering}
             \frame{\includegraphics[width=1\textwidth]{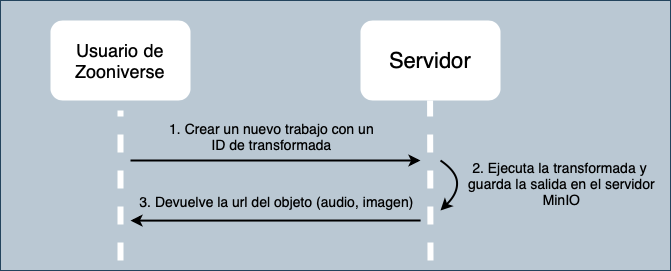}}
             \caption{Zooniverse}
             \label{fig:cap6_server_zooniverseworkflow}
         \end{subfigure}
         \hfill
        \begin{subfigure}[b]{0.8\textwidth}
             \centering
             \captionsetup{justification=centering}
             \frame{\includegraphics[width=1\textwidth]{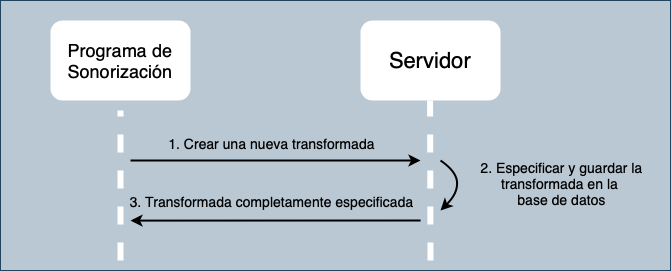}}
             \caption{Programa de sonorización}
             \label{fig:cap6_server_desktopworkflow}
         \end{subfigure}
    \caption{Flujo de trabajo entre el servidor y una API.}
    \label{fig:cap6_server_workflow}
\end{figure}

El flujo de trabajo habitual comienza con la creación y selección de una transformación de sonorización (ver Figura \ref{fig:cap6_server_workflow}). No se ha impuesto un modelado del flujo de trabajo de sonorización como un DAG (gráfico acíclico dirigido), en el que las etapas están conectadas entre sí. En su lugar, se ha adoptado el enfoque sin DAG, que ofrece más flexibilidad en términos de declaraciones que puede ejecutar una \textit{pipeline}, como bucles, condiciones if y bucles de retroalimentación. Al no imponer la construcción de un DAG, se mantiene la posibilidad de interoperar con los administradores de flujo de trabajo populares, como Flyte o Prefect. Este enfoque está más cerca de la forma en que los científicos diseñan sus procesos y análisis. 

La sonorización consiste simplemente en la ejecución de funciones, que pueden llamar a otras funciones. Este enfoque es muy conveniente para los creadores de \textit{pipeline}, pero en términos de ejecución automática y formateo de los datos, la dificultad aumenta. Dos áreas requirieron un significativo trabajo específico:

\begin{itemize}
    \item Para que las salidas intermedias de la \textit{pipeline} estén disponibles, las funciones que representan las etapas de la \textit{pipeline} se pueden decorar con el decorador `exposed' para indicar que sus entradas y salidas se pueden monitorear e incluir en la respuesta del trabajo. El gráfico de dependencia entre las funciones expuestas se obtiene analizando el árbol de sintaxis abstracta del código fuente de la \textit{pipeline}, por ejemplo, sin requerir su ejecución.
    \item La derivación de los esquemas JSON asociados a las entradas y salidas de las funciones expuestas, lo cual se logra inspeccionando las sugerencias de tipo de los argumentos y el tipo de retorno de las funciones. El esquema JSON describe el tipo, las restricciones (como valores mínimos o máximos) y el tipo de medio de contenido de las entradas y salidas.
\end{itemize}

Luego, el identificador de transformación se puede usar para crear un trabajo que ejecutará la transformación contra las entradas especificadas.

\section{Conclusiones}

A lo largo de este capítulo, se han descripto las actualizaciones y mejoras de sonoUno durante estos años, todas ellas en respuesta a los análisis realizados con usuarios presentados en el Capítulo \ref{cap:fg_completo}. El desarrollo de sonoUno, en pos de mantenerlo centrado en el usuario y con sus características de accesibilidad, fue dividido en diferentes etapas: el desarrollo de algoritmos de sonorización (sección \ref{sect:cap6_sonif_reinforce}), la versión de escritorio (sección \ref{sect:cap6_vescritorio}), la versión web (sección \ref{sect:cap6_webcompleta}) y durante el último año, se inicio el diseño de una plataforma unificada que busca integrar las diferentes etapas en un mismo lugar (sección \ref{sect:cap6_server}).

Pese a que los script de sonorización descriptos en la sección \ref{sect:cap6_sonif_reinforce} aún no han sido integrados todos en la versión de escritorio (hasta el momento se ha integrado la sonorización de partículas del LHC), realizan por si mismos el despliegue visual y sonoro de datos. Fue necesario programar estos script de forma separada, debido a que tanto el proceso de visualización como el de sonorización son diferentes, por lo que las opciones para configurar ambos despliegues será diferente. Es por ello que para continuar integrando estos script se necesita realizar un nuevo estudio con usuarios y diseñar una interfaz apropiada para cada tipo de datos.

Aunque la integración de nuevos tipos de datos en la interfaz gráfica de sonoUno requiere un trabajo adicional que se propone como trabajo a futuro, la versión final del programa fue mejorada según los requerimientos y propuestas de personas que la han utilizado. Se ha ejecutado la interfaz de escritorio con datos de rayos cósmicos, curvas de luz, espectros de galaxias, ondas gravitacionales, funciones matemáticas y hasta datos estadísticos descargados de internet en formato tabla. Pudiendo comprobar que la interfaz admite el ingreso y sonorización de gran variedad de datos y de gran tamaño (los datos de ondas gravitacionales que se sonorizaron contienen 1 millón de filas).

Respondiendo también a requerimientos de usuarios, obtenidos de diferentes pruebas y entornos, se realizó el desarrollo de sonoUno en un entorno web. En un inicio, se buscaba que la versión web realizara solamente la sonorización con ajustes básicos, debido a que el entorno web presenta mayores limitaciones que el entorno de escritorio. Sin embargo, la versión final de la web difiere muy poco de la versión de escritorio, no cuenta con la posibilidad de realizar funciones matemáticas con octave, por ejemplo. Se logró realizar una interfaz web que permite el ingreso de datos (muestra además datos precargados), realiza el despliegue visual y la sonorización, permite controlar la reproducción y ajustar parámetros de ambas visualizaciones, posibilitando además la aplicación de funciones matemáticas sencillas: logaritmo, suavizado y buscar picos (recorre la señal buscando los valores máximos).

Finalmente, mantener dos versiones paralelas de un mismo programa es una tarea muy difícil, y prescindir de una de las versiones no es una opción por un tema de accesibilidad (por ejemplo, personas que no cuentan con internet no pueden utilizar la versión web). Debido a ello, y a la gran demanda de poder realizar gran numero de sonorizaciones (salvadas en archivos) en la línea de comandos, se comenzó a trabajar en una versión unificada de sonoUno, que permitiera unificar el código de base de las diferentes versiones de sonoUno. Esta propuesta cuenta con un servidor donde se alojará el código de sonoUno, y las diferentes interfaces (web, de escritorio u otras) deberán acceder a dicho servidor para realizar las tareas de sonorización y visualización. Si bien queda aún mucho camino por recorrer en este tema particular, ya se han comenzado las primeras pruebas con funciones simples.

\chapter{sonoUno como plataforma de sonorización}
\label{cap:sonouno_vfinal}

\section{Introducción}

Incluso cuando las tecnologías actuales presentan el potencial de aumentar la inclusión y las Naciones Unidas ha establecido el acceso digital a la información como un derecho humano, las personas con discapacidad continuamente enfrentan barreras en su profesión. En muchos casos, en ciencias, la falta de herramientas accesibles y centradas en el usuario dejó atrás a los investigadores con discapacidad y no les facilitó realizar investigaciones de primera línea utilizando sus respectivas fortalezas. 

Particularmente, como resultado del estudio de grupos focales (sección \ref{sect:FG_completo}), se concluyó que las personas funcionalmente diversas requieren herramientas para ser autónomas, lo que les permite precisión, certeza, eficacia y eficiencia en su trabajo, logrando una mayor equidad. Continuando, a partir de intercambios con usuarios se detectó la necesidad de herramientas versátiles y robustas, que puedan utilizarse sin necesidad de atravesar innumerables errores de instalación o incompatibilidades de software, que permita al usuario elegir entre una interfaz gráfica o simplemente la funcionalidad que necesita, entre otras.

A partir de la propia experiencia e investigación, plasmada en esta tesis, un programa que reúna las características mencionadas se puede lograr: siguiendo un enfoque de diseño centrado en el usuario como parte integral del desarrollo de software y adaptando los recursos de acuerdo con los objetivos de la investigación. En este capítulo final se detallará cómo cambió desde el inicio el marco de trabajo de sonoUno, hasta llegar a la propuesta de una \textbf{``Plataforma de Sonorización''}.

En función de esta nueva plataforma de sonorización, se detalla la creación y desarrollo de un sitio web dedicado, donde se muestra toda la información pertinente al proyecto, documentación para el usuario, links a las herramientas, una galería con material producido por sonoUno y una sección de noticias. Además, se incluye una sección donde se describe el uso y performance de sonoUno a partir del uso de los principales lectores de pantalla utilizados durante este trabajo (NVDA-Windows, Orca-Ubuntu y VoiceOver-MacOS).

Adicionalmente, se detallan las características y conclusiones de un workshop realizado en julio de 2022, donde se utilizó la versión web de sonoUno y una primer versión de un entrenamiento basado en la sonorización de datos astrofísicos. Este capítulo finaliza con una sección dedicada a las propuestas de líneas de trabajo futuras.

\section{Actualizaciones al marco de trabajo inicial}

En un primer momento, el desarrollo de sonoUno se enfocaba en un software de sonorización de datos astronómicos centrado en el usuario. Se comenzó con la idea de un solo programa multiplataforma, de escritorio y que permitiera el uso del mismo con lectores de pantalla. Sin embargo, luego de alcanzar la primer versión de dicho programa (Capítulo \ref{cap:sonouno_v1}), con toda la documentación organizada, y habiendo realizado múltiples pruebas con usuarios (Capítulo \ref{cap:fg_completo}), se llegó a la conclusión de que era necesario ir más allá de esta primera idea. 

Si bien el desarrollo de un software integrado, multiplataforma (incluyendo la web y dispositivos móviles), acompañado de la investigación pertinente para que dicho programa fuera usable, eficiente e inclusivo, escapa los límites de esta tesis de doctorado, se comenzó con la primera etapa de la propuesta.

Una plataforma de sonorización consiste en brindar una base de programación que permita luego integrar diferentes interfaces, sin necesidad de mayores esfuerzos de programación. De esta forma, desde el punto de vista del diseñador, se podrá integrar una versión de interfaz gráfica para escritorio, web y dispositivos móviles, todas con las mismas funcionalidades y el mínimo esfuerzo de actualización. En cuanto al usuario, se pretende ofrecer una versión que no necesite esfuerzo para su instalación, con la versatilidad de funcionar en diferentes plataformas, contando además con la posibilidad de usar e instalar cada paquete si así lo desea para la integración con desarrollos o actualizaciones propias.

En cuanto a la primer etapa de esta propuesta, se comenzó con la unificación de todo el material y desarrollos existentes, en una página web dedicada a sonoUno (sección \ref{sect:cap7_webinicialcompleta}). Allí se presenta toda la información referente a sonoUno: manuales, links a herramientas y repositorios, galería con ejemplos de despliegue de datos y una sección de noticias. Además, en base a la experiencia previa y a los resultados de un workshop realizado a mediados de 2022 (sección \ref{sect:cap7_workshopjul2022}), se definen los principios que debe seguir el desarrollo, a modo de una hoja de ruta para mantener el desarrollo centrado en el usuario y en pos de una mayor inclusión.

\section{Sitio web de sonoUno}
\label{sect:cap7_webinicialcompleta}

Con la finalidad de reunir la información de sonoUno en un mismo lugar, asegurando la accesibilidad para los usuarios con diferentes estilos sensoriales, se comenzó el desarrollo de una página web dedicada al proyecto \href{https://www.sonouno.org.ar/}{\underline{\textcolor{blue}{sonoUno}}}. Allí se muestra la información del proyecto y se ponen a disposición las diferentes herramientas, desde los manuales e instructivos, hasta una galería de recursos (datos, imágenes, audios) y una sección de noticias.

Dicha página web tiene formato WordPress, con una vista simple; el contenido se muestra en una sola columna sin botones o elementos laterales. En la parte superior derecha, se presenta un botón que permite el cambio de idioma. El color de fondo es blanco con letra negra; luego del título se muestra un banner con todos los logos correspondientes a las instituciones relacionadas con el desarrollo y luego comienza la descripción breve de la herramienta para sonorización.

Al continuar bajando en la inspección de la página y luego de la descripción se encuentran los iconos de cada una de las secciones, con un texto con hipervínculo debajo para poder ingresar a cada una de ellas. Cada sección presenta el mismo formato y diseño que esta página de bienvenida. Cuando se muestra contenido multimedia, se hace en la misma continuidad del texto y se coloca la descripción de cada uno de ellos para que sea descripto por el lector de pantalla.

En cuanto a la navegación, cada sección permite volver a la página de inicio. Además, a medida que se adentra en las secciones se muestra al final de la página la opción de volver al menú anterior, a la sección anterior o a la página principal. Además,  las opciones ``anterior'' y ``siguiente'', permiten navegar entre las secciones internas.

Esta página web se ha probado de forma exitosa con lectores de pantalla y ha sido utilizada por usuarios ciegos y videntes para acceder al material de sonoUno. Una recomendación de los usuarios que se fue repitiendo en el tiempo, fue hacer la página web más atractiva visualmente para los usuarios con posibilidad de detección de formas y colores. En un principio se explicó que la finalidad de la página web era ser accesible para personas con y sin discapacidad, sin embargo, la solicitud se reiteró en cada prueba e intercambio. Fue por ello que se decidió realizar una versión paralela, siguiendo diseños web más sofisticados.

Con tal fin, se creó una nueva página web (\href{https://reinforce.sonouno.org.ar/}{\underline{\textcolor{blue}{link}}}), que recibe al usuario con dos opciones: (1) intérprete de lectores de pantalla (conduce a la página web descripta anteriormente); (2) intérprete gráfico (lleva a la nueva versión). La información que se presenta en ambos diseños es la misma, lo único que cambia es la forma de desplegar la página principal y los primeros niveles de contenido. Esta nueva versión cuenta con un fondo negro con letra blanca e hipervínculos en verde, un menú con listas desplegables en la parte superior, los agradecimientos con cada uno de los logos institucionales, la sección de noticias con imágenes y nuevamente el menú al final junto al mail de contacto.

Las secciones internas de la galería, donde se presenta el material multimedia generado a partir de datos astronómicos obtenidos de grandes bases de datos o facilidades astrofísicas como Pierre Auger, siguen visualizándose en la página web original: amigable a lectores de pantalla, con fondo blanco y letra negra.

\subsection{Manuales de Instalación y para el Usuario}

Estos manuales e instructivos son documentación preparada para el usuario con la finalidad de que puedan utilizar la herramienta. Los instructivos de instalación se encuentran en formato word y pdf para descargar, en castellano e inglés. Se ha elaborado un instructivo por cada sistema operativo (Windows, MacOS y Linux: Ubuntu), y de estos, uno por idioma disponible hasta el momento.

Los manuales de usuario, además de la opción pdf para descargar, se encuentra su contenido disponibles en la misma página web. Al acceder al icono de Manual de Usuario, la persona se encuentra con la opción para descargar y un índice de contenido. Al navegar por el índice de contenido puede ir recorriendo las diferentes secciones del manual. De esta forma una persona puede acceder al contenido sin realizar descargas adicionales, o tener que estar buscando el archivo en su computadora; puede utilizarse con lectores de pantalla y el texto en la página web está disponible actualmente en castellano, inglés, francés, alemán, italiano, griego y portugués.

De manera permanente se actualizan tanto el Manual de Instalación del software para los usuarios que optan por la versión de escritorio del mismo, como el Manual de Usuario, que detalla de manera exhaustiva las capacidades de este desarrollo. Ambos Manuales están accesibles en el Apéndice \ref{ap:manuals} y en la web de sonoUno en las siguientes direcciones: 

\begin{itemize}
    \item \href{https://www.sonouno.org.ar/installation-manuals/}{\underline{\textcolor{blue}{Manual de Instalación}}}
    
    \item \href{https://www.sonouno.org.ar/usermanual/}{\underline{\textcolor{blue}{Manual de Usuario}}}
\end{itemize}

En las secciones siguientes, de presentan y describen los contenidos del Manual de Usuario, Manual de Instalación y repositorio de GitHub, entre otros recursos.

\subsubsection{Introducción descriptiva del sonoUno}

Tal como se describió en los capítulos precedentes, sonoUno utiliza como lenguaje de desarrollo Python, que se ha actualizado a lo largo de los años. En este momento es posible correr este software usando Python 3.10. Se ha aplicado un diseño modular, que permitió el trabajo colaborativo, cuyos resultados se detallaron en secciones previas de este documento. SonoUno ahora es multiplataforma, probado en windows 10, Ubuntu (16.04, 18.04, 20.4 y 22.4), CentOS 7 y Mac (Mojave, Catalina, Big Sur y Monterrey); el equipo de desarrollo trabaja continuamente para mantener este beneficio. Como ya se describió, el objetivo principal de SonoUno es permitir al usuario abrir archivos de datos (extensión txt o csv), mostrar la trama y sonorizar dichos datos. Actualmente, la sonorización se realiza por variación de tono y los ajustes permiten cambiar el volumen y el timbre. 

\subsection{Acceso a sonoUno web y al repositorio GitHub}

Siguiendo con las secciones de la página web de sonoUno, dos de ellas están dedicadas a dar acceso a la herramienta web y al repositorio de GitHub. El sonoUno web (disponible en castellano e inglés por el momento), descripto en la sección \ref{sect:cap6_webcompleta} del Capítulo \ref{cap:reinforce}, es la versión web del software de sonorización sonoUno.

Por otro lado, GitHub es una plataforma de desarrollo colaborativo, que permite la interacción de diferentes programadores, con control de versiones y almacenamiento de diferentes tipos de archivos. Esta plataforma ha sido la elegida para almacenar el código fuente de todo el desarrollo, para lo cual se ha definido una organización llamada ``sonoUno Team'' donde se agrupan todos los repositorios particulares existentes, que se describirán en la siguiente subsección.

\subsubsection{Repositorios en GitHub}

Se debe tener en cuenta que los repositorios en GitHub de las herramientas son trabajos en proceso, algunos con versiones estables que ya han sido presentadas a usuarios (como es el caso de las versiones de escritorio y web de sonoUno), otros en versiones de prueba (como es la primer etapa del servidor sonoUno) y otros donde recién se está iniciando el desarrollo (como es el caso de los entrenamientos en sonorización, tema que cuenta actualmente con una beca doctoral otorgada por CONICET para iniciar el trabajo de investigación en Abril de 2023). 

Cabe destacar que de forma paralela se está utilizando el repositorio \href{https://zenodo.org/}{\underline{\textcolor{blue}{Zenodo}}} para publicar las versiones estables de cada herramienta, esta plataforma presenta relación con GitHub y permite publicar y actualizar cada `release' que se hace (por ejemplo se incluye el repositorio correspondiente a la última actualización de sonoUno versión \href{https://doi.org/10.5281/zenodo.7717725}{\underline{\textcolor{blue}{escritorio}}}). Además de lo mencionado, Zenodo ofrece un DOI para el repositorio y adicionar publicaciones, material audiovisual, presentaciones y conjuntos de datos, entre otros. 

A continuación se describirán los repositorios que contiene la organización ``sonoUno Team'' en GitHub:

\begin{itemize}
    \item \textbf{sonoUno}

        Es el repositorio público principal, fue abierto en mayo del año 2019 luego de lograr la primer versión de sonoUno (descripta en el Capítulo \ref{cap:sonouno_v1}). Contiene el control de versiones desde entonces a la actualidad, la última versión que se encuentra en este repositorio contiene las actualizaciones descriptas en el Capítulo \ref{cap:reinforce}, sección \ref{sect:cap6_vescritorio}.

        Este repositorio cuenta con una descripción de bienvenida, los logos y agradecimiento, y una guía resumida para su instalación. En la carpeta sonoUno es donde se encuentran los script de programación, manuales, instructivos y datos de ejemplo para poder probar el programa luego de instalarlo.

    \item \textbf{sonouno\_web}

        En este caso el repositorio es privado, contiene el desarrollo de la página web del software. Se muestra junto con el código la descripción básica para poder ejecutarlo. El desarrollo y funcionalidades de la página web fueron descriptas en la sección \ref{sect:cap6_webcompleta}.

    \item \textbf{streamUnolib/sonoUnolib}

        Con el inicio de la propuesta de unificación de sonoUno para generar un código base, se tomó la decisión de tener repositorios dedicados a las partes más generales de sonoUno (serían las librerías raíz que luego se utilizarán en los desarrollos específicos). Los repositorios streamUnolib y sonoUnolib tienen las librerías base de sonoUno, por el momento se ha incluido la primer parte de la librería de sonido, que se utiliza actualmente en los script de sonorización de datos de grandes facilidades astrofísicas. Estos repositorios son de acceso público, cabe destacar que la librería ``sonounolib'' que se encuentra en el repositorio ``sonoUnolib'', está disponible también en Pypi (el repositorio oficial de librerías de Python; puede accederse \href{https://pypi.org/project/sonounolib/}{\underline{\textcolor{blue}{aquí}}}), lo que permite que pueda instalarse fácilmente con el comando `pip'.

    \item \textbf{sonoUno-muongraphy/sonoUno-images/sonoUno-lhc}

        Dado que las sonorizaciones particulares de cada tipo de datos (en el caso de esta disertación datos de partículas y una primer aproximación a la sonorización de imágenes) tienen un desarrollo propio, referente a cada tipo de datos, se ha destinado un repositorio para cada desarrollo de sonorización. Esta decisión se basa en que en el diseño del servicio de sonorización se plantea ofrecer cada uno de estos servicios de forma individual.

        De esta forma, ``sonoUno-moungraphy'' contiene la sonorización de datos de muongrafía presentada en la sección \ref{sect:cap6_muon_sonif}. En cuanto a ``sonoUno-images'', incluye la primer versión de sonorización de imágenes descripta en la sección \ref{sect:cap6_imgsonif}. Por último, el repositorio ``sonoUno-lhc'' aloja la sonorización de partículas del gran colisionador de hadrones explicada en la sección \ref{sect:cap6_lhc_sonif}. Los tres repositorios mencionados son de acceso público.

    \item \textbf{sonoUno-server}

        Este repositorio contiene el servidor que ofrecerá los algoritmos de sonorización como servicio. Contiene al inicio una descripción del proyecto, las herramientas que se utilizan y los pasos a seguir para poder utilizar las transformaciones necesarias para realizar y obtener una sonorización. 

    \item \textbf{test\_gui}

        Este proyecto en particular, es un repositorio privado que contiene el código de una interfaz gráfica sencilla programada en wxPython, con la finalidad de comunicarse con el servidor de sonoUno, permitir enviar un archivo de datos y obtener la sonorización para reproducirla. Los resultados fueron prometedores, se pudo obtener el archivo de sonorización del servidor y reproducirlo. El problema actual a resolver para poder utilizar el servidor en las interfaces de sonoUno será la sincronización entre el despliegue visual y sonoro.

    \item \textbf{training\_web}

        Este repositorio ha sido creado para contener el código relacionado con la interfaz web de los entrenamientos, la cual comenzó a desarrollarse como un trabajo final de grado de la carrera de Bioingeniería de la Universidad de Mendoza. Por el momento el repositorio es privado, pero en cuanto se logre la primer versión del mismo con su pertinente publicación, el repositorio será publico, respondiendo siempre a los principios del grupo de trabajo de sonoUno.
    
\end{itemize}

\subsection{Galería y Noticias}

Con la finalidad de mostrar los resultados, eventos y noticias en las cuales está referenciado sonoUno se generaron las secciones ``Galería'' y ``Noticias'' en la página web. La sección noticias por el momento tiene un diseño lineal presentando primero las últimas noticias, se prevé realizar un índice cuando las noticias crezcan en número, para permitir a los usuarios una navegación más sencilla y eficaz.

La sección ``Galería'' está organizada en subsecciones para facilitar la identificación de los datos. En un primer nivel se diferencia entre: datos astronómicos, datos astrofísicos, funciones matemáticas y misceláneas. Dentro de la sección de datos astronómicos se encuentran datos de galaxias, curvas de luz y estrellas variables. La sección de datos astrofísicos contiene datos de glitches, partículas del LHC, muongrafía y rayos cósmicos. Con respecto a la sección de funciones matemáticas muestra una función lineal (creciente y decreciente), cuadrática y seno. Por último, se agregó una sección misceláneas que contiene datos que no pertenecen a las secciones anteriores, contiene un ejemplo de sonorización de datos estadísticos de acceso público.

\section{Uso de sonoUno con lectores de pantalla}
\label{sect:cap7_lectorespantalla}

Durante todas las etapas de los desarrollos de sonoUno, las herramientas han sido probadas con los lectores de pantalla de acceso gratuito de cada sistema operativo (ver sección \ref{sect:lectores_pantalla}). Se buscó siempre mejorar la forma de comunicación y dentro de las posibilidades que ofrecía el código de programación del desarrollo, se trató de mantener una comunicación efectiva. Las pruebas se realizaron siempre con los tres lectores de pantalla, en cada sistema operativo, con la misma versión de sonoUno, para poder realizar comparaciones e inferir cuales eran problemas del programa y cuales propios del lector de pantalla utilizado.

El sistema operativo que más complicaciones presentó en su uso con lector de pantalla fue Ubuntu, con el lector de pantalla Orca. El mismo es un lector de pantalla gratuito, de código abierto, flexible y ampliable. Este lector de pantalla está desarrollado principalmente para infraestructuras de asistencia para Linux y Solaris, por lo que se encontraron algunos problemas al utilizarlo con la interfaz gráfica programada en wxPython. Si bien Orca describe la interfaz de usuario y permite su navegación, la comunicación con el usuario es bastante limitada, esto sumado a problemas de compatibilidad entre wxPython y Orca, hacen que la navegación en sonoUno con el lector de pantalla en Ubuntu no sea del todo efectiva.

En cuanto al sistema operativo que mejor resuelto tiene las ayudas tecnológicas, es MacOS, con su lector de pantalla Voice Over; se debe tener en cuenta que es un software privado. En este caso, el lector de pantalla logra acceder a las descripciones de los elementos y comunica/recuerda al usuario en cada iteración como activar el elemento o continuar al siguiente. Un detalle que se encontró con Voice Over a diferencia de los otros dos, fue que al interactuar con barras deslizables avanza en saltos de determinada magnitud, no punto a punto. Sin embargo, sonoUno presenta la opción de ingresar valores por teclado para que el usuario pueda evitar el uso de barra deslizable si así lo desea. Se puede acceder a un video demostrativo del uso de este lector de pantalla en el siguiente \href{https://youtu.be/Pc_K3uGa2mw}{\underline{\textcolor{blue}{link}}}.

Por último, el lector de pantalla NVDA para Windows, es un lector de pantalla de acceso libre y gratuito desarrollado por NV Access. Se considera que es un punto medio entre los lectores de pantalla descriptos anteriormente, logra detectar todos los elementos de la interfaz, pero presenta complicación en acceder a las descripciones de los elementos. Un aspecto que se consideró a favor es que permite el uso de las barras deslizables, punto por punto. No expone en cada iteración las teclas que se deben ir presionando, pero al igual que para Orca, dicha información está disponible en la documentación. Se puede abrir un video demostrativo con el siguiente \href{https://youtu.be/458_sWX-uw4}{\underline{\textcolor{blue}{link}}}.

Algo importante a destacar es que en el caso de los tres lectores de pantalla se respetan las teclas de navegación estándar, por ejemplo con la tecla `ALT' se avanza en secciones (en la interfaz de escritorio, se avanza elemento a elemento; en el caso de web avanza por elementos de texto e imágenes). Para el uso de la web de sonoUno (tanto la que contiene la información como la herramienta) se utilizó el protocolo ARIA para que el contenido sea accesible para las tecnologías de asistencia. En cuanto a la experiencia al probar los lectores de pantalla fue similar a la experiencia descripta para la versión de escritorio. Se dejan a disposición dos videos demostrativos: \href{https://youtu.be/OCnK_NkYXIU}{\underline{\textcolor{blue}{NVDA}}} y \href{https://youtu.be/rj2ELDwuM40}{\underline{\textcolor{blue}{Voice Over}}}.

Es necesario destacar, que no todos los sitios web tienen la precaución de utilizar ARIA o tomarse el trabajo de probar sus sitios con diferentes lectores de pantalla, por lo que no todos los sitios web son accesibles a tecnologías de asistencia. Este es el caso de la mayoría de las bases de datos, incluso varias bases de datos astronómicas. Esto lleva a que una de las preocupaciones de este proyecto y sus lineamientos futuros sea el acceso de personas con discapacidad a las diferentes bases de datos. Este acceso para personas con discapacidad debería poder resolverse sin necesidad de limitarlos generando un desarrollo nuevo con datos reducidos, ni parches que funcionan por un determinado período de tiempo; debería seguir el mismo lineamiento de sonoUno, buscando formas compatibles con los despliegue actuales y exigiendo el uso de los protocolos y normas de desarrollo vigentes.

\section{Curso Internacional de entrenamiento REINFORCE 2022}
\label{sect:cap7_workshopjul2022}

Los proyectos que incluyen Ciencia Ciudadana generan espacios para que los ciudadanos que desean participar en investigaciones fundamentales en un marco de trabajo innovador, puedan hacerlo de manera remota. Es así que todos pueden contribuir como co-creadores de propuestas y como parte de los grupos de investigación en instituciones y proyectos científicos de frontera.

Estos proyectos comparten los datos con el fin de acelerar o mejorar los descubrimientos, al usar la mayor cantidad posible de datos almacenados en grandes bases de datos, lo que además permite mejorar los sistemas de detección y hasta descubrir fallas instrumentales. Es esta aproximación entre la contribución ciudadana y la investigación científica, una nueva manera de comunicar ciencia pero a la vez, involucrar a las personas mas allá del reconocimiento de los temas de actualidad en investigación en el mundo. Teniendo en cuenta un diseño centrado en el usuario, aun con grupos no especialistas, este modo de hacer ciencia es fundamental para el testeo de nuevas herramientas destinadas a poblaciones con dificultades para el acceso a los datos, ya sea por desconocimiento de los temas, o por discapacidades.

En este marco de referencia, REINFORCE propuso en 2022 un Curso de capacitación Internacional con objetivos claros, relacionados con sus cuatro demostradores (nuevas partículas en el Gran Colisionador de Hadrones, muones cósmicos, ``ruido'' en el detector de ondas gravitacionales y detectores de neutrinos sumergidos en el Mediterráneo):

\begin{itemize}
    \item Colaborar con investigadores para optimizar las grandes infraestructuras de investigación en física para comprender y eliminar el ``ruido'' que obstaculiza descubrimientos potenciales. 
     \item Descubrir cómo la ciencia ciudadana puede ayudar a conectar la investigación y la sociedad. 
     \item Recibir capacitación en los campos de astronomía de ondas gravitacionales, física de alta energía, astronomía de neutrinos y física de rayos cósmicos a través de la ciencia ciudadana y actividades educativas desarrolladas por expertos líderes en investigación de física de frontera y educación científica. 
     \item Realizar visitas virtuales a grandes infraestructuras de investigación.
     \item Aprender a utilizar datos reales para realizar investigaciones propias. 
     \item Ayudar a dar forma al futuro de los planes de estudio en el campo de la física moderna.
\end{itemize}

En el caso particular de la sonorización, una actividad transversal a todos los demostradores:

\begin{itemize}
    \item Aprender a utilizar el sonoUno.
    \item Aplicar las funcionalidades del software para el estudio de datos específicos, como los de los demostradores de REINFORCE.
    \item Recibir entrenamiento para la detección de rasgos en los datos a partir del sonido.
    \item Evaluar el potencial de recursos para análisis multimodal de los datos.
\end{itemize}

Particularmente, se describirán en la próxima subsección los aspectos de la participación en el mencionado workshop relacionados con: la interfaz web de sonoUno desarrollada en esta tesis y un entrenamientos en sonorización, estos últimos son parte de los lineamientos futuros que se desprenden de esta tesis.

\subsection{Descripción de la actividad}

El workshop de sonorización y curso de entrenamiento realizados durante una escuela de verano organizada por el consorcio REINFORCE (\href{http://reinforce.ea.gr/international-training-course/}{\underline{\textcolor{blue}{link}}}) contó con dos encuentros, separados en dos días, dos horas el primer día y una hora el segundo. En primer lugar se presentaron los trabajos que están siendo desarrollados por el grupo de trabajo de sonoUno (importancia de los desarrollos y actividades para la inclusión y la equidad, análisis multimodal de datos, versión de escritorio y web de sonoUno, y scripts de sonorización). Luego se realizaron tres actividades: uso de la interfaz web, uso de los script de sonorización para demostradores en REINFORCE y entrenamiento.

Si bien durante esta tesis no se realizó una investigación enfocada en entrenamientos para el uso de sonorización, una de las conclusiones a la cual se llegó fue la necesidad de los mismos. En este sentido durante el año 2022 se comenzó a trabajar dentro del grupo de trabajo con una línea de investigación enfocada en el estudio de la percepción y la elaboración de entrenamientos. En base a lo mencionado, durante la escuela de verano organizada por REINFORCE en Grecia, en Julio del 2022, además de probar la interfaz web de sonoUno, se realizó el primer entrenamiento en sonorización con docentes e investigadores. A continuación se describirá brevemente el entrenamiento realizado y los resultados, todo ello presenta influencia directa en los lineamientos futuros que se desprenden de esta disertación.

\subsubsection{Entrenamientos}

La percepción es un proceso que requiere una gran cantidad de procesamiento mental que proporciona los medios por los cuales se crea el concepto que uno tiene del entorno y que ayuda a aprender e interactuar con él. La recopilación de estudios previos a lo largo de la historia ha llevado a la conclusión de que el rendimiento auditivo mejora cuando se combina con estímulos visuales y viceversa. Teniendo en cuenta la consideración anterior, se utilizaron las dos vías sensoriales mencionadas con la intención de realizar una prueba de entrenamiento multisensorial con el objetivo de lograr la familiarización de los participantes con este tipo de técnicas en detección de señales. 

Con la finalidad de programar un entrenamiento que permita presentar las señales visuales y auditivas a los participantes, permitiendo la identificación de la misma y guardando dicha interacción del usuario, se utilizó el software PsychoPy \citep{peirce2019}. Dicho software ha sido diseñado para la creación de experimentos en ciencias del comportamiento (como son la psicología, neurociencia, lingüística, entre otras) que permite el control espacial preciso y sincronizado de distintos estímulos. Dicho programa proporciona dos tipos de interfaz con distinta funcionalidad, permitiendo al usuario elegir entre diseñar sus experimentos de forma gráfica o programarlos en lenguaje Python. PsychoPy está disponible para cualquier sistema operativo, no siendo esto una limitación como muchos otros software presentan. Finalmente, posee una comunidad, que comparte sus experimentos en línea, con la posibilidad de descargar el código para, de ser necesario, modificarlo, pudiendo adaptar el experimento a nuevas necesidades.

Durante el inicio del año 2022 se comenzó a realizar pruebas con entrenamientos prototipo dentro del grupo de trabajo, donde se testeo la interfaz con los integrantes del grupo para detectar posibles problemas. Durante dichas pruebas se concluyó la necesidad de desplegar toda la información de forma visual y auditiva (no se había contemplado esto para los textos en la primer aproximación) y la necesidad de dar más tiempo al usuario para responder \citep{sabi2022}. En el caso particular del workshop que se describe aquí, se realizó el diseño del entrenamiento en sonorización con datos de los demostradores de REINFORCE, sonorizados con los script de SonoUno. Se obtuvieron de dichos programas las imágenes y archivos de sonido con los cuales fue diseñado el entrenamiento.

En cuanto al entrenamiento, se llevó a cabo en dos sesiones de complejidad creciente, ambas estuvieron conformadas por tres bloques, una por tipo de datos: (1) tres tipos de glitches; (2) dos eventos del LHC; (3) cuatro eventos de muongrafía (dos con presencia de muón y dos sin). La sesión del primer día fue destinada a que conocieran los datos y pudieran clasificarlos, presentando un ejemplo de cada tipo. En el caso del segundo día se presentaron dos eventos de cada tipo de datos, ubicados aleatoriamente en cada sección.

El segundo día, luego del entrenamiento se le solicitó a los participantes que completaran una encuesta sobre la versión web de sonoUno y el entrenamiento realizado. En la sección \ref{sect:cap7_summerschool_resultados} se detallarán los resultados de dicha encuesta, pero es destacable que los participantes habían mejorado su desempeño en la detección de señales desplegadas de forma multisensorial (vista y audición en este caso).

\subsection{Evaluación de desempeño de usuarios}
\label{sect:cap7_summerschool_resultados}

\begin{figure}[!ht]
    \centering
    \includegraphics[width=1\textwidth]{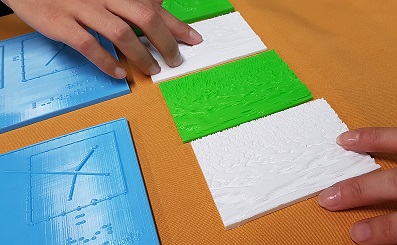}
    \caption{Impresiones 3D de imágenes de partículas e imágenes de Glitches.}
    \label{fig:cap7_impresion3D}
\end{figure}

Durante el workshop se trabajó con todos los participantes en diferentes aspectos de la exploración multisensorial a los datos, se mostró las herramientas de sonoUno, el proceso de sonorización y también se puso a disposición de los participantes impresiones 3D de las mismas imágenes presentadas en las prácticas con los script de sonorización (ver Figura \ref{fig:cap7_impresion3D}).

Con la finalidad de evaluar la utilidad tanto de la actividad, como de las herramientas y del entrenamiento, se realizó al final del último encuentro una encuesta sobre estas temáticas. Dicha encuesta constó de cinco secciones: (1) antecedentes de los participante; (2) acceso a sonoUno web; (3) usabilidad de sonoUno web; (4) demostradores de REINFORCE; (5) documentación y conclusiones. A continuación se describirán los resultados obtenidos, siendo separados entre las secciones descriptas. Se debe tener en cuenta que más allá de que hay secciones enfocadas en sonoUno versión web, los participantes respondieron considerando la totalidad de las herramientas presentadas.

\subsubsection{(1) Antecedentes de los participante}

Si bien del workshop participaron más personas, fueron trece (13) los que respondieron a la encuesta de manera completa. Dado que el workshop tuvo lugar en Grecia, aproximadamente el 50\% de los participantes (6 participantes) era local, los otros 7 residen en diferentes lugares (Austria, Kosovo, 2 de Países Bajos, Rumanía, Servia y Suecia), pero la totalidad de participantes residen en Europa. 

\begin{figure}[!ht]
    \centering
    \includegraphics[width=1\textwidth]{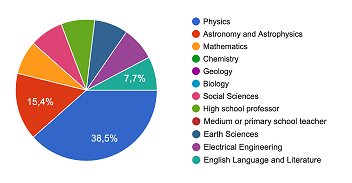}
    \caption{Profesiones de los participantes del Workshop.}
    \label{fig:cap7_resultencuesta_background}
\end{figure}

En cuanto a la profesión, y por tanto background de los participantes en el entrenamiento, se puede observar la diversidad entre los participantes en la Figura \ref{fig:cap7_resultencuesta_background}, donde las porciones del gráfico de torta más pequeñas corresponden todas al 7,7\% (1 participante cada una; profesiones: matemática, ciencias sociales, profesor de secundaria, ciencias de la tierra, ingeniería eléctrica e inglés y literatura), la siguiente en tamaño de color rojo al 15,4\% (2 participantes; profesión: astronomía y astrofísica) y por último la más grande al 38,5\% (3 participantes; profesión: física). Particularmente, para el estado académico de los participantes, tres eran estudiantes, dos estudiantes graduados, dos graduados de doctorado, dos científicos senior o profesor y cuatro respondieron `Otro'.

Se les consultó por la herramienta matemática que mayormente utilizan para aplicar funciones o transformaciones a los datos, obteniendo: que la mayoría (7 participantes) no utiliza ninguna, cuatro de ellos utiliza Python, los otros cuatro utilizan diferentes herramientas (Mathlab, Wolfram Mathematica, LoggerPro y Audacity) y ninguno de ellos utiliza Octave (se resalta este último punto debido a que sonoUno escritorio presenta una conexión con este programa).

Se incluyó una pregunta para saber si conocían sobre el despliegue multisensorial de datos; tres participantes respondieron que conocían del tema antes del workshop y diez que conocieron el tema luego de terminar el workshop. Además, en la siguiente pregunta indicaron que ninguno de ellos había utilizado una herramienta de sonorización anteriormente. Sin embargo, cuando se les consultó a continuación por un dispositivo que hubieran utilizado con un programa de sonorización, once de los participantes respondieron indicando: siete de ellos una computadora portátil, tres de ellos una computadora personal y uno una aplicación móvil. Probablemente la pregunta anterior consideraron que consultaba por el uso de una herramienta de sonorización analizando datos en ciencias, como lo hecho durante el workshop.

\subsubsection{(2) Acceso al material}

La totalidad de los participantes que respondieron la encuesta utilizan el sistema operativo Windows. Un detalle a tener en cuenta en esta sección es que algunas preguntas se respondieron teniendo en cuenta el uso de la versión de escritorio de sonoUno y los script de sonorización de datos. Fue así desde el principio donde luego de establecer que diez de los participantes necesitaron ayuda para utilizar el programa, explicaron que su respuesta se basaba en la dificultad de instalar las librerías de Python necesarias y el uso de la línea de comandos (muchos no estaban familiarizados con esta herramienta).

\begin{figure}[!ht]
    \centering
    \includegraphics[width=1\textwidth]{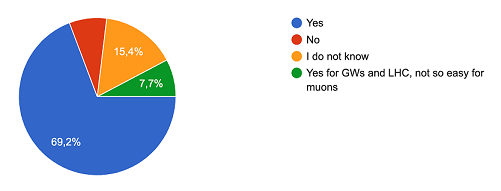}
    \caption{Respuesta a la pregunta: ¿Pudiste extraer información de los datos proporcionados usando sonorización?.}
    \label{fig:cap7_resultencuesta_12}
\end{figure}

En cuanto a la sonorización, aproximadamente el 77\% de los participantes expresaron haber podido extraer información de los datos sonorizados, solo el 7\% de ellos encontró dificultad en la sonorización de Muones (ver Figura \ref{fig:cap7_resultencuesta_12}). En general, sobre el acceso al programa y a los datos, los participantes comentaron la necesidad de ayuda o un período de reconocimiento, pero luego de eso lo encontraron `fácil', `bueno', `efectivo' (esas fueron algunas de las palabras que utilizaron para describirlo).

\begin{figure}[!ht]
    \centering
    \includegraphics[width=1\textwidth]{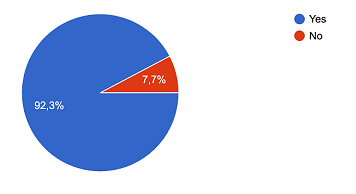}
    \caption{Respuesta a la pregunta: ¿Pudo abrir un conjunto de datos de una manera eficiente en el tiempo?}
    \label{fig:cap7_resultencuesta_16}
\end{figure}

Varios participantes reportaron aspectos que les gustaría modificar sobre los entrenamientos, como por ejemplo: saltar algunas ventanas como la introducción, poder volver a escuchar la sonorización, tener una sesión donde solo esté el sonido sin la imagen, tener retroalimentación por voz sobre la respuesta. En cuanto a la interfaz web un participante respondió que eso fue lo mejor. La Figura \ref{fig:cap7_resultencuesta_16} resalta que casi la totalidad de participantes pudo abrir los archivos de datos sin problemas y en un tiempo eficiente.

\subsubsection{(3) Usabilidad de las herramientas}

En cuanto a la usabilidad de la técnica y del programa, el 76.9\% de los participantes (10 de 13) respondieron que consideran que la sonorización mejoraría su propio trabajo, el 23.1\% restante respondió que no sabían si tendría efecto, ningún participante consideró que la técnica no sería útil. Entre los comentarios que ofrecieron sobre la sonorización de datos, mencionaron mayor eficiencia, una perspectiva diferente, mayor acceso, más divertido y resaltaron la necesidad de conocer más sobre el sonido. Esto último estaría relacionado con la necesidad de profundizar en estudios de percepción y preparar y planificar cursos de entrenamiento.

\begin{figure}[!ht]
    \centering
    \includegraphics[width=1\textwidth]{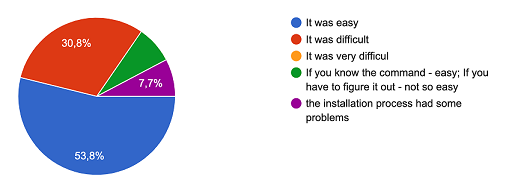}
    \caption{Respuesta a la pregunta: ¿Qué tan problemático fue el proceso de convertir los datos en sonido?}
    \label{fig:cap7_resultencuesta_21}
\end{figure}

Dentro de las preguntas de usabilidad, se les consultó que tan problemático fue el proceso de convertir datos a sonido: se encontró que aproximadamente el 50\% de ellos tuvo problemas o no les resultó fácil el proceso (ver Figura \ref{fig:cap7_resultencuesta_21}). Esto es un indicativo de que si bien los scripts son una buena opción para las personas que trabajan todos los días con los datos o manejan gran cantidad de ellos, no es así para todas las personas.

\subsubsection{(d) Entrenamiento}

\begin{figure}[ht!]
    \centering
        \begin{subfigure}[b]{0.49\textwidth}
             \centering
             \captionsetup{justification=centering}
             \includegraphics[width=1\textwidth]{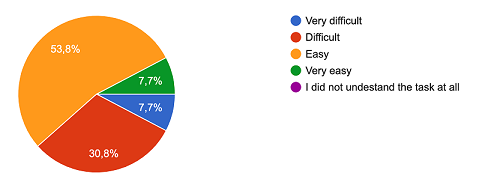}
             \caption{Sonorización de Glitches}
             \label{fig:cap7_resultencuesta_GW1}
         \end{subfigure}
         \hfill
        \begin{subfigure}[b]{0.49\textwidth}
             \centering
             \captionsetup{justification=centering}
             \includegraphics[width=1\textwidth]{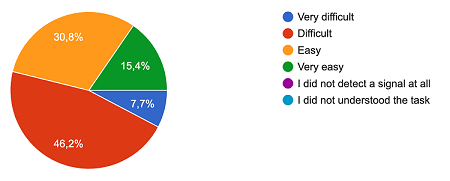}
             \caption{Detección de Glitches}
             \label{fig:cap7_resultencuesta_GW2}
         \end{subfigure}
         \hfill
        \begin{subfigure}[b]{0.49\textwidth}
             \centering
             \captionsetup{justification=centering}
             \includegraphics[width=1\textwidth]{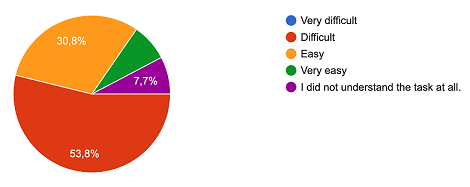}
             \caption{Sonorización de partículas}
             \label{fig:cap7_resultencuesta_LHC1}
         \end{subfigure}
         \hfill
        \begin{subfigure}[b]{0.49\textwidth}
             \centering
             \captionsetup{justification=centering}
             \includegraphics[width=1\textwidth]{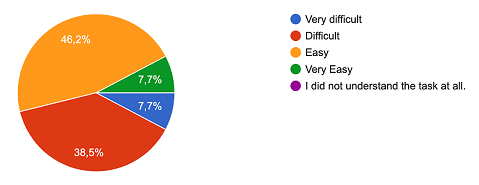}
             \caption{Detección de partículas}
             \label{fig:cap7_resultencuesta_LHC2}
         \end{subfigure}
         \hfill
        \begin{subfigure}[b]{0.49\textwidth}
             \centering
             \captionsetup{justification=centering}
             \includegraphics[width=1\textwidth]{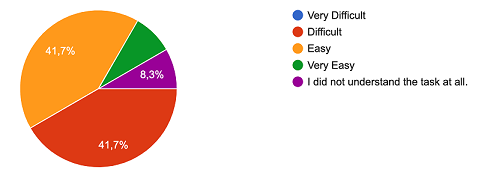}
             \caption{Sonorización de muongrafía}
             \label{fig:cap7_resultencuesta_muon1}
         \end{subfigure}
         \hfill
        \begin{subfigure}[b]{0.49\textwidth}
             \centering
             \captionsetup{justification=centering}
             \includegraphics[width=1\textwidth]{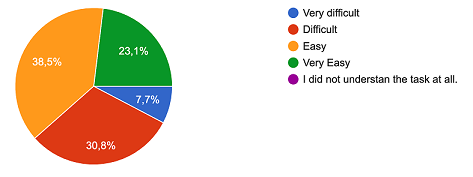}
             \caption{Detección de muones}
             \label{fig:cap7_resultencuesta_muon2}
         \end{subfigure}
    \caption{Preguntas sobre scripts de sonorización y entrenamiento}
    \label{fig:cap7_resultencuesta_entrenamiento}
\end{figure}

En esta sección se les consultó sobre el tipo de datos que se estaba sonorizando en cada sección y luego sobre: la dificultad al ejecutar el script propio de los datos (Figuras \ref{fig:cap7_resultencuesta_GW1}, \ref{fig:cap7_resultencuesta_LHC1} y \ref{fig:cap7_resultencuesta_muon1}) y la dificultad a la hora de detectar el tipo de patrón en los datos (Figuras \ref{fig:cap7_resultencuesta_GW2}, \ref{fig:cap7_resultencuesta_LHC2} y \ref{fig:cap7_resultencuesta_muon2}).

En cuanto a la dificultad de utilizar el script de sonorización, se puede observar en la Figura \ref{fig:cap7_resultencuesta_entrenamiento}, imágenes de la izquierda, que la sonorización de Glitches fue la que resultó más fácil aún cuando reportaron tener problemas para instalar la librería OpenCV. En cuanto a los scripts de partículas del LHC y muongrafía, uno de los participantes reportó no entender la tarea en ambos casos. De estos últimos script el de sonorización de partículas les resultó más difícil que el de muongrafía.

Por otro lado, hablando de la detección de patrones, también se encontró diferencias en la dificultad entre los tipos de datos. Les resultó más fácil la detección de muones que utiliza un arreglo de notas de piano donde se debe encontrar relación ascendente o descendente entre las notas. La sonorización que presentó mayor dificulta fue la de los Glitches, la cual se genera a partir de ondas seno puras que responden a la frecuencia de la señal, con una frecuencia máxima predefinida de 1600 Hz. En cuanto a la sonorización de partículas del LHC utiliza notas de piano pero en un arreglo definido por el script representando diferentes patrones; en este caso las respuestas se encontraron cercanas al 50\% entre difícil o muy difícil y fácil o muy fácil.

Es destacable que ambos aspectos deben reforzarse con más estudios para validar los resultados obtenidos, sin embargo, es notable la preferencia por los sonidos más relacionados a la música o los instrumentos, considerados más placenteros. Deberá estudiarse si esto se mantiene o cambia luego de que la persona realiza un entrenamiento relacionado específicamente con sonorización de datos y detección de rasgos especiales en las señales, y no solo la producción de sonido.

\subsubsection{(e) Documentación y conclusiones}

\begin{figure}[ht!]
    \centering
        \begin{subfigure}[b]{0.49\textwidth}
             \centering
             \captionsetup{justification=centering}
             \includegraphics[width=1\textwidth]{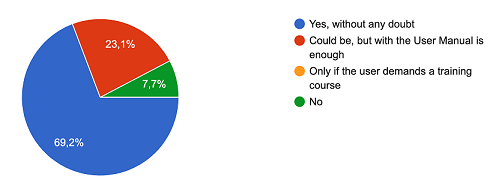}
             \caption{Necesidad de entrenamiento}
             \label{fig:cap7_resultencuesta_30}
         \end{subfigure}
         \hfill
        \begin{subfigure}[b]{0.49\textwidth}
             \centering
             \captionsetup{justification=centering}
             \includegraphics[width=1\textwidth]{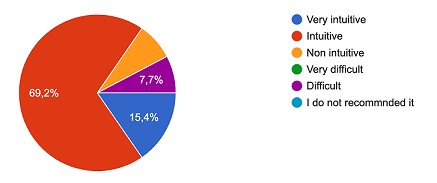}
             \caption{Operabilidad de sonoUno web}
             \label{fig:cap7_resultencuesta_31}
         \end{subfigure}
         \hfill
        \begin{subfigure}[b]{0.49\textwidth}
             \centering
             \captionsetup{justification=centering}
             \includegraphics[width=1\textwidth]{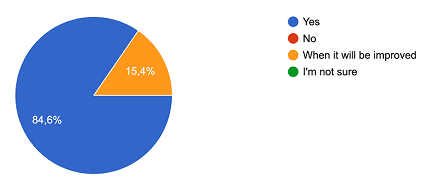}
             \caption{¿Recomendarías sonoUno?}
             \label{fig:cap7_resultencuesta_32}
         \end{subfigure}
         \hfill
        \begin{subfigure}[b]{0.49\textwidth}
             \centering
             \captionsetup{justification=centering}
             \includegraphics[width=1\textwidth]{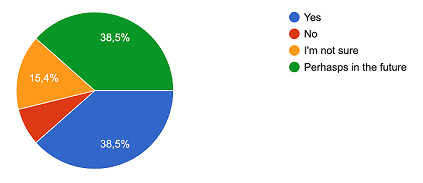}
             \caption{¿Lo usarías en tu investigación?}
             \label{fig:cap7_resultencuesta_33}
         \end{subfigure}
    \caption{Preguntas de cierre sobre el uso de sonoUno}
    \label{fig:cap7_resultencuesta_finalremarks}
\end{figure}

Si bien se comentó anteriormente que la mayoría de los participantes en el Workshop de Sonorización había necesitado asistencia para utilizar las herramientas, nueve de los participantes (69.2\%) reportó no haber utilizado el manual de usuario del programa. La totalidad de participantes que utilizaron el manual, expresaron una opinión positiva sobre el mismo. 

Todos los participantes consideraron que la herramienta cumplió con sus expectativas, les pareció interesante, un enfoque nuevo, divertido, extensible, fácil de utilizar una vez que está instalado y se está familiarizado con el recurso.

La Figura \ref{fig:cap7_resultencuesta_finalremarks} muestra las respuestas a cuatro preguntas con opciones. Es destacable que la mayoría de los participantes considera que es necesario un curso de entrenamiento, lo que refuerza los hallazgos hechos en el Capítulo \ref{cap:fg_completo} de esta tesis. El 85\% de los participantes (11) encontró la interfaz intuitiva, lo que refuerza la importancia de un desarrollo centrado en el usuario desde el inicio. La totalidad de los participantes recomendaría sonoUno a un colega, mientras que el 15\% expresa que lo haría una vez que esté mejorada la interfaz, a partir de los comentarios vertidos en la encuesta. 

Finalmente, la mayoría de los participantes en la encuesta utilizaría sonoUno en su investigación, la persona que expresó que no lo utilizaría, lo justificó mencionando que no realiza análisis de datos en su actividad cotidiana. Los que expresaron no estar seguros, reportaron que fue porque no sabrían como incluirlo en su entorno de trabajo, dado que aún es una herramienta muy nueva.

\subsection{Conclusiones}

El workshop de sonorización y el curso de entrenamiento REINFORCE 2022, permitió obtener importantes resultados y confirmaciones a descubrimientos previos (aquellos mencionados en el Capítulo \ref{cap:fg_completo}) y evaluar la performance de una audiencia frente a la nueva herramienta tras un simple y breve entrenamiento. El mencionado aquí fue el primer encuentro donde se presenta tanto el programa sonoUno como las imágenes obtenidas de sonoUno impresas en 3D, haciendo referencia al despliegue multisensorial de datos que se propone desde el inicio de esta tesis como un nuevo enfoque para el análisis de datos científicos.

Todos los participantes de este workshop, consideraron que el despliegue multisensorial de datos es un enfoque novedoso, y lo utilizarían en sus investigaciones. El cuestionamiento que aflora es como utilizar esta nueva técnica en entornos donde aún no es utilizada por la mayoría de los investigadores. Lo cual se corresponde con una de las preguntas realizada por una de las participantes del grupo focal realizado en 2019: ¿cómo podré discutir sobre los datos con mis pares si ellos no utilizan sonorización? (se ha parafraseado la pregunta).

Se considera que dicho problema puede deberse al desconocimiento que se tiene aún sobre las técnicas de sonorización y la potencialidad que presentan. Esto sumado a una falta de estudios de percepción, que aclaren cómo la sonorización  es interpretada por el cerebro humano, dando a su vez confiabilidad a la técnica.

Además de estos estudios de percepción, cursos de entrenamiento, la propuesta de su uso en niveles educativos como primaria y secundaria ayudarían a reducir el desconocimiento y la falta de familiarización que se tiene con la técnica de sonorización. 

Todo lo mencionado hasta el momento se vincula solo con personas que no presentan una discapacidad visual; ha resultado evidente que la técnica de sonorización mejora la integración de personas con discapacidad en el estudio de la ciencia (ver Capítulo \ref{cap:fg_completo}, específicamente las secciones \ref{sect:fg_conclusiones} y \ref{sect:uso_canarias}).

Es destacable también cómo el enfoque multimodal ayuda al entendimiento de los contenidos conceptuales y científicos, permitiendo explorar un mismo fenómeno por diferentes vías sensoriales. Los participantes de este workshop expresaron que tener las representaciones de los datos tanto de forma visual, como sonora y táctil, fue de mucha utilidad. En este sentido, se refuerza la idea de incluir este tipo de propuestas en los niveles educativos iniciales, ya que en primera instancia, la aproximación multisensorial puede ser abordada de manera lúdica, para luego progresar en el uso de las facultades de percepción complejas del ser humano.

\section{Lineas de trabajos y propuestas futuras}
\label{sect:cap7_lineastrabajo}

Si bien, al iniciar esta investigación sobre el acceso, uso y exploración efectiva de datos astronómicos solo se planteaba el desarrollo de una interfaz de sonorización de datos, se hizo evidente, durante las pruebas con usuarios, la necesidad de redefinir esta primera propuesta. Es así que actualmente, y tras varias actualizaciones,  sonoUno cuenta con: 

\begin{enumerate}
    \item una versión escritorio que además de las funciones comunes con la versión web, presenta conexión con Octave para el análisis de datos con funciones matemáticas; esta es una consecuencia del intercambio con usuarios profesionales, astrónomos y físicos, que trabajan cotidianamente con análisis de datos;
    \item una versión web que permite la sonorización de datos, manipulación de configuraciones de sonido y gráfico, y aplicación de algunas funciones matemáticas sin instalación de librerías y contando solo con conexión a Internet. Además, permite el ingreso de sonido para ser graficado y sonorizado nuevamente; este desarrollo surge como consecuencia de ciertas dificultades que se encontraron a la hora de instalación del software en el caso de usuarios no especialistas o no formados en temas de computación y brinda un acceso inmediato a la herramienta, aún sin grandes recursos locales, pues internet está disponible para la mayor parte de las poblaciones, asegurando acceso al software, más allá de los recursos locales;
    \item un script de sonorización de imágenes que convierte cualquier imagen a sonido, utilizando una sumatoria de intensidades de gris por columna; una demanda para poder incorporar imágenes de cualquier tipo, y no solo datos en archivos organizados en columnas;
    \item un script de sonorización de partículas del LHC, con una novedosa sonorización de partículas donde se seleccionaron cuidadosamente los parámetros de sonido para que representen cada rasgo particular de la partícula en cuestión;
    \item un script de sonorización de imágenes de muongrafía, donde se diseño la sonorización para que sea perceptible por el oído si existe o no relación entre los depósitos de energía de cada capa del detector;
    \item una página web propia de sonoUno que contiene las herramientas, documentación (manuales e instructivos), galería y una sección de noticias. Vale la pena recordar que una falla importante en los recursos de sonorización que han sido estudiados como parte de esta investigación, es la falta de documentación disponible para el usuario o persona interesada en el recurso.
\end{enumerate}

Desde el inicio y durante el desarrollo de cada una de las versiones del programa y los scripts derivados, se ha mantenido los mismos lineamientos: código de acceso gratuito, multiplataforma, probado con usuarios y documentado tanto para el usuario como para el desarrollador.

Dado que el campo de la sonorización de datos a nivel mundial está en una etapa de expansión donde año a año el número de herramientas crece de forma exponencial \citep{zanella2022}, ha sido una prioridad mantener comunicaciones científicas durante todos los años de desarrollo, mostrando los avances de manera permanente y prácticamente en el momento en que se producían. Dicho esfuerzo junto con la trayectoria del espacio de trabajo donde se desarrolló la tesis, permitió que sonoUno se posicionada como un software de sonorización de datos a nivel mundial. Se lo ha reconocido en diferentes reuniones de investigación científica donde se ha invitado al grupo de trabajo para participar como expositores, siendo una de estas ``The Audible Universe'' en sus dos encuentros (\href{https://www.lorentzcenter.nl/the-audible-universe.html}{\underline{\textcolor{blue}{2021}}} y \href{https://www.lorentzcenter.nl/the-audible-universe-2.html}{\underline{\textcolor{blue}{2022}}}). Vale la pena mencionar que este evento está dedicado a las herramientas de sonorización aplicadas en astronomía, busca establecer un marco de trabajo común y lineamientos específicos para el desarrollo de estos recursos.

Del primer encuentro decantó, entre otras cosas mencionadas en el Capítulo \ref{cap:estado_del_arte}, la estadística de crecimiento de software de sonorización en estos últimos años. Además, luego de esa primera edición, se reconoció a sonoUno como software para investigación \citep{zanella2022}. En Diciembre de 2022 tuvo lugar el segundo encuentro, de carácter híbrido. Se realizaron pruebas con diferentes software de sonorización y sonoUno fue evaluado junto a STRAUSS (descripto en la sección \ref{sect:desktop_astronomia} del Capítulo \ref{cap:estado_del_arte}); los participantes provenientes de diferentes especialidades, países, culturas, probaron los software y pudieron compararlos. Dentro de la comparación se reconoció la diferencia considerable entre STRAUSS y sonoUno que se basa en el enfoque que presenta cada uno. STRAUSS es un paquete dedicado a la sonorización de datos, centrado en la producción del sonido. Mientras, sonoUno está enfocado en el acceso a los datos por parte de personas con y sin discapacidad, proveyendo una interfaz gráfica y dedicado a mejorar la accesibilidad tanto a la interfaz como a los datos. Durante dos jornadas, con sesiones de grupos de discusión se intercambiaron ideas, sugerencias, percepciones sobre las formas de diseño y evaluación, en la búsqueda de proponer mejoras a cada herramienta.

\begin{figure}[ht!]
    \centering
        \begin{subfigure}[b]{1\textwidth}
             \centering
             \captionsetup{justification=centering}
             \includegraphics[width=1\textwidth]{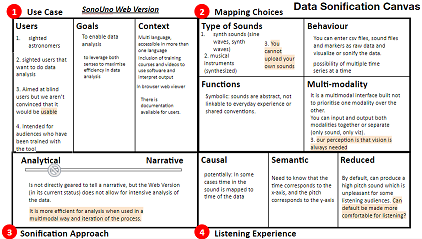}
             \caption{Basado en la versión actual}
             \label{fig:cap7_au2_canvanow}
         \end{subfigure}
         \hfill
        \begin{subfigure}[b]{1\textwidth}
             \centering
             \captionsetup{justification=centering}
             \includegraphics[width=1\textwidth]{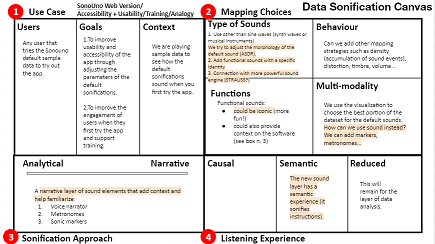}
             \caption{Diseño de nuevas características}
             \label{fig:cap7_au2_canvaproposal}
         \end{subfigure}
    \caption{Canvas dedicados al diseño de mejoras en herramientas de sonorización, elaborados por los organizadores del ``Audible Universe 2'' y completado por los participantes en la sesión del Grupo D.}
    \label{fig:cap7_au2_canva}
\end{figure}

La experiencia de trabajar en grupos interdisciplinarios sobre el desarrollo de herramientas de sonorización, donde se pensó tanto en los datos a sonorizar como en el usuario objetivo de dicha herramienta, fue enriquecedora. La forma de trabajo fue similar a la que se ha utilizado por el equipo de trabajo de sonoUno desde el principio y se utilizaron tablas y recursos colaborativos para poder visualizar todos la misma plantilla de evaluación final (por ejemplo a través de Google Drive). En la Figura \ref{fig:cap7_au2_canva} se pueden observar la plantilla utilizada para la etapa de diseño con sonoUno, por el grupo donde participó la Dra. Beatriz García. La Figura \ref{fig:cap7_au2_canvanow} describe la herramienta sonoUno en su versión actual y la Figura \ref{fig:cap7_au2_canvaproposal} describe las mejoras que se le realizarían según dicho grupo de trabajo, luego de haber utilizado la herramienta el día anterior y tras la discusión de sus funcionalidades. Se agrega, además, la tabla trabajada durante el diseño de nuevas características para mostrar la forma de trabajo que se utiliza al momento de diseñar nuevas implementaciones, antes de comenzar la tarea de implementación. Como se indica en los cuadros de la Figura \ref{fig:cap7_au2_canva}, se debe tener en cuenta: (1) el caso de uso objetivo, quién lo utilizará y en que contexto; (2) las opciones de mapeo como tipo de sonido, comportamiento, multimodalidad; (3) si el discurso o la sonorización tiende a ser más narrativa o analítica; (4) cual será la experiencia del usuario, ¿cómo interpretará la sonorización?.

Durante la sesión de evaluación se diseñó el experimento que se utilizaría para realizar el testeo de la nueva versión de la herramienta con usuarios. Se debió definir las variables, el método, el estímulo y cómo sería realizado. Se contaba con las mismas características que el diseño de grupo focal presentado en el Capítulo \ref{cap:fg_completo}.

La participación en esta reunión internacional permitió dar a conocer entre colegas las últimas novedades de sonoUno, confirmar las técnicas de desarrollo que ya se han utilizado en la herramienta en cuestión y conocer el estatus de la sonorización de datos como campo de investigación. Todos los asistentes en la segunda edición del Audible Universe coinciden en que los siguientes pasos a seguir están relacionados con la percepción y los entrenamientos en sonorización.

Mencionado lo anterior, se continuará con una descripción sobre las propuestas futuras para la integración de todas las herramientas desarrolladas durante esta disertación. En conjunto, se describirán nuevas propuestas que han surgido dentro de las conclusiones parciales de esta tesis y se están comenzando a trabajar por el equipo de sonoUno (una de ellas son los entrenamientos en sonorización).

\subsection{Integración de los scripts de REINFORCE en sonoUno}

Se plantea como trabajo a futuro integrar los scripts de sonorización de datos descriptos en el Capítulo \ref{cap:reinforce}, sección \ref{sect:cap6_sonif_reinforce}, al software sonoUno. Dada la particularidad en el diseño de cada una de las sonorizaciones de datos, se requiere un rediseño completo de la interfaz gráfica para su integración, definiendo qué parámetros de sonido se podrían conservar y cuáles se deberían agregar. Dicha complejidad a su vez requiere nuevas pruebas con usuarios para comprobar la usabilidad, robustez y eficiencia del nuevo diseño antes y después de su implementación.

\begin{figure}[!ht]
    \centering
    \includegraphics[width=0.7\textwidth]{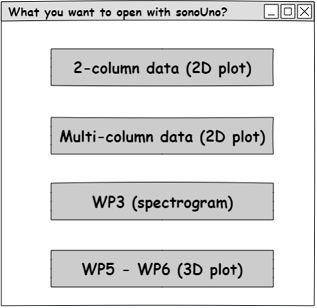}
    \caption{Propuesta de primer interfaz para sonoUno cuando permita el ingreso de diferentes archivos de datos.}
    \label{fig:cap7_sonouno_multidata}
\end{figure}

Se muestra en la Figura \ref{fig:cap7_sonouno_multidata} una propuesta de primer interfaz de sonoUno donde la persona pueda elegir entre los diferentes tipos de datos. Según la elección de la persona, sonoUno desplegaría la interfaz correspondiente sin necesidad de abrir un programa específico para cada tipo de datos. Luego, dentro del programa, en la barra de menú se deberían incluir estas mismas opciones para que el usuario pueda cambiar entre las diferentes versiones según el tipo de datos a sonorizar.

\subsection{Propuesta de plataforma de sonorización}

Teniendo en cuenta todas las herramientas que presenta sonoUno a la fecha, descriptas en el inicio de esta sección (\ref{sect:cap7_lineastrabajo}), se hace evidente la necesidad de poder unificar estas herramientas en pos de un mejor mantenimiento y facilidad de acceso.

En cuanto al usuario, se diseñó la página web de sonoUno que presenta toda la información y se describe en la sección \ref{sect:cap7_webinicialcompleta}. En lo que hace a unificar la interfaz web de sonorización con la de escritorio, se encontró la complejidad de que wxPython no es compatible con el despliegue web y Angular no es compatible con despliegue de escritorio. Debido a que se debe cambiar de librería o entorno de desarrollo, se plantea esta unificación como un trabajo a futuro. Una primer aproximación que se puede explorar es la posibilidad de utilizar Flutter, un entorno de desarrollo de código abierto creado por Google para aplicaciones móviles. Flutter describe en su \href{https://flutter.dev/?gclid=Cj0KCQiA8aOeBhCWARIsANRFrQHauyaonoegWcR8roNCnWSsNMEvjF_xye31sYiiPULEa67PIC0VRioaAq2wEALw_wcB&gclsrc=aw.ds}{\underline{\textcolor{blue}{página web}}} ser multiplataforma, cumpliendo con uno de los mayores objetivos de este desarrollo, en teoría podría utilizarse para desplegar el programa como una aplicación de escritorio, web y como una aplicación móvil; siendo la primer plataforma de sonorización de datos en presentar esta versatilidad. 

En base a esta propuesta de plataforma de sonorización, donde se permitirá acceder a un desarrollo desde diferentes dispositivos y con diferentes usos (recordemos que en la sección \ref{sect:cap6_server} se planteó el uso de las librerías de sonoUno para un servicio de sonorización), se decidió dejar planteados los principios de desarrollo que decantan de la presente disertación:

\begin{itemize}
    \item En cuanto a la interfaz gráfica de usuario:
        \begin{itemize}
            \item Mantener el desarrollo centrado en el usuario;
            \item Para todo cambio que requiera una modificación considerable de los elementos, se debe evaluar su diseño con usuarios previo a la implementación y en una versión beta antes de su lanzamiento oficial;
        \end{itemize}
    \item En cuanto a la programación:
        \begin{itemize}
            \item sonoUno debe continuar siendo de código abierto, multiplaforma, con un despliegue multimodal y debe asegurar el buen funcionamiento con las ayudas tecnológicas (como ser lectores de pantalla);
            \item El diseño será modular, cada nueva función se integrará al módulo correspondiente. Se deberá seguir el diseño modular actual o realizar las modificaciones necesarias al mismo, documentando y justificando el proceso;
            \item Cada funcionalidad debe estar debidamente documentada;
        \end{itemize}
    \item En cuanto a la técnica de sonorización:
        \begin{itemize}
            \item Para nuevos conjuntos de datos que requieran un algoritmo de sonorización nuevo, se deberán seguir los siguientes pasos de desarrollo:
                \begin{itemize}
                    \item Búsqueda bibliográfica para conocer si hay aproximaciones previas a la sonorización de ese tipo de datos en particular;
                    \item Describir las características que se pretender representar mediante sonido: ¿qué información se trasmite actualmente al representar dichos datos? ¿en que forma sensorial se representan los datos al usuario actualmente? ¿qué información se desea transmitir de los datos mediante sonido?
                    \item Corresponder dichas características de los datos con parámetros de sonorización: por ejemplo, si se quiere representar distancia, puede relacionarse con la duración del sonido.
                    \item Realizar pruebas para comprobar si dicha sonorización cumple con las expectativas;
                    \item Si el punto anterior no se cumple, se debe volver al punto 3, si se cumple se puede continuar;
                    \item Publicar el nuevo algoritmo de sonorización, realizar ajustes de ser necesario y comenzar con entrenamientos.
                \end{itemize}
            \item Se diseñarán entrenamientos para capacitar a las personas en el uso de la sonorización, mejorar el uso de la misma para el análisis de datos astrofísicos e investigar la efectividad y eficiencia del despliegue multisensorial.
        \end{itemize}
    \item En cuanto a las herramientas para el usuario:
        \begin{itemize}
            \item Se mantendrán actualizados los manuales de usuario e instructivos de instalación;
            \item Se diseñarán tutoriales y entrenamientos para el uso de sonoUno, a los cuales se tendrá acceso a través de la página oficial de \href{https://www.sonouno.org.ar/}{\underline{\textcolor{blue}{sonoUno}}};
            \item Se mantendrán canales de comunicación con las personas que estén utilizando sonoUno para poder brindar un soporte efectivo.
        \end{itemize}
\end{itemize}

\subsection{sonoUno multiplataforma}

Como se mencionó anteriormente y con la finalidad de hacer énfasis en esta tarea, se busca que sonoUno se pueda acceder desde diferentes plataformas. Además de poder desplegarse en diferentes sistemas operativos (MacOS, Windows y Ubuntu) y la web, se plantea la posibilidad de poder utilizarlo desde una app para dispositivos móviles.

\subsection{Despliegue multisensorial}

El software de sonorización de datos sonoUno desde un principio propuso el uso de dos sentidos (visión y audición) para el análisis de datos, basados en estudios que demuestran la mejora en la percepción cuando se utiliza más de un sentido. Como una de las propuestas futuras dentro del grupo de trabajo de sonoUno se encuentra la integración de la percepción háptica al despliegue multisensorial del programa, los primeros pasos relacionados con esta propuesta se dieron durante dos workshop de sonorización realizados en el año 2022 (uno de ellos se presentó en la sección \ref{sect:cap7_workshopjul2022}) donde se expusieron impresiones 3D de las gráficas producidas por sonoUno para los demostradores de REINFORCE. Por otro lado, también se ha mostrado como el uso de recursos táctiles también es de utilidad para el trabajo con curvas de luz de estrellas variables, espectros, o funciones matemáticas simples, recursos que pueden ser combinados con la percepción e identificación de los datos correspondientes, sonorizados.

\subsection{Entrenamientos}

En búsqueda de profundizar lo explicado en la sección \ref{sect:cap7_workshopjul2022} sobre entrenamientos en sonorización y respondiendo a una de las temáticas que más resonó en los intercambios con usuarios, se comenzó con el desarrollo de entrenamientos en sonorización dentro del grupo de trabajo de sonoUno. Recientemente, una de las integrantes presentó su tesina de grado sobre el desarrollo de una interfaz web para poder realizar entrenamientos en sonorización, y actualmente se encuentra trabajando con una beca doctoral CONICET para poder realizar la investigación en el tema, lo que asegura la continuidad de esta propuesta y su realización en el tiempo del doctorado.

\section{Conclusiones}

Durante este último capítulo, dedicado a definir la integración de las herramientas desarrolladas en una propuesta de \textbf{``Plataforma de sonorización''}, se ha logrado mostrar: el desarrollo de la página web de sonoUno (sección \ref{sect:cap7_webinicialcompleta}), su uso con lectores de pantalla (sección \ref{sect:cap7_lectorespantalla}), una primera propuesta de  entrenamiento en sonorización realizado en el marco del Curso Internacional 2022 de REINFORCE (sección \ref{sect:cap7_workshopjul2022}) y las propuestas a futuro que surgen directamente de la investigación llevada a cabo en esta tesis (sección \ref{sect:cap7_lineastrabajo}).

En cuanto a la primer etapa de integración de sonoUno se ha logrado una página web con información y ejemplos de aplicación del uso de sonoUno, con diferentes datos astronómicos y astrofísicos. Dicha página web ha sido presentada y utilizada en diferentes eventos durante los últimos años.

Por último, es destacable que las propuestas de trabajos futuros que se presentaron en este capítulo son nuevas líneas de investigación que se proyectan de la realización de esta tesis. Algunas de ellas, como los entrenamientos, ya se encuentran en desarrollo y presentan los primeros resultados.

\chapter*{Conclusión}
\addcontentsline{toc}{chapter}{Conclusión}

A partir del inicio de esta tesis doctoral en 2017, se ha evidenciado un crecimiento exponencial de programas y técnicas de sonorización de datos astronómicos y astrofísicos, entre otros, marcando el pulso de estos tiempos en que la aproximación multimodal al estudio de la naturaleza se perfila como tema de alto impacto en el ámbito científico y social. Sin embargo, el enfoque de la investigación realizada y el permanente compromiso con la inclusión y el diseño centrado en el usuario, ha posibilitado que sonoUno logre diferenciarse de sus pares y que el equipo de trabajo sea reconocido en el ámbito científico internacional como un referente \citep{zanella2022}.

Respecto de las técnicas de sonorización, si bien existen algunos antecedentes antes de los '90, la mayoría de los desarrollos y publicaciones que evidencian su uso en ciencia y divulgación son posteriores a esa fecha. De acuerdo a \citet{zanella2022}, los últimos cinco años igualó en porcentaje de desarrollos nuevos, a los 20 años previos. Esto muestra la necesidad que existe por parte de los científicos, de nuevas técnicas de despliegue de datos, reforzando la premisa de que un despliegue multisensorial permite un mejor análisis de los mismos.

Teniendo en cuenta que un despliegue multimodal permite, además, el acceso a los datos para personas con diferentes estilos sensoriales (por ejemplo, personas ciegas, a las cuales el actual despliegue de datos deja excluidas), es que se complementa en esta tesis la investigación de nuevas técnicas de visualización con el desarrollo de herramientas centradas en el usuario y accesibles desde el inicio. De esta forma se logró una herramienta que no solo mejora el actual despliego de los datos, sino que además, permite la inclusión de personas con diversidad funcional en el ámbito científico.

En cuanto a los objetivos planteados inicialmente:

\begin{itemize}
	\item Investigar sobre nuevas modalidades de acceso a los datos astronómicos y astrofísicos, promoviendo la inclusión en los espacios científicos y buscando mejorar la forma de despliegue actual;
	\item Producir herramientas y documentación orientada a mejorar el acceso a los datos promoviendo la inclusión;
\end{itemize}

Se logró desarrollar nuevas modalidades de acceso a los datos mediante la sonorización, pudiendo mostrar el despliegue gráfico de la misma forma que se muestra en otros programas o bases de datos, siendo acompañado dicho despliegue por la novedad de la sonorización de los datos. Se debe tener en cuenta que todas las sonorizaciones producidas en el marco de esta tesis doctoral representan fielmente los datos, no se les aplica ningún tipo de ``maquillaje'' para que suene bonito, lo que permitió mantener claro que es un desarrollo que busca insertarse en el ámbito científico. Las sonorizaciones de tablas de dos o más columnas se basaron en las sonorizaciones realizadas por el software xSonify. No obstante, las sonorizaciones de partículas fueron diseñadas desde cero como parte de esta tesis.

Las investigaciones realizadas con usuarios y otros grupos de investigación, sobre el uso y la accesibilidad del software sonoUno para el análisis de datos, han sido favorables, demostrando que el programa presenta un marco de trabajo centrado en el usuario y en la inclusión. En base a estos intercambios también se han confeccionado guías de trabajo y preparado documentación sobre recomendaciones de accesibilidad que se encuentran disponibles en la \href{https://www.sonouno.org.ar/}{\underline{\textcolor{blue}{web de sonoUno}}} (sección \ref{sect:recomendaciones_publicadas}). Todas estas actividades han sido acompañadas por análisis de normativas vigentes para mantener un marco de trabajo lo más completo y actualizado posible.

En base a los continuos intercambios con usuarios, se materializó la necesidad de que el desarrollo fuera más allá de una simple interfaz gráfica/sonora de escritorio. Así se comenzó con el desarrollo de una interfaz gráfica web que alcanzó un alto nivel de funcionalidades, igualando y mejorando en algunos aspectos al inicial desarrollo de escritorio. El proyecto sonoUno y en particular la investigación de esta tesis, lograron mantener y llevar a cabo sus objetivos iniciales, promoviendo la inclusión en el ámbito científico y convirtiéndose en un referente de sonorización para el análisis multisensorial de datos. 

Si bien sonoUno cuenta con varios casos de uso en donde se trabaja en educación y divulgación en el ámbito científico (ver sección \ref{sect:casos_uso}), adicionalmente se encuentran otros casos de uso como su implementación en Raspberry Py, realizado por \href{https://github.com/Physicslibrary/SonoUno-Raspberry-Pi}{\underline{\textcolor{blue}{Hartwell Fong}}}, y se puede afirmar que sigue siendo un desarrollo enmarcado en el ámbito científico. Así mismo, otro ejemplo que se conoció en los últimos meses fue el proyecto \href{https://stephenserjeant.github.io/sounds-of-bears/}{\underline{\textcolor{blue}{The Sounds of BEARS}}}, donde en su página web incluyen videos que muestran el uso de sus datos en la versión web de sonoUno.

Como se ha mostrado, desde el inicio se ha trabajado con datos reales, descargados desde grandes bases de datos (ver sección \ref{sect:cap6_vescritorio}) o proporcionados por grandes facilidades astrofísicas (ver sección \ref{sect:cap6_sonif_reinforce}). De las encuestas realizadas en los últimos años, la mayoría de los científicos que han utilizado el software estarían dispuestos/as a utilizarlo en sus investigaciones, siempre que el programa se mantenga en desarrollo y continúe mejorando. En base a esta última premisa, resaltó la necesidad de entrenamiento, para una mejor y más profunda comprensión sobre esta nueva modalidad de despliegue multisensorial de datos.

Es así que la línea de investigación que se presenta en este trabajo, deja establecida la posibilidad de mejora continua del software (ver sección \ref{sect:cap7_lineastrabajo}) pero, además sienta las bases para trabajos futuros en diferentes líneas de investigación consideradas estratégicas en la República Argentina, como por ejemplo: 

\begin{itemize}
	\item Análisis de percepción: durante el curso internacional de entrenamiento realizado en el marco de la escuela de verano 2022 del proyecto REINFORCE (ver sección \ref{sect:cap7_workshopjul2022}), se realizó una encuesta donde se evidenció la necesidad de entrenamientos, entre otras cosas. Si bien dicho evento se enfocó mayormente en las sonorizaciones de nuevos datos, como las partículas del LHC, la necesidad de entrenamiento no se limita a los nuevos diseños de sonorizaciones. Durante todo el desarrollo de esta tesis y a lo largo de diferentes encuestas, la necesidad de entrenamiento tanto en el uso de la técnica de sonorización, como en el uso de la herramienta sonoUno, quedaron de manifiesto a través de los comentarios de los usuarios. Incluso durante el último workshop de sonorización (\textit{``The Audible Universe 2''}) se discutió sobre la necesidad de entrenamientos para instruir a las personas en el uso del despliegue multisensorial de datos. Es importante destacar que relacionado a este tema, se ha presentado en la Carrera de Bioingeniería de esta Universidad un trabajo final de grado titulado \textit{``Desarrollo de entrenamientos para el análisis multisensorial de datos''}, donde se describen los primeros pasos en el desarrollo y puesta en marcha de entrenamientos específicos, y la elaboración de una web propia para el despliegue de dichos entrenamientos. Adicionalmente, la misma estudiante dentro del grupo de investigación donde se desarrolló esta tesis, comenzó en abril del 2023 su tesis doctoral con beca interna CONICET, titulada \textit{``Investigación interdisciplinaria de la efectividad de los entrenamientos en sonorización de datos astronómicos en el marco de la astrofísica multimodal''}.
	
	\item Nuevas técnicas de diseño centrado en el usuario: entre ellas se puede mencionar las interfaces que se adapten al usuario, que aprendan de la interacción con el mismo e inteligencia artificial, entre otras. Si bien se encuentran numerosos esfuerzos por alcanzar interfaces centradas en el usuario y que promuevan la inclusión, pocos de ellos se mantienen a largo plazo, ya sea por falta de fondos o por la dificultad de encontrar una técnica que se acomode a todos. Si bien existen estándares que describen buenas prácticas para abordar la accesibilidad, no se logra evidenciar un consenso sobre la mejor forma de desplegar interfaces humano-computadora que cumplan con los criterios de inclusión y sean atractivas para la mayoría de los usuarios.
	
	\item Nuevas técnicas de sonorización: si bien se ha avanzado mucho en materia de sonorización, falta mucho para lograr dar respuesta a los numerosos estilos de visualización de datos presentes en la actualidad, como por ejemplo, la sonorización de imágenes. Esto, sin perder de vista que lo que se busca es dar sentido a los datos a través de la sonorización: no se propone sonorizar un despliegue visual, sino representar un conjunto de datos a través de sonorización para poder analizarlos utilizando las diferentes entradas sensoriales del organismo humano (por ejemplo, no se utilizaría la misma técnica de sonorización para representar la imagen de una galaxia que la de un Glitch, eso es debido a que son conceptos diferentes e incluso su representación gráfica se origina de forma diferente).
	
	\item Tecnologías asistivas e Inteligencia Artificial: durante el desarrollo de esta tesis se tuvieron en cuenta los lectores de pantalla de acceso libre y gratuito de cada sistema operativo, pero no son las únicas tecnologías asistivas disponibles. Dentro de este marco, se deberían incluir todas las tecnologías asistivas disponibles analizando la factibilidad de su uso con la herramienta sonoUno, como sería el caso de Alexa, teclados accesibles (cambia su diseño de acuerdo a la necesidad), tableros de comunicación, reconocimiento de voz o gestual, por mencionar algunas. Sin duda este trabajo excede los límites de esta tesis y es por ello que representa una de las líneas de trabajo a futuro que se desprende de esta investigación.
\end{itemize}

Las posibilidades de contribución en las mencionadas lineas de investigación constituyen un legado de alto impacto en el marco de una sociedad que pugna por la equidad, la inclusión y la sostenibilidad de estos principios. En base a las recomendaciones de la Naciones Unidas y basándonos en el \textit{``Estudio Nacional sobre el Perfil de las Personas con Discapacidad''} es evidente que se necesita tomar acción a largo plazo para asegurar la equidad y reforzando el principio de que las diversidades funcionales que presentan las personas no deben ser un limitante para su desarrollo profesional. 

Los objetivos globales de equidad e inclusión requieren de marcos de referencia que emergen de estudios formales, que puedan ser utilizados como facilitadores para la remoción de las barreras que impiden el crecimiento y progreso social.

\addcontentsline{toc}{chapter}{Bibliografía}
\bibliography{biblio}

\chapter*{Índice de links}
\addcontentsline{toc}{chapter}{Índice de links}

\noindent Página 31 (link) (visitado en Abril de 2023)

\url{https://www.w3.org/standards/webdesign/accessibility} \\

\noindent Página 39 (aquí) (visitado en Abril de 2023)

\url{https://lweb.cfa.harvard.edu/sdu/} \\

\noindent Página 41 (sonoUno)  (visitado en Abril de 2023)

\url{https://www.sonouno.org.ar/reinforce-demonstrators/} \\

\noindent Página 47 (link)  (visitado en Abril de 2023)

\url{https://youtu.be/OCnK\_NkYXIU} \\

\noindent Página 60 (link) (visitado en Abril de 2023)

\url{https://www.audiouniverse.org/} \\

\noindent Página 64 (página web) (visitado en Abril de 2023)

\url{https://ccrma.stanford.edu/~cc/sonify/} \\

\noindent Página 64 (WebChucK IDE)  (visitado en Abril de 2023)

\url{https://github.com/ccrma/webchuck} \\

\noindent Página 94 (WCAG 2.0) (visitado en Abril de 2023)

\url{https://www.w3.org/TR/WCAG20/} \\

\noindent Página 119 (Python) (visitado en Abril de 2023)

\url{https://www.python.org/} \\

\noindent Página 119 (wxPython) (visitado en Abril de 2023)

\url{https://www.wxpython.org/} \\

\noindent Página 119 (Matplotlib) (visitado en Abril de 2023)

\url{https://matplotlib.org/} \\ \\

\noindent Página 120 (Numpy) (visitado en Abril de 2023)

\url{https://numpy.org/} \\

\noindent Página 120 (Pandas) (visitado en Abril de 2023)

\url{https://pandas.pydata.org/} \\

\noindent Página 120 (Mingus) (visitado en Abril de 2023)

\url{https://bspaans.github.io/python-mingus/} \\

\noindent Página 120 (FluidSynth) (visitado en Abril de 2023)

\url{https://www.fluidsynth.org/} \\

\noindent Página 138 (SDSS J151806.13+424445.0) (visitado en Abril de 2023)

\url{http://skyserver.sdss.org/dr15/en/tools/quicklook/summary.aspx?} \\

\noindent Página 138 (Pierre Auger) (visitado en Abril de 2023)

\url{https://labdpr.cab.cnea.gov.ar/opendata/data.php} \\

\noindent Página 140 (galaxia) (visitado en Abril de 2023)

\url{https://www.sonouno.org.ar/wp-content/uploads/sites/9/2021/12/SDSS-J151806.13424445.0_flux_fmax2003.wav} \\

\noindent Página 140 (completo) (visitado en Abril de 2023)

\url{https://drive.google.com/file/d/1skXWhcQvocmXJ_7n5RyEWSPQVKm2r7f5/view?usp=sharing} \\

\noindent Página 140 (recorte) (visitado en Abril de 2023)

\url{https://www.sonouno.org.ar/wp-content/uploads/sites/9/2021/10/auger_public_2016_11_12_range1285973511-1294550092_sound.wav} \\

\noindent Página 146 (Sloan Digital Sky Survey) (visitado en Abril de 2023)

\url{http://skyserver.sdss.org/dr15/en/tools/quicklook/summary.aspx?ra=179.8036794828000&dec=-00.5238033275900} \\

\noindent Página 165 (Sensing the Dynamic Universe) (visitado en Abril de 2023)

\url{https://lweb.cfa.harvard.edu/sdu/} \\

\noindent Página 166 (Joe Palmo) (visitado en Abril de 2023)

\url{https://github.com/joepalmo/sonoUno} \\ \\ \\

\noindent Página 166 (web) (visitado en Abril de 2023)

\url{https://pweb.cfa.harvard.edu/research/sensing-dynamic-universe} \\

\noindent Página 166 (aquí) (visitado en Abril de 2023)

\url{https://lweb.cfa.harvard.edu/sdu/} \\

\noindent Página 167 (Figura 5.1 - Cefeida) (visitado en Abril de 2023)

\url{https://lweb.cfa.harvard.edu/sdu/cepheids.html} \\

\noindent Página 167 (Figura 5.1 - Binaria Eclipsante) (visitado en Abril de 2023)

\url{https://lweb.cfa.harvard.edu/sdu/eclipsingbinaries.html} \\

\noindent Página 167 (nota) (visitado en Abril de 2023)

\url{https://astronomiayeducacion.org/taller-de-introduccion-a-la-sonificacion/?cn-reloaded=1&cn-reloaded=1} \\

\noindent Página 167 (web) (visitado en Abril de 2023)

\url{https://astronomiayeducacion.org/taller-2-de-sonificacion-descubriendo-el-universo/?cn-reloaded=1} \\

\noindent Página 167 (Twitter) (visitado en Abril de 2023)

\url{https://twitter.com/aaecastro/status/1527991640924868608} \\

\noindent Página 168 (Figura 5.2 - Cefeida) (visitado en Abril de 2023)

\url{https://lweb.cfa.harvard.edu/sdu/cepheids.html} \\

\noindent Página 168 (Figura 5.2 - Binaria Eclipsante)  (visitado en Abril de 2023)

\url{https://lweb.cfa.harvard.edu/sdu/eclipsingbinaries.html} \\

\noindent Página 169 (Figura 5.4 - noticia) (visitado en Abril de 2023)

\url{https://astronomiayeducacion.org/taller-2-de-sonificacion-descubriendo-el-universo/?cn-reloaded=1} \\

\noindent Página 171 (REINFORCE) (visitado en Abril de 2023)

\url{https://reinforceeu.eu/} \\

\noindent Página 173 (Zooniverse) (visitado en Abril de 2023)

\url{https://www.zooniverse.org/} \\

\noindent Página 173 (Zooniverse) (visitado en Abril de 2023)

\url{https://www.zooniverse.org/projects/reinforce/gwitchhunters} \\

\noindent Página 173 (link) (visitado en Abril de 2023)

\url{https://www.sonouno.org.ar/glitch-1126409678-84375/} \\

\noindent Página 173 (KM3NeT) (visitado en Abril de 2023)

\url{https://www.km3net.org/} \\

\noindent Página 173 (link) (visitado en Abril de 2023)

\url{https://www.zooniverse.org/projects/reinforce/deep-sea-explorers} \\

\noindent Página 173 (link) (visitado en Abril de 2023)

\url{https://youtu.be/pkiGdZu5gEo} \\

\noindent Página 174 (Figura 6.1(a) - link) (visitado en Abril de 2023)

\url{https://www.zooniverse.org/projects/reinforce/deep-sea-explorers/classify} \\

\noindent Página 174 (Figura 6.1(b) - bioacústica 1)  (visitado en Abril de 2023)

\url{https://www.zooniverse.org/projects/reinforce/deep-sea-explorers} \\

\noindent Página 174 (Figura 6.1(b) - link) (visitado en Abril de 2023)

\url{https://youtu.be/pkiGdZu5gEo} \\

\noindent Página 174 (Zooniverse) (visitado en Abril de 2023)

\url{https://www.zooniverse.org/projects/reinforce/new-particle-search-at-cern} \\

\noindent Página 176 (Figura 6.3 - link) (visitado en Abril de 2023)

\url{https://youtu.be/EYhcdyO2w2I} \\

\noindent Página 177 (Figura 6.4 - link) (visitado en Abril de 2023)

\url{https://youtu.be/MwZ0_EEuD_g} \\

\noindent Página 178 (Figura 6.5 - link a video) (visitado en Abril de 2023)

\url{https://www.sonouno.org.ar/glitch-1126409678-84375/} \\

\noindent Página 179 (link) (visitado en Abril de 2023)

\url{https://www.sonouno.org.ar/glitch-1126409678-84375/} \\

\noindent Página 179 (link) (visitado en Abril de 2023)

\url{https://youtu.be/x6Abb2Ekb8s} \\

\noindent Página 179 (Figura 6.6 - link) (visitado en Abril de 2023)

\url{https://youtu.be/x6Abb2Ekb8s} \\

\noindent Página 179 (link) (visitado en Abril de 2023)

\url{https://youtu.be/pkiGdZu5gEo} \\

\noindent Página 183 (link) (visitado en Abril de 2023)

\url{https://youtu.be/XMaYIJkJIHg} \\

\noindent Página 186 (existencia de muón) (visitado en Abril de 2023)

\url{https://youtu.be/269lsyC_bGg} \\

\noindent Página 186 (caso dudoso donde no existe muón) (visitado en Abril de 2023)

\url{https://youtu.be/1nDW8XGV1PQ} \\

\noindent Página 187 (Sloan Digital Sky Survey) (visitado en Abril de 2023)

\url{https://www.sdss4.org/} \\

\noindent Página 187 (Pierre Auger) (visitado en Abril de 2023)

\url{https://opendata.auger.org/} \\

\noindent Página 187 (ASAS-SN Variable) (visitado en Abril de 2023)

\url{https://asas-sn.osu.edu/variables} \\

\noindent Página 188 (link aquí) (visitado en Abril de 2023)

\url{https://www.sonouno.org.ar/comic-rays-2005-to-2017/} \\

\noindent Página 190 (página web) (visitado en Abril de 2023)

\url{https://www.sdss.org/collaboration/} \\

\noindent Página 190 (link) (visitado en Abril de 2023)

\url{https://skyserver.sdss.org/dr18/VisualTools/navi} \\

\noindent Página 190 (vista rápida) (visitado en Abril de 2023)

\url{https://skyserver.sdss.org/dr18/VisualTools/quickobj} \\

\noindent Página 190 (galería) (visitado en Abril de 2023)

\url{https://www.sonouno.org.ar/galaxy-sdss-j115845-43-002715-7/} \\

\noindent Página 191 (aquí) (visitado en Abril de 2023)

\url{https://www.sonouno.org.ar/galaxy-sdss-j115845-43-002715-7/} \\ \\

\noindent Página 192 (Figura 6.10(a) - video) (visitado en Abril de 2023)

\url{https://www.sonouno.org.ar/wp-content/uploads/sites/9/2022/05/J115845.43-002715.7_video.mp4} \\

\noindent Página 192 (Figura 6.10(b) - sonido) (visitado en Abril de 2023)

\url{https://www.sonouno.org.ar/wp-content/uploads/sites/9/2021/12/SDSS-J115845.43-002715.7_flux_fmax2003.wav} \\

\noindent Página 192 (Figura 6.10(c) - sonido) (visitado en Abril de 2023)

\url{https://www.sonouno.org.ar/wp-content/uploads/sites/9/2021/12/SDSS-J115845.43-002715.7_bestfit_fmax2003.wav} \\

\noindent Página 192 (Figura 6.10(d) - sonido) (visitado en Abril de 2023)

\url{https://www.sonouno.org.ar/wp-content/uploads/sites/9/2021/12/SDSS-J115845.43-002715.7_skyflux_fmax2003.wav} \\

\noindent Página 193 (Figura 6.11 - link) (visitado en Abril de 2023)

\url{https://youtu.be/5TTdcgeDaPw} \\

\noindent Página 193 (página web) (visitado en Abril de 2023)

\url{https://visitantes.auger.org.ar/} \\

\noindent Página 193 (link) (visitado en Abril de 2023)

\url{https://labdpr.cab.cnea.gov.ar/opendata/} \\

\noindent Página 193 (datos) (visitado en Abril de 2023)

\url{https://labdpr.cab.cnea.gov.ar/opendata/data.php} \\

\noindent Página 194 (Figura 6.12(a) - sonido) (visitado en Abril de 2023)

\url{https://www.sonouno.org.ar/wp-content/uploads/sites/9/2021/12/auger2017_sound_fulldata.wav} \\

\noindent Página 194 (Figura 6.12(b) - sonido) (visitado en Abril de 2023)

\url{https://www.sonouno.org.ar/wp-content/uploads/sites/9/2021/12/auger2017\_sound\_1124471483-1125034238.wav} \\

\noindent Página 194 (link) (visitado en Abril de 2023)

\url{https://www.sonouno.org.ar/comic-rays-2005-to-2017/} \\

\noindent Página 194 (link) (visitado en Abril de 2023)

\url{https://youtu.be/3GR4gcx92kU} \\ \\

\noindent Página 195 (Figura 6.13(a) - base de datos) (visitado en Abril de 2023)

\url{https://asas-sn.osu.edu/variables/753bdd73-38a7-5e43-b6c0-063292c7f28d} \\

\noindent Página 195 (Figura 6.13(b) - base de datos) (visitado en Abril de 2023)

\url{https://asas-sn.osu.edu/variables/dfa51488-c6b7-5a03-abd4-df3c28273250} \\

\noindent Página 195 (Figura 6.13(c) - sonido) (visitado en Abril de 2023)

\url{https://www.sonouno.org.ar/wp-content/uploads/sites/9/2023/01/CG-Cas-Cepheid.csv_sound.wav} \\

\noindent Página 195 (Figura 6.13(d) - sonido) (visitado en Abril de 2023)

\url{https://www.sonouno.org.ar/wp-content/uploads/sites/9/2023/01/RW-Phe-Eclipsing-Binary.csv_sound.wav} \\

\noindent Página 195 (ASAS-SN) (visitado en Abril de 2023)

\url{https://asas-sn.osu.edu/variables} \\

\noindent Página 195 (link) (visitado en Abril de 2023)

\url{https://asas-sn.osu.edu/variables/753bdd73-38a7-5e43-b6c0-063292c7f28d} \\

\noindent Página 195 (link) (visitado en Abril de 2023)

\url{https://asas-sn.osu.edu/variables/dfa51488-c6b7-5a03-abd4-df3c28273250} \\

\noindent Página 196 (link) (visitado en Abril de 2023)

\url{https://github.com/sonoUnoTeam/sonoUno/blob/master/sonoUno/sonify_bash_lightcurve.py} \\

\noindent Página 207 (link) (visitado en Abril de 2023)

\url{https://pypi.org/project/sonounolib/} \\

\noindent Página 207 (link) (visitado en Abril de 2023)

\url{https://pchanial.gitlab.io/sonouno-library/} \\

\noindent Página 208 (Figura 6.21 - link) (visitado en Abril de 2023)

\url{https://pchanial.gitlab.io/sonouno-library/demo_pyscript.html} \\

\noindent Página 211 (link) (visitado en Abril de 2023)

\url{https://doi.org/10.5281/zenodo.7717567} \\

\noindent Página 211 (GitHub) (visitado en Abril de 2023)

\url{https://github.com/sonoUnoTeam/sonoUno-server} \\

\noindent Página 211 (link) (visitado en Abril de 2023)

\url{http://api.sonouno.org.ar/redoc} \\

\noindent Página 217 (sonoUno) (visitado en Abril de 2023)

\url{https://www.sonouno.org.ar/} \\

\noindent Página 217 (link) (visitado en Abril de 2023)

\url{https://reinforce.sonouno.org.ar/} \\

\noindent Página 218 (Manual de Instalación) (visitado en Abril de 2023)

\url{https://www.sonouno.org.ar/installation-manuals/} \\

\noindent Página 218 (Manual de Usuario) (visitado en Abril de 2023)

\url{https://www.sonouno.org.ar/usermanual/} \\

\noindent Página 219 (Zenodo) (visitado en Abril de 2023)

\url{https://zenodo.org/} \\

\noindent Página 219 (escritorio) (visitado en Abril de 2023)

\url{https://doi.org/10.5281/zenodo.7717725} \\

\noindent Página 220 (aquí) (visitado en Abril de 2023)

\url{https://pypi.org/project/sonounolib/} \\

\noindent Página 222 (link) (visitado en Abril de 2023)

\url{https://youtu.be/Pc_K3uGa2mw} \\

\noindent Página 223 (link) (visitado en Abril de 2023)

\url{https://youtu.be/458_sWX-uw4} \\

\noindent Página 223 (NVDA) (visitado en Abril de 2023)

\url{https://youtu.be/OCnK_NkYXIU} \\

\noindent Página 223 (Voice Over) (visitado en Abril de 2023)

\url{https://youtu.be/rj2ELDwuM40} \\

\noindent Página 225 (link) (visitado en Abril de 2023)

\url{http://reinforce.ea.gr/international-training-course/} \\ \\

\noindent Página 235 (2021) (visitado en Abril de 2023)

\url{https://www.lorentzcenter.nl/the-audible-universe.html} \\

\noindent Página 235 (2022) (visitado en Abril de 2023)

\url{https://www.lorentzcenter.nl/the-audible-universe-2.html} \\

\noindent Página 239 (página web) (visitado en Abril de 2023)

\url{http://flutter.dev/} \\

\noindent Página 240 (sonoUno) (visitado en Abril de 2023)

\url{https://www.sonouno.org.ar/} \\

\noindent Página 243 (web de sonoUno) (visitado en Abril de 2023)

\url{https://www.sonouno.org.ar/} \\

\noindent Página 243 (Hartwell Fong) (visitado en Abril de 2023)
 
\url{https://github.com/Physicslibrary/SonoUno-Raspberry-Pi} \\
 
\noindent Página 243 (The Sounds of BEARS)  (visitado en Abril de 2023)

\url{https://stephenserjeant.github.io/sounds-of-bears/} \\

\appendix
\chapter{Documentos relativos al análisis normativo ISO}

Se incluyen en este apéndice el paper publicado y los documentos relacionados con el análisis normativo ISO 9241-171:2008. En detalle se encontrará primero el paper, luego una copia del Anexo C de la norma mencionada, que contiene la tabla con la cual se realizó el análisis. A continuación se pondrán a disposición todas las planillas excel obtenidas por evaluador y por programa, para finalizar con el documento obtenido del grupo 2 con todo el detalle del análisis realizado por los dos grupos.

\section{Artículos publicados sobre el análisis con la norma ISO 9241-171:2008}
\label{ap:A_papers}

Se incluyen a continuación dos proceedings presentados en reuniones científicas (relacionados con el análisis de la normativa) y un paper que contiene el análisis ISO completo.

En primer lugar se incluye el proceeding presentado en el XXI Congreso Argentino de Bioingeniería (SABI-2017), el cual contiene el primer análisis realizado con la norma ISO 9241-171:2008. A mediados de 2017, también se realizó una presentación en el Simposio Internacional de Educación en Astronomía y Astrobiología (ISE2A) que se incluye aquí en segundo lugar. Por último, se adjunta el paper con la descripción del análisis completo actualizado en el año 2020 y publicado en la revista \textit{``International Journal of Sociotechnology and Knowledge Development''}.

\includepdf[pages=-,scale=0.8]{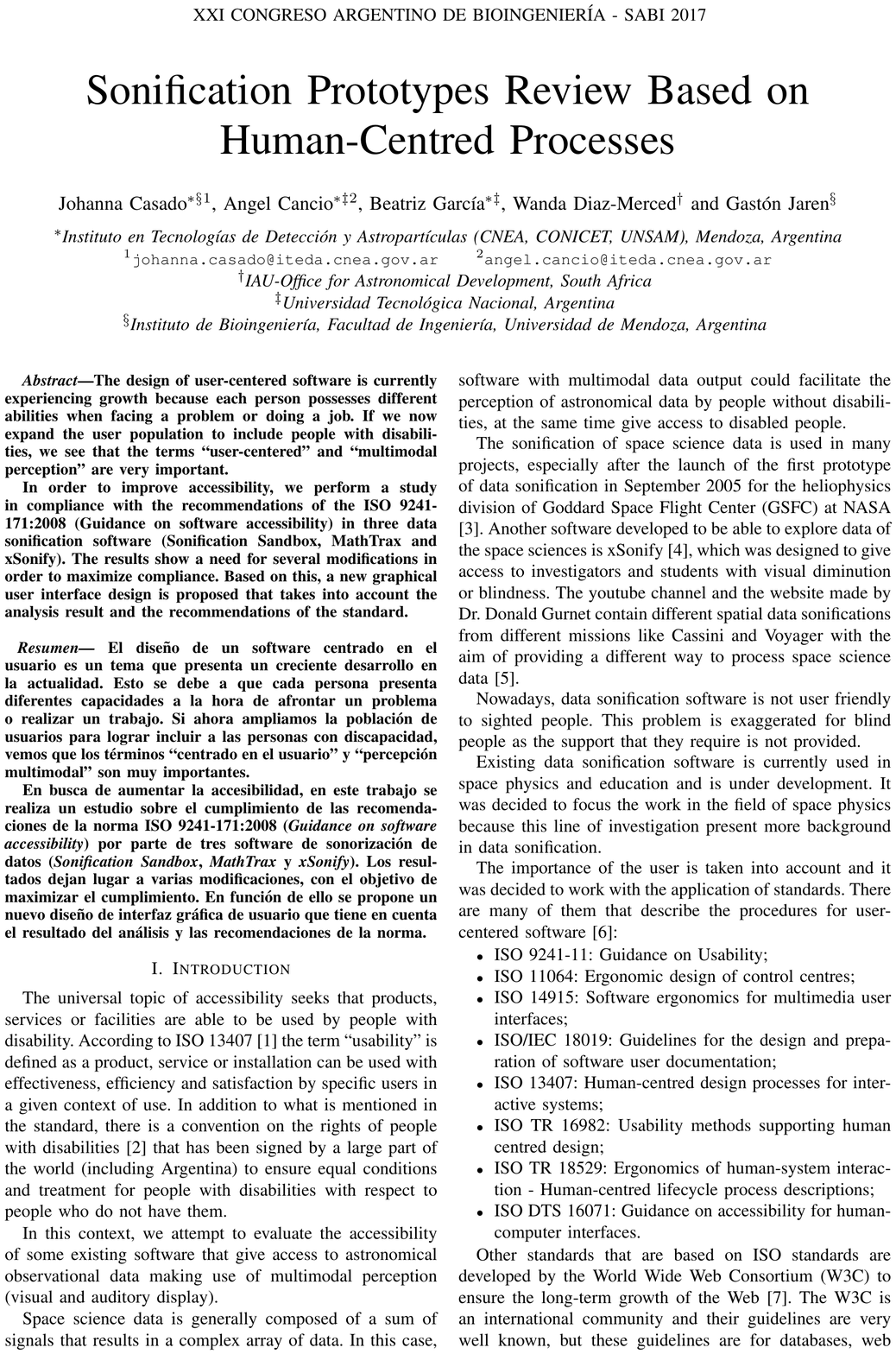}

\includepdf[pages=-,scale=0.8]{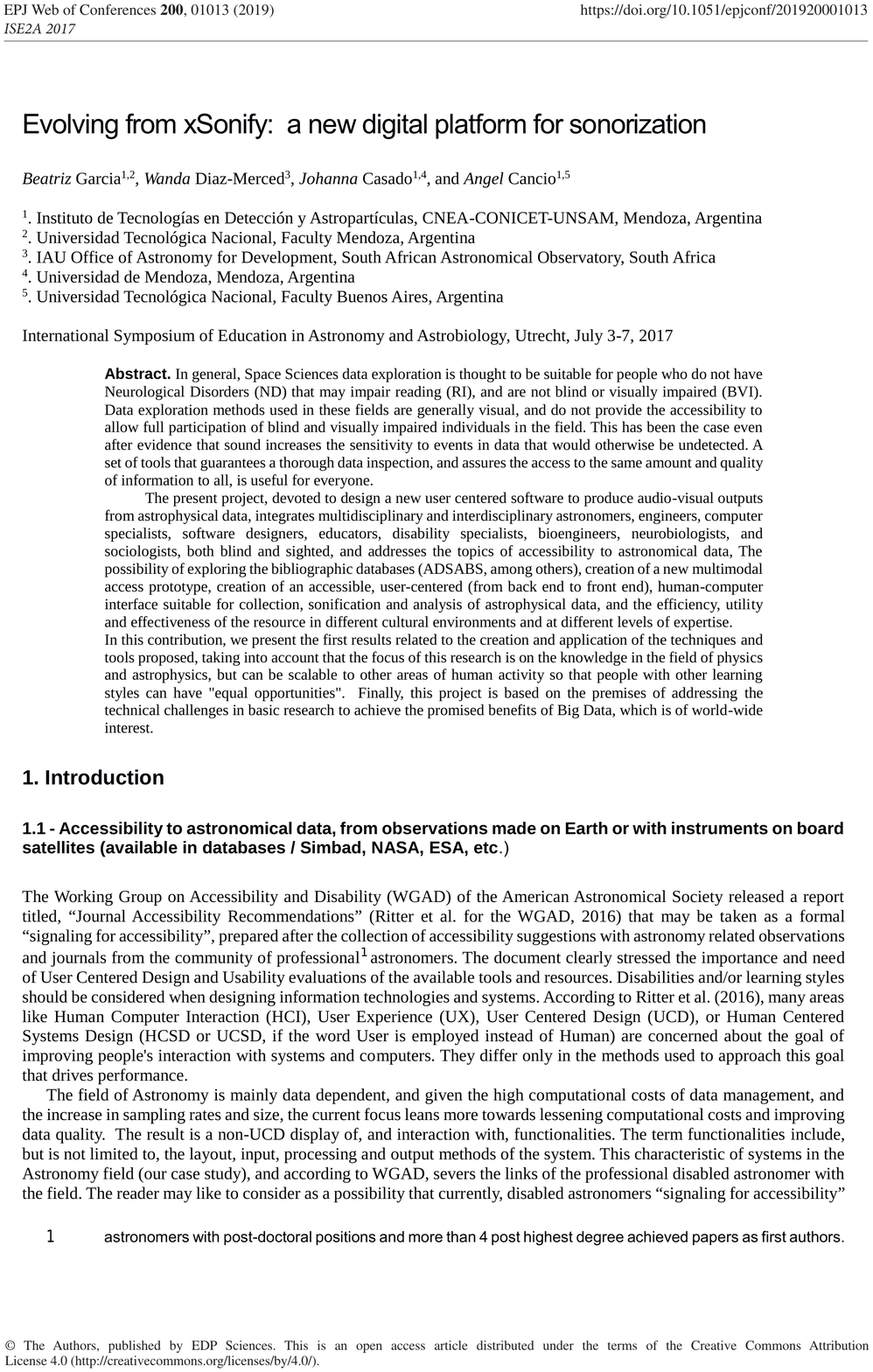}

\includepdf[pages=-,scale=0.8]{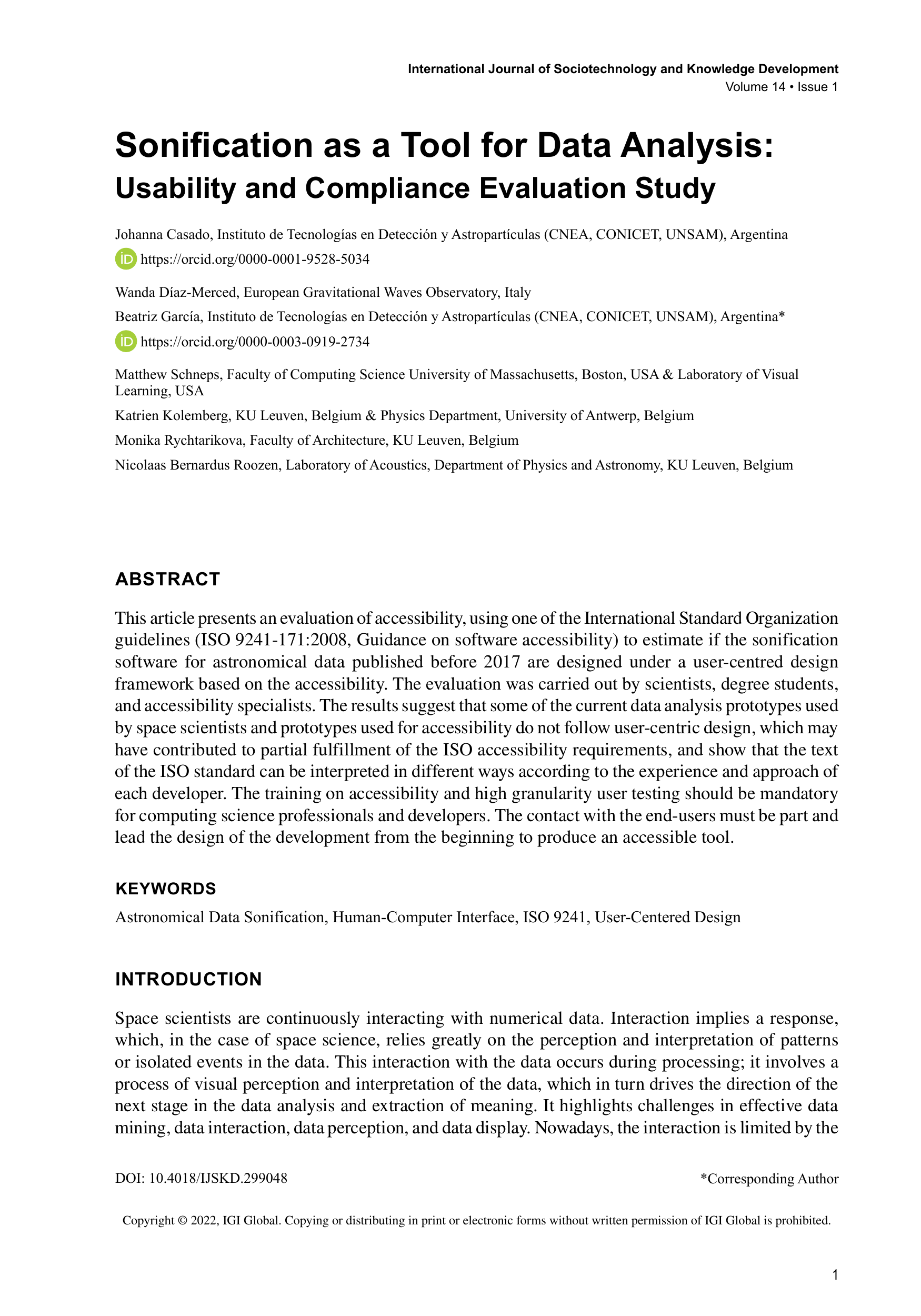}

\section{Copia del Anexo C de la norma ISO 9241-171:2008}
\label{ap:A_anexoc_enblanco}

Se incluye a continuación las 9 hojas del Anexo C de la norma ISO sin completar.

\includepdf[pages=-,scale=0.8]{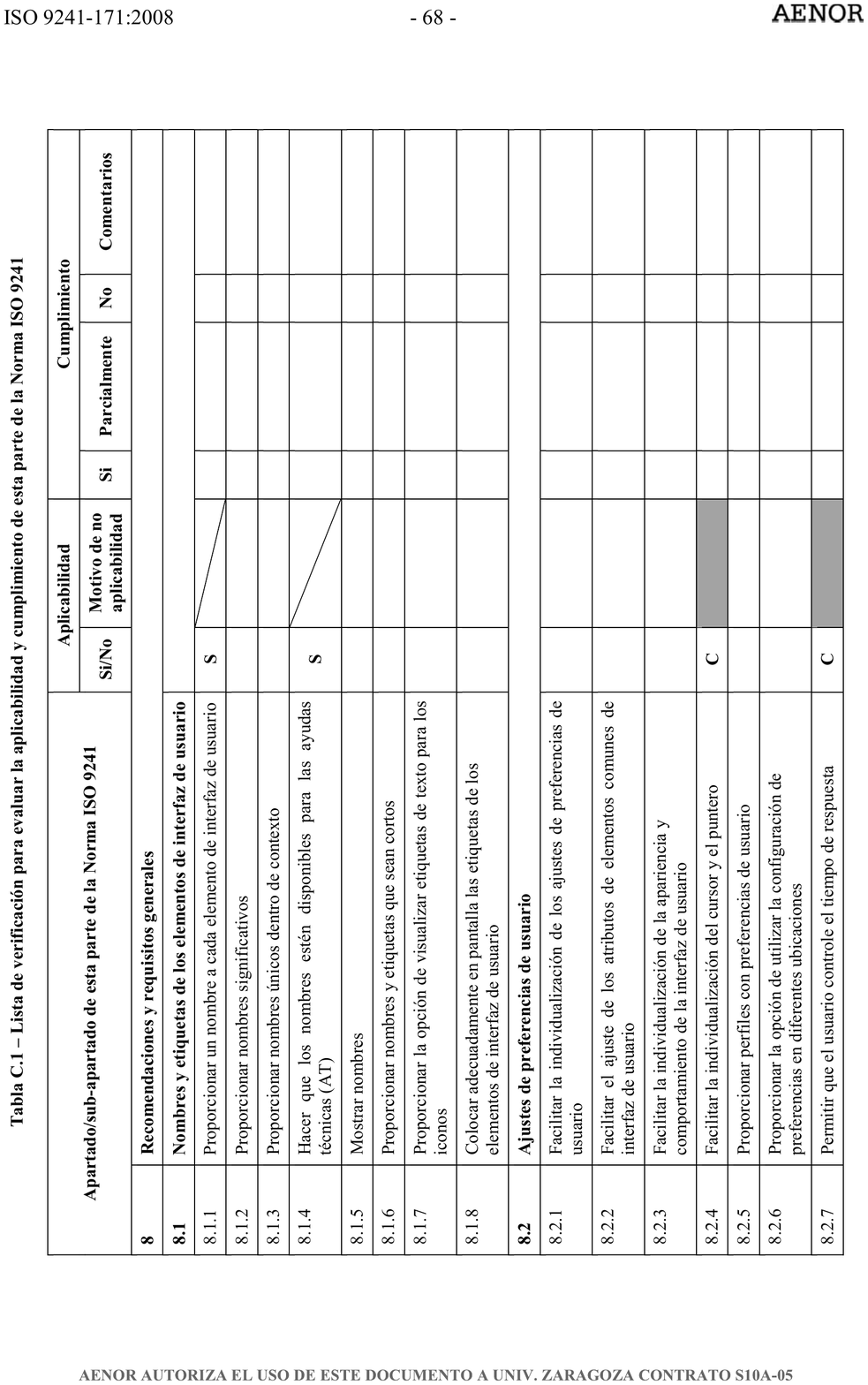}

\section{Planillas excel obtenidas por evaluador y por programa}
\label{ap:A_anexoc_planillasXeval}

Aquí se incluirán las planillas obtenidas por evaluador, y del mismo evaluador, separando por programa.

\subsection{Planillas del Evaluador 1}

\begin{enumerate}
    \item Software MathTrax en MacOS
    \item Software xSonify en MacOS
    \item Software Sonification Sandbox en MacOS
    \item Software MathTrax en Windows
    \item Software xSonify en Windows
    \item Software Sonification Sandbox en Windows
\end{enumerate}


\includepdf[pages=-,scale=0.8]{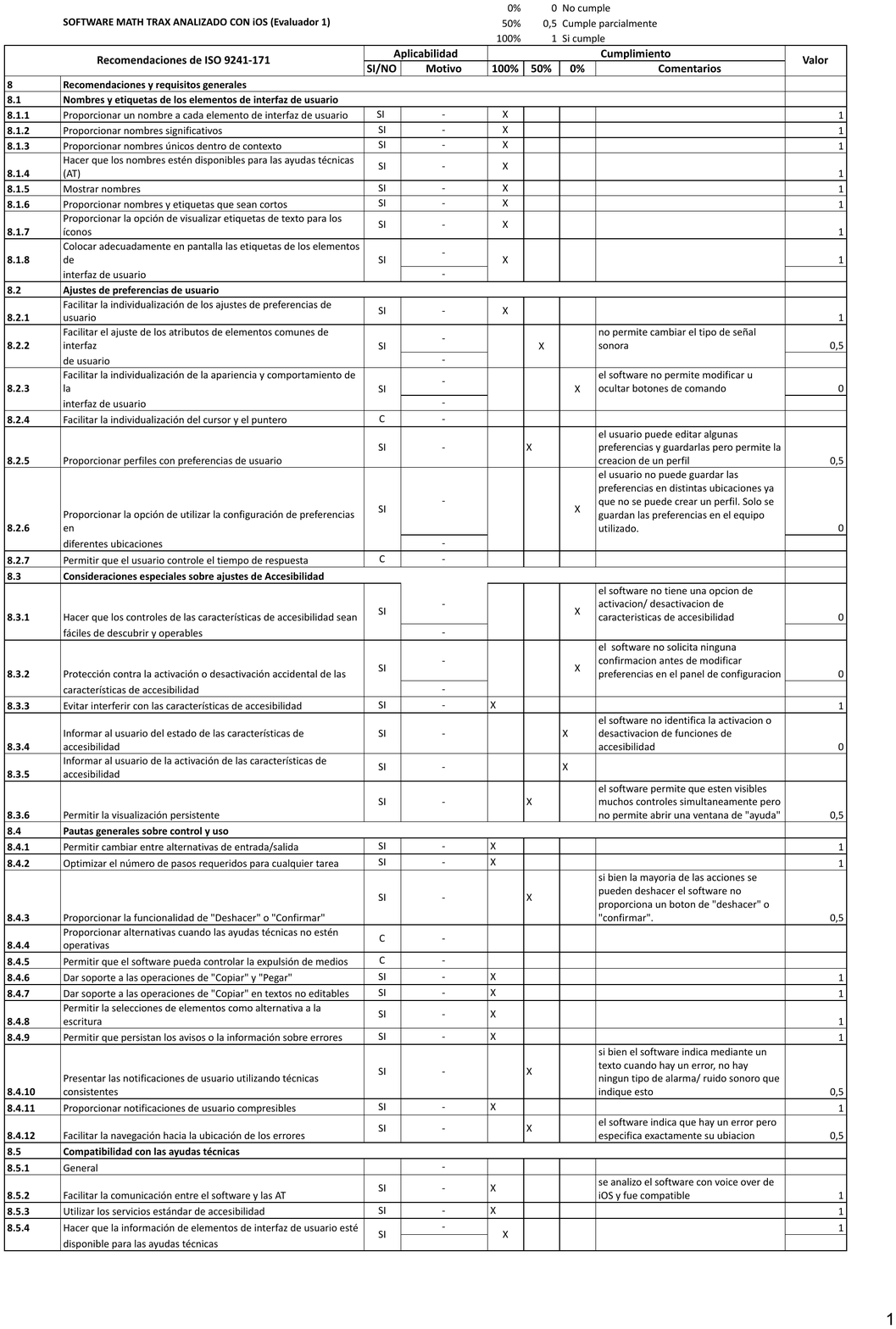}
\includepdf[pages=-,scale=0.8]{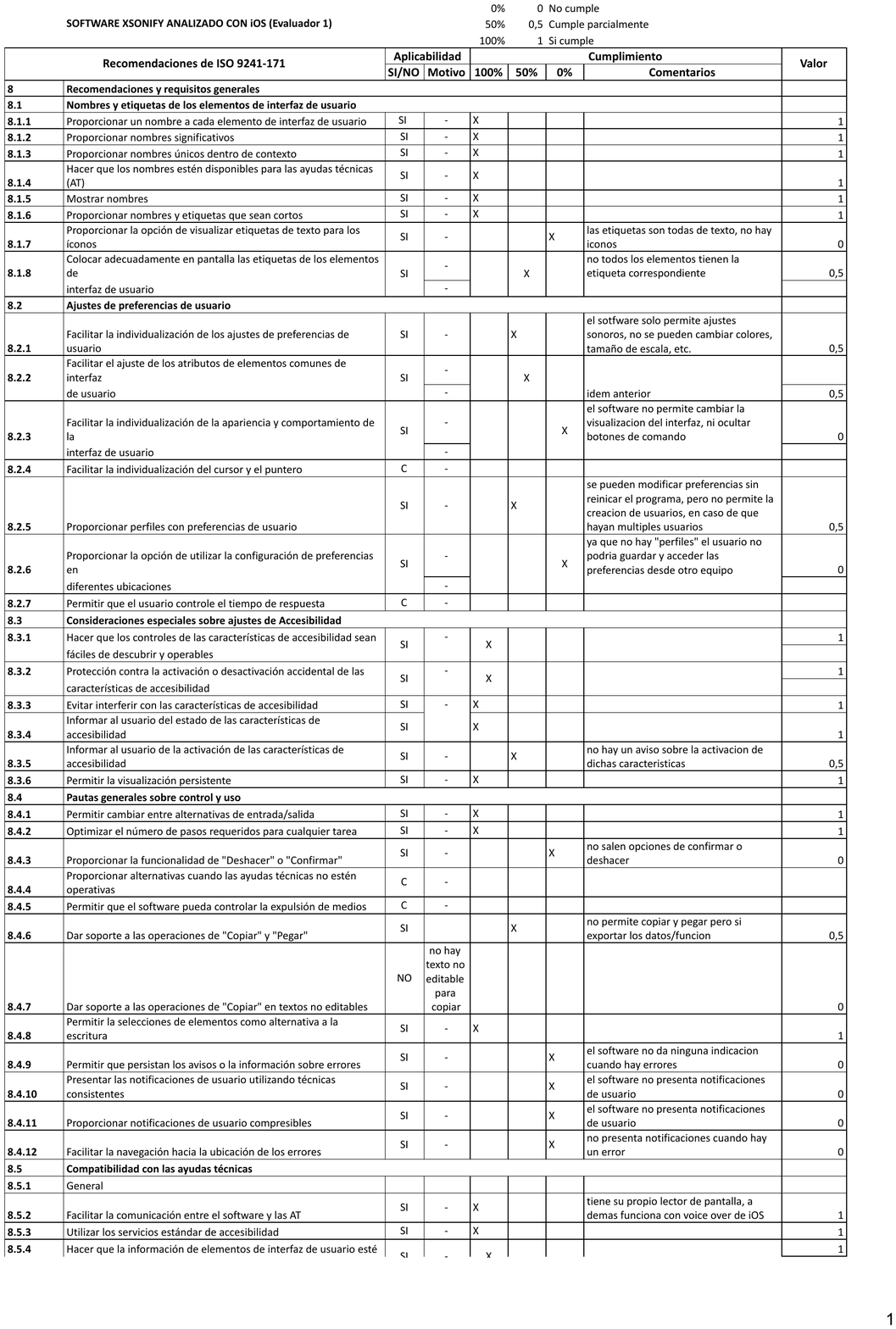}
\includepdf[pages=-,scale=0.8]{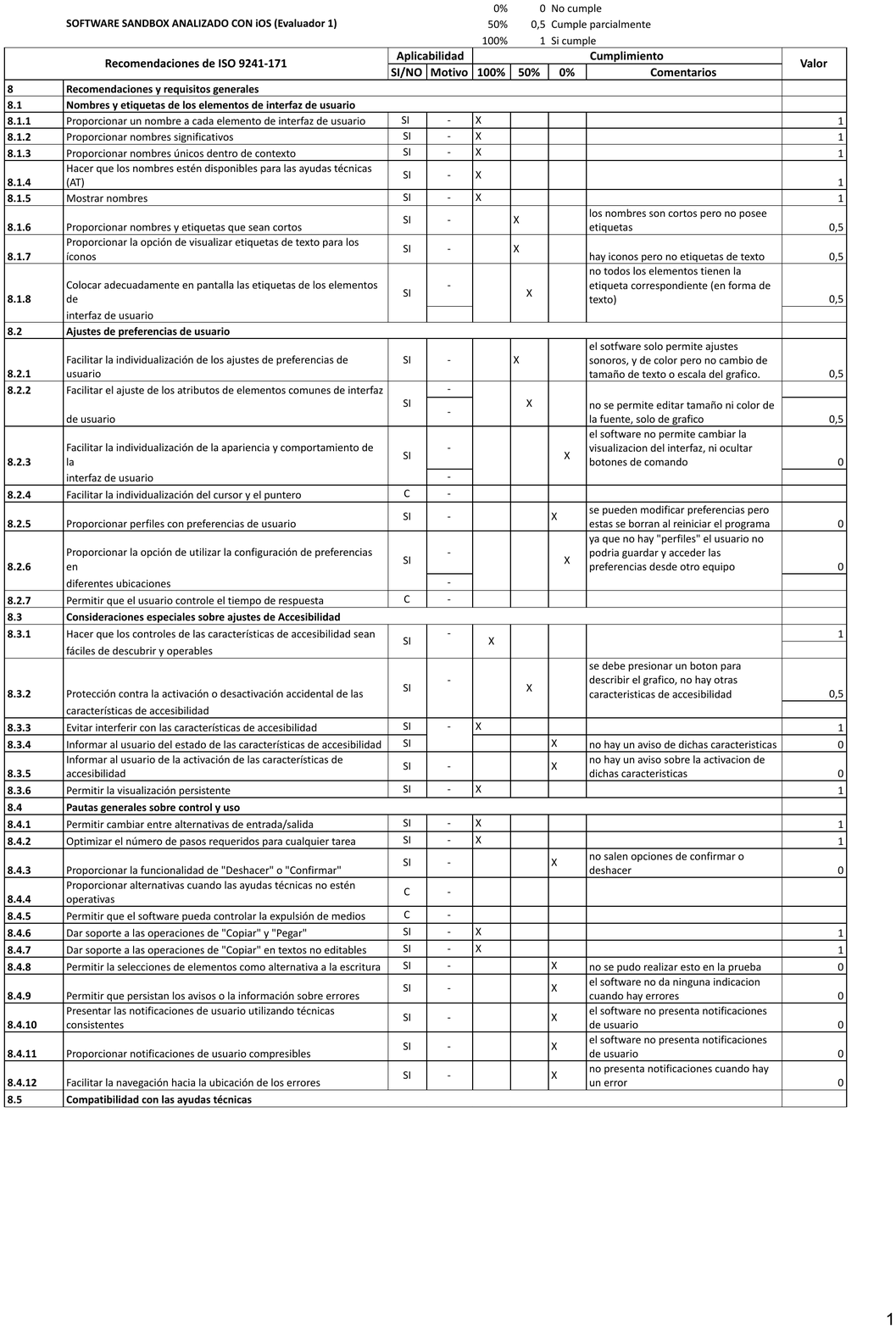}
\includepdf[pages=-,scale=0.8]{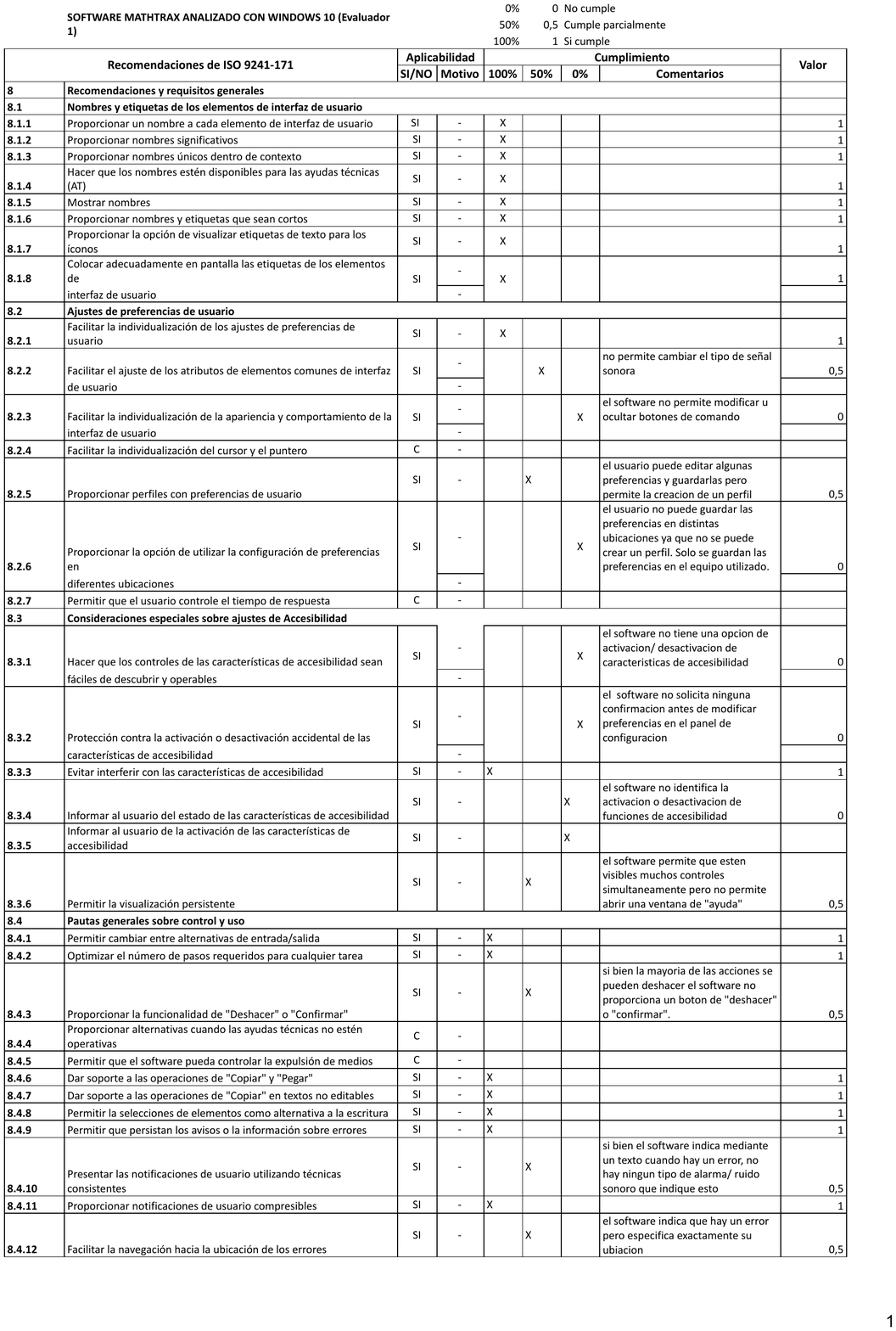}
\includepdf[pages=-,scale=0.8]{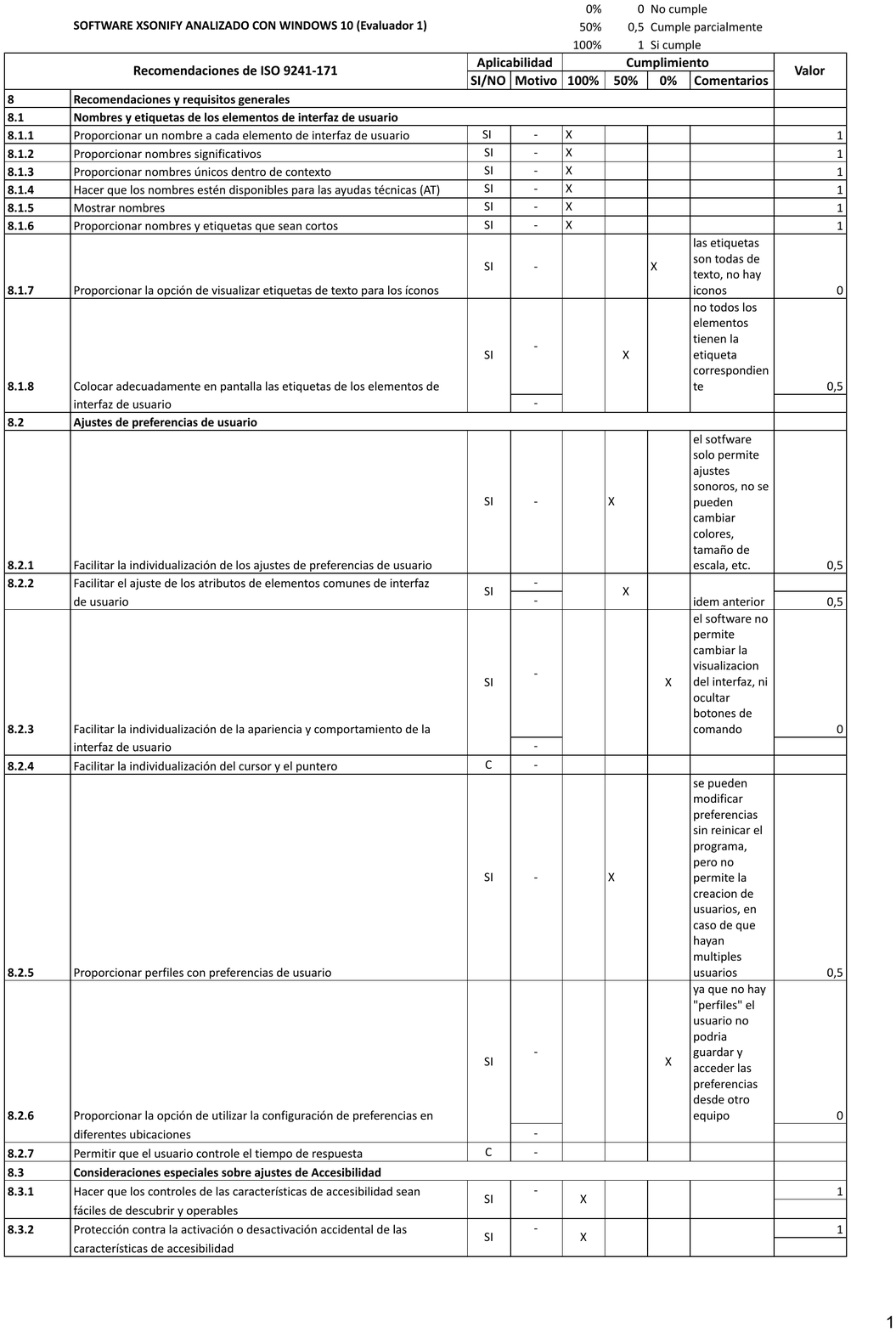}
\includepdf[pages=-,scale=0.8]{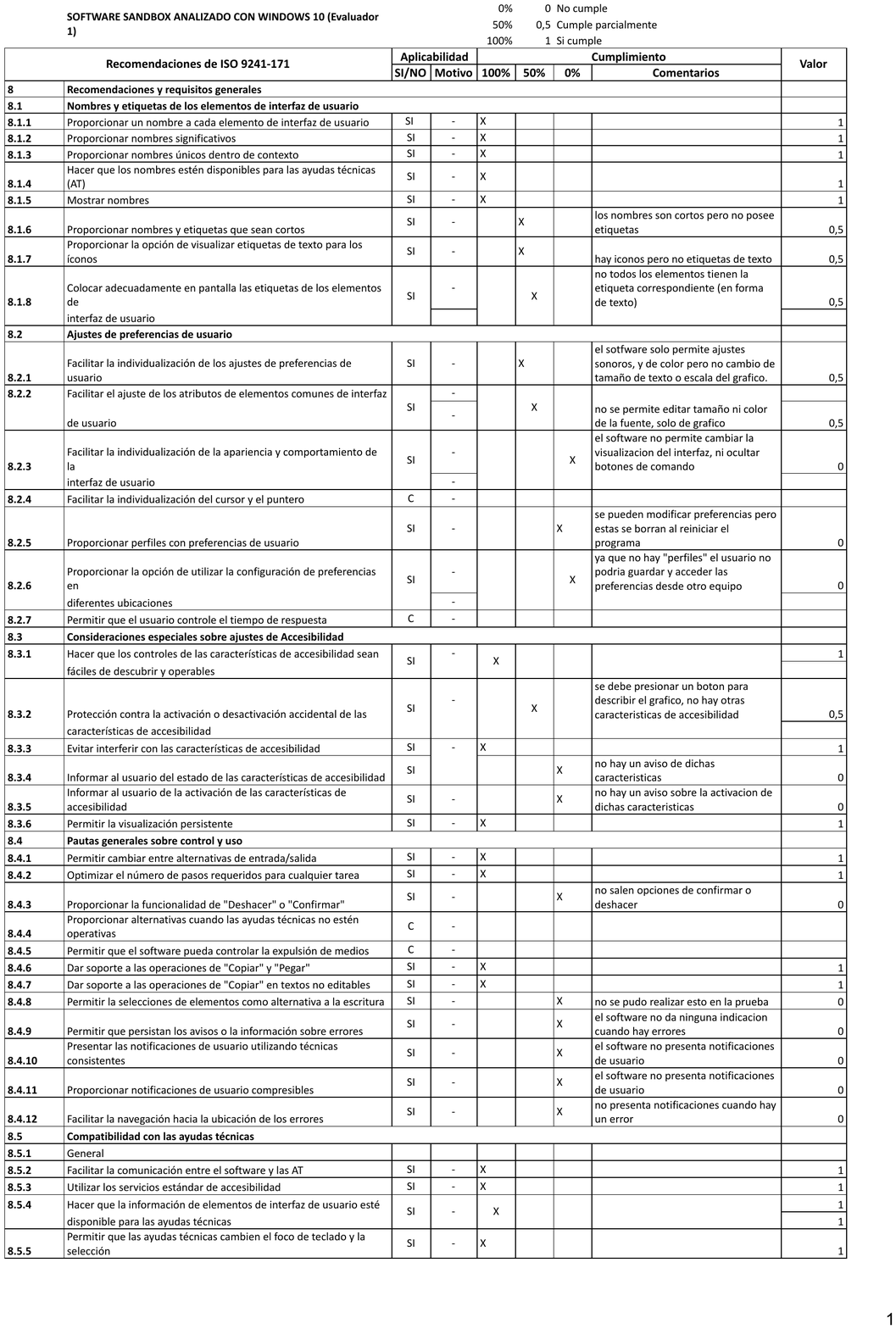}


\subsection{Planillas del Evaluador 2}

\begin{enumerate}
    \item Software MathTrax en MacOS
    \item Software xSonify en MacOS
    \item Software Sonification Sandbox en MacOS
    \item Software MathTrax en Windows
    \item Software xSonify en Windows
    \item Software Sonification Sandbox en Windows
\end{enumerate}


\includepdf[pages=-,scale=0.8]{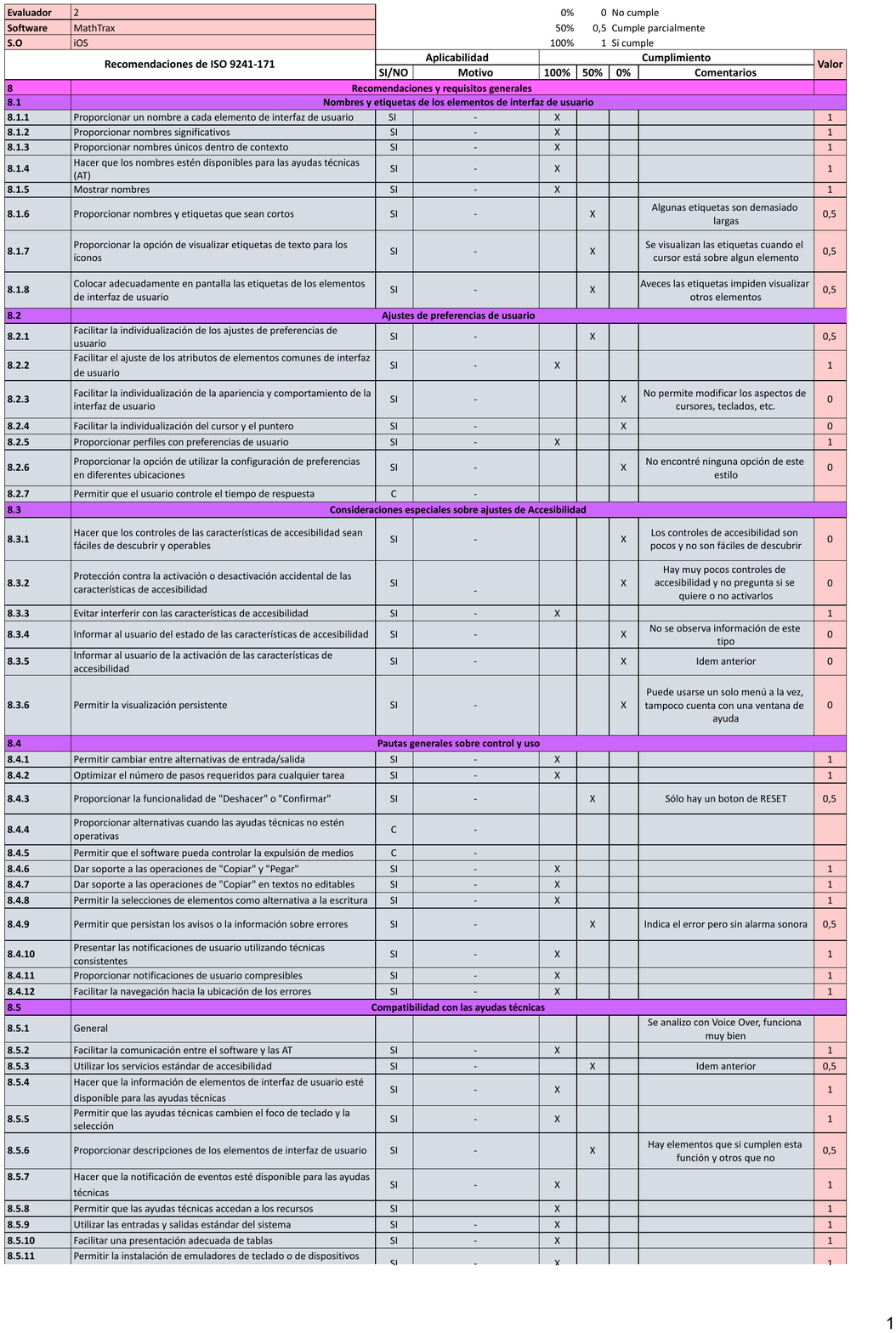}
\includepdf[pages=-,scale=0.8]{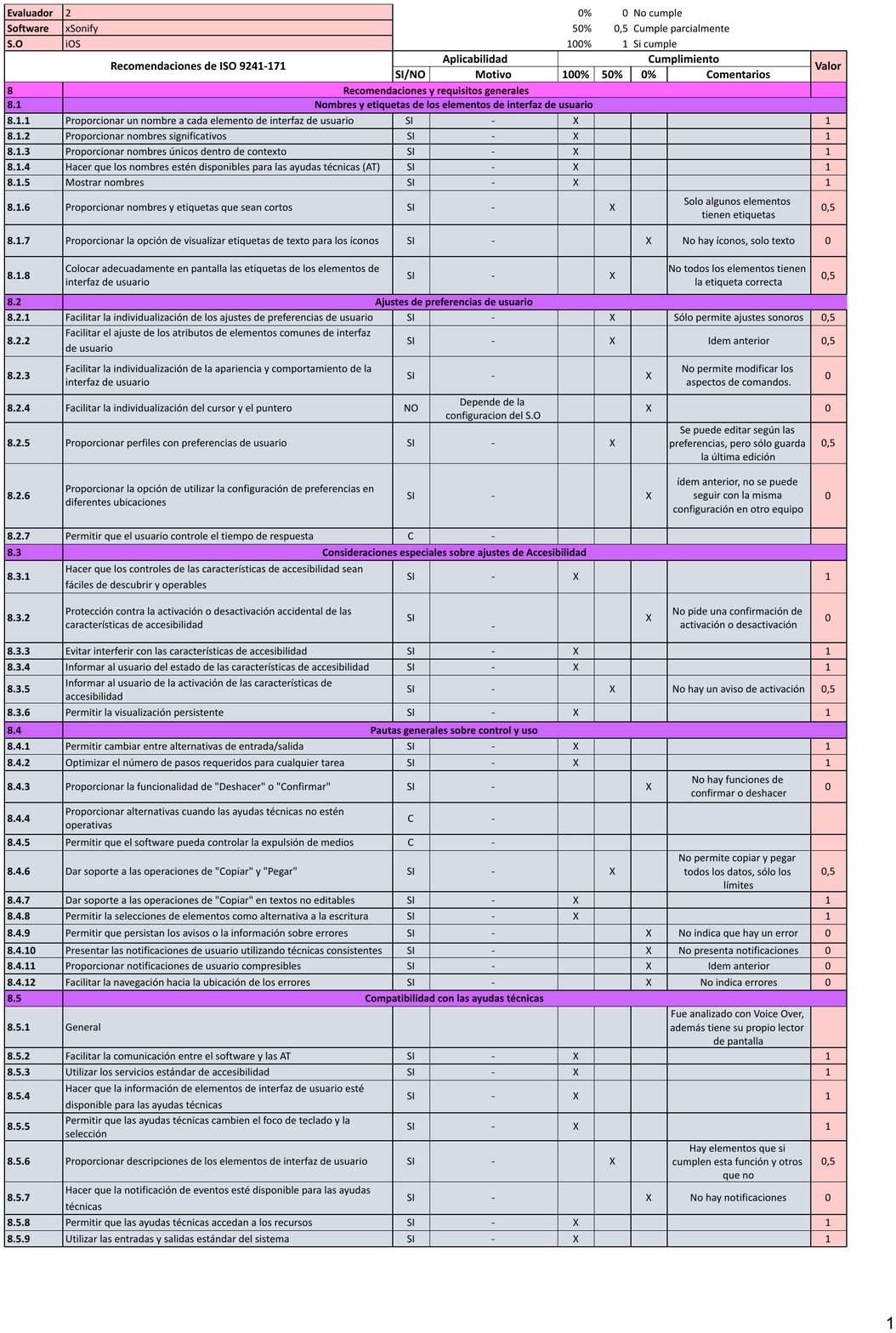}
\includepdf[pages=-,scale=0.8]{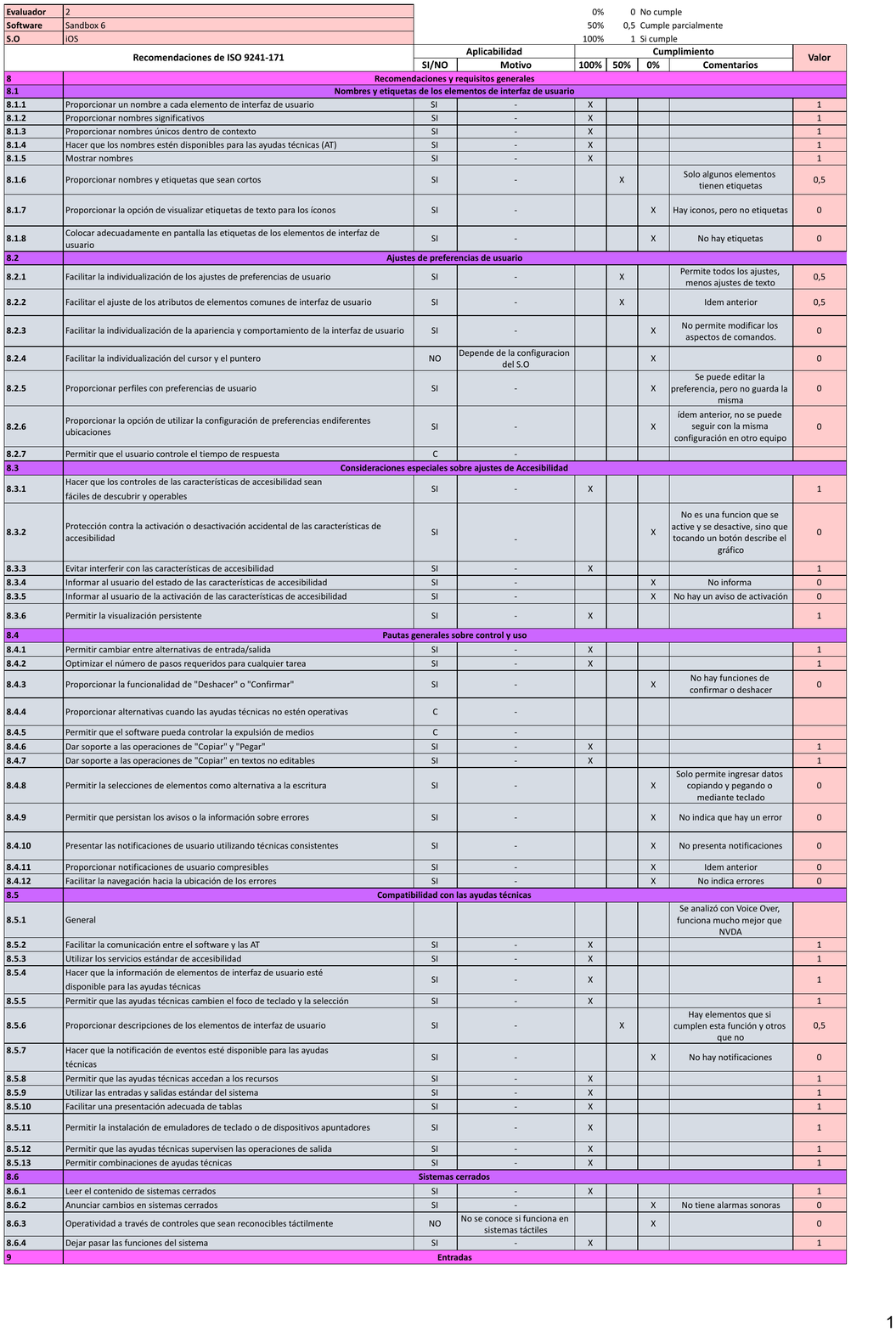}
\includepdf[pages=-,scale=0.8]{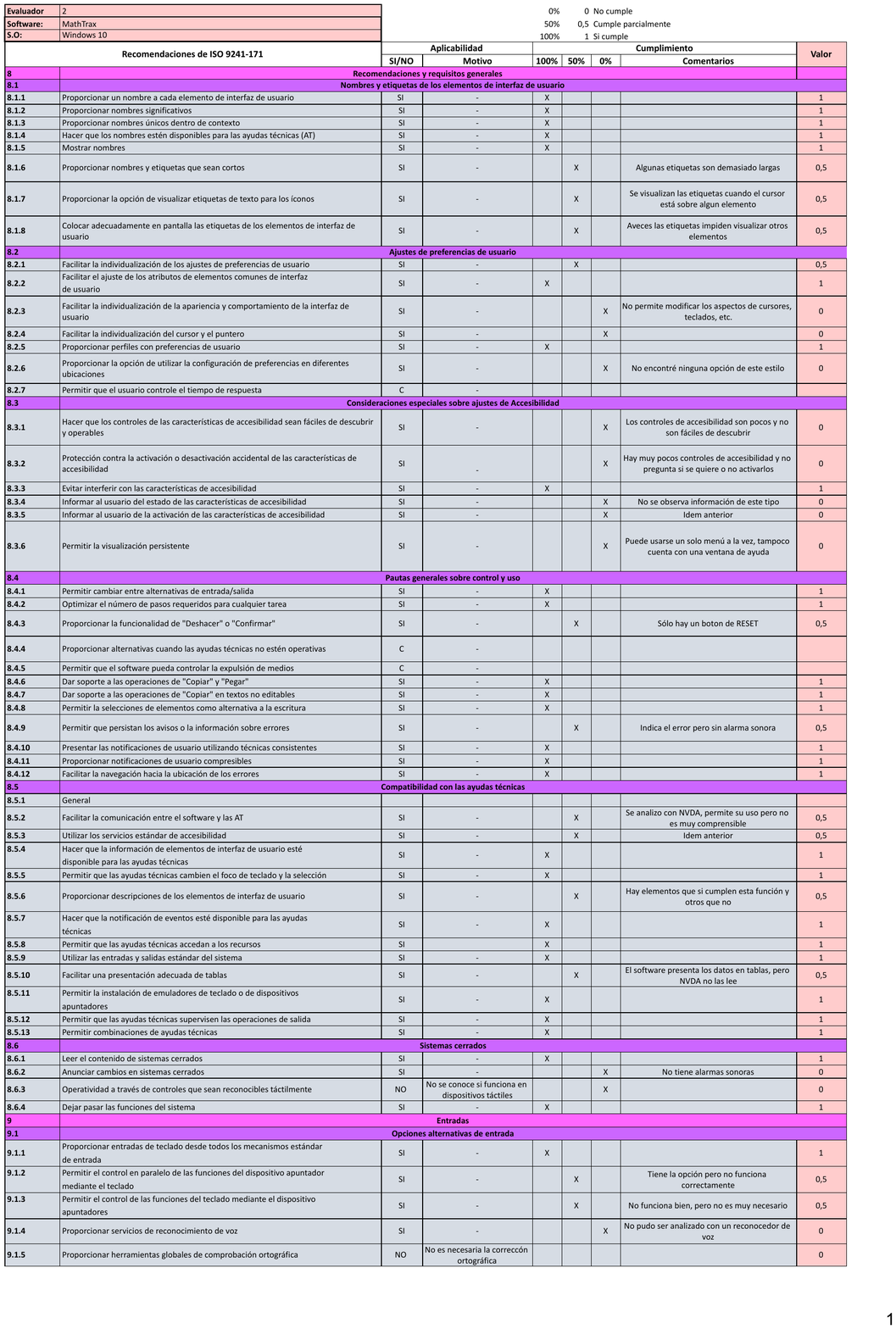}
\includepdf[pages=-,scale=0.8]{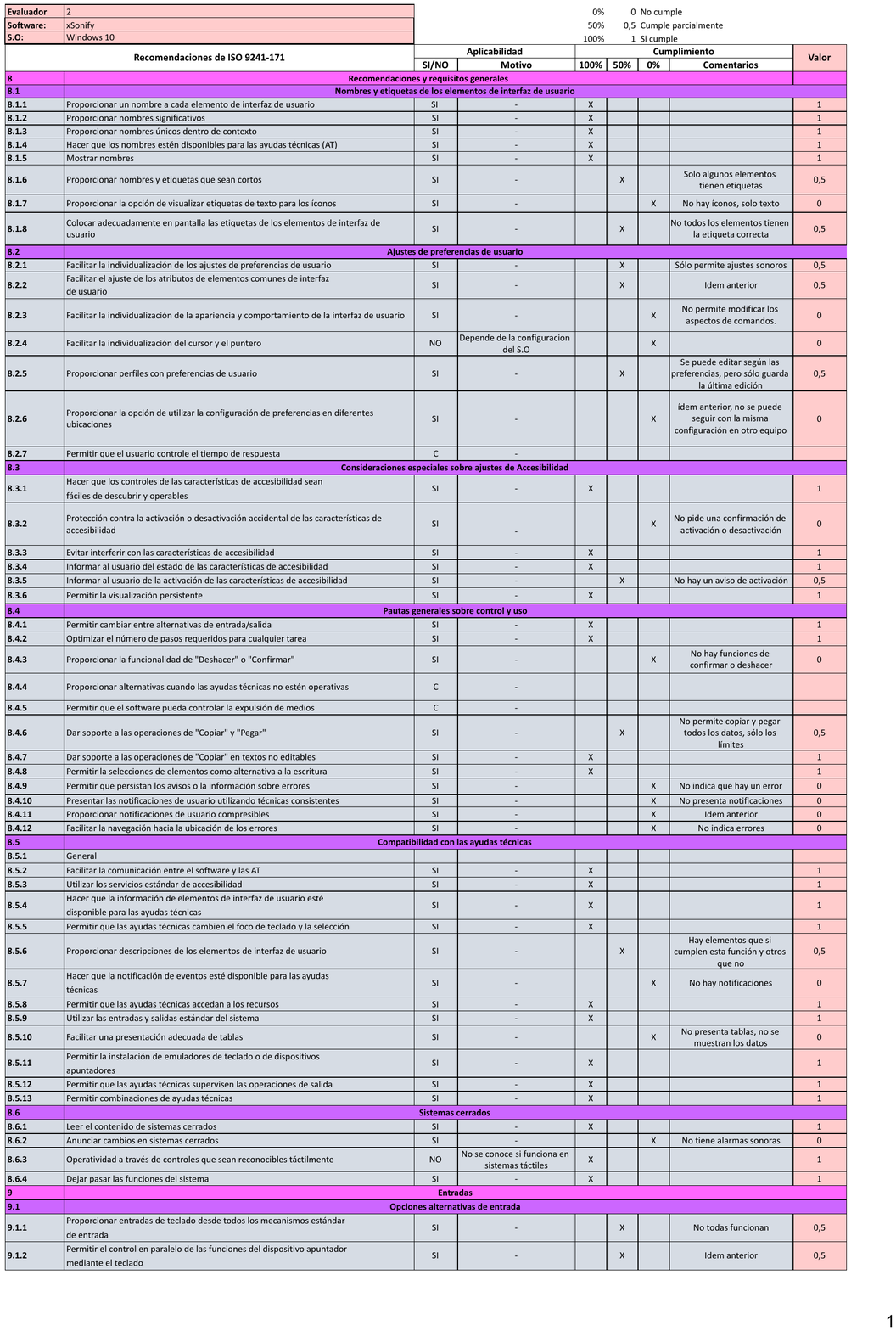}
\includepdf[pages=-,scale=0.8]{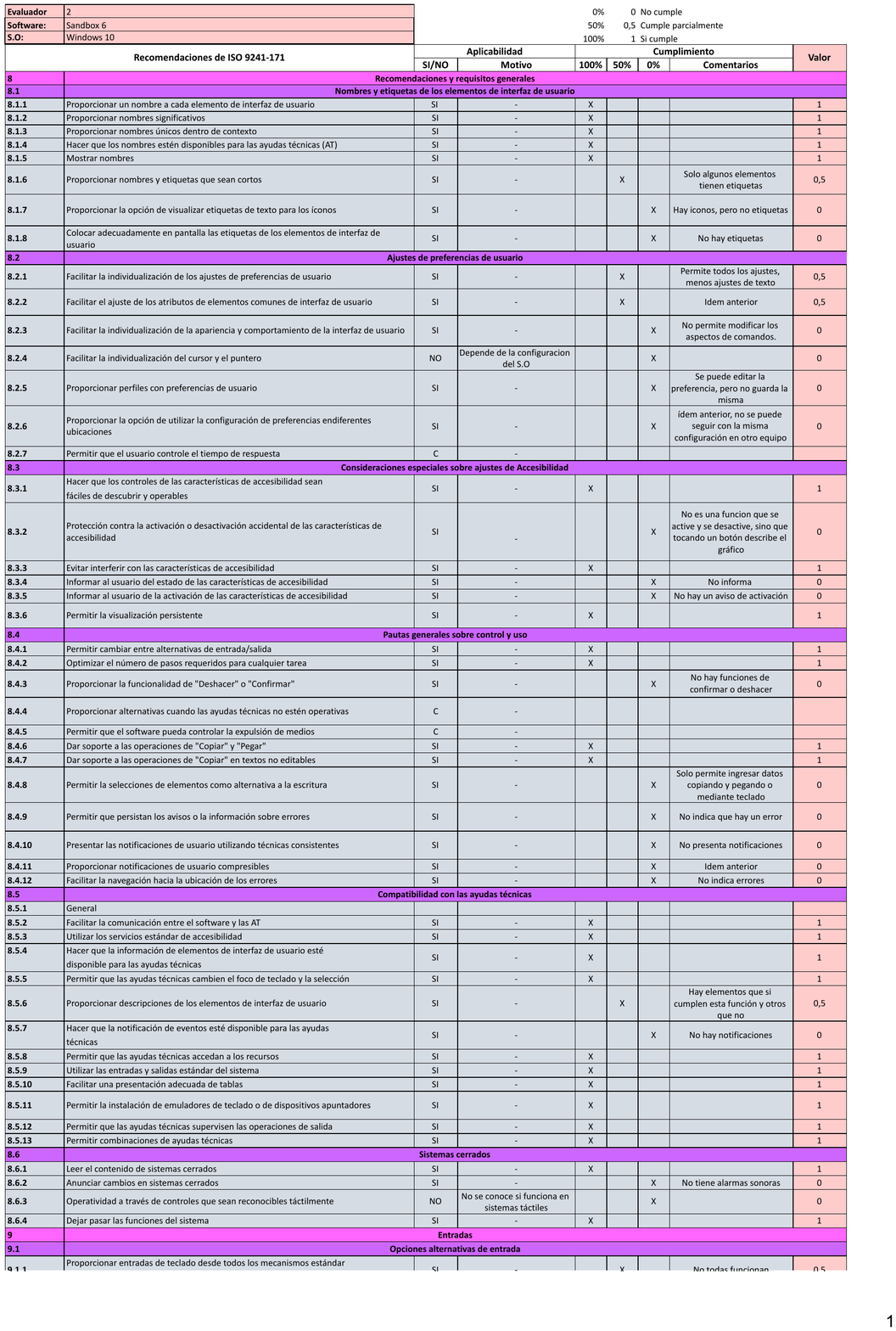}


\subsection{Planillas del Evaluador 3}

\begin{enumerate}
    \item Software MathTrax en MacOS
    \item Software xSonify en MacOS
    \item Software Sonification Sandbox en MacOS
    \item Software MathTrax en Windows
    \item Software xSonify en Windows
    \item Software Sonification Sandbox en Windows
\end{enumerate}


\includepdf[pages=-,scale=0.8]{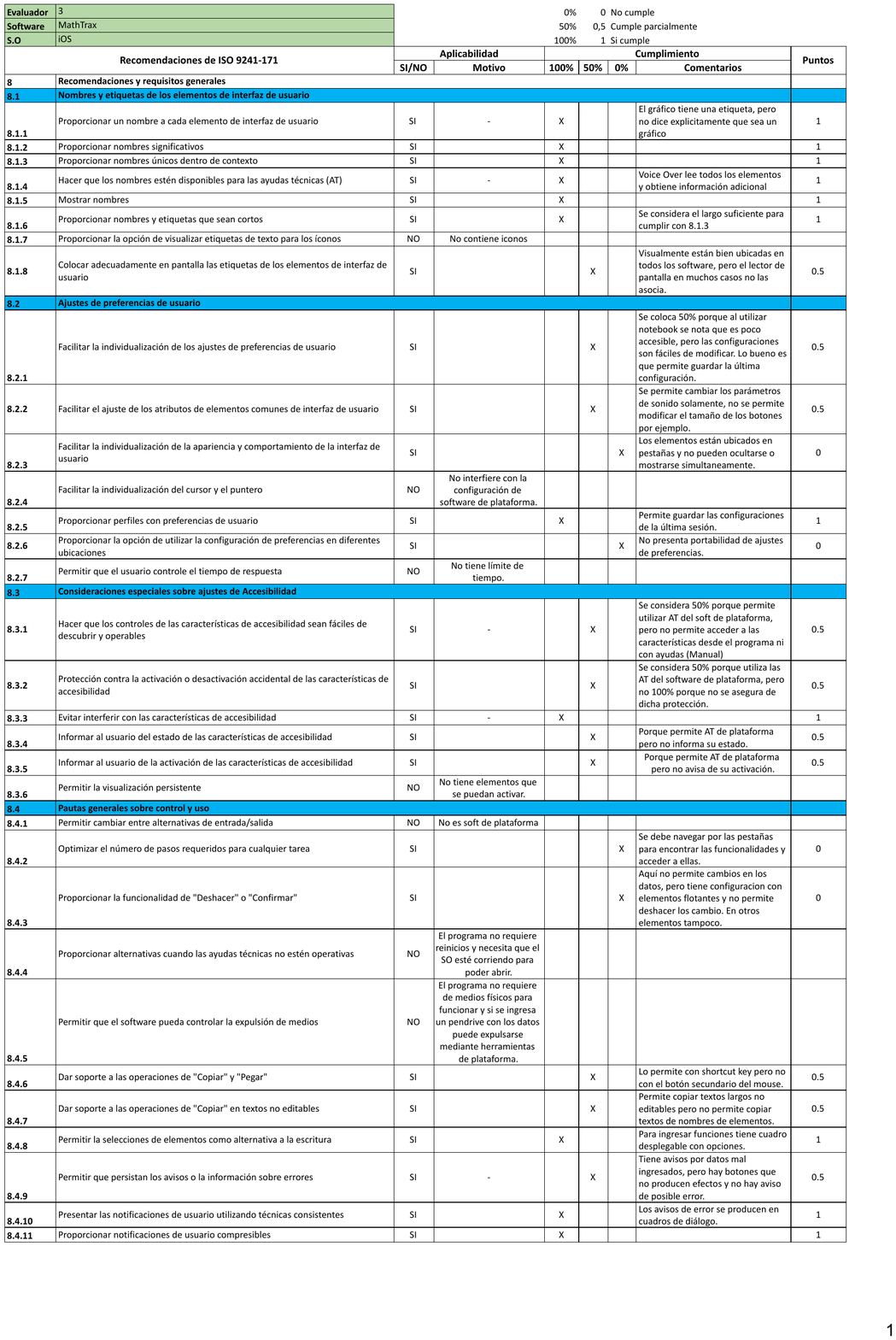}
\includepdf[pages=-,scale=0.8]{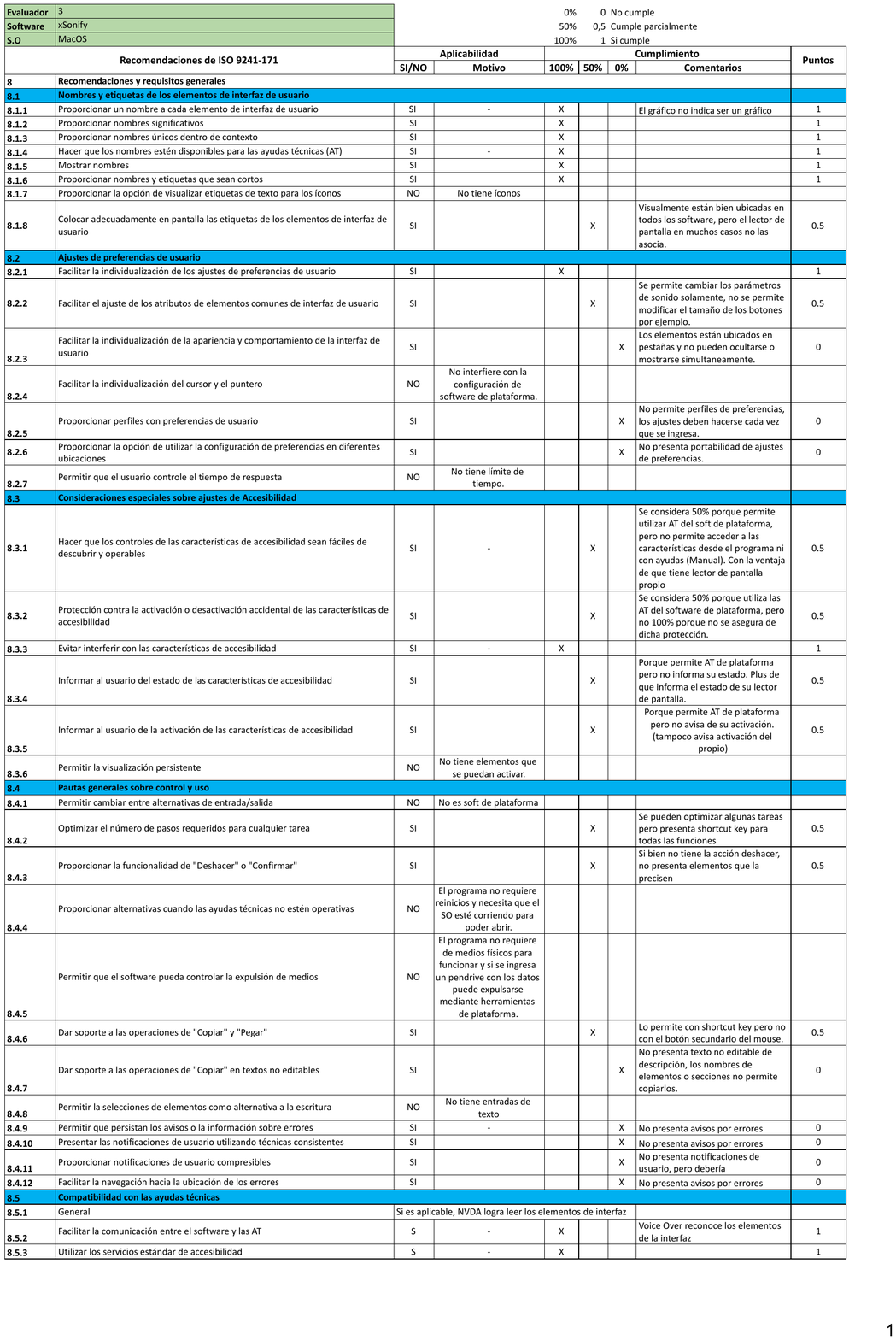}
\includepdf[pages=-,scale=0.8]{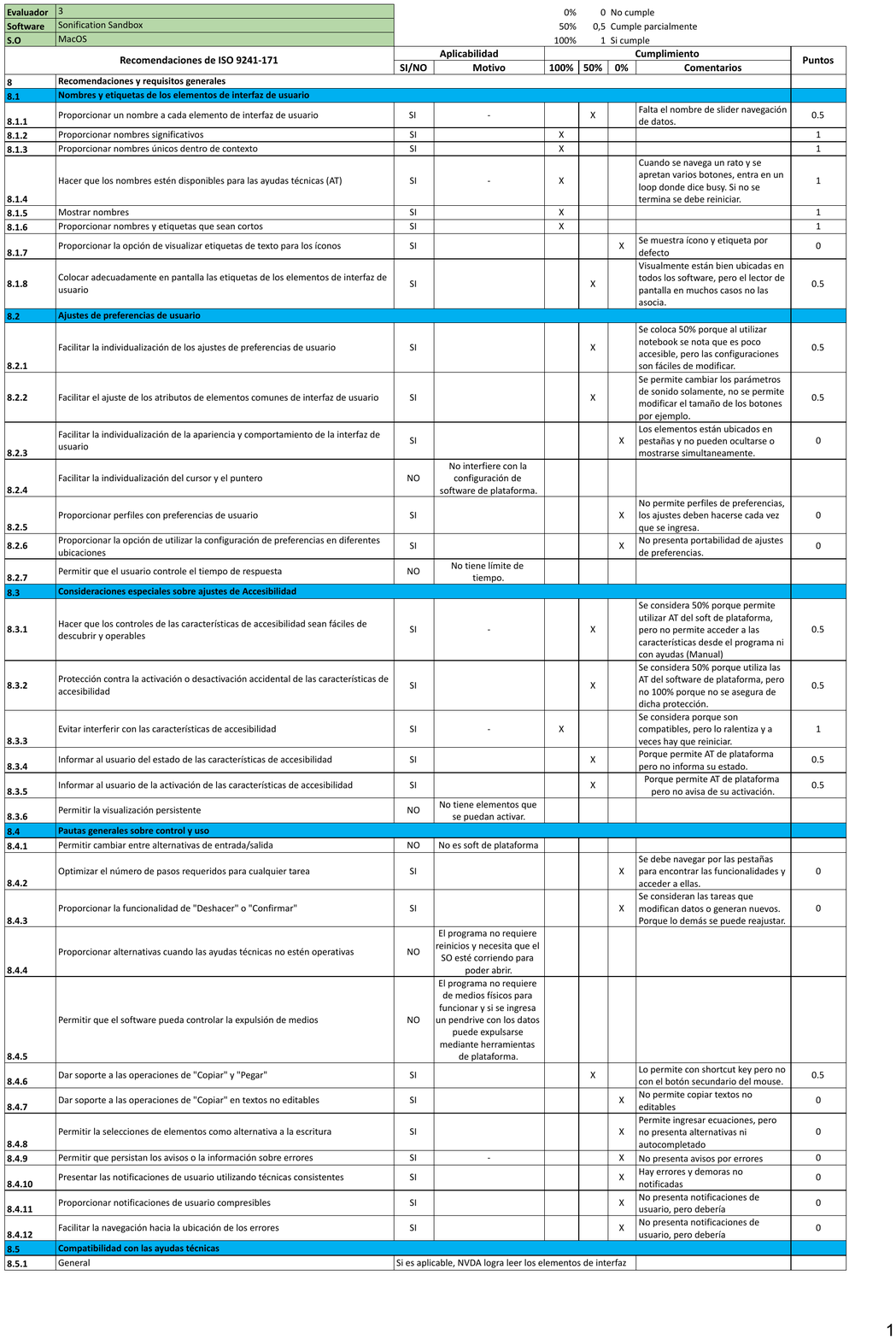}
\includepdf[pages=-,scale=0.8]{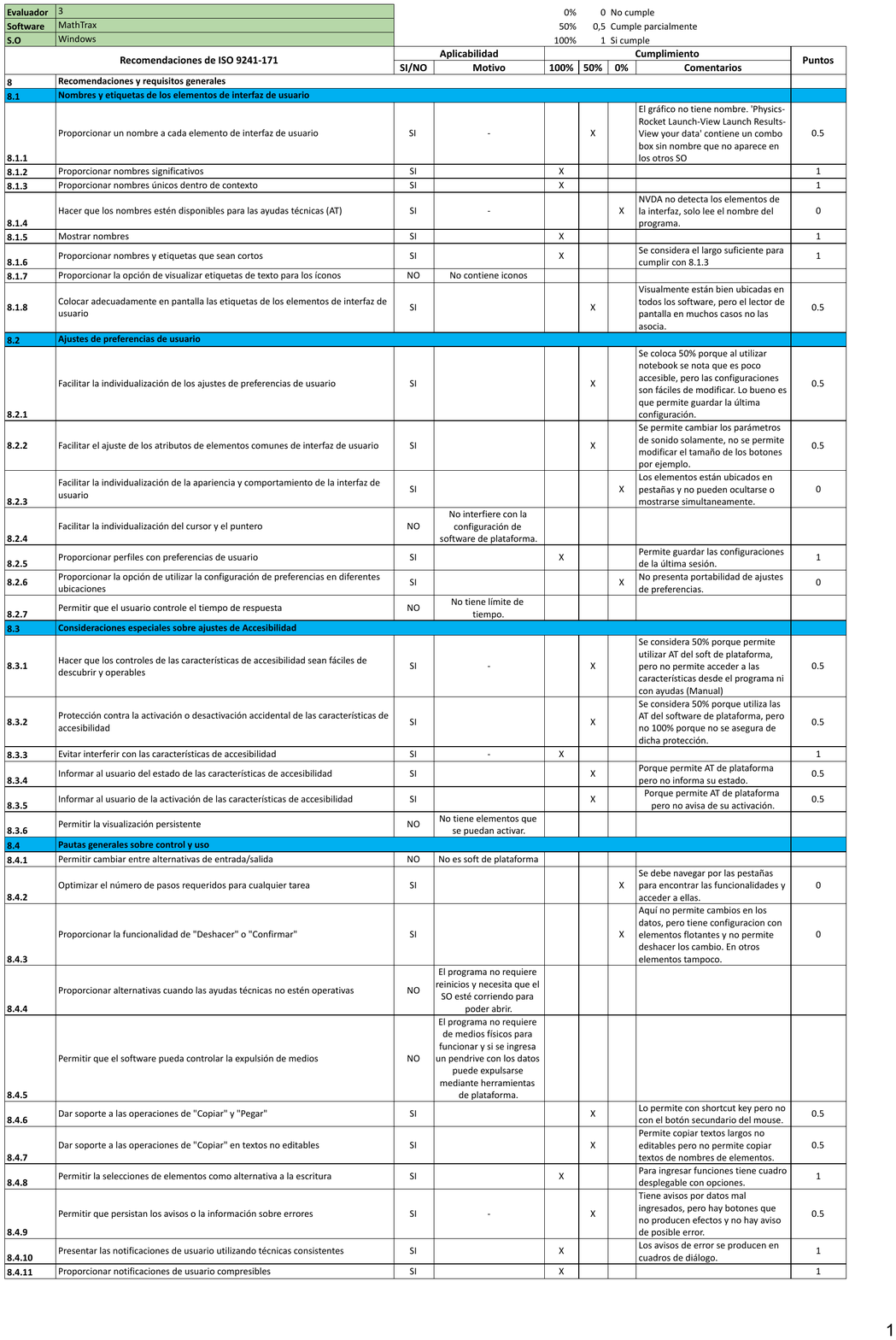}
\includepdf[pages=-,scale=0.8]{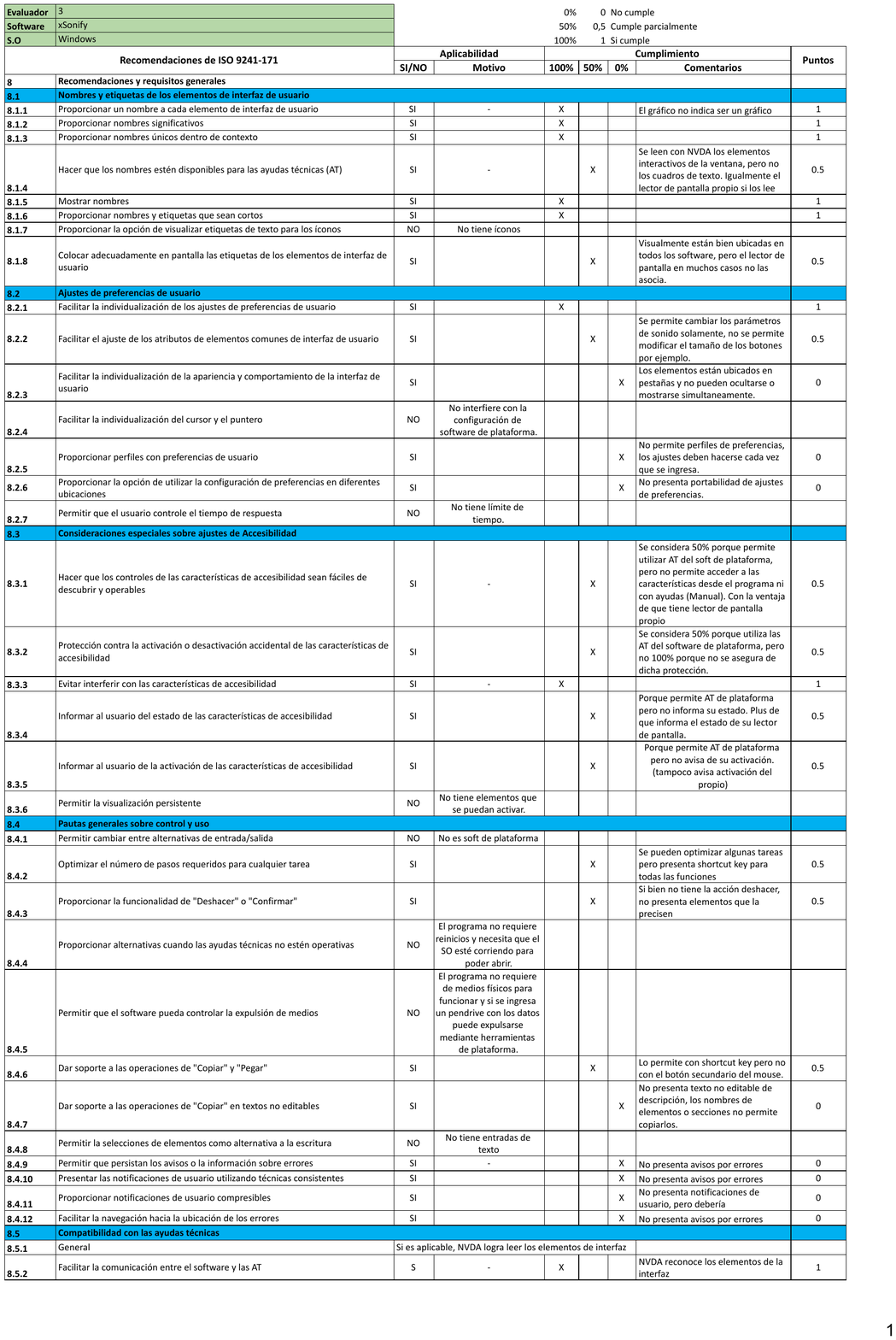}
\includepdf[pages=-,scale=0.8]{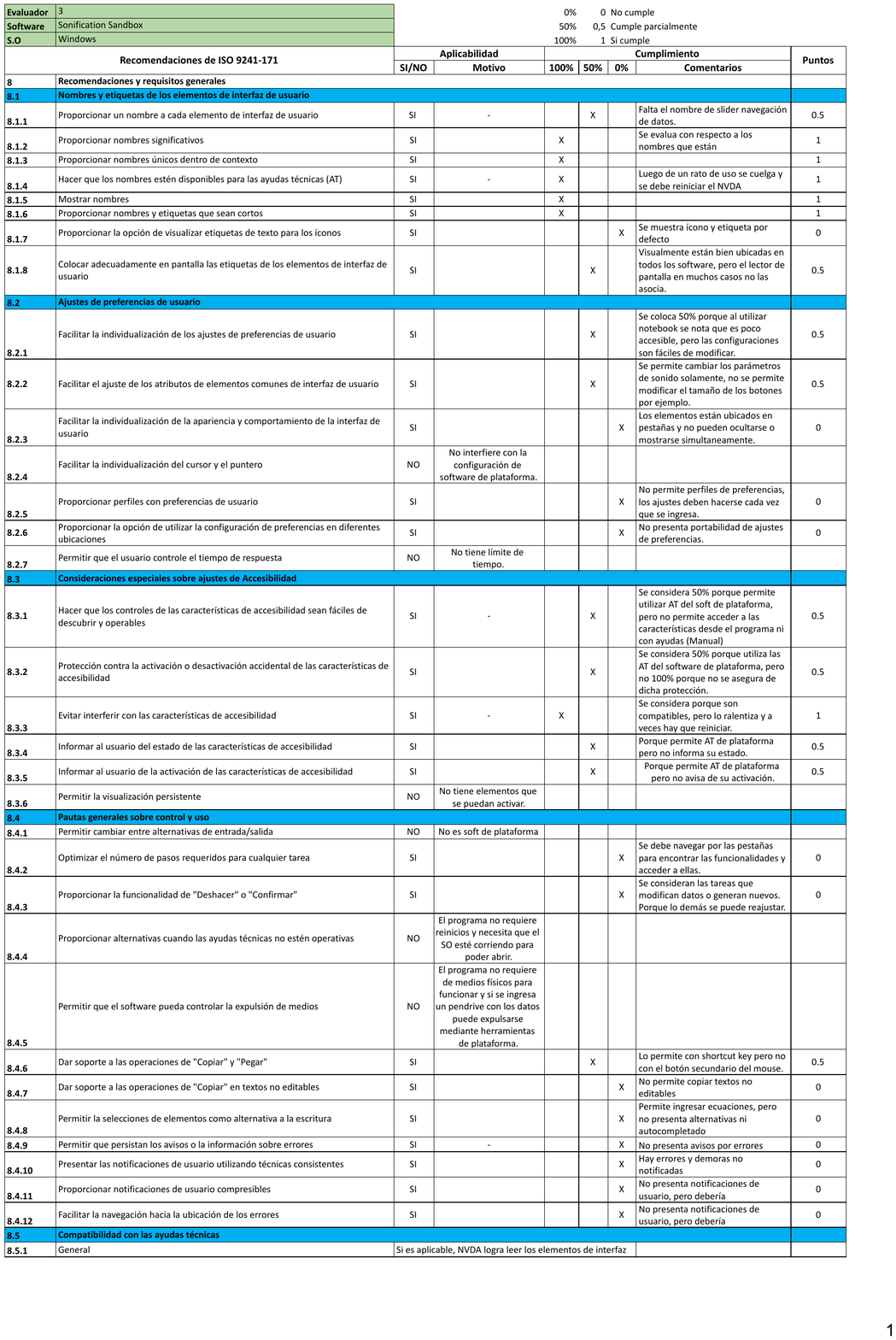}


\subsection{Planillas del Evaluador 4}

\begin{enumerate}
    \item Software MathTrax en MacOS
    \item Software xSonify en MacOS
    \item Software Sonification Sandbox en MacOS
    \item Software MathTrax en Windows
    \item Software xSonify en Windows
    \item Software Sonification Sandbox en Windows
\end{enumerate}


\includepdf[pages=-,scale=0.8]{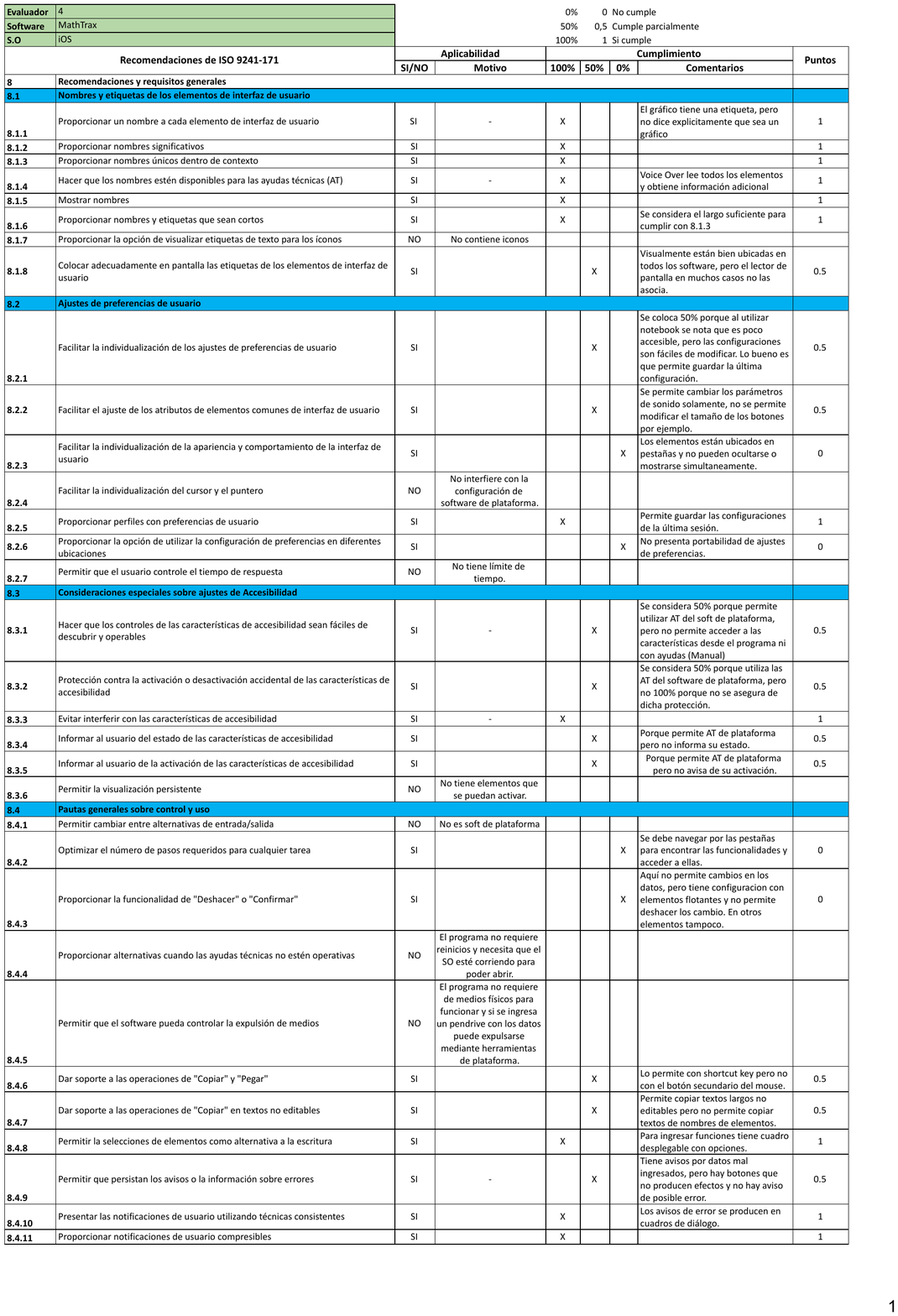}
\includepdf[pages=-,scale=0.8]{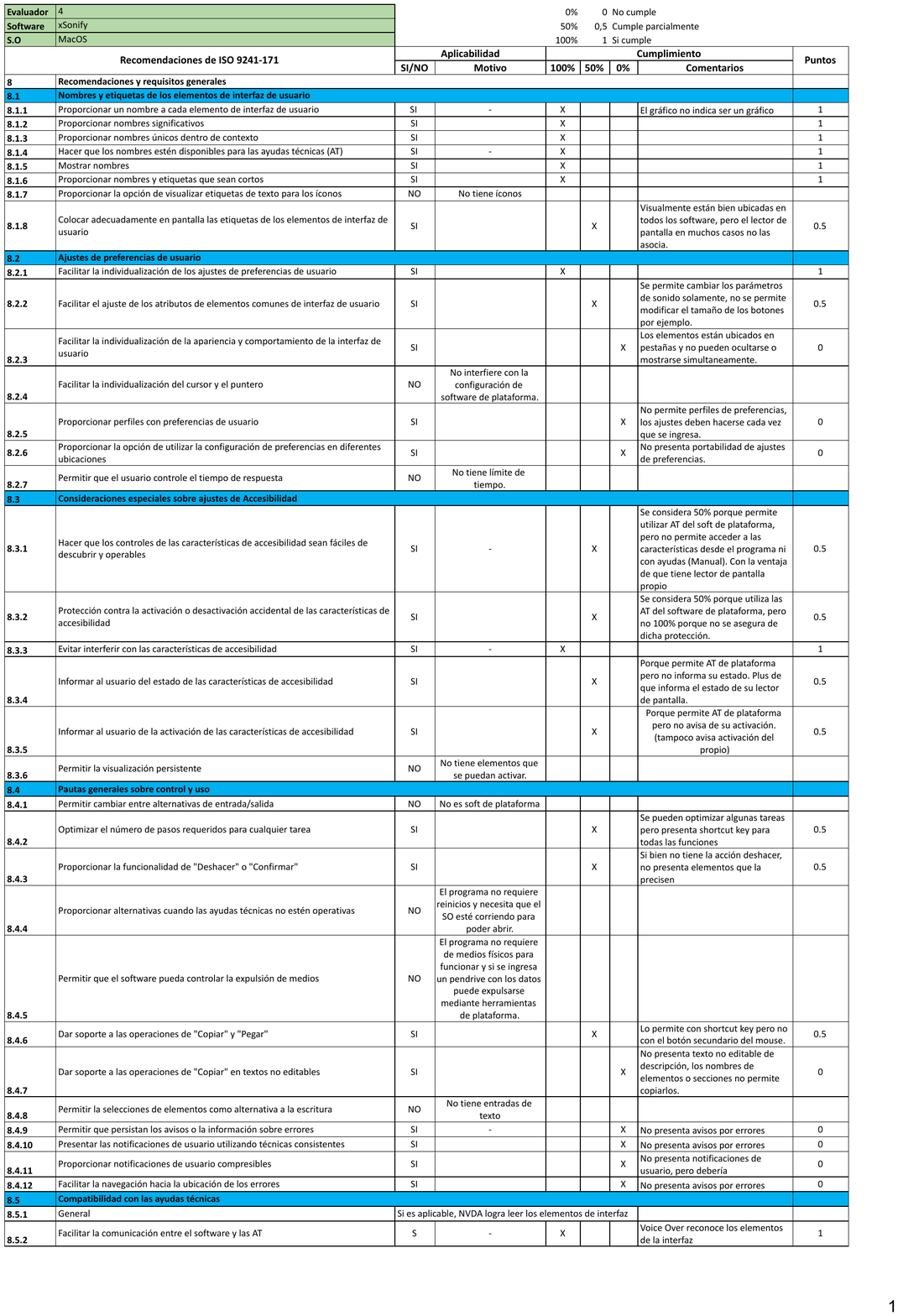}
\includepdf[pages=-,scale=0.8]{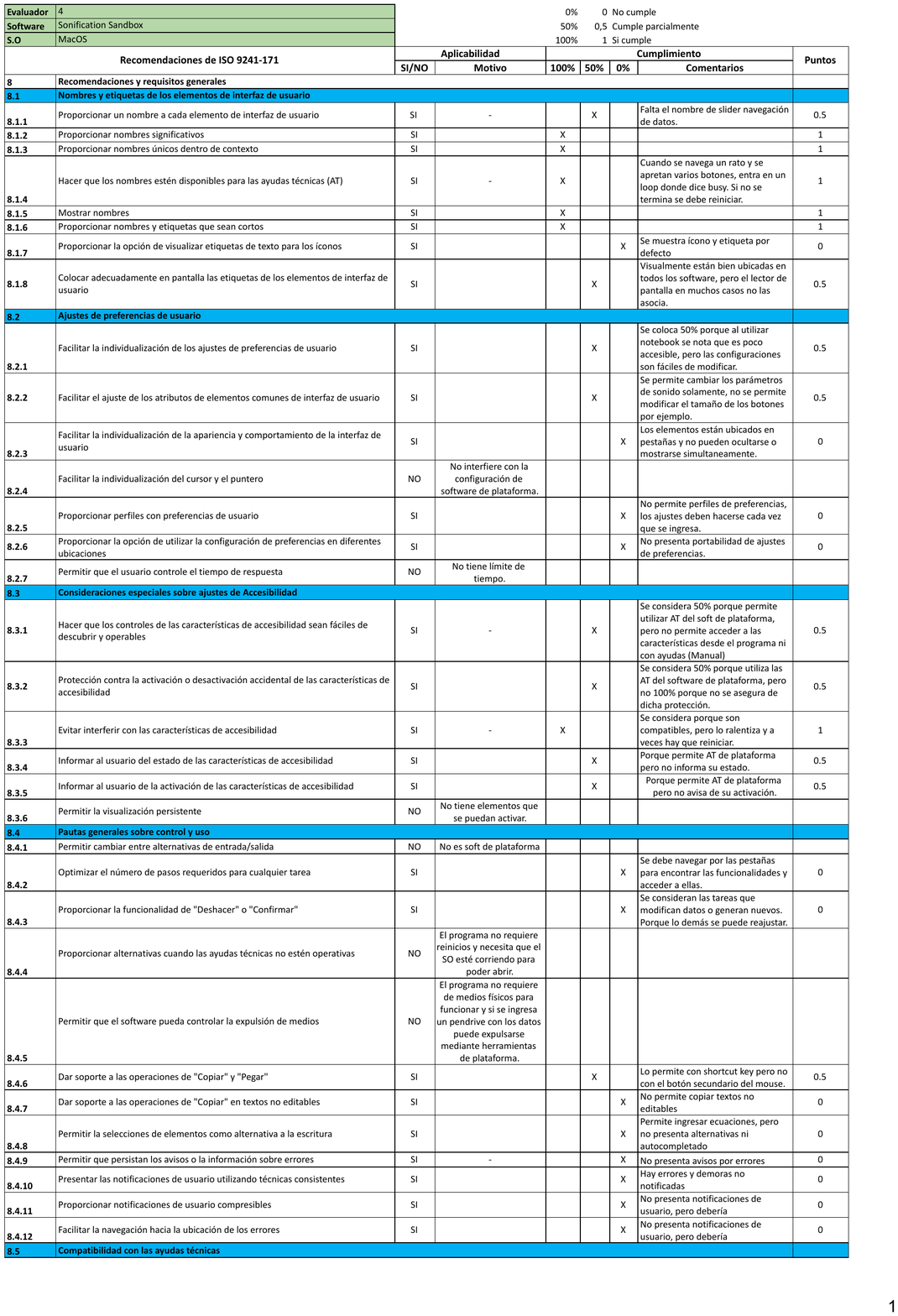}
\includepdf[pages=-,scale=0.8]{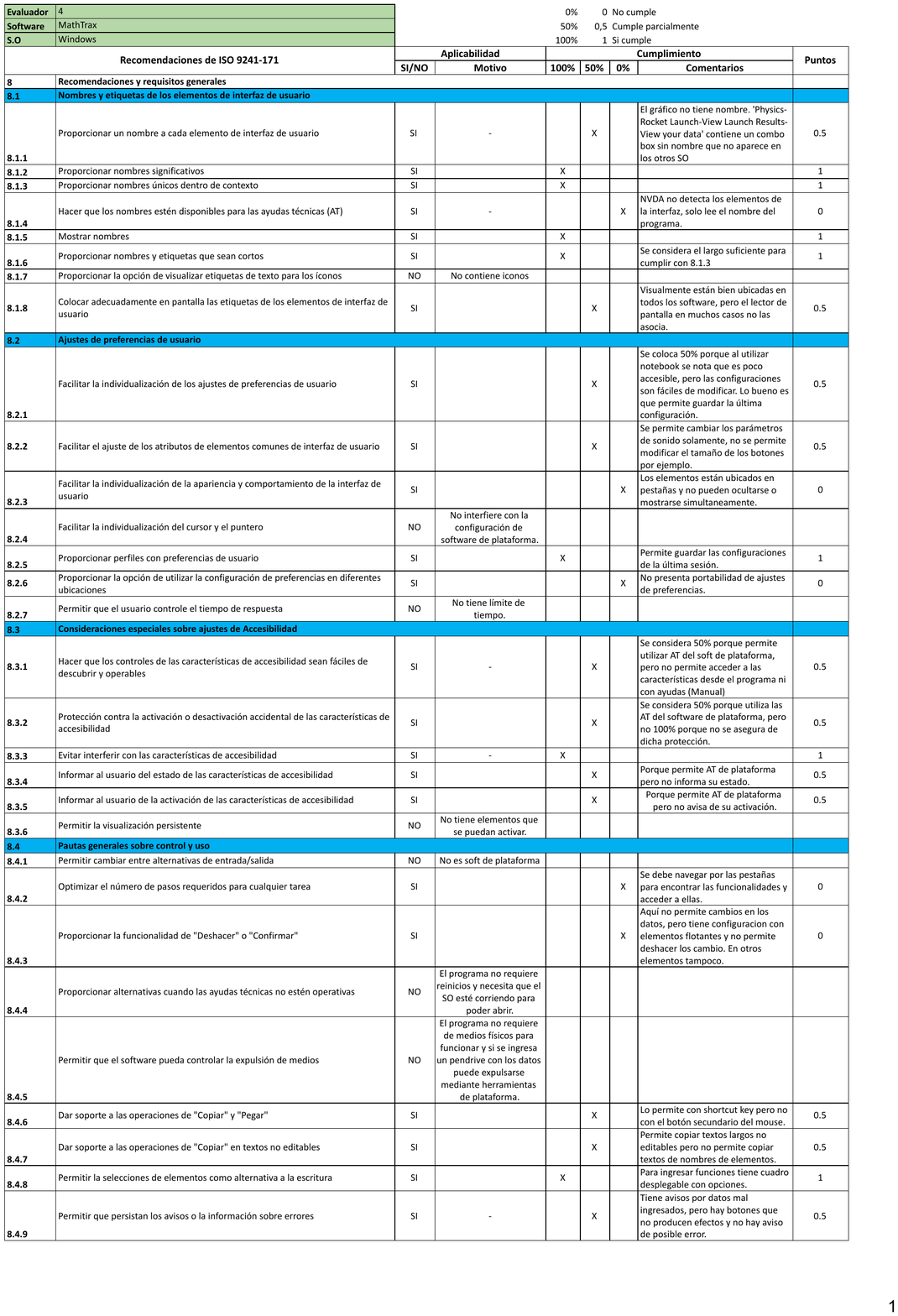}
\includepdf[pages=-,scale=0.8]{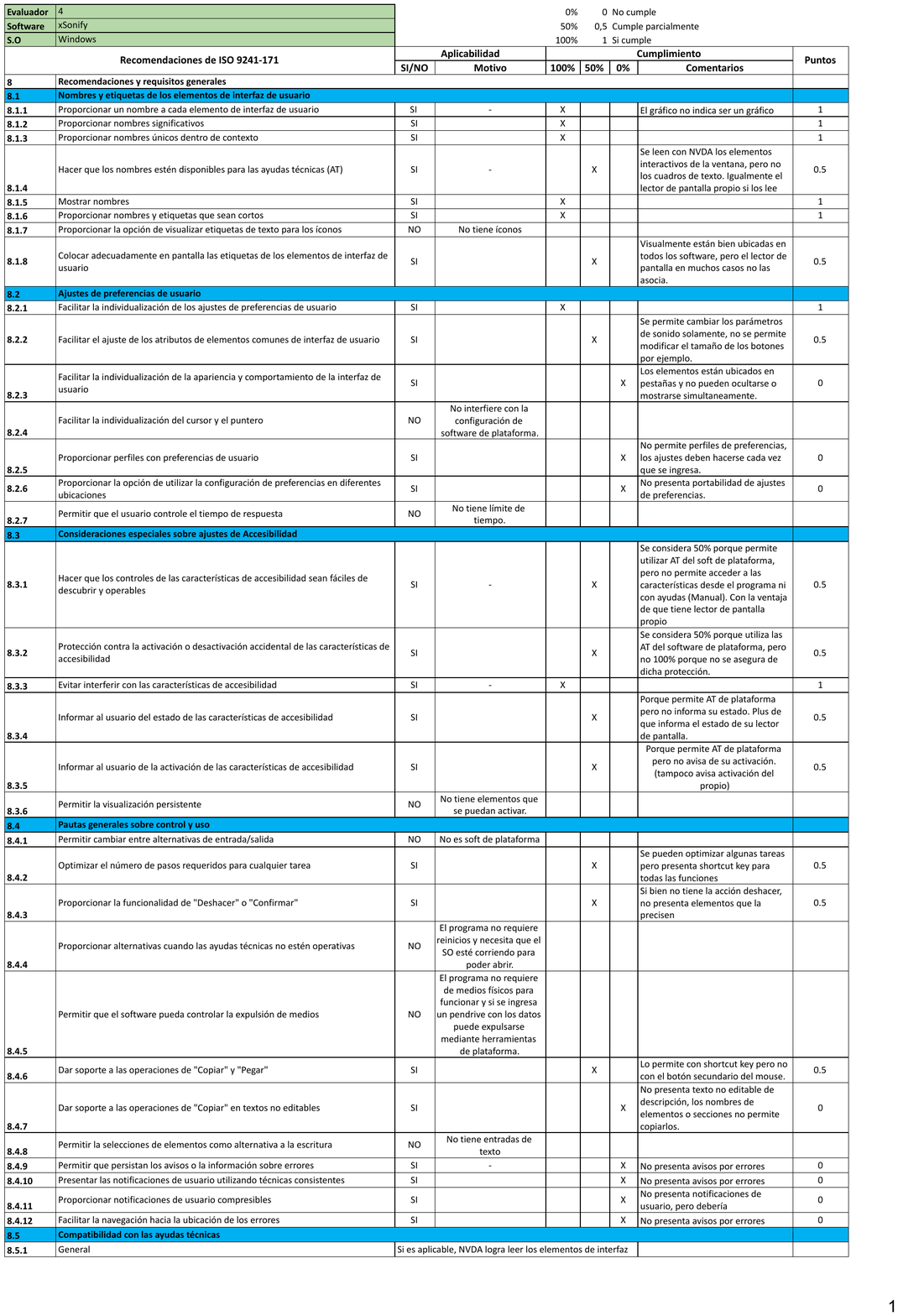}
\includepdf[pages=-,scale=0.8]{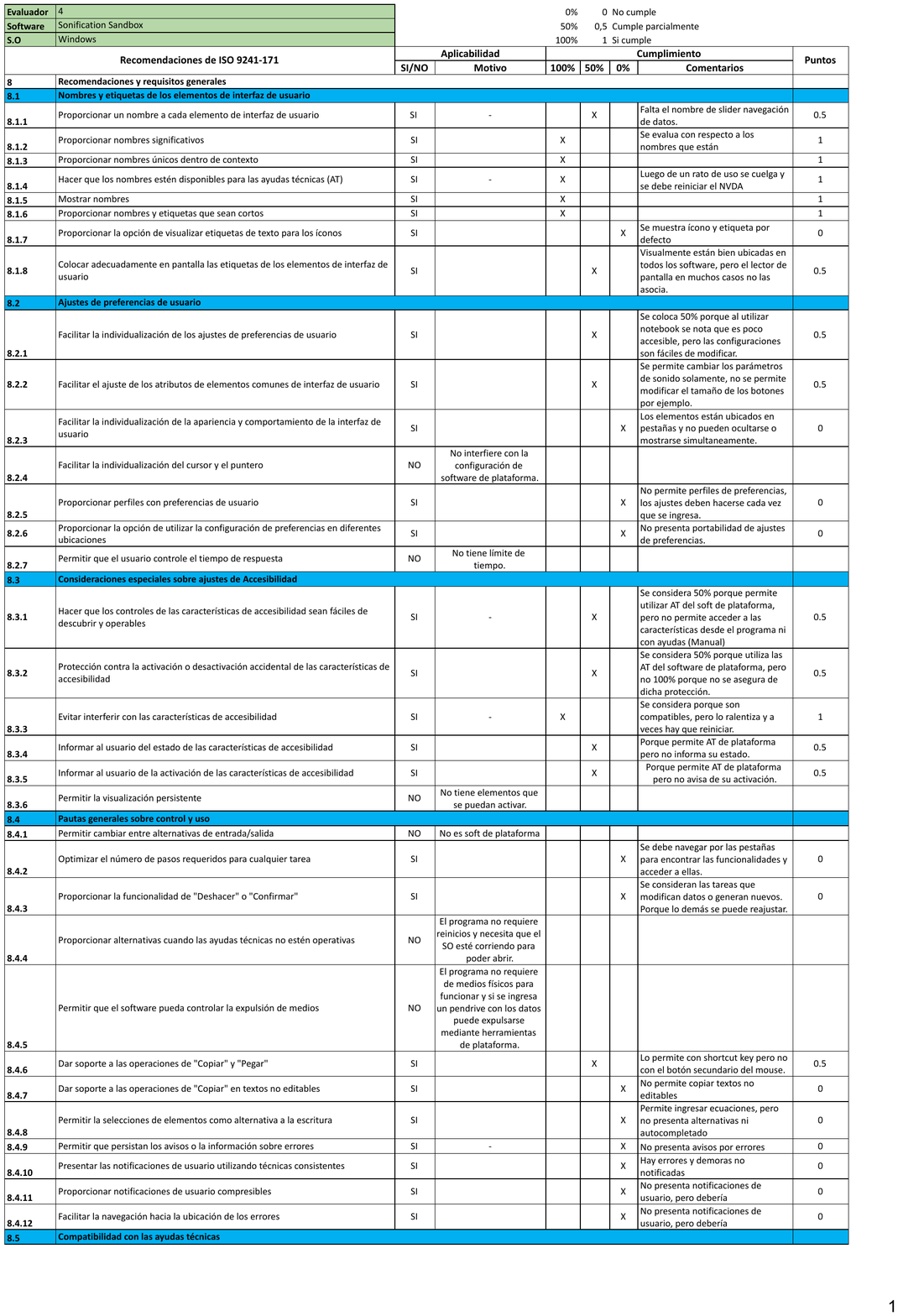}


\subsection{Planillas del Evaluador 5}

\begin{enumerate}
    \item Software MathTrax en Windows
    \item Software xSonify en Windows
    \item Software Sonification Sandbox en Windows
\end{enumerate}


\includepdf[pages=-,scale=0.8]{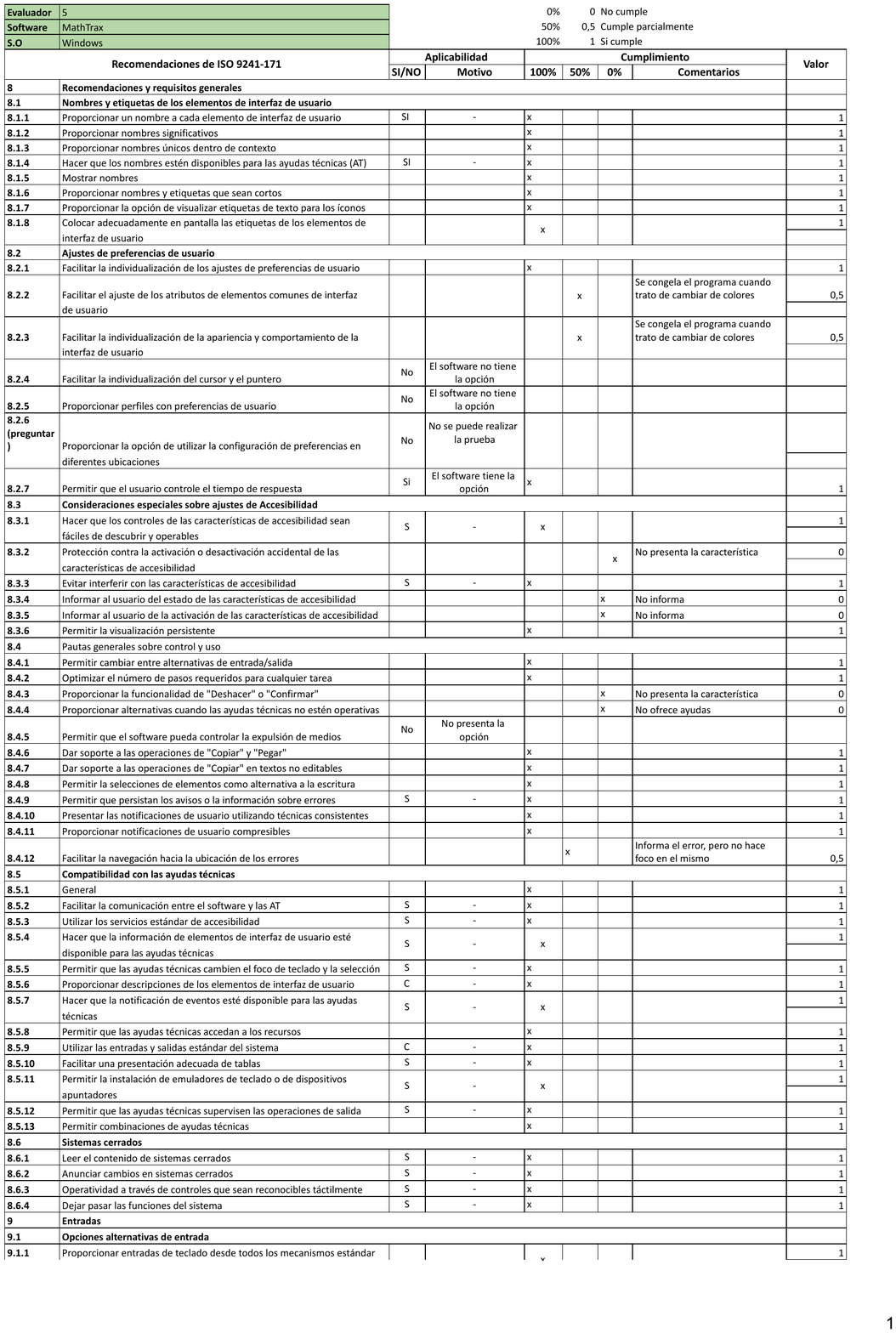}
\includepdf[pages=-,scale=0.8]{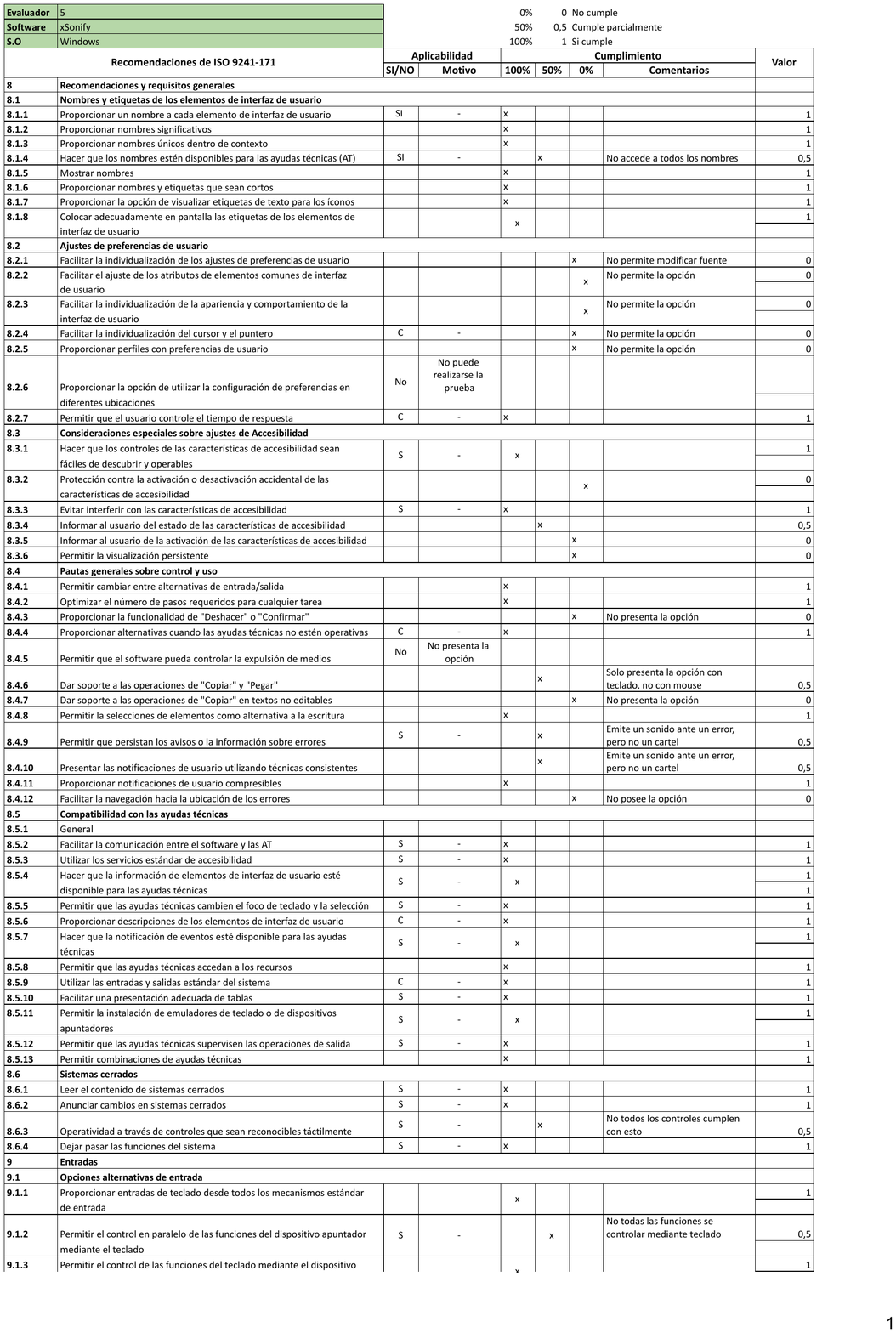}
\includepdf[pages=-,scale=0.8]{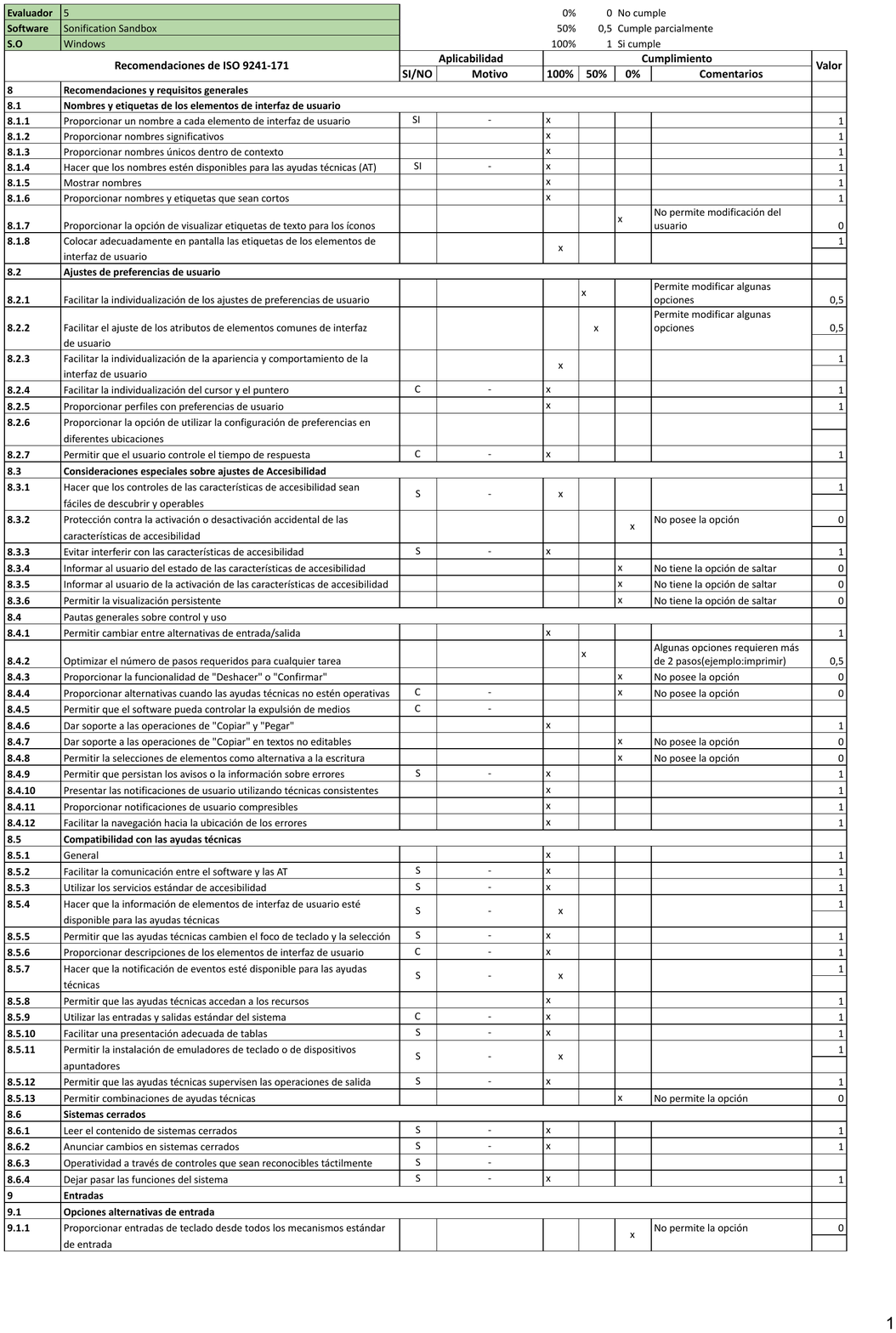}


\section{Documento recibido con el análisis normativo del grupo 2}
\label{ap:A_anexoc_grupo2}

Aquí se incluye el documento pdf con el detalle del análisis del grupo 1 y las anotaciones realizadas por el grupo 2.

\includepdf[pages=-,scale=0.8]{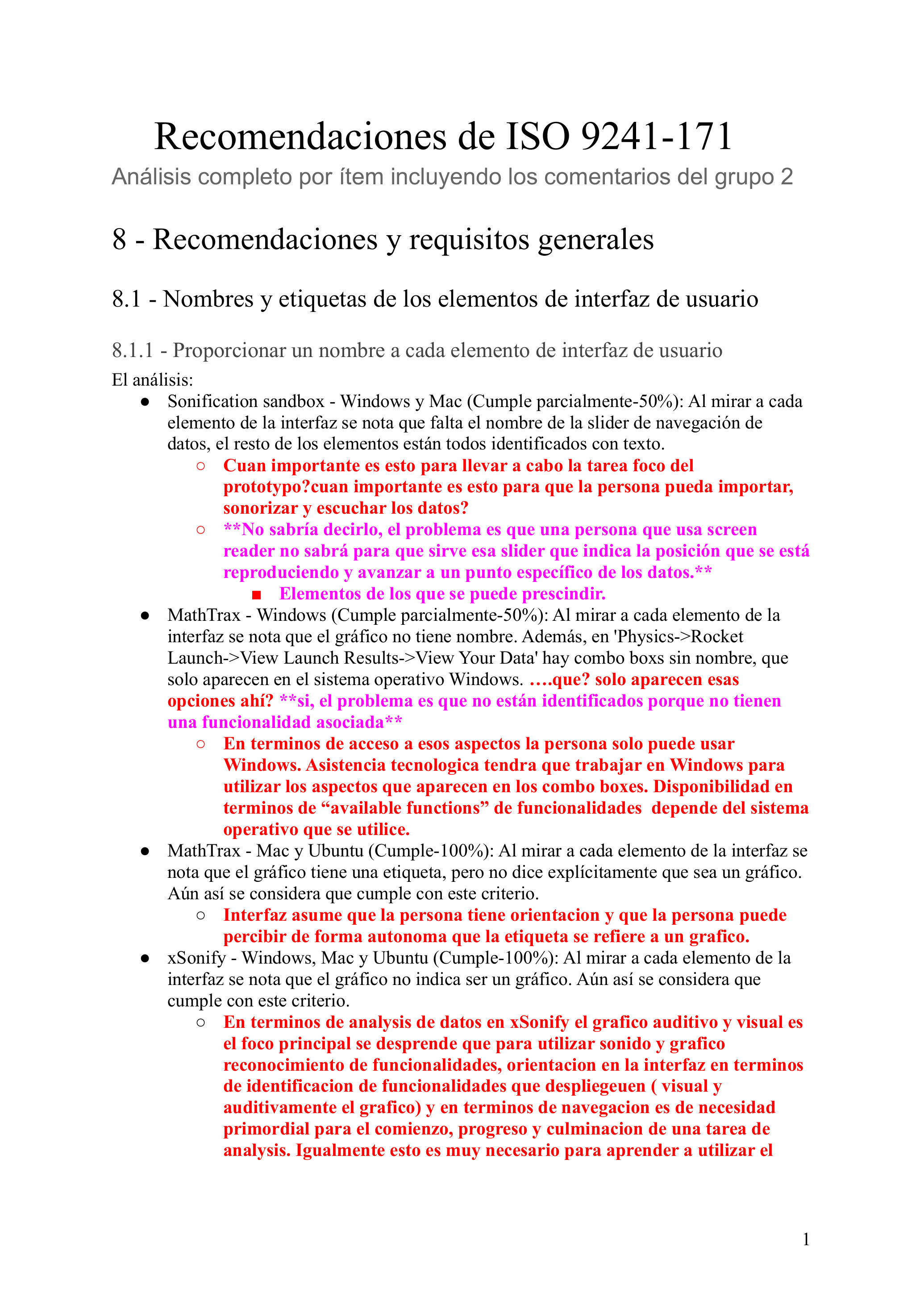}

\chapter{Documentos relativos al análisis normativo web}
\label{ap:web_completo}

Este apéndice incluye un proceeding publicado sobre el análisis de la normativa web y los documentos obtenidos de los evaluadores web, con los cuales se testearon las tres bases de datos: SDSS, ADS y SIMBAD. Por simetría con la forma de despliegue de los datos en las tablas de la sección \ref{sect:w3c_seccioncompleta} se muestran los pdf obtenidos por evaluador, listando cada una de las bases de datos.

\section{Proceeding publicado sobre el análisis de accesibilidad web}
\label{ap:B_proceeding}

A continuación se incluye el proceeding presentado en el año 2017 en la Reunión anual de la Asociación Argentina de Astronomía, el cual incluye el análisis de accesibilidad web realizado a las bases de datos astronómicas SDSS, ADS y SIMBAD.

\includepdf[pages=-,scale=0.8]{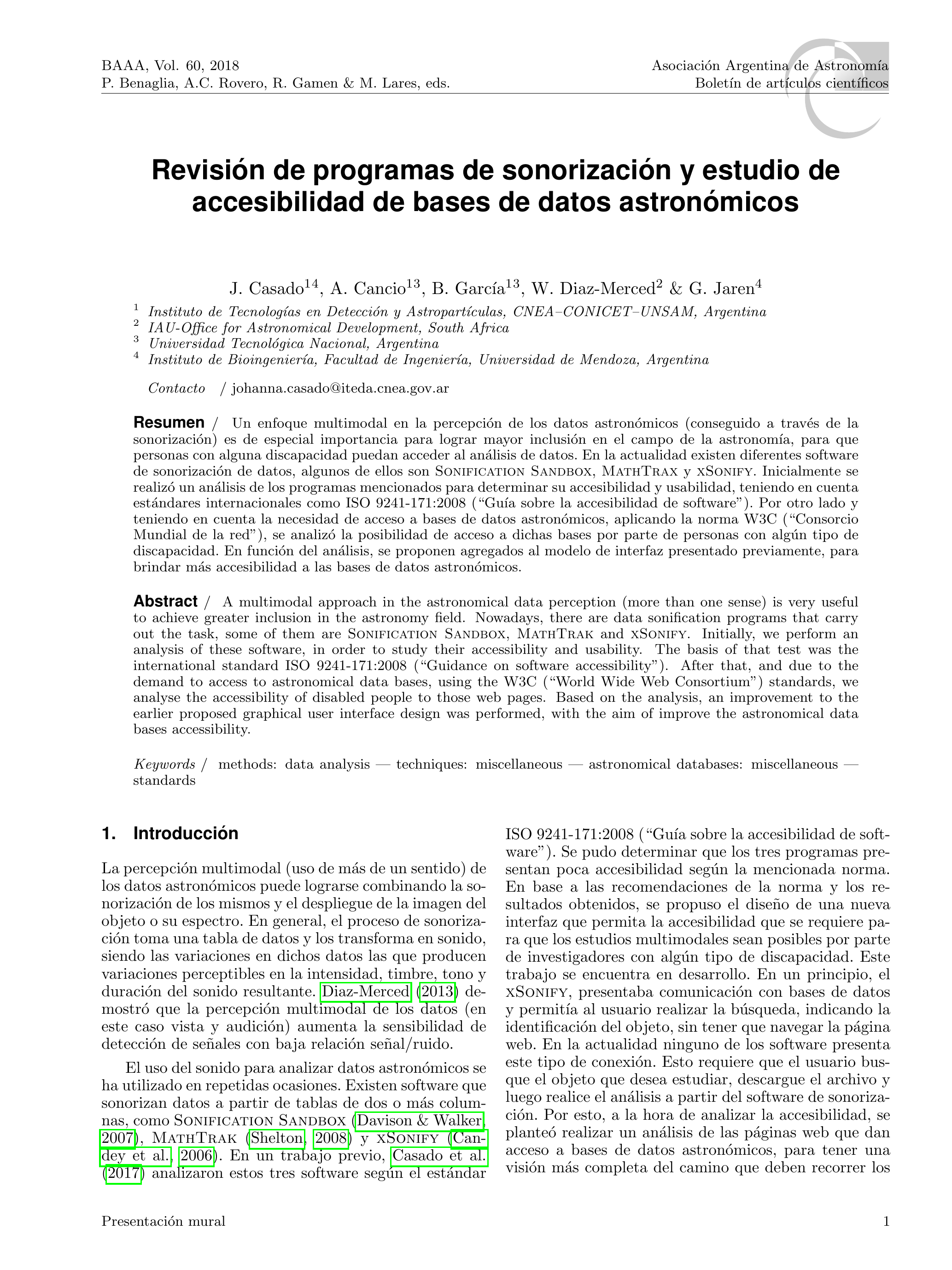}

\section{Informes obtenidos de ACHECKER}
\label{ap:achecker}

A continuación se mostraran los informes obtenidos de la aplicación web ACHECKER en el siguiente orden: SIMBAD, SDSS y ADS.

\includepdf[pages=-,scale=0.8]{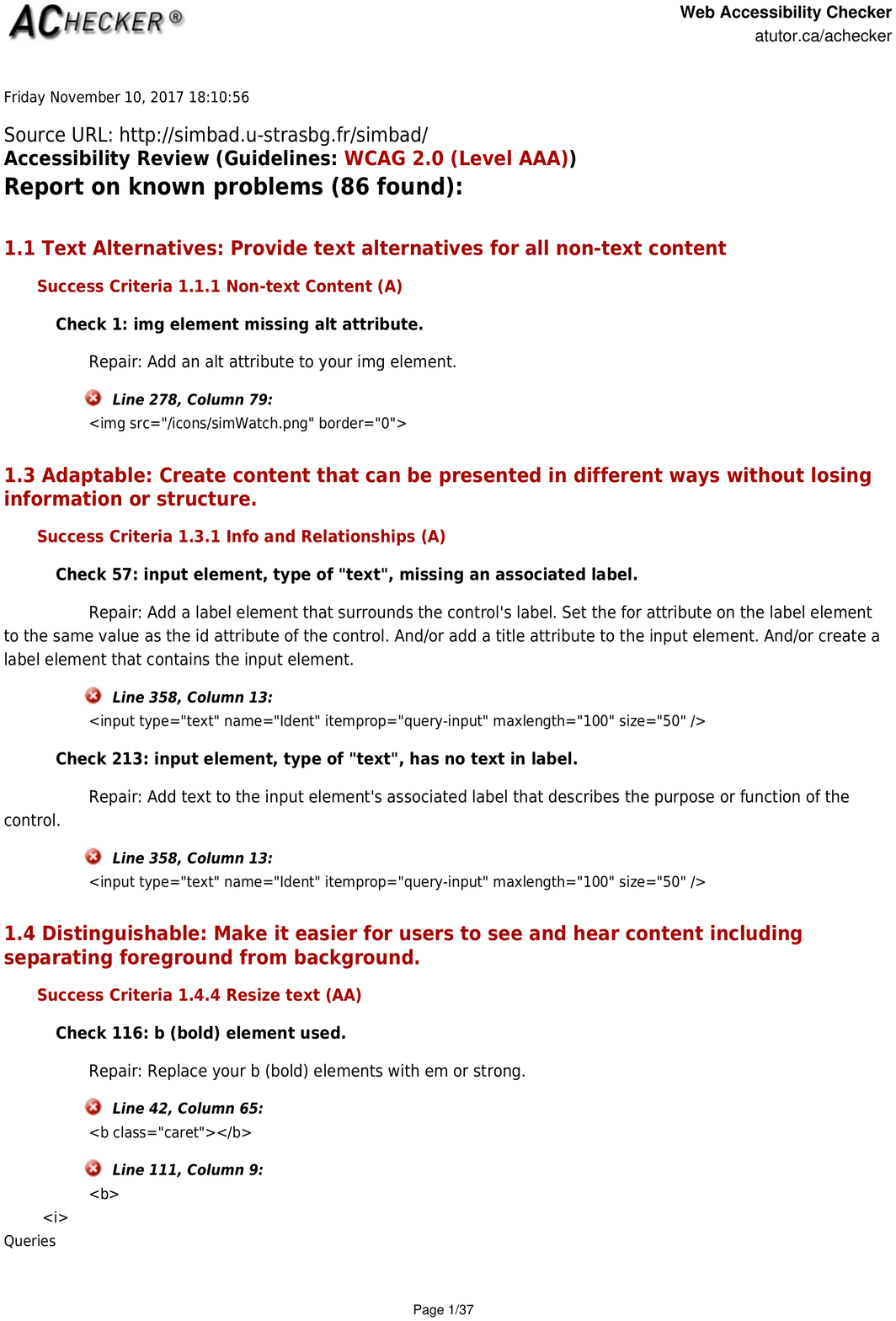}
\includepdf[pages=-,scale=0.8]{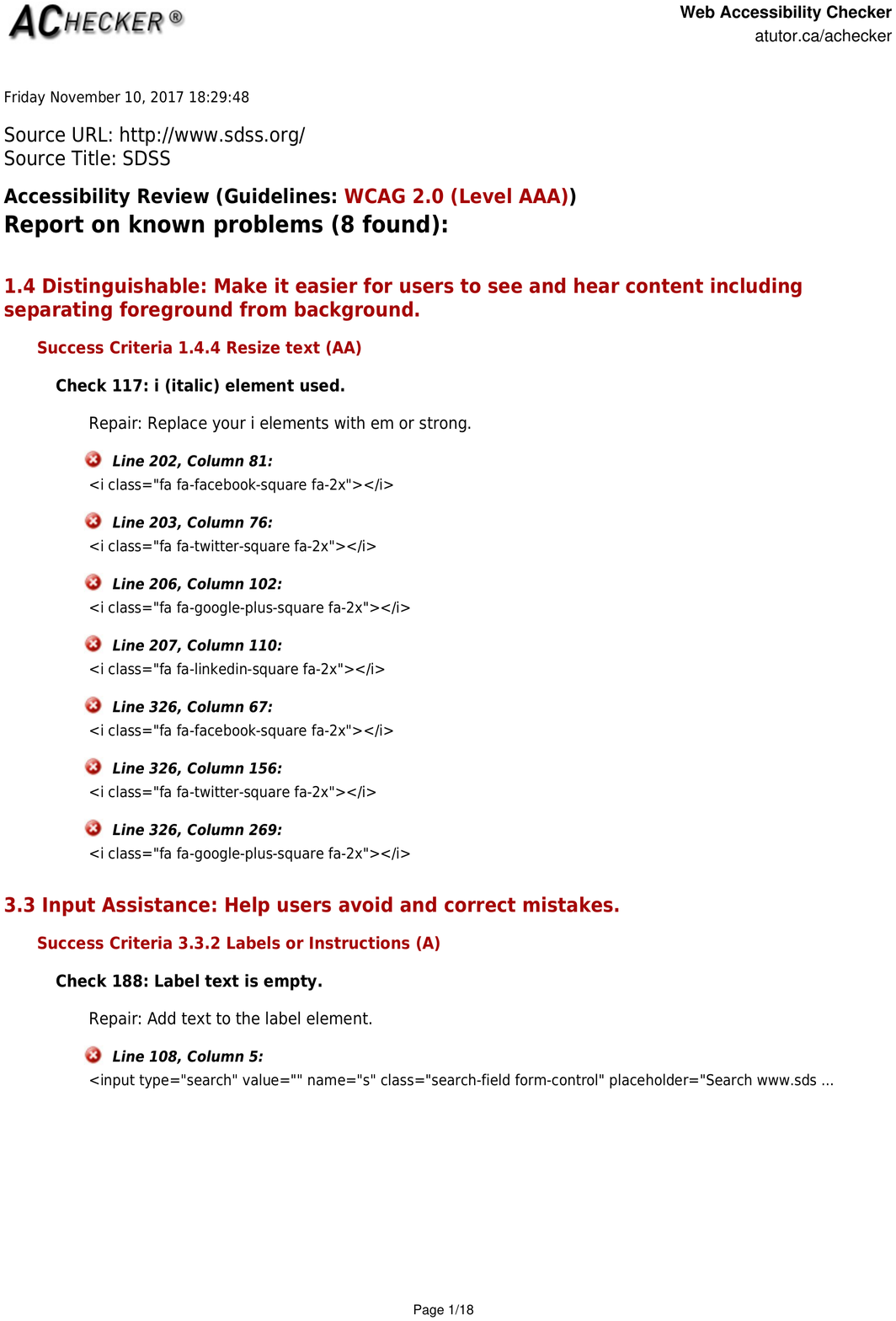}
\includepdf[pages=-,scale=0.8]{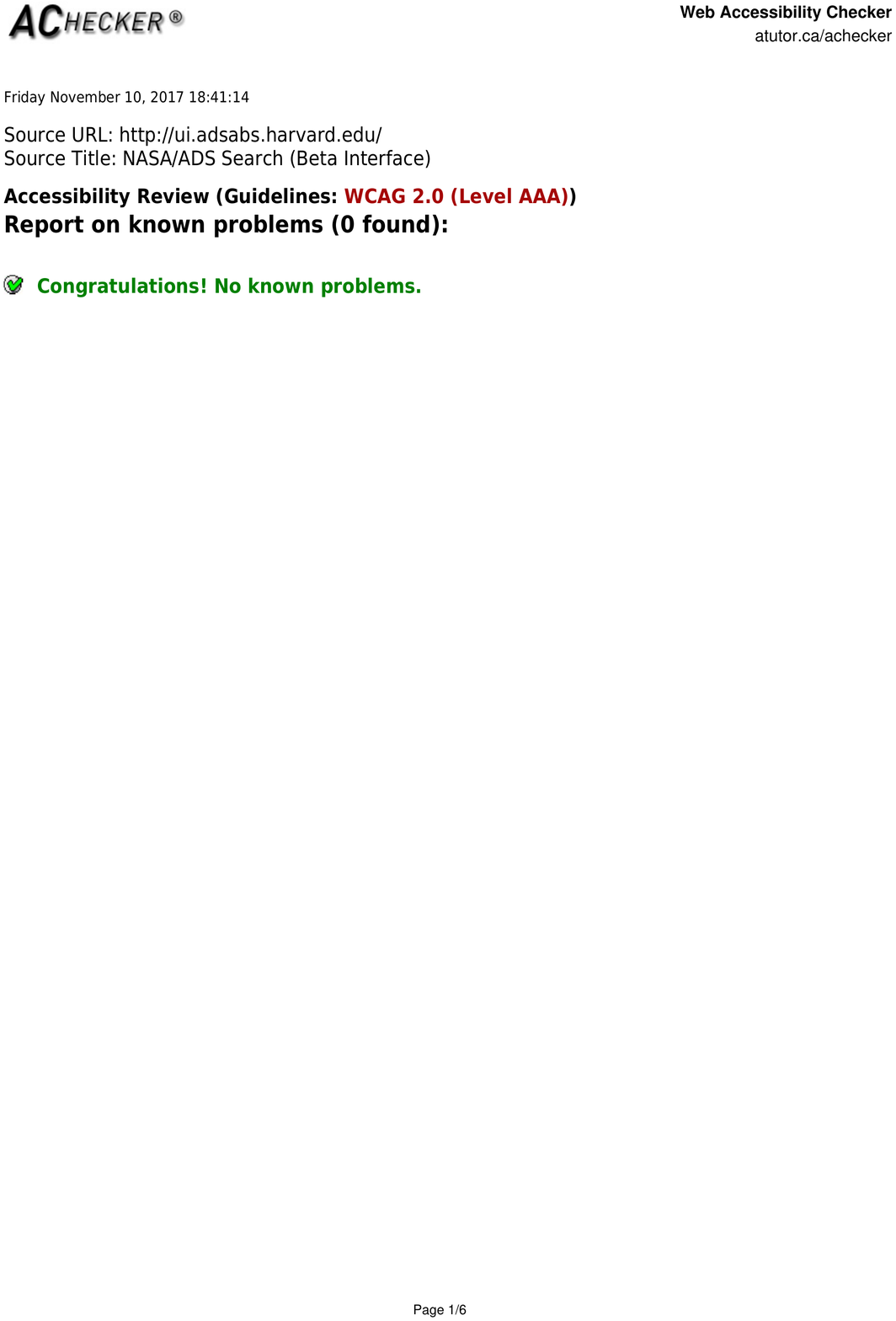}

\section{Informes obtenidos de FAE 2.0}
\label{ap:fae}

A continuación se mostraran los informes obtenidos de la aplicación web FAE 2.0 en el siguiente orden: SIMBAD, SDSS y ADS.

\includepdf[pages=-,scale=0.8]{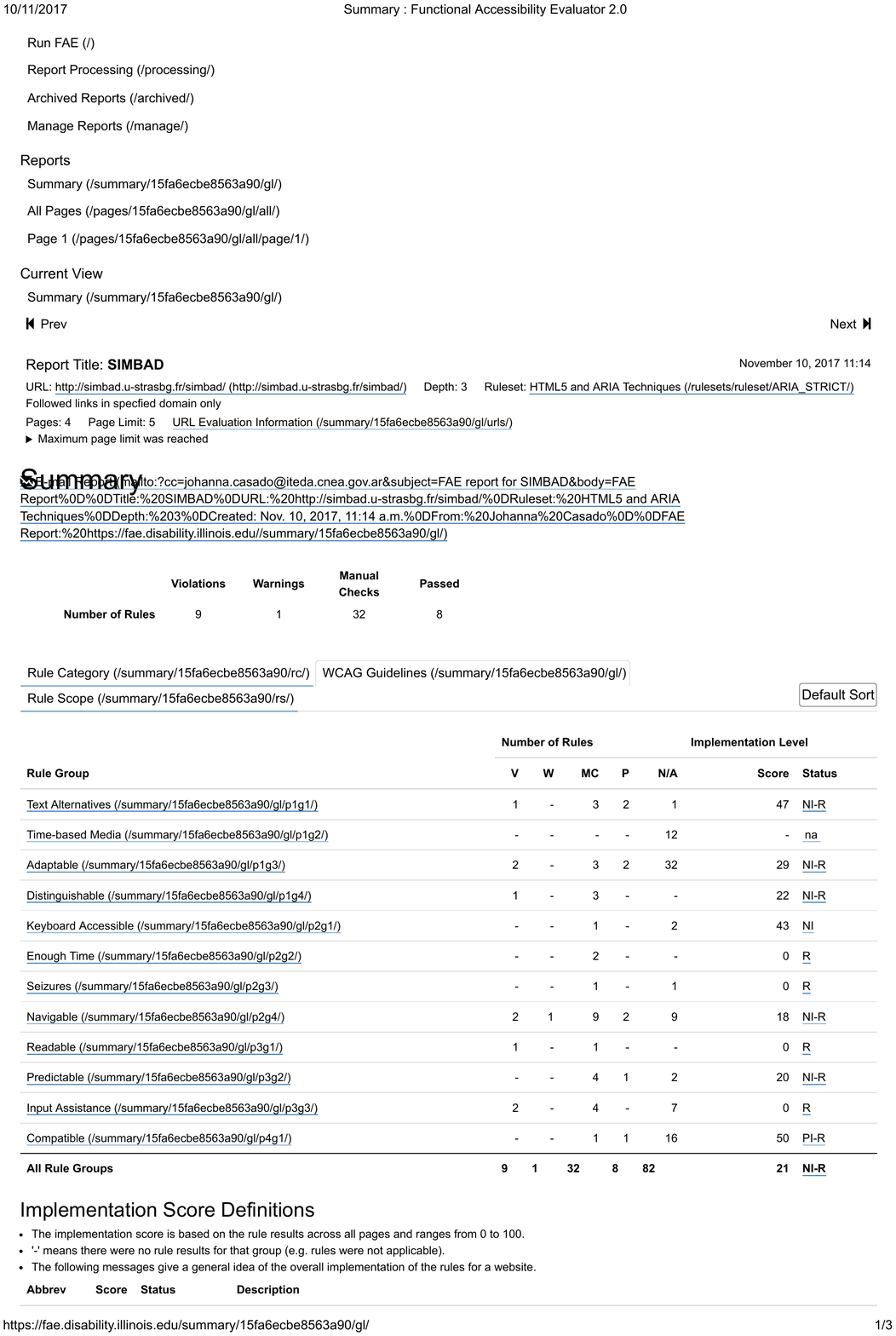}
\includepdf[pages=-,scale=0.8]{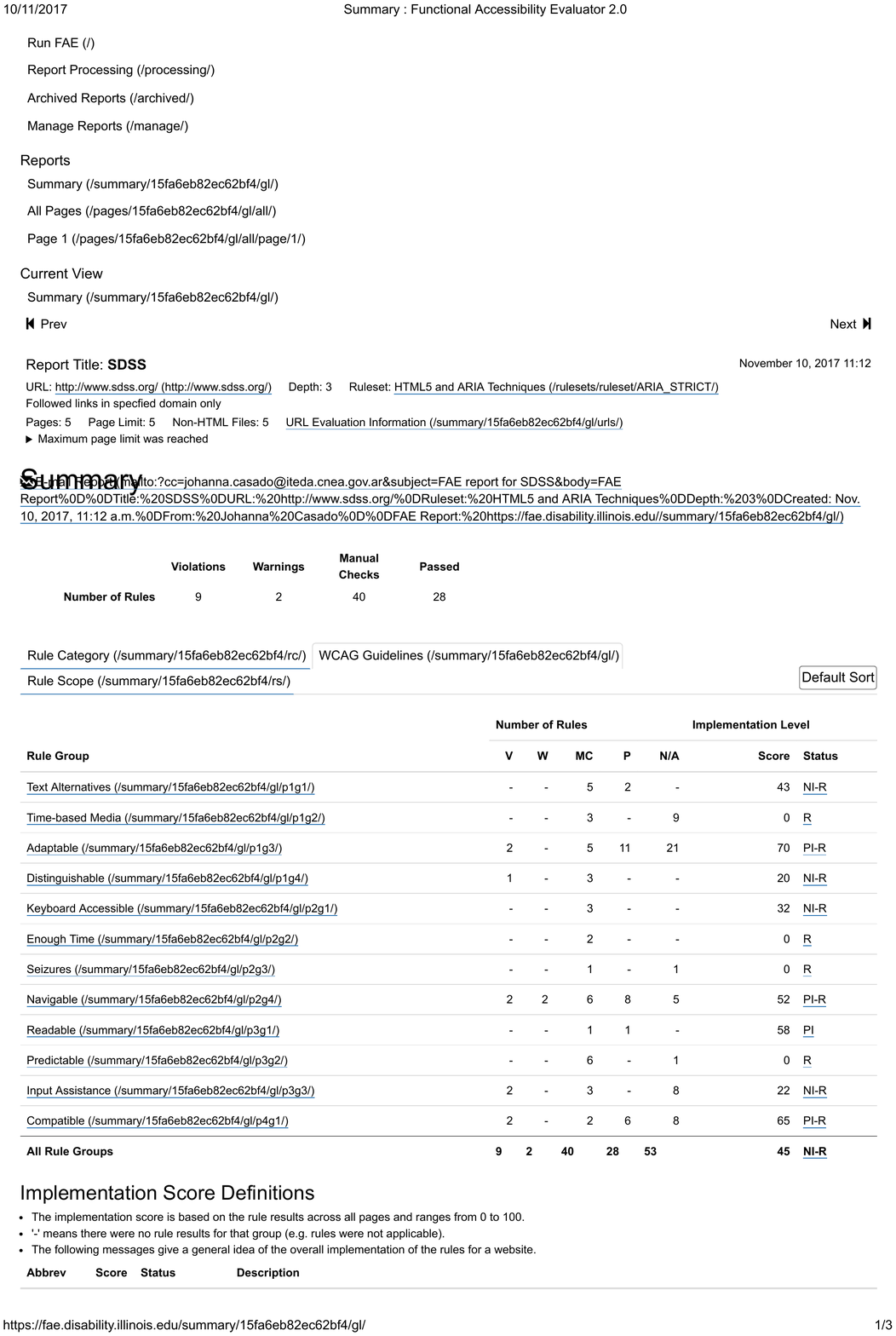}
\includepdf[pages=-,scale=0.8]{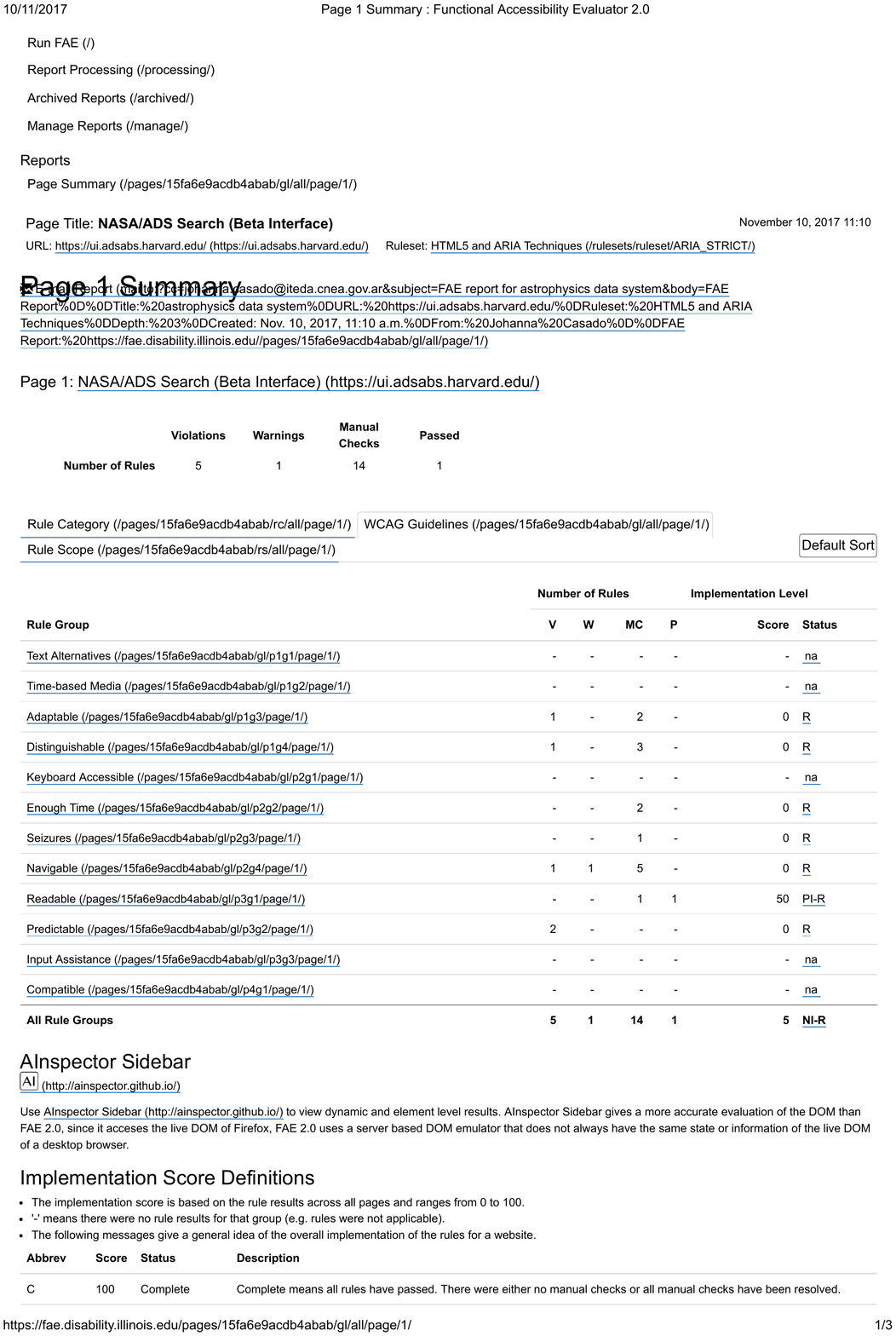}

\section{Informes obtenidos de EXAMINATOR}
\label{ap:examinator}

A continuación se mostraran los informes obtenidos de la aplicación web EXAMINATOR en el siguiente orden: SIMBAD, SDSS y ADS.

\includepdf[pages=-,scale=0.8]{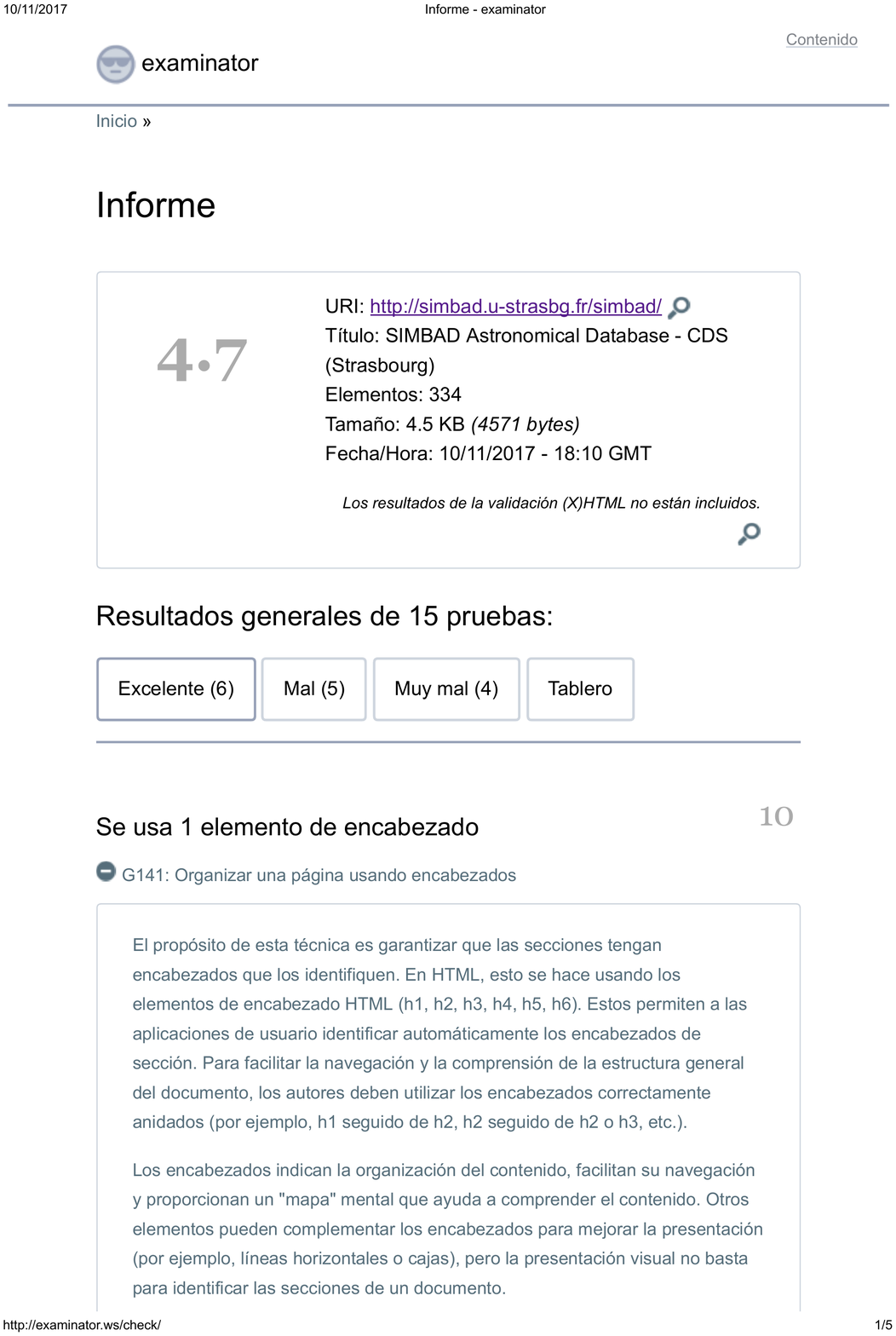}
\includepdf[pages=-,scale=0.8]{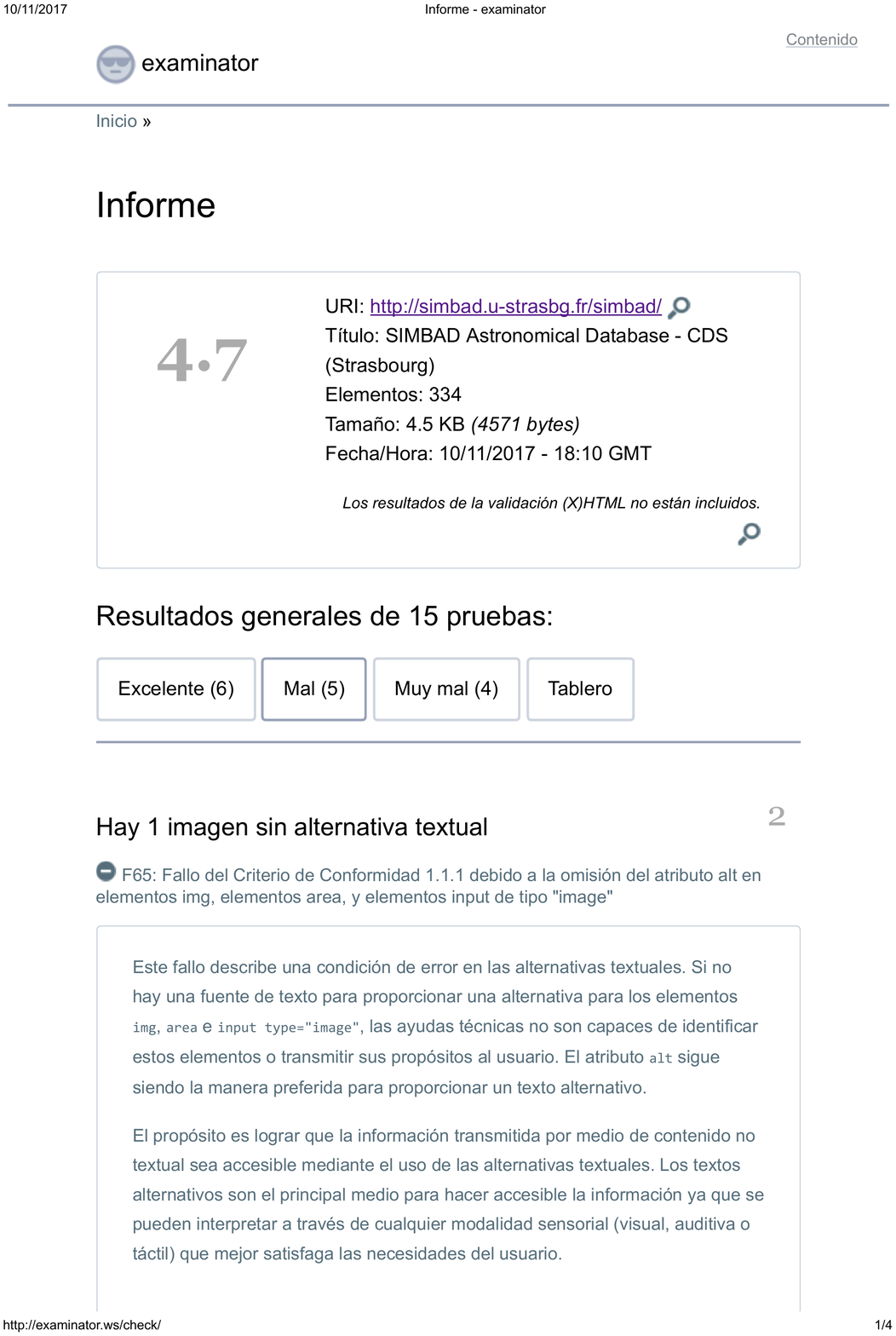}
\includepdf[pages=-,scale=0.8]{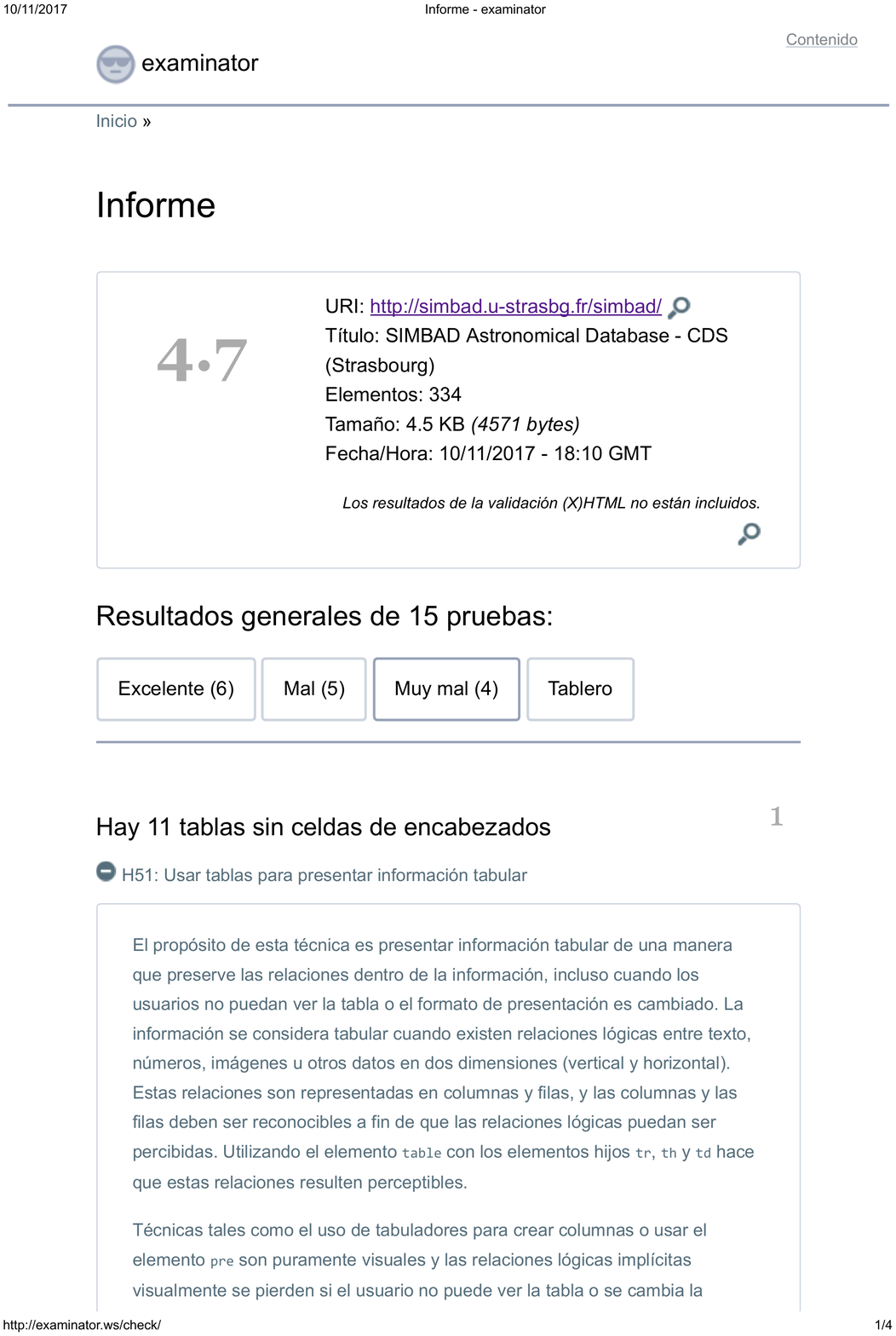}
\includepdf[pages=-,scale=0.8]{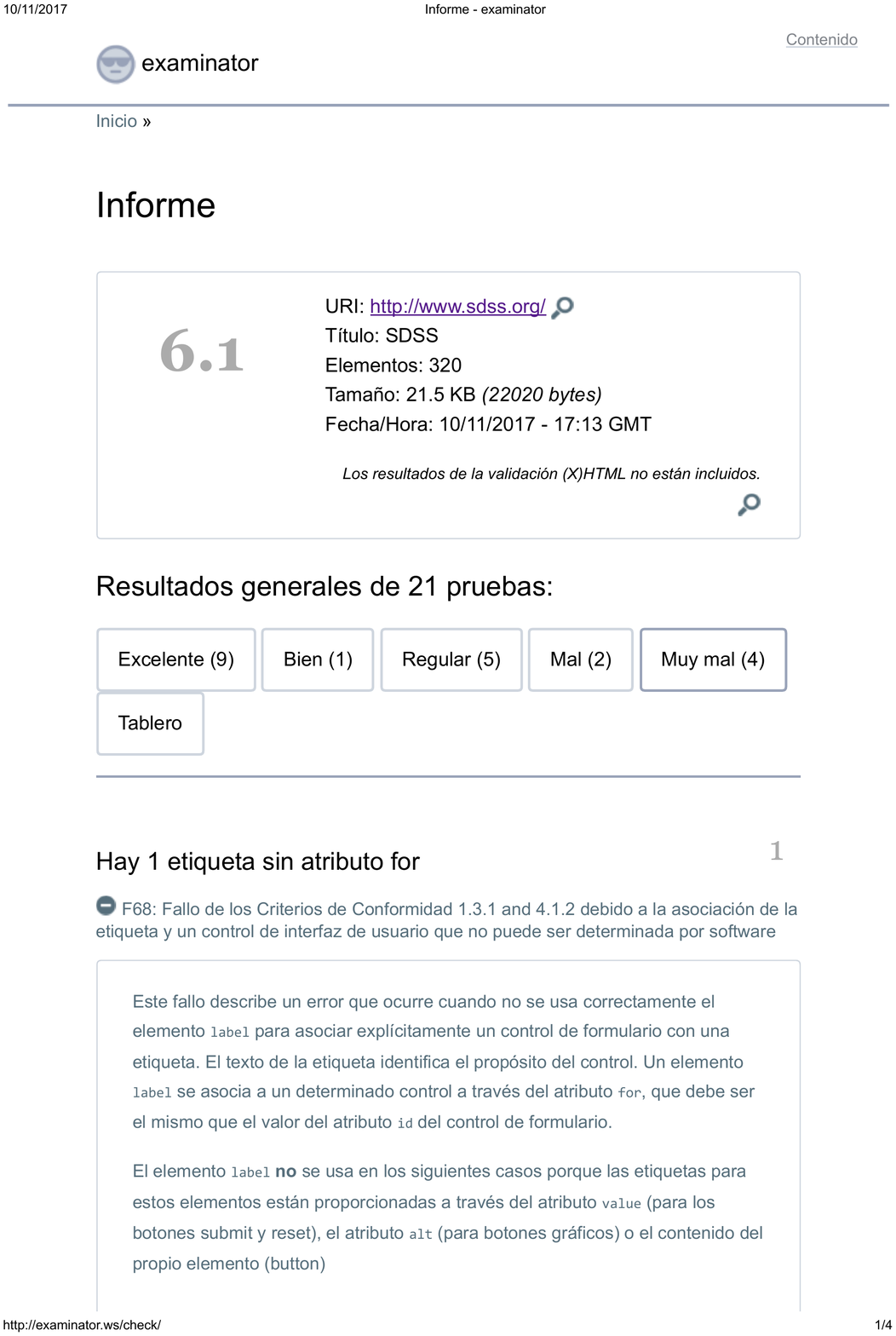}
\includepdf[pages=-,scale=0.8]{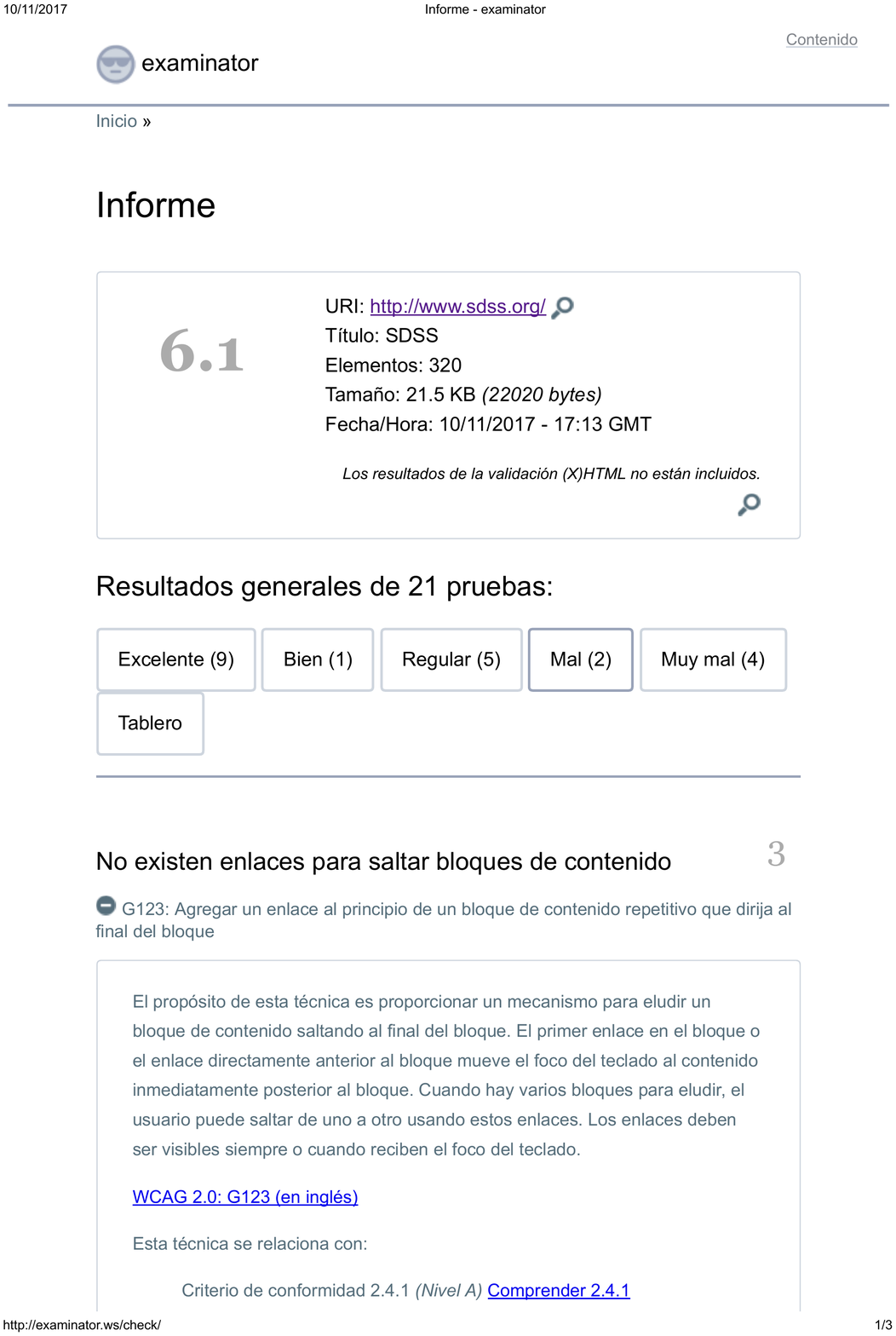}
\includepdf[pages=-,scale=0.8]{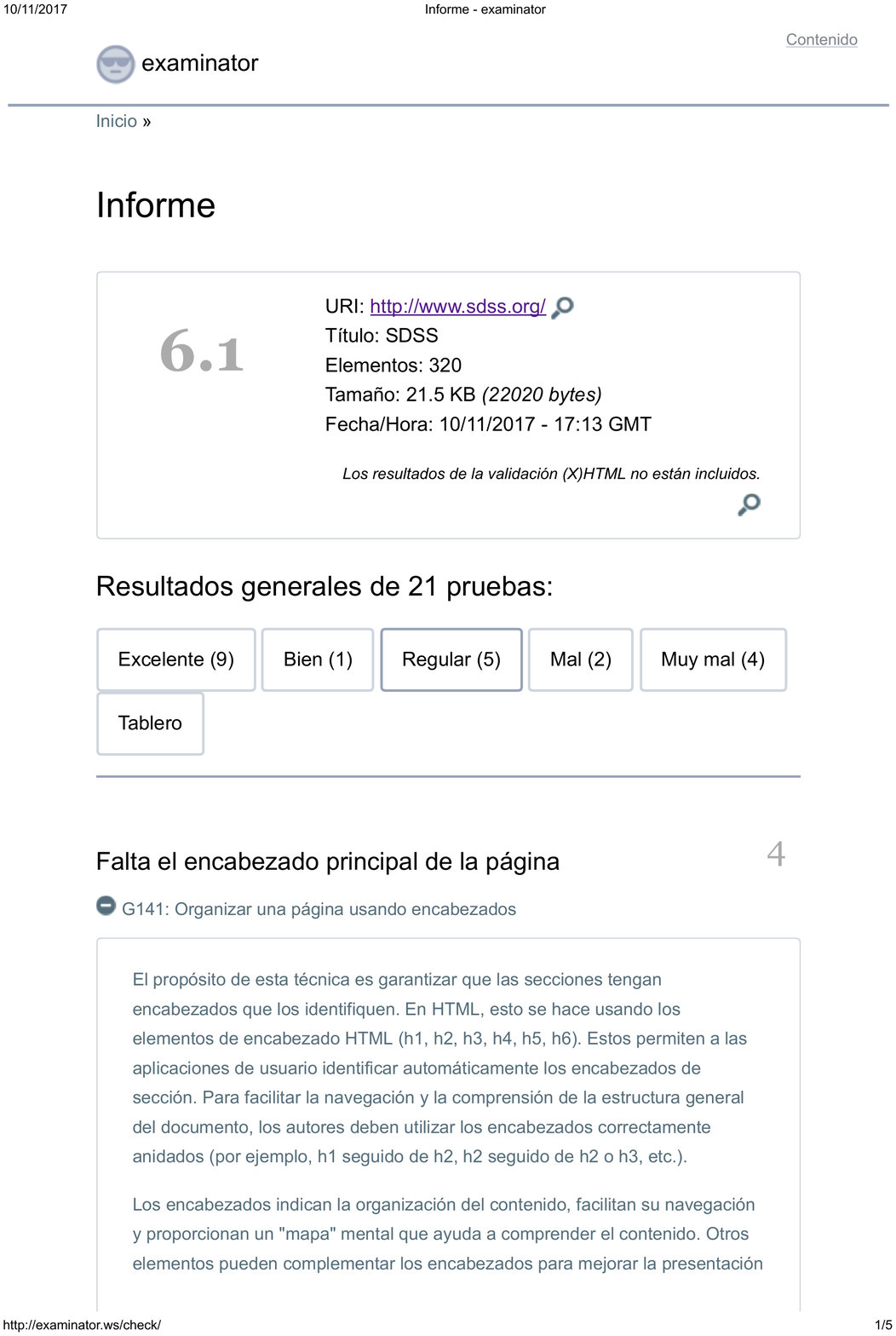}
\includepdf[pages=-,scale=0.8]{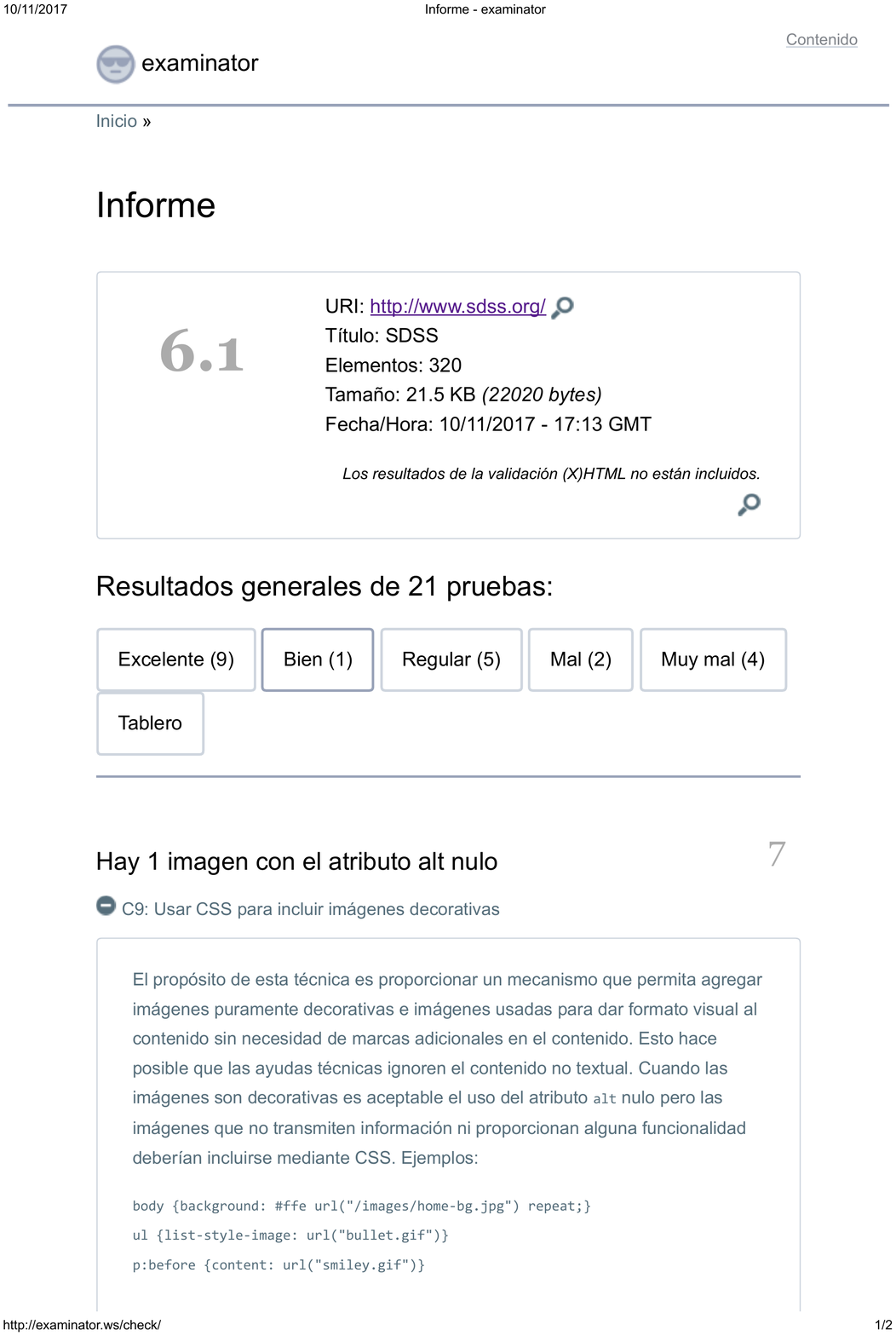}
\includepdf[pages=-,scale=0.8]{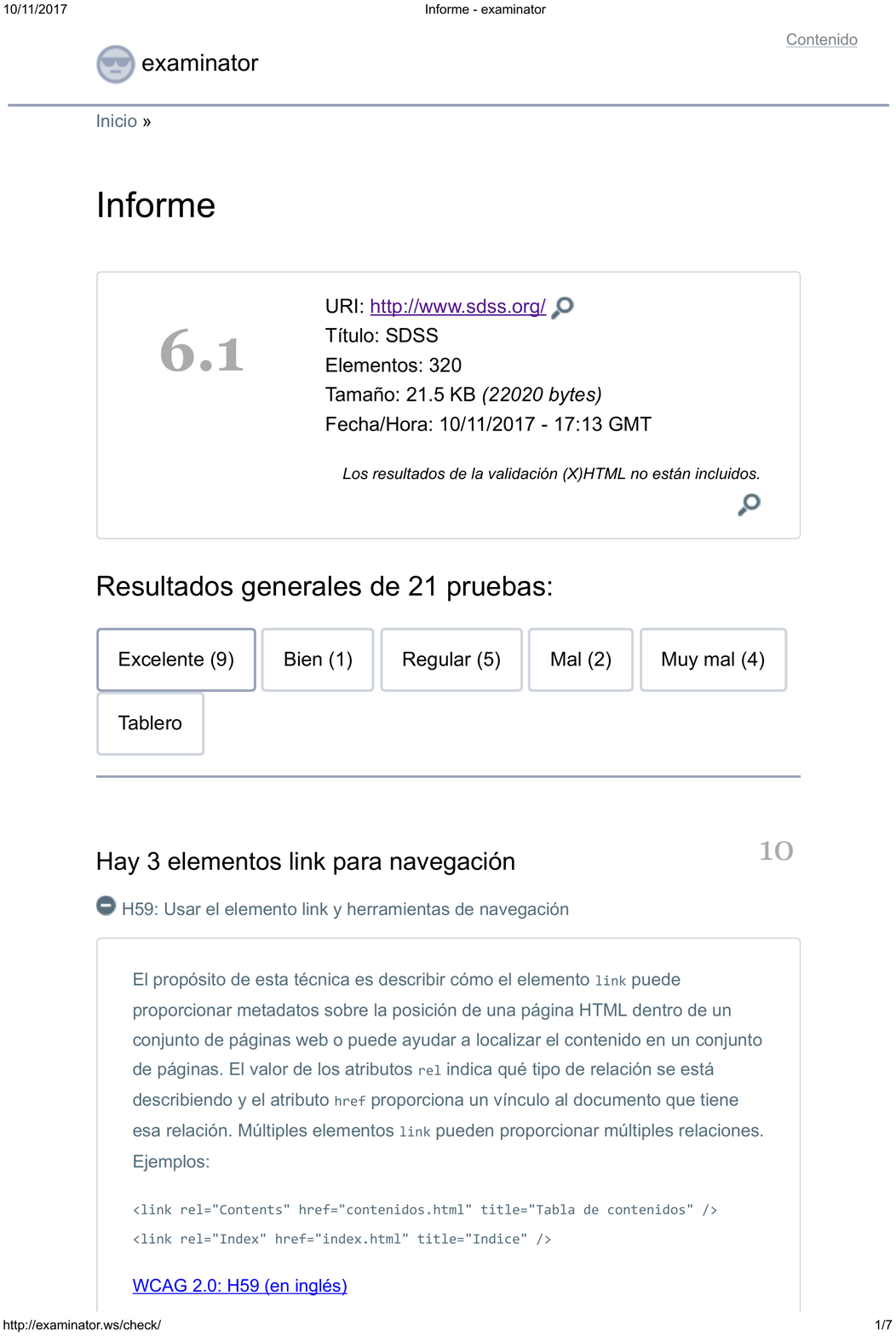}
\includepdf[pages=-,scale=0.8]{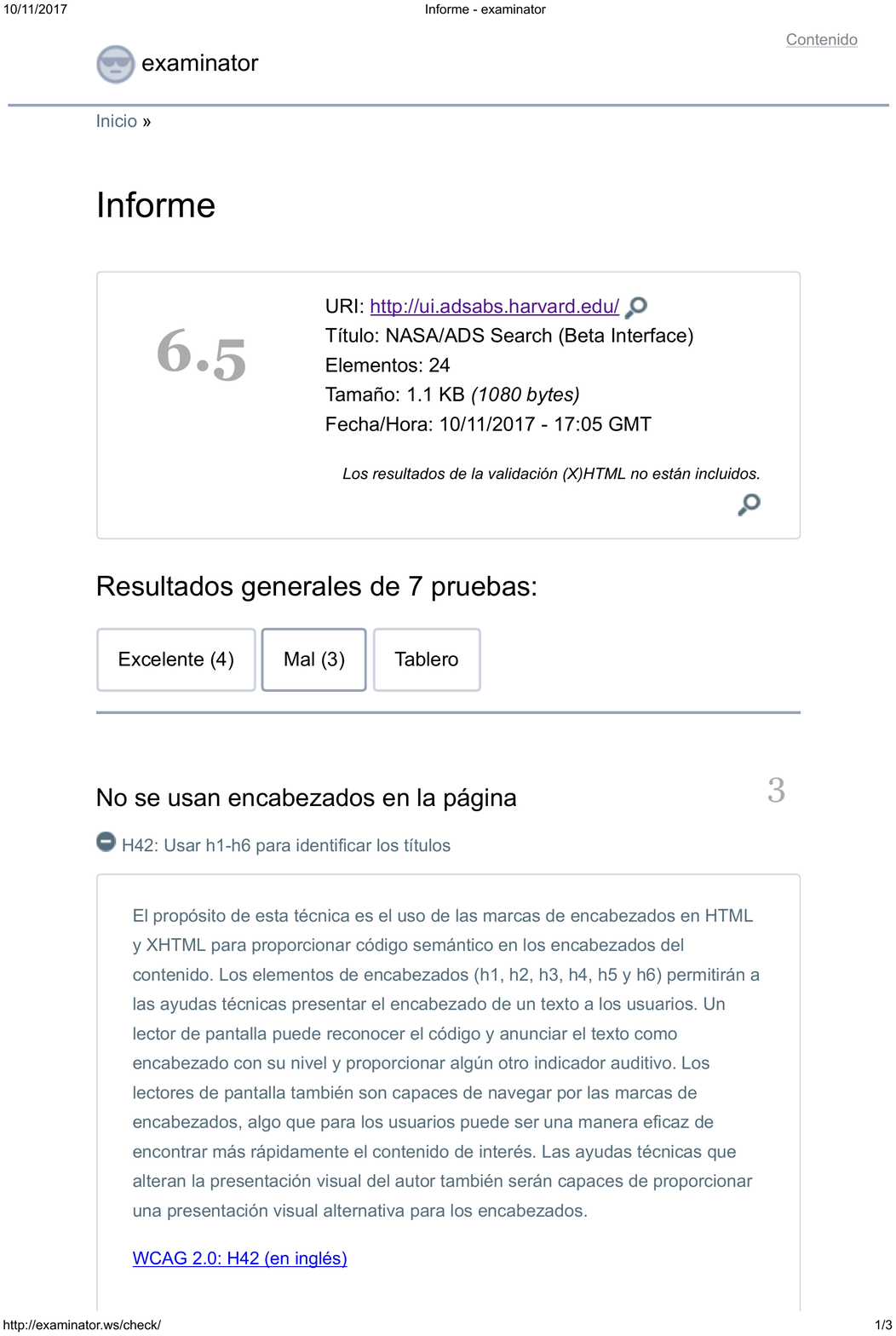}
\includepdf[pages=-,scale=0.8]{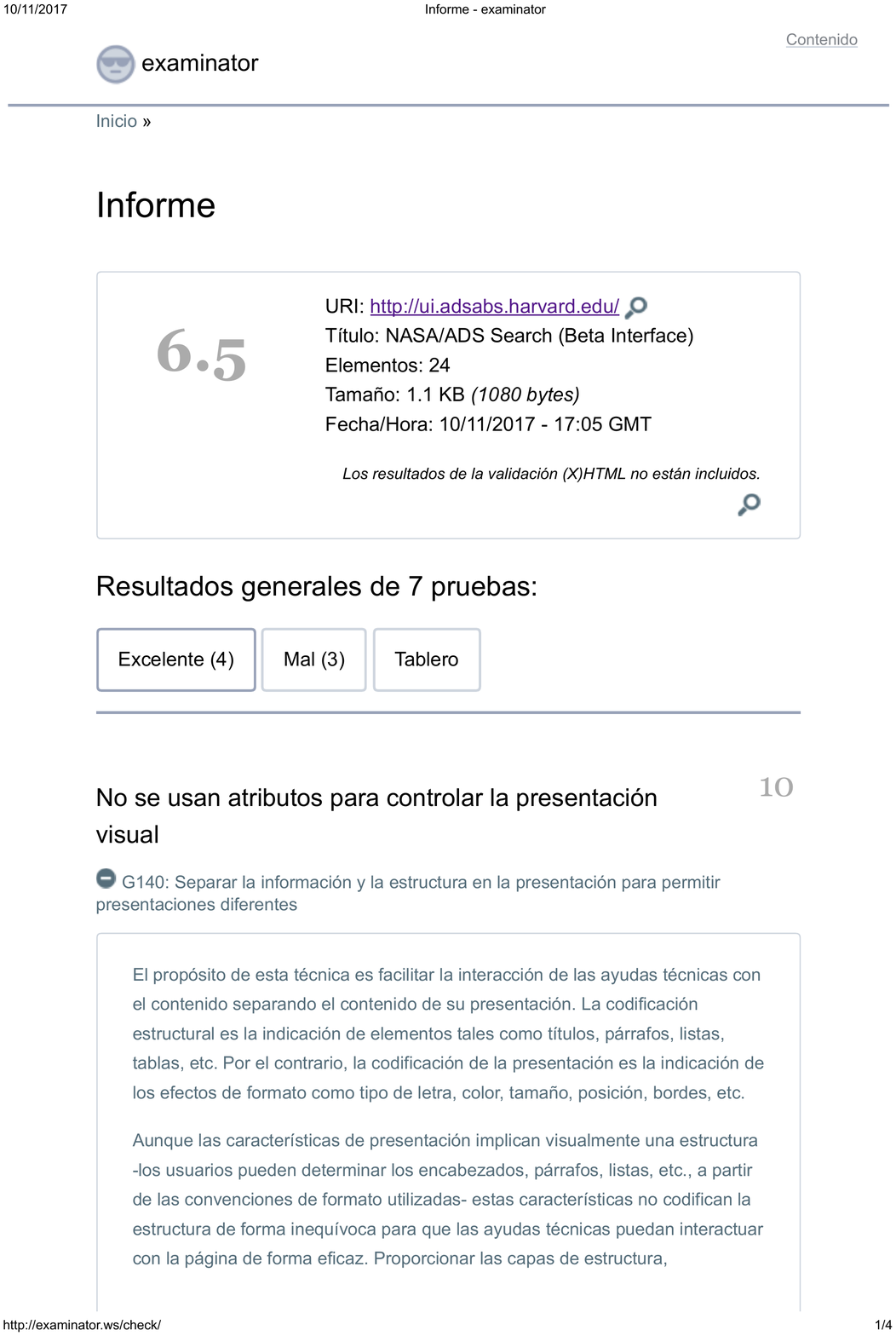}

\chapter{Publicaciones en congreso que hacen referencia al software sonoUno}
\label{ap:pub_sonouno_desktop}

Se agregan en este Apéndice los papers publicados en congreso que describen el desarrollo de sonoUno en su versión escritorio. El primero fue publicado en el Tercer Workshop de Difusión y Enseñanza de la Astronomía (\href{http://sion.frm.utn.edu.ar/WDEAIII/wp-content/uploads/2019/03/WDEA-III-2a-Circular_MARZO-2019.pdf}{\underline{\textcolor{blue}{WDEA III}}}), realizado en San Juan a mediados de 2019; el segundo trabajo que se adjunta aquí, fue presentado a fines de 2019 en el Simposio de la Unión Astronómica Internacional (IAUS385).

\includepdf[pages=-,scale=0.8]{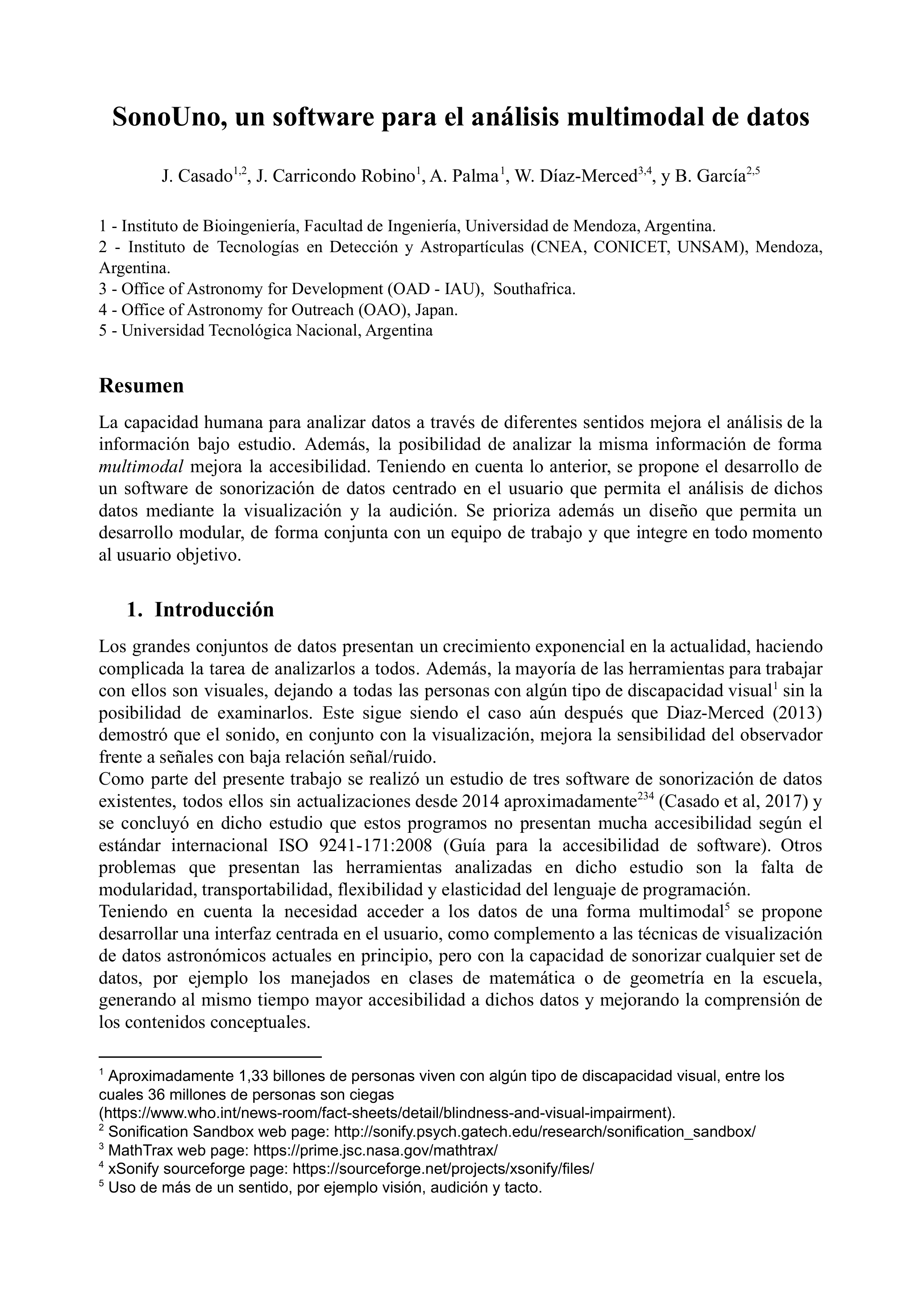}
\includepdf[pages=-,scale=0.8]{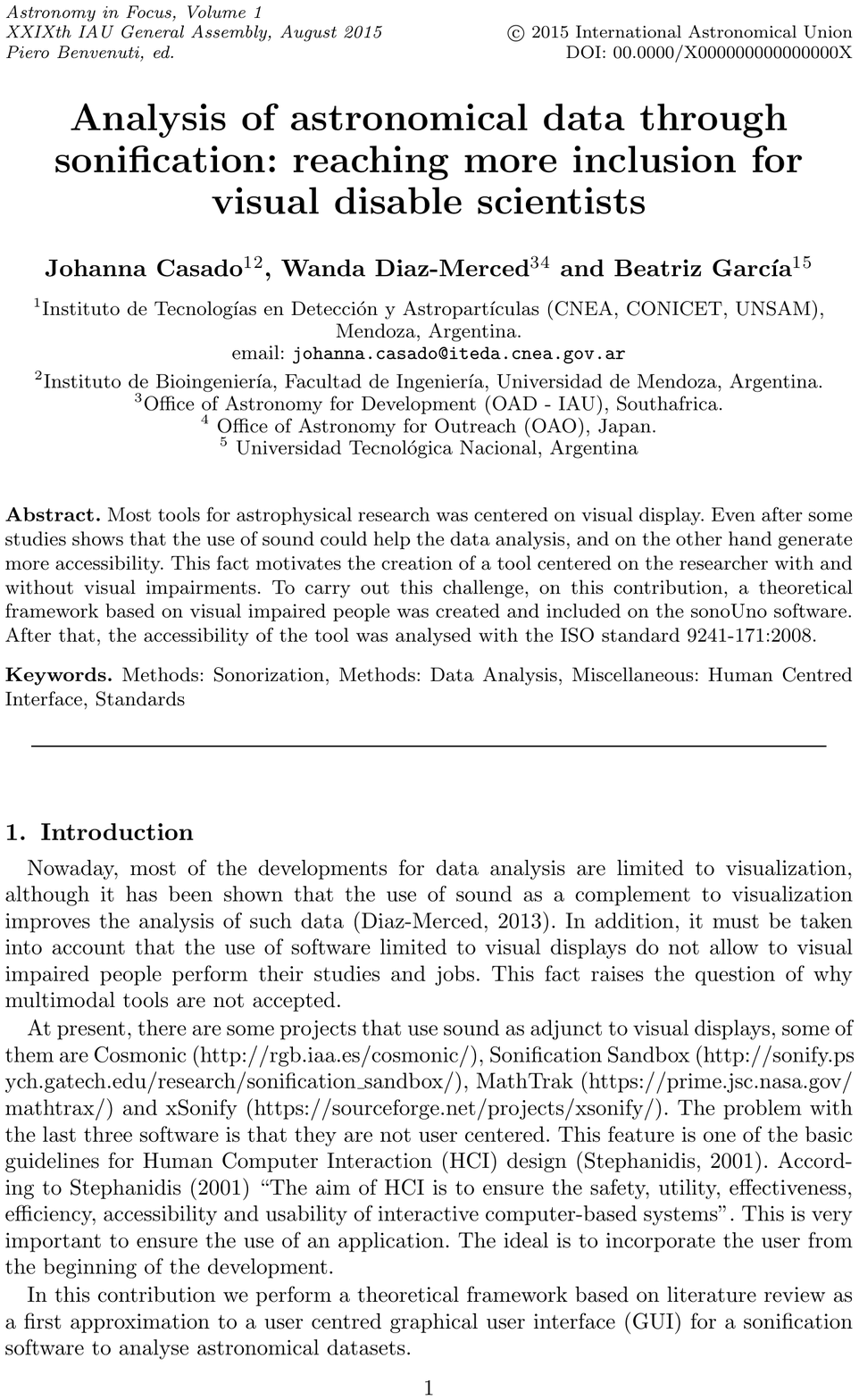}

\chapter{Publicación y documentos relacionados al análisis de grupo focal}
\label{ap:FG}

Se agrega en este Apéndice el paper publicado que describe el trabajo realizado en la Universidad de Southampton, con un grupo focal, con el objetivo de analizar usabilidad del programa desarrollado. Dicha publicación fue realizada en la revista \textit{``American Journal of Astronomy and Astrophysics''}. Este trabajo en particular ha sido descripto en esta tesis en la sección \ref{sect:FG_completo} del Capítulo \ref{cap:fg_completo}.

Adicionalmente, se agregan los archivos pdf que contienen la encuesta que se utilizó como guía para la entrevista de grupo focal y la lista de tareas que se le solicitó realizar a los participantes.

\includepdf[pages=-,scale=0.8]{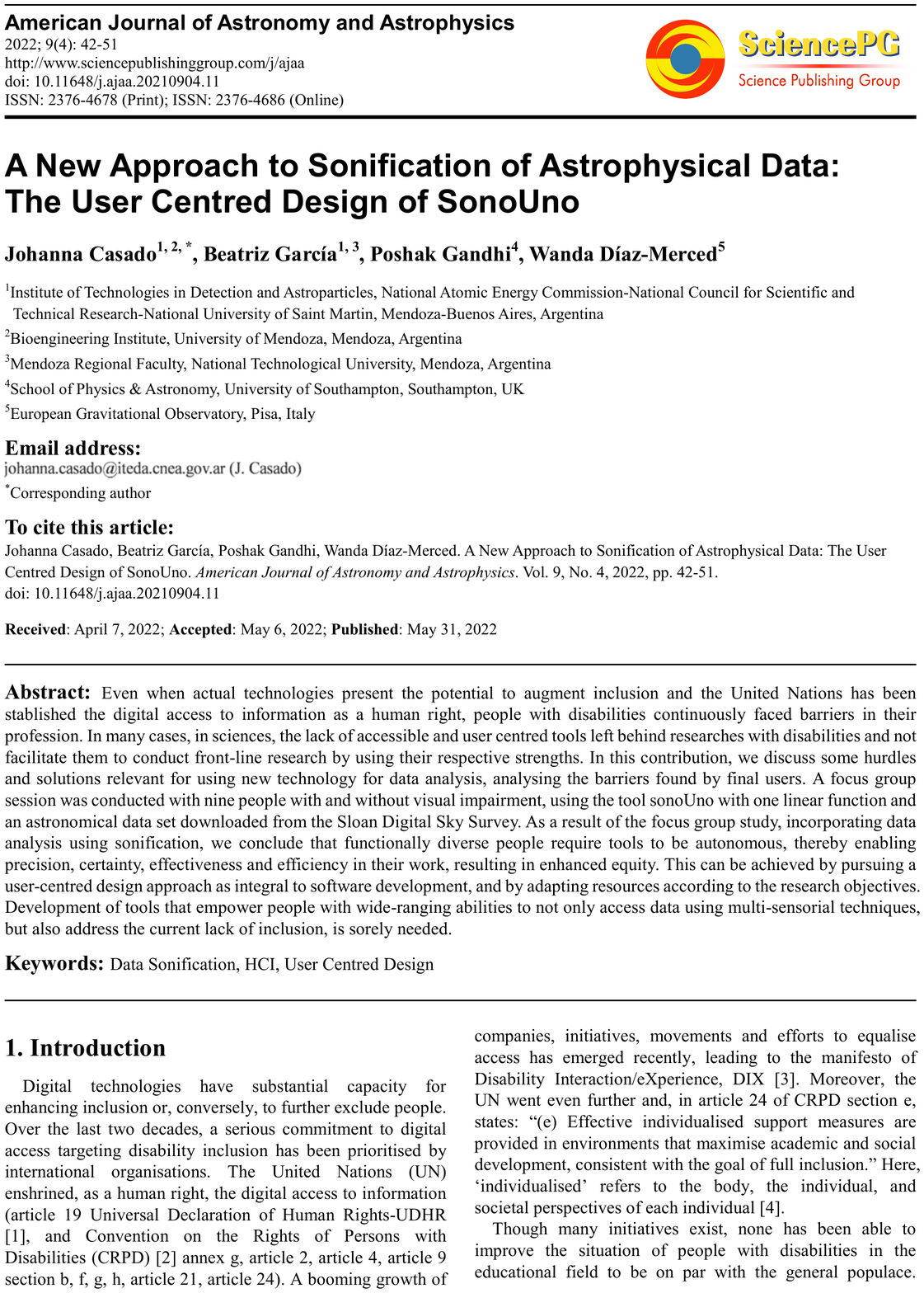}
\includepdf[pages=-,scale=0.8]{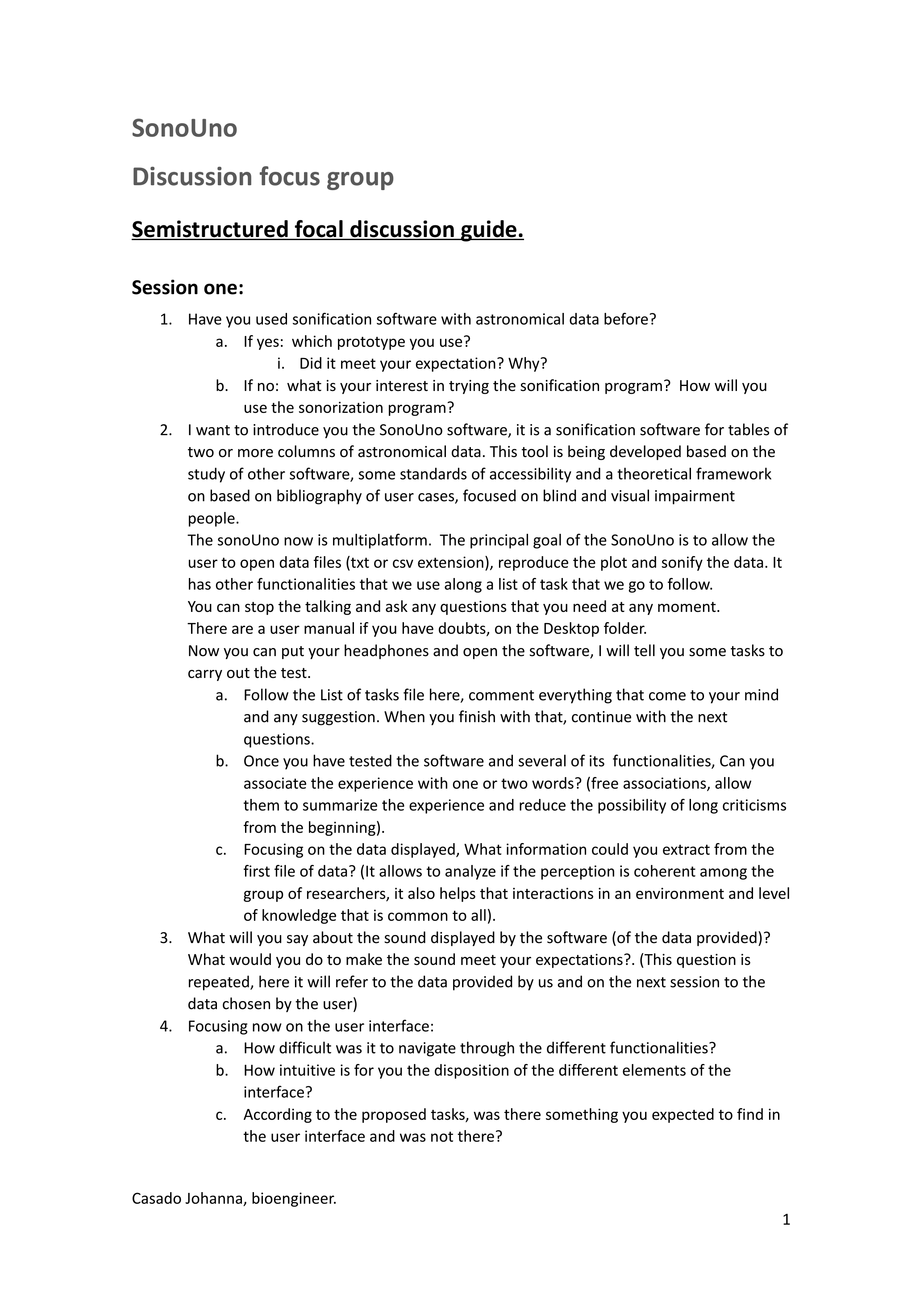}
\includepdf[pages=-,scale=0.8]{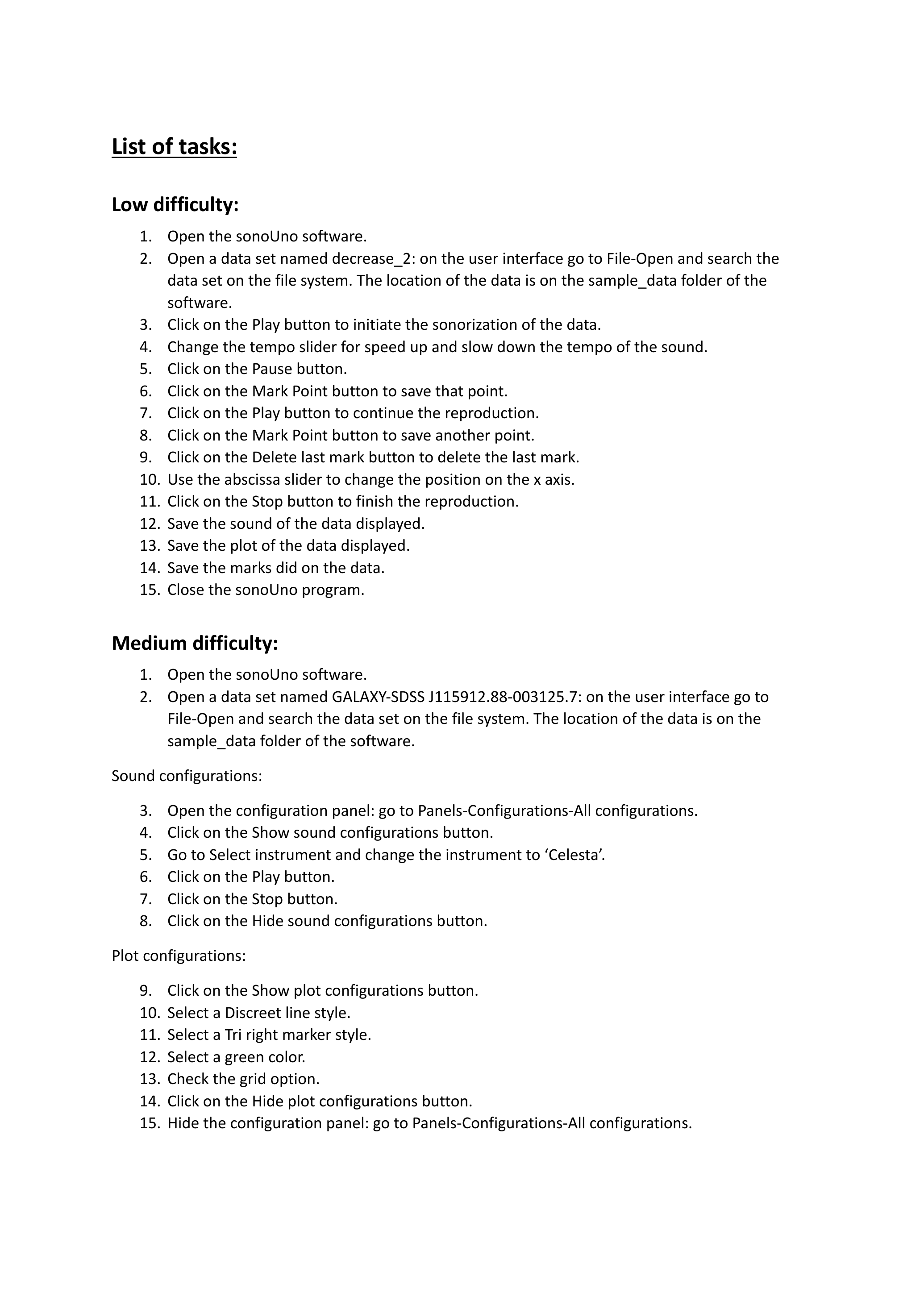}

\chapter{Publicaciones y cuerpo de mails enviados a los usuarios}
\label{ap:pub-y-mails}

Se agrega en este Apéndice el proceeding publicado que describe las primeras conclusiones de las pruebas realizadas con usuarios, las actualizaciones realizadas al programa sonoUno y los inicios de la versión web de sonoUno. Adicionalmente, se agregan dos archivos pdf que contienen el texto del mail enviado a los contactos para que probaran las nuevas versiones de sonoUno.

\includepdf[pages=-,scale=0.8]{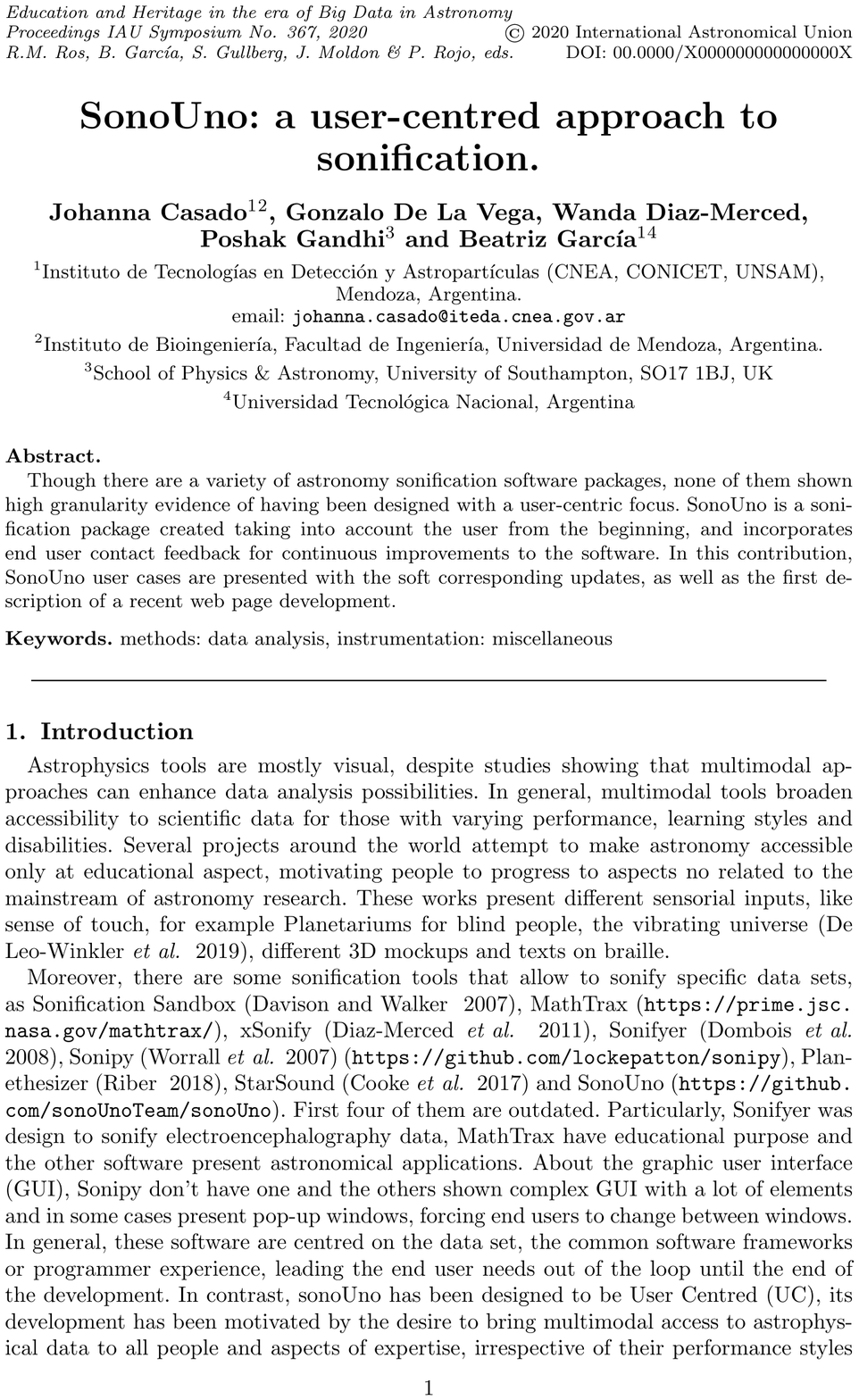}
\includepdf[pages=-,scale=0.8]{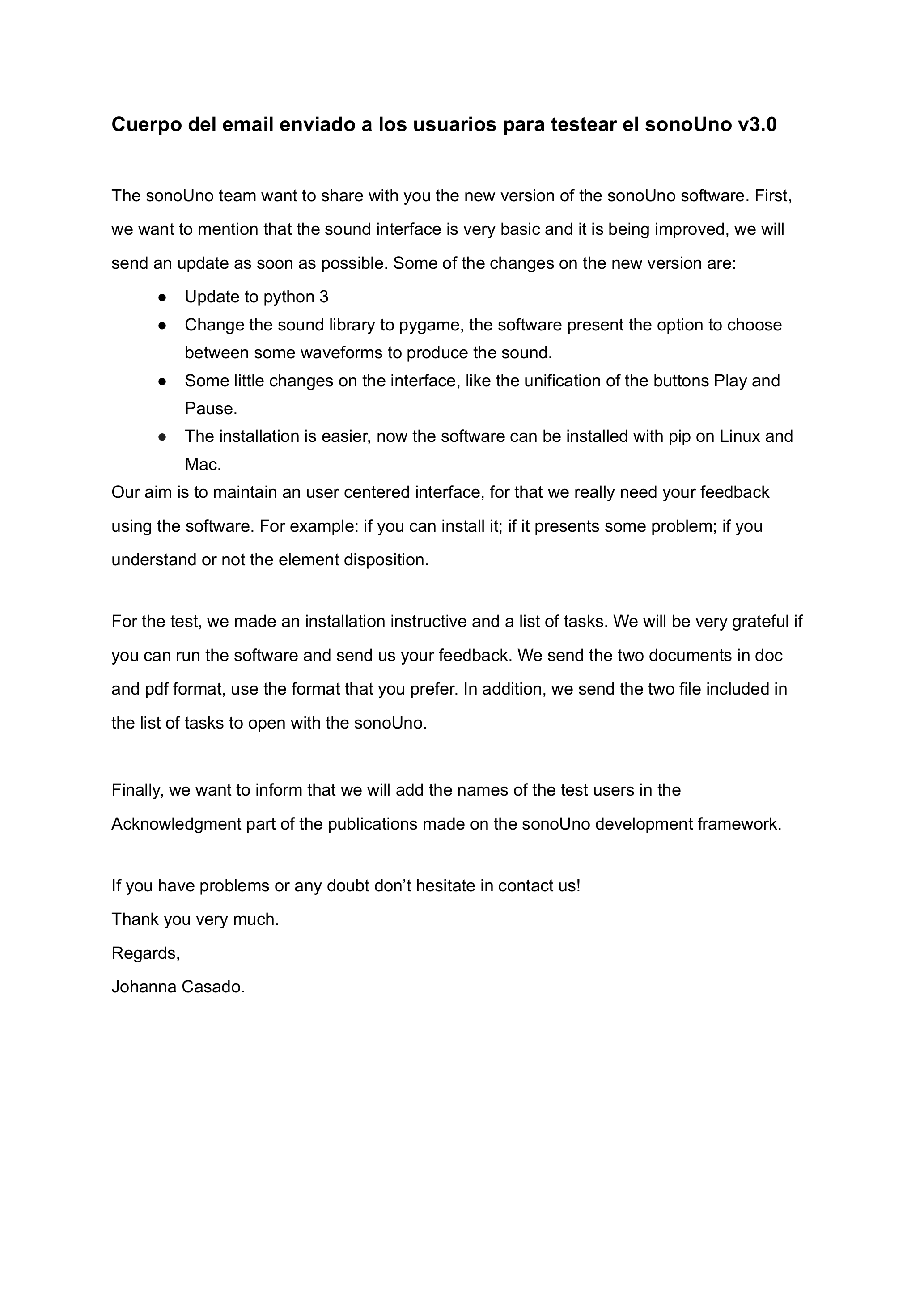}
\includepdf[pages=-,scale=0.8]{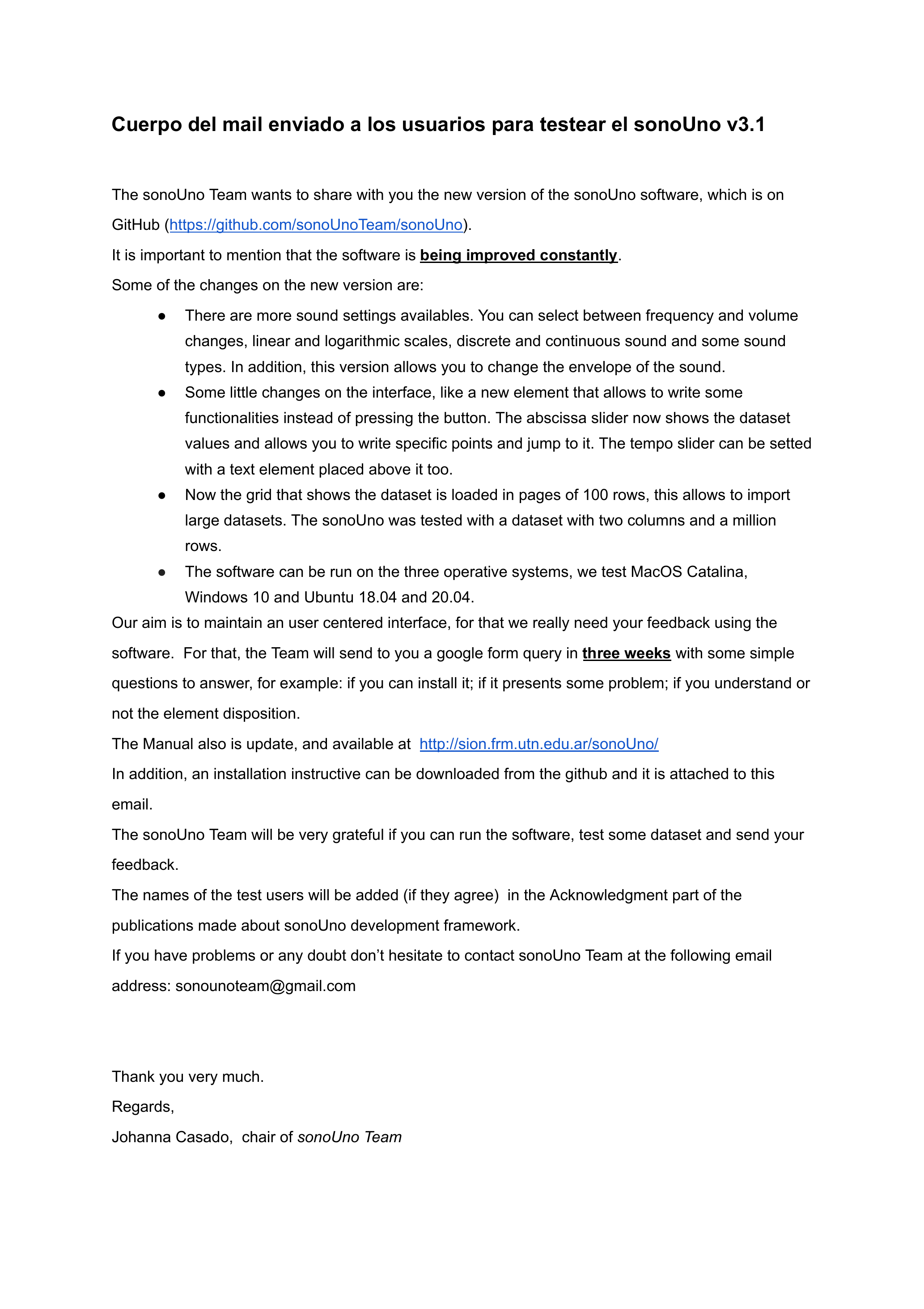}

\chapter{Publicaciones sobre la herramienta}
\label{ap:pub-webyreinforce}

Se agrega en este Apéndice tres proceedings publicados de los cuales: uno describe el proyecto sonoUno en el marco de ciencia ciudadana; y los otros dos describen diferentes partes del desarrollo de la página web de sonoUno.

El primer paper que se adjunta es el relacionado con ciencia ciudadana, fue publicado en la Reunión Internacional de Interfaz Humano Computadora (HCII 2021). El segundo proceeding que se agrega fue publicado en la Revista Mexicana de Astronomía y Astrofísica, se presentó el trabajo en el \textit{2nd Workshop on Astronomy Beyond the Common Senses for Accessibility and Inclusion (2022)}. Por último, el tercer trabajo se presentó en la Reunión Internacional de Interfaz Humano Computadora del año pasado (HCII 2022).

\includepdf[pages=-]{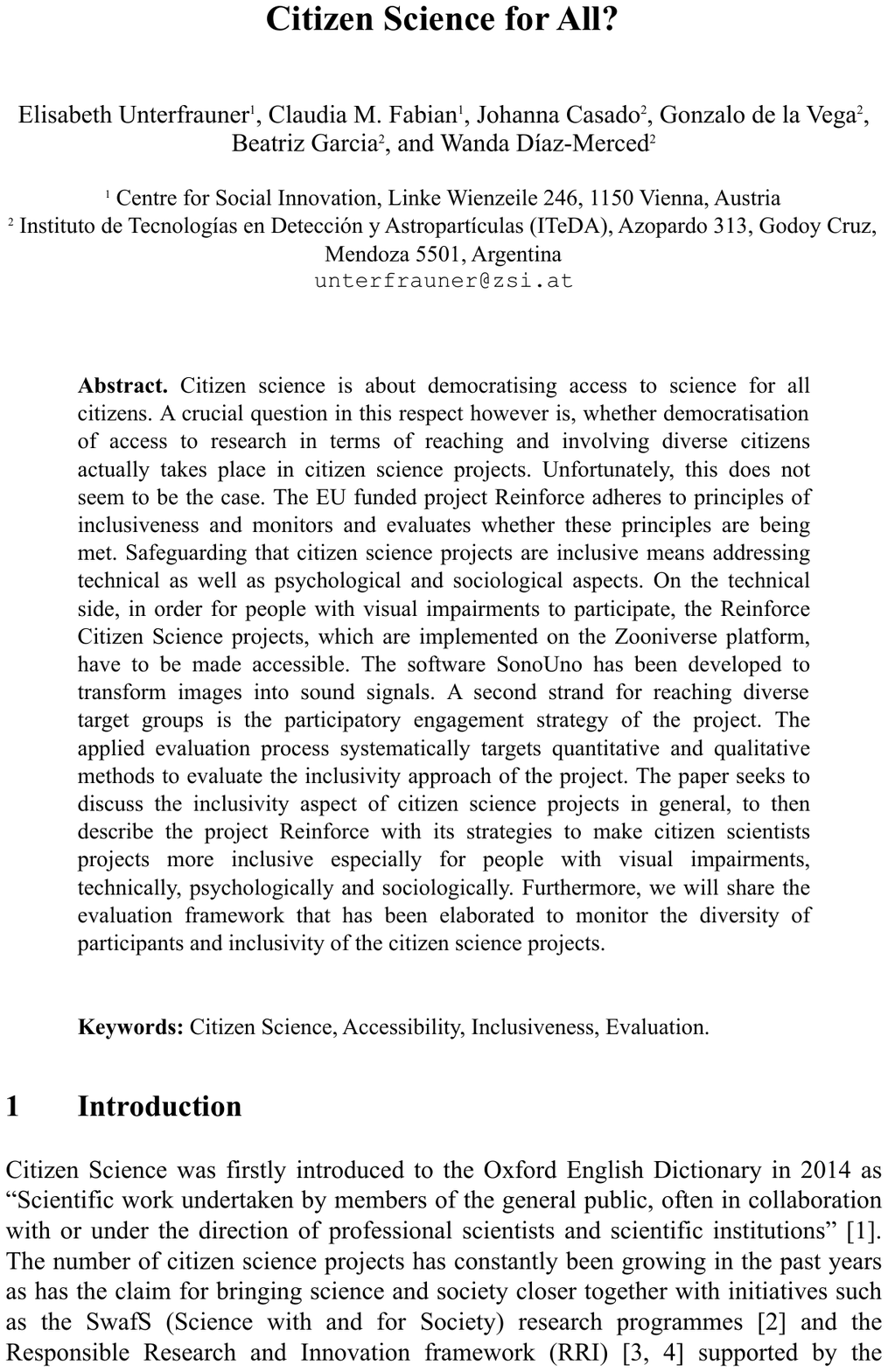}

\includepdf[pages=-,scale=0.8]{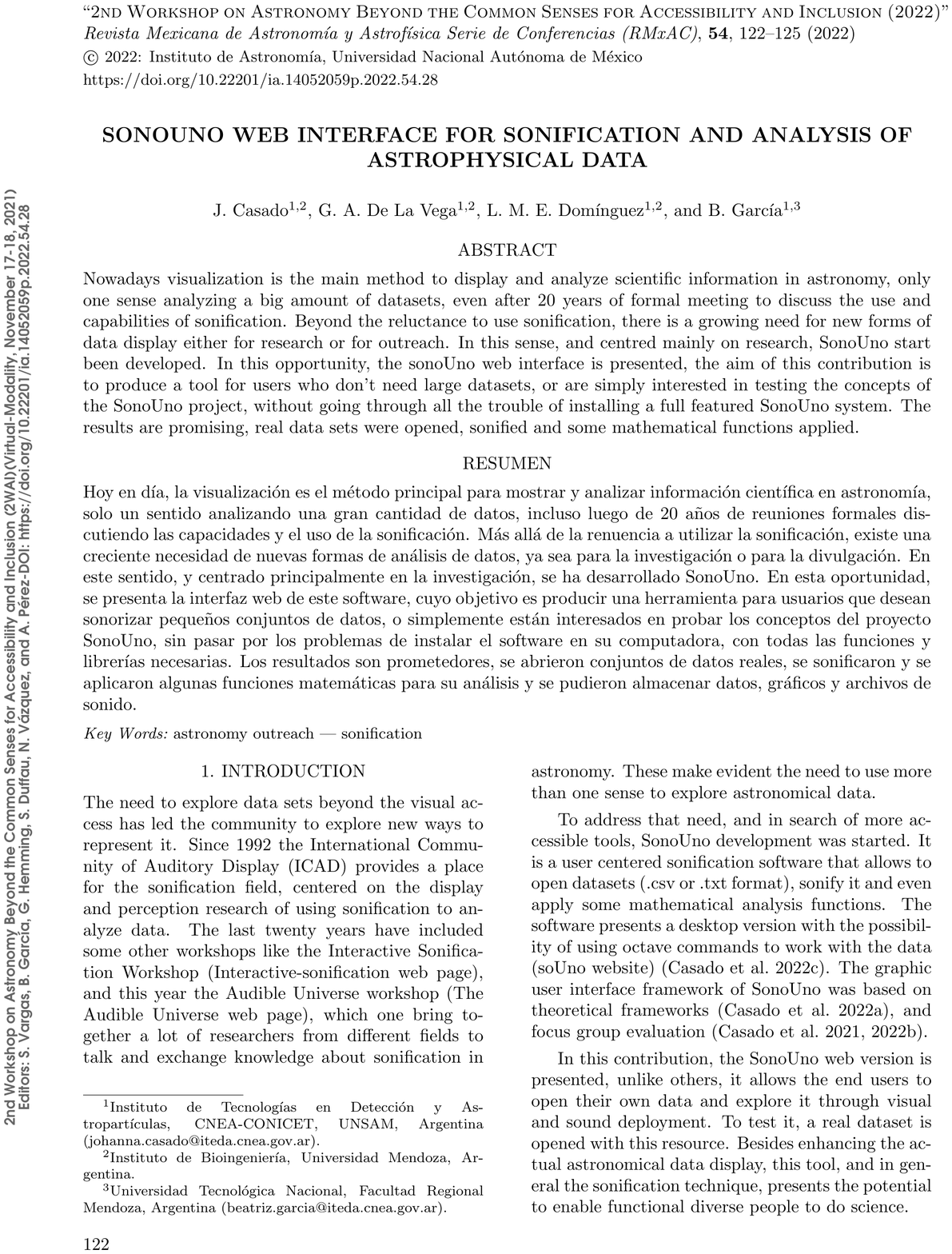}

\includepdf[pages=-,scale=0.8]{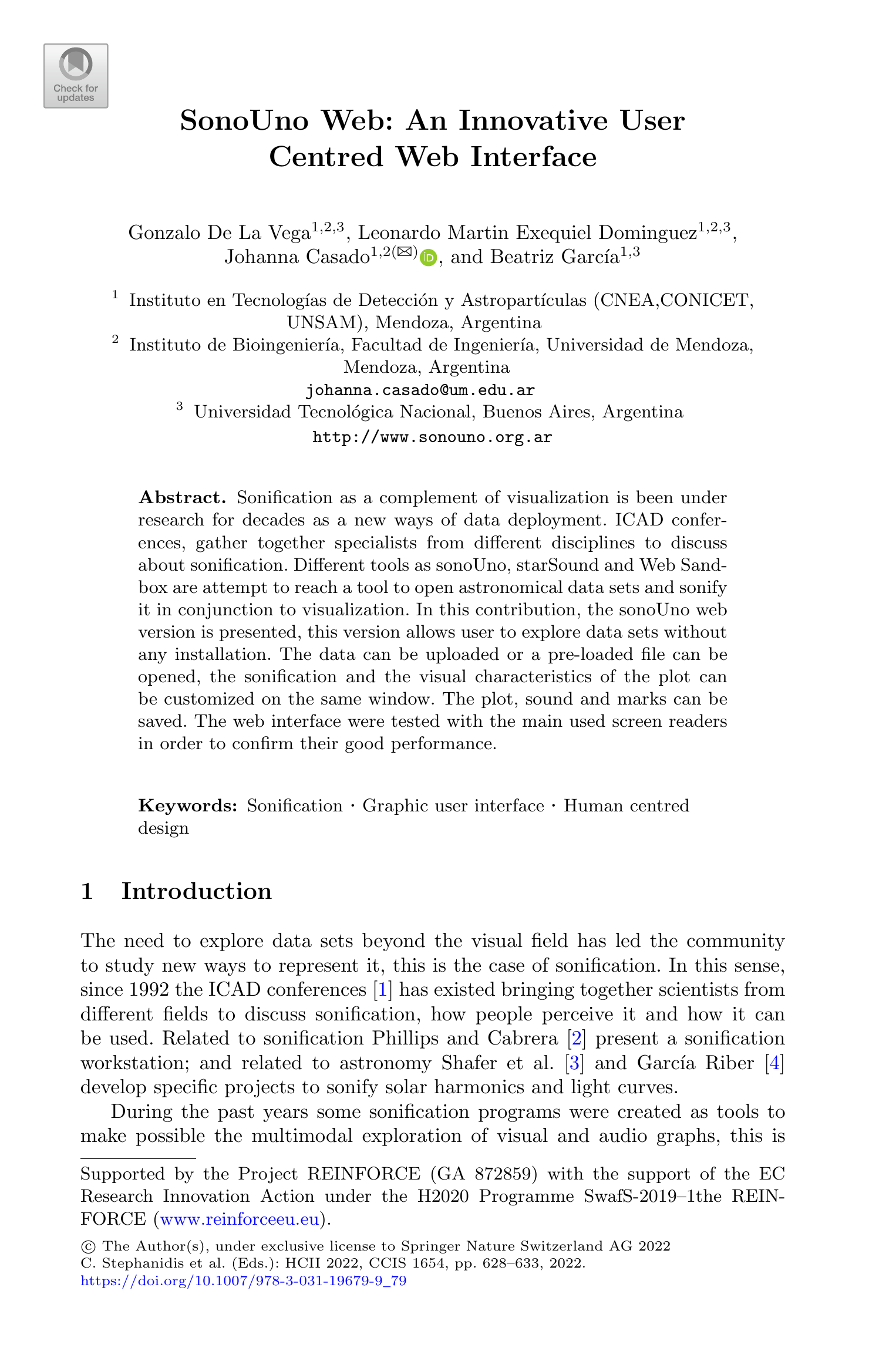}

\chapter{Manuales e Instructivos}
\label{ap:manuals}

Se agrega en este Apéndice el Manual de Usuario del programa sonoUno y los instructivos de instalación. El orden en el cual se encuentran los documentos es el siguiente:

\begin{enumerate}
	\item Manual de Usuario
	\item Instructivo de instalación en Windows
	\item Instructivo de instalación en MacOS - Ingles
	\item Instructivo de instalación en MacOS - Castellano
	\item Instructivo de instalación en Ubuntu
\end{enumerate}

\includepdf[pages=-,scale=0.8]{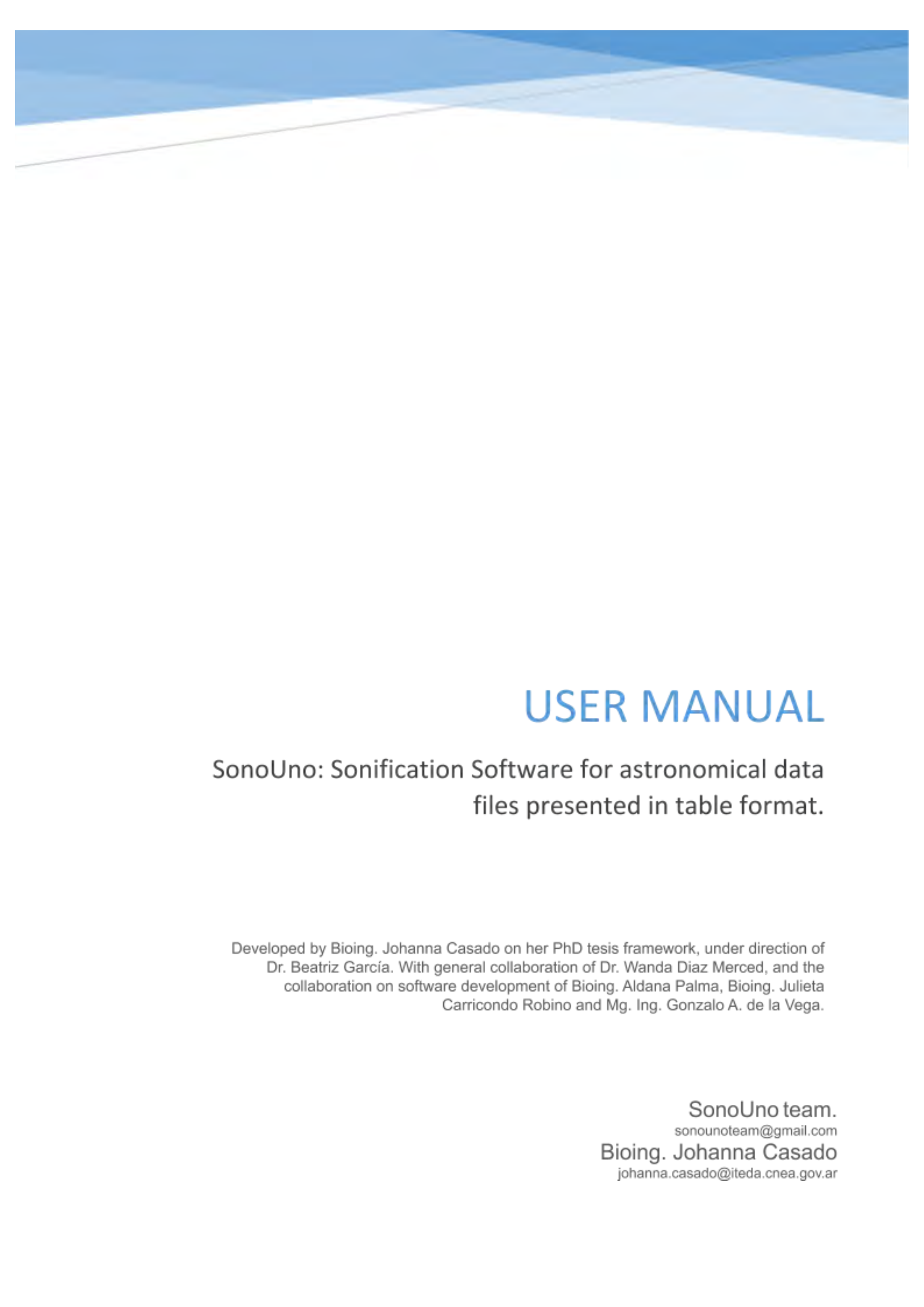}

\includepdf[pages=-,scale=0.8]{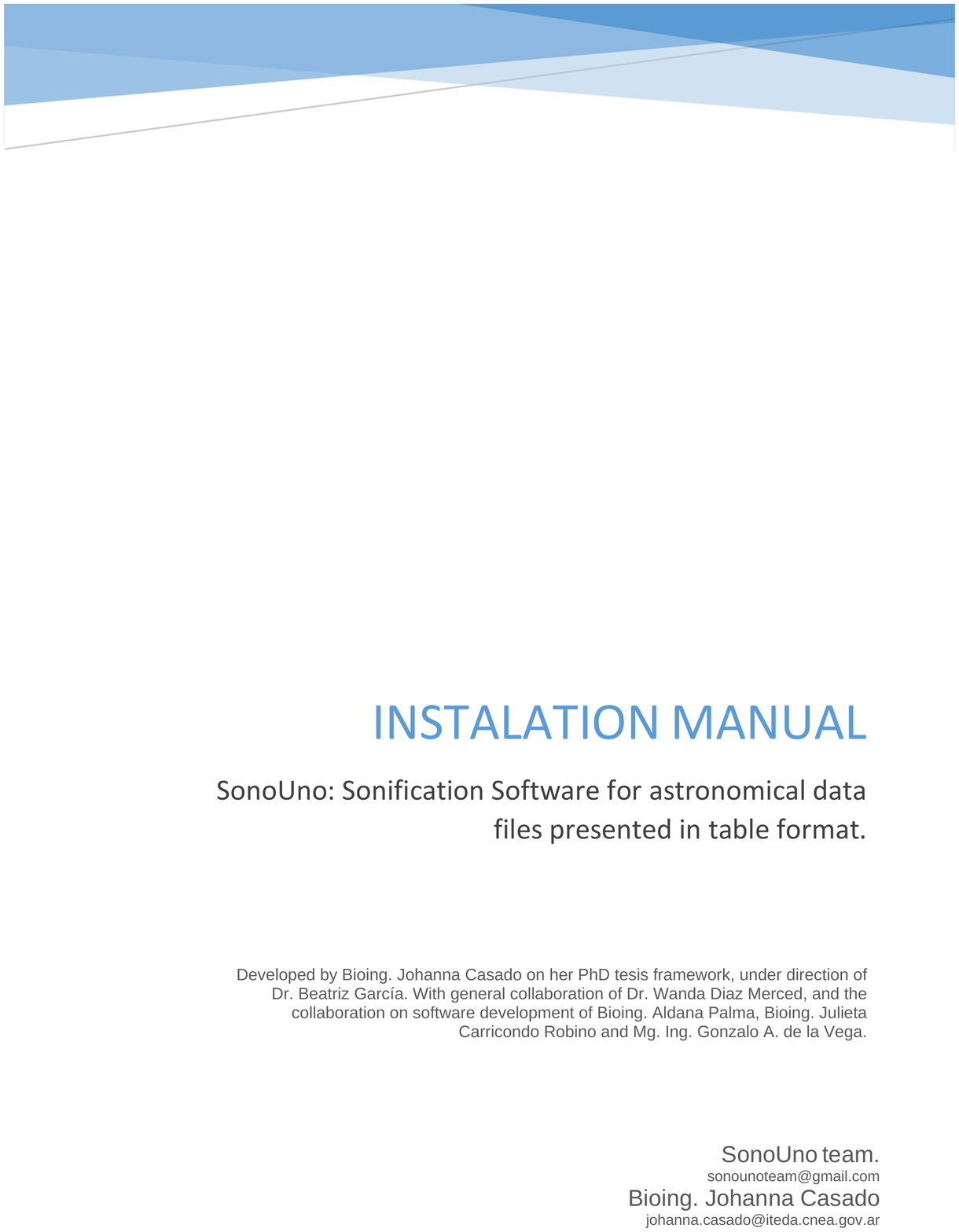}

\includepdf[pages=-,scale=0.8]{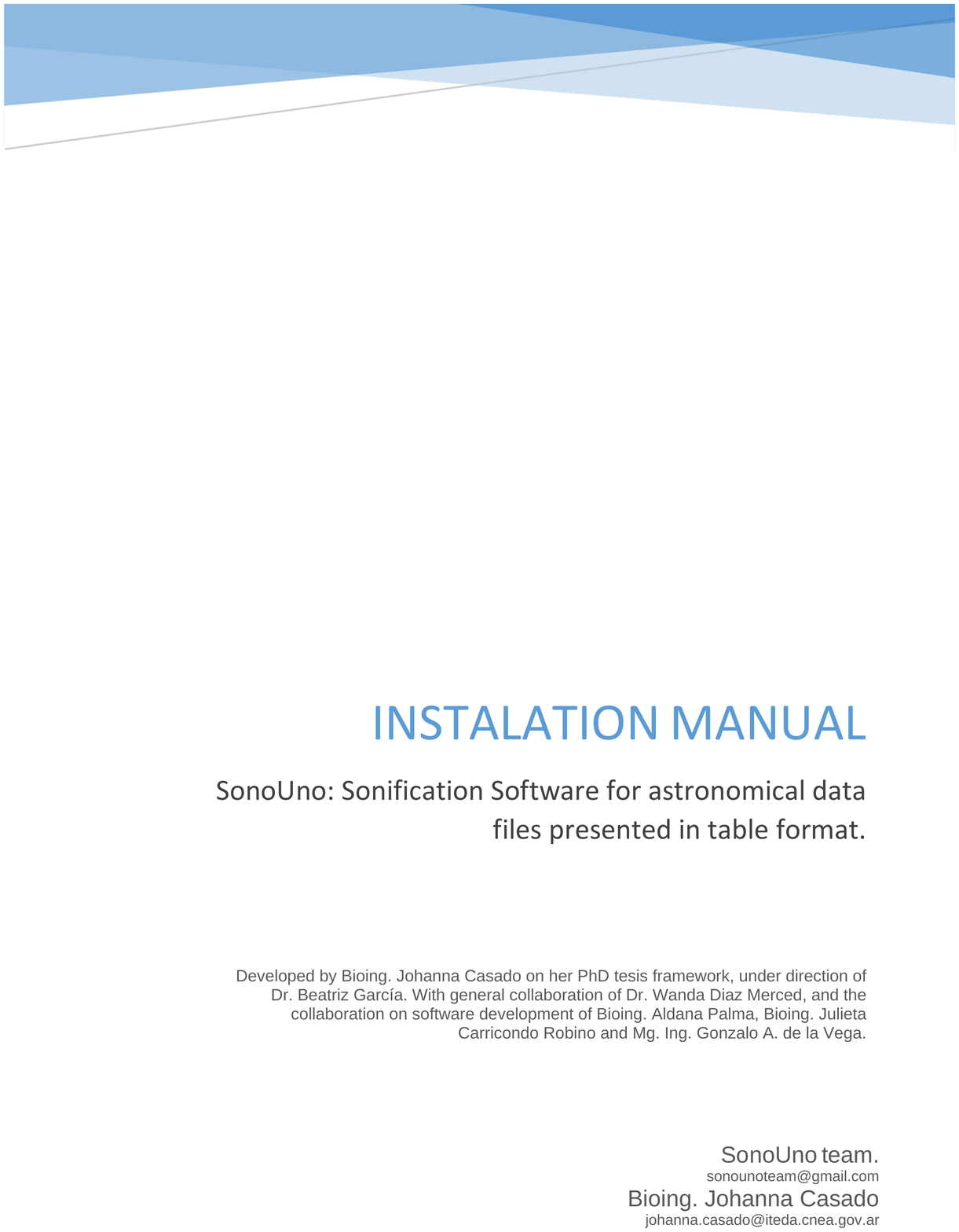}

\includepdf[pages=-,scale=0.8]{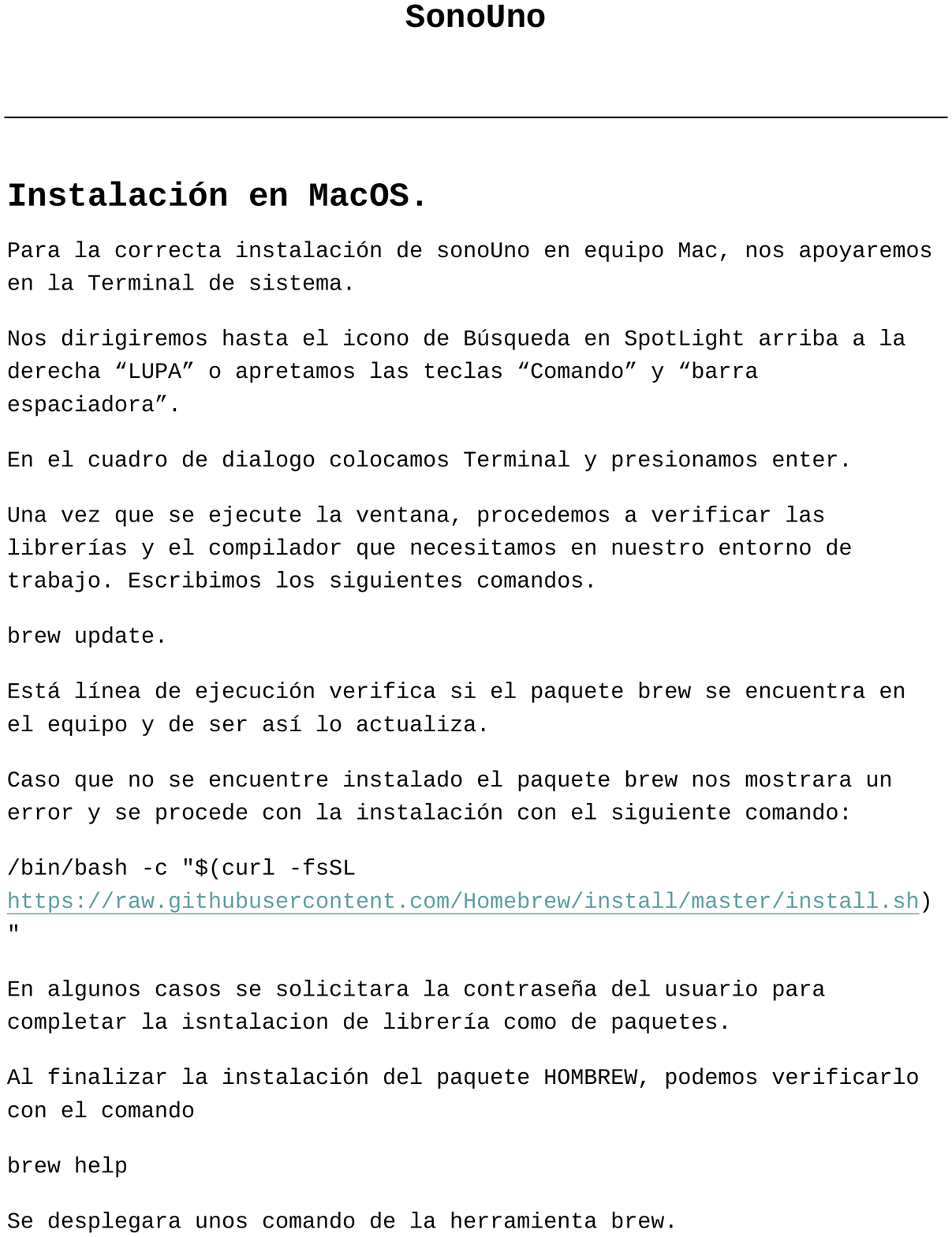}

\includepdf[pages=-,scale=0.8]{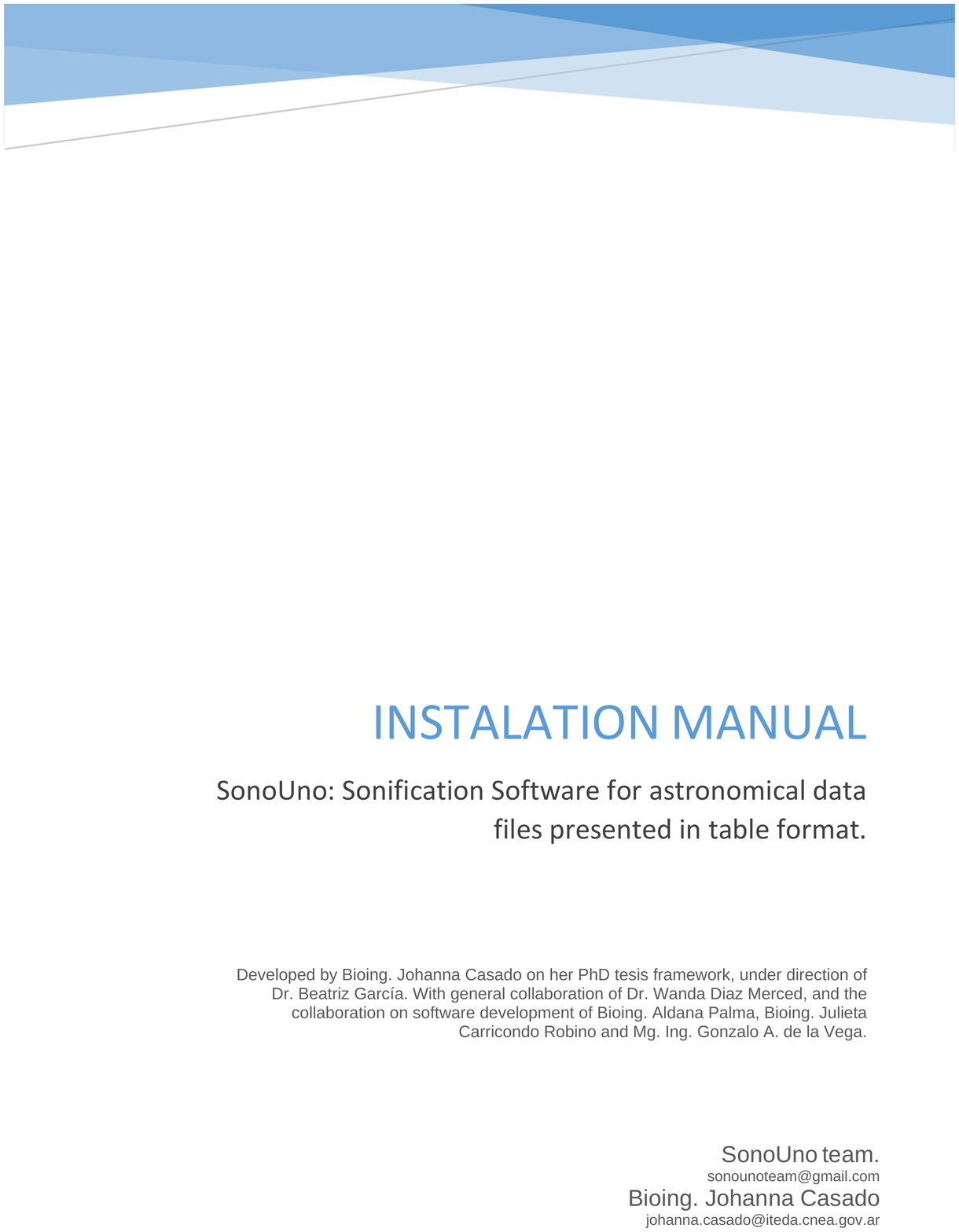}

\end{document}